%% file: main.tex
\documentclass[12pt,preprint]{aastex-thesis}  

\usepackage{epsfig}  
\usepackage{epstopdf}  
\usepackage{graphics}  
\usepackage{natbib}  
\usepackage{iondefs} 

\begin{document}

\doublespace

\pagenumbering{roman}  

\pagestyle{empty}  

\include{title}  

\pagestyle{plain} 

\include{approval}  

\include{dedica}  

\include{ackno}  

\include{vita}  

\doublespace

\include{abstract}  

\singlespace

\tableofcontents  
\newpage

\listoftables  
\addcontentsline{toc}{section}{LIST OF TABLES}  
\newpage
\addcontentsline{toc}{section}{LIST OF FIGURES}  
\listoffigures  

\doublespace

\clearpage
\resetcounters  

\pagenumbering{arabic}  

\input{chp1-intro}  

\clearpage
\resetcounters

\input{chp2-gl436}  

\clearpage
\resetcounters

\input{chp3-wasp12}

\clearpage
\resetcounters
\input{chp4-kepsec}

\clearpage
\resetcounters

\input{chp5-keplmb}

\clearpage
\resetcounters

\input{chp6-keplmb2}

\clearpage
\resetcounters

\input{chp7-ld}

\clearpage
\resetcounters

\input{chp8-sim3}

\clearpage

\pagestyle{empty}  
\begin{center}
APPENDICES  
\end{center}

\addtocontents{toc}{\protect\contentsline{}{}{}}
\appendix  

\clearpage
\pagestyle{plain}  
\resetcounters

\input{appendixA}

\clearpage
\resetcounters

\input{appendixB}

\clearpage
\resetcounters

\input{appendixC}

\clearpage
\resetcounters

\input{appendixD}

\clearpage

\singlespace
\addcontentsline{toc}{section}{REFERENCES}  
\bibliographystyle{apj}  
\bibliography{AstroRefs.bib}  
\end{document}

%% file: title.tex
\begin{center}
FUNDAMENTAL PARAMETERS OF EXOPLANETS\\AND THEIR HOST STARS\\
\bigskip
BY\\
\bigskip
JEFFREY LANGER COUGHLIN, B.S., M.S.
\end{center}
\vspace{1.0in}
\begin{center}
A dissertation submitted to the Graduate School\\
\bigskip
in partial fulfillment of the requirements\\
\bigskip
for the degree\\
\bigskip
Doctor of Philosophy
\end{center}
\bigskip
\begin{center}
Major Subject: Astronomy
\end{center}
\vspace{1.0in}
\begin{center}
New Mexico State University\\
\bigskip
Las Cruces New Mexico\\
\bigskip
September 2012
\end{center}

%% file: approval.tex
\noindent ``Fundamental Parameters of Exoplanets and Their Host Stars" a dissertation prepared by Jeffrey L. Coughlin in partial fulfillment of the requirements for the degree, Doctor of Philosophy, has been
approved and accepted by the following:

\singlespace

\bigskip
\begin{flushleft}
\hrulefill
\newline
Linda Lacey
\newline
Dean of the Graduate School

\vspace{0.5in}

\hrulefill
\newline
Thomas E. Harrison
\newline
Chair of the Examining Committee

\vspace{0.5in}

\hrulefill
\newline
Date
\vspace{0.5in}
\newline
Committee in charge:
\end{flushleft}

\doublespace

\indent Dr. Thomas E. Harrison, Chair\\
\indent Dr. Nancy J. Chanover\\
\indent Dr. Jon Holtzman\\
\indent Dr. Daniel P. Dugas\\

%% file: dedica.tex
\begin{center}
DEDICATION
\end{center}

I dedicate this work to my mother, Marcia Langer, the best Mom in the whole universe.

%% file: ackno.tex
\begin{center}
ACKNOWLEDGMENTS
\end{center}

Over the last 5 years I have learned that no person is ever successful completely on their own. Whether it be friends, family, or mentors, we rely on those people for support in our endeavors, and I've been extremely fortunate to have a generous dose of all three in my life.

I would like to thank the National Science Foundation for their generous support through the Graduate Research Fellowship Program. Thanks as well to NASA and the $Kepler$ Mission for support though their Guest Observing Program, and to the New Mexico Space Grant Research Consortium.

Tom, thanks for being both a mentor and a friend, for helping to reign in my Astronomy ADD, and enabling me to publish so much and be successful in this field. I apologize to you and Joni for the chocolate handprints, but I regret nothing. Jon, thank you for taking me on as a student early, showing me the ropes on the 1 meter, and for all the encouragement. Nancy, thanks for all the support, award nominations, and planetary group treats. Ofelia and Lorenza, thanks for battling the university bureaucracy on my behalf, and often at your own peril. Dr. Dugas, thanks for showing me how cool geology can be; I've never looked at a mountain, river, or valley the same since taking your class. Mercedes, thank you for taking a chance on working with a first year graduate student five years ago, and for being an amazingly supportive collaborator and friend. And to my undergraduate advisors Rick and Horace back at Emory, thank you for all the time you spent with me in undergrad; I could not have been better prepared for graduate school thanks to you two.

To all the graduate students here, thanks for keeping the department a friendly and upbeat place to work. Special thanks to Ryan, Maria, Chas, Jillian, Kyle, Leland, Liz, Adam, Cat, Mike, Nick, Roberto, and Glenn for extra support, listening to all my rants, and simply all the fun times. Joni, thanks for letting me know about all the hidden fun in Cruces, and thanks to you and Tom for hosting such awesome parties. Jom, please never haunt my dreams again. Herbert, my dear office plant, I shall miss your massive tentacle-like vines; thanks for making the office greener, and for not eating Sebastian.

Chris, Josie, Jennifer, Vince, Hazel, Ken, Stephanie, and Charlie, i.e., the Krazy Kutacs, thank you so much for letting me become one of those crazy Kutacs over the past few years, and for letting me use your kitchen tables as desks, where much of this thesis was written.

Grampa Ed and Grammy Joan, thank you for remembering every single holiday, birthday, and special occasion; you are always in my thoughts and in my heart. Uncle Marc, Aunt Cathy, Maggie, Marc, Molly, Melissa, and Eva, thank you for your overly generous contributions to ``the beer fund'', and especially for your overwhelming love and moral support. Aunt Judy, thank you for being a spirit of confidence and self-worth that I will carry with me always.

Julie, thank you for always making time to hang out during my last-minute Tucson visits and for all the support. To Jon, thanks for all the fun times hanging out and saving my squishy scout. To Ryan, thank you for all the nerdy midnight talks over a six pack and for being a great roommate and friend; I don't think I'll ever find a better one. To Maria, thanks for being crazy...I mean, a crazy awesome seriously good friend and for all the trips to Cafe W. To Jamie, although I lament at not being able to write this thesis next to you at Java Monkey, it has still felt like you've been there every step of the way over the past five years.

My brother J.P. and sister-in-law Jenn, thank you for all the love and support, the fun times, and for simply always being there for me. I realized in writing this I've known you Jenn for almost exactly as long as I've been working on this thesis, and what a great 5$\frac{1}{2}$ years it's been! One of these days J.P., I'm going to make up those 31.5 seconds.

My partner Nicholas, thanks for all the time you've spent patiently listening to me ramble on about the various problems I was facing, for the unending encouragement and optimism you've shown me, and for simply making every day since we met brighter than the last. Persistence $is$ everything.

Finally, my greatest thanks goes to my mother, Marcia, for all her love and dedication to my education. I would not have gotten here today without all your love, support, and encouragement.

%% file: vita.tex
\begin{center}
 VITA
\end{center}

\begin{center}
 EDUCATION
\end{center}

\begin{tabular}{ll}
2007-2009 &  M.S., Astronomy\\
 & New Mexico State University, with Honors\\           
 & Las  Cruces, New Mexico, USA\\
2003-2007 &  B.S., Physics and Astronomy\\
 & Emory University, \emph{Summa cum Laude}\\
 & Atlanta, Georgia, USA          
\end{tabular}

\medskip

\begin{center}
AWARDS AND GRANTS
\end{center}
\begin{tabular}{ll}
2012 & NMSU Astronomy Murrell Award for Professional Development\\
2009-2012 & NSF Graduate Research Fellowship\\
2009 & NASA Graduate Student Research Fellowship (Declined)\\
2009 & NMSU Astronomy ZIA Award for Outstanding Research\\
2008 \& 2009 & New Mexico Space Grant Graduate Research Fellowship
\end{tabular}

\medskip

\begin{center}
PROFESSIONAL ORGANIZATIONS
\end{center}
\begin{tabular}{l}
American Astronomical Society (\& Division for Planetary Sciences)\\
Sigma Pi Sigma National Physics Honors Society
\end{tabular}

\medskip

\singlespace

\begin{center}
PUBLICATIONS 
\end{center}

\hangindent=0.5in \noindent \underline{Coughlin, J.L.} and L\'opez-Morales, M., 2012, The Astrophysical Journal, 750, 100. \textit{Modeling Multi-Wavelength Stellar Astrometry. III. Determination of the Absolute Masses of Exoplanets and Their Host Stars}

\hangindent=0.5in \noindent \underline{Coughlin, J.L.} and L\'opez-Morales, M., 2012, The Astronomical Journal, 143, 39. \textit{A Uniform Search for Secondary Eclipses of Hot Jupiters in Kepler Q2 Lightcurves}

\hangindent=0.5in \noindent Harrison, T.E., \underline{Coughlin, J.L.}, Ule, N.M., and L\'opez-Morales, M., 2012, The Astronomical Journal, 143, 4. \textit{Kepler Cycle 1 Observations of Low Mass Stars: New Eclipsing Binaries, Single Star Rotation Rates, and the Nature and Frequency of Starspots}

\hangindent=0.5in \noindent \underline{Coughlin, J.L.}, L\'opez-Morales, M., Harrison, T.E., Ule, N., and Hoffman, D. 2011, The Astronomical Journal, 141, 78. \textit{Low-Mass Eclipsing Binaries in the Initial Kepler Data Release}

\hangindent=0.5in \noindent \underline{Coughlin, J.L.}, Harrison, T.E., and Gelino, D. 2010, The Astrophysical Journal, 723, 1351. \textit{Modeling Multi-wavelength Stellar Astrometry. II. Determining Absolute Inclinations, Gravity Darkening Coefficients, and Spot Parameters of Single Stars with SIM Lite}

\hangindent=0.5in \noindent \underline{Coughlin, J.L.}, Harrison, T.E., Gelino, D., Hoard, D., Ciardi, D., Benedict, F., Howell, S., McArthur, B., and Wachter, S. 2010, The Astrophysical Journal, 717, 776. \textit{Modeling Multi-wavelength Stellar Astrometry. I. SIM Lite Observations of Interacting Binaries}

\hangindent=0.5in \noindent L\'opez-Morales, M., \underline{Coughlin, J.L.}, Sing, D.K., Burrows, A., Apai, D., Rogers, J.C., Spiegel, D.S., and Adams, E.R. 2010, The Astrophysical Journal, 716, 36. \textit{Day-side $z'$-band Emission and Eccentricity of WASP-12b}

\hangindent=0.5in \noindent Holtzman, J., Harrison, T.E., \& \underline{Coughlin, J.L.} 2010, Advances in Astronomy, 46. \textit{The NMSU 1m Telescope at Apache Point Observatory}

\hangindent=0.5in \noindent Tucker, R.S., Sowell, J.R., Williamon, R.M., and \underline{Coughlin, J.L.} 2009, The Astronomical Journal, 137, 2949. \textit{Orbital Solutions and Absolute Elements of the Eclipsing Binary MY Cygni}

\hangindent=0.5in \noindent \underline{Coughlin, J.L.}, Stringfellow, G., Becker, A., L\'opez-Morales, M., Mezzalira, F., and Krajci, T. 2008, The Astrophysical Journal, 689L, 149. \textit{New Observations and a Possible Detection of Parameter Variations in the Transits of Gliese 436b}

\hangindent=0.5in \noindent \underline{Coughlin, J.L.}, Dale, H.A., and Williamon, R.M. 2008, The Astronomical Journal, 136, 1089. \textit{Long-term Photometric Analysis of the Active W Uma-type System TU Boo}

\hangindent=0.5in \noindent Hoffman, D.I., Harrison, T.E., \underline{Coughlin, J.L.}, et al. 2008, The Astronomical Journal, 136, 1067. \textit{New Beta Lyrae and Algol Candidates from the Northern Sky Variability Survey}

\hangindent=0.5in \noindent \underline{Coughlin, J.L.} and Shaw, J.S. 2007, SARA, 1, 7C. \textit{Seven New Low-Mass Eclipsing Binaries}

\hangindent=0.5in \noindent \underline{Coughlin, J.L.} 2007, Emory Undergraduate Research Journal. \textit{A Spotlight on Starlight}

\hangindent=0.5in \noindent \underline{Coughlin, J.L.} 2007, Undergraduate Thesis, Emory University. \textit{Observations and Models of Eclipsing Binary Stars}

\medskip
\begin{center}
CONFERENCE PROCEEDINGS
\end{center}

\hangindent=0.5in \noindent \underline{Coughlin, J. L.}, L\'opez-Morales, M., Harrison, T. E., Ule, N., Hoffman, D. I. 2011, in ASP Conf. Ser. 448, 16$^{th}$ Cambridge Workshop on Cool Stars, Stellar Systems, and the Sun, ed. C. Johns-Krull (Seattle, WA:ASP), 121. \textit{New Low-Mass Eclipsing Binaries from Kepler}

\medskip

\begin{center}
FIELD OF STUDY
\end{center}
\begin{flushleft}
Major Field: Extrasolar Planets \& Eclipsing Binary Stars
\end{flushleft}

%% file: abstract.tex
\begin{center}
ABSTRACT
\end{center}

\bigskip

\begin{center}
FUNDAMENTAL PARAMETERS OF EXOPLANETS\\AND THEIR HOST STARS\\
\bigskip
BY\\
\bigskip
JEFFREY LANGER COUGHLIN, B.S., M.S.
\end{center}

\bigskip

\begin{center}
Doctor of Philosophy\\
New Mexico State University\\
Las Cruces, New Mexico, 2012\\
Dr. Thomas E. Harrison, Chair
\end{center}

\bigskip
\bigskip


For much of human history we have wondered how our solar system formed, and whether there are any other planets like ours around other stars. Only in the last 20 years have we had direct evidence for the existence of exoplanets, with the number of known exoplanets dramatically increasing in recent years, especially with the success of the $Kepler$ mission. Observations of these systems are becoming increasingly more precise and numerous, thus allowing for detailed studies of their masses, radii, densities, temperatures, and atmospheric compositions. However, one cannot accurately study exoplanets without examining their host stars in equal detail, and understanding what assumptions must be made to calculate planetary parameters from the directly derived observational parameters.

In this thesis, I present observations and models of the primary transits and secondary eclipses of transiting exoplanets from both the ground and $Kepler$ in order to better study their physical characteristics and search for additional exoplanets. I then identify, observe, and model new eclipsing binaries to better understand the stellar mass-radius relationship and stellar limb-darkening, compare these observations to the predictions of stellar models, and attempt to define to what extent these fundamental stellar characteristics can impact derived planetary parameters. I also present novel techniques for the direct determination of exoplanet masses and stellar inclinations via multi-wavelength astrometry, the ground-based photometric observation of stars at sub-millimagnitude precision, the reduction of $Kepler$ photometry from pixel-level data, the extraction of radial velocities from spectroscopic observations, and the automatic identification, period analysis, and modeling of eclipsing binaries and transiting planets in large datasets.

%% file: chp1-intro.tex
\begin{singlespace}
\section{\MakeUppercase{Introduction}}
\label{chap1}
\end{singlespace}

\subsection{The Rise of Exoplanets}

\begin{figure}[ht]
\centering
\epsfig{width=\linewidth,file=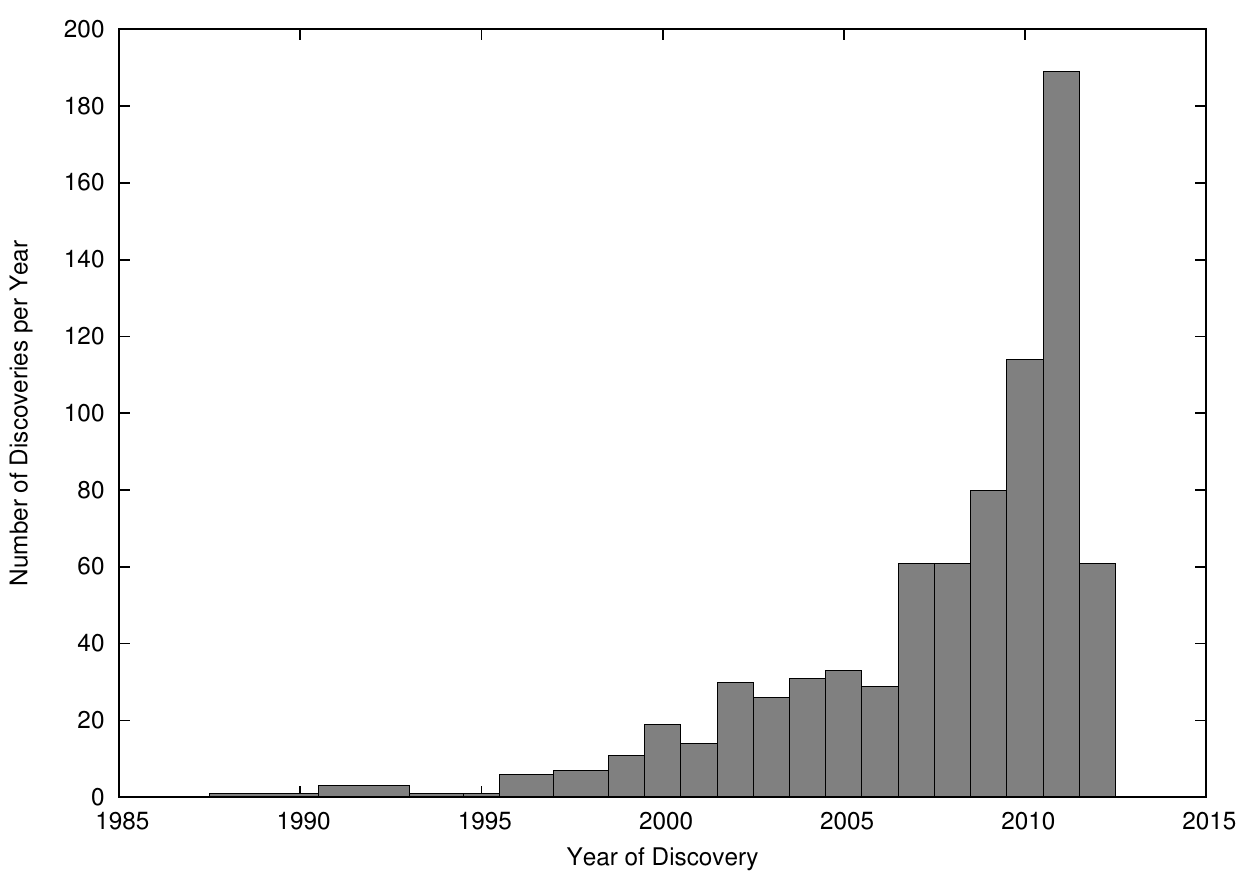}
\caption[Histogram of the number of exoplanet discoveries per year]{Histogram of the number of exoplanet discoveries per year, with data obtained from \citet{Schneider2012}. The number of discovered exoplanets has been increasing roughly exponentially since the first discoveries $\sim$20 years ago. The number of discoveries for the current year, 2012, is incomplete at the time of this writing, but is expected to surpass the number from 2011.}
\label{dischistfig}
\end{figure}

The idea that planets could exist outside our solar system, orbiting stars other than our own sun, has been around for at least hundreds, if not thousands, of years \citep[e.g.,][]{Lucretius58BC,Bruno1584}. However, only in the last $\sim$20 years have we been able to find and study such planets. The first reported, (and subsequently confirmed), detection of an exoplanet was by \citet*{Campbell1988} of a 1.6 $M_{\rm Jup}$, (1.6 times Jupiter's mass), planet in a 2.5 year orbit around the sub-giant $\gamma$~Cephei~A via the observed the velocity variations of the host star. In 1992, \citet{Wolszczan1992} announced the discovery of two planets only a couple times the mass of Earth around a millisecond pulsar via their perturbations to the pulse arrival times. In 1995 \citet{Mayor1995} found the first planetary companion to a main-sequence star, a 0.47 $M_{\rm Jup}$ planet in a 4.2 day orbit around the Sun-like star 51 Pegasi, via radial velocity observations. Since these initial discoveries, the number of exoplanets has increased exponentially due to a combination of increased scientific interest and technological feasibility, as shown in Figure~\ref{dischistfig}. At the time of this writing, there are 778 confirmed exoplanets known, in a total of 624 individual stellar systems \citep{Schneider2012}, with thousands of additional exoplanet candidates awaiting confirmation.

In order to perform fundamental science, it is not enough to simply discover exoplanets; we must thoroughly investigate and characterize the fundamental parameters of each one. Only with a large sample of well-studied planets can we start to answer such fundamental questions as:

\begin{itemize}
 \item How do solar systems and planets form and evolve?
 \item Are planets common around other stars in our galaxy and are they similar to or different from the planets in our solar system?
 \item How do planetary atmospheres function under a range of temperatures and compositions?
 \item Do other planets have conditions capable of supporting life or perhaps even show telltale signs of life?
 \item Are there any planets like Earth around the nearest stars that we can aspire to one day travel?
\end{itemize}

\noindent Unfortunately, the process of determining the fundamental parameters of exoplanets that we need to answer these questions is not straightforward, and often requires a thorough knowledge of the planet, its host star, and the broader astrophysical processes that govern these objects.

\subsection{Detection and Characterization Methods}

There are many methods employed to detect and characterize the fundamental parameters of exoplanets. In this section we briefly summarize each one, and what information can be directly determined via each one.

\subsubsection{Radial Velocity}

The amount of light that a star emits varies with the wavelength of light emitted, and is referred to as a stellar spectrum. An extrasolar planet and its host star orbit a common center of mass, such that both the planet and star move and complete an orbit once an orbital period. The star's orbital size and velocity will be many orders of magnitude smaller than that of the planet, precisely by the ratio of the planet and host star's masses. However, the star is many orders of magnitude brighter than the planet, and thus is typically the only visible component of the system. As the star moves, its spectrum is shifted with respect to wavelength via the Doppler Effect, and thus the velocity of the star, in the direction of the observer on Earth, can be directly measured. The directly determined parameters are the period, eccentricity, and longitude of periastron of the planetary orbit, and the velocity amplitude of the star.

\subsubsection{Transiting Planets}

\begin{figure}[ht]
\centering
\includegraphics[width=\linewidth]{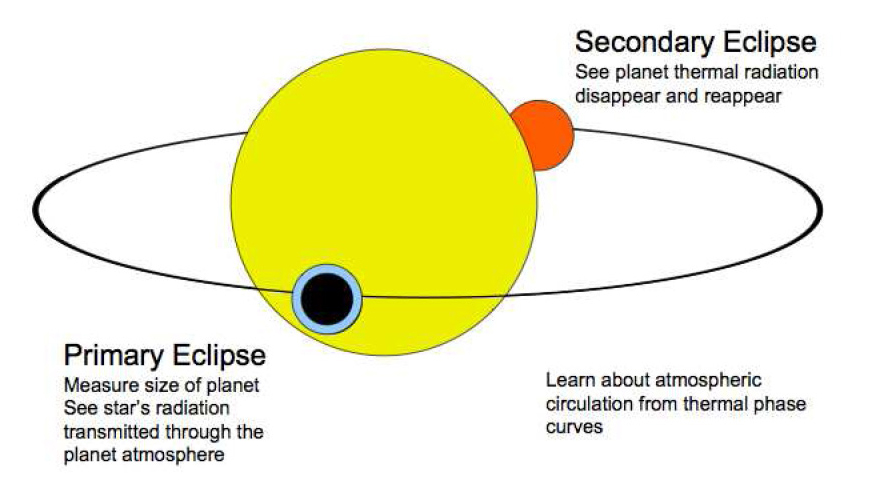}
\caption[Illustration of the primary and secondary eclipses of an exoplanet]{Illustration of the primary and secondary eclipses of an exoplanet. Figure by S. Seager.}
\label{transschemfig}
\end{figure}

If the inclination of an extrasolar planet's orbit is very close to edge-on as viewed from Earth, then the planet will pass in front of its host star once every orbit. In this case, a measurable drop in flux from the system occurs as the planet blocks out light from the host star, and is called the primary transit. This drop in flux is directly proportional to the square of the ratio of the exoplanet's and host star's radii, i.e., the fractional area of the stellar surface that is obscured by the planet. Also during primary transit, some of the star's light will pass through, but only be partially absorbed by, the very upper reaches of the planetary atmosphere, imparting extra absorption features unto the stellar spectrum that are due to the planetary atmosphere. When the planet passes behind the star half an orbit later, there is a much smaller drop in flux that corresponds to the luminosity ratio between the planet and host star, referred to as the secondary eclipse. An illustration is shown in Figure~\ref{transschemfig}. Thus, unlike non-transiting planets, in principle one can determine the mass, radius, density, temperature, and even atmospheric composition of the planet. The detailed study of these characteristics allows the direct testing of various extrasolar planetary atmosphere models, whose predictions as to the existence of key molecular species, general circulation patterns, temperature, and the variation thereof with scale height, can vary widely \citep[e.g.,][]{Cooper2006,Fortney2006,Tinetti2007,Burrows2008a,Burrows2008b,Showman2008,Showman2009,Spiegel2009}. As well, determining atmospheric temperature and composition can provide insights into planetary formation and migration, of which many competing models also exist.

Assuming the star and planet are uniformly illuminated spheres, the directly determined parameters from the light curve primary transit are the period of the system, the fractional sum of the radii, (i.e., the sum of the planet and stellar radii divided by the semi-major axis of the orbit), the ratio of the planetary to stellar radii, and the inclination of the orbit. If the secondary eclipse is also observed in the light curve, one may also directly measure the ratio of the planetary to stellar surface brightness, and the eccentricity and longitude of periastron of the orbit. Via mathematical re-arrangements of these directly determined parameters, \citet{SeagerOrnelas2003} found that the host star's mean stellar density is directly determined from the primary transit light curve alone, and \citet{Southworth2007} found that the exoplanet's surface gravity is directly determinable with both a primary transit light curve and single-line radial velocity curve.

\subsubsection{Transit Timing and Parameter Variations}

In addition to measuring the depth of a transit, the timing and duration of the transit can also be measured. If there is another planet in the system, or a moon around the transiting planet, it will exert gravitational perturbations on the transiting planet. These perturbations will manifest as changes in the timing, duration, and depth of the observed transits, and can in certain cases directly yield the mass, orbital period, and semi-major axis of the non-transiting planet \citep{Agol2005,Holman2005}. Furthermore, if the second planet or the moon also transits, the masses, radii, orbital periods, and semi-major axes of all three components can be directly determined from the light curve alone \citep{Kipping2010b,Carter2011}.

\subsubsection{Microlensing}

According to General Relativity, any object with mass will warp the space-time surrounding it, and deflect the path that light takes when it travels close to the object. For objects as massive as the Sun and other stars, this effect can be significant, and in fact deflections of stellar positions near the Sun during a solar eclipse were among the first confirmations of the theory of General Relativity. A star, and even a planet, is thus capable of acting like a giant lens, magnifying the light from distant sources. In our galaxy, the amount of deflection provided by a star, and thus its focal length, are of the order such that stars approximately halfway between us and the galactic center are at the right distance to act as a lens, provided that Earth, the intervening system that acts as the lens, and another distant star line up exactly right. Astronomers have monitored a very large number of stars towards the center of the galaxy, and have detected several microlensing events. In these cases, a primary magnification event is seen that increases the light observed by several orders of magnitude, as the three components slowly drift into alignment, and the intervening lens star focuses the light from the distant star onto the Earth. On top of this primary event, one or more smaller magnification events are often seen, which is due to planetary companions of the lens star also acting as, albeit smaller, lenses. The amplitude of these events directly measures the masses of the host star and its planets, as well as the orbital separation between them.

\subsubsection{Astrometry}

As discussed with the radial velocity technique, both the planet and host star move over an orbital period around their common center of mass. By precisely measuring the position of a star on the sky, the on-sky projected motion of the star can be measured. This directly yields the projected distance between the star and barycenter of the system, as well as the period, eccentricity, and longitude of periastron of the orbit. Nearly always these measurements also directly determine the distance to the system via geometric parallax, and thus the physical distance between the star and barycenter is also directly determined.

\subsubsection{Direct Imaging}

In all of the above techniques, the presence and properties of extrasolar planets are deduced via the perturbations they induce upon their host stars. However, it is possible to directly image and detect an extrasolar planet by taking a very long exposure of the system while using a mask to block light from the host star. Typically, adaptive optics are also employed in order to achieve spatial resolution better than the projected separation of the planet and star. If detected, the relative brightness of the exoplanet compared to the host star can be directly determined, as well as the period, eccentricity, and longitude of periastron of the orbit if multiple images are taken over a significant fraction of the planet's orbit.

\subsection{Inferred Stellar and Planetary Parameters from Directly Determined Quantities}

Of the techniques discussed above, the radial velocity technique has by far been the most productive technique for finding planets and planet candidates over the past 20 years. However, the amount of information yielded by this technique alone is severely limited. In order to calculate a mass for the planet, one must assume a mass for the host star, as well as an inclination for the planetary orbit. While the former can be estimated based on the stellar spectrum, the latter is impossible to determine from radial velocity observations alone, and thus one can only truly determine lower limits to the planetary mass. Astrometry is a bit more useful, as it directly yields the inclination of the orbit and the star-barycenter distance, and thus with an assumed mass for the star, a mass for the planet and a semi-major axis for its orbit can be calculated.

The transit technique has been the second-most productive to-date, with 240 planets known to transit in 206 individual stellar systems at the time of this writing \citep{Schneider2012}, and is expected to rapidly leap ahead as the most productive technique given the thousands of planet candidates recently announced via the $Kepler$ mission \citep{Batalha2012}. Most confirmed transiting planets have radial velocity observations of the host star taken as well. Although a transiting planet can yield much more information than a non-transiting planet, there are some very important caveats, namely that most of the planetary parameters of interest are inherently dependent on the assumed stellar parameters. Although the inclination is directly determinable from the light curve, since only the radial velocity of the star is known we must assume a mass for the host star in order to calculate a mass for the exoplanet. Since the fractional radii of the star and planet are directly determined from the light curve, if we assume a mass for the star (and either assume a mass for the planet or treat it as negligible compared to the star) we can combine that with the directly determined period to calculate the semi-major axis of the system, and thus a radius for the star and exoplanet. Alternatively, one may assume a radius for the host star, and use the directly determined ratio of radii to calculate a radius for the planet, as is often the case in the absence of radial velocity observations. In practice, since the mean stellar density and the planetary surface gravity are directly determined, one would choose values of mass and radius for the host star that would reproduce the mean stellar density and planetary surface gravity values within the observational errors. The planetary temperature can be calculated from the directly determined ratio of the surface brightness if one knows the wavelength bandpass of observation, assumes a spectral energy distribution over the bandpass for both the planet and star, and assumes a temperature for the star. Generally the stellar spectral energy distribution and temperature of the host star can be directly determined from high-resolution spectroscopic observations. Of course, in order to utilize this technique, the planet has to have the fortuitous alignment that it does transit as seen from Earth, which is quite rare.

\begin{figure}[ht]
\centering
\includegraphics[width=0.75\linewidth]{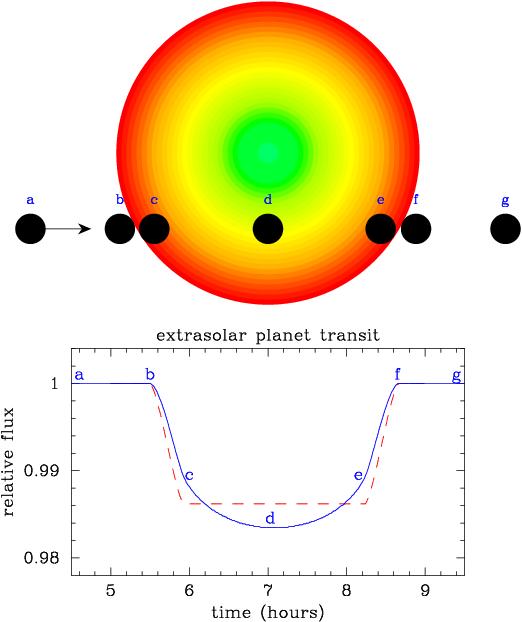}
\caption[Illustration of a planetary transit and the resultant light curve for both a stellar disk of uniform brightness and one that includes limb-darkening]{Illustration of a planetary transit and the resultant light curve for both a stellar disk of uniform brightness (dashed line) and one that includes limb-darkening (solid line).}
\label{transitillustration}
\end{figure}

An assumption we made in the above statements is that the stellar disk is uniformly bright, which turns out to be a poor assumption. Stellar limb-darkening is the phenomenon that stars are brighter towards the center of their observed disks, and darker towards their edges, or limbs, and can be quite significant when examining exoplanet transit curves (see Section~\ref{ldintro} for a complete explanation of the effect). Figure \ref{transitillustration} is an illustration of a primary transit, and the resulting light curve one would observe for both a star with a constant brightness distribution (dashed line) and one that has brightness variation across its surface due to limb-darkening (solid line). The surface brightness distribution of the host star due to limb-darkening will affect the directly determined planetary parameters above, and thus it is important to know if limb-darkening can be directly determined from transit curves, and if not, how well we understand and can characterize it for various stars.

There are a few very recent advances as well that provide more information than we have previously been able to obtain. \citet{Snellen2010}, \citet*{Rodler2012}, and \citet{Brogi2012} were just recently able to directly detect the radial velocity of an exoplanet in the combined spectrum, but only for very special cases. In these cases one is able to determine the velocities of both the planet and star, and if the inclination is known, are thus able to directly determine the absolute masses of both the planet and star. With high-precision light curves, such as those being produced by the $Kepler$ mission, one is also able to detect several effects that occur over an entire orbit. First, the star and planet are not point sources, and thus distort each other through gravitational tidal interactions. As the planet raises a tide on the star, it increases the emitting surface area of the star, thus increasing its total observed luminosity in the light curve. By measuring this photometric signature, it becomes possible to estimate the mass of the planet by assuming the mass and density profile of the star, or visa versa. Second, light emitted by the star may be reflected off the planet's day side, causing a similar photometric signature over the course of an orbit. If we assume a certain albedo for the planet, this technique can be used to determine the inclination of the orbit, and thus the mass of the planet if combined with radial velocity observations. Third and finally, it is possible to directly determine the velocity of the star from the light curve alone, without spectroscopic radial velocity measurements at all, via an effect known as photometric beaming. Due to special relativity, the brightness of an object is magnified in the direction of its motion, as seen relative to an observer, as photons emitted from the object are preferentially emitted in the direction of motion. \citet{LoebGaudi2003} first realized this effect could be applied to exoplanets, with the stellar flux change quantified as

\begin{equation}
 F = F_{0}\left(1+4\frac{v_{r}}{c}\right)
\end{equation}

\noindent where $F$ is the observed flux, $F_{0}$ is the flux of the object at rest, $v_{r}$ is the radial velocity of the object, and $c$ is the speed of light. Thus, if this effect can be measured in a light curve, the velocity of the star is directly determined, and can be applied to determine other quantities of interest just as the radial velocity determined from traditional spectroscopic observations would be used.

\subsection{Thesis Goals and Format}

Observations of the fundamental parameters of exoplanets and their host stars are critical to performing fundamental science and answering long sought questions regarding planetary formation, evolution, and uniqueness. The accuracy of these observations also depends on critically understanding the interplay between the stellar and planetary parameters, and the assumptions that must be made to extract those parameters. Thus, the questions this thesis aims to answer are:

\begin{enumerate}
 \item What can be learned from high-precision observations of exoplanet transits?
 \item What are the atmospheric properties of exoplanets, how much do they vary planet-to-planet, and how do they change as a function of temperature?
 \item How accurately are we able to estimate the fundamental parameters of stars, how much variation exists star-to-star, and what are the implications for the study of planets around them?
 \item What new techniques might we employ in the future to better discover and characterize exoplanets?
\end{enumerate}

In Chapter~\ref{chap2} we present high-precision observations of the primary transit of the Neptune-mass exoplanet Gliese 436b in an effort to better characterize its orbit and search for additional low-mass planets in the system, originally published as \citet{Coughlin2008}. In Chapter~\ref{chap3} we detect and measure the secondary eclipse of exoplanet Wasp-12b via ground-based observations in the $z'$-band to probe its atmosphere, originally published as \citet{LopezMorales2010}. In Chapter~\ref{chap4} we search for and detect the secondary eclipses of numerous hot Jupiters in $Kepler$ data, allowing for a statistical study of hot Jupiter atmospheres, originally published as \citet{Coughlin2012a}. In Chapters~\ref{chap5} and \ref{chap6} we identify new low-mass eclipsing binaries in the $Kepler$ field, obtain follow-up observations, and model them to accurately measure their masses and radii in an attempt to better understand the long-standing discrepancy between predicted and observed radii of low-mass stars, of which Chapter~\ref{chap5} is published as \citet{Coughlin2011}. In Chapter~\ref{chap7} we present observational measurements of the limb-darkening coefficients of main-sequence stars in the $Kepler$ bandpass in an effort to test model predictions and examine star-to-star variation. Finally, in Chapter~\ref{chap8} we present a novel theoretical technique that can directly measure the masses of exoplanets utilizing multi-wavelength astrometry, published as \citet{Coughlin2012b}. We also note that Appendices~\ref{epdmappendsec} and \ref{agaappendix} were originally published as appendices to \citet{Coughlin2011}, and that Appendices~\ref{sim1appendix} and \ref{sim2appendix} were originally published as \citet{Coughlin2010a} and \citet*{Coughlin2010b}.

%% file: chp2-gl436.tex
\begin{singlespace}
\section[\MakeUppercase{New Observations and a Possible Detection of Parameter Variations in the Transits of\\Gliese~436\MakeLowercase{b}}]{\MakeUppercase{New Observations and a Possible Detection of Parameter Variations in the Transits of Gliese~436\MakeLowercase{b}}}
\label{chap2}
\end{singlespace}

\subsection{Introduction}

Gliese 436 is an M-dwarf (M2.5V) with a mass of 0.45 M$_{\sun}$ and hosts the extrasolar planet Gliese 436b, which is a Neptune-sized planet with a mass of 23.17 M$_{\earth}$ \citep{Torres2007}. Gliese 436b was first discovered via radial-velocity (RV) variations by \citet{Butler2004}, who also searched for a photometric transit, but failed to detect any signal greater than 0.4\%. It was thus a surprise when \citet{Gillon2007b} reported the detection of a transit with a depth of 0.7\%, implying a planetary radius of 4.22 R$_{\earth}$ \citep{Torres2007} and thus a composition similar to Uranus and Neptune.  In addition, both \citet{Deming2007} and \citet{Maness2007} calculated that the significant eccentricity of the orbit, e = 0.15, coupled with its short period of $\sim$2.6 days, should result in circularization timescales of $\sim$10$^{8}$ years, which contrasts with the old age of the system at $\gtrsim$6$\times$10$^{9}$ years. The existence of one or more additional planets in the system could be responsible for perturbations to Gliese 436b's orbit, and thus result in the observed peculiarities. We considered this possibility right after the initial publication of \citet{Gillon2007b}, and began an intensive campaign to observe the photometric transits of Gliese 436b in order to search for variations indicative of orbital perturbations.

\citet{Ribas2008a} reported the possible detection of a $\sim$5 M$_{\earth}$ companion in the Gliese 436 system located near the outer 2:1 resonance of Gliese 436b via analysis of all the RV data compiled to date. Theoretically this planet would be perturbing Gliese 436b so as to increase its orbital inclination at a rate of $\sim$0.1 deg yr$^{-1}$, and thus its transit depth and length, so that the non-detection by \citet{Butler2004} and the observed transit of \citet{Gillon2007b} were compatible. Since the RV detection of this second planet had a significant false-alarm probability of $\sim$20\%, \citet{Ribas2008a} proposed that confirmation could be achieved through 2008 observations of Gliese 436b's transits, which would show a lengthening of transit duration by $\sim$2 minutes compared to the \citet{Gillon2007b} data. As well, transit-timing variations (TTVs) of several minutes should also be detectable by observing a significant number of transits.

\citet{Alonso2008} reported a lack of observed inclination changes and TTV evidence for the second planet, based on a comparison of a single $H$-band light curve obtained in March 2008 to 8$\micron$ data taken with Spitzer 254 days earlier \citep{Gillon2007a,Deming2007}. This result, combined with additional radial velocity measurements that contradicted the proposed period of the second planet, drove \citet{Ribas2009} to retract their claim of the companion at IAU Symposium 253. However, \citet{Shporer2009} presented multiple light curves obtained in May 2007, and could not rule out TTVs on the order of a minute. While the planet specifically proposed by \citet{Ribas2008a} most likely does not exist,  \citet{Ribas2009} makes a strong case that a second planet is still needed to explain the peculiarities of Gliese 436b, and most likely exists in a non-resonant configuration where no strong TTVs are induced. Amateur astronomers have been diligent in observing Gliese 436b since it's initial transit discovery, and thus along with this data, published data, and our own data, we are able to present a thorough analysis of the TTVs, inclination, duration, and depth of the transit changes in the Gliese 436 system. We present our observations in Section~\ref{gl436obssec}, our modeling and derivation of parameters in Section~\ref{gl436modelsec}, and explore the observed TTVs and parameters of the system over time in Section~\ref{gl436parsec}.

\subsection{Observations}
\label{gl436obssec}

\begin{figure}
\centering
\epsfig{width=0.85\linewidth,file=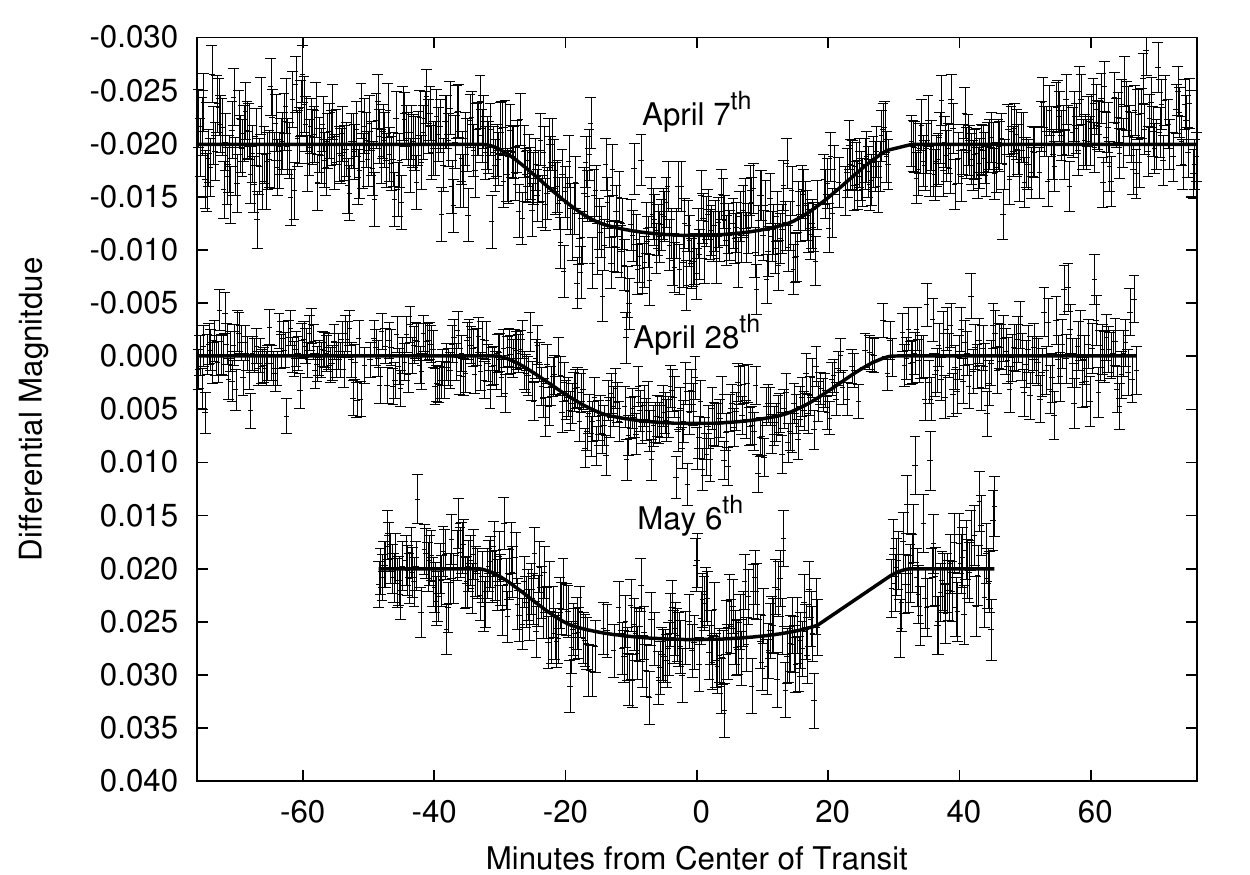}
\epsfig{width=0.85\linewidth,file=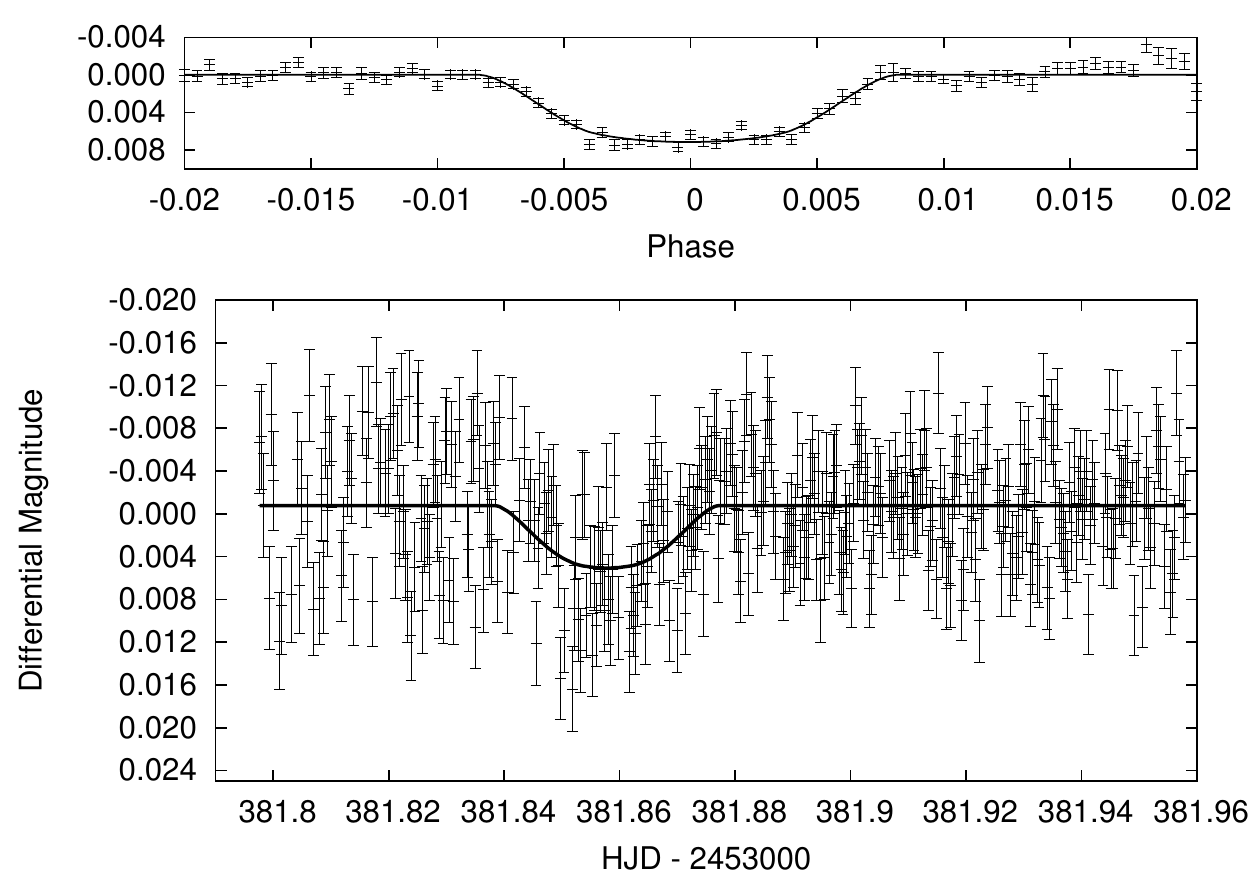}
\caption[APO 3.5m and 1m light curves of Gliese 436b]{\emph{Top}: The $V$-band light curves observed with the APO 3.5-meter in 2008 with model fits. \emph{Middle}: The 2008 3.5-meter data combined, phased, and binned in increments of 0.0005 phase. \emph{Bottom}: The transit observed by the NMSU 1-meter telescope on the night of January 11 2005 UT. A 3-sigma clip has been applied, and is shown with a model fit for which the radii were fixed.}
\label{gl436lcfig}
\end{figure}

We observed Gliese 436 (11h 42m 11s, +26$\degr$ 42$\arcmin$ 24$\arcsec$ J2000) in the $V$ filter on the nights of April 7, April 28, and May 6 2008 UT with the 3.5-meter telescope at Apache Point Observatory (APO). We used SPIcam, a backside-illuminated SITe 2048$\times$2048 CCD with 2$\times$2 binning, resulting in a plate scale of 0.28$\arcsec$/pixel, and sub-framed to a field of view of 4.8$\arcmin$ by 0.56$\arcmin$ to decrease readout time. We applied typical overscan, bias, and flat-field calibrations. For photometric reduction we used the standard IRAF task PHOT, with the aperture selected as a constant multiple of the Gaussian-fitted FWHM of each image to account for any variable seeing. We performed differential photometry with respect to the star USNO 1167-0208653 (2MASS ID 175252970) located at 11h 42m 12.08s, +26$\degr$ 46$\arcmin$ 07.45$\arcsec$ J2000. This star has $V$~=~10.82 and color $V$-$I$~=~1.48, compared to Gliese 436 which has $V$~=~10.68, and color $V$-$I$~=~1.70. In the error bar computation, we account for both standard noise from the photometry, as well as due to scintillation following equation 10 of \citet{Dravins1998}. Having obtained at least 30 minutes of data on each side of the transit, we subtracted a linear fit for all data outside of transit vs. airmass to account for any differential reddening. Resulting individual data points have errors ranging from 1.5 to 2.8 mmag, which agrees with the rms of the residuals from the model fits, and a typical cadence of about 17 seconds. We have searched for correlated noise on the timescale of ingress and egress, via the technique of \citet{Pont2006}, but only find a statistically significant amount for the night of April 7, measured to be 0.11 mmag. The three transits are shown in Figure~\ref{gl436lcfig}. 

We also carried out accompanying observations with the New Mexico State University (NMSU) 1-meter telescope at APO, in the $V$ filter on the night of April 7 2008 UT, and in the I filter on the night of April 28 2008 UT. A 2048$\times$2048 E2V CCD was used with 1x1 binning and sub-framing, resulting in a field of view of 8.0$\arcmin$ square and a plate scale of 0.47$\arcsec$/pixel, and we applied the aforementioned standard calibration and photometric extraction techniques. We performed ensemble photometry with respect to the USNO star that was used as the 3.5m reference, as well as BD+27 2046 ($V$~=~10.64, $V$-$I$~=~0.44), and another star at 11h42m00s, +26$\degr$45$\arcmin$56$\arcsec$ J2000 ($V$~=~12.81, $V$-$I$~=~1.46). Resulting typical errors on individual points range from 3 to 5 mmag with a typical cadence of about 12 seconds.

The NMSU 1-meter telescope can also function as a robotic telescope, and is used intermittently to photometrically monitor stars with known radial-velocity discovered planets to search for transits \citep*{Holtzman2010}. A search of the 1-meter archives revealed that it observed Gliese 436 on the night of January 11 2005 UT, during which a transit should have occurred, according to the precise ephemeris for Gliese 436b that is now available by incorporating the many observed transits in 2007 and 2008. At the time, this 1-meter program depended on visual inspection of automatically generated photometry and plots. For this night, the plot had large temporal and brightness ranges, and thus the tiny transit was easily missed visually. However, carefully inspecting the region constrained by the ephemeris, as well as re-performing the photometry to maximize signal-to-noise, we find a transit signature within a minute of that predicted by the ephemeris with reasonable width and depth, as shown in Figure~\ref{gl436lcfig}. Individual data points have an error of about 4 mmag, a cadence of 30 seconds, and we do not detect any correlated noise with any level of significance.

We also conducted observations on the nights of April 28 and May 13 2008 UT using a 24$"$ telescope located at the Sommers-Bosch Observatory (SBO) on the University of Colorado at Boulder campus, using an I filter. These observations also used a windowed chip and an exposure time to maximize signal-to-noise without saturating, and have comparable temporal resolution to the 3.5m and 1m telescopes due to a shorter readout time. As well, we used an unfiltered 11$"$ telescope at Cloudcroft, NM (CC) with a SBIG ST-7E CCD and 2$\times$2 binning on May 6 2008 UT, with a resulting cadence of about 25 seconds. We have also gathered all the amateur data currently available, $\sim$15 light curves, on the system as compiled by Bruce Gary (http://brucegary.net/AXA/GJ436/gj436.htm).

\subsection{Modeling and Derivation of Parameters}
\label{gl436modelsec}

We used the JKTEBOP code \citep{Southworth2004a,Southworth2004b} to model all the transit light curves, both our own and previously published, in a consistent and uniform manner. \citet{Southworth2008} has performed an exhaustive analysis of fourteen transiting planets using the JKTEBOP code, and shows it compares well with results reported elsewhere. JKTEBOP offers the advantage of incorporating a Levenberg-Marquardt optimization algorithm, improved limb darkening treatments, and extensive error analysis routines, which are critical for confirming any trends in the system.

For each transit curve, we solved for the ratio of radii ($k$ = $R_{p}$/$R_{s}$), the orbital inclination ($i$), the time of mid-transit ($T_{0}$), and a scale factor that defines the normalized value of the out-of-transit flux in the light curves. In order to obtain reasonable results for the scale of the system for all data sets, the sum of the radii ($R_{s}$ + $R_{p}$) was set to that found by \citet{Torres2007}. We also fixed the eccentricity to a value of 0.15 and the longitude of periastron to 343$\degr$ as given by \citet{Deming2007} and \citet{Mardling2008}. We used a quadratic limb-darkening law with coefficients taken from \citet{Claret2000b} for T$_{eff}$ = 3500K, log(g) = 4.5, V$_{t}$ = 2.0 km s$^{-1}$, and [M/H] = 0.0, for the appropriate filters. In the case of the Spitzer 8$\micron$ data, we used the coefficients as determined by \citet{Gillon2007a}. From each fit, still assuming a constant sum of radii, we were thus also able to calculate the individual star and planet radii, as well as the depth and width of transit. In order to rule out any potential correlations in derived planet size and inclination, we then re-modeled all data with the same procedure, but also fixing $k$, and thus the star and planet sizes, to that found by \citet{Torres2007}. This generally produced similar results, but for the noisier data sets achieved more consistent results. Parameters from both techniques are shown in Table~\ref{modeldattab}.

In order to obtain robust errors, we ran 10,000 Monte-Carlo simulations for each data set and performed a residual-permutation analysis \citep{Jenkins2002} to investigate temporally correlated noise. In both cases, the previously fixed parameters, as well as the limb-darkening coefficients, were allowed to vary so that their individual uncertainties would be taken into account in the derived parameter uncertainties. For each Monte Carlo simulation, random Gaussian noise with amplitude equal to the given error bars, or in the absence thereof the standard deviation of the residual scatter from the best-fit solution, was added to each data point and the curve re-fitted with random perturbations applied to the initial parameter values. This ensured a detailed exploration of the parameter space and parameter correlations. However, this Monte Carlo technique will underestimate errors for certain parameters in the presence of temporally correlated noise, which can result from trends in seeing, extinction, focus, or other atmospheric or telescope related phenomena \citep{Southworth2008}. The residual-permutation method takes the residuals of the best-fit model, shifts them to the next data point, and finds a new solution. The residuals are shifted again, a new fit is found, and the process repeats as many times as there are datapoints. Thus, there is a distribution of fitted values similar to the Monte Carlo technique, but any temporal trends will have been propagated around the light curve, and thus taken into account. For our final errors we adopt the larger value found between the two methods, although for the majority of parameters and data sets the two methods agree quite well.

In total we modeled 28 light curves, (16 professional and 12 amateur), covering 19 separate transit events over a baseline of nearly 3.3 years.

\begin{deluxetable}{rcccccccc}
\rotate
\tablewidth{0in}
\tabletypesize{\scriptsize}
\tablecaption{Parameters derived from using the JKTEBOP code with 1$\sigma$ errors}
\tablecolumns{9}
\tablehead{Epoch & Source & Filter\tablenotemark{c} & $T_{\rm min}$ & Inclination & $R_{\star}$ & $R_{p}$ & Depth & Width\\ & & & (HJD-2450000) & ($\degr$) & ($R_{\sun}$) & ($R_{\earth}$) & (mmag) & (minutes)}
\startdata
\cutinhead{Ratio of Radii $k$ Allowed to Vary}
-318 & NMSU 1m & V & 3381.85584$\pm$0.00179 & 86.02$\pm$0.23 & 0.446$\pm$0.046 & 6.19$\pm$4.81 & 7.05$\pm$1.35 & 47.0$\pm$7.1\\
0 & \citet{Gillon2007b}\tablenotemark{b} & V & 4222.61617$\pm$0.00060 & 86.38$\pm$0.18 & 0.463$\pm$0.016 & 4.32$\pm$0.24 & 6.98$\pm$0.43 & 60.1$\pm$1.6\\
1 & \citet{Shporer2009}\tablenotemark{b} & None & 4225.26052$\pm$0.00089 & 86.43$\pm$0.17 & 0.463$\pm$0.015 & 4.33$\pm$0.27 & 7.16$\pm$0.81 & 61.4$\pm$2.5\\
1 & \citet{Shporer2009}\tablenotemark{b} & V & 4225.26050$\pm$0.00072 & 86.35$\pm$0.17 & 0.462$\pm$0.016 & 4.47$\pm$0.25 & 7.31$\pm$0.46 & 59.2$\pm$1.9\\
9 & \citet{Shporer2009}\tablenotemark{b} & R & 4246.41012$\pm$0.00079 & 86.27$\pm$0.18 & 0.456$\pm$0.014 & 5.10$\pm$0.66 & 9.07$\pm$0.87 & 56.7$\pm$2.6\\
22 & \citet{Gillon2007a}  & 8$\micron$ & 4280.78219$\pm$0.00011 & 86.34$\pm$0.16 & 0.464$\pm$0.016 & 4.23$\pm$0.16 & 7.46$\pm$0.10 & 59.8$\pm$1.1\\
110 & Gregor Srdoc\tablenotemark{a}  & R & 4513.43393$\pm$0.00174 & 86.10$\pm$0.24 & 0.457$\pm$0.030 & 5.04$\pm$2.91 & 7.11$\pm$1.29 & 49.8$\pm$6.5\\
110 & Tonny Vanmunster\tablenotemark{a}  & R & 4513.44404$\pm$0.00247 & 87.13$\pm$0.30 & 0.461$\pm$0.016 & 4.57$\pm$0.45 & 9.63$\pm$1.67 & 77.8$\pm$4.9\\
112 & Bruce Gary\tablenotemark{a}  & R & 4518.72999$\pm$0.00278 & 86.04$\pm$0.34 & 0.443$\pm$0.087 & 6.55$\pm$9.38 & 8.78$\pm$2.37 & 47.0$\pm$11.7\\
113 & Gregor Srdoc\tablenotemark{a}  & R & 4521.37338$\pm$0.00130 & 87.27$\pm$0.30 & 0.459$\pm$0.015 & 4.81$\pm$0.30 & 11.06$\pm$1.13 & 80.0$\pm$4.2\\
115 & James Roe\tablenotemark{a}  & V & 4526.65995$\pm$0.00124 & 86.03$\pm$0.27 & 0.447$\pm$0.036 & 6.05$\pm$3.34 & 7.47$\pm$1.55 & 48.4$\pm$9.0\\
115 & Joao Gregorio\tablenotemark{a}  & V & 4526.65972$\pm$0.00130 & 87.09$\pm$0.28 & 0.468$\pm$0.016 & 3.80$\pm$0.25 & 6.50$\pm$0.66 & 77.1$\pm$4.0\\
117 & Richard Schwartz\tablenotemark{a}  & V & 4531.94399$\pm$0.00222 & 86.18$\pm$0.20 & 0.461$\pm$0.015 & 4.51$\pm$0.65 & 6.37$\pm$1.67 & 53.4$\pm$6.5\\
118 & \citet{Alonso2008} & H & 4534.59611$\pm$0.00014 & 86.39$\pm$0.17 & 0.463$\pm$0.016 & 4.32$\pm$0.17 & 7.74$\pm$0.11 & 61.1$\pm$0.9\\
127 & Manuel Mendez\tablenotemark{a} & R & 4558.38849$\pm$0.00173 & 86.60$\pm$0.24 & 0.466$\pm$0.016 & 3.99$\pm$0.33 & 6.60$\pm$0.87 & 66.6$\pm$5.6\\
129 & NMSU 1m & V & 4563.67937$\pm$0.00257 & 86.45$\pm$0.33 & 0.467$\pm$0.019 & 3.87$\pm$0.61 & 6.34$\pm$1.12 & 61.5$\pm$8.0\\
129 & APO 3.5m & V & 4563.67968$\pm$0.00051 & 86.44$\pm$0.17 & 0.459$\pm$0.015 & 4.73$\pm$0.28 & 8.60$\pm$0.44 & 61.8$\pm$2.2\\
132 & James Roe\tablenotemark{a}  & B & 4571.61844$\pm$0.00107 & 88.60$\pm$0.62 & 0.455$\pm$0.015 & 5.24$\pm$0.24 & 14.84$\pm$0.83 & 95.5$\pm$3.6\\
137 & NMSU 1m & I & 4584.83301$\pm$0.00117 & 86.55$\pm$0.19 & 0.449$\pm$0.015 & 5.89$\pm$0.51 & 14.61$\pm$1.72 & 65.2$\pm$3.6\\
137 & APO 3.5m & V & 4584.83084$\pm$0.00035 & 86.32$\pm$0.16 & 0.464$\pm$0.015 & 4.20$\pm$0.20 & 6.36$\pm$0.25 & 58.3$\pm$1.2\\
137 & SBO 24$"$ & I & 4584.82868$\pm$0.00166 & 86.63$\pm$0.21 & 0.448$\pm$0.015 & 5.95$\pm$0.50 & 15.23$\pm$2.13 & 67.3$\pm$3.8\\
137 & Bruce Gary\tablenotemark{a} & R & 4584.82876$\pm$0.00087 & 86.51$\pm$0.18 & 0.463$\pm$0.015 & 4.32$\pm$0.25 & 7.41$\pm$0.58 & 64.0$\pm$2.4\\
138 & Manuel Mendez\tablenotemark{a} & R & 4587.47754$\pm$0.00170 & 86.91$\pm$0.28 & 0.462$\pm$0.016 & 4.45$\pm$0.35 & 8.83$\pm$1.13 & 73.9$\pm$4.9\\
140 & CC 11$"$ & None & 4592.76123$\pm$0.00140 & 86.25$\pm$0.17 & 0.463$\pm$0.015 & 4.38$\pm$0.48 & 6.64$\pm$1.03 & 56.0$\pm$3.4\\
140 & APO 3.5m & V & 4592.76281$\pm$0.00084 & 86.50$\pm$0.17 & 0.465$\pm$0.015 & 4.12$\pm$0.21 & 6.71$\pm$0.36 & 63.5$\pm$3.4\\
140 & SBO 24$"$ & I & 4592.76202$\pm$0.00177 & 86.55$\pm$0.26 & 0.453$\pm$0.017 & 5.43$\pm$0.84 & 12.25$\pm$1.50 & 65.3$\pm$4.8\\
143 & SBO 24$"$ & I & 4600.69795$\pm$0.00118 & 85.88$\pm$0.24 & 0.425$\pm$0.067 & 8.52$\pm$6.54 & 6.75$\pm$1.08 & 42.0$\pm$8.5\\
146 & James Roe\tablenotemark{a}  & V & 4608.62470$\pm$0.00107 & 86.32$\pm$0.23 & 0.454$\pm$0.015 & 5.31$\pm$0.67 & 9.86$\pm$0.55 & 58.4$\pm$6.3\\
\nodata & 3.5m Data Combined & V & \nodata & 86.39$\pm$0.16 & 0.463$\pm$0.015 & 4.39$\pm$0.22 & 7.25$\pm$0.31 & 60.6$\pm$1.3\\
\cutinhead{Star and Planet Radii Fixed by Fixing k}
-318 & NMSU 1m & V & 3381.85596$\pm$0.00212 & 86.15$\pm$0.17 & 0.464$\pm$0.016 & 4.23$\pm$0.28 & 5.61$\pm$0.63 & 52.5$\pm$3.2\\
0 & \citet{Gillon2007b}\tablenotemark{b} & V & 4222.61617$\pm$0.00062 & 86.40$\pm$0.16 & 0.464$\pm$0.016 & 4.23$\pm$0.28 & 6.74$\pm$0.69 & 60.5$\pm$2.6\\
1 & \citet{Shporer2009}\tablenotemark{b} & None & 4225.26049$\pm$0.00094 & 86.45$\pm$0.19 & 0.464$\pm$0.016 & 4.23$\pm$0.30 & 6.86$\pm$0.76 & 61.8$\pm$3.7\\
1 & \citet{Shporer2009}\tablenotemark{b} & V & 4225.26050$\pm$0.00076 & 86.39$\pm$0.16 & 0.464$\pm$0.016 & 4.23$\pm$0.30 & 6.65$\pm$0.72 & 59.9$\pm$2.1\\
9 & \citet{Shporer2009}\tablenotemark{b} & R & 4246.41009$\pm$0.00103 & 86.37$\pm$0.17 & 0.464$\pm$0.017 & 4.23$\pm$0.29 & 6.66$\pm$0.70 & 59.6$\pm$3.6\\
22 & \citet{Gillon2007a}  & 8$\micron$ & 4280.78219$\pm$0.00011 & 86.34$\pm$0.16 & 0.464$\pm$0.016 & 4.23$\pm$0.30 & 7.47$\pm$1.00 & 59.5$\pm$1.0\\
110 & Gregor Srdoc\tablenotemark{a} & R & 4513.43416$\pm$0.00191 & 86.19$\pm$0.18 & 0.464$\pm$0.016 & 4.23$\pm$0.28 & 5.92$\pm$0.74 & 54.0$\pm$4.3\\
110 & Tonny Vanmunster\tablenotemark{a} & R & 4513.44424$\pm$0.00386 & 87.29$\pm$0.54 & 0.464$\pm$0.016 & 4.23$\pm$0.28 & 8.20$\pm$1.11 & 77.5$\pm$9.0\\
112 & Bruce Gary\tablenotemark{a} & R & 4518.73038$\pm$0.00358 & 86.20$\pm$0.25 & 0.464$\pm$0.017 & 4.23$\pm$0.31 & 6.00$\pm$1.03 & 55.7$\pm$7.3\\
113 & Gregor Srdoc\tablenotemark{a}  & R & 4521.37312$\pm$0.00244 & 87.34$\pm$0.39 & 0.464$\pm$0.016 & 4.23$\pm$0.31 & 8.32$\pm$1.26 & 80.1$\pm$6.3\\
115 & James Roe\tablenotemark{a} & V & 4526.66055$\pm$0.00302 & 86.14$\pm$0.27 & 0.464$\pm$0.015 & 4.23$\pm$0.28 & 5.56$\pm$1.30 & 51.7$\pm$9.1\\
115 & Joao Gregorio\tablenotemark{a} & V & 4526.65996$\pm$0.00101 & 86.65$\pm$0.19 & 0.464$\pm$0.016 & 4.23$\pm$0.29 & 7.42$\pm$0.93 & 67.3$\pm$3.4\\
117 & Richard Schwartz\tablenotemark{a} & V & 4531.94392$\pm$0.00198 & 86.26$\pm$0.26 & 0.464$\pm$0.016 & 4.23$\pm$0.31 & 5.95$\pm$1.08 & 56.1$\pm$7.5\\
118 & \citet{Alonso2008} & H & 4534.59610$\pm$0.00014 & 86.40$\pm$0.16 & 0.464$\pm$0.016 & 4.23$\pm$0.28 & 7.42$\pm$0.98 & 61.1$\pm$1.0\\
127 & Manuel Mendez\tablenotemark{a} & R & 4558.38809$\pm$0.00164 & 86.51$\pm$0.21 & 0.464$\pm$0.016 & 4.23$\pm$0.23 & 7.00$\pm$0.76 & 64.2$\pm$4.7\\
129 & NMSU 1m & V & 4563.67966$\pm$0.00252 & 86.38$\pm$0.23 & 0.464$\pm$0.016 & 4.23$\pm$0.28 & 6.63$\pm$0.79 & 60.1$\pm$5.3\\
129 & APO 3.5m & V & 4563.67971$\pm$0.00116 & 86.53$\pm$0.18 & 0.464$\pm$0.016 & 4.23$\pm$0.28 & 7.06$\pm$0.78 & 64.1$\pm$3.2\\
132 & James Roe\tablenotemark{a} & B & 4571.61831$\pm$0.00467 & 88.62$\pm$1.02 & 0.464$\pm$0.016 & 4.23$\pm$0.28 & 9.25$\pm$1.27 & 93.3$\pm$6.5\\
137 & NMSU 1m & I & 4584.83373$\pm$0.00379 & 86.84$\pm$0.78 & 0.464$\pm$0.016 & 4.23$\pm$0.27 & 7.70$\pm$1.19 & 72.6$\pm$14.5\\
137 & APO 3.5m & V & 4584.83084$\pm$0.00036 & 86.32$\pm$0.16 & 0.464$\pm$0.016 & 4.23$\pm$0.29 & 6.45$\pm$0.62 & 58.0$\pm$1.8\\
137 & SBO 24$"$ & I & 4584.82787$\pm$0.00912 & 86.64$\pm$0.73 & 0.464$\pm$0.016 & 4.23$\pm$0.28 & 7.05$\pm$1.45 & 67.8$\pm$17.1\\
137 & Bruce Gary\tablenotemark{a} & R & 4584.82874$\pm$0.00100 & 86.53$\pm$0.18 & 0.464$\pm$0.016 & 4.23$\pm$0.28 & 7.07$\pm$0.76 & 64.3$\pm$2.8\\
138 & Manuel Mendez\tablenotemark{a} & R & 4587.47761$\pm$0.00204 & 87.00$\pm$0.31 & 0.464$\pm$0.016 & 4.23$\pm$0.29 & 7.92$\pm$0.99 & 75.0$\pm$5.3\\
140 & CC 11$"$ & None & 4592.76119$\pm$0.00142 & 86.27$\pm$0.15 & 0.464$\pm$0.016 & 4.23$\pm$0.29 & 6.24$\pm$0.66 & 56.5$\pm$2.9\\
140 & APO 3.5m & V & 4592.76248$\pm$0.00093 & 86.47$\pm$0.17 & 0.464$\pm$0.016 & 4.23$\pm$0.29 & 6.94$\pm$0.68 & 62.6$\pm$2.9\\
140 & SBO 24$"$ & I & 4592.76090$\pm$0.00430 & 86.71$\pm$0.32 & 0.464$\pm$0.016 & 4.23$\pm$0.28 & 7.57$\pm$1.06 & 69.5$\pm$7.8\\
143 & SBO 24$"$ & I & 4600.69668$\pm$0.00171 & 86.08$\pm$0.17 & 0.464$\pm$0.016 & 4.23$\pm$0.30 & 5.54$\pm$0.71 & 50.4$\pm$3.6\\
146 & James Roe\tablenotemark{a} & V & 4608.62542$\pm$0.00508 & 86.40$\pm$0.68 & 0.464$\pm$0.016 & 4.23$\pm$0.28 & 6.53$\pm$2.03 & 63.8$\pm$19.6\\
\nodata & 3.5m Data Combined & V & \nodata & 86.43$\pm$0.16 & 0.464$\pm$0.016 & 4.23$\pm$0.28 & 6.82$\pm$0.67 & 61.7$\pm$2.7
\enddata
\tablenotetext{a}{Amateur Observer with data obtained from Bruce Gary. http://brucegary.net/AXA/GJ436/gj436.htm}
\tablenotetext{b}{Data were digitized from published plot}
\tablenotetext{c}{Johnson-Cousins System}
\tablecomments{All errors are 1$\sigma$}
\label{modeldattab}
\end{deluxetable}

\subsection{Transit Timing and Eclipse Variations}
\label{gl436parsec}

\begin{figure}
\centering
\epsfig{width=\linewidth,file=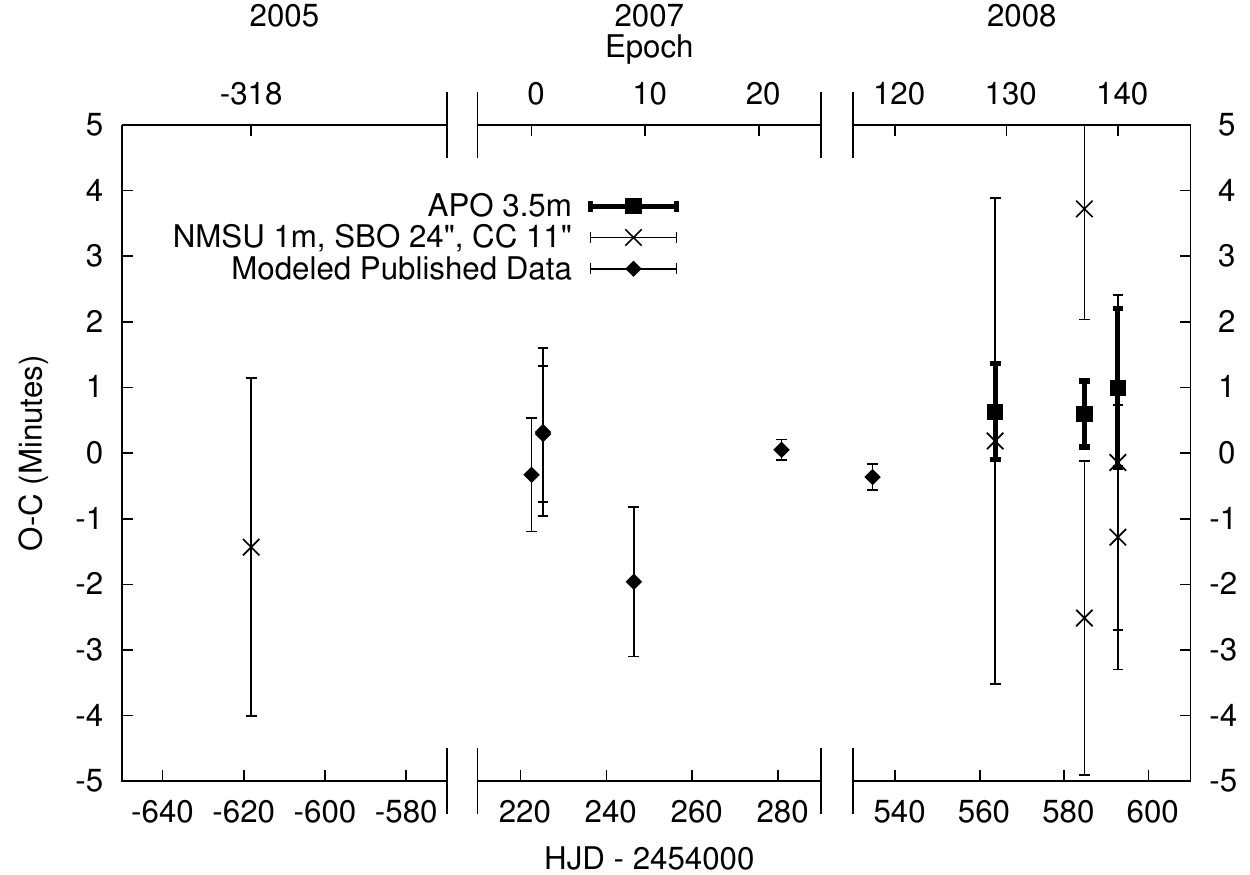}
\caption[O-C diagram for all professional times of minima]{O-C diagram for all professional times of minima.}
\label{ocfig}
\end{figure}

Using the derived time of minima in Table~\ref{modeldattab} for all the data when allowing $k$ to vary, we derive a new linear, error-weighted ephemeris of  T$_{c}$(HJD) = 2454222.6164(1) + 2.643897(2)$\cdot$E, where the parentheses indicate the amount of uncertainty in the last digit, and E is the epoch with E = 0 the initial transit discovery of \citet{Gillon2007b}. Using this ephemeris, we then compute an observed minus calculated (O-C) diagram for the time of transit center, as shown in Figure~\ref{ocfig}. We have currently excluded the amateur data from the plot due to much larger error bars, so that the high-precision data points can be seen clearly. We have examined the TTVs and various subsets thereof using a phase dispersion minimization technique \citep{Stellingwerf1978}, but do not find any periods with statistical significance. Examining the best data, specifically the previously published data and our 3.5-meter observations, there is a standard deviation of 52 seconds. Assuming a sinusoidal TTV trend, we can then rule out any TTVs with amplitude greater than $\sim$1 minute.

\begin{figure}[ht]
\centering
\epsfig{width=\linewidth,file=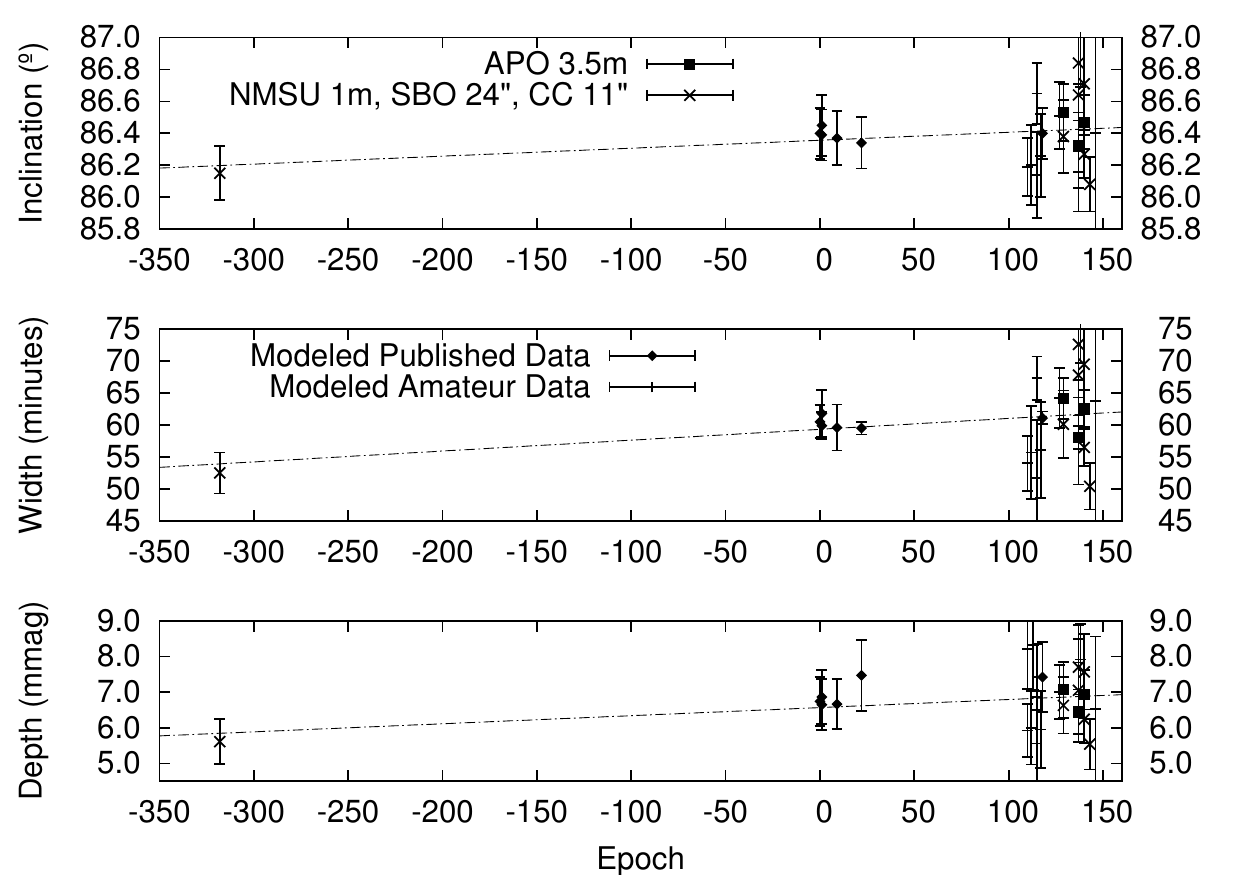}
\caption{Measured inclination, width, and depth of transit over time for all data, with the star and planet radii fixed.}
\label{trendfig}
\end{figure}

We have searched for any trends in derived inclination, width, and depth of transit over time via error-weighted least-squares linear regression. In addition, we have also performed 10,000 Monte Carlo simulations for each fit, where Gaussian noise with amplitude equal to each point's error bars was added in each iteration and the data re-fitted, with resulting 1$\sigma$ parameter distributions giving robust errors. The two methods agree to within 1\% for all values. As mentioned in Section~\ref{gl436modelsec}, we modeled all the light curves by both allowing the ratio of radii to vary as well as fixing it, and thus we list the values for each set. Performing fits to all the data, we have a tentative detection of increasing inclination, transit width, and transit depth with time, as shown in Table~\ref{trendtab}. We present these fits with the actual data derived when fixing the radii in Figure~\ref{trendfig}. As a precaution against any bias being introduced by the much larger number of data points at later epochs, we decided to separately bin the 2005, 2007, and 2008 data using an error-weighted mean, and re-fit the three resulting data points for each modeling method. As shown in Table~\ref{trendtab}, the values agree very well with those derived when not binning the data.


\begin{deluxetable}{lccc}
\tablewidth{0pt}
\tablecaption{Trends in derived inclination, width, and depth of transit over time}
\tablecolumns{4}
\tablehead{Data Set & deg yr$^{-1}$ & min yr$^{-1}$ & mmag yr$^{-1}$}
\startdata
\cutinhead{Variable Radius}
All & 0.120$\pm$0.062 & 3.43$\pm$1.01 &  0.28$\pm$0.16\\
Binned & 0.126$\pm$0.061 & 3.53$\pm$0.97 & 0.26$\pm$0.14\\
No 2005 & 0.092$\pm$0.099 & 3.10$\pm$1.10 & 0.29$\pm$0.17\\
\cutinhead{Fixed Radius}
All & 0.069$\pm$0.051 & 2.36$\pm$0.84 &  \ 0.32$\pm$0.20\\
Binned & 0.071$\pm$0.050 & 2.37$\pm$0.81 & \ 0.32$\pm$0.19\\
No 2005 & 0.020$\pm$0.099 & 1.68$\pm$1.29 & -0.01$\pm$0.42\\
\enddata
\label{trendtab}
\end{deluxetable}

The trends are moderately dependent on the single 2005 transit data point, which greatly extends the temporal baseline, and as such we are cautious about any claims. Resulting temporal trends when removing the 2005 data point are also shown in Table~\ref{trendtab}. Although while removing the 2005 data point significantly weakens the claim of a variation of inclination with time, the trend of increasing width still holds. Also of interest is that at a rate of 0.120 deg yr$^{-1}$, as derived from our fit to all the data fitted with a variable radius, the JKTEBOP program yields an increase in transit width of 4.36 min yr$^{-1}$, and depth of 0.544 mmag yr$^{-1}$, which are in agreement with our observed trends, and thus are self-consistent. As well, the measured rate of inclination change is compatible with the $\sim$0.1 deg yr$^{-1}$ required to make congruent the non-detection of \citet{Butler2004} and the observed transit of \citet{Gillon2007b}. Extending the measurement baseline a couple years into the future will confirm or negate this result.

\subsection{Discussion and Conclusion}

We have presented a total of ten new primary transit light curves of Gliese 436b, three of which come from the 3.5-meter telescope at APO, and one of which is from the NMSU 1-meter in January 2005. We have collected and uniformly modeled all available professional and amateur light curves, and searched for any trends in transit timing, width of transit, and depth of transit variations. We find statistically significant, self-consistent trends that are compatible with the perturbation of Gliese 436b by a planet with mass $\lesssim$ 12 M$_{\earth}$ in a non-resonant orbit with semi-major axis $\lesssim$ 0.08 AU. This conclusion is based on the numerical simulations of \citet[][see Fig. 1]{Ribas2008a} who constrain the mass and semi-major axis of the theoretical second planet by examining which configurations could produce the observed orbital perturbations while still remaining undetected by the existing radial-velocity data. From our analysis, we infer a non-resonant orbit based on a lack of detected TTVs with amplitude $\gtrsim$ 1 minute. We stress that our measured trends are moderately dependent on our 2005 data, and thus subsequent high-precision observations over the next few years need to be carried out to confirm or refute this trend. If confirmed, it would be strong evidence for the first extrasolar planet discovered via orbital perturbations to a transiting planet. Also, we would like to note that although \citet{Alonso2008} had previously limited the rate of inclination change to 0.03$\pm$0.05 deg/yr, they did so only by measuring the change in width between the 2007 Spitzer observations and their own 2008 $H$-band data, which they found to be 0.5$\pm$1.2 minutes. Via Table~\ref{modeldattab}, we find the difference in transit width between the two observations to be 1.5$\pm$1.4 minutes, which is in agreement with our derived inclination and width values, and is a more reliable result due to using full model fits with proper limb-darkening coefficients. With respect to the amateur observations, although they are numerous, the very small depth of the transit makes it a challenge for most small aperture systems, resulting in very large uncertainties in $i$ and $T_{0}$. Also, while amateur observers are aware of the importance of precision timing, we of course cannot examine each of their observing set-ups, and thus one must be aware of the possibility, although small, of systemic time offsets on a given night when interpreting their data.

Very recently, \citet{Stevenson2012} found preliminary evidence for two additional low-mass, sub-Earth-sized transiting planets in both interior and exterior orbits to Gliese 436b via direct observations of their transits in recently obtained $Spitzer$ data. Although it is not clear if these planets could be responsible for the parameter variations we observe due to their very low masses, it does provide evidence that the Gliese 436 system is populated by multiple planets, increasing the likelihood that a second planet massive enough to induce the observed variations exists in the system.

%% file: chp3-wasp12.tex
\begin{singlespace}
\section[\MakeUppercase{Day-side \MakeLowercase{z}$'$-band emission and Eccentricity\\of WASP-12\MakeLowercase{b}}]{\MakeUppercase{Day-side \MakeLowercase{z}$'$-band emission and Eccentricity of WASP-12\MakeLowercase{b}}}
\label{chap3}
\end{singlespace}

\subsection{Introduction} \label{sec:intro}

The transiting hot Jupiter WASP-12b, discovered by \cite{Hebb2009}, has many notable characteristics. With a mass of 1.41 $\pm$ 0.10 $M_{\rm Jup}$ and a radius of 1.79 $\pm$ 0.09 $R_{\rm Jup}$, WASP-12b was the planet with the second largest radius reported at discovery, and the sixth largest transiting planet known at the time of this writing \citep{Schneider2012}. It is also one of the most heavily irradiated planets known, with an incident stellar flux at the substellar point of over 9$\times10^{9}$ $erg$ $cm^{-2}$ $s^{-1}$. In addition, model fits to its observed radial velocity and transit light curves suggest that the orbit of WASP-12b is slightly eccentric. All these attributes make WASP-12b one of the best targets to test current irradiated atmosphere and tidal heating models for exoplanets.

In irradiated atmosphere model studies WASP-12b is an extreme case even in the category of highly irradiated gas giants. Such highly irradiated planets are expected to show thermal inversions in their upper atmospheric layers \citep{Burrows2008a}, although the chemicals responsible for such inversions remain unknown. TiO and VO molecules, which can act as strong optical absorbers, have been proposed \citep{Hubeny2003,Fortney2008}, but \citet{Desert2008} claim that the concentration of those molecules in planetary atmospheres is too low ($< 10^{-3} - 10^{-2}$ times solar) to cause thermal inversions. \citet{Spiegel2009} argue that TiO needs to be at least half the solar abundance to cause thermal inversions, and very high levels of macroscopic mixing are required to keep enough TiO in the upper atmosphere of planets.  $S_{2}$, $S_{3}$ and HS compounds have also recently been suggested and then questioned as causes of the observed thermal inversions \citep{Zahnle2009}.

In the case of tidal heating, detailed models are now being developed \citep[e.g.,][]{Bodenheimer2003,Miller2009,Ibgui2009,Ibgui2010,Ibgui2011} to explain the inflated radius phenomenon observed in hot Jupiters, of which WASP-12b, with a radius over 40\% larger than predicted by standard models, is also an extreme case. All models assume that the planetary orbits are slightly eccentric, and directly measuring those eccentricities is key not only to test the model hypotheses, but also to obtain information about the planets' core mass and energy dissipation mechanisms \citep[see][]{Ibgui2010}.

We present the detection of the eclipse of WASP-12b in the $z'$-band (0.9 $\mu$m), which gives the first measurement of the atmospheric emission of this planet, and the first direct estimation of its orbital eccentricity. This is also only the second detection of an exoplanet secondary eclipse at $\lambda$ $<$ 1 $\mu m$ from ground-based observations, (the first was by \citet{Sing2009a} with combined data from 6.5 and 8-meter telescopes), while we employ only a 3.5-meter telescope. Section~\ref{sec:obs} summarizes the observations and analysis of the data. In Section~\ref{sec:model} we compare the emission of the planet to models. The results are discussed in Section~\ref{sec:sum}.

\subsection{Observations and Analyses}
\label{sec:obs}

We monitored WASP-12 [RA = 06:30:32.794, Dec = +29:40:20.29 (J2000), $V$ = 11.7] during two eclipses, and under photometric conditions, on February 19 and October 18 2009 UT.  An additional attempt on October 30 2009 UT was lost due to weather. The data were collected with the SPICam instrument on the ARC's 3.5-meter telescope at Apache Point Observatory, using a SDSS z$^{\prime}$ filter with an effective central wavelength of $\sim$0.9 $\mu$m. SPICam is a backside-illuminated SITe TK2048E 2048x2048 pixel CCD with 24 micron pixels, giving an unbinned plate scale of 0.14 arc seconds per pixel and a field of view of 4.78 arc minutes square. The detector, cosmetically excellent and linear through the full A/D converter range, was binned 2x2, which gives a gain of 3.35 e$^{-}$/ADU, a read noise of 1.9 DN/pixel, and a 48 second read time.

On February 19 we monitored WASP-12 from 3:00 to 3:28 UT and from 3:54 to 7:10 UT, losing coverage between 3:28 and 3:54 UT when the star reached a local altitude greater than 85$\degr$, the soft limit of the telescope at that time. These observations yielded 1.20 hours of out-of-eclipse and 2.45 hours of in-eclipse coverage, at airmasses between 1.005--1.412. On October 18 we extended the altitude soft limit of the telescope to 87$\degr$ and covered the entire eclipse from 7:05 to 12:45 UT, yielding 2.73 hours of out-of-eclipse and 2.93 hours of in-eclipse coverage, with airmasses between 1.001--1.801. In both nights we defocused the telescope to a FWHM of $\sim$2$\arcsec$ to reduce pixel sensitivity variation effects, and also to allow for longer integration times, which minimized scintillation noise and optimized the duty cycle of the observations. Pointing changed by less than (x,y)=(4,7) pixels in the October 18 dataset, and by less than (x,y)=(3,12) pixels on February 19, with the images for this second night suffering a small gradual drift in the $y$ direction throughout the night. Integration times ranged from 10 to 20 seconds.  Taking into account Poisson, readout, and scintillation noise, the photometric precision on WASP-12 and other bright stars in the images ranged between 0.07--0.15\% per exposure on February 19, and between 0.05--0.09\% per exposure on October 18.

The field of view of SPICam was centered at RA = 06:30:25, Dec = +29:42:05 (J2000) and included WASP-12 and two other isolated stars at RA = 06:30:31.8, Dec = +29:42:27 (J2000) and RA = 06:30:22.6, Dec = +29:44:42 (J2000), with apparent brightness and $B-V$ and $J-K$ colors similar to the target.  Each night's dataset was analyzed independently and the results combined in the end. The timing information was extracted from the headers of the images and converted into Heliocentric Julian Days using the IRAF task {\it setjd}, which has been tested to provide sub-second timing accuracy.

We corrected each image for bias level and flatfield effects using standard IRAF routines. Dark current was negligible. DAOPHOT-type aperture photometry was performed in each frame. We recorded the flux from the target and the comparison stars over a wide range of apertures and sky background annuli around each star. We used apertures between 2 and 35 pixels in one-pixel steps during a first preliminary photometry pass, and 0.05 pixel steps in the final photometric extraction. To compute the sky background around each star we used variable width annuli, with inner radii between 35 and 60 pixels sampled in one-pixel steps. The best aperture and sky annuli combinations were selected by identifying the most stable, (i.e., minimum standard deviation), differential light curves between each comparison and the target at phases out-of-eclipse\footnote{We had to iterate on the out-of-eclipse phase limits after finding that the eclipse was centered at $\phi$ = 0.51. Out-of-eclipse was finally defined as phases $\phi$$<$0.45 and $\phi$$>$0.57.}. In the February 19 data, the best photometry results from an aperture radius of 14.7 pixels for both the target and the comparison stars, and sky annuli with a 52-pixel inner radius and 22-pixel wide. For the October 18 data, 17.9 pixel apertures and sky annuli with a 45-pixel inner radius and 22-pixel wide produce the best photometry.

The resultant differential light curves between the target and each comparison contain systematic trends that can be attributed to either atmospheric effects, such as airmass, seeing, or sky brightness variations, or to instrumental effects, such as small changes in the location of the stars on the detector. Systematics can also be introduced by instrumental temperature or pressure changes, but those parameters are not monitored in SPICam. We modeled systematics for each light curve by fitting linear correlations between each parameter (airmass, seeing, sky brightness variations, and target position) and the out-of-eclipse portions of the light curves. All detected trends are linear and there are no apparent residual color difference effects. The full light curves are then de-trended using those correlation fits. In the October 18 dataset, airmass effects are the dominant systematic, introducing a linear baseline trend with an amplitude in flux of 0.07\%. The February 19 dataset also shows systematics with seeing and time with a total amplitude of also 0.07\%. The systematics on this night were modeled using only the after-eclipse portion of the light curve, and we consider this dataset less reliable that the October 18 one. The 18 pre-ingress images collected between 3:00 and 3:38 UT suffer from a $\sim$50 pixel position shift with respect to the rest of the images collected that night, which cannot be modeled using overall out-of-eclipse systematics. We chose not to use those points in the final analysis. Correlations with the other parameters listed above are not significant in any of the two datasets.

\begin{figure}
\epsfig{width=\linewidth, file=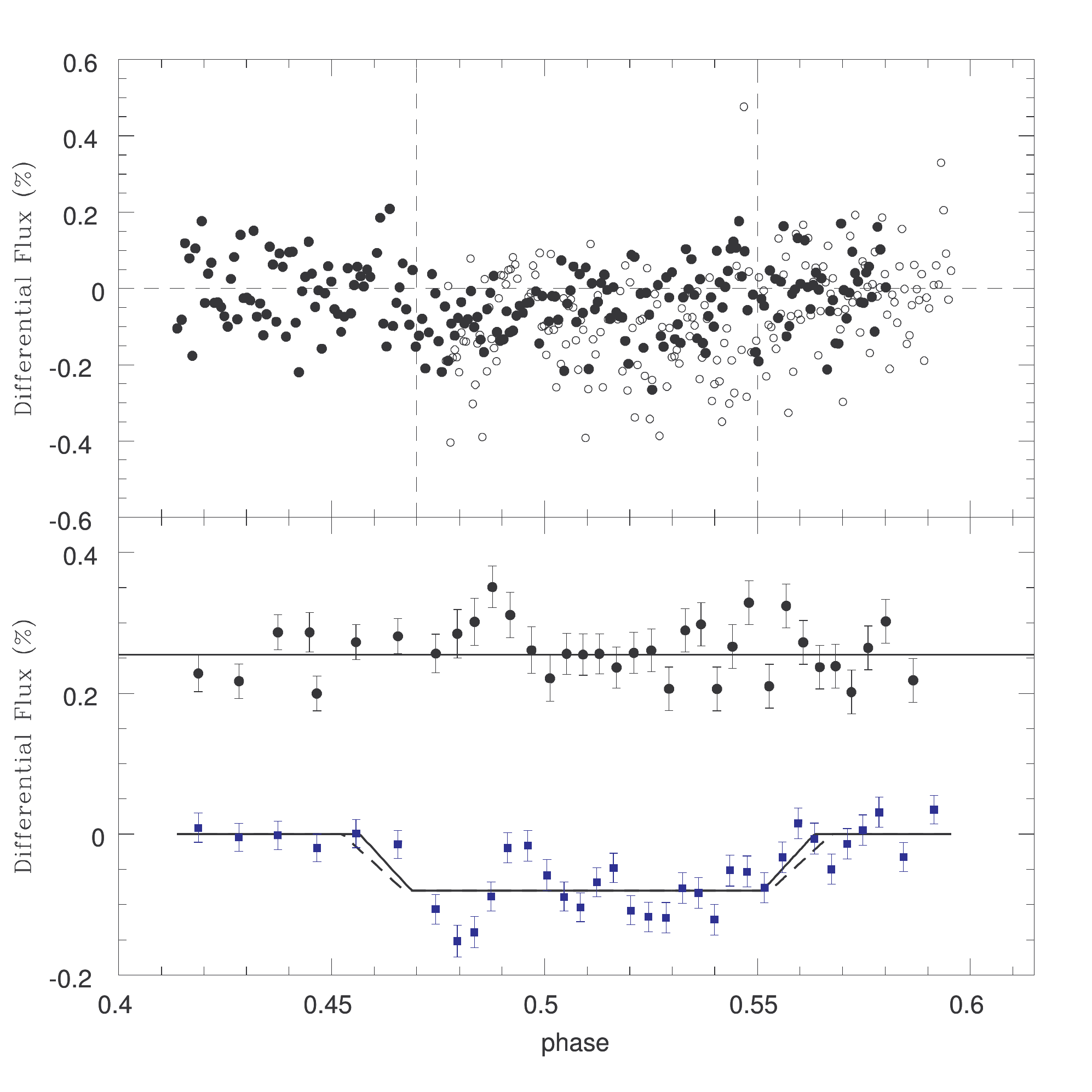}
\caption[Secondary eclipse light curve for WASP-12b]{{\it Top}: De-trended differential light curves. Open and filled dots show, respectively, the Feb 19 and Oct 18 UT 2009 data. The vertical dashed lines denote the start and end of total eclipse. {\it Bottom}: Combined light curves binned by a factor of 12. Blue squares correspond to WASP-12 and black dots to the differential light curve of the two comparison stars. The best fit models are shown as solid lines (for $e=0$) and dashed lines (for $e=0.057$). Both models produce the same depth and center phase, but the $e=0.057$ model lasts 11.52 minutes longer. We attribute the flux jumps between phases 0.475 and 0.5 to unremoved systematics. Notice that, although the systematics appear in both curves, the trends in each curve are not correlated in phase.}
\label{fig:LC}
\end{figure}

Finally, we produce one light curve per night by combining the de-trended light curves of each comparison. The light curves are combined applying a weighted average based on the Poisson noise of the individual light curve points. The result is illustrated in Figure~\ref{fig:LC}. The out-of-eclipse scatter of the combined light curves is 0.11\% for the February 19 data and 0.09\% for the October 18 data. De-trending significantly improves the systematics, but some unidentified residual noise sources remain, which we have not been able to fully model.

\subsubsection{Eclipse detection and error estimation}

The two-night combined light curve contains 421 points between phases 0.413 and 0.596, based on the \cite{Hebb2009} ephemerides. To establish the presence of the eclipse and its parameters, we fit the data to a grid of models generated using the {\it JASMINE} code, which combines the \citet{Kipping2008} and \citet{MandelAgol2002} algorithms to produce model light curves in the general case of eccentric orbits. The models do not include limb darkening, (which is not important for secondary eclipse observations), and use as input parameters the orbital period, stellar and planetary radii, argument of the periastron, orbital inclination, stellar radial velocity amplitude, and semi-major axis values derived by \citet{Hebb2009}.  The eccentricity is initially assumed to be $e$=0, which produces models with a total eclipse duration of 2.808 hours. The best fit model is found by $\chi^{2}$ minimization, with the depth, the central phase of the eclipse, and the out-of-eclipse differential flux as free parameters. 

First we fit the individual night light curves to ensure the eclipse signal is present in each dataset. The February 19 data give an eclipse depth of 0.100 $\pm$ 0.023\%, while the derived eclipse depth for the October 18 data is 0.068 $\pm$ 0.021\%. The central phases are $\phi$=0.510 for the first eclipse and $\phi$ = 0.508 for the second. We assume the difference in depth is due to systematics we have not been able to properly model. The incomplete eclipse from February 19 might seem more prone to systematics, but our inspection of both datasets does not reveal stronger trends in that dataset. We therefore combined the data from both nights, weighting each light curve based on its out-of-eclipse scatter.

The result of the combined light curve analysis is the detection of an eclipse with a depth of 0.082 $\pm$ 0.015\% and centered at orbital phase $\phi$ = 0.51, as shown in Figure~\ref{fig:LC}. The reduced $\chi^{2}$ of the fit is 0.952. The error in the eclipse depth is computed using the equation $\sigma_{depth}^2=\sigma_w^2/N+\sigma_r^2$, where $\sigma_w$ is the scatter per out-of-eclipse data point and $\sigma_r^2$ describes the red noise.  The $\sigma_r$ is estimated with the binning technique by \citet{Pont2006} to be 1.5$\times10^{-4}$ when binning on timescales up to the ingress and egress duration of about 20 minutes.

We investigate to what extent the uncertainties in the system's parameters affect our eclipse depth and central phase results. Varying the impact parameter, planet-to-star ratio, and scale of the system by 1$\sigma$ of the reported values in \cite{Hebb2009}, the measured eclipse depth changes only by 0.004\% or 0.27$\sigma_{depth}$, while the central phase remains unchanged. Our result is therefore largely independent of the adopted system parameters.

\begin{figure}[ht!]
\epsfig{width=\linewidth, file=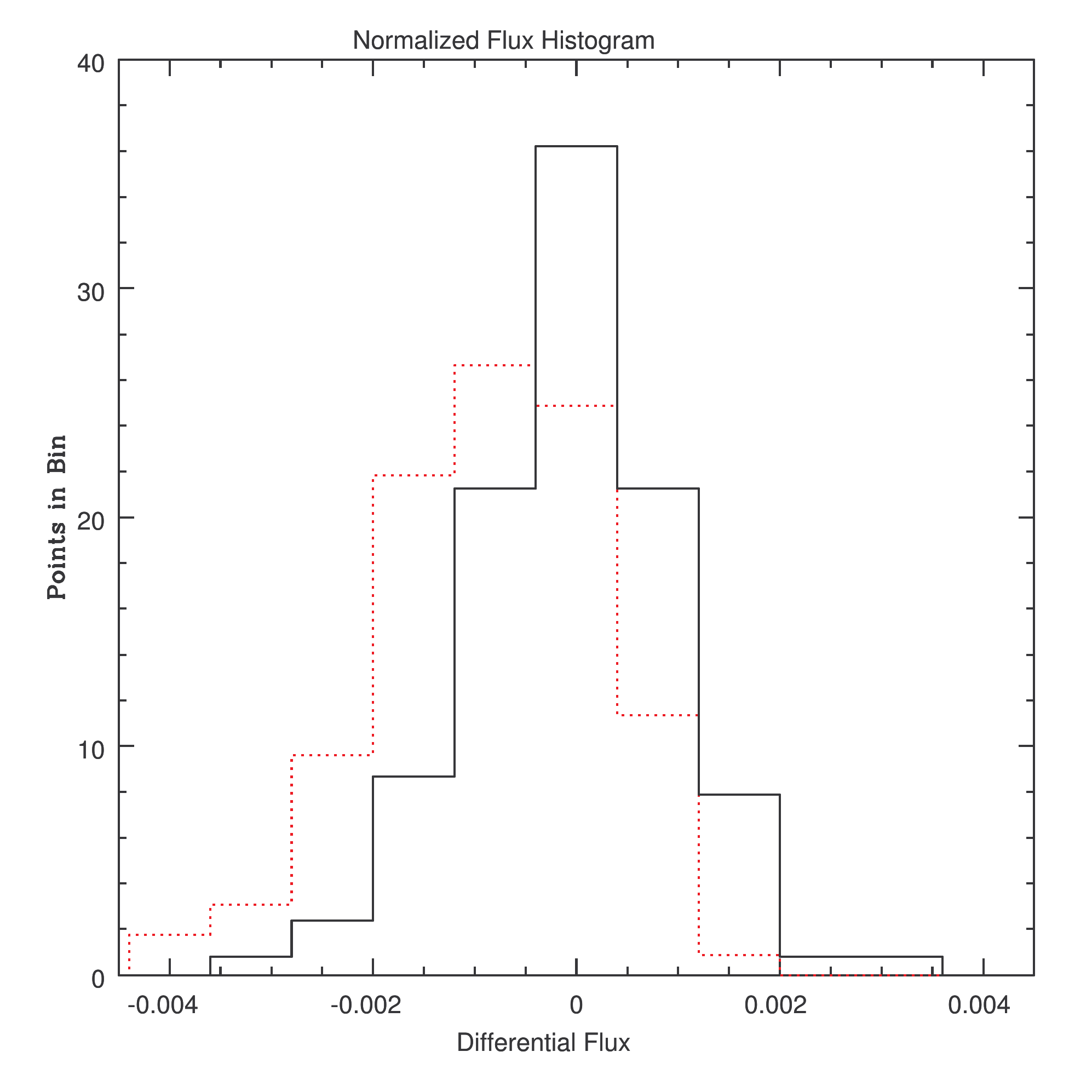}
\caption[Histograms of the in and out-of-eclipse flux distributions]{Normalized flux histograms of the in-eclipse (dotted red line) and out-of-eclipse (solid line) portions of the WASP-12 light curve in Figure~\ref{fig:LC}. The bin width is 0.00082 in differential flux, coincident with the detected eclipse depth.}
\label{fig:histo}
\end{figure}

We perform several more tests to confirm the eclipse detection in a manner similar to previously reported eclipse results \citep{Deming2005,Sing2009a,Rogers2009}. From the average of the 125 out-of-eclipse light curve data points versus the 228 in-eclipse points (only points where the planet is fully eclipsed, adopting $\phi$=0.51 as the central eclipse phase), we measure an eclipse depth of 0.080 $\pm$ 0.015\%. We further check the detection by producing histograms of the normalized light curve flux distribution in the in-eclipse and out-of-eclipse portions of the light curve. The result, illustrated in Figure~\ref{fig:histo}, shows how the flux distribution of in-eclipse points is shifted by 0.00082 with respect to the out-of-eclipse flux distribution, centered at zero. We also fit the light curve with the JKTEBOP code \citep{Southworth2004a,Southworth2004b}, and use Monte-Carlo, prayer-bead and bootstrapping analyses to estimate the errors, obtaining errors on the eclipse depth of $\pm$0.011\%, $\pm$0.008\%, and $\pm$0.011\%, and errors on the central phase of $\pm$0.0021, $\pm$0.0026, and $\pm$0.0018, for the three methods, respectively.

Adopting the largest error estimates from all of these analysis techniques, we derive final values of a depth and central phase of 0.082 $\pm$ 0.015\% (5.5$\sigma$) and 0.5100 $\pm$ 0.0026 (3.8$\sigma$).

\subsubsection{Eccentricity}

The eccentricity $e$ of WASP-12b was calculated from the measured central phase shift value using Eq. 6 from \citet{Wallenquist1950},

\begin{equation} 
e\cos\omega = \frac{\pi}{P}\frac{(t_{2}-t_{1}-P/2)}{1+\csc^{2}i},
\end{equation}

\noindent where $P$, $i$ and $\omega$ are, respectively, the orbital period, inclination, and periastron angle of the system, and $t_{2}$ - $t_{1}$ is the time difference between transit and eclipse. In our case $t_{2}$ - $t_{1}$ = 0.51$P$. Using the values of $P$, $i$ and $\omega$ from \citet{Hebb2009}, we derive an $e$ = 0.057 $\pm$ 0.015, which agrees with the non-zero eccentricity result reported by these authors. This eccentricity can be in principle explained if 1) the system is too young to have already circularized, 2) there are additional bodies in the system pumping the eccentricity of WASP-12b, 3) the tidal dissipation factor $Q^{'}_{P}$ \citep{Goldreich1963} of WASP-12b is several orders of magnitude larger than Jupiter's, estimated to be between $6\times10^{4}$ and $2\times10^{6}$ \citep{YoderPeale1981}, or 4) the orbit is really circular but there is a wavelength-dependent brightness variation across the surface of the planet that would shift the center of the eclipse, as suggested for HD 189733b by \citet{Swain2010}.

\subsection{Comparison with atmospheric models} 
\label{sec:model}

\begin{figure}[ht]
\epsfig{angle=90, width=\linewidth, file=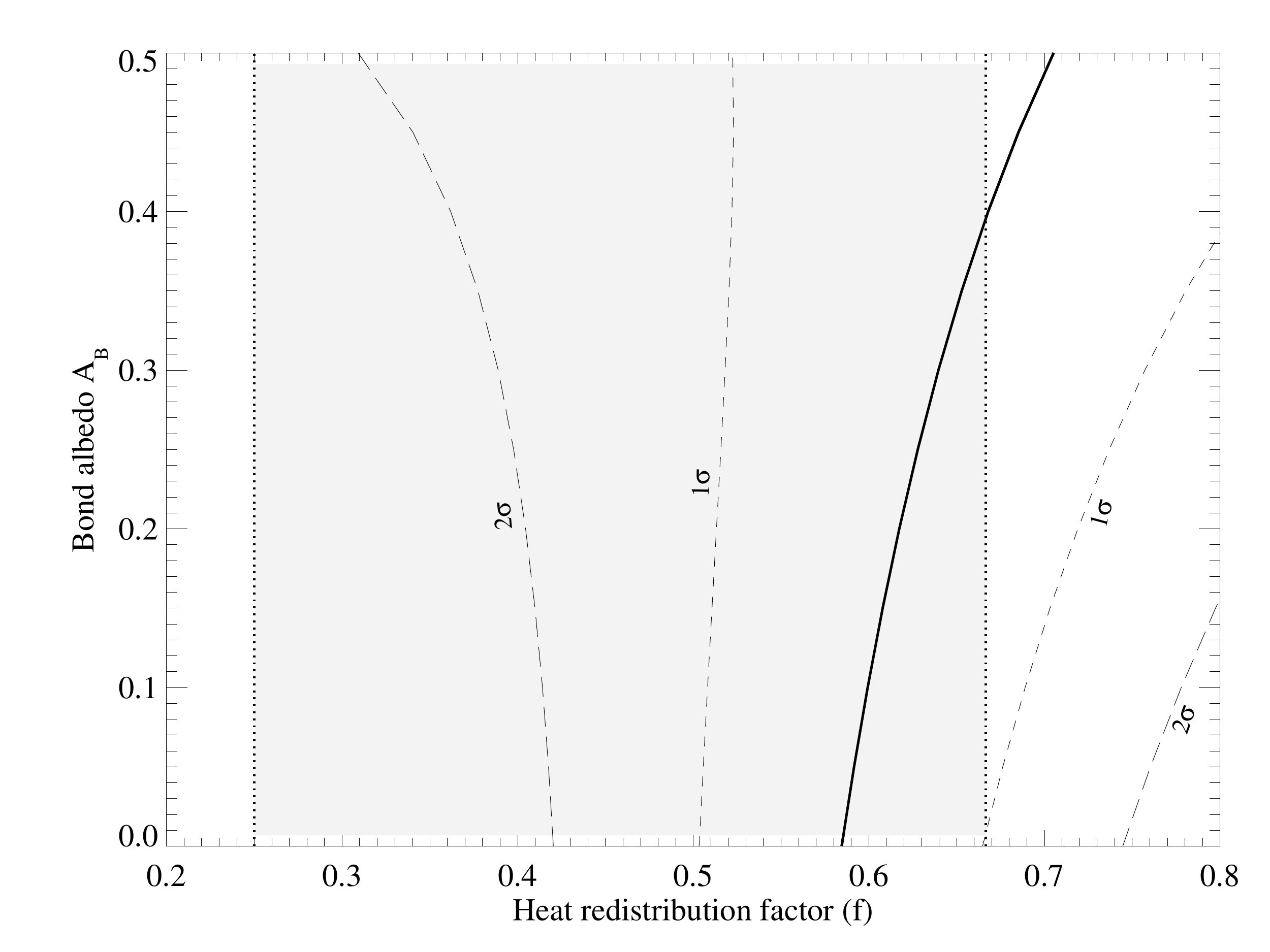}
\caption[Values of $A_{B}$ and $f$ that reproduce the observed $z'$-band eclipse depth of WASP-12b]{Values of $A_{B}$ and $f$ that reproduce the observed $z'$-band eclipse depth of WASP-12b, assuming the planet emits as a blackbody. The shaded area highlights the region of allowed $f$ values ($1/4 - 2/3$). The short and long dashed lines delimit, respectively, the $1\sigma$ and $2\sigma$ confidence regions of the result.}
\label{fig:fvsab}
\end{figure}

\begin{figure}[ht!]
\epsfig{width=\linewidth, file=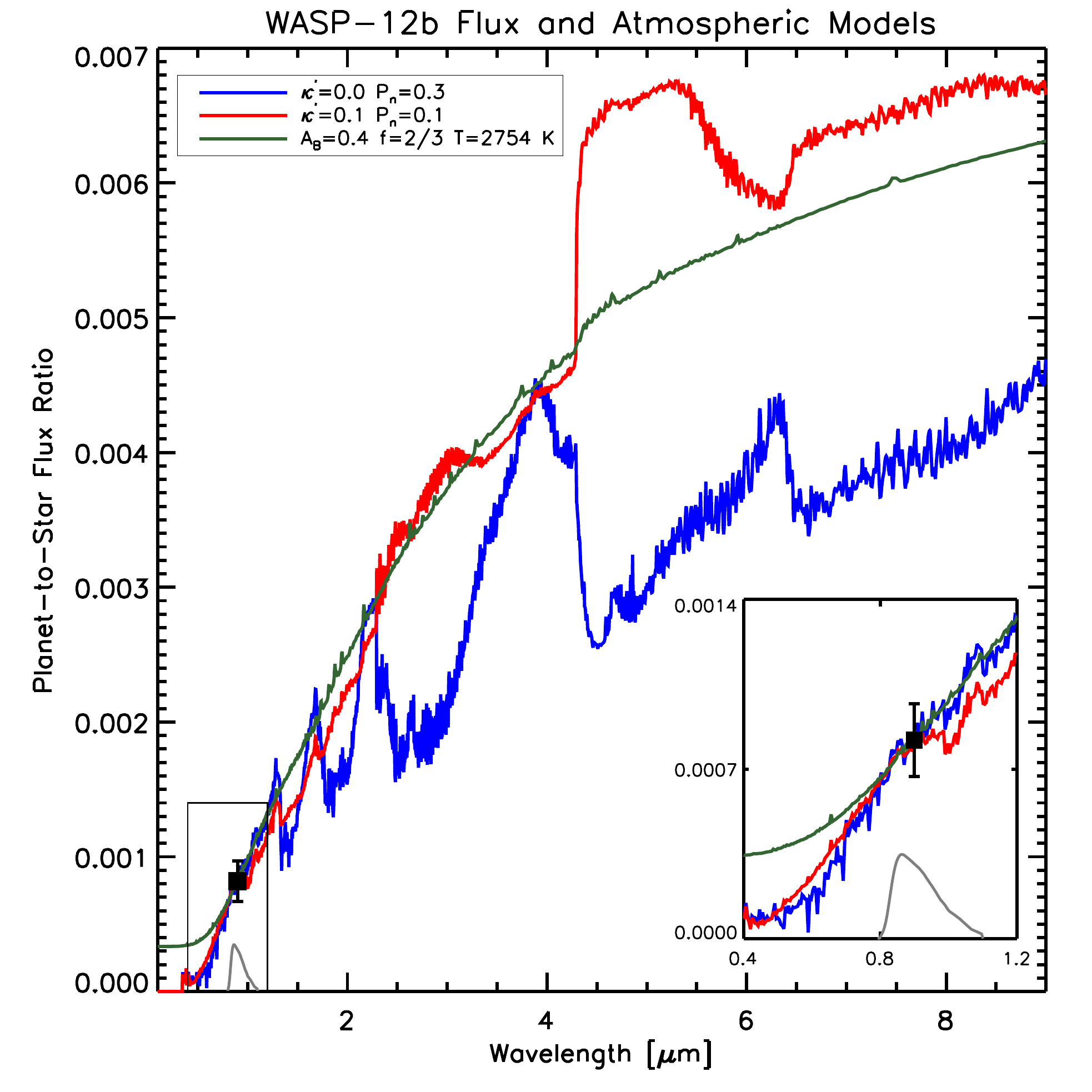}
\caption[Comparison of the eclipse depth (planet-to-star flux ratio) to models]{Comparison of the eclipse depth, (planet-to-star flux ratio), shown as the black square with errorbars, to models. The green line shows the $A_{B}$ = 0.4, $f=2/3$ blackbody model from Figure~\ref{fig:fvsab}. The blue and red lines show, respectively, the best fit model for an atmosphere with no extra absorber, and with an extra absorber of opacity $\kappa'$ between 0.43 and 1.0 $\mu$m. The black thin line at the bottom indicates the SPICam plus SDSS $z'$-band filter response. See Section~\ref{sec:model} for more details.}
\label{fig:models}
\end{figure}

We compare the observed $z'$-band flux of WASP-12b to simple blackbody models and to expectedly more realistic radiative-convective models of irradiated planetary atmospheres in chemical equilibrium, following the same procedure described in \citet{Rogers2009}. The results are shown in Figures~\ref{fig:fvsab} and \ref{fig:models}.

In the simplistic blackbody approximation, a 0.082 $\pm$ 0.015\% deep eclipse corresponds to a $z'$-band brightness temperature of $T_{z'}$ = 3028 $\pm$ 105 $K$, slightly lower than the planet's equilibrium temperature of $T_{p}$ = 3129 $K$ assuming zero Bond albedo ($A_{B}$ =0) and no energy re-radiation ($f$ = $\frac{2}{3}$) \citep[see][]{LopezMoralesSeager2007}. However, when the thermal and reflected flux of the planet are included, different combinations of $A_{B}$ and $f$ can yield the same eclipse depth, as illustrated in Figure~\ref{fig:fvsab}. From that figure we can constrain the energy redistribution factor to $f$ $\ge$ 0.585 $\pm$ 0.080, but the albedo is not well constrained.  Assuming a maximum $A_{B}$ $\le$ 0.4, the temperature of the day-side of WASP-12b is $T_{p}$ $>$ 2707 $K$. 

The more realistic atmospheric models are derived from self-consistent coupled radiative transfer and chemical equilibrium calculations, based on the models described in \citet{Sudarsky2000,Sudarsky2003}, \citet{Hubeny2003} and \citet{Burrows2005,Burrows2006,Burrows2008a} \citep[see][for details]{Rogers2009}.  We generate models with and without thermal inversion layers, by adding an unidentified optical absorber between 0.43 and 1.0 $\mu$m, with different level of opacity $\kappa'$. The opacity of the absorber varies parabolically with frequency, with a peak value of $\kappa'$ = 0.25 $cm^{2}$ $gr^{-1}$. As Figure~\ref{fig:models} shows, models with and without extra absorbers produce similar fits to the observed $z'$-band flux. The best model without absorber has a $P_{n} = 0.3.$\footnote{$P_{n} = 0$ and $P_{n} = 0.5$ correspond, respectively, to $f=2/3$ and $f=1/4$, however there is not a well-defined $P_{n}-f$ relation for intermediate values since the physical models account for atmospheric parameters (e.g. pressure, opacity) in a way different than blackbody models.} The best model with an extra absorber has a $P_{n} = 0.1$ and $\kappa_{e}$ = 0.1 $cm^{2}$ $gr^{-1}$. Observations at other wavelengths are necessary to further constrain the models.

\subsection{Discussion and Conclusions} 
\label{sec:sum}

This first detection of the eclipse of WASP-12b agrees with the slight eccentricity of the planet's orbit found by \citet{Hebb2009}, and places initial constraints to its atmospheric characteristics. We note though that detections of the secondary eclipse in the near-infrared $J$, $H$, and $K$-bands \citep{Croll2011}, and in the mid-infrared with $Spitzer$ \citep{Campo2011}, published after our detection, did not find any significant eccentricity. The most likely explanation is that a bright spot exists on the surface of the planet that is prominent in the $z'$-band, but not at longer wavelengths.

The presence of other bodies in the system can be tested via radial velocity or transit timing variation observations, although the current RV curve by \citet{Hebb2009} shows no evidence of additional planets, unless they are in very long orbits.

One would expect that if extra absorbers are present in the upper atmosphere of the planet in gaseous form, they might give rise to thermal inversion layers.  However, as Figure~\ref{fig:models} illustrates, the observed 0.9 $\mu$m eclipse depth can be fit equally well by a model without extra absorbers. Additional observations at longer wavelengths, specially longer than $\sim$4.0 $\mu$m, will break that model degeneracy. Observations at wavelengths below $\sim$0.6 $\mu$m will also better constrain $A_{B}$. Indeed, after our detection was published, \citet{Madhusudhan2011} combined our measurement with those at 1.2, 1.6, 2.1, 3.6, 4.5, 5.8, and 8 $\mu$m to determine the planet is extremely rich in methane.

%% file: chp4-kepsec.tex
\begin{singlespace}
\section{\MakeUppercase{A Uniform Search for Secondary Eclipses of Hot Jupiters in \emph{Kepler} Q2 Lightcurves}}
\label{chap4}
\end{singlespace}

\subsection{Introduction}

\label{introsec}

Measuring the secondary eclipses of transiting exoplanets at optical wavelengths is a powerful tool for probing their atmospheres, in particular their albedos, brightness temperatures, and energy redistribution factors. The $Kepler$ mission has recently uncovered over a thousand new transiting planet candidates \citep{Borucki2011}, which provide an unprecedented and uniform sample of high photometric precision light curves among which secondary eclipse signals can be detected.

In the past decade, many surprising discoveries regarding the atmospheric properties of hot Jupiters have been made. For example, many hot Jupiters appear to have temperature inversions, with numerous proposed explanations, but no definitive evidence for exactly which physical processes are involved \citep{Hubeny2003,Fortney2006,Burrows2007,Fortney2008,Spiegel2009,Zahnle2009,Knutson2010,Madhusudhan2010}. Other results have found that the atmospheric composition of different planets vary significantly, or that they present a wide range of heat circulation efficiencies between their day and night sides \citep[see][and references therein]{Baraffe2010}.

Most of the observations yielding to those discoveries have been done in the mid-infrared (3.6 - 24 $\mu$m) with $Spitzer$. Observations at shorter wavelengths are more scarce, especially in the visible, but most of them point towards the predominance of very low geometric albedo, ($A_g$ $<$ 0.3 at the 3$\sigma$ level upper limits), atmospheres in hot Jupiters \citep{Charbonneau1999,Leigh2003a,Leigh2003b,Rodler2008,Rodler2010,Rowe2008,Alonso2009a,Alonso2009b,Alonso2010,Snellen2009,Christiansen2010,Welsh2010,KippingBakos2011a,KippingSpiegel2011,Desert2011a,Desert2011b,Langford2011}, in contrast to the $A_g$ $\approx$ 0.5 albedos observed in the colder gas giants in our Solar System \citep{Karkoschka1994}. Those results are in fair agreement with early theoretical models \citep[e.g.,][]{Marley1999,Sudarsky2000,Seager2000}, which predict significant absorption of the incident stellar radiation in the visible by sodium and potassium, followed by re-emission in the infrared. Other molecules, such as TiO, VO, and HS, have also been suggested as possible strong optical absorbers \citep[e.g.,][]{Hubeny2003,Fortney2008,Zahnle2009}.

However, three recent studies suggest higher geometric albedos for two planets. \citet{Berdyugina2011} have published a value of $A_g$ = 0.28$\pm$0.16 for HD~189733b via polarized reflected light\footnote{We note that \citep{Wiktorowicz2009} reported a non-detection of polarized light from this planet, and placed an upper limit to the polarimetric modulation of the exoplanet at $\Delta$P $<$ 7.9$\times$10$^{-5}$.}, while \citet{KippingBakos2011a} and \citet{Demory2011} respectively suggest albedos of $A_g$ = 0.38$\pm$0.12 and $A_g$ = 0.32$\pm$0.03 for Kepler-7b based on measurements of the emission of the planet with $Kepler$ during secondary eclipse. Some plausible explanations for such high albedos include Rayleigh scattering and the presence of clouds or hazes in the atmospheres of those planets \citep{Demory2011}. Also, in the case of the hottest planets, some amount of thermal emission could be contributing to the measured emission levels in the reddest edge of the observed visible wavelength windows (e.g., at $\lambda \sim$ 0.8 $\mu$m).

Although new theoretical work is necessary to determine the cause of apparently high albedos in some hot Jupiters, the key answer to whether these results are typical or not relies on more observations, since the current discussions are based on a statistically insufficient sample of only three planets. The purpose of this work is to significantly increase that sample by searching for the emission of hot Jupiters among the publicly available $Kepler$ light curves of planet candidates reported by \citet{Borucki2011}. Given the photometric precision of the $Kepler$ data and the wavelength coverage of the $Kepler$ passband (0.4 - 0.9 $\mu$m), these datasets provide unprecedented quality data to detect the secondary eclipses of those planets in the visible and statistically determine the albedos of hot Jupiters. Furthermore, as \citet{Borucki2011} do no explicitly state how they modeled their light curves or obtained their parameters, a re-modeling of the data will perform an independent test on the methods they employed.

In addition to providing estimations of the planetary albedo, measuring the timing and duration of the secondary eclipse, when coupled with the primary eclipse, can directly measure the orbital eccentricity of a system \citep[e.g.,][]{Knutson2007b}. Also, if there is a significant flux contrast between the day and night side of the planet, one may be able to measure the varying amount of emitted light by the planet in the light curve, and directly measure the day-to-night contrast ratio \citep{Harrington2006,Knutson2007b}. Even a robust upper limit on the eclipse depth can narrow the range of possible planetary albedos and yield useful information on the statistics of exoplanetary albedos.

In Section~\ref{datasec} we present our target selection criteria and describe how we reprocess the $Kepler$ light curves from the pixel-level data. In Section~\ref{modelsec} we describe how we model the data using the JKTEBOP code, and obtain robust errors on all parameters while accounting for potential systematic noise. We present our derived physical parameters of both the planet candidates and their host stars in Section~\ref{paramsec}, and in Section~\ref{trendssec} we examine possible trends in our results. In Section~\ref{indivsec} we discuss individual candidates of interest, and finally in Section~\ref{concsec} we summarize our findings and examine possible future directions for the study of this sample.

\subsection{Observational Data}
\label{datasec}

The first step of our analysis consisted of selecting a set of planet candidates suitable for secondary eclipse detection among the 1,235 planet candidates published by \citet{Borucki2011}. We made a pre-selection of potentially detectable objects using the planetary and stellar parameters listed in Table 2 of \citet{Borucki2011}, choosing only those systems with $P$ $<$ 5 days and $R_p$ $>$ 0.5 $R_{J}$, after estimating that planets with longer orbital periods and smaller radii are too cool, too small, or too far away from their host star to produce deep enough secondary eclipse signals to be detectable by $Kepler$, even in the most extreme albedo conditions (i.e., $A_g$ = 1.0). Our secondary eclipse depth estimations also account for the amount of stellar irradiation, given the effective temperature of the stars reported by \citet{Borucki2011}. The result is a list of 76 candidates.

The next step consisted of an inspection of the $Kepler$ light curves of those 76 targets. The analysis of \citet{Borucki2011} uses the first four months of $Kepler$
observations, which include quarters Q0, Q1, and Q2. However, significantly discrepant systematic noise patterns exist between the light curves from different quarters, which result in additional noise when all the data are combined. Therefore, we decided to use only the data from Q2, which alone contains continuous 90-day observation coverage and is well suited for our search.

We modeled two different light curves for each target: the Presearch Data Conditioned (PDC) light curve, and our own generated light curves that we produced using the pixel-level data and our own photometric pipeline. In the remainder of the paper we refer to this second analysis as the CLM pipeline, (for \citet{Coughlin2012a}). As detailed in \citet{Jenkins2010a}, the first step in creating the PDC light curves was correcting the pixel level data for bias pattern noise, dark current, gain, non-linearity, cosmic rays, shutter smearing, pixel-to-pixel sensitivity, and other pixel-level effects. The calibrated pixels were then run through a Photometric Analysis (PA) that measures and subtracts background flux, and sums up pixels within a photometric aperture for each star, creating the PA light curve. The size of those apertures are defined such that it is supposed to maximize the mean signal-to-noise for each star. The PA light curves were then subjected to Pre-Search Data Conditioning which attempts to remove systematic effects due to temperature, focus, pointing, and other effects by correlating with ancillary engineering data. The PDC module also corrects for any sudden jumps in the data, for example due to sudden pointing changes or pixel sensitivities due to cosmic ray hits, as well as removes excess flux in the photometric aperture due to crowding.

During our inspection of the PDC data we noted that, despite the thorough analysis detailed by \citet{Jenkins2010a}, many of the PDC light curves produced by the $Kepler$ PDC pipeline still contain significant systematic trends at a level of couple percent variation, an effect that can significantly hinder the detection of secondary eclipses and phase brightness variations. Upon thorough examination of the pixel-level data, PA, and PDC light curves, we concluded that the majority of the trends correlate with and are due to the 0.1-0.5 pixel centroid position drift experienced by the majority of the target stars each quarter. This drift is principally due to Differential Velocity Aberration (DVA), where the amount of stellar aberration introduced by the spacecraft's velocity varies over the large field of view, resulting in the shifting of stellar positions on the detector as large as 0.6 pixels over a 90 day period \citep{Jenkins2010b}. Spacecraft pointing error only accounts for 0.05 pixels of the total movement \citep{Jenkins2010b}. This drift in the stellar position causes light from the wings of each star's point spread function to enter and leave the optimal photometric aperture at different rates, resulting in a flux variance of several percent over each quarter. To remove those effects we re-analyzed the $Kepler$ pixel-level data using the CLM pipeline.

The CLM photometric pipeline starts with the calibrated pixel-level data, with background flux removed from each pixel. As the majority of target stars are well-isolated, we simply summed up the flux for every pixel that was downloaded from the spacecraft for each star. We find that this removes the majority of long-term systematic variations due to DVA, usually producing light curves with significantly less systematic noise than the similarly produced PA light curves. Usually the only time summing up all the pixels produced more systematic noise is when there was significant crowding in the field by comparably bright stars, but we find this only affects a small fraction of the selected transiting planet candidates, and note that crowding can still significantly affect PA photometry as well. 

Even after minimizing the amount of light variation within the aperture, pixel-to-pixel sensitivity, both in spectral response as well as quantum efficiency, and intra-pixel variations, still produced significant systematics. We cut out areas of significant systematic variation, which principally occur around BJD 2455015 and BJD 2455065, due to a safe mode event and a pointing tweak. Then, we performed a correlation-based Principal Component Analysis (PCA) \citep{Murtagh1987} on the pixel level data, and subtract out the first three PCA components, thus removing the majority of major systematic noise, which is still principally correlated with the large position drift. We have checked and verified that this does not remove or significantly modify the transit signal for any system. We then fit a B\'ezier curve \citep{Kahaner1989} to the data, performed a 3-sigma rejection, re-fit a B\'ezier curve, and then divided by the fit. This iterative procedure does an excellent job of removing any possible remaining low-frequency systematic features in the data without removing or affecting the transits or real high-frequency stellar variations. We then, for only a few systems, removed one or two points that were extremely significant outliers. For one system, KOI 433.01, we removed a single transit that was clearly from another long-period companion in the system. For both the $Kepler$ PDC and CLM light curves, we finally subtracted a linear trend to ensure the light curves were completely flat before modeling (see Section~\ref{modelsec}).

We would like to note that, as a possible technique of removing systematic noise, we also attempted to directly solve for a pixel mask that would account for pixel-to-pixel variation in quantum efficiency and spectral response. Every image in the time series was multiplied by this pixel mask, whose values ranged from 0 to 1, and then the pixel fluxes summed to produce a corrected light curve. We used an asexual genetic algorithm, similar to that presented in \citet{Coughlin2011}, to solve for the values of each pixel in the pixel mask that produced corrected light curves with a minimum amount of systematic noise, defined via various methods. We found that the technique was very successful at removing nearly all systematic noise from the light curves. However, depending on the minimization criteria selected, we found that the algorithm was prone to over-correct the light curve, and remove features due to real astrophysical phenomena. As well, even when it did appear to remove the systematic trends and not the real astrophysical signatures, it was difficult to tell, unlike with the PCA analysis, whether or not the solution had a real physical basis. Thus, we decided not to employ this technique in our analysis. However, with more work or a better understanding of the systematic noise sources, it might become a viable means for removing systematic noise from $Kepler$, and possibly other, light curves.

\begin{figure}
\centering
\epsfig{width=0.75\linewidth,file=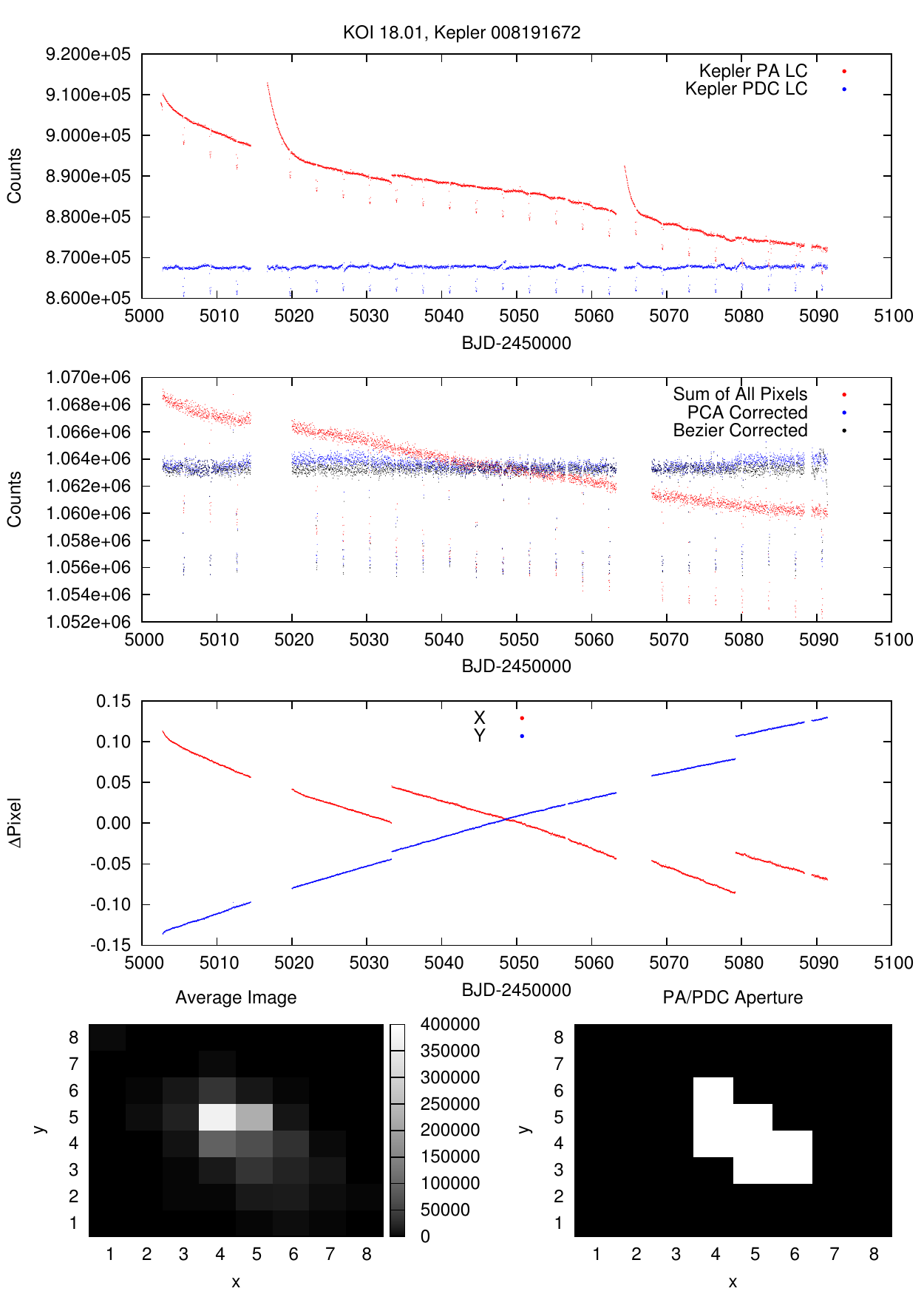} 
\caption[Plots of the light curves, centroid positions, pixel-level images, and photometric apertures used in the $Kepler$ PA and PDC reduction, for all initial 76 candidate systems]{\small Plots of the light curves, centroid positions, pixel-level images, and photometric apertures used in the $Kepler$ PA and PDC reduction, for all initial 76 candidate systems. The top panel for each system shows the Q2 $Kepler$ PA and PDC light curves. The next panel shows our CLM pipeline reduced light curves, including those from simply summing all the pixels in each frame, applying PCA correction, and then applying a Bezier correction, with the most severe systematic noise regions cut out. The next panel shows the flux-weighted relative centroid movement in both X and Y over Q2, using all pixels in the frame, again with the most severe systematic noise regions cut out. The bottom left panel is the average image of all the frames over the quarter. The bottom right panel shows the photometric aperture used in the $Kepler$ PA and PDC light curve reduction, where only white pixels were counted and summed. Only a single plot, Figure~\ref{candlcs}.7, is shown in the text for guidance. Figures~\ref{candlcs}.1-\ref{candlcs}.76 are available in the online version of the Astronomical Journal under \citet{Coughlin2012a}.}
\label{candlcs}
\end{figure}

In Figure~\ref{candlcs} we plot the $Kepler$ PA and PDC light curves, our CLM light curves, (including those from simply summing up all the pixels in each frame, applying the PCA correction, and then applying the Bezier correction), the centroid positions, averaged pixel-level image, and the aperture used in the $Kepler$ PA and PDC photometry, for each of the 76 candidate systems. Of the original 76 candidates, 36 were deemed unmodelable based on their PDC light curves, due to either strong systematics or intrinsic stellar variability with amplitudes on the order of, or greater than, the depth of the transits. The 36 discarded systems were KOI 1.01, 17.01, 20.01, 127.01, 128.01, 135.01, 183.01, 194.01, 203.01, 208.01, 214.01, 217.01, 254.01, 256.01, 552.01, 554.01, 609.01, 667.01, 767.01, 823.01, 882.01, 883.01, 895.01, 981.01, 1152.01, 1176.01, 1177.01, 1227.01, 1285.01, 1382.01, 1448.01, 1452.01, 1540.01, 1541.01, 1543.01, and 1546.01. In the case of our newly generated CLM pipeline light curves, 26 candidates turned out to be unmodelable, most of them due to stellar variability, as in the case of the PDC light curves above, or due to blends in the images resulting in significant light contamination of the target light curves. The 26 systems discarded in this case were KOI 102.01, 135.01, 194.01, 199.01, 208.01, 256.01, 552.01, 554.01, 609.01, 823.01, 882.01, 883.01, 895.01, 931.01, 961.02, 961.03, 981.01, 1152.01, 1177.01, 1227.01, 1285.01, 1382.01, 1448.01, 1452.01, 1540.01, and 1546.01. Thus, in total, there are 21 targets that have no modelable light curve from either analysis, (nearly all due to intrinsic stellar variability), 35 systems that have modelable light curves from both the $Kepler$ PDC data and our CLM reduction, and 55 systems that have at least one modelable light curve from either the $Kepler$ PDC data or our CLM analysis. In Table~\ref{kepsec-tab1} we present the $Kepler$ Object of Interest (KOI) number, $Kepler$ ID number, and host stars' $Kepler$ magnitude, effective temperature, surface gravity, and metallicity from the $Kepler$ Input Catalog of each of the 55 modelable candidates.

\begin{deluxetable}{rccccrcccc}
  \rotate
  \tablewidth{0pt}
  \tabletypesize{\scriptsize}
  \tablecaption{Modeled Systems and Their Host Star Properties}
  \tablecolumns{10}
  \tablehead{KOI & \emph{Kepler} ID & m$_{\rm kep}$ & $T_{\star}$ & $log~g$ & [Fe/H] & $M_{\star,\rm KIC}$ & $R_{\star, \rm KIC}$ & $M_{\star,\rm ISO}$ & $R_{\star, \rm ISO}$\\ & & & (K) & (cgs) & & ($M_{\sun}$) & ($R_{\sun}$) & ($M_{\sun}$) & ($R_{\sun}$)}
  \startdata
  \input{kepsec-tab1.tex}
  \enddata
  \label{kepsec-tab1}
\end{deluxetable}

We now highlight a few systems to illustrate the different types of systematic and stellar noise in the $Kepler$ data, and differences between the PDC and CLM light curves. For KOI 17.01, (see Figure~\ref{candlcs}.6), the PA and PDC light curves both show a $\sim$1.7\% systematic variation over the quarter, as the PA/PDC aperture does not encompass many pixels in the wing of the PSF that contain significant signal, and the star experiences a $\sim$0.4 pixel drift over the quarter. In contrast, the raw pixel-summed CLM light curve shows only a $\sim$0.54\% systematic variation, and the PCA-corrected and Bezier-corrected CLM light curves show virtually no remaining systematic noise. For KOI 102.01 and 199.01, (see Figures~\ref{candlcs}.11 and \ref{candlcs}.22), there is a close companion star that causes the CLM photometry to produce much worse light curves than the PDC data. In the case of KOI 102.01, significant systematics are introduced in the CLM light curve from the movement of the companion in and out of the frame, given the $\sim$0.3 pixel drift over the quarter, and as well the extra third light causes the transits of the primary to be damped out. In the case of KOI 199.01, the companion is an eclipsing binary, and its light curve is imposed on top of that of the transiting system. Significant systematics are still present in the PA data in these two cases, but much less so than the CLM data, and they appear to be removed in the PDC data. In the cases of KOI 256.01 and KOI 1452.01, (see Figures~\ref{candlcs}.32 and \ref{candlcs}.71), the stars exhibit clear high-frequency variations at the same level as the transits, possibly due to stellar pulsation or rapid rotation and star spots. Note that the CLM pipeline does not remove the stellar signal because it is high-frequency and intrinsic to the system, and also that for KOI 256.01 additional long-term systematic noise is present in the $Kepler$ PA and PDC data due to the small aperture they employ and the $\sim$0.35 pixel drift, but does not exist in the CLM data.

\subsection{Light Curve Modeling}
\label{modelsec}

We used the JKTEBOP eclipsing binary modeling code \citep{Southworth2004a,Southworth2004b}, which is based on the EBOP code \citep{Etzel1981,Popper1981}, to model both the $Kepler$ PDC light curves and our own CLM pipeline light curves for the 55 modelable systems. In short, JKTEBOP\footnote{For more information on JKTEBOP, see http://www.astro.keele.ac.uk/jkt/codes/jktebop.html} models the projection of each star as a biaxial ellipsoid and calculates light curves by numerical integration of concentric ellipses over each star, and is well-suited to modeling detached eclipsing binaries or transiting extrasolar planets. We modeled each light curve first fixing the eccentricity to $e=0$, and then leaving it as a free parameter. The reason for leaving $e$ as a free parameter is that, even though systems with $P$ $<$ 5 days are generally expected to be circularized, additional bodies in the system or other evolutionary effects can perturb their orbits. Indeed, at the time of this writing, $\sim$36\% of currently known transiting planets with $P$ $<$ 5 days have a measured non-zero eccentricity \citep{Schneider2012}. Therefore, since we are performing a blind search for secondary eclipses, restricting the search to only circular orbits might result in detection biases. The results between fixing $e=0$ and letting it vary can sometimes vary significantly, as shown at the end of this section.

For both cases of either fixing $e=0$ or letting it vary, we also simultaneously solved for the orbital period of the system, $P$, time of primary transit minimum, $T_{0}$, the inclination of the orbit, $i$, e$\cdot$cos($\omega$) and e$\cdot$sin($\omega$), where $\omega$ is the longitude of periastron, the planet-to-star surface brightness ratio, $J$, the sum of the fractional radii, $r_{sum}$, the planet-to-star radii ratio, $k$, and the out-of-eclipse (baseline) flux. We note for clarity that the relation between $J$, $k$, and the planet-to-star luminosity ratio, $L_{r}$, is $L_{r}$ = $k^{2}J$. To account for any potential brightness variations with phase, we also multiply the planet's luminosity, $L_{p}$, by a factor of one plus a sinusoidal curve, so that

\begin{equation}
  L_{p}(T) = L_{p} + A_{L_{p}} \cdot sin\left(\frac{2\pi(T-T_{0})}{P} - \frac{\pi}{2}\right)
\end{equation}

\noindent where $L_{p}(T)$ is the planet's luminosity at a given observed time, $T$, and $A_{L_{p}}$, for which we solve, is the amplitude of the sinusoidal curve. Note that we have fixed the period of this sine wave to the orbital period of each system, and fixed the reference zero phase so that the maximum amplitude peak coincides with the center of the secondary eclipse. Although there has been at least one case of a measured planetary brightness phase curve having its maximum offset from secondary eclipse in the infrared \citep{Knutson2007b}, many optical observations indicate planetary brightness phase curve maxima coincident with the secondary eclipse \citep{Borucki2009,Snellen2009,Welsh2010,Bonomo2011}. We note that a value of $A_{L_{p}}$ = 0.0 implies no brightness variations with phase. A value of $A_{L_{p}}$ = 0.2 implies the planet is 20\% brighter at phase 0.5, when the day-side is visible, and 20\% fainter at phase 1.0, when the night-side is visible, compared to phases 0.25 and 0.75. A value of $A_{L_{p}}$ = 1.0 implies a perfectly dark night-side. Negative values of $A_{L_{p}}$ would imply a brighter night-side than day-side, which is not physically expected, but allowed for in the code so as not to introduce any bias towards positive values of $A_{L_{p}}$. Note also that $J$ is allowed to be both positive and negative so as not to introduce any bias towards positive values of $J$, and thus false detections. 

In both cases we assumed a quadratic limb-darkening law for the stars, and fixed coefficients to the values found by \citet{Sing2010} for the $Kepler$ bandpass, using the estimated stellar effective temperatures, surface gravities, and metallicities. We also set the values of the gravity darkening coefficients to those derived by \citet{Claret2000a}, based on the effective temperature of the stars. Even though we fix the limb and gravity darkening coefficients, they are computed from stellar models and have an associated uncertainty when compared to reality. As the choice of these coefficients can affect the determination of other system parameters, their uncertainty must be taken into account in the error analysis. \citet{Claret2008} determined this uncertainty to be $\sim$10\%, and thus we allowed the values of the limb and gravity darkening coefficients to vary over a range of $\pm$10\% during the error estimation analysis, described below.

Finally, as pointed out by \citet{Kipping2010a}, sparse sampling times, such as the 29.4244 minute sampling of the $Kepler$ long-cadence data, can significantly alter the morphological shape of a transit light curve and result in erroneous planetary parameter estimations if the effect is not taken into account. Thus, we instructed JKTEBOP to integrate the models over 29.4244 minutes, composed of 10 separate sub-intervals centered on each observed data point, to account for this effect. 

We derived error estimates using three error analysis techniques implemented in JKTEBOP: Monte Carlo, Bootstrapping, and Residual Permutation, but chose to adopt the parameter errors estimated by this last technique as it has been shown to best account for the effect of systematic noise in transit light curves \citep{Jenkins2002}. While Monte Carlo and Bootstrapping tend to underestimate errors in the presence of systematic noise, those two techniques have traditionally been chosen over Residual Permutation because in the latter one can only refit the data as many times as available data points. This poses a problem for most ground based transit light curves, which typically have only a couple hundred points, but for $Kepler$ Q2 data, which contains almost 5000 data points over a 90-day interval for the long-cadence data, and nearly 30 times more for the short-cadence data, the method is not statistically limited and therefore best suited to derive robust errors.

\begin{figure}
\centering
\epsfig{width=0.85\linewidth,file=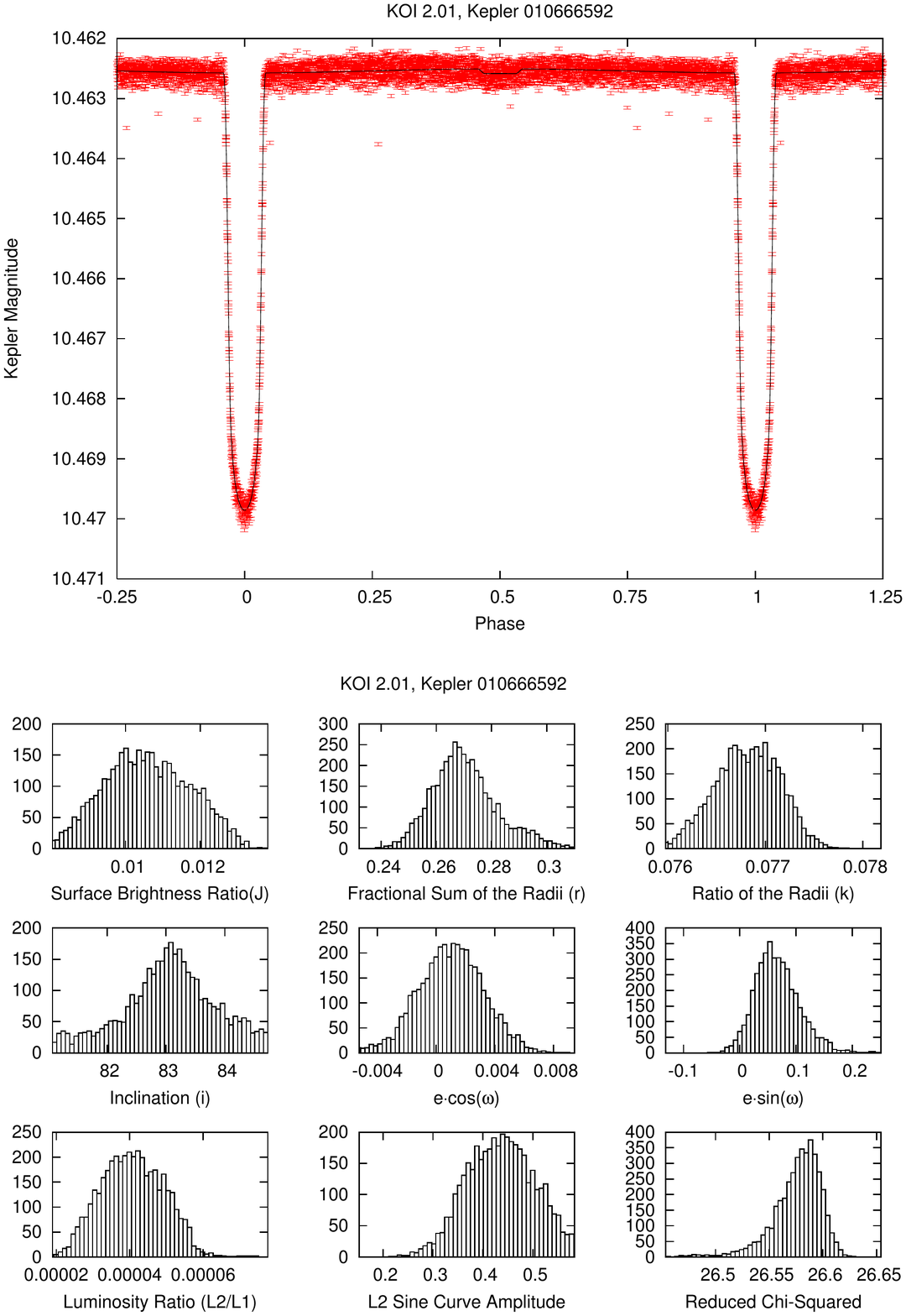}
\caption[Plots of the phased light curves of the 40 systems produced from the $Kepler$ PDC photometric pipeline]{Plots of the phased light curves of the 40 systems produced from the $Kepler$ PDC photometric pipeline, shown with our best model fits, allowing eccentricity to vary, and histograms of the resulting parameter distributions from the error analysis. Only the first plot, Figure~\ref{keplcs}.1, is shown in the text for guidance. Figures~\ref{keplcs}.1-\ref{keplcs}.40 are available in the online version of the Astronomical Journal under \citet{Coughlin2012a}.}
\label{keplcs}
\end{figure}

\begin{figure}
\centering
\epsfig{width=0.85\linewidth,file=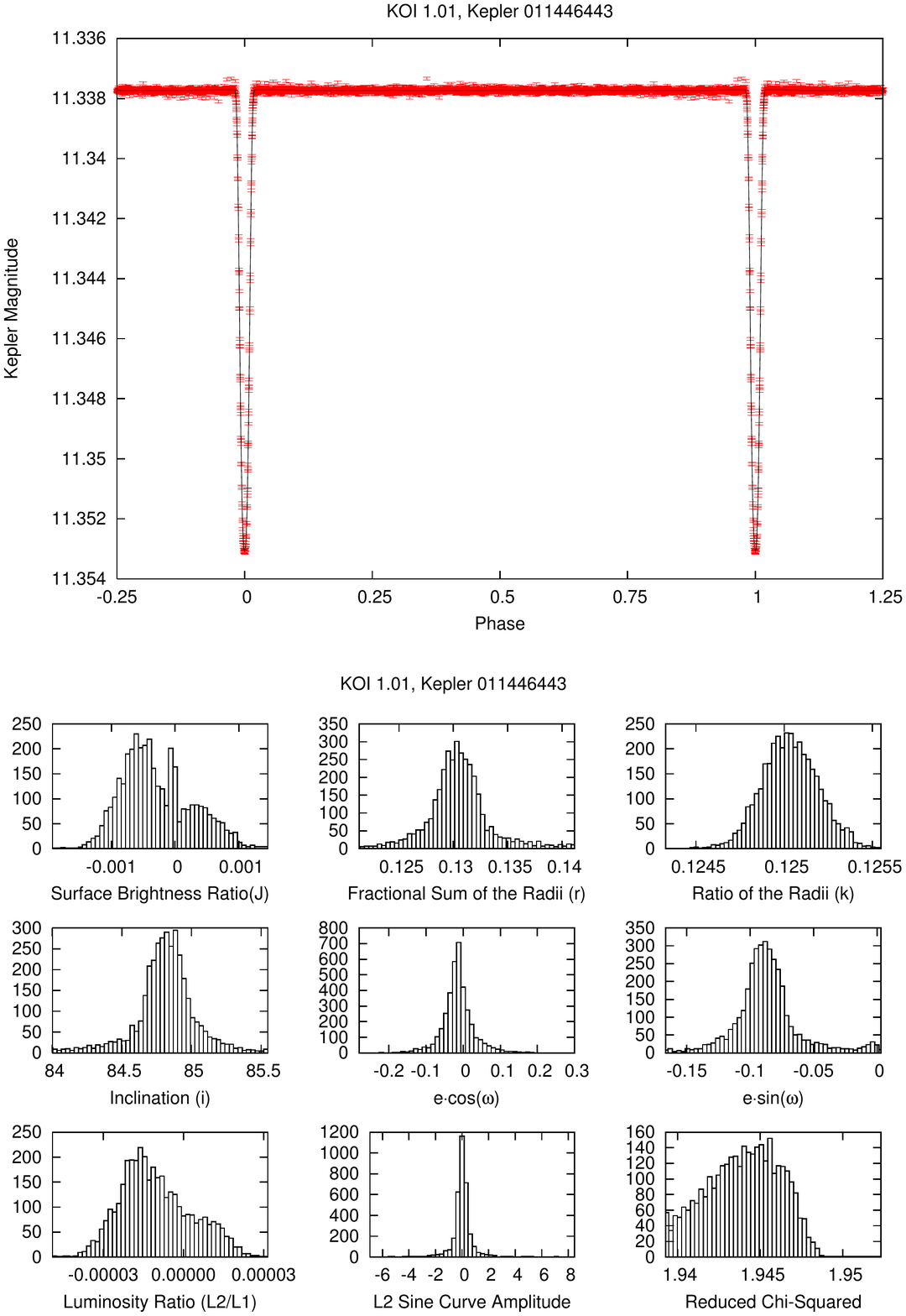}
\caption[Plots of the phased light curves of the 50 systems produced using our CLM photometric pipeline]{Plots of the phased light curves of the 50 systems produced using our CLM photometric pipeline, shown with our best model fits, allowing eccentricity to vary, and histograms of the resulting parameter distributions from the error analysis. Only the first plot, Figure~\ref{ourlcs}.1, is shown in the text for guidance. Figures~\ref{ourlcs}.1-\ref{ourlcs}.50 are available in the online version of the Astronomical Journal under \citet{Coughlin2012a}.}
\label{ourlcs}
\end{figure}

In Figure~\ref{keplcs} we plot the resulting phased light curves, with the corresponding best-fit model light curve when allowing eccentricity to vary, along with histograms of the parameter distributions from the error analysis, for the 40 modelable systems with $Kepler$ PDC light curves. In Figure~\ref{ourlcs} we do the same, but for the 50 modelable systems with CLM light curves. In Table~\ref{modelresultstab} we list the median values for all the modeling parameters, for both sets of light curves, and for both fixing $e=0$ and allowing it to vary, along with their determined asymmetric, Gaussian, 1$\sigma$ errors.

\begin{deluxetable}{rcccccccccc}
  \rotate
  \tablewidth{0pt}
  \tabletypesize{\scriptsize}
  \setlength{\tabcolsep}{5pt} 
  \tablecaption{Modeling Results: Median Values and Associated 1$\sigma$ Uncertainties}
  \tablecolumns{11}
  \tablehead{KOI & $J$ & $r_{\rm sum}$ & $k$ & $i$ & $e$cos$w$ & $e$sin$w$ & $P$ & $T_{0}$ & $A_{L_{p}}$ & $\chi^{2}_{red}$\\ & & & & ($\degr$) & & & (Days) & (BJD-2450000) & & }  \startdata
  \cutinhead{PDC Light Curve With Eccentricity Fixed to Zero}
  \input{kepsec-tab2a.tex}
\enddata
  \label{modelresultstab}
\end{deluxetable}

As a result of our light curve modeling we find, when fixing $e=0$, nine secondary eclipse detections at the 1-2$\sigma$ level, three detections at the 2-3$\sigma$ level and four detections at the $>$3$\sigma$ level in the PDC light curves. In the CLM light curves, we find 11 secondary eclipse detections at the 1-2$\sigma$ level, four detections at the 2-3$\sigma$ level and four detections at the $>$3$\sigma$ level. In the case of allowing eccentricity to vary we find 18 detections at 1-2$\sigma$ level, three detections at 2-3$\sigma$ level, and four detections at $>$3$\sigma$ level in the PDC light curves. In the CLM light curves we find 10 detections at 1-2$\sigma$ level, 10 detections at 2-3$\sigma$ level, and five detections at $>$3$\sigma$ level. Each set of results has been used independently in the statistical study of candidate emission parameters described in Sections~\ref{paramsec} and \ref{trendssec}. Examining both sets of light curves, and both $e=0$ and $e$ allowed to vary, we find 16 systems with 1-2$\sigma$, 14 systems with 2-3$\sigma$, and six systems with $>$3$\sigma$ confidence level secondary eclipse detections in at least one light curve. It is more difficult to quantify the number of systems that have certain level detections among multiple light curves and eccentricity constraints, given that not all systems had both PDC and CLM light curves and that eccentric systems may not be detected in the non-eccentric model, and is best left to the discussion of individual systems in Section~\ref{indivsec}. Additionally, examining the 35 systems that had both modelable PDC and CLM light curves, we find that for the PDC light curves the average reduced $\chi^{2}$ value is 4.86, while for the CLM light curves it is 2.48, and that on average, each system's CLM light curve has a 27\% lower reduced $\chi^{2}$ value compared to the PDC light curve.

We also note that there were significant detections of negative $J$ values for some systems. When fixing eccentricity to zero, for the PDC light curves, we find three detections of negative $J$ at the 1-2$\sigma$ level, but none at higher significance. In the CLM light curves, we find four detections of negative $J$ at the 1-2$\sigma$ level, but none at higher significance. In the case of allowing eccentricity to vary, for the PDC light curves, we find four detections of negative $J$ at the 1-2$\sigma$ level, and one detection at the 2-3$\sigma$ level, but none at higher significance. In the CLM light curves we find four detections of negative $J$ at the 1-2$\sigma$ level, one detection at the 2-3$\sigma$ level, and one detection at the $>$3$\sigma$ level. Since there is no known physical mechanism to increase the flux of the system when the planet passes behind the host star, these detections are obviously spurious. Since there is no bias towards or preference for positive or negative $J$ values in the modeling code, and assuming that the $Kepler$ data does not suffer from systematics that preferentially result in either decrements or increments in flux that span expected secondary eclipse durations, statistically speaking we must have as many false detections of positive $J$ values, or secondary eclipses, for as many detections of negative $J$ values we have, per each confidence interval. Thus, when fixing eccentricity to zero, for PDC light curves, we estimate our false alarm probabilities as 33\% for 1-2$\sigma$ detections, and 0\% for $>$2$\sigma$ detections. For CLM light curves, we estimate a 36\% false alarm probability for 1-2$\sigma$ detections, and 0\% for $>$2$\sigma$ detections. When allowing eccentricity to vary, for PDC light curves we estimate false alarm probabilities of 22\% for 1-2$\sigma$ detections, 33\% for 2-3$\sigma$ detections, and 0\% for $>$3$\sigma$ detections. For the CLM light curves, we estimate a 40\% false alarm probability for 1-2$\sigma$ detections, 10\% for 2-3$\sigma$ detections, and 20\% for $>$3$\sigma$ detections. Although we are dealing with small number statistics and the uncertainties on the determined false alarm probabilities are large, we note that these roughly agree with what we would statistically expect for each confidence interval quoted, i.e., a 1$\sigma$ detection has a formal 31.73\% false alarm probability by definition, though allowing eccentricity to vary does appear to induce false detections at $\sim$1.5 times greater frequency. Combining all the results from each light curve type and eccentricity parameter, we can generalize our false alarm probabilities to 31\%, 10\%, and 6\% for the 1-2$\sigma$, 2-3$\sigma$, and $>$3$\sigma$ confidence intervals respectively.

\subsection{Derivation of Stellar and Planetary Parameters}
\label{paramsec}

The secondary eclipse detections presented in the previous section allow, for the first time, to make a statistically significant analysis of the emission properties of exoplanet candidates at visible wavelengths, specifically in the $\sim$0.4 - 0.9 $\mu m$ $Kepler$ passband.

In this section we first revise the parameters of the host stars necessary to derive the physical properties of the planets and then compute physical and atmospheric parameters for each planet candidate, i.e., the brightness, equilibrium, and maximum effective temperatures, radii, and albedos, using both the originally reported and revised stellar parameter values. A detailed statistical study of the properties for the planets based on those parameters is presented in Section~\ref{trendssec}.

\subsubsection{Stellar Parameters}

The $Kepler$ Input Catalog (KIC) provides estimates of the effective temperature, surface gravity, and radius for all the host stars in our sample. Those parameters have been derived from a combination of broad and narrow-band photometry \citep{Brown2011}, although it has been also recognized that some of the parameter values in the KIC might contain significant errors.  As explained in detail by \citet{Brown2011}, the majority of approximately Sun-like stars in the KIC have effective temperatures that only disagree by 200 K or less from the temperature values of a control sample derived by other methods. However, for stars significantly more massive or less massive than the Sun, i.e., with $T_{\star} \gtrsim$ 9000 K and $T_{\star} \lesssim$ 4000 K, where $T_{\star}$ is the effective temperature of the star, the temperatures in the KIC can suffer from large systematics and are not reliable. As well, in the case of the derived stellar radii, $R_{\star}$, the values reported in the KIC are derived from statistical relations between the values obtained for $T_{\star}$, the surface gravity of the stars, $log~g$, and the luminosity, $L_{\star}$ (see sections 7 and 8 of \citet{Brown2011}, for more details). Therefore, if any of those parameters are systematically off, (such as $log~g$, which has an associated error of $\pm$0.4 dex), the values derived for $R_{\star}$ will be erroneous.

All the host stars in our sample have KIC effective temperature estimates between 4000 and 9000 K, so we have assumed that those values are accurate within the errors. From those temperatures we recomputed the radius and mass of the stars via interpolation of up-to-date stellar evolutionary models by \citet{Bertelli2008} for $M_{\star} \leq 1.4 M_{\sun}$ and \citet{Siess2000} for $M_{\star} >$ 1.4 $M_{\sun}$. In the models we have assumed that all the stars are nearly coeval, with an age of $\sim$1 Gyr and therefore on the main sequence, have abundances similar to the Sun, i.e., Z = 0.017, Y = 0.26, and have a mixing length of $\alpha$ = 1.68. 

The errors in those parameters have been estimated by recomputing the mass and radius of each star 10,000 times, each time adding random Gaussian noise to the underlying variables and examining the 1$\sigma$ spread of the resulting distribution. In the error estimations using the KIC values, we assumed an error of $\pm$0.4 dex for $log~g$, as reported by \citet{Brown2011}. In the error estimations using the model isochrones we assume an error of $\pm$200 K for $T_{\star}$, as reported by \citet{Brown2011}. The values for the mass and radius of each star as computed from the KIC, labeled ``KIC'', and via interpolation of the stellar isochrones, labeled ``ISO'', along with their estimated errors, are listed in Table~\ref{kepsec-tab1}.

\subsubsection{Planetary Parameters}

\begin{figure}
\centering
\begin{tabular}{c}
  \epsfig{width=0.75\linewidth,file=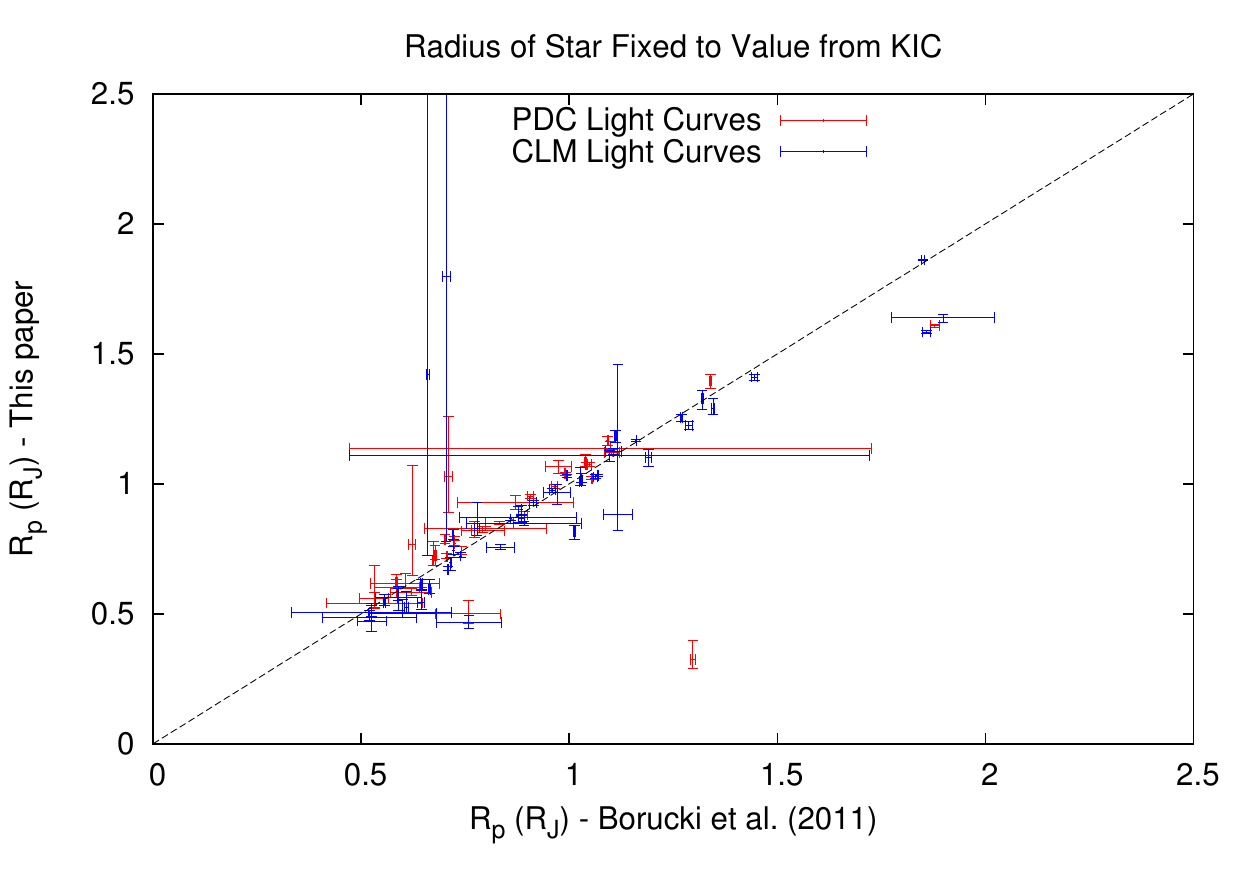}\\
  \epsfig{width=0.75\linewidth,file=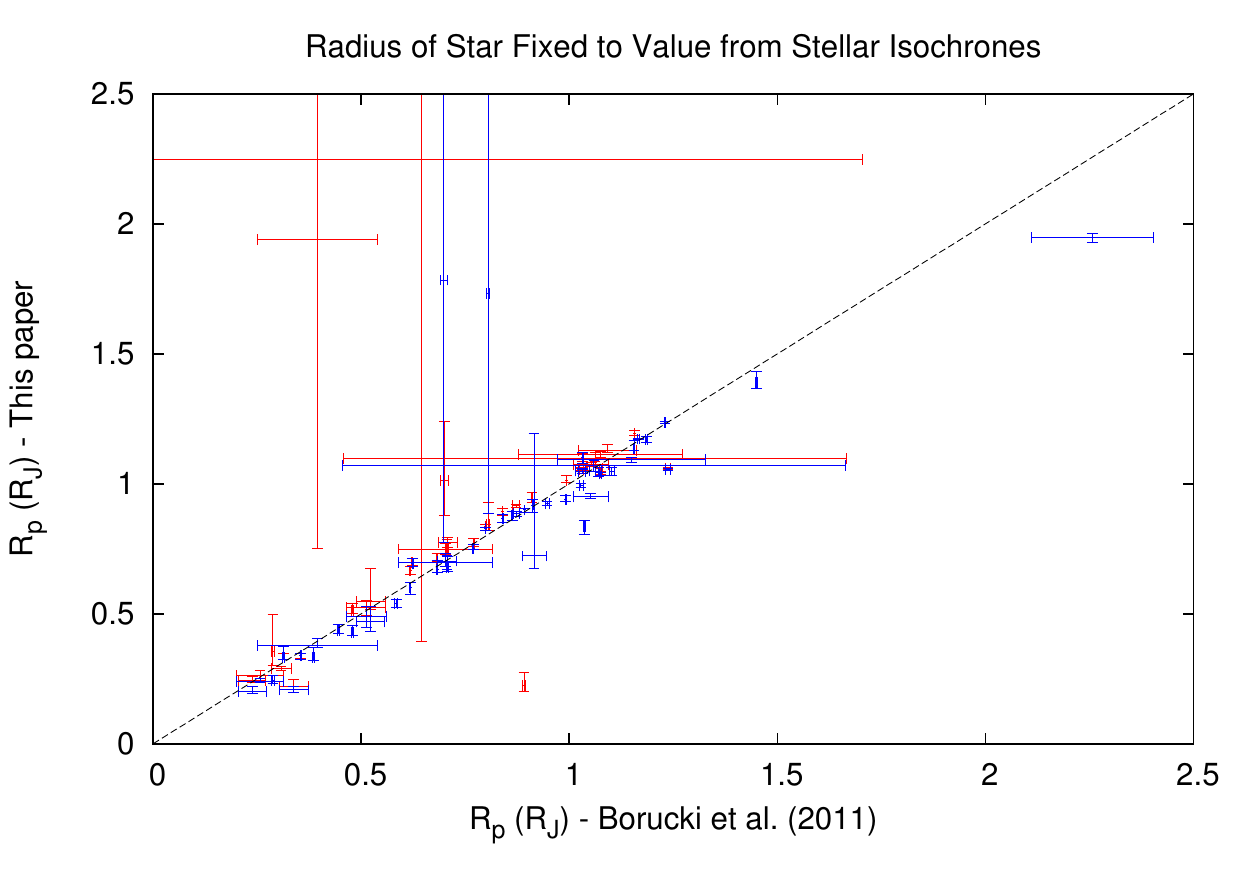}
\end{tabular}
\caption[Comparison of the values for the planetary radius as given by \citet{Borucki2011} to those we derived]{Comparison of the values for the planetary radius as given by \citet{Borucki2011} and derived in this paper. Red and blue symbols correspond, respectively, to the results from the PDC and CLM light curves. The errorbars are computed assuming a fixed stellar radius, but taking into account the errors on the value of the ratio of the radii. In the top panel the stellar radius has been set to its value in the KIC, while in the bottom panel the stellar radius is computed via stellar isochrones from its given $T_{\star}$ value in the KIC. The dashed line delineates an expected 1:1 correlation.}
\label{borcompfig}
\end{figure}

Given the stellar parameters and their associated errors, we proceeded with calculating physical parameters for each planet. From the orbital period of the system and the mass of the star, we calculated the semi-major axis of the planets via Newton's version of Kepler's Third Law. The radius of each planet, $R_{p}$, was calculated from the stellar radius and the value of the radius ratio derived in Section~\ref{modelsec}. The obtained $R_{p}$ values are compared with those reported by \citet{Borucki2011} in Figure~\ref{borcompfig} by multiplying the derived value of the ratio of the radii from each study, including results from both the PDC and CLM data for ours, by the radius of the host star derived via both the KIC and stellar isochrones. Except for a handful of outliers, most values of the planetary radii derived via different parameter estimations seem to agree with the \citet{Borucki2011} results within $\sim$5\%. We note though that the radii of individual planet candidates can be significantly affected depending on whether their stellar radii are derived from the KIC or stellar isochrones, on average $\sim$20\%.

We also calculated the sub-stellar equilibrium temperature, maximum effective temperature, and brightness temperature of each planet following the same equations as in \citet{Cowan2011}, who themselves draw upon \citet{Hansen2008} and \citet{Burrows2008a}. We calculated the equilibrium temperature of each planet at its sub-stellar point, $T_{0}$, using $T_{\star}$ and the semi-major axis of the system, $a$, via,

\begin{equation}
  T_{0} = T_{\star}\cdot(R_{\star}/a)^{0.5}
\end{equation}

\noindent We note that this expression assumes non-significant eccentricity effects on the heating of the planet by the star. 

We calculated the maximum effective temperature of the planet, $T_{\epsilon=0}$, using the equation

\begin{equation}
  T_{\epsilon=0} = (2/3)^{\frac{1}{4}}\cdot T_{0}
\end{equation}

\noindent assuming no albedo or heat recirculation.

Finally, to calculate the measured brightness temperature of each planet, $T_{b}$, we assume both the planet and star emit like blackbodies and compute $T_{b}$ by integrating their fluxes over the $Kepler$ passband from the equation

\begin{equation}
\label{Jeq}
  J = \frac{\int t_{\lambda}\cdot\lambda^{-5}\cdot(exp(\frac{hc}{\lambda kT_{b}})-1)^{-1} \cdot d\lambda}{\int t_{\lambda}\cdot\lambda^{-5}\cdot(exp(\frac{hc}{\lambda kT_{\star}})-1)^{-1} \cdot d\lambda}
\end{equation}

\noindent where $\lambda$ is a given wavelength, $t_{\lambda}$ is the net transmission of the telescope and detector at a given wavelength, $h$ is Planck's constant, $c$ is the speed of light, and $k$ is Boltzmann's constant.

In the cases where significant sinusoidal variations were detected in the light curves, i.e., cases where we found a positive value for $A_{L_{p}}$ with a detection level of at least 2$\sigma$, we treated the effect as real and accounted for it in the determination of the day-side flux of the planet by multiplying the values of $J$ in Eq.~\ref{Jeq} by (1+$A_{L_{p}}$). The only systems where significant sinusoidal variations were detected are KOI 2.01 and KOI 13.01, for both the PDC and CLM data, and both fixing eccentricity to zero and letting it vary. For KOI 1541.01 a significant sinusoidal variation was also found, but only for the CLM data and when allowing eccentricity to vary. Notice that, if real, the observed amplitude of sinusoidal variations can be either due to a significant albedo, and thus a varying amount of reflected light with phase, a significant temperature difference between the day and night sides of the planets, i.e., very little heat redistribution, and therefore varying amounts of emitted light with phase, or photometric beaming. It is possible that sinusoidal systematic signals could mask as significant phase variations, but in order to be detected as such by the residual-permutation error analysis we employed, the systematic feature would have to have a stable amplitude and period over the course of the 90-day observations, and have the period and phase of maximum amplitude coincide with the orbital period and secondary eclipse phase of the planet, which we deem unlikely.

In the above temperature calculations we have assumed no albedo and therefore all the observed planetary fluxes are due to thermal emission. However, the atmosphere of the planets can contain clouds or hazes which would reflect at least part of the incident stellar light. To account for those effects, we can estimate the different contribution amounts of reflected light to the measured planet-to-star surface brightness ratio, as 

\begin{equation}
\label{albeq}
  F_{a} = \frac{A \cdot R_{\star}^{2}}{a^{2}}
\end{equation}

\noindent where $A$ is the geometric albedo of the planet in the integrated $Kepler$ passband. Assuming different values of $A$ between 0.0 (no albedo) and 1.0 (purely reflective atmosphere), we can subtract the resultant value of $F_a$ from $J$ in order to remove the reflected light contribution from the measurements of the eclipse depths before computing the $T_{b}$ of the planet that accounts for the remaining, thermally emitted, light. Furthermore, given the measured surface brightness ratio and Equation~\ref{albeq}, we can determine the maximum possible geometric albedo of the planet in the $Kepler$ wavelength range, $A_{max}$, by assuming that all of the detected emission is solely due to reflected light. Setting $F_{a}$ = $J$ and solving for $A$, we obtain the expression

\begin{equation}
\label{maxalbeq}
 A_{max} = \frac{a^{2}J}{R_{\star}^{2}}
\end{equation}

Finally, we have computed robust errors for all the derived quantities, i.e.,  $T_0$, $T_{\epsilon = 0}$, $T_b$, $a$, $R_p$, $A_{max}$, and also $T_b / T_0$ (see next section) by re-calculating all their values 10,000 times, each time adding random Gaussian noise with amplitudes equal to the errors of the underlying quantities $J$, $P$, $k$, $T_*$, $M_*$, and $R_*$. The resulting median values of each parameter and their asymmetric Gaussian 1$\sigma$ errors are listed in Table~\ref{kepsec-tab3} along with the detection significance of the secondary eclipse, $\sigma_{sec}$, and the luminosity ratio of the system, $L_{r}$ = $L_{p}$/$L_{\star}$, for both the PDC and CLM light curves, both letting eccentricity vary and fixing it to zero, and both using the stellar parameters derived from the KIC and via stellar isochrones. Negative values of $\sigma_{sec}$ mean that a negative value of $J$ was found, i.e., an increase of light at secondary eclipse, instead of the expected decrease. We deem those results unphysical, but note we can still use them to establish upper limits for the depth of the eclipse, and estimate the fraction of spurious eclipse detections in our analysis, as already described at the end of Section~\ref{modelsec}.

\begin{deluxetable}{rrcccccccc}
  \rotate
  \tablewidth{0pt}
  \tabletypesize{\scriptsize}
  \tablecaption{Derived System Parameters and Associated 1$\sigma$ Uncertainties}
  \tablecolumns{10}
  \tablehead{KOI & $L_{r}$ & $\sigma_{sec}$ & $T_{\rm 0}$ & $T_{\epsilon=0}$ & $T_{\rm b}$\tablenotemark{a} & $T_{\rm b}$/$T_{0}$ & $R_{\rm p}$ & $a$ & $A_{\rm max}$\\ & (ppm) & & (K) & (K) & (K) & & ($R_{\rm Jup}$) & (AU) & } 
  \startdata
  \cutinhead{PDC Light Curve with Eccentricity Fixed to Zero and Stellar Parameters from the KIC}
  \input{kepsec-tab3a.tex}
\enddata
  \tablenotetext{a}{Derived assuming $A$ = 0.0}
  \label{kepsec-tab3}
\end{deluxetable}

\subsection{Statistical Properties of the Secondary Eclipse Emissions}
\label{trendssec}

\begin{figure}[h!]
\centering
\begin{tabular}{cc}
   \epsfig{width=0.475\linewidth,file=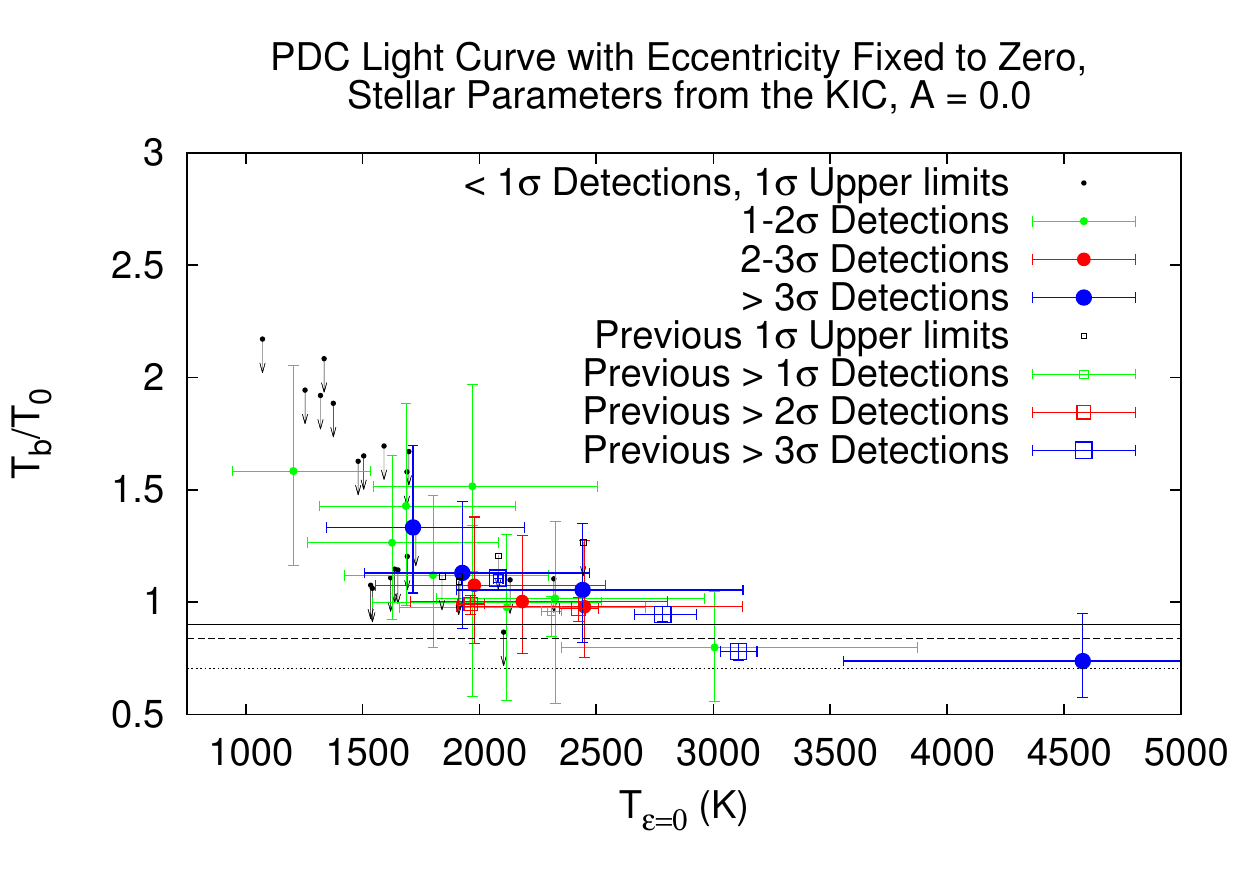} &
   \epsfig{width=0.475\linewidth,file=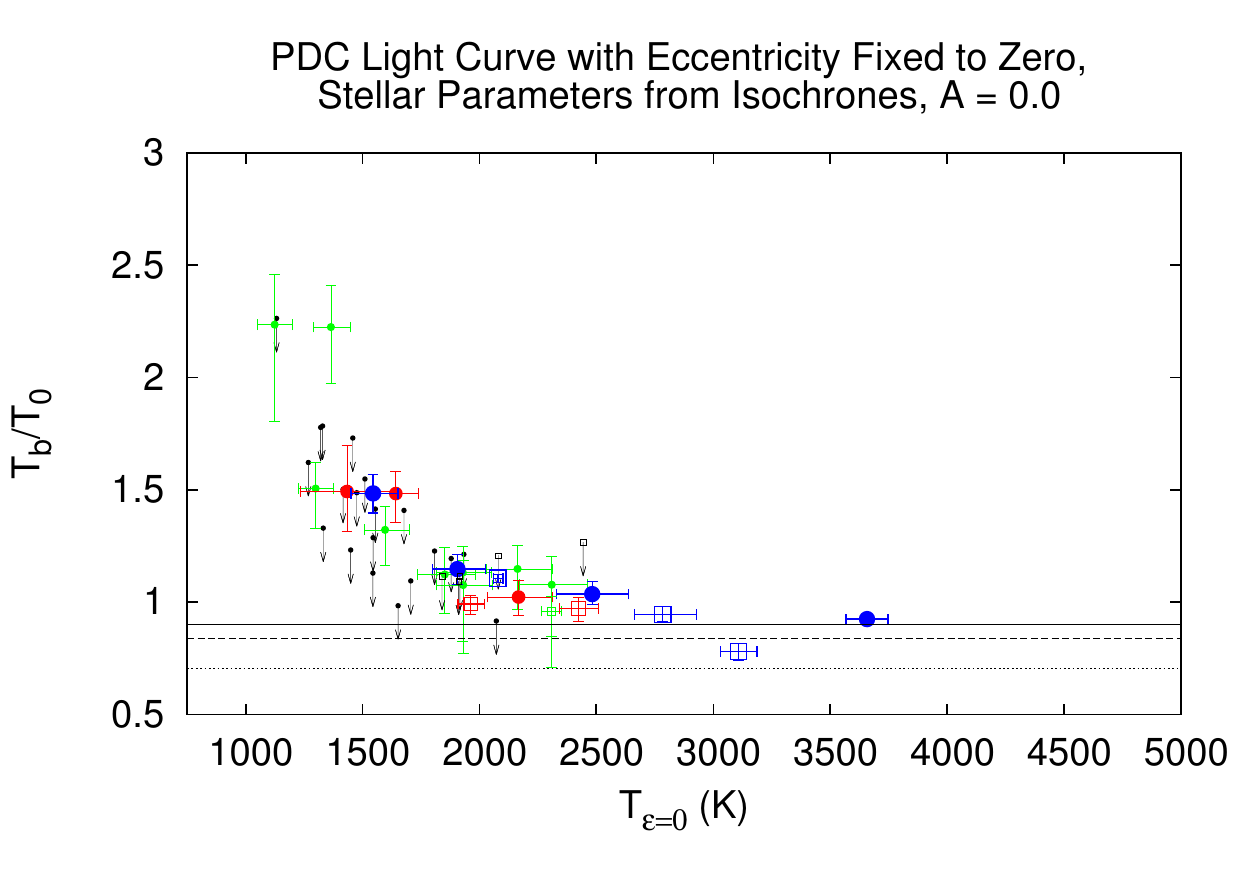} \\
   \epsfig{width=0.475\linewidth,file=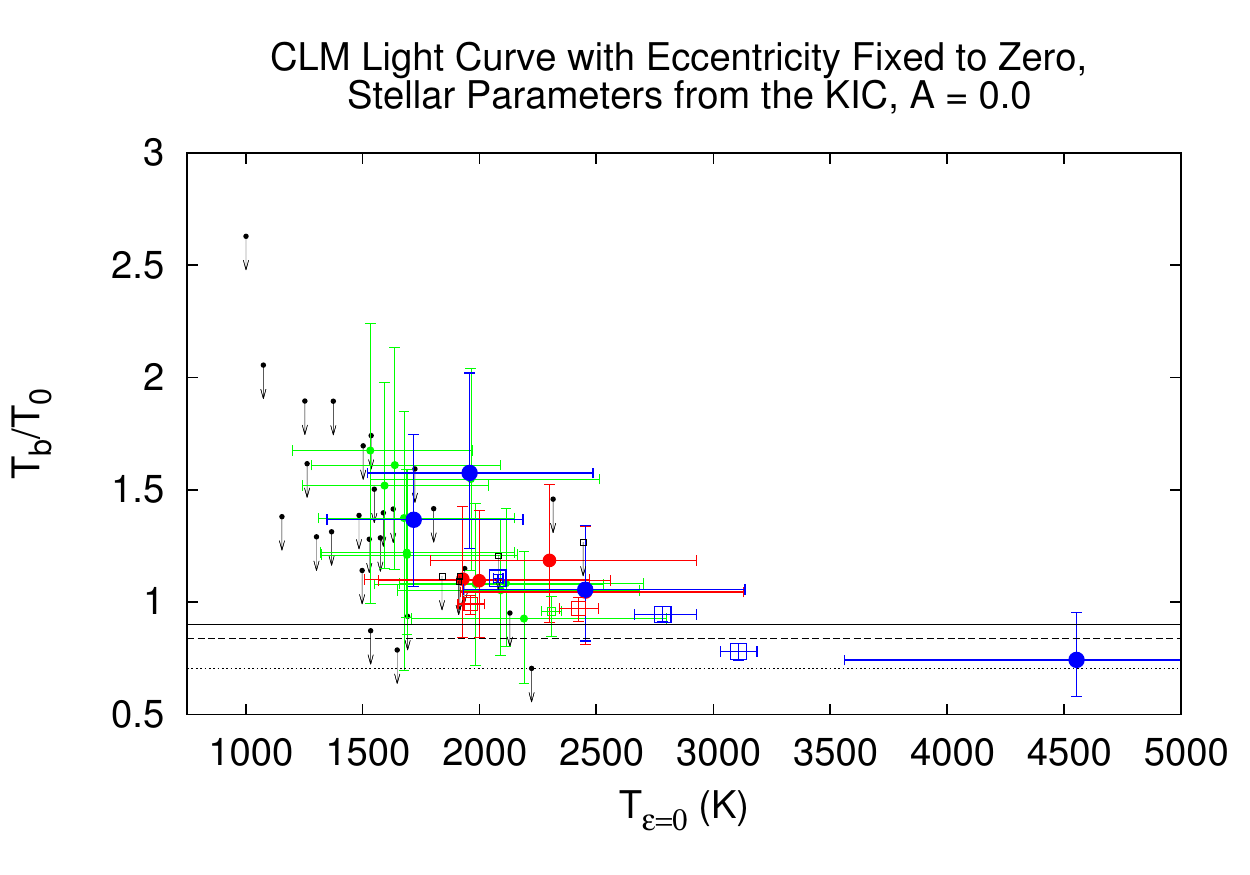} &
   \epsfig{width=0.475\linewidth,file=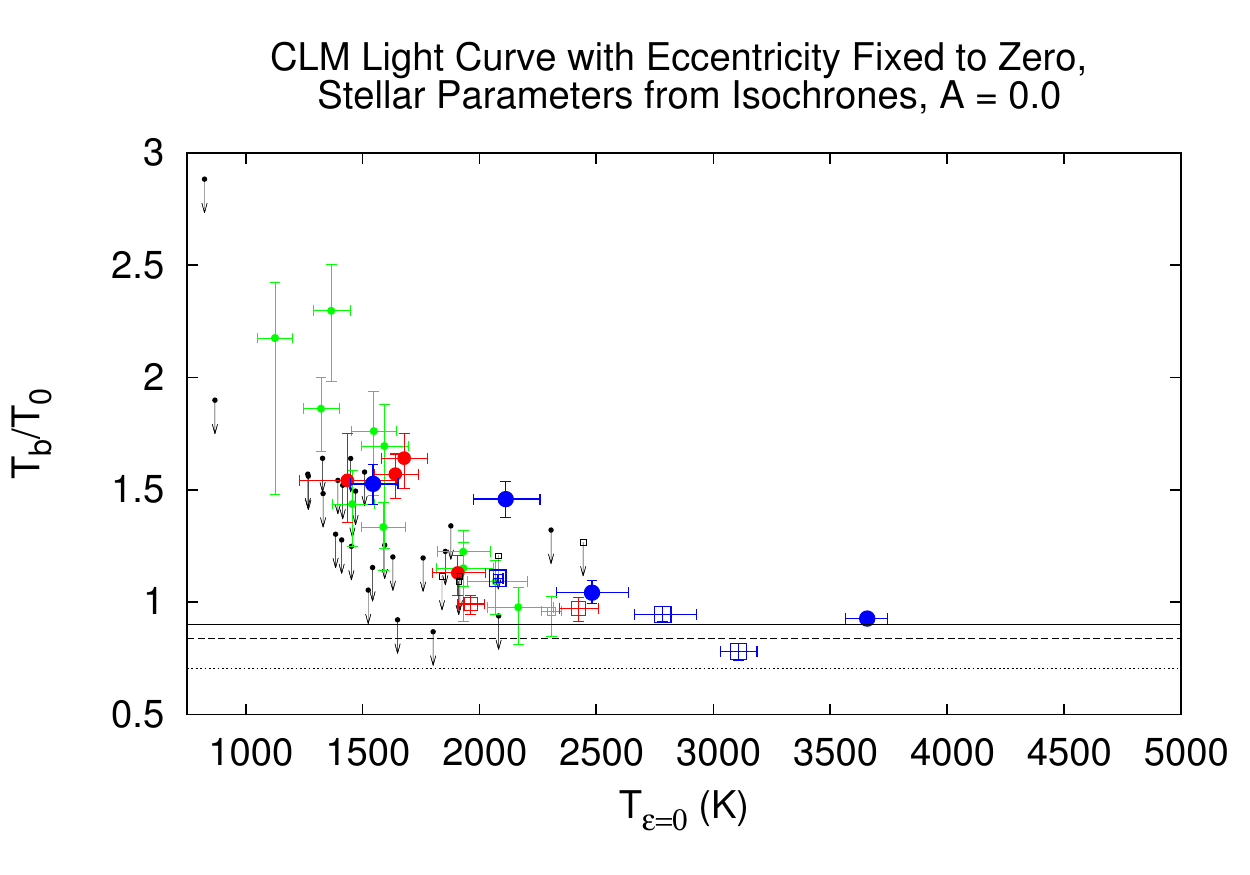} \\
\end{tabular}
\caption[Plots of the effective day side temperature ratio versus the maximum effective day side temperature when fixing eccentricity to zero]{Plots of the effective day side temperature ratio versus the maximum effective day side temperature when fixing eccentricity to zero. The values obtained when deriving stellar parameters from the KIC are shown in the left column, while values obtained when deriving stellar parameters from stellar isochrones are shown in the right column. Values obtained when using the $Kepler$ PDC light curves are shown in the first row, while values obtained when using the CLM pipeline are shown in the second row. Solid circles correspond to $Kepler$ systems modeled in this paper, while open squares are previously published detections or upper limits of exoplanet secondary eclipses at optical wavelengths. All errors are 1$\sigma$. The x-axis errorbars are not shown for the $<$1$\sigma$ detections for clarity. The solid, dashed, and dotted black lines in each figure correspond to the expected temperature ratio assuming no heat recirculation, a uniform day-side temperature, and a uniform planetary temperature respectively.}
\label{trfig}
\end{figure}

\begin{figure}[ht]
\centering
\begin{tabular}{cc}
   \epsfig{width=0.475\linewidth,file=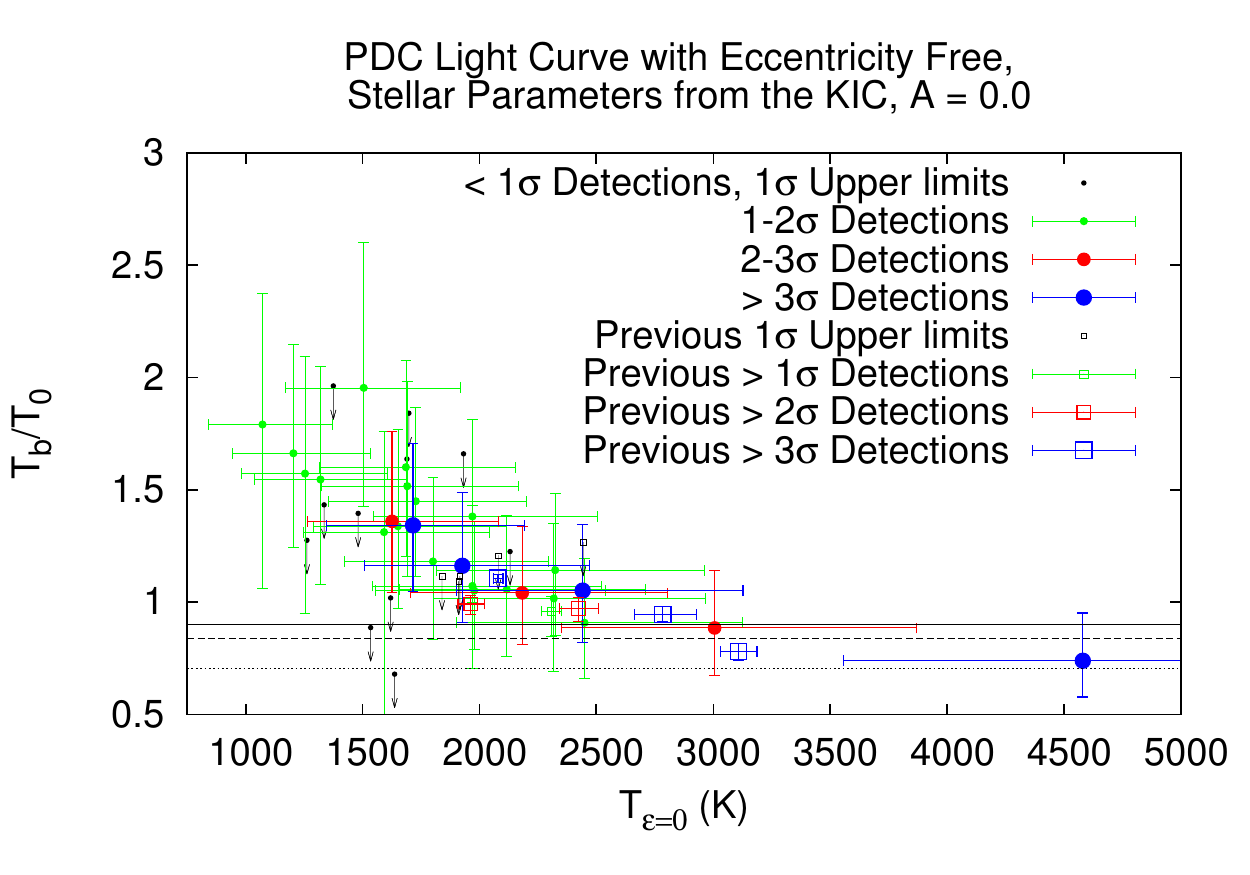} &
   \epsfig{width=0.475\linewidth,file=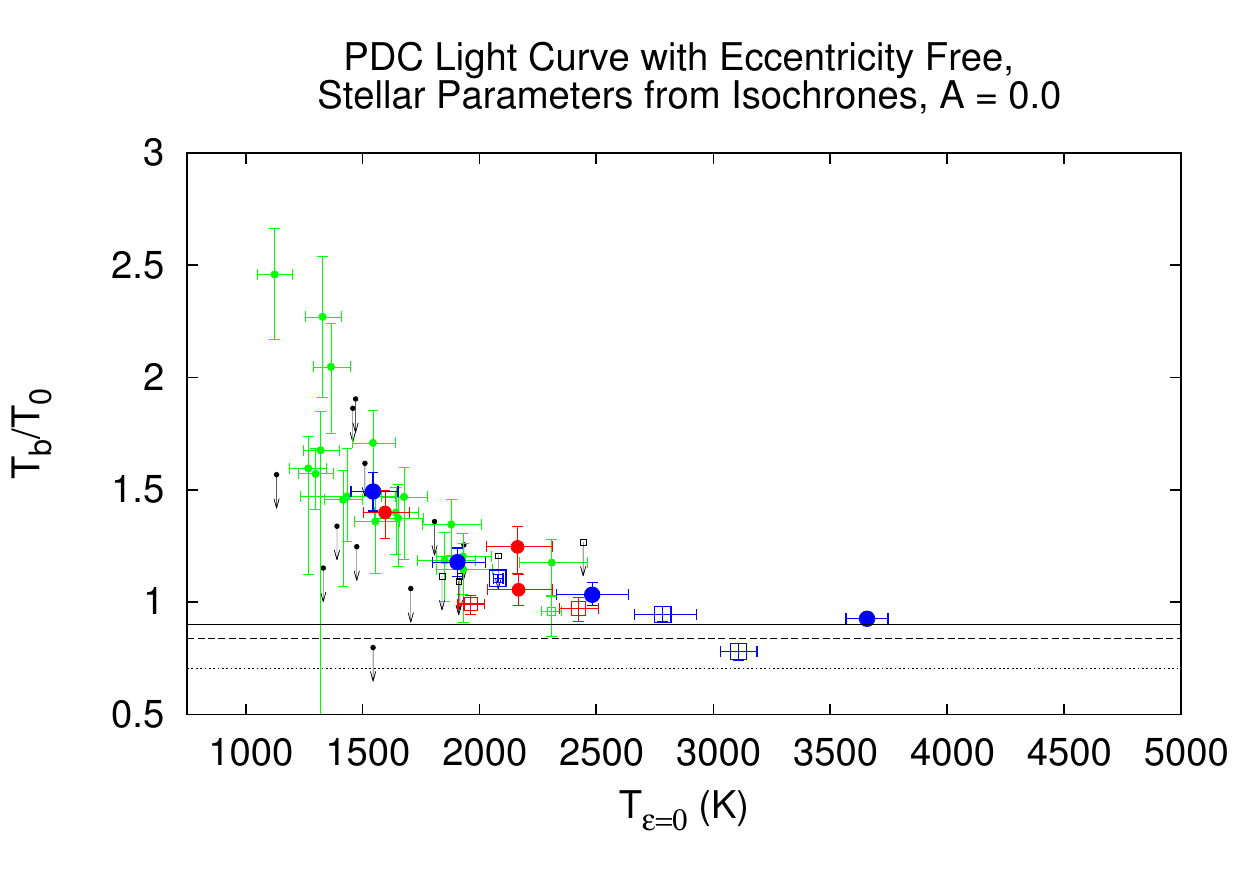} \\
   \epsfig{width=0.475\linewidth,file=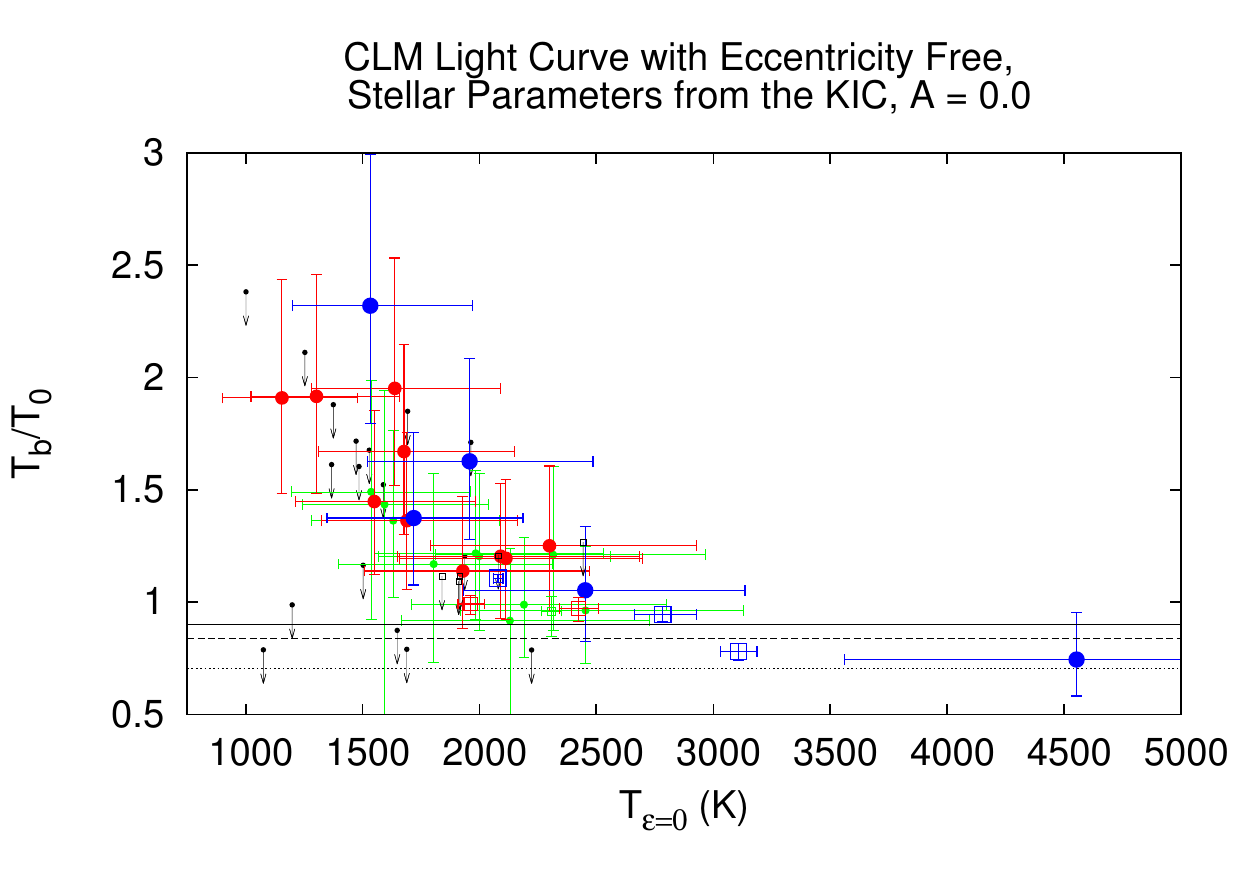} &
   \epsfig{width=0.475\linewidth,file=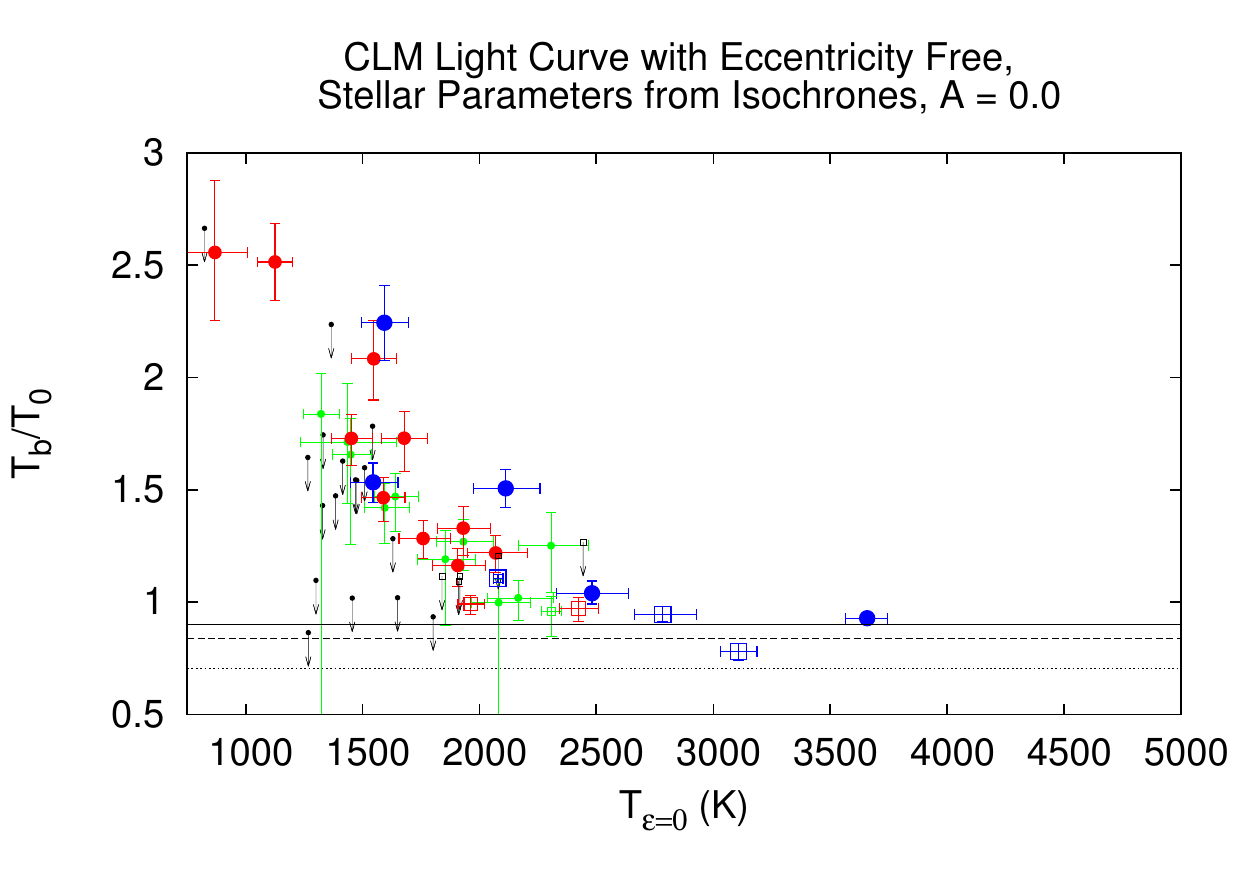} \\
\end{tabular}
\caption[Similar to Figure~\ref{trfig}, but with eccentricity allowed to vary]{Similar to Figure~\ref{trfig}, but with eccentricity allowed to vary.}
\label{trfig2}
\end{figure}

Following \citet{Cowan2011}, we plot in Figure~\ref{trfig} the dimensionless $Kepler$ passband day-side brightness temperature ratio of each planet candidate in our sample, $T_{b}$/$T_{0}$, versus their maximum expected day-side temperature, $T_{\epsilon = 0}$, for the case of eccentricity fixed to zero. The different panels in the figure correspond to the results from the PDC and the CLM light curves, and both using the stellar parameters derived from the KIC and via stellar isochrones. In all the panels we have assumed zero albedo, which is equivalent to assuming that all the emission from the planet is thermal. (The case of non-zero albedos is considered below.) The 1-2$\sigma$, 2-3$\sigma$, and $>$3$\sigma$ detections, and 1$\sigma$ upper limits of the $<$1$\sigma$ detections, are represented by solid circles of different colors and sizes. In addition, the open squares correspond to planets with previously published secondary eclipse detections in the optical, i.e., CoRoT-1b \citep{Alonso2009b}, CoRoT-2b \citep{Alonso2009a}, Hat-P-7b \citep{Welsh2010}, Kepler-5b \citep{KippingBakos2011a}, Kepler-7b \citep{KippingBakos2011a,Demory2011}, and OGLE-TR-56b \citep{Sing2009b,Adams2011}, as well as previously published secondary eclipse upper limits, i.e., HD209458 \citep{Rowe2008}, TrES-2b \citep{KippingBakos2011b}, and Kepler-4b, Kepler-6b, and Kepler-8b \citep{KippingBakos2011a}. Each point is shown with its 1$\sigma$ x and y-axis errorbars, except for the $<$ 1$\sigma$ detection upper limits, where the x-axis errorbars are omitted for clarity. Finally, the three horizontal lines in each plot indicate the expected values of $T_{b}$/$T_{0}$ for no energy redistribution, i.e., $f = 2/3$, a uniform day-side temperature, i.e., $f = 1/2$, and a uniform planetary temperature, i.e., $f = 1/4$ \citep[see][]{LopezMoralesSeager2007,Cowan2011}. In Figure~\ref{trfig2} we reproduce the same plots but with eccentricity allowed to vary.

Unless there is some extra emission at optical wavelengths that is not being accounted for, all the planets should lie below the $f = 2/3$ lines in Figures~\ref{trfig} and \ref{trfig2}. However, it is immediately apparent that the vast majority of candidates lie above that line. In addition, there appears to be a trend of increasing  $T_b / T_0$ with decreasing $T_{\epsilon = 0}$, with some possible detections approaching 2.5 times the maximum expected brightness temperature, i.e., nearly 40 times more flux than expected. All the planets with previously published secondaries also appear to follow the same trend, although they all have $T_{\epsilon = 0}$ $\gtrsim$ 2000 K. This trend will be discussed in more detail below.

We have explored several possible explanations for such large observed emissions at visible wavelengths: 1) A bias in the determination of stellar and planetary parameters or the secondary eclipse detection efficiency, 2) high albedos, which would make reflected light a major contributor to the planetary emission, 3) very large amounts of non-LTE or other thermal emission at optical wavelengths, 4) the presence of a significant source of internal energy generation within the planet, and 5) some of the candidates are in fact very low mass stars or brown dwarf companions, or background eclipsing binary blends.

{\it Potential~Biases}: The determined stellar parameters of $T_{\star}$, $M_{\star}$, and $R_{\star}$ can have significant uncertainties, as noted in Section~\ref{paramsec}, and indeed can vary notably when taken from the KIC or computed from stellar isochrones based on $T_{\star}$. The values of those parameters are intimately tied to the derivation of the planetary parameters $T_{\epsilon = 0}$, $T_{0}$, $T_{b}$, $a$, $R_{p}$, $F_{a}$, and $A_{max}$, and that must be kept in mind when interpreting any possible results. For example, the stellar isochrones assume the stars are main-sequence, and thus if a host star is really a sub-giant or otherwise evolved, the stellar flux at the planet's surface would be underestimated. This would in turn cause an underestimated value of both $T_{\epsilon = 0}$ and $T_0$, and thus potentially overestimated values of $T_b / T_0$ at lower $T_{\epsilon = 0}$. However, it is unlikely that a large fraction of the examined systems are significantly evolved, and sub-giants would likely show telltale variations in their light curves. Given this, and that as far as we know there is no other preferential bias in the determination of the stellar parameters, we would not expect this problem to systematically influence the results presented in Figures~\ref{trfig} and \ref{trfig2}.

When examining secondary detection efficiency, we note that the derived upper limits on the secondary eclipse depths are roughly at the same level as the noise of the $Kepler$ data. That level of noise is consistent among the set of $Kepler$ light curves we have examined. Thus, as we search the data for planets with lower values of $T_{\epsilon =0}$ and $T_{0}$, the corresponding upper limit for $T_{b}$/$T_{0}$ naturally increases, as the expected eclipse depth decreases while the noise level remains the same. This introduces an artificial trend of higher  $T_{b}$/$T_{0}$ upper limits as $T_{\epsilon =0}$ decreases, which can be seen in Figures~\ref{trfig} and \ref{trfig2}. Similarly, when examining the $\sim$1$\sigma$ detections, about $\sim$32\% of the detections will be statistically spurious, as exemplified in the data and already discussed in Section~\ref{modelsec}. We thus expect an artificial trend of higher $T_{b}$/$T_{0}$ values with decreasing values of $T_{\epsilon = 0}$ for $\sim$32\% of the $\sim$1$\sigma$ detections. However, when we move to the $\gtrsim$ 2$\sigma$ detections, the expected rate of spurious detections drop to $\lesssim$ 5-10\%. The data in Figures~\ref{trfig} and \ref{trfig2} still reveal a significant trend of increasing $T_{b}$/$T_{0}$ values with decreasing $T_{\epsilon = 0}$ in the $\gtrsim$~2$\sigma$ detections, (including as well previous secondary eclipse detections from the literature), so we conclude this trend is real and due to either increasing albedos or emission features in the visible as the atmospheric temperature of the planets decrease. The hypothesis of high albedos is further discussed below. As for emission features, a literature search on this topic does not reveal any known physical processes that would produce this effect, so further theoretical work might be necessary.

\begin{figure}
\centering
\begin{tabular}{cc}
  \epsfig{width=0.45\linewidth,file=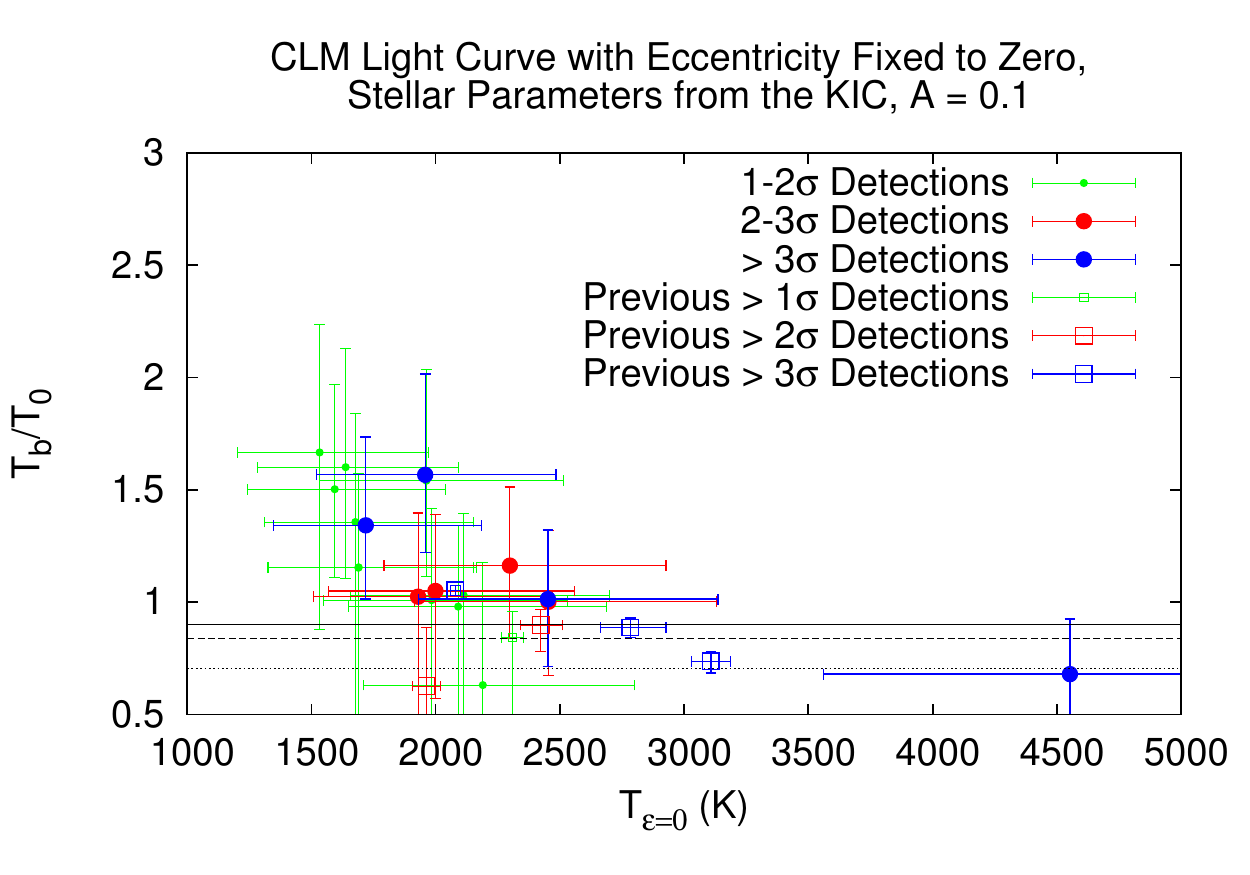} &
  \epsfig{width=0.45\linewidth,file=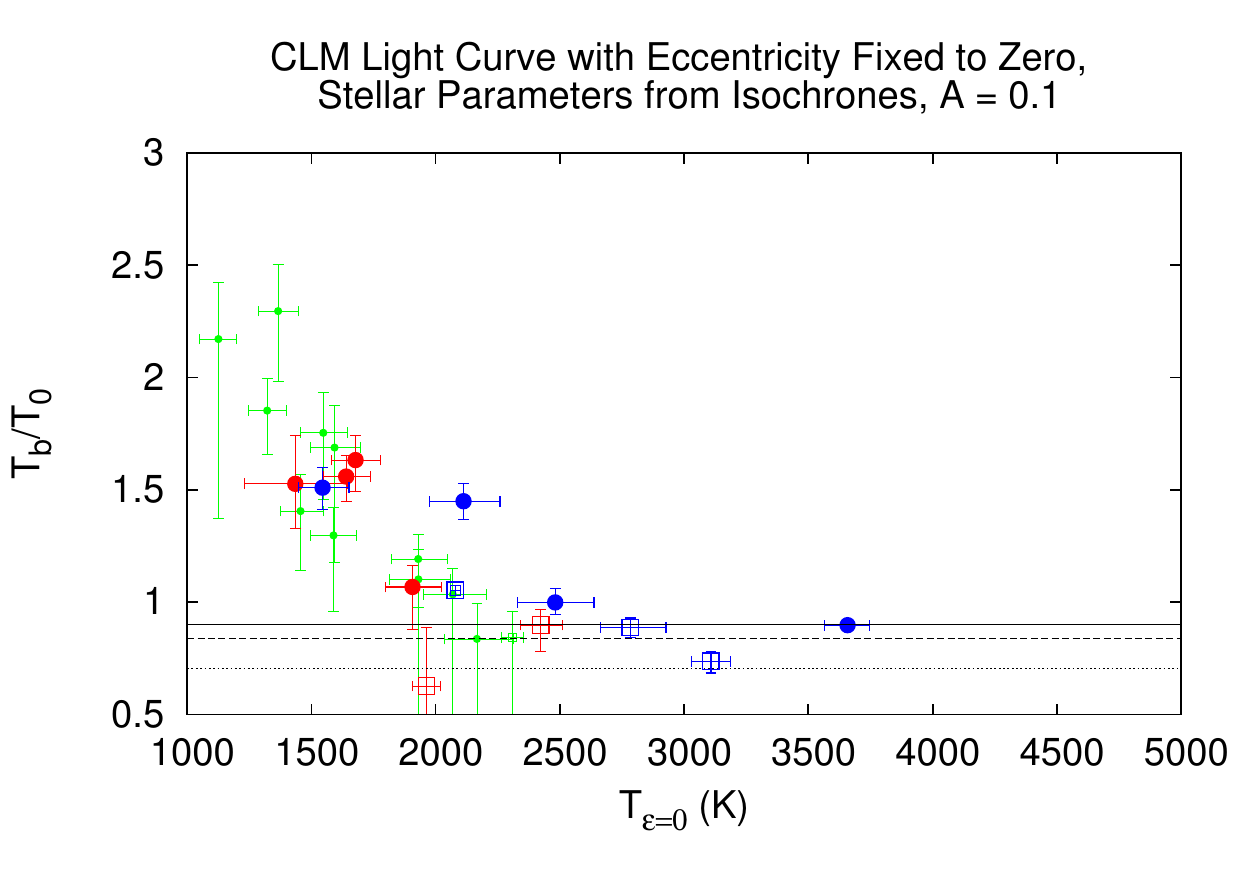} \\
  \epsfig{width=0.45\linewidth,file=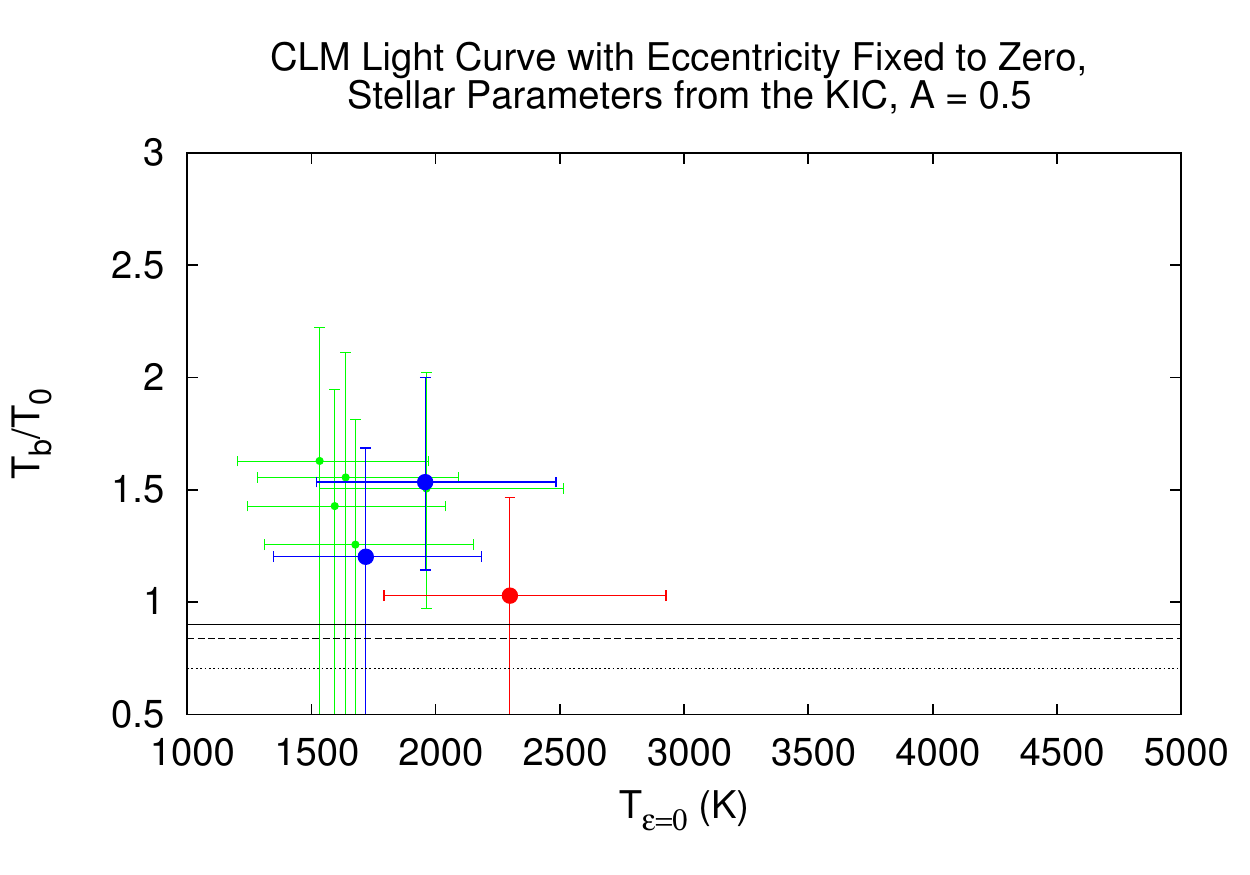} &
  \epsfig{width=0.45\linewidth,file=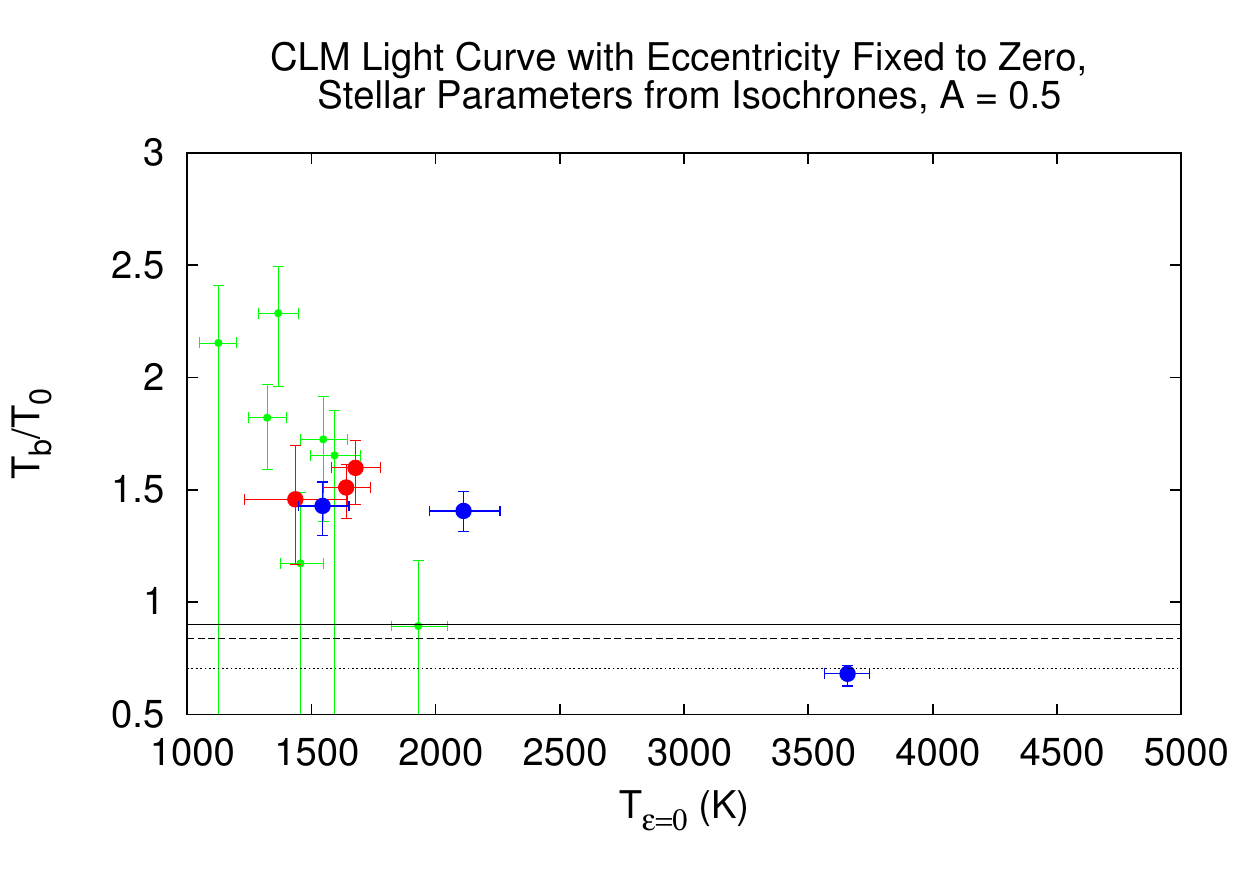} \\
  \epsfig{width=0.45\linewidth,file=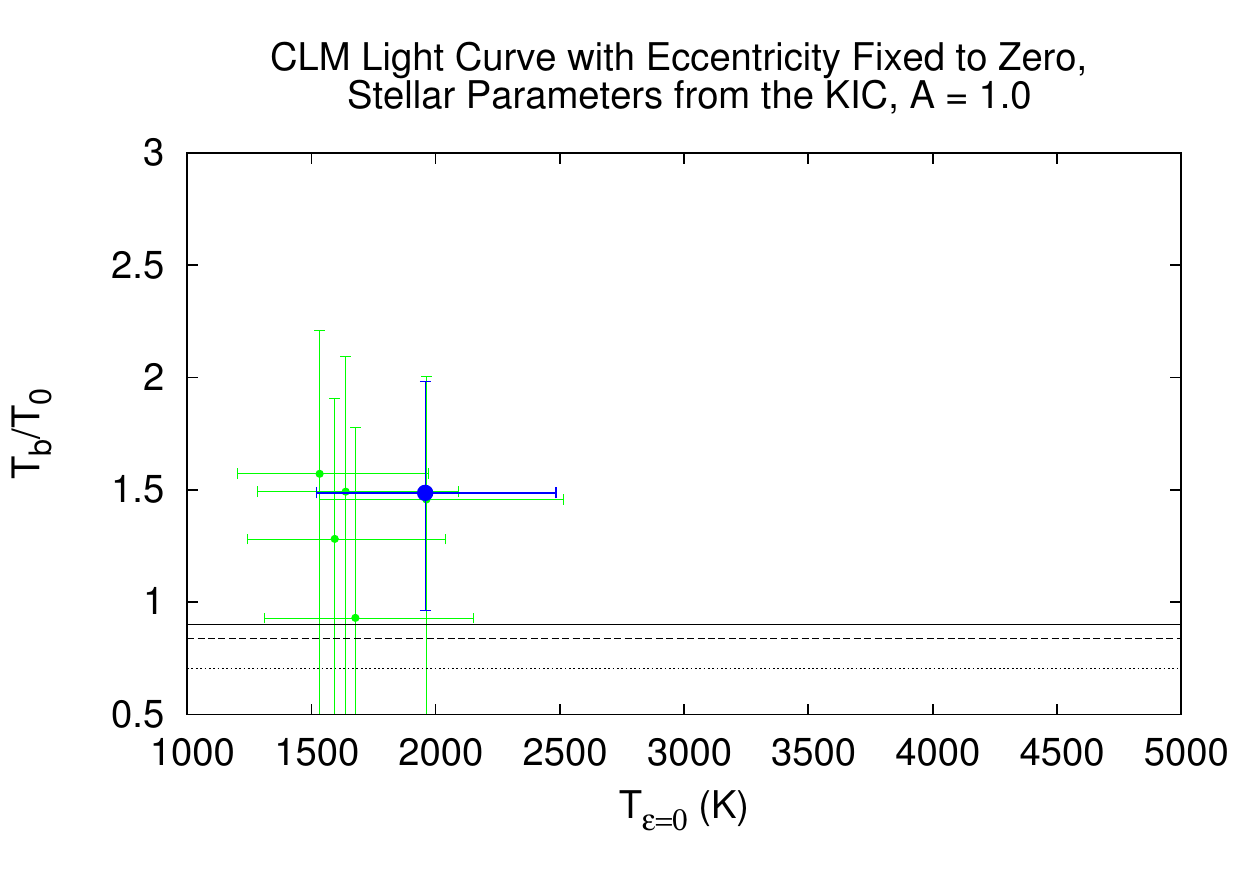} &
  \epsfig{width=0.45\linewidth,file=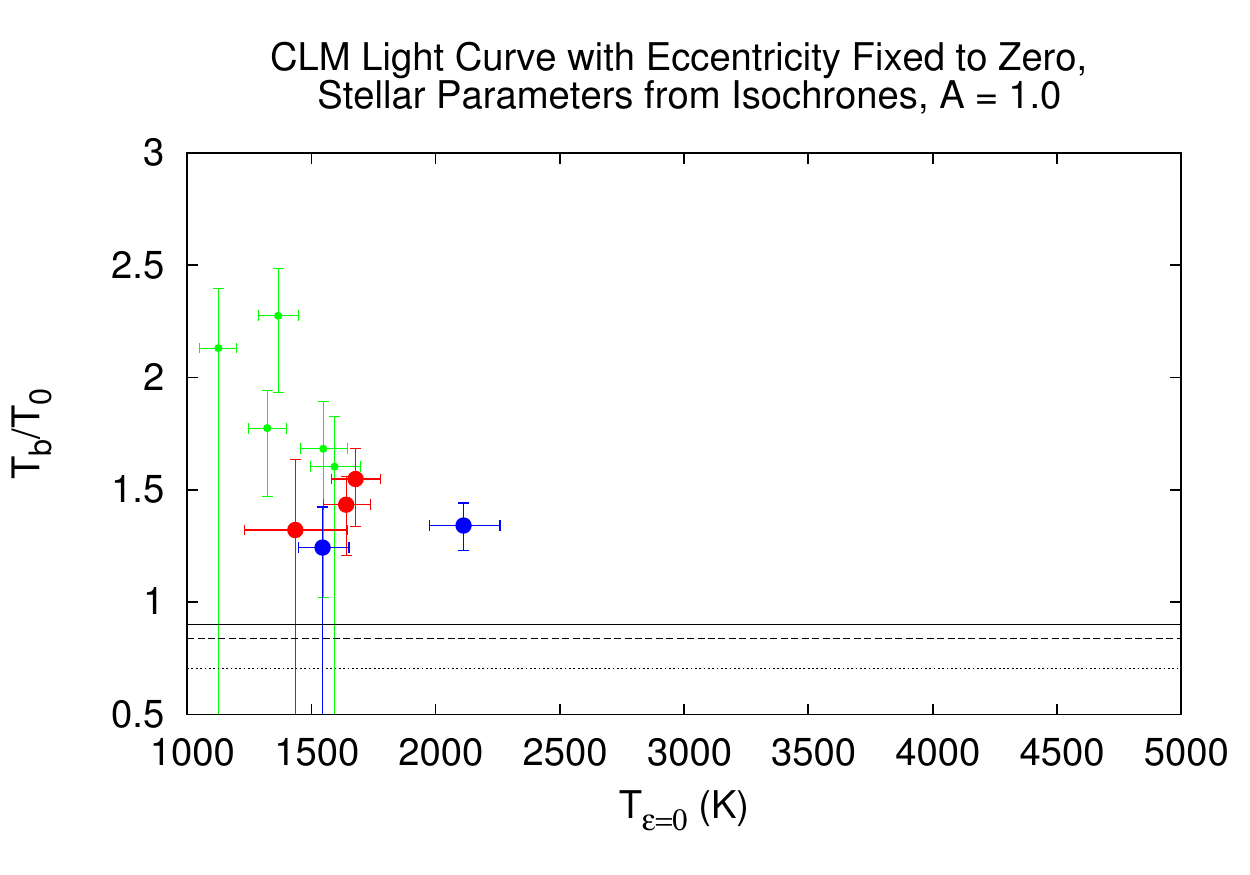} \\
\end{tabular}
\caption[Plots of the effective day side temperature ratio versus the maximum effective day side temperature for albedos of 0.1, 0.5, and 1.0]{Plots of the effective day side temperature ratio versus the maximum effective day side temperature for, from top to bottom, albedos of 0.1, 0.5, and 1.0, using the CLM pipeline light curves and assuming no eccentricity. The values obtained when deriving stellar parameters from the KIC are shown in the left column, while values obtained when deriving stellar parameters from stellar isochrones are shown in the right column. Solid circles correspond to $Kepler$ systems modeled in this paper, while open squares are previously published detections or upper limits of exoplanet secondary eclipses at optical wavelengths. The solid, dashed, and dotted black lines in each figure correspond to the expected temperature ratio assuming no recirculation, a uniform day-side temperature, and a uniform planetary temperature respectively. Systems that disappear from the plots when moving from low to high albedos can be fully explained by reflected light, while systems that still remain at A = 1.0 present excess emission that cannot be explained solely by reflected light.}
\label{albfig}
\end{figure}

{\it High~Albedos}: As mentioned in the introduction, some recent studies suggest high albedos for some known hot giant planets \citep[e.g.][]{Berdyugina2011,KippingBakos2011a,Demory2011}. To examine the possibility that the excess flux observed in our list of secondary eclipse candidates is due to albedo, we have recomputed the expected normalized brightness temperature $T_{b}$/$T_{0}$ of each candidate when assuming increasing values of the albedo. The reflected light contribution is computed using equation~\ref{albeq} and subtracted from the total flux measured for each object. $T_{b}$ is then recomputed using the remaining flux, assuming it is solely due to thermal emission. The results are shown in Figure~\ref{albfig} for three different albedos, $A$ = 0.1, 0.5, and 1.0, using the secondary eclipse depths measured from the CLM pipeline light curves for $e = 0$, (the eclipse depths measured from the other light curves described in Section~\ref{datasec} give similar results).

As expected, $T_{b}$/$T_{0}$ decreases as the albedo increases, and many of the points in Figure~\ref{albfig} go below the $T_{b}$/$T_{0}$ = (2/3)$^{1/4}$, i.e., $A$ =0 and $f$ = 2/3, limit once high enough albedos are assumed. The emission of 53\% of the planet candidates can be interpreted as a combination of reflected light and thermal emission when we assume a geometric albedo of $A$ = 0.5, set $e$ = 0, and derive stellar parameters from the KIC, though only 31\% when deriving stellar parameters from stellar isochrones. Those levels of reflectivity might indicate the presence of clouds, haze, or Rayleigh scattering in the atmosphere of those planets, with a general trend of increasing albedo with decreasing planetary temperature responsible for the trend of increasingly excess emission at lower planetary temperatures. We note, however, that a significant number of points, 40\% and 63\% when deriving stellar parameters from the KIC and stellar isochrones respectively, with $e$ = 0, still remain above $T_{b}$/$T_{0}$ = (2/3)$^{1/4}$, even if we assume perfectly reflecting planets, i.e. $A$ = 1.0. Many of these remaining systems are at low values of $T_{\epsilon = 0}$, and some of those points are $\gtrsim$ 2$\sigma$ detections, so there is still excess emission and a correlation of $T_{b}$/$T_{0}$ with decreasing $T_{\epsilon = 0}$ that cannot be explained by reflected light, and needs to be explained in some other way.

{\it Non-LTE~or~Other~Thermal~Emission}: We have used the brightness temperature parameter $T_{b}$ to represent the amount of thermal emission from each planet, assuming that the planets emit as blackbodies. In that case, if the emission of the planet yields a $T_{b}$ larger than that predicted by $f$ = 2/3 and $A$ = 0, that emission is above the so-called equilibrium temperature. However, the atmospheres of exoplanets do not necessarily emit as blackbodies, and some spectral models of hot Jupiters \citep[e.g.,][]{Fortney2008} predict significantly higher emission levels in the optical region covered by the $Kepler$ bandpass ($\sim$ 0.4 - 0.9 $\mu m$) compared to blackbody approximations. The emission spectrum of a planet will depend strongly on its atmospheric composition and Temperature-Pressure (T-P) profile. Although some models including the presence of strong absorbers, such as TiO and VO, have been proposed \citep{Hubeny2003,Fortney2008,Burrows2008a}, uncertainties in the T-P profiles and the lack of previous observational data limit the reliability of those models. In addition, and as already mentioned above, there is currently no model that predicts the increase of $T_{b}$/$T_{0}$ for decreasing $T_{\epsilon = 0}$ observed in Figures~\ref{trfig} and \ref{trfig2}.

Another possibility is that the large amount of stellar irradiation these planets receive induces resonant and non-resonant fluorescent transitions by exciting the chemical species present in their upper atmospheres. Fluorescence has been measured in Solar System objects, i.e., Titan, Saturn and Jupiter, at IR, UV, and X-ray wavelengths \citep[e.g.,][] {Yelle2003,LopezValverde2005,Cravens2006,Lupu2011}, and possibly in the NIR for the exoplanet HD189733b \citep{Swain2010}, but we could not find any observations of or theoretical work on non-LTE/fluorescence emission in the 0.4 - 0.9$\mu m$ wavelength range covered by $Kepler$. However, it is expected that non-LTE emission lines in exoplanetary atmospheres will not be significantly broadened by collisions, and will appear instead as sharp emission features (Martin-Torres, priv. comm.). Considering the amount of energy required to produce these fluorescent emissions, it is unlikely that, given their narrow emission range, they would be luminous enough to significantly increase the measured emission levels over the very wide $Kepler$ bandpass above the expected LTE emissions observed in Figures~\ref{trfig} and \ref{trfig2}.

{\it Significant~Internal~Energy~Generation}: Our own Jupiter radiates about 1.6 times as much energy as it receives from the Sun. The additional heat source is generally attributed to either residual heat left over from the initial Solar System nebula collapse, or ongoing slow contraction of the planet's core. However, if the planet candidates in our list were undergoing a similar internal energy generation process, $T_b / T_0$ would only reach up to about 1.6$^{1/4}$ = 1.12, not high enough to explain the emission of objects in our sample with $T_{\epsilon =0}$ $\lesssim$ 1500 K.

\begin{figure}[ht!]
\centering
\begin{tabular}{cc}
  \epsfig{width=0.475\linewidth,file=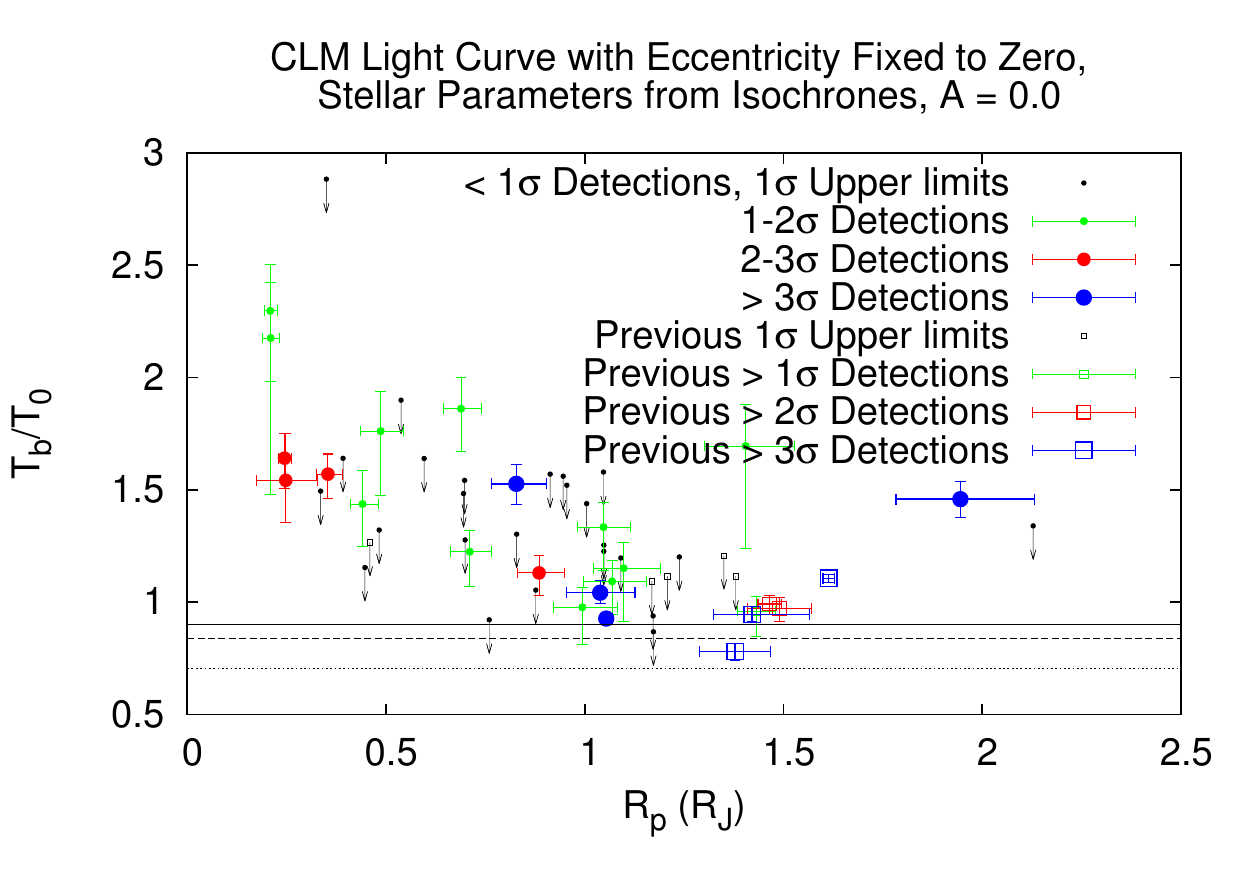}&
  \epsfig{width=0.475\linewidth,file=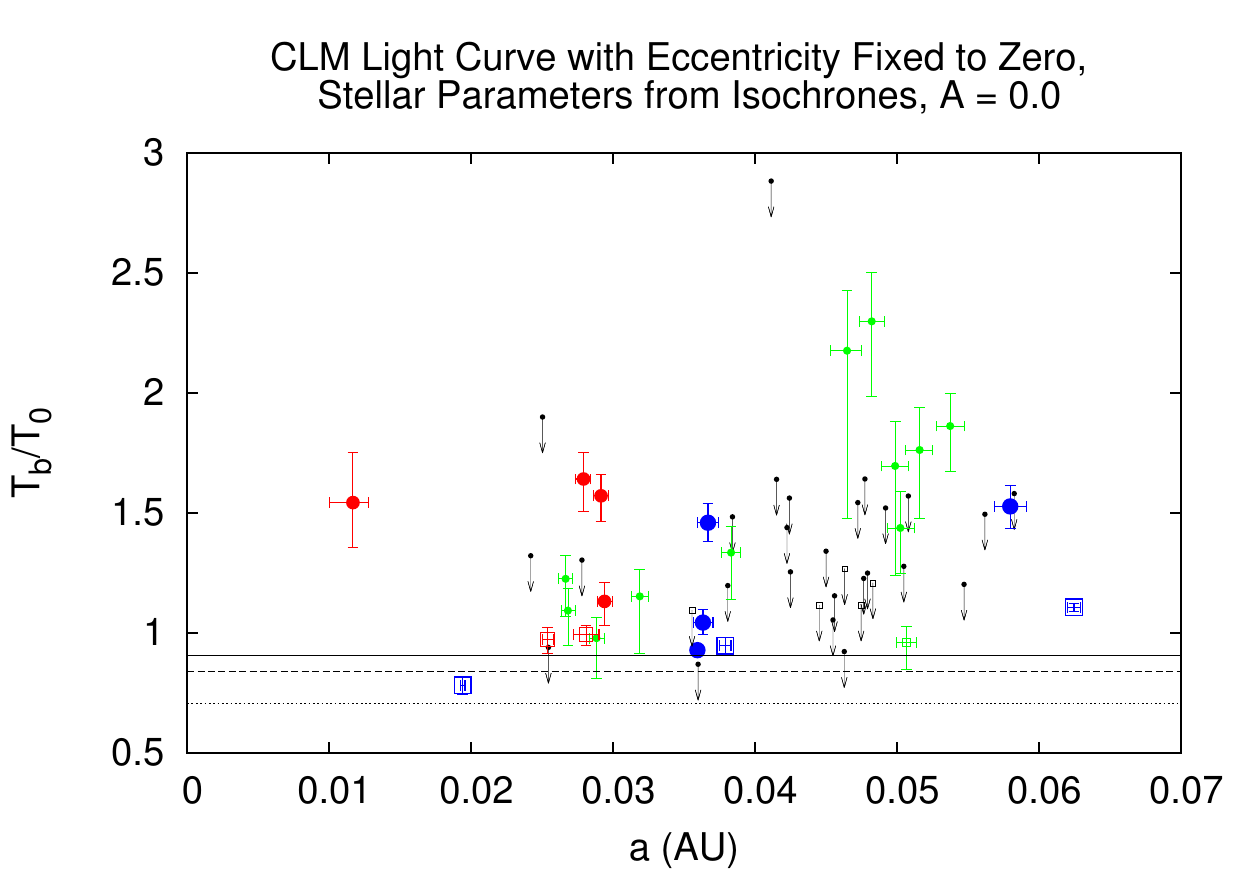}\\
  \epsfig{width=0.475\linewidth,file=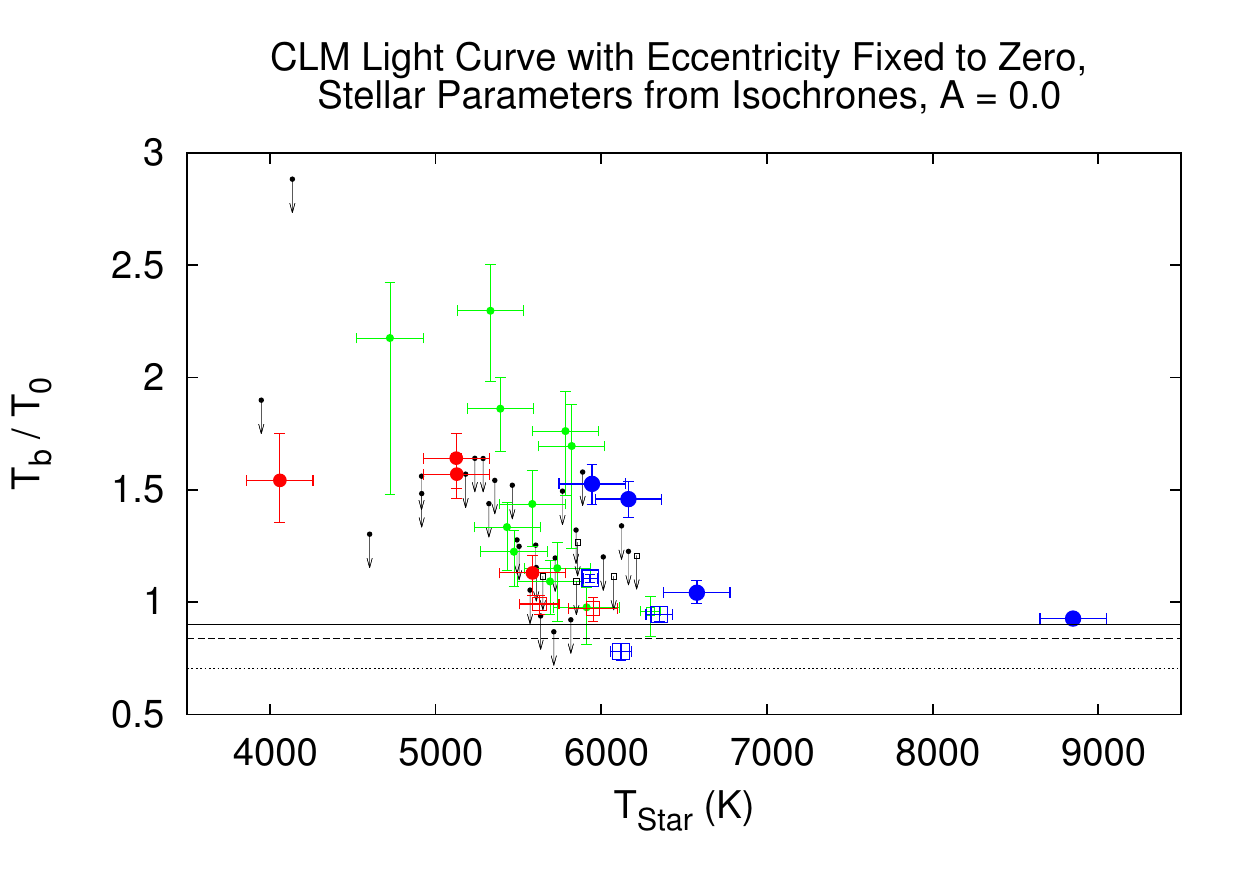}&
  \epsfig{width=0.475\linewidth,file=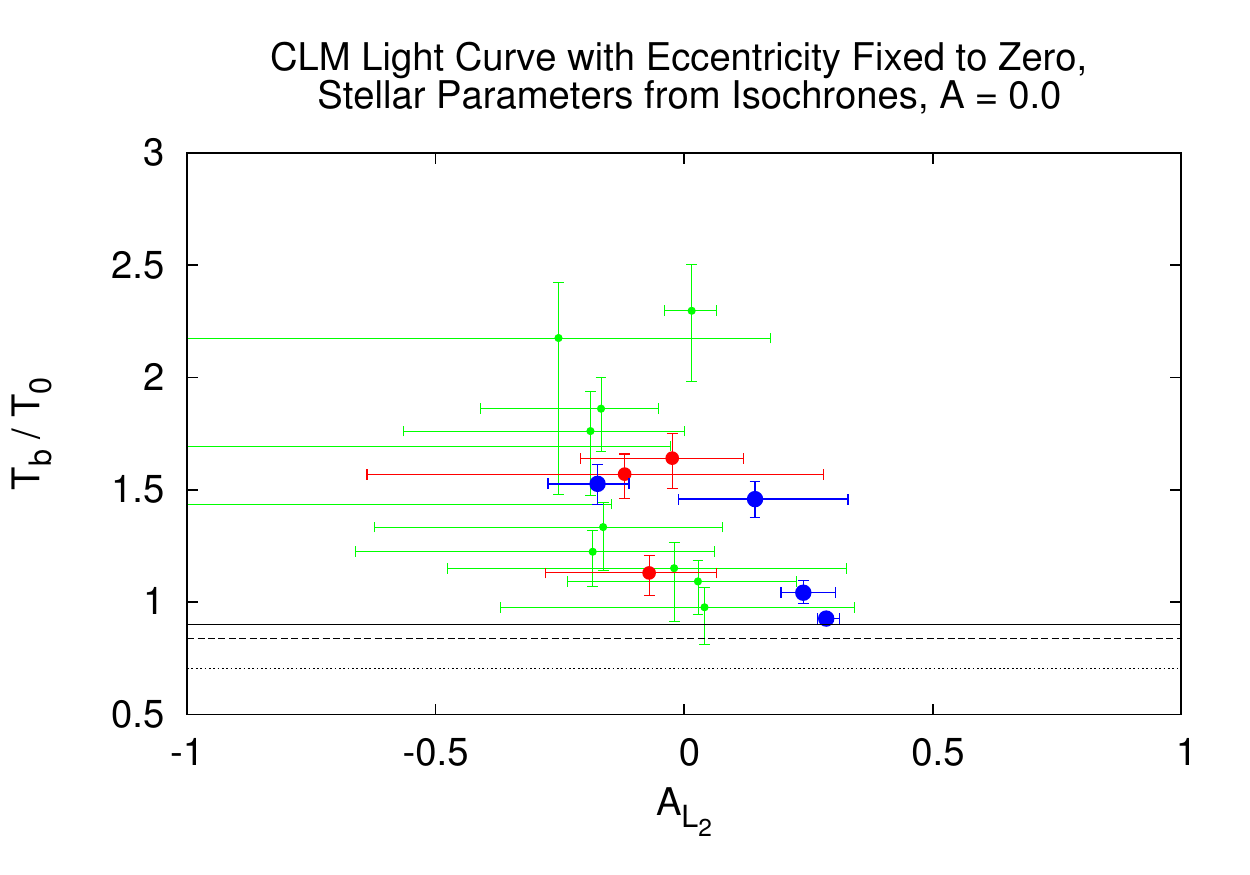}\\
\end{tabular}
\caption[Plots of the effective day side temperature ratio versus planetary radii, semi-major axis, stellar effective temperature, and out-of-eclipse luminosity variation]{Plots of the effective day side temperature ratio versus the radius of the planet (top-left), the semi-major axis of the system (top-right), the effective temperature of the star (bottom-left), and the amplitude of the sine curve applied to the planet's luminosity (bottom-right), using the CLM light curves, assuming no eccentricity, and deriving stellar parameters from stellar isochrones. Results for the PDC light curves are not shown, but produce similar results. Solid circles correspond to $Kepler$ systems modeled in this paper, while open squares are previously published detections or upper limits of exoplanet secondary eclipses at optical wavelengths. The x-axis errorbars are not shown for the $<$1$\sigma$ detections for clarity. The solid, dashed, and dotted black lines in each figure correspond to the expected temperature ratio assuming no recirculation, a uniform day-side temperature, and a uniform planetary temperature respectively.}
\label{rtrendfig}
\end{figure}

{\it Low~Mass~Stars,~Brown~Dwarfs,~or~Blends}: The last possibility we consider is that some of the objects in the list are in fact brown-dwarfs or low-mass stars, but this assumption also possesses some problems. Exoplanet search results over the years have revealed what appears to be a ``brown dwarf desert'' within orbital separations from the host star of $a$ $\lesssim$ 5 AU, for solar type stars \citep[see e.g.,][]{Grether2006}. However, the discovery of CoRoT-3b, a 21.7 $M_{Jup}$ object orbiting at a separation of only 0.057 AU around an F3-type star, has opened some debate about whether this object is really a brown dwarf, or if planets more massive than the defined Deuterium burning limit can form around stars more massive than the Sun. To test this idea we plot in Figure~\ref{rtrendfig} the measured $T_{b}$/$T_{0}$ of each planet candidate versus the effective temperature of the star, $T_{\star}$. We see no clear correlation between the temperature of the host star and an excess brightness of the planet candidates, and an error-weighted linear fit does not yield a statistically significant slope. We also plot in Figure~\ref{rtrendfig} the values for $T_{b}$/$T_{0}$ versus $R_{p}$, $a$, and $A_{L_{p}}$, but do not see any significant linear correlations either.

\begin{figure}[ht!]
\centering
\begin{tabular}{cc}
  \epsfig{width=0.475\linewidth,file=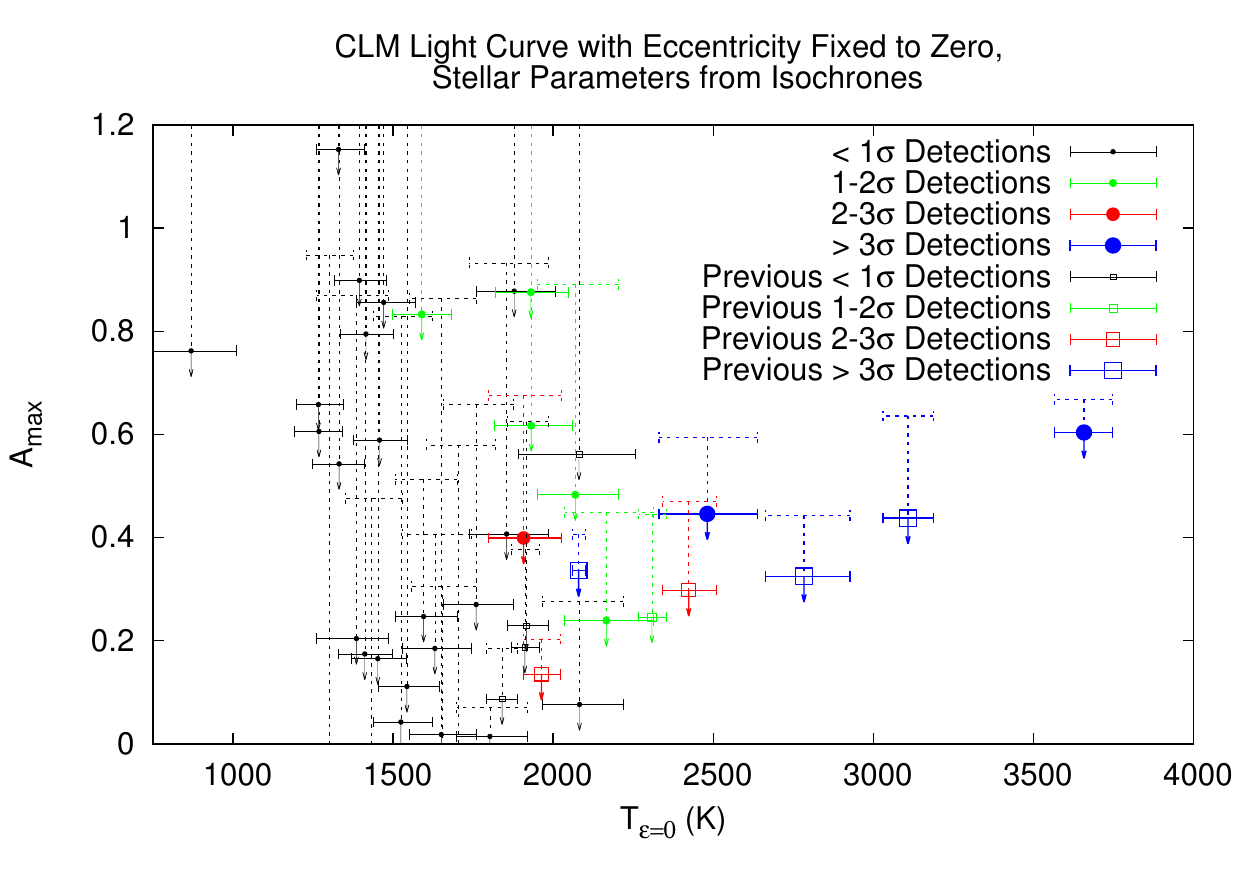}&
  \epsfig{width=0.475\linewidth,file=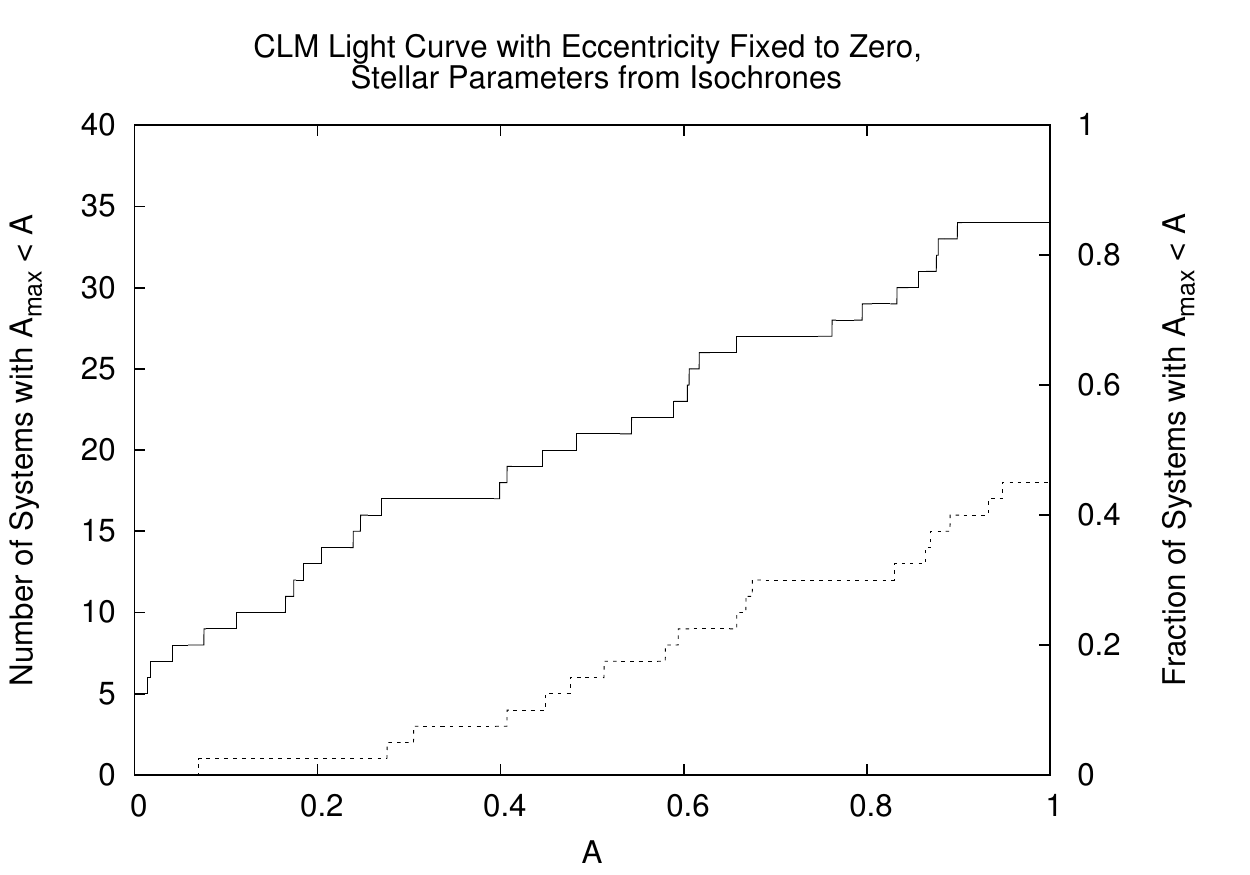}\\
  \epsfig{width=0.475\linewidth,file=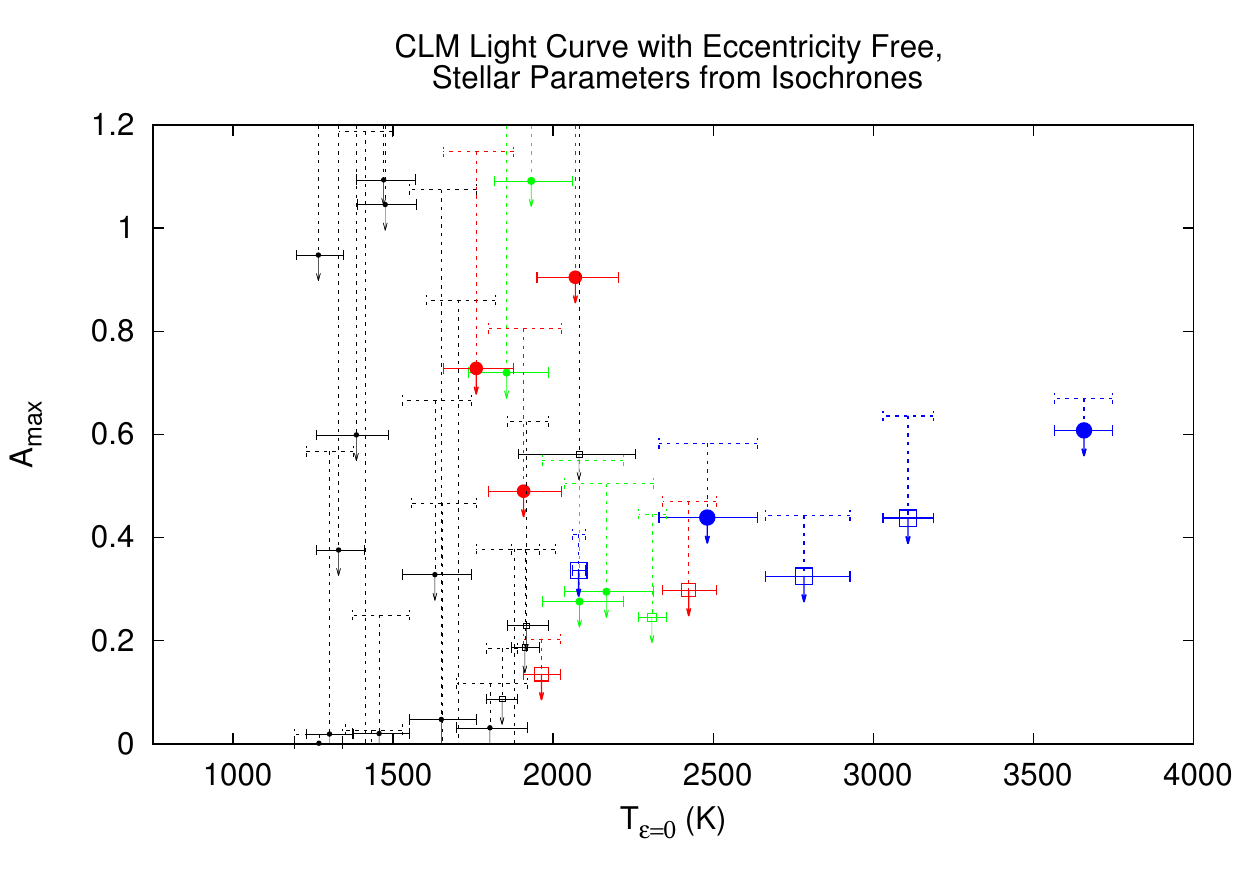}&
  \epsfig{width=0.475\linewidth,file=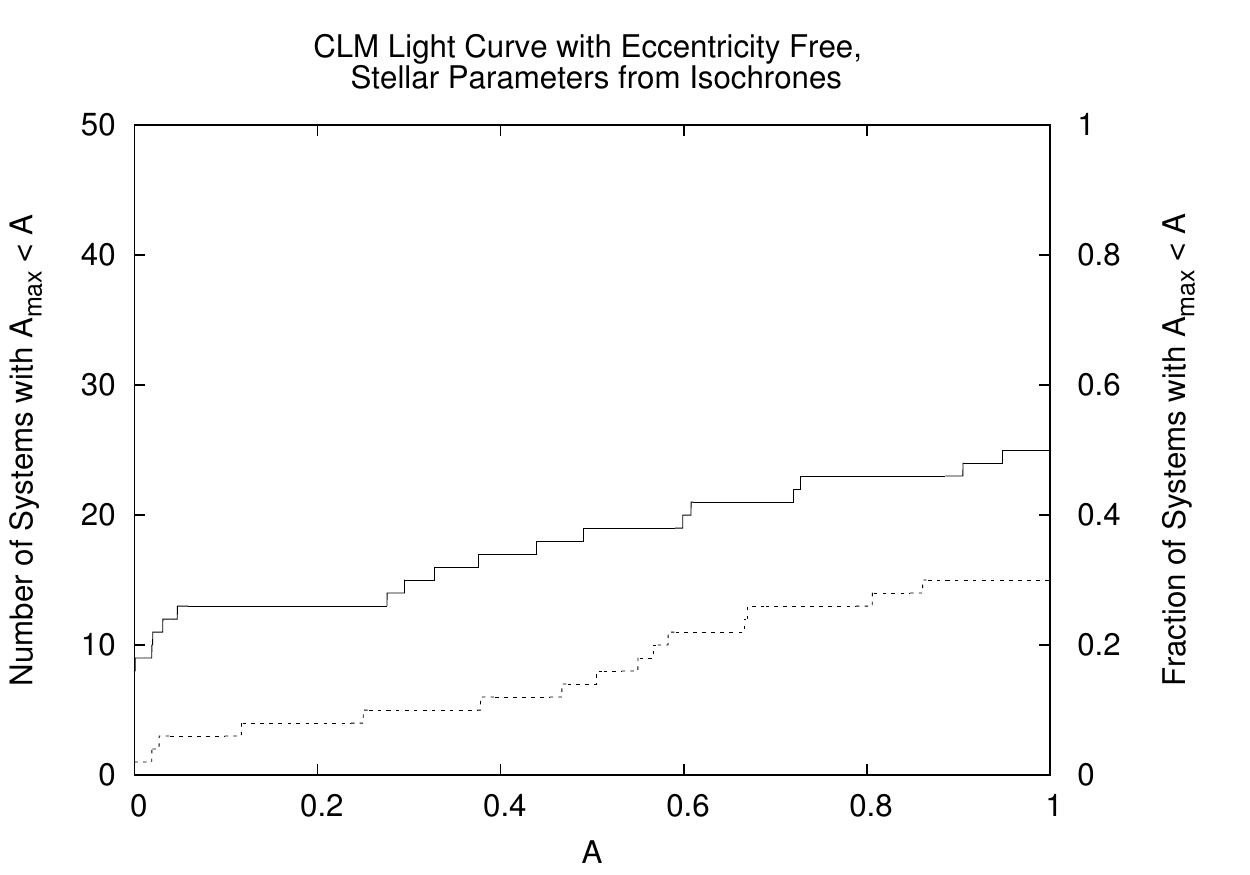}\\
\end{tabular}
\caption[Plots of the 1$\sigma$ and 3$\sigma$ upper limits on the maximum possible albedo]{Left Column: Plots of the 1$\sigma$ and 3$\sigma$ upper limits, delineated by solid and dashed lines respectively, on the maximum possible albedo values versus the maximum effective day side temperature when deriving stellar parameters from stellar isochrones, using the CLM pipeline light curves. Right Column: Plots of the cumulative number of systems and total fraction of all systems modeled in this paper that have their upper limit of $A_{max}$ below a given value of $A$, for both 1$\sigma$ and 3$\sigma$ upper limits. Values obtained when fixing eccentricity to zero are shown in first row, while values obtained when allowing eccentricity to vary are shown in the second row. Solid circles correspond to $Kepler$ systems modeled in this paper, while open squares are previously published detections or upper limits of exoplanet secondary eclipses at optical wavelengths. The results when using the $Kepler$ PDC curves are not plotted, but are very similar to the presented CLM light curves.}
\label{maxalbfig}
\end{figure}

We also utilize the upper limits on possible secondary eclipses we derived to examine potential trends in $A_{max}$ as computed using Eq.~\ref{maxalbeq}. In Figure~\ref{maxalbfig} we plot both the 1$\sigma$ and 3$\sigma$ upper limits, delineated by solid and dashed lines respectively, on the values of $A_{max}$ versus $T_{\epsilon = 0}$ for both fixing eccentricity to zero and allowing it to vary, for the CLM light curves and deriving stellar parameters from isochrones. We also include the values for previously published detections and upper limits of secondary eclipses in the optical wavelength regime. Also in Figure~\ref{maxalbfig} we plot the cumulative number of systems, and total fraction of all systems, that were modeled in this paper and that have their upper limit of $A_{max}$ below a given value of $A$, for both 1$\sigma$ and 3$\sigma$ upper limits. As can be seen, when fixing eccentricity to zero, we can generally obtain constraints on the maximum possible albedo for $\sim$85\% of the $Kepler$ systems at the 1$\sigma$ confidence level, and $\sim$45\% of systems at the 3$\sigma$ confidence level. When letting eccentricity vary, we can only constrain $A_{max}$ for $\sim$50\% and $\sim$30\% of the $Kepler$ systems at the 1$\sigma$ and 3$\sigma$ confidence levels respectively. However, comparing the $T_{\epsilon = 0}$ values of the previously published planets to that of the $Kepler$ candidates, we find we have significantly increased the number of systems with constrained albedos in the $T_{\epsilon = 0} \lesssim$ 2000 K regime. As can be seen, many of the systems in this temperature regime appear to have maximum possible albedos below 0.3 at the 1$\sigma$ confidence level, thus confirming previous findings of low albedos for hot Jupiters at optical wavelengths, and indicating that such low albedos may be common down to planetary temperatures of 1200~K.

\subsection{Properties of Some Individual Objects}
\label{indivsec}

In the previous section we analyzed the secondary eclipse detections as a set, in an attempt to find common characteristics of the sample. We have carefully examined each individual system, and in this section present and discuss the most interesting objects in more detail.

\subsubsection{KOI 1.01 / TrES-2b}

KOI 1.01 is also known as TrES-2b, and was discovered to be a transiting planet by \citet{ODonovan2006} before the $Kepler$ mission was launched. \citet{KippingBakos2011b} determined an upper limit to the eclipse depth of 37ppm at the 1$\sigma$ level, and 73 ppm at the 3$\sigma$ level, based on short-cadence Q0 and Q1 $Kepler$ data, thus limiting the geometric albedo to $A_{g}$ $<$ 0.146 at 3$\sigma$ confidence. \citet{KippingSpiegel2011} have recently published the detection of phase curve variations with an amplitude of 6.5$\pm$1.9 ppm using Q0-Q2 short cadence data, but do not detect the secondary eclipse itself, calculating that any secondary eclipse measurement must have an uncertainty of $\sim$13 ppm. If this variation is due to reflected light, then \citet{KippingSpiegel2011} calculate the albedo of the planet as A$_{g}$ = 0.0253$\pm$0.0072.

Using the CLM pipeline light curves, and fixing eccentricity to zero, we determined a secondary eclipse depth of -3.9$^{+8.1}_{-8.0}$ ppm, thus yielding upper limits on the eclipse depth of 4.2 and 20.4 ppm for 1$\sigma$ and 3$\sigma$ confidence levels respectively. We determined a value for the maximum possible albedo, $A_{max}$, using stellar values from the KIC, of -0.004$^{+0.010}_{-0.019}$, yielding upper limits on the albedo of 0.006 and 0.026 for 1$\sigma$ and 3$\sigma$ confidence levels respectively. If using values from stellar isochrones, we instead determine $A_{max}$ = -0.014$^{+0.028}_{-0.029}$, and thus 1$\sigma$ and 3$\sigma$ upper limits of 0.016 and 0.072, respectively. We did not detect any significant orbital phase variation, although a significant value of the luminosity of the planet must be found in order to produce a significant value of $A_{L_{p}}$ via our modeling technique. Given a difference in the calculated planet-to-star luminosity ratio between our measurements and those of \citet{KippingSpiegel2011} of 10.4$\pm$8.3 ppm, our results do not conflict with those of \citet{KippingSpiegel2011} at a confidence level greater than 1.25$\sigma$.  

We also note that the error on individual points in the Q2 long cadence data is 43 ppm, and thus given the predicted 77.1 minute occultation duration, the 29.4244 minute cadence of $Kepler$ long cadence data, the 88.7 days of coverage, and the 2.47 day orbital period of the system, we calculate that one could detect the secondary eclipse of the planet to a 1$\sigma$ precision of 4.5 ppm. This is in agreement with our 1$\sigma$ upper limit, although our formal 1$\sigma$ errors on the eclipse depth are twice as large, likely due to remaining systematics that were accounted for in the residual-permutation error analysis. However, with better systematic noise reduction, and an additional 1-2 quarters of data, the secondary eclipse of this planet could very well be detected to 3$\sigma$ confidence. Future efforts should be directed towards this goal to confirm the phase signal found by \citet{KippingSpiegel2011}, and ensure it is not due to remaining systematics in the $Kepler$ lightcurves or intrinsic stellar variability.

\subsubsection{KOI 2.01 / HAT-P-7b}

KOI 2.01 is also known as HAT-P-7 and was discovered by \citet{Pal2008} prior to the launch of the $Kepler$ mission. \citet{Borucki2009} detected a secondary eclipse in the Q0 $Kepler$ data of 130$\pm$11 ppm, and a 122 ppm phase variation. \citet{Christiansen2010} determined an independent 3$\sigma$ upper limit of 550 ppm on the secondary eclipse depth at optical wavelengths using the EPOXI spacecraft. Using Q1 $Kepler$ data, \citet{Welsh2010} determined a secondary eclipse depth of 85.8 ppm, a 63.7 ppm phase variation due to reflection from the planet, a 37.3 ppm phase variation due to ellipsoidal distortions in the star induced by tidal interaction between the planet and star, and determined an albedo in the $Kepler$ passband of 0.18, though no errors on the derived quantities were given.

We do not detect any significant orbital eccentricity ($>$3$\sigma$) in KOI 2.01 / HAT-P-7 in either the PDC or the CLM light curves when allowing eccentricity to vary. Fixing eccentricity to zero, for the $Kepler$ PDC light curve, we determine a planet-to-star luminosity ratio of 75.0$^{+9.5}_{-8.1}$ ppm and a value for $A_{L_{p}}$ of 0.421$^{+0.073}_{-0.071}$, and thus a 63.2$^{+13.6}_{-12.6}$ ppm phase variation. Similarly, for the CLM light curve, we derive a planet-to-star luminosity ratio of 77.7$^{+10.3}_{-9.5}$ ppm and a value for $A_{L_{p}}$ of 0.240$^{+0.065}_{-0.045}$, and thus a 37.3$^{+11.2}_{-8.3}$ ppm phase variation. Our determined eclipse depths are very consistent between the CLM and PDC light curves, and within $\sim$1$\sigma$ of the value found by \citet{Welsh2010}, though are not compatible with the value found by \citet{Borucki2009} at the $\sim$4$\sigma$ level. Examining the amplitude of the phase variation of the system, we first point out that there is a $\sim$1.5$\sigma$ difference in the values derived between the PDC and CLM light curves, and likely is a result of the different methods employed to remove systematic noise. While the high-frequency signal of the secondary eclipse was not affected, the low-frequency signal of the phase variation, with a period of $\sim$2.2 days, was much more easily distorted. It is not clear which measurement on the phase variation is more valid, though the CLM pipeline light curves yields a $\chi^{2}_{red}$ of 7.6, versus a value of 26.5 for the PDC data. This result should highlight the level of care that needs to be taken when examining and interpreting phase variations and other low-frequency signals in $Kepler$ data.

Finally, utilizing stellar isochrones for the mass and radius determination of the host star, we determine 3$\sigma$ upper limits to the maximum albedo of the planet of 0.556 and 0.594 for the PDC and CLM light curves respectively, in agreement with the values determined by \citet{Welsh2010}.

\subsubsection{KOI 10.01 / Kepler-8b}

KOI 10.01 is also known as Kepler-8b, and was first discovered to be a transiting planet by \citet{Jenkins2010c}. \citet{KippingBakos2011a} found that the orbit is consistent with a circular orbit, and placed a 3$\sigma$ upper limit on a secondary eclipse of 101.1 ppm, thus constraining the albedo to $<$ 0.63.

We do not detect any significant eccentricity ($>$3$\sigma$) in either the PDC or CLM light curves when allowing eccentricity to vary. Fixing the eccentricity to zero, we place 3$\sigma$ upper limits on the secondary eclipse of the planet at 119 ppm, (14.9$^{+34.7}_{-13.9}$ ppm), and 114 ppm, (19.6$^{+31.6}_{-33.7}$ ppm), for the PDC and CLM light curves respectively. Deriving the parameters for the host star from stellar isochrones yields a 3$\sigma$ limit on the maximum albedo of 0.898 and 0.933 for the PDC and CLM light curves respectively. Thus, we do not provide any additional constraints on the atmosphere of this planet over previous studies.

\subsubsection{KOI 13.01}

KOI 13 was noted by \citep{Borucki2011} to be a double star, and unresolved in the $Kepler$ images due to the $\sim$4$\arcsec$ pixel size. \citet{Szabo2011} recently conducted a thorough analysis of the system with careful detail to isolating the transiting planet candidate in the double star system, and concluded that KOI 13.01 is likely a brown dwarf with a radius of 2.2$\pm$0.1 $R_{J}$. They also concluded that the transit showed an asymmetrical profile, due to the rapidly rotating nature of the host of KOI 13.01, detected a secondary eclipse with a depth of 120$\pm$10 ppm, and did not find any significant orbital eccentricity. More recently, \citet{Shporer2011} determined a mass of 9.2$\pm$1.1 M$_{Jup}$ via photometric beaming, and \citet{Barnes2011} used the asymmetrical profile of the transit to measure a planetary radius of 1.756$\pm$0.014 R$_{\sun}$, thus making it more likely that this object is a massive hot Jupiter. \citet{Mazeh2011} further support this characterization, and additionally measure a secondary eclipse depth of 163.8$\pm$3.8 ppm.

We do not detect any significant eccentricity ($>$3$\sigma$) in either the PDC or CLM light curves when allowing eccentricity to vary, and we can confirm that the asymmetrical transit shape exists in both the PDC and CLM lightcurves. Fixing the eccentricity to zero, we determine a secondary eclipse depth of 124.3$^{+6.9}_{-7.8}$ and 125.6$^{+6.3}_{-8.6}$ ppm for the PDC and CLM light curves respectively, which are statistically consistent with each other and the values found by \citet{Szabo2011}, though is discrepant at the 5$\sigma$ level with the value measured by \citet{Mazeh2011}. We note that KOI 13.01 is the planet with the highest value of $T_{\epsilon=0}$ in our sample, and one of the few that is consistent with a $T_{b}$/$T_{0}$ value of less than 1.0, although we did not take third light into account in our analysis.

\subsubsection{KOI 17.01 / Kepler-6b}

KOI 17.01 is also known as Kepler-6b, and was discovered to be a transiting planet by \citet{Dunham2010}. \citet{KippingBakos2011a} did not find any evidence for a non-circular orbit, and constrained any possible secondary eclipses to less than 51.5 ppm, and thus a geometric albedo less than 0.32, both at 3$\sigma$ confidence. \citet{Desert2011a} was able to use $Kepler$ data from Q0-Q5, of which Q3-Q5 were not yet publicly accessible at the time of writing, to measure a secondary eclipse of 22$\pm$7 ppm, and did not find any evidence for a non-circular orbit. Combining this eclipse measurement with others obtained via $Spitzer$, they determined a geometric albedo of 0.11$\pm$0.04.

We do not detect any significant eccentricity ($>$3$\sigma$), and place a 3$\sigma$ upper limit on any possible secondary eclipse at 34.5 ppm, (-17.1$^{+17.2}_{-22.3}$ pm). This constrains the planetary albedo, at the 3$\sigma$ level, to less than 0.25 and 0.31 when deriving stellar parameters from the KIC and isochrones respectively, consistent with both the values derived by \citet{KippingBakos2011a} and \citet{Desert2011a}.

\subsubsection{KOI 18.01 / Kepler-5b}

KOI 18.01 is also known as Kepler-5b, and was discovered to be a transiting planet by \citet{Koch2010}. \citet{KippingBakos2011a} did not find any evidence for a non-circular orbit, though found weak evidence for a secondary eclipse with a depth of 26$\pm$17 ppm, implying a geometric albedo of 0.15$\pm$0.10. \citet{Desert2011a} detected the secondary eclipse, again using Q0-Q5 data, to greater precision and determined a depth of 21$\pm$6 ppm, which they combined with $Spitzer$ observations to determine a geometric albedo of 0.12$\pm$0.04.

We do not detect any significant eccentricity ($>$3$\sigma$), and do not detect the secondary eclipse in either the PDC or CLM light curves, placing a 3$\sigma$ upper limit on the eclipse depth of 62.9 ppm (-27.4$^{+30.1}_{-33.2}$ ppm) using the CLM light curve. The derived median value and associated 1$\sigma$ uncertainties on the secondary eclipse depth for the PDC data is 0.21$^{+1.4}_{-0.3}$ ppm, in obvious contradiction to the previously mentioned measured eclipse depths. However, the PDC light curve for KOI 18.01 appears to suffer from a high level of systematic noise, and inspection of the parameter distribution histograms for the surface brightness ratio and luminosity ratio reveal them to significantly deviate from a Gaussian shape, having directly derived 2$\sigma$ uncertainties of $^{+36.1}_{-10.7}$ ppm, thus providing a more reasonable 3$\sigma$ upper limit on the eclipse depth of 54.4 ppm for the PDC light curve.

\subsubsection{KOI 20.01 / Kepler-12b}

KOI 20.01 has recently been announced by \citet{Fortney2011} as Kepler-12b, a 1.7 $R_{J}$, 0.43 $M_{J}$ planet orbiting a slightly evolved G0 star at a period of 4.4 days. Using $Kepler$ data from Q0-Q7, of which Q3-Q7 were not publicly accessible at the time of writing, they were able to measure a 31$\pm$8 ppm secondary eclipse, which implies a geometric albedo of 0.14$\pm$0.04 when combined with additional $Spitzer$ observations. They also do not detect any significant orbital eccentricity.

Using our CLM light curves, we derived a 3$\sigma$ upper limit on the eclipse depth of 56.3 ppm, (4.7$^{+17.2}_{-9.7}$ ppm), when fixing eccentricity to zero, implying a 3$\sigma$ upper limit on the maximum possible albedo of the planet of 0.40. These results are in agreement with the values recently found by \citet{Fortney2011}.

\subsubsection{KOI 64.01}

\citet{Borucki2011} noted that KOI 64.01 may be a binary system composed of a F-type primary and a M-type secondary. We do not detect any significant eccentricity ($>$3$\sigma$) in the system, though we do detect marginal evidence for a secondary eclipse in the system. Fixing the eccentricity to zero, we detect a secondary eclipse with a depth of 47.7$^{+25.0}_{-23.3}$ and 75.1$^{+35.1}_{-27.7}$ ppm, (2.0$\sigma$ and 2.7$\sigma$ detections), for the PDC and CLM light curves respectively. Taking the CLM light curve detection as a more reliable measurement, given its $\chi_{red}$ value of 7.3 versus a value of 10.4 for the PDC data, we cannot provide any constraints on the maximum albedo, and in fact even an albedo of 1.0 cannot account for this level of emission. Assuming it has an albedo of 0.0, we calculate a value of $T_{b}$/$T_{0}$ = 1.05$^{+0.29}_{-0.23}$ if deriving stellar values from the KIC, and 1.57$^{+0.09}_{-0.11}$ if using stellar isochrones, which are both above the maximum allowed value of (2/3)$^\frac{1}{4}$ for a planet with no heat redistribution. Thus, this object, if the secondary eclipse detection is real, likely has a significant source of internal energy generation, and certainly may be a brown dwarf or low-mass star. We note however that the effective temperature for the star given in the KIC is 5128 K, which would suggest a K0 spectral type, not F-type. Thus if the host star is a K0 dwarf, the companion, if not a planet, would likely be a brown dwarf, unless the system is composed of a foreground K0 dwarf and a background F+M type eclipsing binary.

\subsubsection{KOI 97.01 / Kepler-7b}

KOI 97.01 is also known as Kepler-7b, and was discovered to be a transiting planet by \citet{Latham2010}. \citet{KippingBakos2011a} found a secondary eclipse depth of 47$\pm$14 ppm, implying a geometric albedo of 0.38$\pm$0.12, and found a day/night flux difference of 17$\pm$9 ppm. \citet{Demory2011} used $Kepler$ data from Q0-Q4, of which Q3 and Q4 were not yet publicly available at the time of writing, to detect a secondary eclipse of 44$\pm$5 ppm, which implied an albedo of 0.32$\pm$0.03. They also detected an orbital phase curve with an amplitude of 42$\pm$4 ppm. Neither study found evidence for significant orbital eccentricity.

We obtain $>$3$\sigma$ detections of the secondary eclipse in both the PDC and CLM light curves, both fixing eccentricity to zero and allowing it to vary, though we do not detect any significant ($>$3$\sigma$) eccentricity in the system. Fixing the eccentricity to zero, we obtain secondary eclipse depths of 53.2$^{+14.1}_{-13.0}$ ppm and 66.1$^{+17.4}_{-17.5}$ ppm for the PDC and CLM light curves respectively, however we find we cannot place any significant limits on the maximum possible albedo. We determine values for $A_{L_{p}}$ of 0.12$\pm$0.24 and -0.17$^{+0.06}_{-0.10}$ for the PDC and CLM light curves respectively, which translate to phase variations of 12.8$^{+25.8}_{-25.7}$ ppm and -22.5$^{+9.9}_{-14.5}$, neither of which is significant at a $>$3$\sigma$ level. 

Both of our values for the eclipse depth are consistent with those obtained by \citet{KippingBakos2011a} at $<$1$\sigma$ discrepancy, and at $<$1.5$\sigma$ with those of \citet{Demory2011}. Upon inspection of the data, it turns out that we are not able to significantly constrain the albedo, when both \citet{KippingBakos2011a} and \citet{Demory2011} were able to, due to the values we adopt for the stellar mass and radius. We determined median values of 1.09 $M_{\sun}$ and 1.28 $R_{\sun}$ using the KIC, and 1.09 $M_{\sun}$ and 1.03 $R_{\sun}$ via stellar isochrones, where the other studies adopted values of $\sim$1.3 $M_{\sun}$ and $\sim$1.9 $R_{\sun}$, as \citet{Latham2010} found this star to be a G-type sub-giant. Unfortunately the KIC did not hint at this star being non-main-sequence, and the stellar isochrones we employ assume the host star is on the main-sequence. Obviously changing the stellar radius by a factor of $\sim$2 greatly impacts the estimate of reflected light, and this should emphasize the connection between the assumed stellar properties and derived planetary properties. Although we do not detect significant phase variations, our obtained values for the PDC light curve are not in conflict with either previously published result at greater than $\sim$1$\sigma$ significance. The value for the CLM light curve does conflict at $>$3$\sigma$ significance, and we attribute it to differences in the light curve processing from the pixel level data. This is another example of how low-frequency signals can change significantly in $Kepler$ data depending on the reduction technique employed.

\subsubsection{KOI 183.01}

We highlight KOI 183.01 due to a possibly significant detection of its secondary eclipse and eccentricity. Using the CLM light curves, we obtain a value for the secondary eclipse depth of 14$^{+42}_{-40}$ ppm when fixing eccentricity to zero, but a value of 125$^{+42}_{-39}$, (3.2$\sigma$), when allowing eccentricity to vary. In the latter case, we measure values of $e\cdot$cos($\omega$) = -0.152$^{+0.009}_{-0.008}$ and $e\cdot$sin($\omega$) = 0.03$^{+0.12}_{-0.14}$, yielding values of $e$ = 0.178$^{+0.075}_{-0.023}$ and $\omega$ = 169$^{+47}_{-34}$ degrees.

We employ the Bayesian Information Criterion (BIC) \citep{Schwarz1978} to determine if allowing the eccentricity to vary, thus adding two more free parameters, provides a statistically significantly better fit to the data than fixing it to zero. Given competing models with different values of $\chi^{2}$, and a different number of free parameters, $k$, the value

\begin{equation}
  BIC = \chi^{2} + k\cdot ln(N)
\end{equation}

\noindent is computed, where $N$ is the number of data points, and the model with the lowest BIC value is the preferred model. Given the 3,720 data points in the light curve, 8 free parameters with a $\chi^{2}$ value of 12499 when fixing the eccentricity to zero, and 10 free parameters with a $\chi^{2}$ value of 12476 when allowing eccentricity to vary, values for the BIC of 12565 and 12558 are obtained for the fixed and free eccentricity models respectively. Given the lower BIC value for the eccentricity free model, and given that the resulting parameter distributions from the residual permutation analysis are well-behaved Gaussian curves and do not show any anomalies (see Figure~\ref{ourlcs}.14), we conclude that the eccentricity free model is preferred and statistically significant.

Adopting values for the host star from stellar isochrones, we compute a maximum possible albedo of 0.52$^{+0.21}_{-0.19}$, or a value of $T_{b}$/$T_{0}$ = 1.28$^{+0.08}_{-0.09}$. These are both reasonable values for a moderately reflective planet, one heated beyond radiative thermal equilibrium via tidal heating due to its significant eccentricity, or likely a combination of the two. Additional $Kepler$ data and other follow-up observations should hopefully confirm this detection.

\subsubsection{KOI 196.01}

KOI 196.01 has recently been confirmed as a 0.49 $M_{J}$, 0.84 $R_{J}$, transiting planet by \citet{Santerne2011} via SOPHIE RV measurements and an analysis of the $Kepler$ light curve. They detect a secondary eclipse with a depth of 64$^{+10}_{-12}$ ppm, with corresponding phase variations, and determine a geometric albedo of 0.30$\pm$0.08. They do not find any significant orbital eccentricity.

We also detect the secondary eclipse with a depth of 77$\pm$24 ppm and 63$^{+33}_{-32}$ ppm in the PDC and CLM light curves respectively, and also do not find any evidence for significant orbital eccentricity, though we do not find any significant orbital phase variation. Via our eclipse depths, and utilizing stellar isochrones, we determine a maximum possible albedo of 0.29$^{+0.10}_{-0.09}$ and 0.26$^{+0.14}_{-0.13}$ for the PDC and CLM light curves respectively. Both of our values for the eclipse depth and geometric albedo are consistent and agree with those determined by \citet{Santerne2011}.

\subsubsection{KOI 202.01}

We highlight KOI 202.01 due to a possible detection of its secondary eclipse, and a robust upper limit on its albedo. We measure the secondary eclipse depth of the system as 69$^{+31}_{-30}$ ppm (2.3$\sigma$) and 46$^{+31}_{-36}$ ppm (1.3$\sigma$), for the PDC and CLM light curves respectively, and do not detect any significant orbital eccentricity nor phase variations. Although these are not significant enough to claim a robust detection, they are certainly interesting enough results to merit further follow-up with additional data, and place robust 3$\sigma$ upper limits on the maximum albedo of the system at 0.46 and 0.43 for the PDC and CLM light curves respectively, when deriving host star parameters from stellar isochrones.

\subsubsection{KOI 203.01 / Kepler-17b}

KOI 203.01 is also known as Kepler-17b, and was first confirmed as a transiting planet by \citet{Desert2011b}. Using $Kepler$ Q0-Q6 observations, of which Q3-Q6 were not yet publicly available at the time of writing, along with follow-up observations from $Spitzer$, they were able to detect a secondary eclipse with a depth of 58$\pm$10 ppm, and thus an albedo of 0.10$\pm$0.02, while finding no evidence for any orbital eccentricity. \citet{Bonomo2011} also detect the secondary eclipse at the 2.5$\sigma$ level using Q1-Q2 observations, after fitting and subtracting 4$^{th}$ order polynomials to 4-day segments of the light curve, with a depth of 52$\pm$21 ppm and consistent with a circular orbit.

We were only able to obtain a 3$\sigma$ upper limit on the secondary eclipse depth of 159 ppm (-15$\pm$58 ppm) using the CLM light curve and fixing eccentricity to zero. This corresponds to 3$\sigma$ upper limits on the albedo of 0.26 and 0.28 when deriving the host star parameters from the KIC and stellar isochrones respectively. Our results are thus statistically consistent with the detections of \citet{Desert2011b} and \citet{Bonomo2011}, although our errorbars are much larger. Examining the data, this star's lightcurve has large out of eclipse variations, on the order of the depth of the primary transit, due to both intrinsic stellar variability and systematics introduced by the star's movement, that rendered the PA and PDC data unmodelable. Our CLM pipeline was able to remove a large amount of this variability, (enough to reliably measure the primary transit), but did not fully remove it all, as illustrated by the $\chi_{red}^{2}$ value of 19.4. Thus, the eclipse signal for this planet is below the noise level for the CLM lightcurve.

\subsubsection{KOI 1541.01}

We highlight this system due to the unusually deep secondary eclipse and high eccentricity we detect at $>$3$\sigma$ confidence. However, the value for the eccentricity, $\sim$0.78, along with the unusually deep eclipse depth of $\sim$1100 ppm, and a large $\chi_{red}^{2}$ value of 28.7, lead us to believe this system suffers from severe systematics which happened to phase together in such a way as to create an artificial eclipse. If the signal is real, then this system must be a background eclipsing binary blend or other similar object.

\subsubsection{KOI 1543.01}

This system is very similar to KOI 1541.01 in that we also obtain a $>$3$\sigma$ detection of a secondary eclipse, but with an unusually high values for the eclipse depth, eccentricity, and $\chi_{red}^{2}$. Inspection of the light curve also reveals this system to contain significant systematics, or else must be a background eclipsing binary.

\subsection{Summary and Conclusions}
\label{concsec}

We have analyzed the $Kepler$ Q2 light curves of 76 hot Jupiter transiting planet candidates using both the $Kepler$ PDC data and the results from our own photometric pipeline for producing light curves from the pixel-level data. Of the 76 initial candidates only 55 have light curves with high enough photometric stability to search for secondary eclipses. For the remaining targets this search is hindered by either intrinsic variability of the host star or residual systematics in the light curve analyses. We have found that significant systematics in the $Kepler$ light curves due to small photometric apertures and large centroid motions hinder analyses if not properly removed or accounted for, and that a re-reduction of the photometry is best done at the pixel-stamp level. We also stress the importance of taking into account how $Kepler$ light curves are produced from the pixel level data when considering any detection of low-frequency signals, as they can vary significantly depending on the technique employed. 

We have also re-determined the stellar and planetary parameters of each system while deriving robust errors that take into account residual systematic noise in the light curves. We detect what appear to be the secondary eclipse signals of $\sim$20-30 of the targets in our list at $> 1 \sigma$ confidence levels, and also derive robust upper limits for the secondary eclipse emission of all the remaining objects in our sample. All of our sample present excess emission compared to what is expected via blackbody thermal emission alone, as well as a trend of increasing excess emission with decreasing expected maximum effective planetary temperature, in agreement with previously reported secondary eclipse detections of hot Jupiters in the optical, which can be attributed to the 'appearance' of increasing albedos with decreasing planetary temperatures. By performing statistical analyses of those results we arrive to the following main conclusions.

\begin{enumerate}

\item Assuming no contribution from reflected light, i.e., A = 0.0, the majority of the detected secondary eclipses reveal thermal emission levels higher than the maximum emission levels expected for planets in local thermodynamical equilibrium.

\item While the extra emission from many of the planets can be accounted for by varying the amount of reflected light, the emission from  $\sim$50\% of the detected objects ($>$1$\sigma$) can not be accounted for even when assuming perfectly reflective planets (i.e., A = 1.0). These planets must either have much higher thermal emission in the optical compared to a blackbody, as predicted by theoretical models, have very large non-LTE optical emission features, have underestimated host star masses, radii, or effective temperatures, or are in fact false-positives and not planets but rather brown dwarfs, very low-mass stars, or stellar blends. Follow-up observations of these systems are necessary to confirm this conclusion. The most outstanding potentially false positive systems are KOI 64.01, 144.01, 684.01, 843.01, 1541.01, and 1543.01. Given that this is 6 of the 55 systems that we modeled, or 11\% of the sample, it appears to agree with the expected $\sim$10\% false positive rate of the initial 1,235 candidates estimated by \citet{Morton2011}. We note that \citet{Santerne2012} found a potential false positive rate for $Kepler$ hot Jupiters as high as 35\% from radial-velocity observations of KOIs, which could certainly be compatible with our data if we count more of the systems with unusually deep eclipses as false positives.

\item Although we do not identify a sole cause of the observed trend of increasing excess planetary emission with decreasing expected maximum effective planetary temperature, the hypothesis of increasing planetary albedo with decreasing planetary temperature is able to explain many of the systems. This would be physically plausible as the upper atmospheres of Jupiter-like planets transition from very low albedos at high temperatures, as observed for hot Jupiters, to higher albedos at lower temperatures, as observed for cool Jupiters in our own solar system. We note that further observations via $Kepler$ and multi-wavelength ground-based facilities of both the planetary candidates and their host stars are still needed to fully explain this trend.

\item From the emission upper limits placed on planet candidates for which we do not detect secondary eclipse signals, we conclude that a significant number, at least 30\% at the 1$\sigma$ level, of those targets must have very low-albedos, ($A_{g}$ $<$ 0.3), which is a result that is consistent with the majority of previous observations and early theoretical hot Jupiter model predictions. All previous observational upper limits had been placed on hot Jupiters with expected atmospheric effective temperatures higher than $\sim$1650 K. Our results extend that temperature limit to planets with expected effective temperatures higher than 1200 K and can help further establish what chemicals play a critical role in the atmospheric properties of hot Jupiters.

\item From the inspection of individual targets we conclude that the majority of our secondary eclipse depths for candidates with previously published eclipse detections are consistent with the results from those other studies. We note that several of those other studies have access to $Kepler$ data from quarters after Q2, while our analysis is limited to just the public Q2 data, and therefore our results have larger detection errorbars in some cases.

\item Our results are based on only one quarter of $Kepler$ data, but 12-24 quarters should become available in the future. Therefore, we expect that future studies of these targets will be able to improve our secondary eclipse detections and upper limits by factors of 3-5, or much greater if noise systematics in the light curves can be further reduced.

\end{enumerate}

%% file: kepsec-tab1.tex
1.01 & 011446443 & 11.338 & 5713 & 4.14 & -0.139 & 1.133$^{_{+1.726}}_{^{-0.681}}$ & 1.496$^{_{+1.400}}_{^{-0.727}}$ & 1.019$^{_{+0.060}}_{^{-0.053}}$ & 0.943$^{_{+0.074}}_{^{-0.060}}$ \\
2.01 & 010666592 & 10.463 & 6577 & 4.32 & 0.000 & 1.351$^{_{+2.059}}_{^{-0.812}}$ & 1.336$^{_{+1.251}}_{^{-0.649}}$ & 1.316$^{_{+0.078}}_{^{-0.075}}$ & 1.363$^{_{+0.115}}_{^{-0.113}}$ \\
5.01 & 008554498 & 11.665 & 5766 & 4.04 & 0.116 & 1.176$^{_{+1.720}}_{^{-0.706}}$ & 1.748$^{_{+1.574}}_{^{-0.846}}$ & 1.035$^{_{+0.060}}_{^{-0.055}}$ & 0.962$^{_{+0.077}}_{^{-0.065}}$ \\
10.01 & 006922244 & 13.563 & 6164 & 4.44 & -0.128 & 1.110$^{_{+1.718}}_{^{-0.666}}$ & 1.056$^{_{+1.002}}_{^{-0.510}}$ & 1.162$^{_{+0.074}}_{^{-0.066}}$ & 1.133$^{_{+0.109}}_{^{-0.093}}$ \\
13.01 & 009941662 & 9.958 & 8848 & 3.93 & -0.141 & 1.813$^{_{+2.807}}_{^{-1.088}}$ & 2.454$^{_{+2.329}}_{^{-1.186}}$ & 1.989$^{_{+0.055}}_{^{-0.053}}$ & 1.616$^{_{+0.021}}_{^{-0.022}}$ \\
17.01 & 010874614 & 13.000 & 5724 & 4.47 & 0.000 & 0.908$^{_{+1.353}}_{^{-0.538}}$ & 0.911$^{_{+0.842}}_{^{-0.443}}$ & 1.022$^{_{+0.060}}_{^{-0.054}}$ & 0.947$^{_{+0.075}}_{^{-0.061}}$ \\
18.01 & 008191672 & 13.369 & 5816 & 4.46 & 0.000 & 0.922$^{_{+1.447}}_{^{-0.551}}$ & 0.942$^{_{+0.862}}_{^{-0.454}}$ & 1.050$^{_{+0.062}}_{^{-0.059}}$ & 0.981$^{_{+0.081}}_{^{-0.071}}$ \\
20.01 & 011804465 & 13.438 & 6012 & 4.47 & -0.161 & 1.069$^{_{+1.637}}_{^{-0.639}}$ & 0.999$^{_{+0.933}}_{^{-0.483}}$ & 1.110$^{_{+0.071}}_{^{-0.061}}$ & 1.059$^{_{+0.100}}_{^{-0.080}}$ \\
64.01 & 007051180 & 13.143 & 5128 & 3.94 & -0.341 & 1.180$^{_{+1.737}}_{^{-0.694}}$ & 1.956$^{_{+1.729}}_{^{-0.936}}$ & 0.868$^{_{+0.048}}_{^{-0.047}}$ & 0.786$^{_{+0.046}}_{^{-0.042}}$ \\
97.01 & 005780885 & 12.885 & 5944 & 4.27 & 0.052 & 1.086$^{_{+1.705}}_{^{-0.649}}$ & 1.279$^{_{+1.170}}_{^{-0.616}}$ & 1.089$^{_{+0.067}}_{^{-0.060}}$ & 1.031$^{_{+0.094}}_{^{-0.077}}$ \\
102.01 & 008456679 & 12.566 & 5919 & 3.90 & -0.358 & 1.216$^{_{+1.862}}_{^{-0.727}}$ & 2.058$^{_{+1.922}}_{^{-0.994}}$ & 1.081$^{_{+0.067}}_{^{-0.061}}$ & 1.020$^{_{+0.092}}_{^{-0.075}}$ \\
127.01 & 008359498 & 13.938 & 5570 & 4.53 & 0.174 & 1.043$^{_{+1.581}}_{^{-0.634}}$ & 0.921$^{_{+0.850}}_{^{-0.436}}$ & 0.980$^{_{+0.056}}_{^{-0.053}}$ & 0.897$^{_{+0.066}}_{^{-0.055}}$ \\
128.01 & 011359879 & 13.758 & 5718 & 4.18 & 0.362 & 1.139$^{_{+1.700}}_{^{-0.688}}$ & 1.443$^{_{+1.284}}_{^{-0.699}}$ & 1.020$^{_{+0.061}}_{^{-0.054}}$ & 0.945$^{_{+0.076}}_{^{-0.061}}$ \\
144.01 & 004180280 & 13.698 & 4724 & 4.00 & 0.241 & 1.098$^{_{+1.680}}_{^{-0.662}}$ & 1.744$^{_{+1.568}}_{^{-0.838}}$ & 0.767$^{_{+0.053}}_{^{-0.055}}$ & 0.694$^{_{+0.049}}_{^{-0.052}}$ \\
183.01 & 009651668 & 14.290 & 5722 & 4.71 & -0.141 & 1.012$^{_{+1.536}}_{^{-0.608}}$ & 0.734$^{_{+0.678}}_{^{-0.353}}$ & 1.022$^{_{+0.062}}_{^{-0.054}}$ & 0.947$^{_{+0.077}}_{^{-0.062}}$ \\
186.01 & 012019440 & 14.952 & 5826 & 4.56 & 0.021 & 1.059$^{_{+1.660}}_{^{-0.632}}$ & 0.877$^{_{+0.819}}_{^{-0.414}}$ & 1.052$^{_{+0.063}}_{^{-0.058}}$ & 0.984$^{_{+0.083}}_{^{-0.070}}$ \\
188.01 & 005357901 & 14.741 & 5087 & 4.73 & 0.255 & 0.893$^{_{+1.353}}_{^{-0.542}}$ & 0.671$^{_{+0.619}}_{^{-0.318}}$ & 0.859$^{_{+0.046}}_{^{-0.048}}$ & 0.778$^{_{+0.043}}_{^{-0.042}}$ \\
195.01 & 011502867 & 14.835 & 5604 & 4.50 & -0.188 & 1.058$^{_{+1.580}}_{^{-0.640}}$ & 0.968$^{_{+0.862}}_{^{-0.469}}$ & 0.988$^{_{+0.059}}_{^{-0.051}}$ & 0.907$^{_{+0.071}}_{^{-0.053}}$ \\
196.01 & 009410930 & 14.465 & 5585 & 4.51 & 0.096 & 1.029$^{_{+1.574}}_{^{-0.620}}$ & 0.937$^{_{+0.842}}_{^{-0.450}}$ & 0.984$^{_{+0.056}}_{^{-0.053}}$ & 0.901$^{_{+0.067}}_{^{-0.054}}$ \\
199.01 & 010019708 & 14.879 & 6214 & 4.60 & 0.104 & 1.081$^{_{+1.641}}_{^{-0.650}}$ & 0.863$^{_{+0.797}}_{^{-0.415}}$ & 1.182$^{_{+0.076}}_{^{-0.070}}$ & 1.161$^{_{+0.113}}_{^{-0.100}}$ \\
201.01 & 006849046 & 14.014 & 5491 & 4.45 & 0.187 & 1.039$^{_{+1.628}}_{^{-0.620}}$ & 0.985$^{_{+0.920}}_{^{-0.465}}$ & 0.959$^{_{+0.053}}_{^{-0.052}}$ & 0.875$^{_{+0.059}}_{^{-0.052}}$ \\
202.01 & 007877496 & 14.309 & 5912 & 4.44 & 0.120 & 1.093$^{_{+1.667}}_{^{-0.654}}$ & 1.052$^{_{+0.952}}_{^{-0.516}}$ & 1.080$^{_{+0.064}}_{^{-0.060}}$ & 1.019$^{_{+0.088}}_{^{-0.075}}$ \\
203.01 & 010619192 & 14.141 & 5634 & 4.49 & 0.041 & 1.056$^{_{+1.592}}_{^{-0.636}}$ & 0.980$^{_{+0.915}}_{^{-0.473}}$ & 0.996$^{_{+0.060}}_{^{-0.051}}$ & 0.915$^{_{+0.072}}_{^{-0.054}}$ \\
204.01 & 009305831 & 14.678 & 5287 & 4.48 & -0.104 & 0.966$^{_{+1.447}}_{^{-0.577}}$ & 0.950$^{_{+0.854}}_{^{-0.460}}$ & 0.905$^{_{+0.054}}_{^{-0.045}}$ & 0.821$^{_{+0.054}}_{^{-0.042}}$ \\
214.01 & 011046458 & 14.256 & 5322 & 4.44 & 0.018 & 1.022$^{_{+1.510}}_{^{-0.612}}$ & 1.015$^{_{+0.890}}_{^{-0.488}}$ & 0.916$^{_{+0.052}}_{^{-0.048}}$ & 0.832$^{_{+0.054}}_{^{-0.046}}$ \\
217.01 & 009595827 & 15.127 & 5504 & 4.72 & 0.220 & 0.975$^{_{+1.448}}_{^{-0.584}}$ & 0.706$^{_{+0.649}}_{^{-0.330}}$ & 0.963$^{_{+0.054}}_{^{-0.053}}$ & 0.879$^{_{+0.060}}_{^{-0.053}}$ \\
229.01 & 003847907 & 14.720 & 5608 & 4.37 & 0.219 & 1.069$^{_{+1.616}}_{^{-0.645}}$ & 1.124$^{_{+1.041}}_{^{-0.533}}$ & 0.990$^{_{+0.056}}_{^{-0.052}}$ & 0.909$^{_{+0.067}}_{^{-0.054}}$ \\
254.01 & 005794240 & 15.979 & 3948 & 4.54 & 0.234 & 0.530$^{_{+0.810}}_{^{-0.322}}$ & 0.650$^{_{+0.602}}_{^{-0.307}}$ & 0.347$^{_{+0.168}}_{^{-0.118}}$ & 0.319$^{_{+0.124}}_{^{-0.090}}$ \\
356.01 & 011624249 & 13.807 & 5124 & 4.07 & -0.503 & 1.111$^{_{+1.623}}_{^{-0.665}}$ & 1.593$^{_{+1.466}}_{^{-0.750}}$ & 0.869$^{_{+0.048}}_{^{-0.049}}$ & 0.787$^{_{+0.045}}_{^{-0.043}}$ \\
412.01 & 005683743 & 14.288 & 5584 & 4.28 & -0.011 & 1.093$^{_{+1.667}}_{^{-0.654}}$ & 1.275$^{_{+1.153}}_{^{-0.625}}$ & 0.983$^{_{+0.058}}_{^{-0.053}}$ & 0.901$^{_{+0.068}}_{^{-0.054}}$ \\
421.01 & 009115800 & 14.995 & 5181 & 4.32 & -0.075 & 1.016$^{_{+1.500}}_{^{-0.610}}$ & 1.155$^{_{+1.053}}_{^{-0.547}}$ & 0.880$^{_{+0.051}}_{^{-0.047}}$ & 0.797$^{_{+0.049}}_{^{-0.042}}$ \\
433.01 & 010937029 & 14.924 & 5237 & 4.37 & 0.375 & 0.986$^{_{+1.476}}_{^{-0.589}}$ & 1.080$^{_{+0.971}}_{^{-0.523}}$ & 0.894$^{_{+0.052}}_{^{-0.046}}$ & 0.810$^{_{+0.052}}_{^{-0.041}}$ \\
611.01 & 006309763 & 14.022 & 6122 & 4.55 & -0.132 & 1.085$^{_{+1.599}}_{^{-0.645}}$ & 0.914$^{_{+0.852}}_{^{-0.429}}$ & 1.149$^{_{+0.069}}_{^{-0.066}}$ & 1.115$^{_{+0.102}}_{^{-0.092}}$ \\
667.01 & 006752502 & 13.826 & 4135 & 4.57 & 0.000 & 0.607$^{_{+0.917}}_{^{-0.364}}$ & 0.681$^{_{+0.597}}_{^{-0.326}}$ & 0.501$^{_{+0.126}}_{^{-0.165}}$ & 0.431$^{_{+0.114}}_{^{-0.121}}$ \\
684.01 & 007730747 & 13.831 & 5331 & 3.96 & 0.113 & 1.174$^{_{+1.743}}_{^{-0.704}}$ & 1.870$^{_{+1.718}}_{^{-0.873}}$ & 0.917$^{_{+0.053}}_{^{-0.048}}$ & 0.833$^{_{+0.054}}_{^{-0.045}}$ \\
760.01 & 011138155 & 15.263 & 5887 & 4.62 & 0.010 & 1.060$^{_{+1.611}}_{^{-0.634}}$ & 0.840$^{_{+0.801}}_{^{-0.403}}$ & 1.072$^{_{+0.063}}_{^{-0.061}}$ & 1.008$^{_{+0.087}}_{^{-0.076}}$ \\
767.01 & 011414511 & 15.052 & 5431 & 4.44 & 0.023 & 1.026$^{_{+1.615}}_{^{-0.625}}$ & 1.007$^{_{+0.943}}_{^{-0.485}}$ & 0.946$^{_{+0.050}}_{^{-0.052}}$ & 0.862$^{_{+0.053}}_{^{-0.052}}$ \\
801.01 & 003351888 & 15.001 & 5472 & 4.39 & 0.182 & 1.041$^{_{+1.590}}_{^{-0.632}}$ & 1.080$^{_{+1.000}}_{^{-0.510}}$ & 0.955$^{_{+0.054}}_{^{-0.054}}$ & 0.871$^{_{+0.059}}_{^{-0.054}}$ \\
809.01 & 003935914 & 15.530 & 5690 & 4.48 & -0.385 & 1.047$^{_{+1.518}}_{^{-0.630}}$ & 0.967$^{_{+0.877}}_{^{-0.454}}$ & 1.012$^{_{+0.061}}_{^{-0.053}}$ & 0.934$^{_{+0.075}}_{^{-0.059}}$ \\
813.01 & 004275191 & 15.725 & 5357 & 4.73 & -0.285 & 0.955$^{_{+1.428}}_{^{-0.576}}$ & 0.695$^{_{+0.641}}_{^{-0.328}}$ & 0.924$^{_{+0.052}}_{^{-0.050}}$ & 0.839$^{_{+0.054}}_{^{-0.047}}$ \\
830.01 & 005358624 & 15.224 & 4915 & 4.90 & 0.155 & 0.797$^{_{+1.176}}_{^{-0.478}}$ & 0.528$^{_{+0.481}}_{^{-0.250}}$ & 0.819$^{_{+0.047}}_{^{-0.052}}$ & 0.742$^{_{+0.041}}_{^{-0.049}}$ \\
838.01 & 005534814 & 15.311 & 5794 & 4.48 & -0.095 & 1.049$^{_{+1.596}}_{^{-0.619}}$ & 0.987$^{_{+0.913}}_{^{-0.471}}$ & 1.043$^{_{+0.059}}_{^{-0.057}}$ & 0.972$^{_{+0.078}}_{^{-0.068}}$ \\
840.01 & 005651104 & 15.028 & 4916 & 4.39 & -0.091 & 0.936$^{_{+1.379}}_{^{-0.556}}$ & 1.023$^{_{+0.953}}_{^{-0.481}}$ & 0.818$^{_{+0.048}}_{^{-0.052}}$ & 0.742$^{_{+0.042}}_{^{-0.049}}$ \\
843.01 & 005881688 & 15.270 & 5784 & 4.40 & 0.203 & 1.093$^{_{+1.650}}_{^{-0.655}}$ & 1.109$^{_{+0.972}}_{^{-0.531}}$ & 1.041$^{_{+0.060}}_{^{-0.057}}$ & 0.969$^{_{+0.079}}_{^{-0.068}}$ \\
897.01 & 007849854 & 15.257 & 5734 & 4.46 & 0.270 & 1.066$^{_{+1.597}}_{^{-0.637}}$ & 1.024$^{_{+0.934}}_{^{-0.497}}$ & 1.025$^{_{+0.060}}_{^{-0.056}}$ & 0.951$^{_{+0.075}}_{^{-0.064}}$ \\
908.01 & 008255887 & 15.113 & 5391 & 4.25 & 0.128 & 1.050$^{_{+1.591}}_{^{-0.636}}$ & 1.279$^{_{+1.197}}_{^{-0.618}}$ & 0.933$^{_{+0.052}}_{^{-0.051}}$ & 0.849$^{_{+0.053}}_{^{-0.050}}$ \\
913.01 & 008544996 & 15.198 & 5463 & 4.75 & -0.281 & 0.967$^{_{+1.521}}_{^{-0.589}}$ & 0.688$^{_{+0.644}}_{^{-0.332}}$ & 0.953$^{_{+0.052}}_{^{-0.053}}$ & 0.869$^{_{+0.057}}_{^{-0.053}}$ \\
931.01 & 009166862 & 15.272 & 5714 & 4.78 & 0.319 & 1.016$^{_{+1.547}}_{^{-0.617}}$ & 0.685$^{_{+0.624}}_{^{-0.330}}$ & 1.020$^{_{+0.061}}_{^{-0.053}}$ & 0.944$^{_{+0.075}}_{^{-0.061}}$ \\
961.02 & 008561063 & 15.920 & 4188 & 4.56 & 0.000 & 0.612$^{_{+0.888}}_{^{-0.368}}$ & 0.671$^{_{+0.609}}_{^{-0.315}}$ & 0.536$^{_{+0.117}}_{^{-0.157}}$ & 0.461$^{_{+0.114}}_{^{-0.123}}$ \\
961.03 & 008561063 & 15.920 & 4188 & 4.56 & 0.000 & 0.626$^{_{+0.955}}_{^{-0.377}}$ & 0.679$^{_{+0.621}}_{^{-0.325}}$ & 0.539$^{_{+0.115}}_{^{-0.155}}$ & 0.463$^{_{+0.113}}_{^{-0.123}}$ \\
1176.01 & 003749365 & 15.715 & 4601 & 4.69 & 0.376 & 0.746$^{_{+1.115}}_{^{-0.453}}$ & 0.648$^{_{+0.614}}_{^{-0.309}}$ & 0.736$^{_{+0.053}}_{^{-0.078}}$ & 0.663$^{_{+0.053}}_{^{-0.083}}$ \\
1419.01 & 011125936 & 15.507 & 5848 & 4.46 & -0.292 & 1.075$^{_{+1.614}}_{^{-0.645}}$ & 0.999$^{_{+0.920}}_{^{-0.474}}$ & 1.059$^{_{+0.065}}_{^{-0.059}}$ & 0.992$^{_{+0.087}}_{^{-0.072}}$ \\
1459.01 & 009761199 & 15.692 & 4060 & 4.40 & 0.098 & 0.646$^{_{+1.001}}_{^{-0.389}}$ & 0.830$^{_{+0.781}}_{^{-0.393}}$ & 0.444$^{_{+0.138}}_{^{-0.159}}$ & 0.384$^{_{+0.115}}_{^{-0.113}}$ \\
1541.01 & 004840513 & 15.189 & 6164 & 4.53 & 0.068 & 1.100$^{_{+1.685}}_{^{-0.657}}$ & 0.954$^{_{+0.850}}_{^{-0.465}}$ & 1.163$^{_{+0.074}}_{^{-0.068}}$ & 1.134$^{_{+0.108}}_{^{-0.095}}$ \\
1543.01 & 005270698 & 14.985 & 5821 & 4.54 & -0.240 & 1.044$^{_{+1.587}}_{^{-0.618}}$ & 0.912$^{_{+0.870}}_{^{-0.433}}$ & 1.052$^{_{+0.061}}_{^{-0.059}}$ & 0.983$^{_{+0.080}}_{^{-0.071}}$ \\

%% file: kepsec-tab2a.tex
   2.01 & $ $ 0.0108$^{_{+0.0013}}_{^{-0.0011}}$ & 0.254$^{_{+0.005}}_{^{-0.005}}$ & 0.0769$^{_{+0.0004}}_{^{-0.0004}}$ & 83.92$^{_{+0.7045}}_{^{-0.5848}}$ & $ $ 0.000$^{_{+0.000}}_{^{-0.000}}$ & $ $ 0.000$^{_{+0.000}}_{^{-0.000}}$ & 2.204732$^{_{+3.1\textrm{e-}06}}_{^{-3.0\textrm{e-}06}}$ & 4954.35796$^{_{+0.00013}}_{^{-0.00013}}$ & $ $ 0.421$^{_{+0.07}}_{^{-0.07}}$ & 26.5\\
   5.01 & $ $-0.0088$^{_{+0.0088}}_{^{-0.0097}}$ & 0.123$^{_{+0.010}}_{^{-0.013}}$ & 0.0356$^{_{+0.0009}}_{^{-0.0011}}$ & 83.60$^{_{+0.7978}}_{^{-0.5911}}$ & $ $ 0.000$^{_{+0.000}}_{^{-0.000}}$ & $ $ 0.000$^{_{+0.000}}_{^{-0.000}}$ & 4.780381$^{_{+6.7\textrm{e-}05}}_{^{-6.7\textrm{e-}05}}$ & 4965.97227$^{_{+0.00119}}_{^{-0.00121}}$ & $ $ 0.300$^{_{+0.72}}_{^{-0.35}}$ & 2.69\\
  10.01 & $ $ 0.0014$^{_{+0.0033}}_{^{-0.0013}}$ & 0.151$^{_{+0.004}}_{^{-0.004}}$ & 0.0938$^{_{+0.0006}}_{^{-0.0007}}$ & 84.68$^{_{+0.3495}}_{^{-0.2975}}$ & $ $ 0.000$^{_{+0.000}}_{^{-0.000}}$ & $ $ 0.000$^{_{+0.000}}_{^{-0.000}}$ & 3.522511$^{_{+1.5\textrm{e-}05}}_{^{-1.6\textrm{e-}05}}$ & 4954.11838$^{_{+0.00043}}_{^{-0.00041}}$ & $ $-5.752$^{_{+4.04}}_{^{-64.5}}$ & 4.31\\
  13.01 & $ $ 0.0248$^{_{+0.0013}}_{^{-0.0015}}$ & 0.294$^{_{+0.005}}_{^{-0.006}}$ & 0.0659$^{_{+0.0002}}_{^{-0.0003}}$ & 79.79$^{_{+0.5070}}_{^{-0.4079}}$ & $ $ 0.000$^{_{+0.000}}_{^{-0.000}}$ & $ $ 0.000$^{_{+0.000}}_{^{-0.000}}$ & 1.763589$^{_{+2.3\textrm{e-}06}}_{^{-2.3\textrm{e-}06}}$ & 4953.56510$^{_{+0.00013}}_{^{-0.00014}}$ & $ $ 0.339$^{_{+0.02}}_{^{-0.02}}$ & 15.9\\
  18.01 & $ $ 0.0000$^{_{+0.0001}}_{^{-0.0000}}$ & 0.174$^{_{+0.027}}_{^{-0.005}}$ & 0.0783$^{_{+0.0025}}_{^{-0.0007}}$ & 87.64$^{_{+1.7674}}_{^{-3.6176}}$ & $ $ 0.000$^{_{+0.000}}_{^{-0.000}}$ & $ $ 0.000$^{_{+0.000}}_{^{-0.000}}$ & 3.548460$^{_{+2.4\textrm{e-}05}}_{^{-2.5\textrm{e-}05}}$ & 4955.90081$^{_{+0.00068}}_{^{-0.00064}}$ & $ $-78.785$^{_{+238.}}_{^{-525.}}$ & 9.33\\

%% file: kepsec-tab3a.tex
   2.01 & $ $   75.0$^{_{+9.48}}_{^{-8.13}}$ & $ $9.11 & $ $2702$^{_{+0759}}_{^{-0597}}$ & $ $2441$^{_{+0686}}_{^{-0540}}$ & $ $ 2851$^{_{+0062}}_{^{-0064}}$ & $ $ 1.055$^{_{+0.294}}_{^{-0.232}}$ & 1.022$^{_{+0.955}}_{^{-0.497}}$ & 0.0367$^{_{+0.0132}}_{^{-0.0097}}$ & $ $ 0.38$^{_{+0.65}}_{^{-0.24}}$  \\
   5.01 & $ $  -13.4$^{_{+13.4}}_{^{-14.8}}$ & $ $-0.99 & $ $2139$^{_{+0597}}_{^{-0455}}$ & $ $1932$^{_{+0539}}_{^{-0411}}$ & $ $-2607$^{_{+0827}}_{^{-0256}}$ & $ $-1.159$^{_{+0.539}}_{^{-0.389}}$ & 0.617$^{_{+0.552}}_{^{-0.300}}$ & 0.0586$^{_{+0.0205}}_{^{-0.0154}}$ & $ $-0.40$^{_{+0.39}}_{^{-1.15}}$  \\
  10.01 & $ $   14.9$^{_{+34.7}}_{^{-13.9}}$ & $ $1.07 & $ $1993$^{_{+0546}}_{^{-0420}}$ & $ $1801$^{_{+0493}}_{^{-0380}}$ & $ $ 2319$^{_{+0205}}_{^{-0358}}$ & $ $ 1.121$^{_{+0.355}}_{^{-0.321}}$ & 0.990$^{_{+0.905}}_{^{-0.470}}$ & 0.0471$^{_{+0.0172}}_{^{-0.0124}}$ & $ $ 0.21$^{_{+0.55}}_{^{-0.17}}$  \\
  13.01 & $ $  124.3$^{_{+6.93}}_{^{-7.81}}$ & $ $16.15 & $ $5069$^{_{+1414}}_{^{-1135}}$ & $ $4580$^{_{+1278}}_{^{-1025}}$ & $ $ 3744$^{_{+0055}}_{^{-0056}}$ & $ $ 0.738$^{_{+0.213}}_{^{-0.161}}$ & 1.610$^{_{+1.524}}_{^{-0.779}}$ & 0.0348$^{_{+0.0127}}_{^{-0.0092}}$ & $ $ 0.23$^{_{+0.40}}_{^{-0.14}}$  \\
  18.01 & $ $    0.2$^{_{+1.42}}_{^{-0.27}}$ & $ $0.33 & $ $1827$^{_{+0514}}_{^{-0398}}$ & $ $1651$^{_{+0464}}_{^{-0360}}$ & $ $ 1648$^{_{+0132}}_{^{-0252}}$ & $ $ 0.866$^{_{+0.278}}_{^{-0.267}}$ & 0.749$^{_{+0.703}}_{^{-0.357}}$ & 0.0446$^{_{+0.0158}}_{^{-0.0118}}$ & $ $ 0.01$^{_{+0.03}}_{^{-0.01}}$  \\
  64.01 & $ $   47.7$^{_{+25.0}}_{^{-23.3}}$ & $ $2.05 & $ $2711$^{_{+0748}}_{^{-0605}}$ & $ $2449$^{_{+0676}}_{^{-0547}}$ & $ $ 2702$^{_{+0165}}_{^{-0227}}$ & $ $ 0.980$^{_{+0.294}}_{^{-0.226}}$ & 0.829$^{_{+0.756}}_{^{-0.403}}$ & 0.0324$^{_{+0.0115}}_{^{-0.0084}}$ & $ $ 0.24$^{_{+0.49}}_{^{-0.17}}$  \\

%% file: chp5-keplmb.tex
\begin{singlespace}
\section[\MakeUppercase{Low-Mass Eclipsing Binaries in the Initial\\\emph{Kepler} Data Release}]{\MakeUppercase{Low-Mass Eclipsing Binaries in the Initial \emph{Kepler} Data Release}}
\label{chap5}
\end{singlespace}

\subsection{Introduction}

A double-lined, detached, eclipsing binary (DDEB) is a system that contains two non-interacting, eclipsing stars, in which the spectra of both components can
be clearly seen, allowing for the radial-velocity (RV) of each component to be obtained. In these systems, the mass and radius of each star can be determined with errors usually less than 1-2\%, thus making DDEBs currently the most accurate method of obtaining masses and radii of stars. Models of main-sequence stars with masses similar to or greater than the Sun have been tested over the years using DDEBs. \citet{Popper1980} compiled available masses and radii of DDEB's with accuracies of $\le$ 15\%, up to that date, and found general agreement with stellar models, though stressed the need for more accurate observations and models. \citet{Andersen1991} provided a compilation of all available DDEB systems up to that date, with accuracies $\le$ 2\%, and showed that the masses and radii of these stars were in general agreement with the current stellar evolution models, with any discrepancies attributable to abundance variations. \citet{Torres2010} recently performed a similar review with nearly double the sample of DDEBs. They were able to show the need to include non-classical effects such as diffusion and convection in stellar models, definitively demonstrate the existence of significant structural differences in magnetically active and fast-rotating stars, test theories of rotational synchronization and orbital circularization, and validate General Relativity via apsidal motion rates. However, while observations of DDEBs have enhanced our understanding of stellar structure and evolution for stars with M $\geq$ 1.0 M$_{\sun}$, low-mass, main-sequence (LMMS) stars, (M $<$ 1.0 M$_{\sun}$ and T$_{\rm eff}$ $<$ 5800 K), have not been tested to the same extent.

Although a few systems with late G or early K type components had been studied prior to 2000, \citep[c.f.][and references therein]{Popper1980,Andersen1991,Torres2006,Clausen2009}, only three LMMS DDEBs with late K or M type components were known \citep{Lacy1977,Leung1978,Delfosse1999}. This number had only increased to nine by the beginning of 2007 \citep[cf.][Table 1]{LopezMorales2007}. Despite the fact that the majority of main-sequence stars are low-mass, these stars are both intrinsically fainter, and physically smaller, than their more massive counterparts. Therefore, they have a lower eclipse probability and are harder to discover and study. As outlined by \citet{LopezMorales2007}, analysis of these systems showed that the observed radii for these stars are consistently $\sim$10-20\% larger than predicted by stellar models \citep{Baraffe1998} for 0.3 M$_{\sun}$ $\lesssim$ M $\lesssim$ 0.8 M$_{\sun}$. \citet{Fernandez2009} recently showed this was also likely the case for five M dwarfs in short-period eclipsing systems with an F type primary, though since the systems are only single-lined, the masses could not be determined directly. This discrepancy between the radii derived from models and from observations either reveals a flaw in the stellar models for this mass regime, or is due to differences in metallicity, magnetic activity, or interpretation of the light curve data when star spots are present \citep{Morales2008}. As to this last point, \citet{Morales2010} recently noted that improperly taking polar spots into account in the light curve modeling process may possibly cause the derivation of stellar radii a few percent larger than the true values for some of these systems. Of all of these scenarios, enhanced magnetic activity has been proposed as the principal cause of inflated radii \citep{Chabrier2007,LopezMorales2007,Morales2008}.

If enhanced magnetic activity is the principal cause of the inflated radii, shorter-period binary systems, with the stellar rotation rate enhanced by the revolution of the system, would be expected to show greater activity and thus larger radii than longer-period systems \citep{Chabrier2007}. Binary systems with component masses of 0.5 M$_{\sun}$ are expected to synchronize, and therefore be spun-up, in less than 0.1 Gyr for periods less than 4 days, and in less than 1 Gyr for periods less than 8 days \citep{Zahn1977, Zahn1994}. Thus, the discovery of LMMS DDEBs with P $\gtrsim$ 10 days, where the binary components should have natural rotation rates, is crucial to probing if enhanced rotation due to binarity is the underlying cause of this phenomenon. This theory might be supported by measurements of isolated field M and K dwarf stars via very long baseline interferometry, which \citet{Demory2009} found to match stellar models. However, recently a much larger sample of nearly two dozen isolated M and K dwarf stars finds, for $\sim$80\% of the sample, larger radii than the model predictions for 0.35 $<$ M $<$ 0.65 M$_{\sun}$ \citep{Boyajian2010}, indicating that there are likely multiple causes of inflation at work, or a remaining flaw in the stellar models.

Though several more LMMS DDEB systems have been found since 2007, \citep{Coughlin2007,Shaw2007,Becker2008,Blake2008,Devor2008a,Devor2008b,Shkolnik2008,Hoffman2008,Irwin2009,Dimitrov2010,Shkolnik2010}, there are to-date only seven well-studied systems with 1.0 $<$ P $<$ 3.0 days \citep[][and references therein]{LopezMorales2007,Becker2008,Shkolnik2008}, and only one has a larger period, at P = 8.4 days \citep{Devor2008b}. This is mostly due to the fact that ground-based photometric surveys, such as NSVS, TrES, and OGLE, are either cadence, precision, magnitude, or number limited, and thus not sensitive to long periods. The \emph{Kepler Mission}, with 3 years of constant photometric monitoring of over 150,000 stars with V $\lesssim$ 17, at 30-minute cadence and sub-millimagnitude precision, is the key to discovering a large number of long-period, LMMS DDEBs. 

In this chapter we present the results of our search through all the newly available \emph{Kepler} Q0 and Q1 public data for LMMS DDEBs. Section~\ref{keplmbdatasec} describes the data we use in this chapter. Section~\ref{binaryidentsec} describes our binary identification technique, and Section~\ref{keplmbmodelsec} describes how we model the light curves. Our selection and list of new LMMS DDEBs is presented in Section~\ref{newlmbsec}, and we present new transiting planet candidates in Section~\ref{transsec}. In Section~\ref{lmbmodelcompsec} we compare the new LMMS DDEBs with theoretical models, and conclude with a summary of our results in section~\ref{keplmbconcsec}. Once accurate mass and radius values exist for a large range of both mass and period, our understanding of these objects should substantially improve, and we will be one step closer to extending to the lower-mass regime the advanced study of stellar structure and evolution that sun-like and high-mass stars have been a subject of for some time.

\subsection{Observational Data}
\label{keplmbdatasec}

The data used in our analysis consists of the 201,631 light curves made public by the \emph{Kepler Mission}\footnote{http://kepler.nasa.gov/} as of June 15, 2010 from \emph{Kepler} Q0 and Q1 observations. All light curves can be accessed through the Multi-mission Archive at STScI (MAST)\footnote{http://archive.stsci.edu/kepler/}. The data consist of 51,366 light curves from \emph{Kepler} Q0, (observed from 2009-05-02 00:54:56 to 2009-05-11 17:51:31 UT), and 150,265 light curves from \emph{Kepler} Q1, (observed from 2009-05-13 00:15:49 to 2009-06-15 11:32:57 UT), each at 29.43 minute cadence. Individual light curves for Q0 contain $\sim$470 data points, and for Q1 contain $\sim$1,600 data points. Targets range in \emph{Kepler} magnitude from 17.0 at the faintest, to 5.0 at the brightest.

The \emph{Kepler} team has performed pixel level calibrations, (including bias, dark current, flat-field, gain, and non-linearity corrections), identified and cleaned cosmic-ray events, estimated and removed background signal, and then extracted time-series photometry using an optimum photometric aperture. They have also removed systematic trends due to spacecraft pointing, temperature fluctuations, and other sources of systematic error, and corrected for excess flux in the optimal photometric aperture due to crowding \citep{Vancleve2010}. It is this final, ``corrected'' photometry that we have downloaded for use in our analysis.

\subsection{Eclipsing Binary Identification} 
\label{binaryidentsec}

\citet{Prsa2010} released an initial catalog of eclipsing binary stars they found in the \emph{Kepler} field from the same Q0 and Q1 data we use in this chapter. They first identified EB candidates via \emph{Kepler's} Transit Planet Search (TPS) algorithm, eliminating those targets already identified as exoplanet candidates. To determine the ephemeris of each candidate, they used Lomb-Scargle, Analysis of Variance, and Box-fitting Least Squares periodogram techniques, combined with manual inspection and modification. They then culled, through manual inspection, non-EB candidates, such as pulsating and heavily spotted stars, as well as duplicates due to contamination from nearby stars, and arrive at their final list of 1,832 binaries, which are manually classified as detached, semi-detached, over-contact, ellipsoidal, or unknown. Next, they estimate the principal parameters of each system, (temperature ratio, sum of the fractional radii, e$\cdot$cos($\omega$), e$\cdot$sin($\omega$), and sin(i) for detached systems), via a neural network technique called Eclipsing Binaries via Artificial Intelligence \citep[``EBAI''][]{Prsa2008}. For our search, which focuses on the detection of LMMS DDEBs, we have devised our own DDEB identification technique, which we apply to the Q1 data. We do not use the Q0 data in this part of the analysis to avoid discrepant systematics between the two quarters, which complicate the analysis. 

\begin{figure}[h]
\centering
\epsfig{width=\linewidth,file=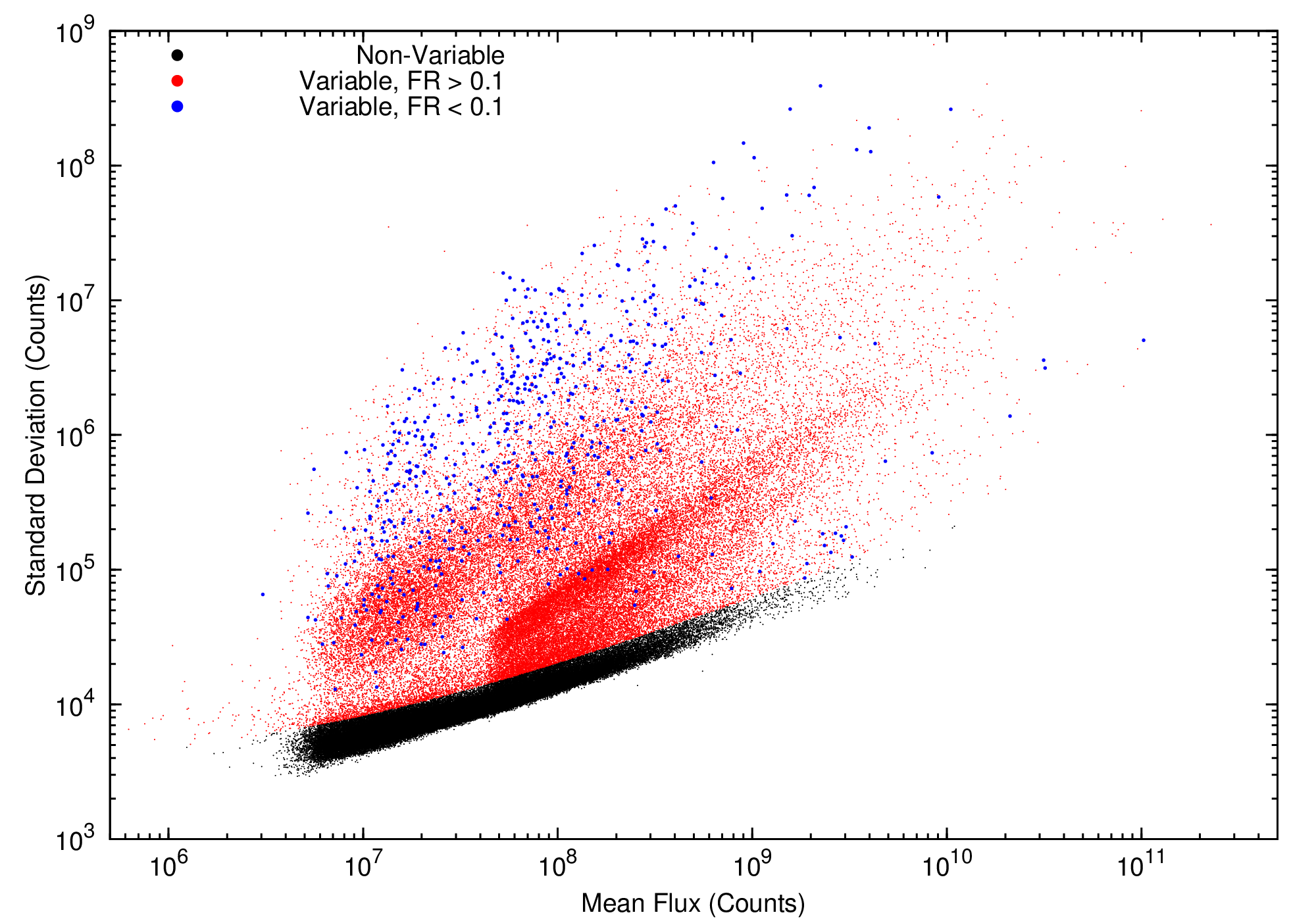}
\caption[Plot of standard deviation versus mean flux for the 150,265 stars in Q1]{Plot of standard deviation versus mean flux for the 150,265 stars in Q1. Black dots represent stars that vary by less than 1-sigma from a best-fit power-law to the data, and thus we classify them as non-variables. Red dots represent variables with a flux ratio greater than 0.1. Blue dots represent variables with a flux ratio less than 0.1, and thus are good candidates to be eclipsing binaries.}
\label{stdevfig}
\end{figure}

Our search consisted of two steps. The first was to identify variable stars, and to do so, we placed a light curve standard deviation limit above which the objects are classified as variables. We first subtracted an error-weighted, linear fit of flux versus time from all data, to remove any remaining linear systematic trends, and then plotted the standard deviation of each light curve versus its average flux and fit a power law. These data are shown in Figure~\ref{stdevfig}, where the black dots correspond to light curves which deviate by less than 1$\sigma$ from the standard deviation versus average flux fit, and we thus classify as non-variable. The colored dots indicate the variable candidates that deviate by more than 1$\sigma$. Next, we used the flux ratio (FR) measurement criterion, which we adapted from the magnitude ratio given in \citet{Kinemuchi2006}, and is defined as 
\begin{equation} \textrm{FR} = \frac{\textrm{maximum flux - median flux}}{\textrm{maximum flux - minimum flux}} \end{equation} 
as a measure of whether or not the variable spends most of its time above (low FR value) or below (high FR value) the median flux value. Perfectly sinusoidal variables have FR $=$ 0.50, pulsating variables, such as RR Lyrae's, have FR $>$ 0.5, and eclipsing binaries have FR $<$ 0.5. As we are principally interested in finding well detached systems with relatively deep, narrow eclipses, which thus have low FR values, we make a further cut of the systems and only examine those variables with FR $<$ 0.1, shown by blue dots in Figure~\ref{stdevfig}.

The second step of the analysis was to determine the orbital period of each candidate. This was done using two independent techniques that are both well-suited for detached eclipsing binary systems. The first is Phase Dispersion Minimization (PDM) \citep{Stellingwerf1978}, which attempts to find the period that best minimizes the variance in multiple phase bins of the folded light curve. This technique is not sensitive to the shape of the light curve, and thus is ideal for non-sinusoidal variables such as detached eclipsing binaries. The downside of this technique is that if strong periodic features exist in the light curve, which do not correspond to the period of eclipses, such as rapidly varying spots, stellar pulsations, or leftover systematics, they can weaken the signal of the eclipse period. We use the latest implementation given by \citet{Stellingwerf2006}, and determine the best three periods via this technique to ensure that the true period is found, and not just an integer multiple, or fraction, thereof. 

The second technique we use is one we invented specifically for detached eclipsing binaries, and call Eclipse Phase Dispersion Minimization (EPDM). The idea behind EPDM is that we want to automatically identify and align the primary eclipses in an eclipsing binary, thus finding the period of the system. To accomplish this, EPDM finds the period that best minimizes the dispersion of the actual phase values of the faintest N points in a light curve, i.e. the very bottom of the eclipses. Since EPDM only selects the N faintest points in a light curve, it is not affected by systematics or periodic features that do not correspond to the period of eclipses, assuming the systematics do not extend below the depth of the eclipses. The technique works for all binary systems with equal or unequal eclipse depths, and transiting planets, both with either zero or non-zero eccentricity. Computationally, EPDM is significantly faster than traditional PDM techniques. For a detailed and illustrative explanation of this new technique, please see Appendix~\ref{epdmappendsec}. We use EPDM to find the three best fit periods for each system as well, for the same reasons as we did with PDM.

\begin{figure}
\centering
\epsfig{width=\linewidth,file=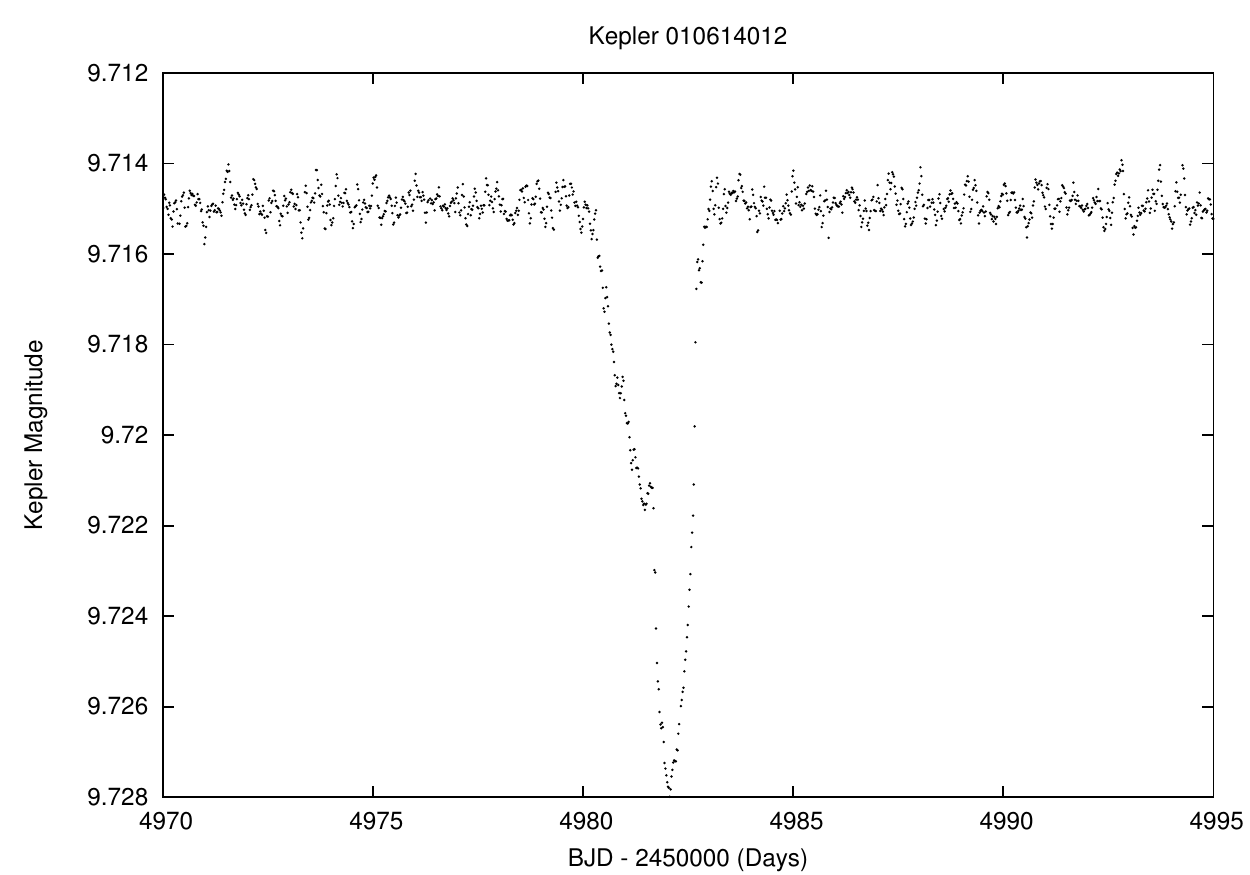}
\caption[Kepler 010614012: An apparent red giant, with an unusual, shallow, eclipse-like feature]{Kepler 010614012: An apparent red giant, (T$_{\rm eff}$ = 4859K, logg = 3.09, [M/H] = -0.64, R$_{\star}$ = 5.71 R$_{\sun}$), with an unusual, shallow, eclipse-like feature.}
\label{Kepler010614012}
\end{figure}

We identify 577 EB candidates in the Q1 data. Of these, 486 are listed by \citet{Prsa2010} as detached eclipsing binaries, and 20 are identified as semi-detached eclipsing binaries. The 71 remaining candidates were manually inspected by examining both the raw and phased light curves at the 6 best periods found via PDM and EPDM. Of these 71 remaining candidates, 48 turned out to be false positives with significantly large, sharp systematic features, and one is an apparent red giant, (Kepler 010614012, T$_{\rm eff}$ = 4859K, $\log{\rm g}$ = 3.086, [M/H] = -0.641, R$_{\star}$ = 5.708 R$_{\sun}$), with an unusual, asymmetrical, eclipse-like feature that lasts for $\sim$3 days with a depth of 1.2\%, shown in Figure~\ref{Kepler010614012}. This does not appear to be a systematic feature due to its very flat out of eclipse baseline, contiguous nature, long duration, and the actual time at which the feature occurs, compared to the majority of other objects with strong systematics. The remaining 22 targets are: two transiting exoplanet candidates contained in the recently released list of 306 candidates by \citet{Borucki2010}, three already published transiting planets, (Kepler-5b, Kepler-6b, and TrES-2b), seven shallow eclipsing systems with primary eclipse depths ranging from 1.4\% to 5.7\%, visible secondary eclipses ranging from 0.05\% to 4.6\%, and periods ranging from 4.7 to 45.3 days, the already published transiting hot compact object Kepler 008823868 \citep{Rowe2010}, a 6.4 day eclipsing binary with T$_{\rm eff}$ = 5893K and eclipse depths of 38.4\% and 12.2\% (Kepler 006182849), and eight transiting exoplanet candidates with transit depths ranging from 0.75\% to 4.9\%, and periods ranging from 2.5 to 24.7 days. For the seven new extremely shallow eclipsing systems, we list their \emph{Kepler} ID numbers, periods, effective temperatures, surface gravities, and primary and secondary eclipse depths in Table~\ref{shallowebtab}, and note they could be of interest for follow-up due to the potential to contain brown dwarf or extremely low-mass secondaries, or even anomalously hot exoplanet companions. Of the eight transiting candidates, only one is listed in the \emph{Kepler} false positive catalog\footnote{http://archive.stsci.edu/kepler/false\_positives.html}, Kepler 011974540. None of them are in the list of the 306 released candidates by \citet{Borucki2010}, nor are among the 400 planetary candidates currently reserved for follow-up observations \citep{Borucki2010}. These will be further discussed in Section~\ref{transsec}.

\begin{deluxetable}{cccccc}
\tablewidth{0pt}
\tabletypesize{\small}
\tablecaption{Period, Effective Temperature, Surface Gravity, and Eclipse Depth Estimates for the Seven New Extremely Shallow Eclipsing Systems}
\tablecolumns{6}
\tablehead{\emph{Kepler} ID & Period & T$_{\rm eff}$ & logg & Pri. & Sec.\\ & (Days) & (K) & & (\%) & (\%)}
\startdata
\input{keplmb-tab1.tex}
\enddata
\tablenotetext{a}{System is listed in the \emph{Kepler} False Positive Catalog as likely to be an EB.}
\tablenotetext{b}{System has non-zero eccentricity.}
\tablenotetext{c}{Period derived assuming zero eccentricity.}
\label{shallowebtab}
\end{deluxetable}

\subsection{Light Curve Modeling}
\label{keplmbmodelsec}

Since the system parameters determined by \citet{Prsa2010} are only estimates and do not incorporate spots, and since we seek to obtain as accurate physical parameters as possible, we modeled each system using a robust global minimization scheme with a commonly used, physically detailed eclipsing binary modeling code. We took all 314 detached eclipsing binaries with T$_{\rm eff}$ $<$ 5500K and that are publicly available, (5 systems are still proprietary), identified from both our search and the \citet{Prsa2010} catalog, combined Q0 and Q1 data if available, and via manual inspection classified systems as double-eclipse (i.e. contained two visible eclipses), single-eclipse (i.e. only contained one eclipse), or as spurious results that were not recognizable as eclipsing systems. (Given the errors in the KIC temperature determination, and to ensure the primary is below 1.0 M$_{\sun}$, we used 5500K as our cutoff, instead of 5800K. As well, the definition of a ``double-lined'' system is one in which the lines of both components are visible in an observed spectrum. Although in general if two eclipses are clearly visible in the photometric light curve, it is likely to be ``double-lined'', this cannot be determined without an actual spectrum. Thus, we use the term ``double-eclipse'' throughout the chapter, with the assumption that when observed spectroscopically, the majority of these systems will be observed as ``double-lined''.) 

We then used the JKTEBOP eclipsing binary modeling program \citep{Southworth2004a,Southworth2004b} to model every double-eclipse eclipsing binary system, of which there were 231, solving for the period, time of primary minimum, inclination, mass ratio, e$\cdot$cos($\omega$), e$\cdot$sin($\omega$), surface brightness ratio, sum of the fractional radii, ratio of the radii, and out of eclipse flux. In addition, we also solved for the amplitude and time of minimum of a sinusoidal term imposed on the luminosity of the primary component, with the period fixed to that of the binary, in order to account for spots. Note that in the JKTEBOP model the mass ratio is only used to determine the amount of tidal deformation of the stars from a pure sphere. Thus, it has no effect on the light curve of long-period systems, which due to their large separations are almost perfectly spherical, but must be included to properly model very short-period systems, where the tidal deformation can have a significant impact on the light curve. We used the quadratic limb darkening law, which works well for late-type stars \citep[e.g.][]{Manduca1977,Wade1985,Claret1990}, with coefficients set to those found by \citet{Sing2010} for the \emph{Kepler} bandpass via interpolation given the systems' effective temperatures, surface gravities, and metallicities as listed in the $Kepler$ Input Catalog (KIC)\footnote{http://archive.stsci.edu/kepler/kepler\_fov/search.php}. We also fixed the gravity darkening exponent based on the effective temperature as prescribed by \citet{Claret2000a}. As any contaminating flux from nearby stars in the photometric aperture has already been compensated for in the \emph{Kepler} pipeline \citep{Vancleve2010}, we set the amount of third light to 0.0. Note that third light might still exist in some systems if there is a background star or tertiary component that is unidentifiable from ground-based surveys, (i.e. less than $\sim$1$\arcsec$ separation), but since third light is usually unconstrained in a single-color light curve, we do not let it vary. If third light existed in a system and was not accounted for, the solution would result in an inclination determination lower than the true value, and therefore an over-estimation of the stellar radius. However, this should only occur in a minority of systems. For a couple of binaries in our list, the light curves absolutely could not be modeled without the inclusion of third light, (i.e. very sharp eclipses with depths of less than 0.01 mag). For these cases only, we let the third light vary, and thus be a non-zero parameter. Additionally, if the effect of spots in a light curve deviates significantly from the adopted sinusoidal shape, it could affect the derived luminosity ratio to a minor extent, but it should not affect the sum of the radii.

In order to model such a large number of systems over such a large solution space, and to ensure we have found the best global solution, we adapted the JKTEBOP code to use a modified version of the asexual genetic algorithm (AGA) described by \citet{Canto2009}, coupled with its standard Levenberg-Marquardt minimization algorithm. Genetic algorithms (GA) are an extremely efficient method of fitting computationally intensive, multi-parameter models over a large and potentially discontinuous parameter space, and thus ideal for this work. For the details of how genetic algorithms work, and the specific changes we made to the \citet{Canto2009} AGA, please see Appendix~\ref{agaappendix}. 

We found that our modified AGA does an excellent job of solving well-behaved light curves, simultaneously varying all 12 aforementioned parameters over the entire range of possible solutions. For some of the systems however, strong systematics and/or variable star spots introduced a significant amount of noise, especially in systems with shallow eclipses, for which it was more difficult to arrive at a robust solution. For these systems we had to manually correct the systematics, often by either eliminating the Q0 or Q1 data, equalizing the base flux levels of Q0 and Q1 data, or subtracting out a quasi-sinusoidal variation in the base flux level due to remaining \emph{Kepler} systematics. When possible we attempted to minimize the amount of manual interference. Hopefully this will become much less of a problem with subsequent data releases. We then re-ran the AGA using a larger initial population until a good solution was found, i.e., both eclipse depths and widths were fit well by visual inspection. Every light curve in the end was visually inspected to be a good fit compared to the scatter of the data points, i.e., appeared to have $\chi_{\rm red}$ = 1.0 when ignoring systematic errors, and the obtained parameters were confirmed to be reasonable when visually inspecting the light curves.

\subsection{New Low-Mass Binary Candidates}
\label{newlmbsec}

In order to identify the main-sequence stars from our list of 231 candidates, and determine the best candidates for follow-up, we employ the following technique to estimate the temperature, mass, and radius of each star using the sum of the fractional radii, $r_{sum}$, period, $P$, the luminosity ratio, L$_{r}$, (which is derived from the surface brightness ratio, $J$, and radii ratio, $k$), obtained from our JKTEBOP models, and the effective temperature of the system, $T_{\rm eff}$, obtained from the KIC, with an estimated error of $\pm$200 K. The value for $T_{\rm eff}$ given in the KIC was determined via interpolation of standard color magnitude relations as determined by ground-based, multi-wavelength photometry \citep{Vancleve2010}. Although in principle one might be able to deconvolve two separate spectral energy distributions from this photometry, in reality given the level of photometric error in the KIC and uncertainty at which binary phase the photometry was obtained, this is untenable. Instead, we assume the stars radiate as blackbodies, and that each star contributes to the determined $T_{\rm eff}$ in proportion to its luminosity. Thus, following our assumption, we obtain the following relation,

\begin{equation}
\label{teffeq}
T_{\rm eff} = \frac{L_{1}T_{1} + L_{2}T_{2}}{L_{1}+L_{2}}
\end{equation}

\noindent where $L_{1}$, $L_{2}$, $T_{1}$ and $T_{2}$  are the luminosities and effective temperatures of star 1 and 2 respectively. Still assuming the stars radiate as blackbodies, the luminosity of each star is proportional to its radius squared and temperature to the fourth power, with the temperature proportional to is surface brightness to the one-fourth power. Thus, we find that the luminosity ratio can be expressed as,

\begin{eqnarray}
\label{lumeq}
L_{r} & = & \frac{L_{1}}{L_{2}} = \frac{r_{1}^{2}T_{1}^{4}}{r_{2}^{2}T_{2}^{4}} = k^{2}T_{r}^{4} = k^{2}\left[\left(\frac{SB_{1}}{SB_{2}}\right)^{1/4}\right]^{4} \nonumber \\ & = & k^{2}\left(J^{\frac{1}{4}}\right)^{4} = k^{2}J
\end{eqnarray}

\noindent where $SB_{1}$ and $SB_{2}$ are the surface brightnesses of star 1 and star 2 respectively, and $r_{1}$ and $r_{2}$ are the fractional radii of star 1 and 2 respectively, defined as $R_{1}/a$ and $R_{2}/a$, where $R_{1}$ and $R_{2}$ are the physical radius of each star, and $a$ is the semi-major axis of, or separation between, the components. Combining equations~\ref{teffeq} and \ref{lumeq} yields the expression,

\begin{equation}
\label{finallumeq}
T_{\rm eff} = \frac{L_{r}T_{1} + T_{2}}{L_{r}+1}
\end{equation}

\noindent which has two known parameters, T$_{\rm eff}$ and L$_{r}$, and two unknown parameters, T$_{1}$ and T$_{2}$. To place a further constraint upon the values of T$_{1}$ and T$_{2}$, we make the assumption that both stars in the binary are on the main-sequence, and employ the mass, temperature, radius, and average of the $V$-band and $R$-band luminosity relations given in \citet{Baraffe1998} for 0.075 $\le$ M $\le$ 1.0 M$_{\sun}$ and in \citet{Chabrier2000} for M $<$ 0.075 M$_{\sun}$, both assuming an age of 5.0 Gyr and [M/H] = 0.0. (We average the $V$ and $R$-band luminosities to obtain a very close approximation to the \emph{Kepler} bandpass.) From these models, for a given value of T$_{1}$, there is only one value of T$_{2}$ which will reproduce the observed value of L$_{r}$. Thus, there only exists one set of unique values for T$_{1}$ and T$_{2}$ that reproduces both the observed T$_{\rm eff}$ and L$_{r}$ values for the system. 

For each T$_{1}$ and T$_{2}$ then, we obtain the absolute masses and radii, (M$_{1}$, M$_{2}$, R$_{1}$, and R$_{2}$), via interpolation from the \citet{Baraffe1998} and \citet{Chabrier2000} models. Then, utilizing Kepler's 3$^{rd}$ law, given the total mass of the system, we calculate the semi-major axis, $a$, via

\begin{equation}
\label{kepeq}
a = (GM_{tot})^{\frac{1}{3}}(\frac{P}{2\pi})^{\frac{2}{3}}
\end{equation}

\noindent where $M_{\rm tot}$ is the total mass of the system, M$_{1}$ + M$_{2}$, and G is the gravitational constant. We then multiply each radius determined above by a constant so that the sum of the fractional radii derived from the JKTEBOP model, $r_{\rm sum}$, is equal to $(R_{1} + R_{2})/a$, the sum of the fractional radii when using the physical values of $M_{1}$, $M_{2}$, $R_{1}$, $R_{2}$, and $P$. This technique is robust because while individual parameters such as $i$, $J$, and $k$ can suffer from degeneracies, especially in systems with shallow eclipses, the values of $r_{sum}$ and $L_{r} = k^{2}J$, which we rely on, are firmly set by the width of the eclipses and the difference in their eclipse depths, respectively.

For clarity, we now illustrate the individual steps of this procedure using the example of an actual system, Kepler 002437452. This system was found to have T$_{\rm eff}$ = 5398 K and L$_{r}$ = 3.90 from the KIC and the JKTEBOP modeling respectively. Now, assuming the stars are main-sequence, one could choose values of T$_{1}$ = 4000 K and T$_{2}$ = 3620 K, and looking up their luminosities from the \citet{Baraffe1998} models, find that the luminosity ratio between two main-sequence stars with temperatures of 4000 K and a 3620 K is 3.90. In this case, the luminosity ratio criterion would be satisfied, but T$_{\rm eff}$ would be $\sim$3922 K, nowhere near the measured value of 5398 K. Similarly, one could choose values of T$_{1}$ = 5400 and T$_{2}$ = 5393, and this would yield T$_{\rm eff}$ = 5398 K, but L$_{r}$ would be 1.01, nowhere near the needed value of 3.90. The unique solution that satisfies both the effective temperature and luminosity ratio constraints is that T$_{1}$ = 5591 and T$_{2}$ = 4647, which yields both T$_{\rm eff}$ = 5398 and L$_{r}$ = 3.90. Now, given these temperatures, interpolating from the \citet{Baraffe1998} models yields values of M$_{1}$ = 0.963 M$_{\sun}$, R$_{1}$ = 0.966 R$_{\sun}$, M$_{2}$ = 0.792 M$_{\sun}$, and R$_{2}$ = 0.783 R$_{\sun}$. Taking the masses, and the period of the system of 14.47184 days, and utilizing Eq.~\ref{kepeq}, we find that the semi-major axis, $a$, would be 30.1 R$_{\sun}$. Dividing the sum of the estimated physical radii by the semi-major axis just calculated, we find a value of 0.058 for the sum of fractional radii. Now, from the JKTEBOP model, this system was measured to have a sum of the fractional radii of 0.084, and so it appears that the current values for the radii are underestimated. Thus, we multiply the radii by a factor of 0.084/0.058 = 1.45, to obtain our final radii values of R$_{1}$ = 1.40 R$_{\sun}$ and R$_{2}$ = 1.13 R$_{\sun}$, with, as above, M$_{1}$ = 0.96 M$_{\sun}$, T$_{1}$ = 5591 K, M$_{2}$ = 0.79 M$_{\sun}$, and T$_{2}$ = 4647 K.

\begin{figure}
\centering
\begin{tabular}{c}
\epsfig{width=0.75\linewidth,file=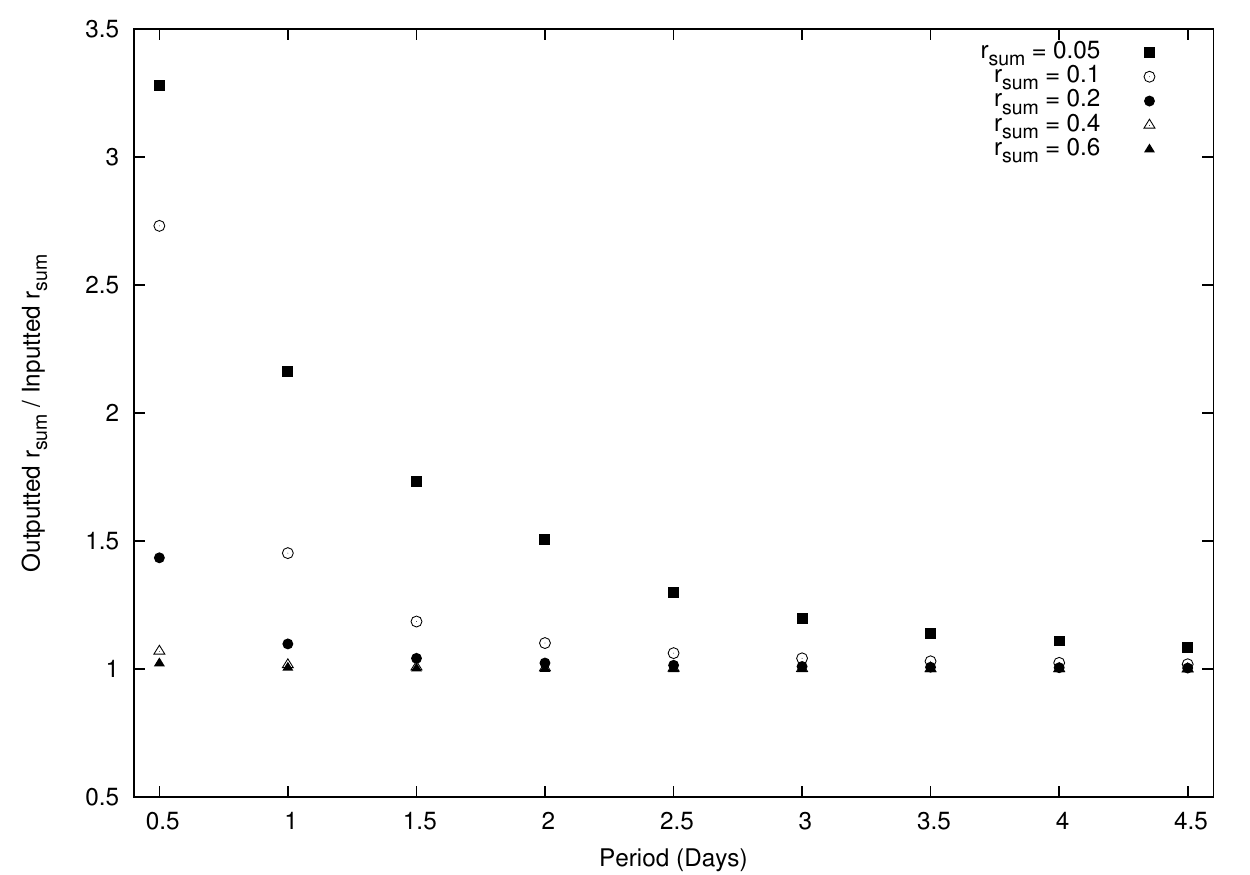} \\
\epsfig{width=0.75\linewidth,file=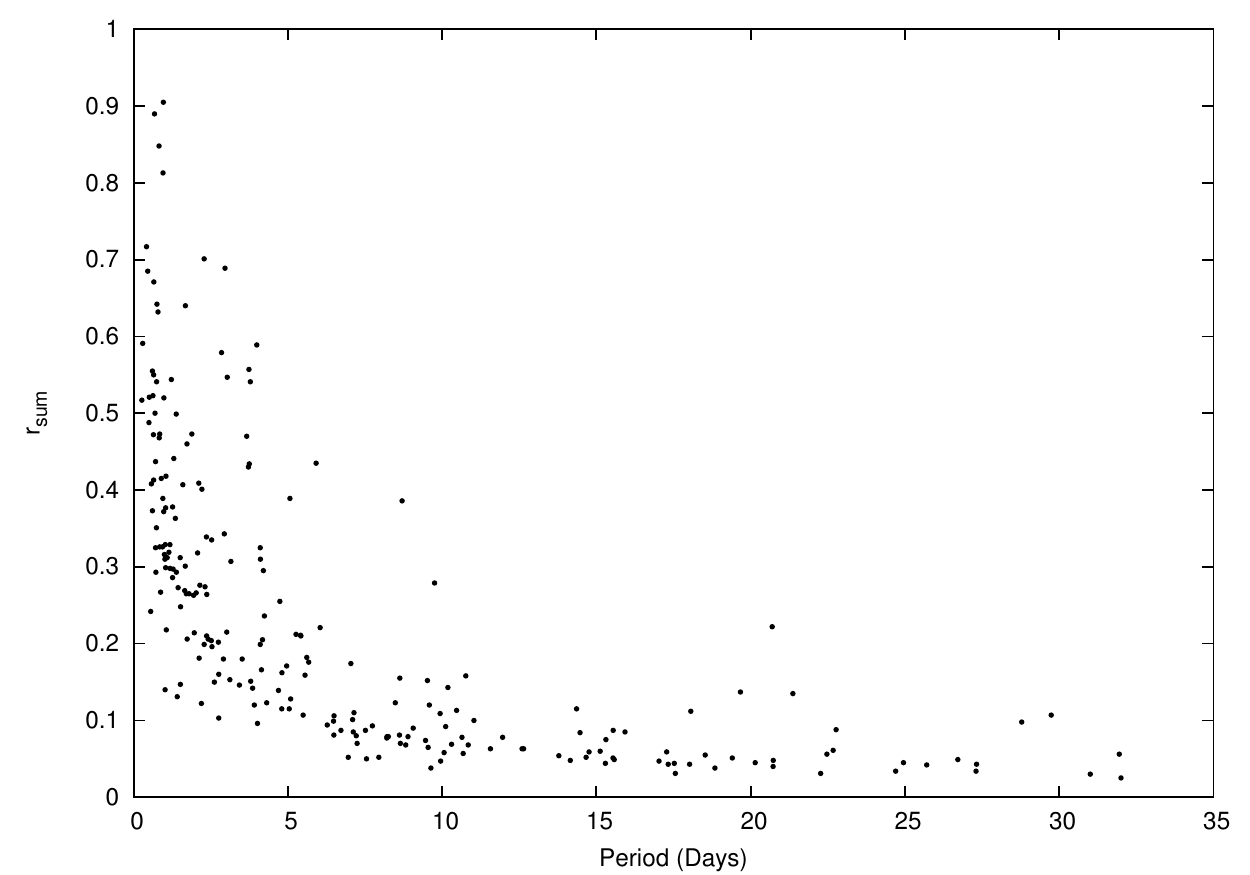} \\
\end{tabular}
\caption[The effect that the 29.43 minute integration time has on the derivation on the sum of the fractional radii at a given period]{Top: The effect that the 29.43 minute integration time has on the derivation on the sum of the fractional radii, r$_{sum}$, at a given period. As can be seen, only very small values of r$_{sum}$ and P yield discrepancies $\gtrsim$ 10\%, for example, combinations of P $<$ 3 days and r$_{sum}$ $<$ 0.05, P $<$ 1.5 days and r$_{sum}$ $<$ 0.1, P $<$ 0.75 days and r$_{sum}$ $<$ 0.2, etc. Bottom: The values of r$_{sum}$ versus period for the binaries we have modeled in this chapter, presented in Table~\ref{ddebcandstab}. Very few of the systems, $\lesssim$ 2\%, in our sample lie in a region where they would be significantly affected by the 29.43 minute integration time.}
\label{parintfig}
\end{figure}

\citet{Kipping2010a} has recently examined the effects of the long, ($\sim$30 minute), integration time of long-cadence \emph{Kepler} observations on transit light curves, and found that it can significantly alter the morphological shape of a transit curve and result in erroneous parameters if not properly taken into account in the modeling procedure. Certainly, eclipsing binaries are also affected by long integration times, namely by a ``smearing'' of the eclipses so that they appear to be shallower and have a longer duration. Qualitatively, this would result in a lower inclination and larger sum of the fractional radii, while the luminosity ratio would remain unchanged, since both eclipse depths are equally affected. To quantitatively investigate the extent to which the long integration could affect the derived parameters, we generated model light curves of a typical eclipsing binary, varying its period and the sum of the fractional radii. We then binned these light curves as if they had a 29.43 minute integration time, and the same number of data points as the Q1 \emph{Kepler} light curves. We then re-solved the light curves without accounting for the integration time, and compared the computed parameters to those used to generate the original light curve. We found that for the long-cadence \emph{Kepler} integration time of 29.43 minutes, only systems with very low values of $r_{sum}$ and P are significantly affected, as can be seen in Figure~\ref{parintfig}. These types of systems are less than 2\% of our sample. Nevertheless, we modified the JKTEBOP program to perform a numerical integration over a given exposure time, as suggested by \citet{Kipping2010a}. We tested our modifications by solving the aforementioned generated light curves, now taking the integration time into account, and successfully retrieved the inputted parameters.

\begin{figure}[ht]
\centering
\epsfig{width=\linewidth,file=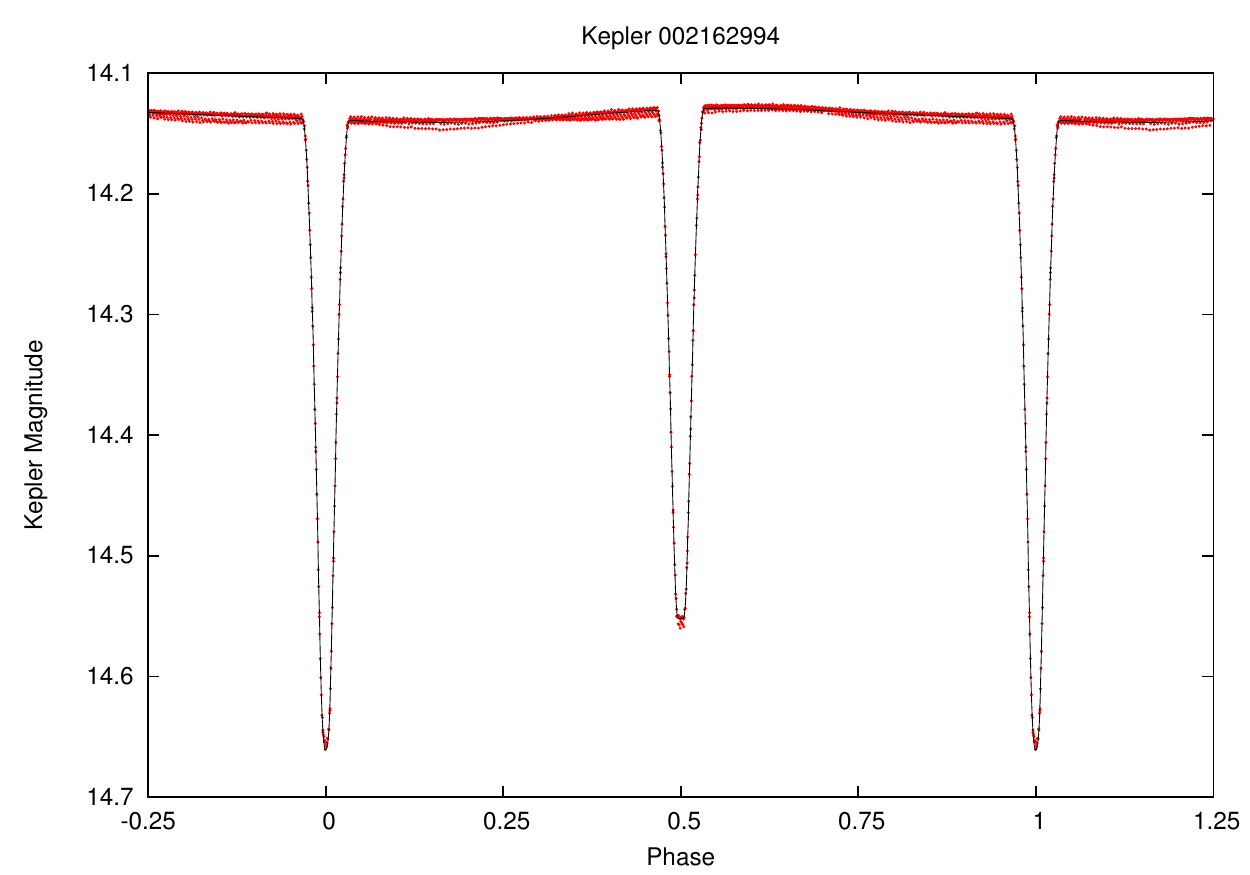}
\caption[Plots of the light curves of the 231 systems modeled with the JKTEBOP code]{Plots of the light curves of the 231 systems modeled with the JKTEBOP code, presented in Table~\ref{ddebcandstab}. Only the first plot, Figure~\ref{lightcurveplots}.1, is shown in the text for guidance. Figures~\ref{lightcurveplots}.1-\ref{lightcurveplots}.231 are available in the online version of the Astronomical Journal under \citet{Coughlin2011}.}
\label{lightcurveplots}
\end{figure}

After estimating the individual mass, radius, and temperature for each component, we re-computed the gravity and limb-darkening coefficients for each individual star, and performed a Levenberg-Marquardt minimization starting from our previously best solutions, taking into account the 29.43 minute integration time. We then repeated the processes of deriving the physical values of the components, interpolating gravity and limb-darkening coefficients, and performing a Levenberg-Marquardt minimization several more times to ensure convergence. The JKTEBOP solutions for all initial 231 candidates are shown in Table~\ref{ddebcandstab}, including the \emph{Kepler} ID number, effective temperature of the system, apparent \emph{Kepler} magnitude, magnitude range of the light curve, period, time of primary minimum, inclination, eccentricity, longitude of periastron, sum of the fractional radii, surface brightness ratio, radii ratio, luminosity ratio, amplitude of the sine curve applied to the luminosity of the primary star to account for spots, and the amount of third light. Although we list the derived surface brightness and radii ratios here, we note again that they are not always reliable on their own, and thus are combined to obtain the luminosity ratio in our analysis via Eq.~\ref{lumeq}. Plots of each of the eclipsing binaries with their model fit are given in Figure~\ref{lightcurveplots}.

\begin{deluxetable}{ccccccccrcccccc}
\tablewidth{0pt}
\rotate
\tabletypesize{\scriptsize}
\tablecaption{Model System Parameters via JKTEBOP for the 231 DDEBs with T$_{\rm eff}$ $<$ 5500 K}
\tablecolumns{15}
\tablehead{\emph{Kepler} ID & T$_{\rm eff}$ & m$_{\rm kep}$ & $\Delta$m$_{\rm kep}$ & Period & T$_{\rm 0}$ & i & e\tablenotemark{a} & $\omega$\tablenotemark{a} & r$_{\rm sum}$ & J & k & L$_{r}$ & L$_{1}$ Sine & L$_{3}$\\ & (K) & & & (Days) & (BJD-2450000) & ($\degr$) & & ($\degr$) & & & & & Amplitude & }
\startdata
\input{keplmb-tab2.tex}
\enddata
\tablenotetext{a}{Although the values for e and $\omega$ are presented in this table for ease of reading, the values of e$\cdot$cos($\omega$) and e$\cdot$sin($\omega$) were actually solved for in the analysis.}
\label{ddebcandstab}
\end{deluxetable}

As a check on the reliability of our analysis technique we took the well-studied low-mass eclipsing binary GU Boo \citep{LopezMorales2005}, and modeled only the $R$-band light curve, (not using the radial velocity curves), via the exact same procedure as stated above in Sections~\ref{keplmbmodelsec} and \ref{newlmbsec}. The only differences were that we used only the $R$-band luminosities from the \citet{Baraffe1998} and \citet{Chabrier2000} models, and an integration time of 2 minutes as stated in \citet{LopezMorales2005}. We used only the period, time of primary minimum, and estimated effective temperature of the system from broadband photometry provided in \citet{LopezMorales2005}, as we did for the systems in our main study. We find T$_{1}$ = 3912 K, M$_{1}$ = 0.61 M$_{\sun}$, R$_{1}$ = 0.62 R$_{\sun}$, T$_{2}$ = 3813 K, M$_{2}$ = 0.57 M$_{\sun}$, and R$_{2}$ = 0.59 via our technique. In comparison, \citet{LopezMorales2005} found with multi-color light curves and radial-velocity curves of the system, values of T$_{1}$ = 3920 K, M$_{1}$ = 0.610 M$_{\sun}$, R$_{1}$ = 0.623 R$_{\sun}$, T$_{2}$ = 3810 K, M$_{2}$ = 0.599 M$_{\sun}$, and R$_{2}$ = 0.620. The values derived from our technique using only a single color light curve are accurate to within a few percent of the very precise values derived from a study using multi-color light and radial-velocity curves, thus validating our technique.

As noted above, \citet{Prsa2010} estimated the parameters of temperature ratio, sum of the fractional radii, e$\cdot$cos($\omega$), e$\cdot$sin($\omega$), and sin(i) for detached systems, via the EBAI technique \citep{Prsa2008}. Before comparing to the parameters obtained by \citep{Prsa2010}, we note that the modeling approach between EBAI and our AGA presented in this chapter have some fundamental differences. EBAI is extremely computationally efficient, but relies on a fitted polynomial to the actual data \citep{Prsa2008}, which is then compared to a neural network training set of 33,235 light curves generated by the Wilson-Devinney code \citep{Wilson1971,Wilson1993}. \citet{Prsa2008} notes that ``...the artificial neural network output is viable for statistical analysis and as input to sophisticated modeling engines for fine-tuning.'' In comparison, the use of our AGA coupled with JKTEBOP is computationally slower, but models each actual data point, obtaining an actual best-fit model while varying all physical parameters of interest over the global solution space. As well, our AGA takes into account the 29.43 minute integration time, while EBAI does not. Thus, although the EBAI technique is excellent for mining large databases, identification of light curve morphology, and obtaining estimates of parameters for statistical studies, it is not intended to model individual light curves as precisely and accurately as possible. Keeping this in mind, comparing the parameters obtained by \citet{Prsa2010} to our solutions for the same systems, we first note a moderate correlation between the sum of radii given by \citet{Prsa2010} and our results, with an average discrepancy of $\sim$20\%. However some of the \citet{Prsa2010} solutions are unphysical, (r$_{sum}$ $<$ 0.0), and visual inspection of the polyfit curves given by \citet{Prsa2010} appears to reveal a systematic underestimation of the eclipse depths. With respect to eccentricity, the parameters presented by \citet{Prsa2010} reveal an unusually large number of eccentric systems, with only 3\% of systems having e $\le$ 0.01, and 11\% of systems with e $\le$ 0.05. In contrast, our parameters show 36\% of systems with e $\le$ 0.01, and 60\% of systems with e $\le$ 0.05, which better matches the large number of systems observed that do not show any offset of secondary eclipse from phase 0.5, and no difference in the eclipse widths, indicative of a circular orbit. There is only a slight correlation between our inclination values and that of \citet{Prsa2010}, but as we previously noted, the \citet{Prsa2010} polyfit curves appear not to fit the eclipse depths well. There is practically no correlation between our values for the surface brightness ratio and EBAI's temperature ratio provided in \citet{Prsa2010}, though \citet{Prsa2010} notes that for detached systems, the ``...eclipse depth ratio is strongly affected by eccentricity and star sizes as well, rendering T$_{2}$/T$_{1}$ a poor proxy for the surface brightness ratio.''

In Table~\ref{lmbmrtab} we list the \emph{Kepler} ID number, orbital period, effective temperature of the system, and the estimated effective temperature, mass and radius of each stellar component for the 95 systems that contain two main-sequence stars, which we define as having a radius less than 1.5 times the \citet{Baraffe1998} and \citet{Chabrier2000} model relationships, and a light curve amplitude of at least 0.1 magnitudes, (suitable for ground-based follow-up and less likely to contain any third light). All of these 95 systems have both stars with masses less than 1.0 M$_{\sun}$. Note that we have ordered Table~\ref{lmbmrtab} such that Star 1 is always the more massive star, regardless if L$_{r}$ was greater or less than 1.0 in Table~\ref{ddebcandstab}. Also note that since we are using $V$+$R$-band luminosities, which best correspond to the \emph{Kepler} bandpass, one cannot always use the simple R$^{2}\cdot$T$^{4}$ relation to derive luminosity ratios from Table~\ref{lmbmrtab} to compare to Table~\ref{ddebcandstab}, since that would correspond to the bolometric luminosity. However, if one takes a system from Table~\ref{lmbmrtab}, looks up the $V$+$R$-band luminosity for each component, based on their mass and temperature, from the \citet{Baraffe1998} and \citet{Chabrier2000} models, and derives a luminosity ratio, this will exactly match the luminosity ratio in Table~\ref{ddebcandstab} from the JKTEBOP models, because the technique defines it as such. These results substantially increase the number of LMMS DDEB candidates in general, and provide 29 new LMMS DDEBs with both components below one solar mass, and at least 0.1 magnitude eclipse depths, in the heretofore unexplored period range of P $>$ 10 days. We further discuss the impact of these systems and comparison to theoretical models in Section~\ref{lmbmodelcompsec}.

\begin{deluxetable}{ccccccccc}
\tablewidth{0pt}
\tabletypesize{\scriptsize}
\tablecaption{Temperature, Mass, and Radius Estimates for the 95 New LMMS DDEB Candidates with Amplitudes $\ge$ 0.1 Magnitudes and Both Masses~$<$~1.0~M$_{\sun}$}
\tablecolumns{9}
\tablehead{\emph{Kepler} ID & Period (Days) & T$_{\rm eff}$(K) & T$_{1}$(K) & T$_{2}$(K) & M$_{1}$(M$_{\sun}$) & M$_{2}$(M$_{\sun}$) & R$_{1}$(R$_{\sun}$) & R$_{2}$(R$_{\sun}$)}
\startdata
\input{keplmb-tab3.tex}
\enddata
\label{lmbmrtab}
\end{deluxetable}

\begin{figure}
\centering
\begin{tabular}{cc}
\epsfig{width=0.7\linewidth,file=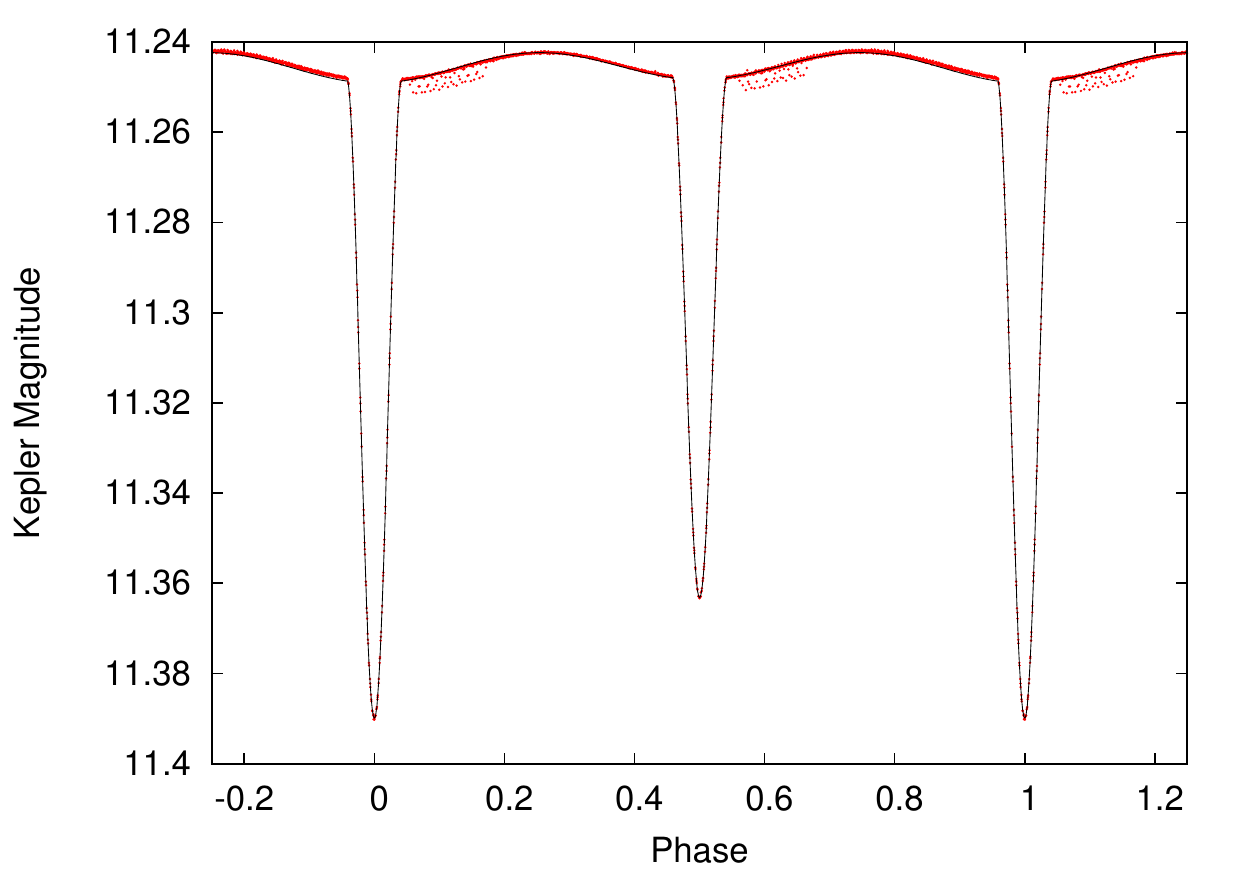} \\
\epsfig{width=0.7\linewidth,file=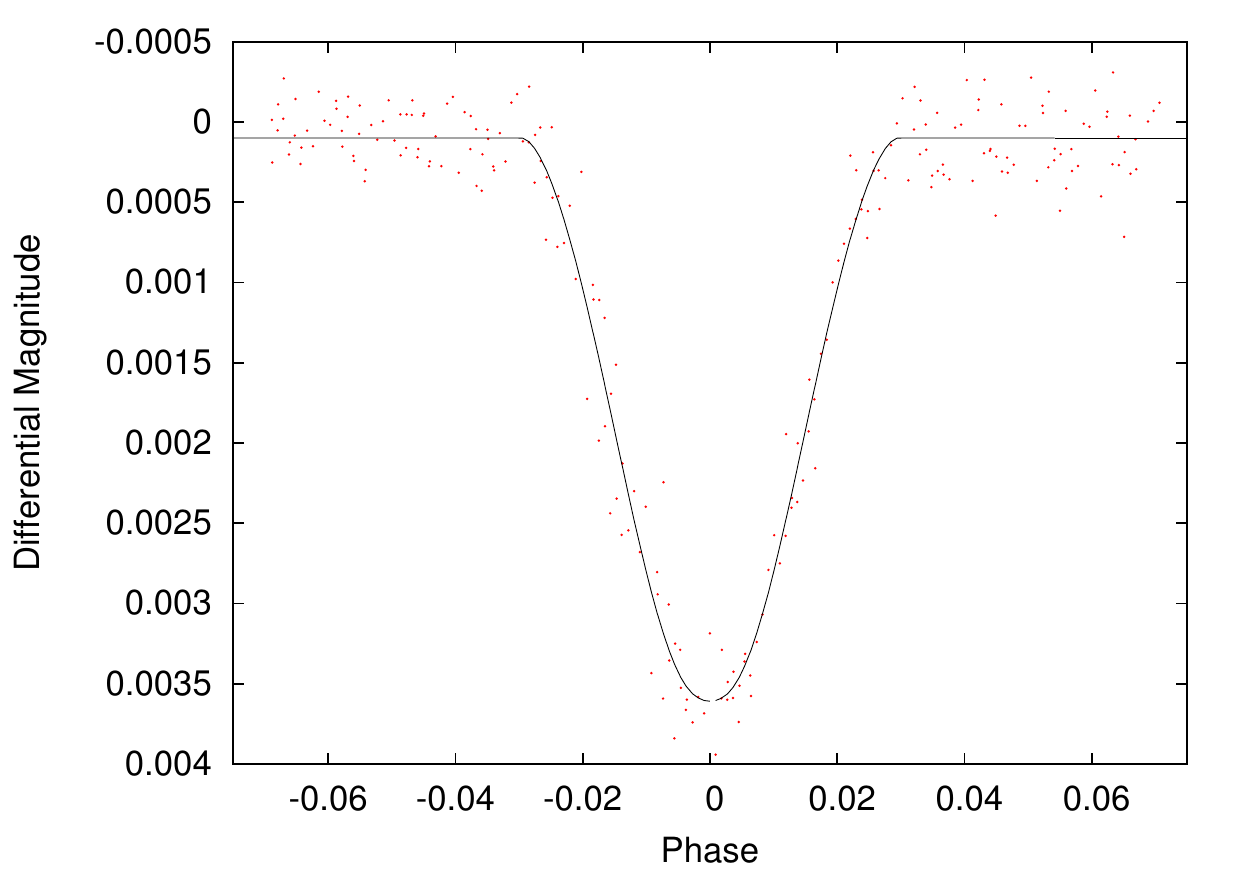} \\
\end{tabular}
\caption[Kepler 004247791. An example of a system which was determined not to be main-sequence]{Kepler 004247791. An example of a system which was determined not to be main-sequence in Section~\ref{newlmbsec}. Top: The light curve phased at its period of 4.100866 days with our best model fit. Given the shallow, wide eclipses for a $\sim$4.1 day period and T$_{\rm eff}$ = 4063K, if this system were main-sequence, it would have a combined mass of 1.28 M$_{\sun}$ and a combined radius of 3.82 R$_{\sun}$. Thus, this system contains one or more evolved stars. Bottom: The model-fit subtracted light curved phased at a period of 2.02484 days, showing a transit-like feature imposed on the light curve of the eclipsing binary. Possible explanations may include, but are certainly not limited to a background eclipsing binary with no visible secondary eclipse at 0.49376 times the orbital period of the foreground binary, a background eclipsing binary with nearly identical primary and secondary eclipses at 0.98752 times the orbital period of the foreground binary, a circumbinary transiting object, or a transiting object around one of the stars in an almost 2:1 resonant orbit with the binary.}
\label{Kepler00424fig}
\end{figure}

In Figure~\ref{Kepler00424fig} we show an example of a system which did not meet the main-sequence criterion, Kepler 004247791, which has T$_{\rm eff}$ = 4063K and a period of 4.100866 days. If this system were main-sequence, via our method, it would have a combined mass of 1.28 M$_{\sun}$ and a combined radius of 3.82 R$_{\sun}$. This can be seen by the wide, shallow eclipses for a system of this period and effective temperature. Thus, this system contains one or two evolved stars. An additional curiosity of this system is a periodic transit-like feature that is superimposed on the eclipsing binary light curve. The transit feature occurs at just slightly less than half the orbital period of the eclipsing binary, so that it is seen twice per every revolution of the eclipsing binary system, occurring at a slightly earlier phase every revolution. We subtract the model fit from the eclipsing binary, and plot the transit feature at its period of 2.02484 days in the right panel of Figure~\ref{Kepler00424fig}. Some possible explanations may include, but are certainly not limited to: 1) a background eclipsing binary with no visible secondary eclipse at 0.49376 times the orbital period of the foreground binary, 2) a background eclipsing binary with nearly identical primary and secondary eclipses at 0.98752 times the orbital period of the foreground binary, 3) a circumbinary transiting object, or 4) a transiting object around one of the stars in an almost 2:1 resonant orbit with the binary. Follow-up multi-color light curves, spectra, and radial velocities will be needed to fully characterize this interesting system.

\subsection{New Transiting Planet Candidates}
\label{transsec}

\begin{figure}[ht]
\centering
\epsfig{width=\linewidth,file=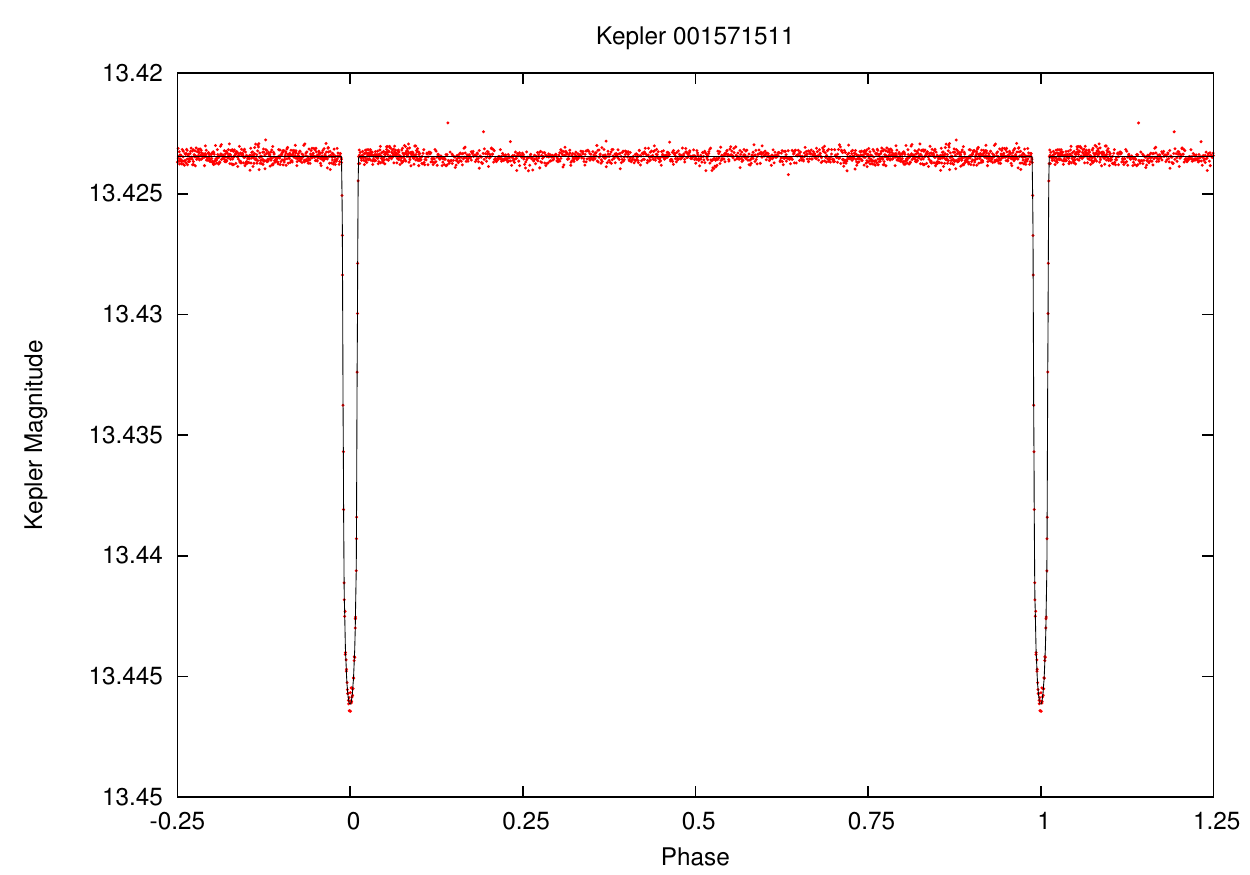}
\caption[Plots of the light curves of the eight transiting planet candidates modeled with the JKTEBOP code]{Plots of the light curves of the eight transiting planet candidates modeled with the JKTEBOP code, presented in Table~\ref{transmrtab}. Only the first plot, Figure~\ref{transcandslcplots}.1, is shown in the text for guidance. Figures~\ref{transcandslcplots}.1-\ref{transcandslcplots}.8 are available in the online version of the Astronomical Journal under \citet{Coughlin2011}.}
\label{transcandslcplots}
\end{figure}

For the eight new transiting planet candidates mentioned in Section~\ref{binaryidentsec}, we combined Q0 and Q1 data, and modeled the transit curves using JKTEBOP, accounting for the 29.43 minute integration time, and using our modified AGA in the same manner described in Section~\ref{keplmbmodelsec}. We assumed zero eccentricity and negligible flux from each planet, and interpolated the limb-darkening and gravity-darkening coefficients via the effective temperature, surface gravity, and metallicity from the relations of \citet{Sing2010} and \citet{Claret2000a}. We then solved for the period, time of primary minimum, inclination, sum of the fractional radii, ratio of the radii, and the out of transit flux level. With this narrowed set of parameters, the AGA proved to be extremely quick and precise, and all fits were confirmed by eye and $\chi^{2}$ values to accurately fit the data. Plots of the transit light curves with model fits are shown in Figure~\ref{transcandslcplots}.

To estimate the physical radius of each transiting exoplanet candidate, we took the value for the radius of the host star from the KIC, and multiplied by the ratio of the radii, $k$, found from the model. In Table~\ref{transmrtab} we list the \emph{Kepler} ID number, apparent \emph{Kepler} magnitude, time of primary minimum, period, effective temperature of the star, inclination, radius of the star, and radius of the exoplanet candidate in both solar radii and Jupiter radii.

\begin{deluxetable}{lccccccccc}
\tablewidth{0pt}
\tabletypesize{\scriptsize}
\tablecaption{Model Parameters for the Eight Transiting Exoplanet Candidates}
\tablecolumns{10}
\tablehead{\emph{Kepler} ID & M$_{\rm kep}$ & T$_{0}$ & P & T$_{\rm eff,\star}$ & i & R$_{\star}$ & R$_{\rm p}$ & R$_{\rm p}$\\ & & (BJD-2454900) & (Days) & (K) & ($\degr$)& (R$_{\sun}$) & (R$_{\sun}$) & (R$_{\rm Jup}$) }
\startdata
\input{keplmb-tab4.tex}
\enddata
\tablenotetext{a}{$ $Listed in the \emph{Kepler} False Positive Catalog as ``velocity measurements indicate eclipsing binary''}
\label{transmrtab}
\end{deluxetable}

As can be seen, the radii for these transiting planet candidates range from 0.56 to 2.1 R$_{\rm Jup}$, with periods between 4.1 and 24.6 days. Only one of these, Kepler 011974540, has been ruled out as a planet from follow-up RV measurements, which are needed for the rest of the candidates to confirm or refute their planetary nature. However, even if these objects turn out not to be planetary mass, they then must be either brown dwarfs or very low-mass stars, which still are valuable finds. In the case of brown dwarfs, these targets would be located within the so-called ``brown dwarf desert'' \citep{McCarthy2004}.

\subsection{Comparison of the New Low-Mass Binary Candidates to Models}
\label{lmbmodelcompsec}

As described in the introduction, one of the current outstanding questions in the study of low-mass stars is whether the inflated radii observed in binaries is caused by their enhanced stellar rotation, and therefore enhanced magnetic activity. We explore this problem in this section using the list of the 95 new LMMS DDEB candidates with estimated individual masses both below 1.0 M$_{\sun}$ and light curve amplitudes greater than 0.1 magnitudes, given in Table~\ref{lmbmrtab}. This sample, for the first time, provides a statistically significant number of systems with orbital periods larger than 10 days. 

\begin{figure}
\centering
\begin{tabular}{cc}
\epsfig{width=0.45\linewidth,file=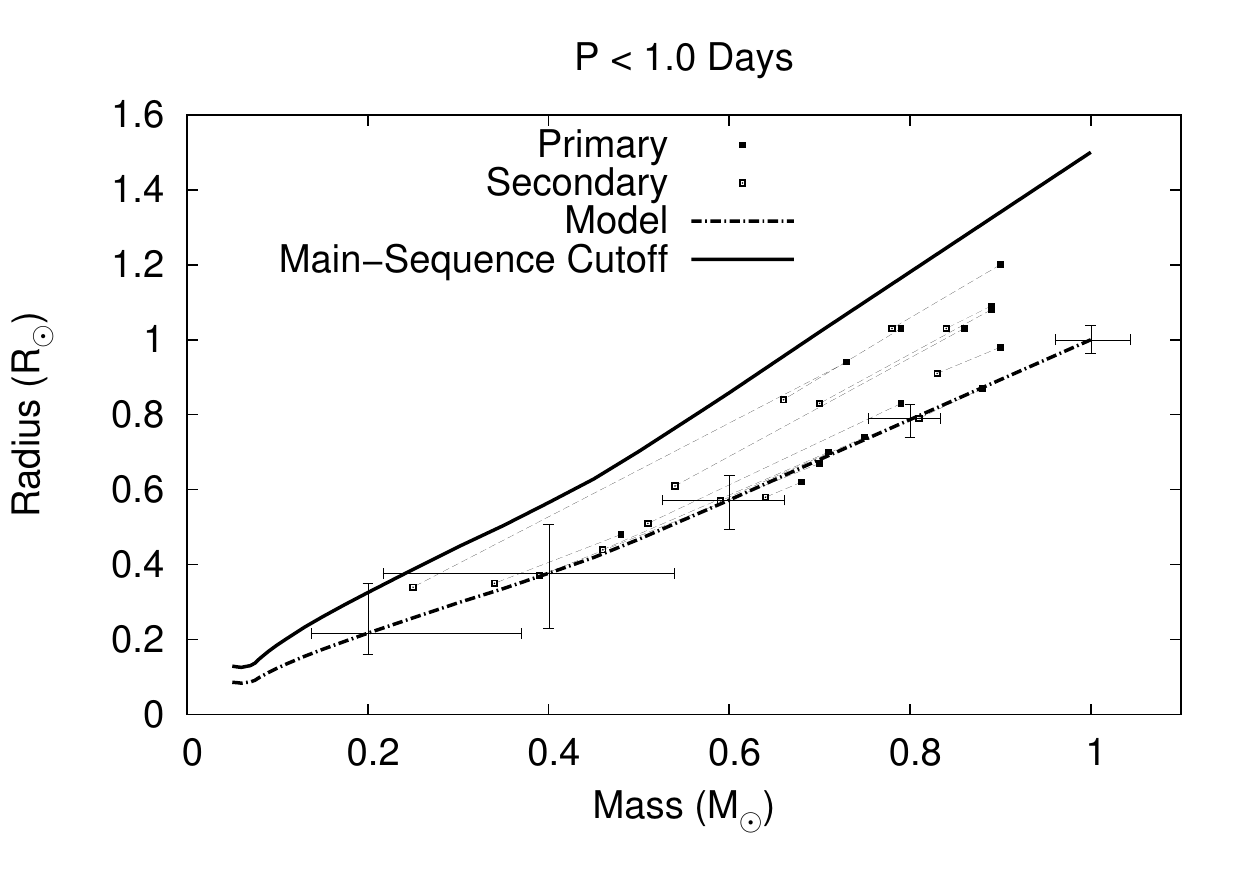} &
\epsfig{width=0.45\linewidth,file=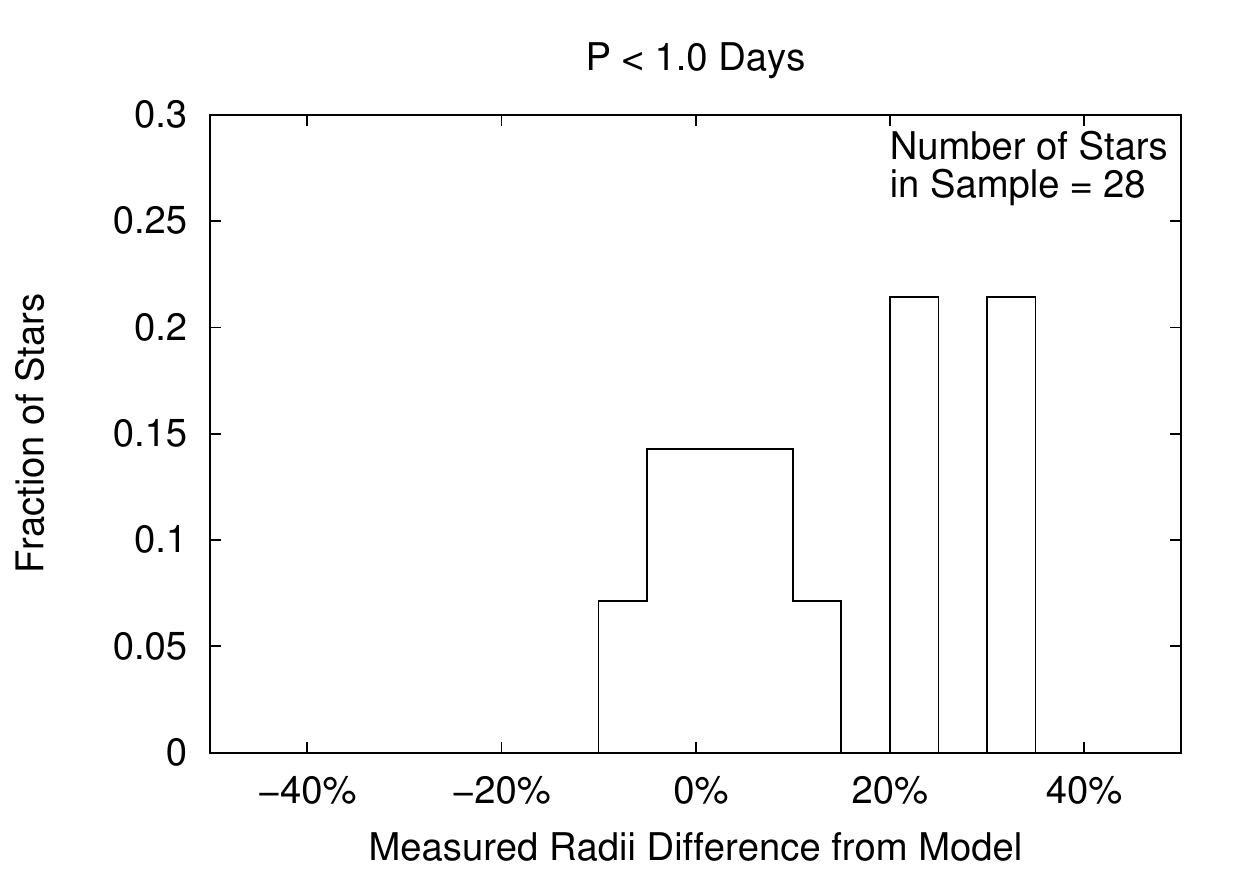} \\
\epsfig{width=0.45\linewidth,file=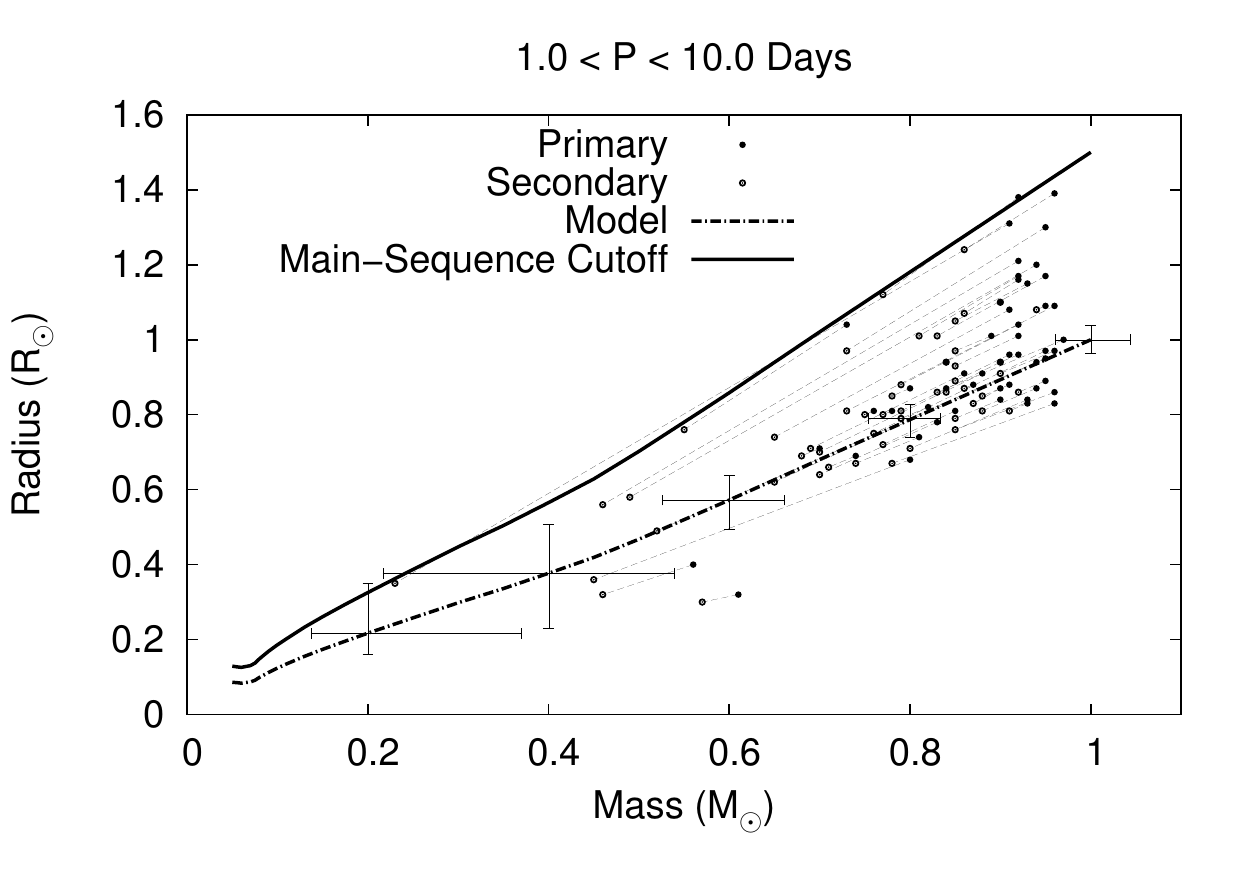} &
\epsfig{width=0.45\linewidth,file=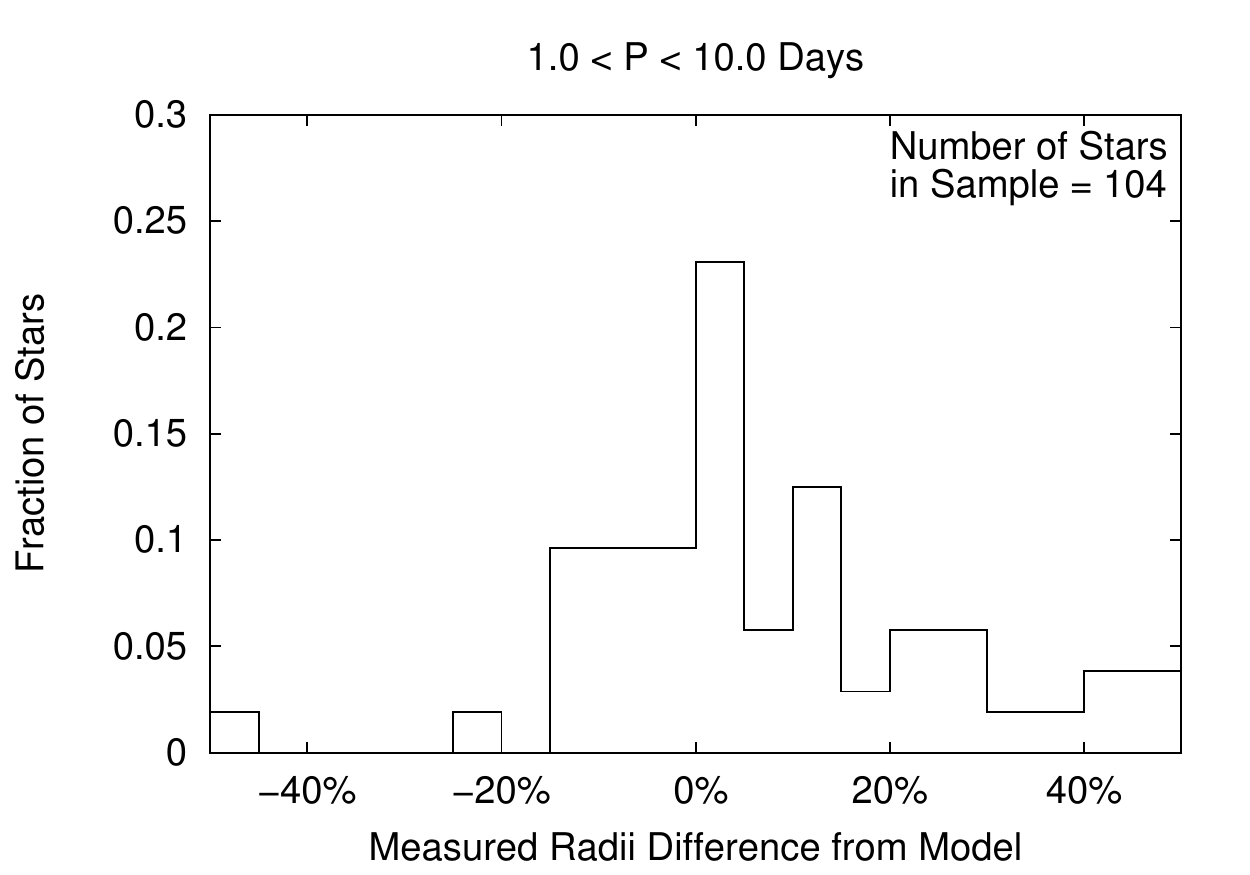} \\
\epsfig{width=0.45\linewidth,file=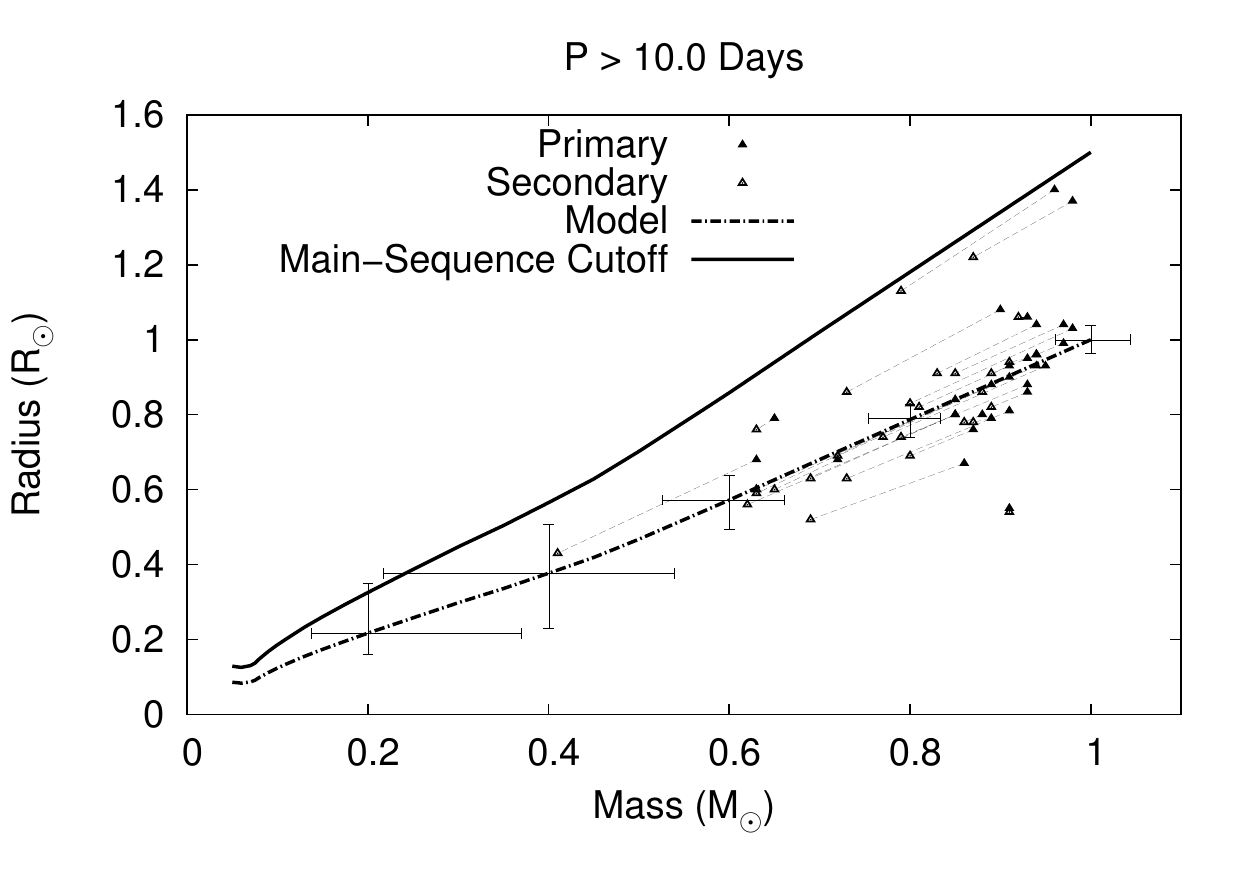} &
\epsfig{width=0.45\linewidth,file=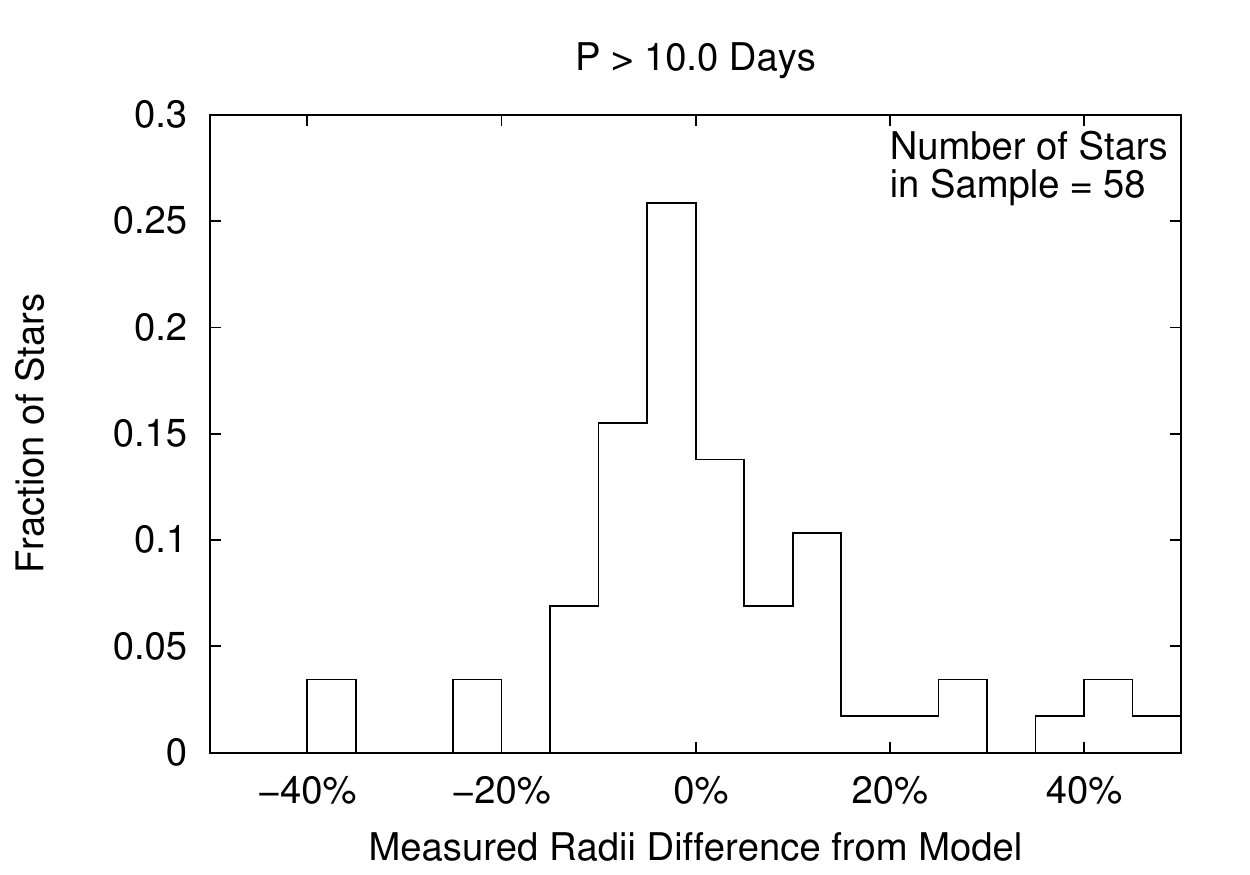} \\
\end{tabular}
\caption[Mass-radius diagrams for each binary with both components $<$ 1.0 M$_{\sun}$ and photometric amplitudes greater than 0.1 mag]{Left: Mass-radius diagrams for each binary with both components $<$ 1.0 M$_{\sun}$ and photometric amplitudes greater than 0.1 mag, as given in Table~\ref{lmbmrtab}, with systems connected by faint lines. The systems are sorted into short-period (P $<$ 1.0 days, top panel), medium-period, (1.0 $<$ P $<$ 10.0 days, middle panel), and long-period groupings, (P $>$ 10.0 days, bottom panel). The theoretical mass-radius relations of \citet{Baraffe1998} for 0.075 M$_{\sun}$ $\le$ M $\le$ 1.0 M$_{\sun}$, and of \citet{Chabrier2000} for M $<$ 0.075 M$_{\sun}$, both for [M/H] = 0.0 and an age of 5.0 Gyr, are over-plotted. The solid line shows the main-sequence cutoff criterion. The error bars indicate the error in mass and radius obtained when interpolating from the mass-temperature-radius relations with an error of 200K. Right: Histograms of the fraction of stars in the sample versus their deviance from the models for each period grouping. As can be seen by both the mass-radius relation plots and the histograms, shorter period binaries in general appear to exhibit larger radii compared to the models than longer period systems.}
\label{mrfig}
\end{figure}

The left-side panels of Figure~\ref{mrfig} show mass-radius diagrams using the mass and radius of each binary star component estimated in Section~\ref{newlmbsec}. The LMMS DDEB candidates have been separated into three categories, with orbital periods P $<$ 1.0 day, 1.0 $<$ P $<$ 10 days, and P $>$ 10 days. Each primary and secondary in a binary pair is traced by a connecting line. We also plot in each panel of Figure~\ref{mrfig} the theoretical mass-radius relation predicted by the \citet{Baraffe1998} models for M $\ge$ 0.075 M$_{\sun}$, and the \citet{Chabrier2000} models for M $<$ 0.075 M$_{\sun}$, both for [M/H] = 0.0, and an age of 5.0 Gyrs. We have also defined a main-sequence cutoff as 1.5 times the theoretical mass-radius relation, which is illustrated by the solid line in each diagram. In the models we have used an $\alpha$ = 1.0 for M $\le$ 0.7 M$_{\sun}$ and interpolated the radius of the models for 0.7 $M_{\sun}$ $<$ M $\le$ 1.0 $M_{\sun}$ by fixing the radius of the 1.0 $M_{\sun}$ model to 1.0 $R_{\sun}$, therefore avoiding the dependence of the stellar radius with $\alpha$ between 0.7 $M_{\sun}$ and 1.0 $M_{\sun}$ \citep{Baraffe1998}. We also include in the mass-radius diagrams estimations of the error in our M and R values at several masses, computed by adding and subtracting 200 K, (the error in the T$_{\rm eff}$ determinations given by the KIC), from a given temperature and interpolating the mass and radius from the theoretical relations. Note that one of the long-period stars, Kepler 008075618, falls well below the main-sequence, with two identical components with M = 0.91 M$_{\sun}$ and R = 0.53 R$_{\sun}$. Inspection of this light curve, coupled with the light curve model, reveals that this system could in fact be a single-lined system at half the listed period.

In the figure, many of the stellar radii of binaries with P $<$ 1.0 days appear to fall above the model predictions, but as the orbital period increases, a larger fraction of the systems appear to have radii that are either consistent with or fall below the models. There certainly is a fair amount of scatter in these data introduced by the large error in the mass and radius estimations, but a histogram analysis of the radius distributions confirms these apparent trends. On the right-side panels of Figure~\ref{mrfig} we show 5\% bin-size histograms representing how many stars have a radius that deviates by a given percentage from the models. The average radius discrepancy is 13.0\%, 7.5\%, and 2.0\% for the short (P $<$ 1.0 days), medium (1.0 $<$ P $<$ 10.0 days), and long-period (P $>$ 10.0 days) systems respectively. Although a full analysis of each system with multi-color light and radial-velocity data is still needed, these preliminary estimates support the hypothesis that binary spin-up is the primary cause of inflated radii in short period LMMS DDEBs.

\subsection{Summary}
\label{keplmbconcsec}

We present 231 new double-eclipse, detached eclipsing binary systems with T$_{\rm eff}$ $<$ 5500 K, found in the Cycle 0 data release of the \emph{Kepler Mission}, and provide their \emph{Kepler} ID, estimated effective temperature, \emph{Kepler} magnitude, magnitude range of the light curve, orbital period, time of primary minimum, inclination, eccentricity, longitude of periastron, sum of the fractional radii, and luminosity ratio. We estimate the masses and radii of the stars in these systems, and find that 95 of them contain two main-sequence stars with both components having M $<$ 1.0 M$_{\sun}$ and eclipse depths of at least 0.1 magnitude, and thus are suitable for ground-based follow-up. Of these 95 systems, 14 have periods less than 1.0 day, 52 have periods between 1.0 and 10.0 days, and 29 have periods greater than 10.0 days. This new sample of low-mass, double-eclipse, detached eclipsing binary candidates more than doubles the number of previously known systems, and extends the sample into the completely heretofore unexplored P $>$ 10.0 day period range for LMMS DDEBs. 

Comparison to the theoretical mass-radius relation models for stars below 1.0 M$_{\sun}$ by \citet{Baraffe1998} show preliminary evidence for better agreement with the models at longer periods, where the rotation rate of the stars is not expected to be spun-up by tidal locking, although, in the absence of radial-velocity measurements, the errors on the estimated mass and radius are still quite large. For systems with P $<$ 1.0 days, the average radius discrepancy is 13.0\%, whereas for 1.0 $<$ P $<$ 10.0 days and P $>$ 10.0 days, the average radius discrepancy is 7.5\% and 2.0\%, respectively. Ground-based follow-up, in the form of radial velocity and multi-wavelength light curves, is needed to derive the mass and radius of each star in each system to $\sim$1-2\%, which we have already begun to acquire. With accurate masses and radii for multiple long-period systems, we should be able to definitively test the hypothesis that inflated radii in low-mass binaries are principally due to enhanced rotation rates.

We also present eight new transiting planet candidates. Only one of them is currently listed in the \emph{Kepler} False Positive Catalog. The remaining candidates require radial-velocity follow-up to confirm or refute their planetary nature. Even if these systems do not turn out to be planets, they then must be brown dwarf or very low-mass, late-type M dwarfs, which would still be a very valuable find. In fact, all false positive planet candidates determined by the \emph{Kepler} team will be of great interest to stellar astrophysics. We also present seven new extremely shallow eclipsing systems, one well detached binary with deep eclipses, and one apparent red giant with an unusual eclipse-like feature. We also highlight a very unusual eclipsing binary system containing at least one evolved star and an additional transit-like feature in the light curve. Finally, the systems that we determined are not main-sequence, and we therefore did not include in the subsequent analysis, should be further studied for valuable science. Accurate mass, radius, and temperature determinations of those systems could yield valuable insights into stellar and binary evolution.

%% file: keplmb-tab1.tex
003098197 & 38.3840\tablenotemark{c} & 5675 & 4.814 & 4.9 & 4.60\\
004178389\tablenotemark{a} & 45.2600\tablenotemark{c} & 5645 & 4.670 & 3.4 & 2.80\\
009016295\tablenotemark{b} & 19.9858 & 5819 & 4.582 & 4.1 & 0.17\\
009071386\tablenotemark{a} & 4.68513 & 6324 & 4.267 & 1.4 & 0.05\\
009838975\tablenotemark{a} & 18.7000 & 5018 & 4.802 & 5.7 & 0.21\\
012017140\tablenotemark{b} & 22.8624 & 6026 & 4.500 & 4.7 & 0.11\\
012504988\tablenotemark{a} & 5.09473 & 5985 & 4.464 & 2.9 & 0.06\\

%% file: keplmb-tab2.tex
002162994 & 5410 & 14.162 & 0.535 & 4.101544 & 5002.545861 & 89.87 &  0.01 & 270 & 0.199 &    0.991 &  0.702 & 0.4888 & 0.008 & 0.00\\
002437452 & 5398 & 16.981 & 0.256 & 14.47184 & 5003.759350 & 87.46 &  0.08 &  90 & 0.084 &    0.641 &   2.47 & 3.905 & 0.011 & 0.00\\
002580872 & 5293 & 14.880 & 0.374 & 15.92672 & 4978.550988 & 87.95 &  0.26 & 102 & 0.084 &     1.14 &   1.25 & 1.774 & 0.014 & 0.00\\
002719873 & 5086 & 15.160 & 0.235 & 17.27953 & 4968.273250 & 87.76 &  0.31 &  90 & 0.059 &    0.633 &   2.64 & 4.425 & 0.007 & 0.00\\
002852560 & 5381 & 15.308 & 0.460 & 11.96119 & 4964.912794 & 88.06 &  0.44 &  41 & 0.079 &     1.04 &  0.986 & 1.008 & 0.000 & 0.00\\
002860788 & 5319 & 14.043 & 0.137 & 5.259798 & 4965.066945 & 82.29 &  0.00 & 268 & 0.212 &    0.876 &  0.561 & 0.2755 & 0.009 & 0.00\\
003003991 & 5366 & 13.926 & 0.115 & 7.244790 & 4964.859062 & 86.88 &  0.28 & 270 & 0.083 &   0.0438 &   11.2 & 5.447 & 0.000 & 0.40\\
003102024 & 5117 & 12.809 & 0.351 & 13.78248 & 4958.697309 & 89.50 &  0.54 & 302 & 0.054 &    0.605 &   1.37 & 1.138 & 0.000 & 0.43\\

%% file: keplmb-tab3.tex
002162994 & 4.102 & 5410 & 5593 & 5038 & 0.96 & 0.86 & 1.39 & 1.24 \\
002437452 & 14.47 & 5398 & 5591 & 4647 & 0.96 & 0.79 & 1.40 & 1.13 \\
002719873 & 17.28 & 5086 & 5246 & 4382 & 0.90 & 0.73 & 1.08 & 0.86 \\
002852560 & 11.96 & 5381 & 5385 & 5378 & 0.93 & 0.92 & 1.06 & 1.06 \\
003003991 & 7.245 & 5366 & 5554 & 4598 & 0.96 & 0.78 & 0.83 & 0.67 \\
003102024 & 13.78 & 5117 & 5160 & 5069 & 0.89 & 0.87 & 0.79 & 0.78 \\
003241344 & 3.913 & 5422 & 5461 & 3688 & 0.94 & 0.52 & 0.94 & 0.49 \\
003241619 & 1.703 & 5165 & 5344 & 4622 & 0.92 & 0.79 & 1.04 & 0.88 \\
003458919 & 0.8920 & 5063 & 5206 & 4254 & 0.89 & 0.70 & 1.08 & 0.83 \\
003730067 & 0.2941 & 4099 & 4158 & 4010 & 0.68 & 0.64 & 0.62 & 0.58 \\
003848919 & 1.047 & 5226 & 5238 & 5214 & 0.90 & 0.90 & 1.10 & 1.10 \\
004049124 & 4.804 & 5349 & 5501 & 4347 & 0.95 & 0.73 & 1.30 & 0.97 \\
004077442 & 0.6929 & 4523 & 4643 & 4094 & 0.79 & 0.66 & 1.03 & 0.84 \\
004346875 & 4.694 & 5339 & 5367 & 3599 & 0.92 & 0.46 & 1.21 & 0.56 \\
004352168 & 10.64 & 5115 & 5281 & 4744 & 0.91 & 0.81 & 0.93 & 0.82 \\
004484356 & 1.144 & 5080 & 5250 & 4636 & 0.90 & 0.79 & 0.94 & 0.81 \\
004540632 & 31.01 & 4818 & 4953 & 4190 & 0.85 & 0.69 & 0.80 & 0.63 \\
004633434 & 22.27 & 4902 & 5041 & 4219 & 0.86 & 0.69 & 0.67 & 0.52 \\
004678171 & 15.29 & 4240 & 4331 & 4048 & 0.72 & 0.65 & 0.68 & 0.60 \\
004773155 & 25.71 & 5447 & 5448 & 5447 & 0.94 & 0.94 & 0.96 & 0.96 \\
004908495 & 3.121 & 4731 & 4791 & 4655 & 0.82 & 0.79 & 0.82 & 0.79 \\
005036538 & 2.122 & 4199 & 4236 & 4155 & 0.70 & 0.68 & 0.71 & 0.69 \\
005080652 & 4.144 & 5344 & 5536 & 4858 & 0.95 & 0.83 & 1.17 & 1.01 \\
005300878 & 1.279 & 4631 & 4667 & 4590 & 0.80 & 0.78 & 0.87 & 0.85 \\
005597970 & 6.717 & 5179 & 5284 & 4060 & 0.91 & 0.65 & 1.08 & 0.74 \\
005731312 & 7.946 & 4658 & 4701 & 3583 & 0.80 & 0.45 & 0.68 & 0.36 \\
005781192 & 9.460 & 5372 & 5546 & 4482 & 0.95 & 0.76 & 0.97 & 0.75 \\
005802470 & 3.792 & 5418 & 5620 & 4859 & 0.97 & 0.83 & 1.00 & 0.86 \\
005871918 & 12.64 & 4021 & 4052 & 3983 & 0.65 & 0.63 & 0.79 & 0.76 \\
006029130 & 12.59 & 5160 & 5201 & 5114 & 0.89 & 0.88 & 0.88 & 0.86 \\
006131659 & 17.53 & 4870 & 4970 & 3972 & 0.85 & 0.63 & 0.84 & 0.59 \\
006449552 & 20.15 & 5357 & 5537 & 4532 & 0.95 & 0.77 & 0.93 & 0.74 \\
006464285 & 0.8436 & 5061 & 5159 & 4923 & 0.89 & 0.84 & 1.09 & 1.03 \\
006466939 & 2.286 & 4920 & 4925 & 4916 & 0.84 & 0.84 & 0.87 & 0.86 \\
006591789 & 5.088 & 5410 & 5560 & 4342 & 0.96 & 0.73 & 1.09 & 0.81 \\
006697716 & 1.443 & 4898 & 5036 & 4215 & 0.86 & 0.69 & 0.91 & 0.71 \\
006706287 & 2.535 & 5182 & 5327 & 4931 & 0.91 & 0.85 & 0.96 & 0.89 \\
006778050 & 0.9458 & 5091 & 5223 & 4872 & 0.90 & 0.83 & 0.98 & 0.91 \\
006841577 & 15.54 & 5478 & 5676 & 4676 & 0.98 & 0.80 & 1.03 & 0.83 \\
006863840 & 3.853 & 5024 & 5050 & 4997 & 0.87 & 0.86 & 0.88 & 0.87 \\
007119757 & 0.7429 & 5072 & 5242 & 4607 & 0.90 & 0.78 & 1.20 & 1.03 \\
007125636 & 6.491 & 4358 & 4417 & 4277 & 0.74 & 0.71 & 0.69 & 0.66 \\
007128918 & 7.119 & 5386 & 5574 & 4968 & 0.96 & 0.85 & 0.86 & 0.76 \\
007129465 & 5.492 & 5182 & 5269 & 5069 & 0.90 & 0.87 & 0.87 & 0.83 \\
007200102 & 14.67 & 5207 & 5390 & 4643 & 0.93 & 0.79 & 0.88 & 0.74 \\
007624297 & 18.02 & 5135 & 5291 & 4352 & 0.91 & 0.73 & 0.81 & 0.63 \\
007670617 & 24.70 & 4876 & 4971 & 3945 & 0.85 & 0.62 & 0.80 & 0.56 \\
007671594 & 1.410 & 3717 & 3773 & 3597 & 0.56 & 0.46 & 0.40 & 0.32 \\
007691527 & 4.800 & 5354 & 5492 & 5138 & 0.94 & 0.88 & 0.87 & 0.81 \\
007749318 & 2.372 & 5211 & 5347 & 4991 & 0.92 & 0.86 & 1.16 & 1.07 \\
007798259 & 1.734 & 4619 & 4735 & 4386 & 0.81 & 0.74 & 0.74 & 0.67 \\
007846730 & 11.03 & 5476 & 5667 & 5079 & 0.98 & 0.87 & 1.37 & 1.22 \\
008075618 & 17.56 & 5288 & 5301 & 5275 & 0.91 & 0.91 & 0.55 & 0.54 \\
008094140 & 0.7064 & 4200 & 4266 & 3598 & 0.71 & 0.46 & 0.70 & 0.44 \\
008296467 & 10.30 & 5316 & 5427 & 5159 & 0.93 & 0.89 & 0.86 & 0.82 \\
008364119 & 7.736 & 5443 & 5581 & 5232 & 0.96 & 0.90 & 0.97 & 0.91 \\
008411947 & 1.798 & 5086 & 5168 & 4980 & 0.89 & 0.85 & 1.01 & 0.97 \\
008580438 & 6.496 & 5307 & 5314 & 3348 & 0.91 & 0.23 & 1.31 & 0.35 \\
008906676 & 8.210 & 5249 & 5436 & 4709 & 0.93 & 0.80 & 0.84 & 0.71 \\
009001468 & 17.33 & 4949 & 5089 & 4676 & 0.87 & 0.80 & 0.76 & 0.69 \\
009029486 & 6.277 & 5368 & 5421 & 5309 & 0.93 & 0.91 & 0.83 & 0.81 \\
009098810 & 8.258 & 5126 & 5240 & 4956 & 0.90 & 0.85 & 0.84 & 0.79 \\
009210828 & 1.656 & 4893 & 4898 & 4888 & 0.84 & 0.84 & 0.94 & 0.94 \\
009284741 & 20.73 & 5085 & 5156 & 4998 & 0.88 & 0.86 & 0.80 & 0.78 \\
009328852 & 0.6458 & 4338 & 4357 & 3375 & 0.73 & 0.25 & 0.94 & 0.34 \\
009346655 & 0.8716 & 4183 & 4232 & 3512 & 0.70 & 0.39 & 0.67 & 0.37 \\
009474485 & 1.025 & 4469 & 4492 & 4444 & 0.76 & 0.75 & 0.81 & 0.80 \\
009639265 & 0.5063 & 5004 & 5147 & 4730 & 0.88 & 0.81 & 0.87 & 0.79 \\
009665503 & 11.57 & 5141 & 5293 & 4321 & 0.91 & 0.72 & 0.90 & 0.69 \\
009714358 & 6.480 & 4825 & 4964 & 4522 & 0.85 & 0.77 & 0.81 & 0.72 \\
009762519 & 7.515 & 5435 & 5528 & 4050 & 0.95 & 0.65 & 0.95 & 0.62 \\
009837578 & 20.73 & 5359 & 5390 & 5327 & 0.93 & 0.91 & 0.95 & 0.94 \\
009934208 & 9.059 & 4258 & 4347 & 3743 & 0.73 & 0.55 & 1.04 & 0.76 \\
009944421 & 7.095 & 5304 & 5348 & 5255 & 0.92 & 0.90 & 0.96 & 0.94 \\
010129482 & 0.8463 & 4558 & 4622 & 3669 & 0.79 & 0.51 & 0.83 & 0.51 \\
010189523 & 1.014 & 5002 & 5143 & 4239 & 0.88 & 0.70 & 0.91 & 0.70 \\
010215422 & 24.40 & 5427 & 5625 & 4944 & 0.97 & 0.85 & 1.04 & 0.91 \\
010264202 & 1.035 & 5207 & 5347 & 4971 & 0.92 & 0.85 & 1.01 & 0.93 \\
010292465 & 1.353 & 5258 & 5417 & 4965 & 0.93 & 0.85 & 1.15 & 1.05 \\
010711646 & 0.7376 & 4339 & 4440 & 3877 & 0.75 & 0.59 & 0.74 & 0.57 \\
010753734 & 19.41 & 5446 & 5603 & 5183 & 0.97 & 0.89 & 0.99 & 0.91 \\
010794242 & 7.144 & 5459 & 5490 & 3633 & 0.94 & 0.49 & 1.20 & 0.58 \\
010979716 & 10.68 & 3932 & 3996 & 3530 & 0.63 & 0.41 & 0.68 & 0.43 \\
010992733 & 18.53 & 5274 & 5457 & 4848 & 0.94 & 0.83 & 1.04 & 0.91 \\
011134079 & 1.261 & 5201 & 5381 & 4732 & 0.92 & 0.81 & 1.17 & 1.01 \\
011233911 & 4.960 & 5193 & 5370 & 4531 & 0.92 & 0.77 & 1.38 & 1.12 \\
011391181 & 8.617 & 5218 & 5288 & 5133 & 0.91 & 0.88 & 0.88 & 0.85 \\
011975363 & 3.518 & 5482 & 5507 & 5457 & 0.95 & 0.94 & 1.09 & 1.08 \\
012004679 & 5.042 & 5432 & 5514 & 5330 & 0.95 & 0.92 & 0.89 & 0.86 \\
012004834 & 0.2623 & 3576 & 3620 & 3468 & 0.48 & 0.34 & 0.48 & 0.35 \\
012356914 & 27.31 & 5368 & 5455 & 4003 & 0.94 & 0.63 & 0.93 & 0.60 \\
012400729 & 0.9317 & 4949 & 5005 & 3715 & 0.86 & 0.54 & 1.03 & 0.61 \\
012418816 & 1.522 & 4583 & 4603 & 4563 & 0.78 & 0.77 & 0.81 & 0.80 \\
012470530 & 8.207 & 4725 & 4863 & 4245 & 0.83 & 0.70 & 0.78 & 0.64 \\
012599700 & 1.018 & 3887 & 3936 & 3816 & 0.61 & 0.57 & 0.32 & 0.30 \\

%% file: keplmb-tab4.tex
001571511 & 13.42 & 68.529019 & 14.02065 & 5804 & 89.28 & 1.08 & 0.14 & 1.43\\
003342592 & 14.92 & 69.190452 & 17.17864 & 5717 & 89.20 & 0.93 & 0.14 & 1.37\\
005372966 & 15.37 & 67.675070 & 9.286422 & 5464 & 88.91 & 0.92 & 0.19 & 1.87\\
006756669 & 15.33 & 65.860125 & 5.851827 & 5353 & 88.34 & 0.90 & 0.16 & 1.59\\
006805146 & 13.21 & 56.568771 & 13.77974 & 6214 & 89.14 & 1.41 & 0.21 & 2.11\\
008544996 & 15.20 & 65.898818 & 4.081488 & 5463 & 87.61 & 1.00 & 0.13 & 1.27\\
011974540$^{a}$ & 13.22 & 65.862352 & 24.67058 & 6507 & 89.53 & 0.69 & 0.06 & 0.56\\
012251650 & 14.76 & 71.657743 & 17.76233 & 4952 & 88.97 & 1.00 & 0.16 & 1.64\\

%% file: chp6-keplmb2.tex
\begin{singlespace}
\section[\MakeUppercase{Mass and Radius Determination of New\\Long-Period Low-Mass Eclipsing Binaries}]{\MakeUppercase{Mass and Radius Determination of New Long-Period Low-Mass Eclipsing Binaries}}
\label{chap6}
\end{singlespace}

\subsection{Introduction}

In Chapter~\ref{chap5} we presented the identification of 231 new double-eclipse, detached eclipsing binary systems with $T_{\rm eff}$ $<$ 5500 K, found in the Cycle 0 data release of the $Kepler$ Mission, and estimated that 95 of them contained two main-sequence components with eclipses of at least 0.1 magnitudes. Of these, 29 have periods greater than 10 days, and thus provide the opportunity to test whether the large radii, compared to stellar model predictions, historically observed in low-mass eclipsing binaries is principally a result of enhanced rotation rates due to being in short-period systems. Additionally, as part of a $Kepler$ Guest Observer program, we found an additional 29.911 day period low-mass eclipsing binary through 90-day monitoring of $Kepler$ field stars \citep{Harrison2012}. Although the existing $Kepler$ light curves provide excellent constraints on the fractional radii of the components, follow-up spectroscopic observations and radial velocity measurements are needed in order to directly determine the absolute masses and temperatures of each component, as well as the scale of each system.

\subsection{Observational Data}

\subsubsection{Photometric Observations}
\label{photobs}

In sections~\ref{datasec} and \ref{binaryidentsec} we noted that $Kepler$ data can contain discrepant systematics from quarter to quarter, as the spacecraft is rotated 90 degrees and each star lands on a different set of pixels. Consequently, we chose to select only a single quarter of data for each binary system of interest for modeling. For each star, we examined all available public data, which spanned up to Q9 at the time of analysis, and selected one of the most recently available quarters which showed to have the least amount of systematic noise in the PA and PDC light curves. Given that the PDC pipeline does a comparatively good job compared to our PCA-based analysis (see Section~\ref{datasec}) for recent quarters of data, we used the PDC data, applying a B\'ezier curve to the out of eclipse light curve to remove any remaining systematics. We used short-cadence (1 min) when available, but the majority of the systems only had long-cadence (30 minute) data available.

\subsubsection{Spectroscopic Observations}

For the radial-velocity observations, longslit spectra were obtained at both the Apache Point Observatory 3.5-meter telescope via the Dual Imaging Spectrograph (DIS), and the Kitt Peak National Observatory (KPNO) 4-meter via the R-C Spectrograph. Typically for short-period eclipsing binaries with P $<$ 1 day, a single night is sufficient to obtain a good radial-velocity curve, assuming that the system is bright. If the system is fainter, one has to make sure that the individual exposure times are not so long that orbital smearing occurs, where the binary moves a significant fraction of its orbit during the exposure. For long-period systems, orbital smearing is not a concern, but it can take weeks of observing to cover an entire orbit. Thus, we decided to employ the strategy of obtaining single radial velocity observations of as many systems as possible in a given night, spreading the observing nights over several months. From our initial list, we observed the brightest stars first, thus allowing us to maximize the number of systems observed.

The observations were carried out over two observing seasons; the first from June - December 2010, and the second from May - September 2011. The DIS spectrograph is able to simultaneously obtain spectra in both blue ($\sim$4400~\AA) and red ($\sim$7500~\AA) channels, but since the majority of the flux of low-mass stars is emitted in the red channel, the blue channel was discarded and only the red channel utilized. The R1200 grating was employed resulting in a resolution of 0.58~\AA~per pixel, or R~$\approx$~10,000. The wavelength range was set to $\sim$5900 - 7100~\AA~for the first observing season, and $\sim$5700 - 6800~\AA~for the second observing season. For the R-C Spectrograph, the KPC-24 grating was used in second order, resulting in a resolution of 0.53~\AA~per pixel, or R~$\approx$~10,000, with the wavelength range set to $\sim$5700 - 6750~\AA. In total, 615 individual stellar spectra were taken.

All raw frames were bias, dark, and flat-field corrected. For the DIS observations, column 1023 is a dead column, and thus those values were replaced via linear interpolation from the two neighboring columns. Due to the long exposure times ($\gtrsim$ 5 minutes) for many systems, there were a significant amount of cosmic rays on the detector for most frames. For all frames, a Laplacian edge algorithm was used to detect and replace cosmic rays via the $lacos\_spec$ IRAF package, (see \citet{Dokkum2001}). This routine was thoroughly examined and observed to be robust in only removing cosmic rays, and not altering any real spectral features.  We used the $apall$ package in IRAF to extract one-dimensional spectra for each image, tracing the apertures to account for curvature on the chip, and subtracting off the night sky background from the science images.

We used lamps containing Helium, Neon, and Argon gas (HeNeAr) in order to produce wavelength calibration frames. For the first observing season of DIS data, one wavelength calibration frame was taken at the start of each night, but for all subsequent DIS observations and all KPNO observations one HeNeAr image was taken directly before or after each science exposure. After identifying the $\sim$20 strong HeNeAr spectral features in each calibration frame, we used a 4$^{th}$ order Legendre polynomial to map the wavelength solution to the pixel values, typically obtaining a dispersion solution with error $\lesssim$~0.025~\AA. We then applied that solution to the respective science frames to transform them to wavelength space. As we typically had sufficiently long exposure times so as to measure the position of the 6300.3086~\AA~OI(D) terrestrial night glow emission line in the sky spectrum, we used this line to correct for any systematic shifts that might have occurred between the science and wavelength calibration frames. We also note that we disabled rotation when using the DIS instrument to minimize any spurious flexure due to rotation on the 3.5 meter Alt-Az telescope.

After wavelength calibration, all science spectra were flattened, (i.e., we normalized each spectrum by its continuum value at each wavelength), by fitting a 20-piece cubic spline fit over 10 iterations, during which points 3$\sigma$ above the fit or 1.5$\sigma$ below the fit were rejected. For each spectrum, the BJD(TT), (the Barycentric Julian Date in Terrestrial Time), was calculated and used for all future time keeping. In order to compensate for the velocity of Earth, the IRAF task $bcvcorr$ was used to correct all observations to the reference frame of the solar system barycenter.

In the end, sufficient data for a radial-velocity curve analysis was collected for 11 systems: Kepler 003102024, Kepler 004352168, Kepler 004773155, Kepler 006029130, Kepler 006131659, Kepler 006431670, Kepler 007846730, Kepler 008296467, Kepler 009284741, Kepler 010753734, and Kepler 010992733.

\subsection{Radial-Velocity Measurement}
\label{rvobs}

Traditionally in binary star work, cross-correlation is used to extract the radial velocities of each component from the observed spectrum. In this technique, the observed spectrum and a reference spectrum are each re-binned to the same linear wavelength scale. The correlation value, (i.e., the sum of the products of the spectrum's and reference's flux values at each wavelength), is computed for all possible pixel shifts, and the maximum correlation value corresponds to the most likely value for the pixel shift, or velocity, of the observed spectrum. When two spectra are visible in the observed spectrum, as is typically the case for binary stars, one observes a double-peaked correlation value, with each peak corresponding to a component of each binary. Using only one reference spectrum in this manner has the significant drawbacks of it tends to underestimate the velocity of each component due to line blending, and often does not reveal the velocities of the secondary component if it is significantly fainter than the primary, or the two stars have similar velocities in a given spectrum.

Thus, \citet{Zucker1994} developed TODCOR, a two-dimensional cross-correlation technique for finding the radial velocities of binary stars. TODCOR performs the cross-correlation with two reference spectra simultaneously, finding the values for the velocity of each reference spectrum, and their relative luminosity ratio, $\alpha$, that results in the highest correlation value. Although this is a significant advantage over one-dimensional cross-correlation, there still exists the drawback that the observed spectrum needs to be re-binned to a linear wavelength scale. This can result in the distortion of spectral lines and thus result in a degradation of accuracy when extracting the component's velocities. Furthermore, it is difficult to assess the formal error of the observations using the correlation value as a goodness-of-fit parameter, and the need for a constant linear wavelength scale can hamper robust error determination methods such as Monte Carlo or Bootstrapping simulations.

Thus, we present here a new method we invented to extract radial velocity values from double-lined spectra. Instead of using cross-correlation, we directly fit reference spectra to each observed spectrum via a traditional standard deviation minimization, which is equivalent to minimizing $\chi^{2}$ assuming all observed points have equal errors. Specifically, for each observed spectrum, we fit for the velocity of each reference spectra as well as $\alpha$, for a total of 3 free parameters. During the fitting, the original observed spectrum is never changed, and thus the original spectrum can have any arbitrary number and distribution of wavelength and flux value pairs. Given a total of $M$ observed spectra, each with $N_{j}$ points, taken at times $t_{j}$, we want to find the values of $V_{1,j}$, $V_{2,j}$, and $\alpha$, i.e., the velocities of reference spectra 1 and 2 at each $t_{j}$ and the relative luminosity of the reference spectra, that minimizes the function

\begin{equation}
\label{rmseq}
  \sum_{j=0}^{M}{\sum_{i=0}^{N_{j}}{\left(F_{0,i} - \frac{\alpha\cdot F_{1}\left(\lambda_{0,i}\cdot(1-\frac{V_{1,j}}{c})\right) + F_{2}\left(\lambda_{0,i}\cdot(1-\frac{V_{2,j}}{c})\right)}{\alpha+1}\right)^{2}}}
\end{equation}

\noindent where $F_{0,i}$ is the flux of an observed spectrum at given wavelength, $\lambda_{0,i}$, $F_{1}$($\lambda$) and $F_{2}$($\lambda$) are the fluxes of reference stars 1 and 2 at a given wavelength, and $c$ is the speed of light. In order to determine $F_{1}$($\lambda$) and $F_{2}$($\lambda$), cubic spline interpolation is used on the original reference spectra.

For reference spectra, we employ the normalized (flattened) synthetic spectra of \citet{Munari2005}, which ensures that no spurious signals are found that correspond to telluric features. The \citet{Munari2005} grid we employ covers models with 3500 $\geq$ $T_{\rm eff}$ $\geq$ 10,000 at steps of 250 K, 0.0 $\geq$ $\log{g}$ $\geq$ 5.0 at steps of 0.5 dex, -2.5 $\geq$ [M/H] $\geq$ 0.5 at steps of 0.5 dex, and 0 $\geq$ $V_{rot}$ $\geq$ 100 km/s in steps of 10 km/s. We fix the value of $\alpha$ to the value of the luminosity ratio found via the light curve modeling, as the wavelength range of the spectra closely matches the $Kepler$ bandpass. Additionally we require that the ratio of blackbodies with temperatures $T_{1}$ and $T_{2}$, the temperatures of reference spectra 1 and 2, when integrated over the $Kepler$ bandpass, match the observed surface brightness ratio from the light curve. In order to ensure we find the global minimum in both selected reference spectra and their velocities, we perform a global grid search looping over all possible combinations of $V_{1,j}$, $V_{2,j}$, $T_{1}$, (with $T_{2}$ set as just mentioned), a common [M/H] for both reference spectra, $\log{g}$ for each reference spectrum, and $V_{rot}$ for each reference star.

To determine robust errors on the derived radial velocities, we employ a Bootstrapping re-sampling method. In Bootstrapping, the original data, which in this case is each observed spectrum, is re-sampled with replacement N times, where N is the number of data points in the original dataset, to create a new dataset. This new dataset is re-fit and new values on the parameters of interest, (in this case the reference velocities), are computed and recorded. The process is repeated a large number of times, which we choose to be 10,000, and thus a median and 1$\sigma$ confidence interval is able to be computed on the resulting distribution. Bootstrapping is advantageous in this case because it does not require any error estimates on the individual data points, (which are difficult to compute for spectroscopic measurements), and it draws on the inherent distribution of the original dataset without any assumptions. 

To estimate errors on the derived temperature of the components, we utilize the spectroscopic quality-of-fit parameter \citep{LopezMoralesBonanos2008,Behr2003}, defined as

\begin{equation}
  z = \sqrt{N}\left(\frac{rms^{2}}{rms^{2}_{min}} - 1\right)
\end{equation}

\noindent where N is the number of data points, $rms^{2}$ is the standard deviation of the fit under consideration, and $rms^{2}_{min}$ is the best fit found. The $z$ parameter is similar to a reduced $\chi^{2}$ in the absence of known errors on the individual points. By definition $z$ = 0 at the best-fit, and where $z$ = 1 represents the 1$\sigma$ confidence interval. Starting from the best-fit value of $T_{1}$, we compute the $rms^{2}$ value for neighboring values of $T_{1}$ via Eq.~\ref{rmseq}, which results in a smoothly varying parabolic-like curve of $z$ versus $T_{1}$. Fitting a parabola to these measurements, we find the value of $T_{1}$ that minimizes $z$, i.e., the best fit value, and the confidence interval where $z$ $<$ 1, i.e., the resulting 1$\sigma$ confidence interval. We compute a value of $T_{2}$ via the value of the surface brightness ratio from the light curve, as described above, propagating the errors on $T_{1}$ and the surface brightness ratio to obtain the error on $T_{2}$.

\subsection{Modeling}

\subsubsection{Light Curve Modeling}
\label{rvmodeltech}

We use the JKTEBOP code \citep{Southworth2004a,Southworth2004b} to model each light curve. As the systems under investigation are well-detached systems, we assume they are perfect spheres and not tidally distorted. As they are all late-type stars, we assume a linear limb-darkening law, (see Section~\ref{ldintro}), and simultaneously solved for the period, time of primary minimum, inclination, e$\cdot$cos($\omega$), e$\cdot$sin($\omega$), surface brightness ratio, sum of the fractional radii, ratio of the radii, out of eclipse flux, and linear limb-darkening coefficient of each star, designated $c_{p}$ and $c_{s}$ for the primary and secondary star respectively. We set gravity darkening coefficients for each star based on their estimated $T_{\rm eff}$ via interpolation from stellar models as prescribed by \citet{Claret2000a}, although we note that this has extremely minimal impact on the resulting light curves due to the wide separation of the components. For the long cadence data, we instructed the JKTEBOP program to perform 10 sub-integrations over the 29.43 minute integration time to account for photometric smearing \citep{Kipping2010a}. The light curves for each system, along with the best-fit model, are shown in Figures~\ref{Kepler003102024}~-~\ref{Kepler010992733}.

To determine robust errors, we first scaled the error bars so that the best fit had a reduced $\chi^{2}$ = 1, as sometimes there were remaining systematics in the light curve, and we did not want to consequently underestimate errors by ignoring this source of noise. We then performed 10,000 Monte Carlo simulations where new datasets were created by adding noise to the original light curves via each point's individual error bars. At each iteration the original best-fit parameters were significantly perturbed, and the resulting light curve re-fit via a Levenberg-Marquardt minimization algorithm. The posterior distributions on all the parameters were then used to compute the median and 1$\sigma$ confidence intervals for all parameters. We list these values for each system in Table~\ref{lmblctab}.

\begin{deluxetable}{cccccccccc}
\tablewidth{0pt}
\rotate
\tabletypesize{\scriptsize}
\tablecaption{Modeling Results of the Long-Period Low-Mass Binary Light Curves}
\tablecolumns{10}
\tablehead{\emph{Kepler} ID & $P$ & $r_{\rm sum}$ & $k$ & $J$ & $i$ & $e$\tablenotemark{a} & $\omega$\tablenotemark{a} & $c_{p}$ & $c_{s}$\\ & (Days) & & & & ($\degr$) & & ($\degr$) & & }
\startdata
\input{lmblcmodel.tab}
\enddata
\label{lmblctab}
\end{deluxetable}

\subsubsection{Radial Velocity Curve Modeling}

In order to model the observed radial velocities of each system, we employ the standard analytical expressions for the radial velocities of two stars in a classic Newtonian orbit \citep{Paddock1913}. The velocities of each star are described by the equations

\begin{equation}
  V_{1} = V_{0} + K_{1}\cdot(\cos{(\omega + \theta(t))} + e\cos{\omega})
  \label{star1eq}
\end{equation}
\begin{equation}
  V_{2} = V_{0} - K_{2}\cdot(\cos{(\omega + \theta(t))} + e\cos{\omega})
  \label{star2eq}
\end{equation}

\noindent where $V_{1}$ and $V_{2}$ are the velocities of stars 1 and 2 respectively, $V_{0}$ is the velocity of the system's center of mass with respect to the solar system barycenter, $K_{1}$ and $K_{2}$ are the semi-major velocity amplitudes of stars 1 and 2 respectively, $e$ is the eccentricity of the system, $\omega$ is the longitude of periastron of the system, and $\theta(t)$ is the true anomaly of the system at a given time, $t$. The true anomaly is related to the orbital phase of the binary, $\phi(t)$, by the following equations

\begin{equation}
 \tan{\left(\frac{\theta(t)}{2}\right)} = \frac{\sqrt{1+e}}{\sqrt{1-e}} \cdot \tan{\left(\frac{E}{2}\right)}
\end{equation}

\begin{equation}
 2\pi\phi(t) = E - e\sin{E}
 \label{transceq}
\end{equation}

\noindent where $E$ is referred to as the eccentric anomaly. We note that Equation~\ref{transceq} is a transcendental equation, and thus $\theta(t)$ must be solved numerically for each $\phi(t)$.

\begin{figure}
\centering
\begin{tabular}{c}
  \epsfig{width=0.9\linewidth,file=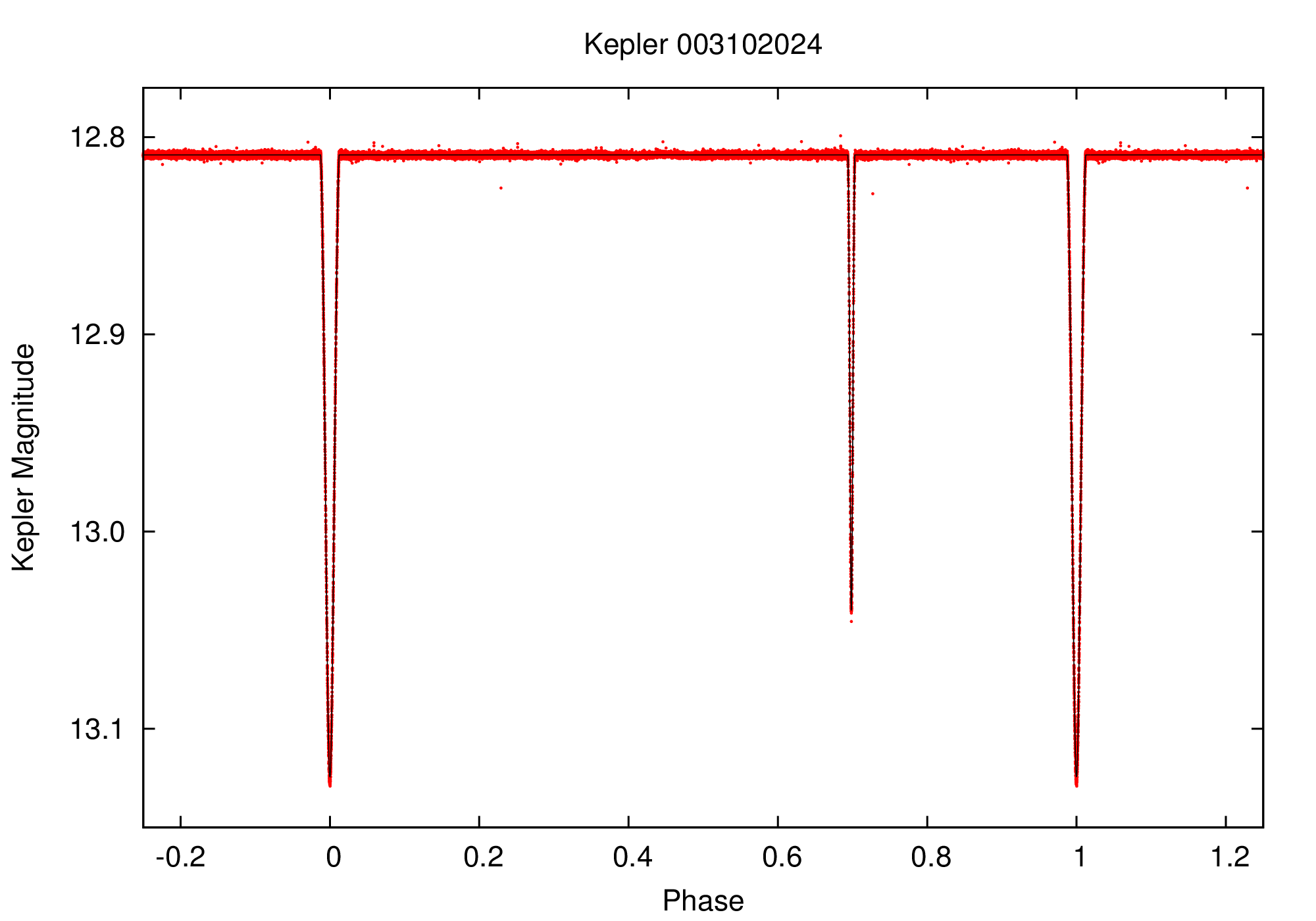}\\
  \epsfig{width=0.9\linewidth,file=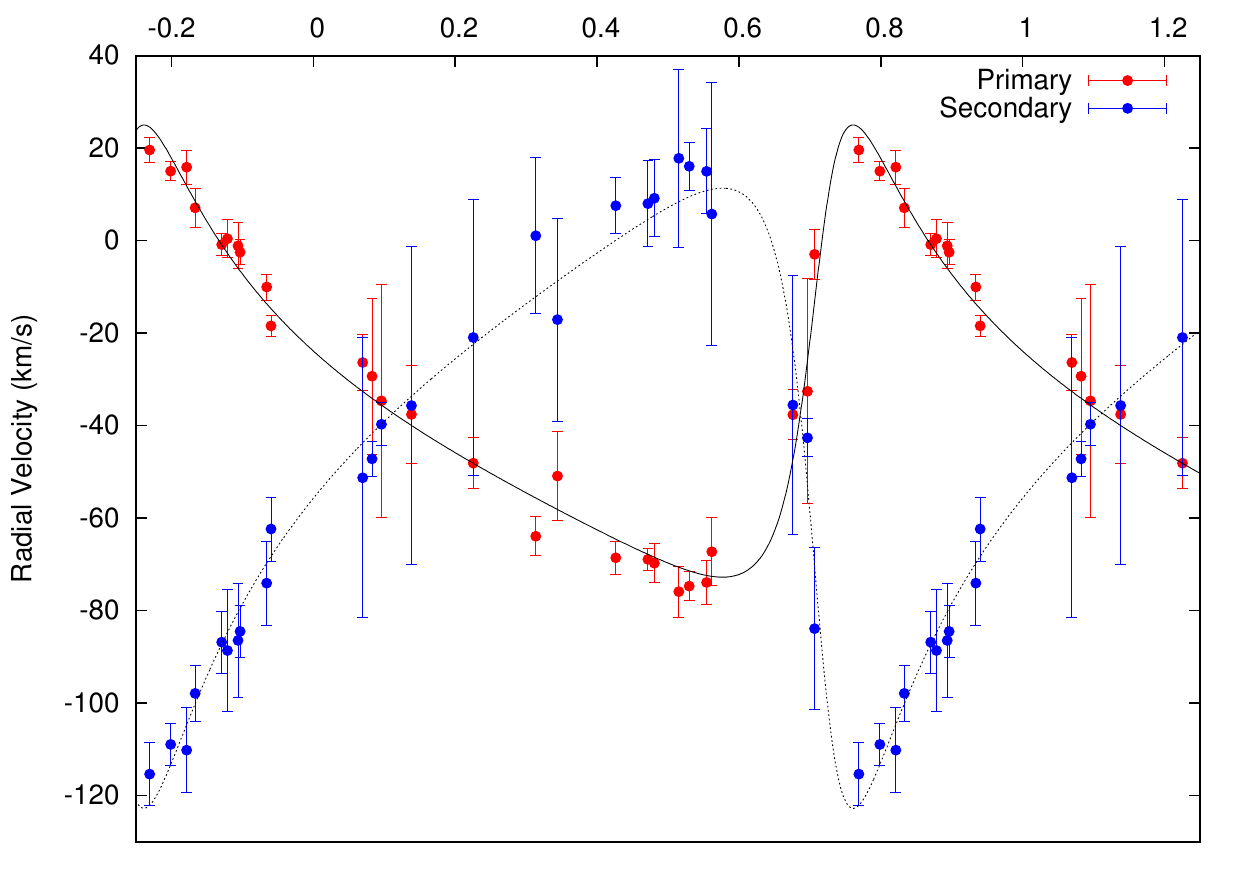}
\end{tabular}
\caption[Light and radial-velocity curves of Kepler 003102024]{$Kepler$ light curve (top panel) and ground-based radial velocity curves (bottom panel) as a function of orbital phase for Kepler 003102024. The best-fit models are shown via black lines.}
\label{Kepler003102024}
\end{figure}

\begin{figure}
\centering
\begin{tabular}{c}
  \epsfig{width=0.9\linewidth,file=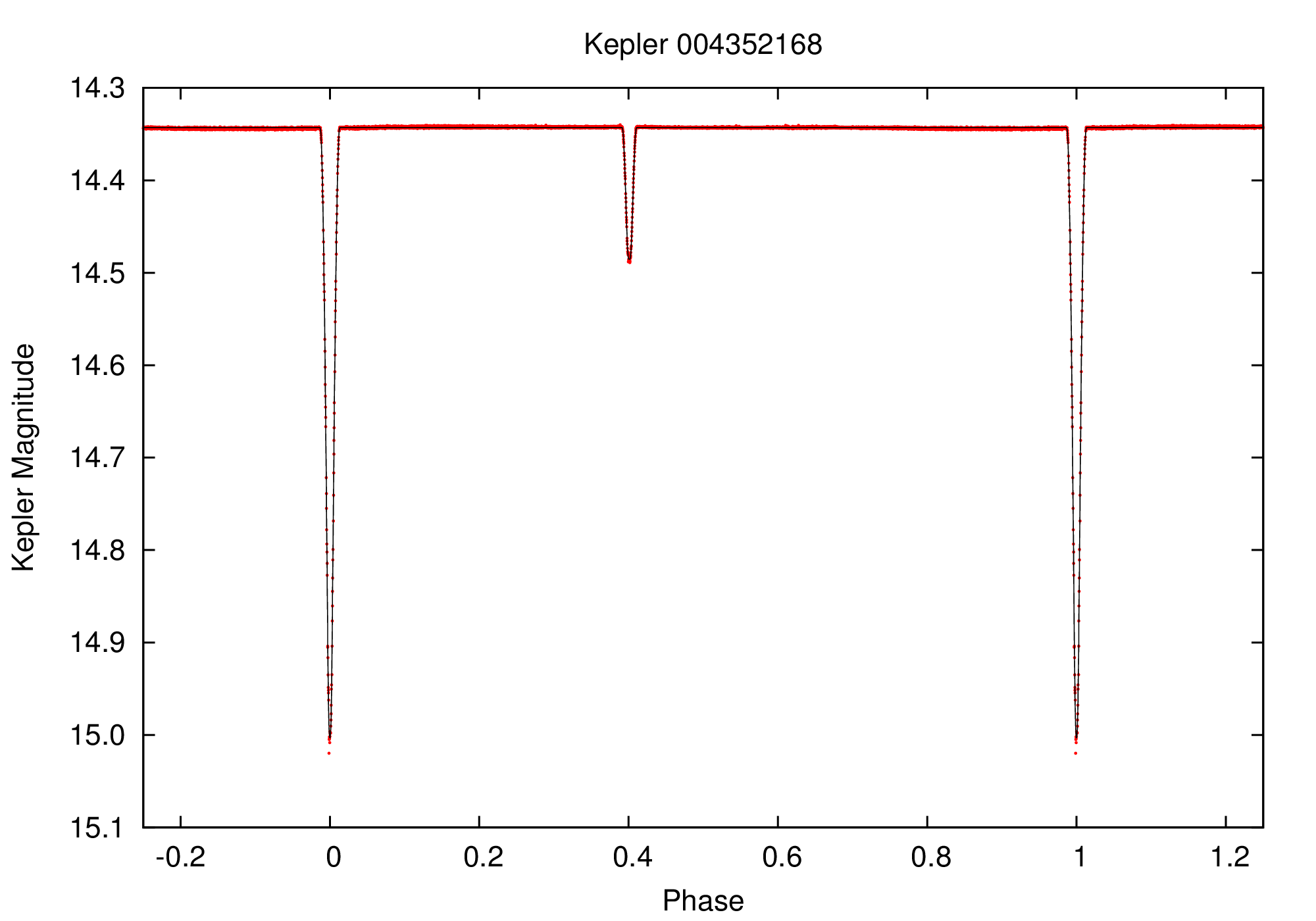}\\
  \epsfig{width=0.9\linewidth,file=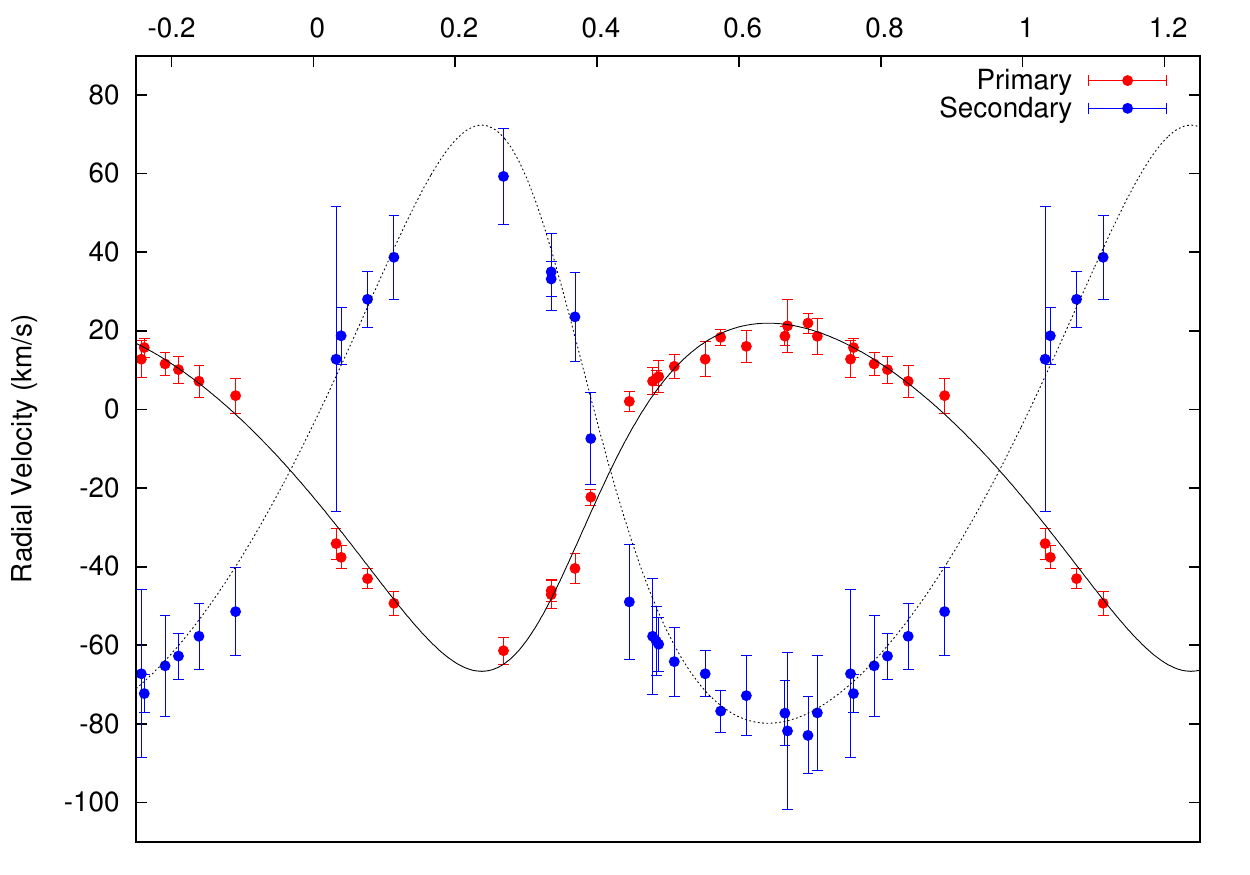}
\end{tabular}
\caption[Light and radial-velocity curves of Kepler 004352168]{$Kepler$ light curve (top panel) and ground-based radial velocity curves (bottom panel) as a function of orbital phase for Kepler 004352168. The best-fit models are shown via black lines.}
\label{Kepler004352168}
\end{figure}

\begin{figure}
\centering
\begin{tabular}{c}
  \epsfig{width=0.9\linewidth,file=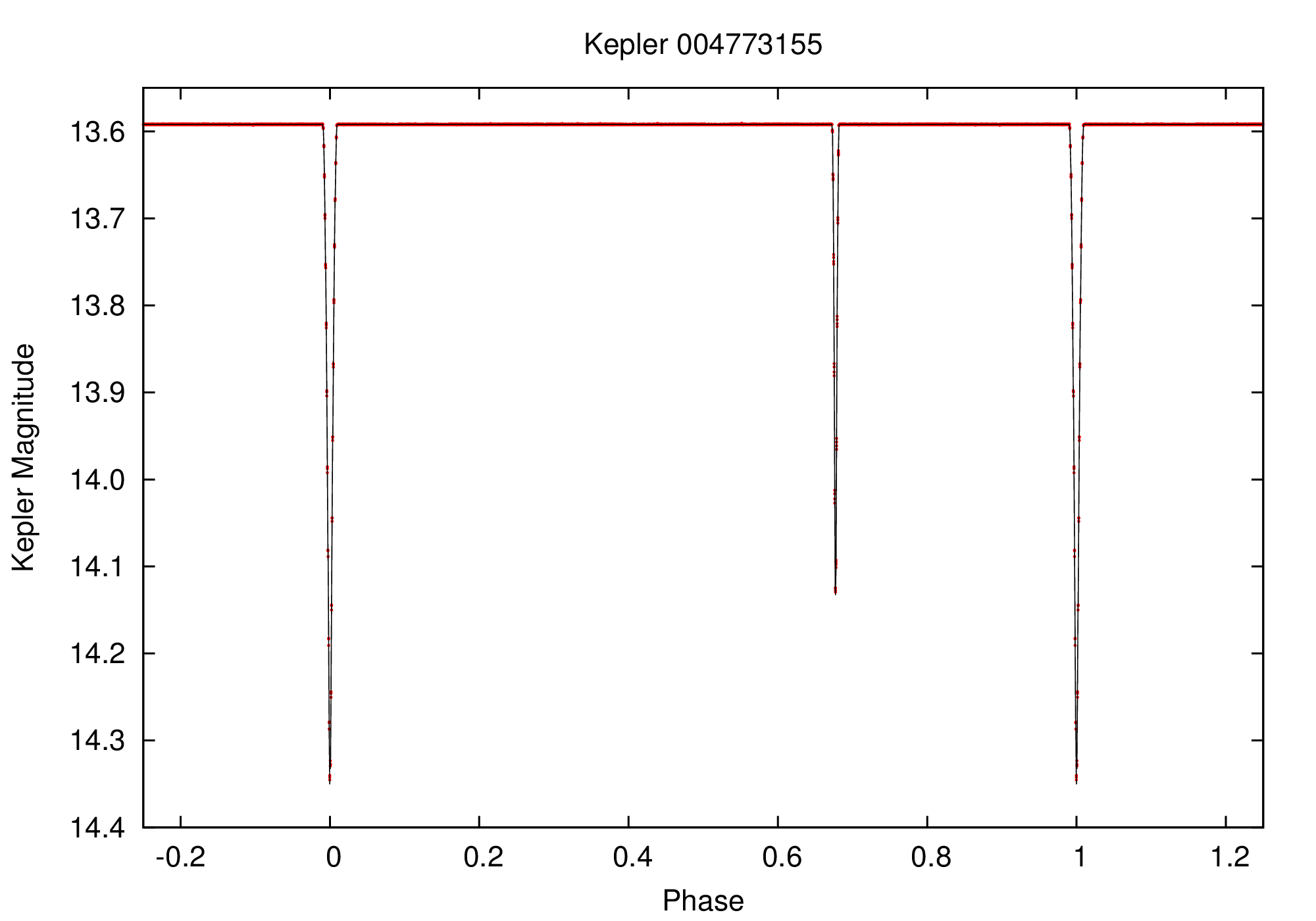}\\
  \epsfig{width=0.9\linewidth,file=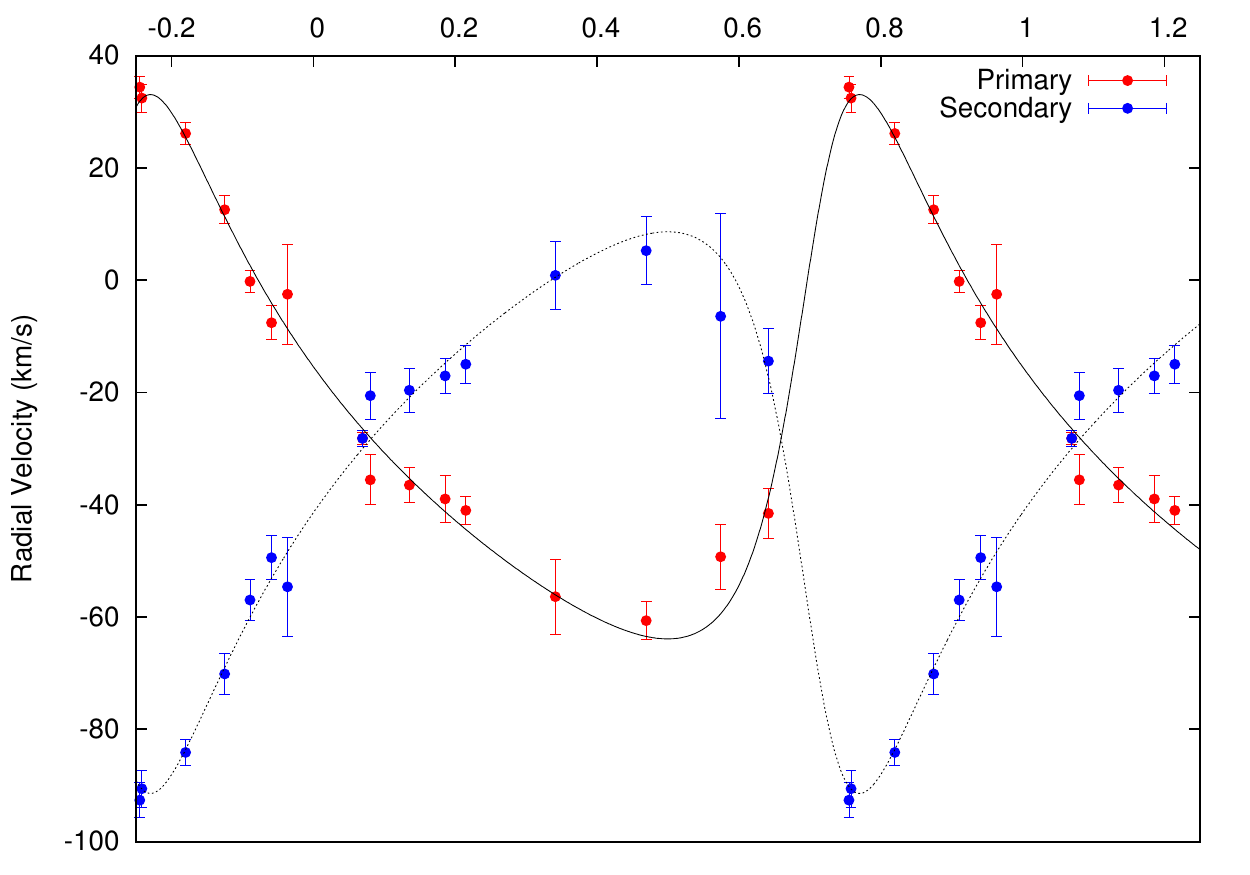}
\end{tabular}
\caption[Light and radial-velocity curves of Kepler 004773155]{$Kepler$ light curve (top panel) and ground-based radial velocity curves (bottom panel) as a function of orbital phase for Kepler 004773155. The best-fit models are shown via black lines.}
\label{Kepler004773155}
\end{figure}

\begin{figure}
\centering
\begin{tabular}{c}
  \epsfig{width=0.9\linewidth,file=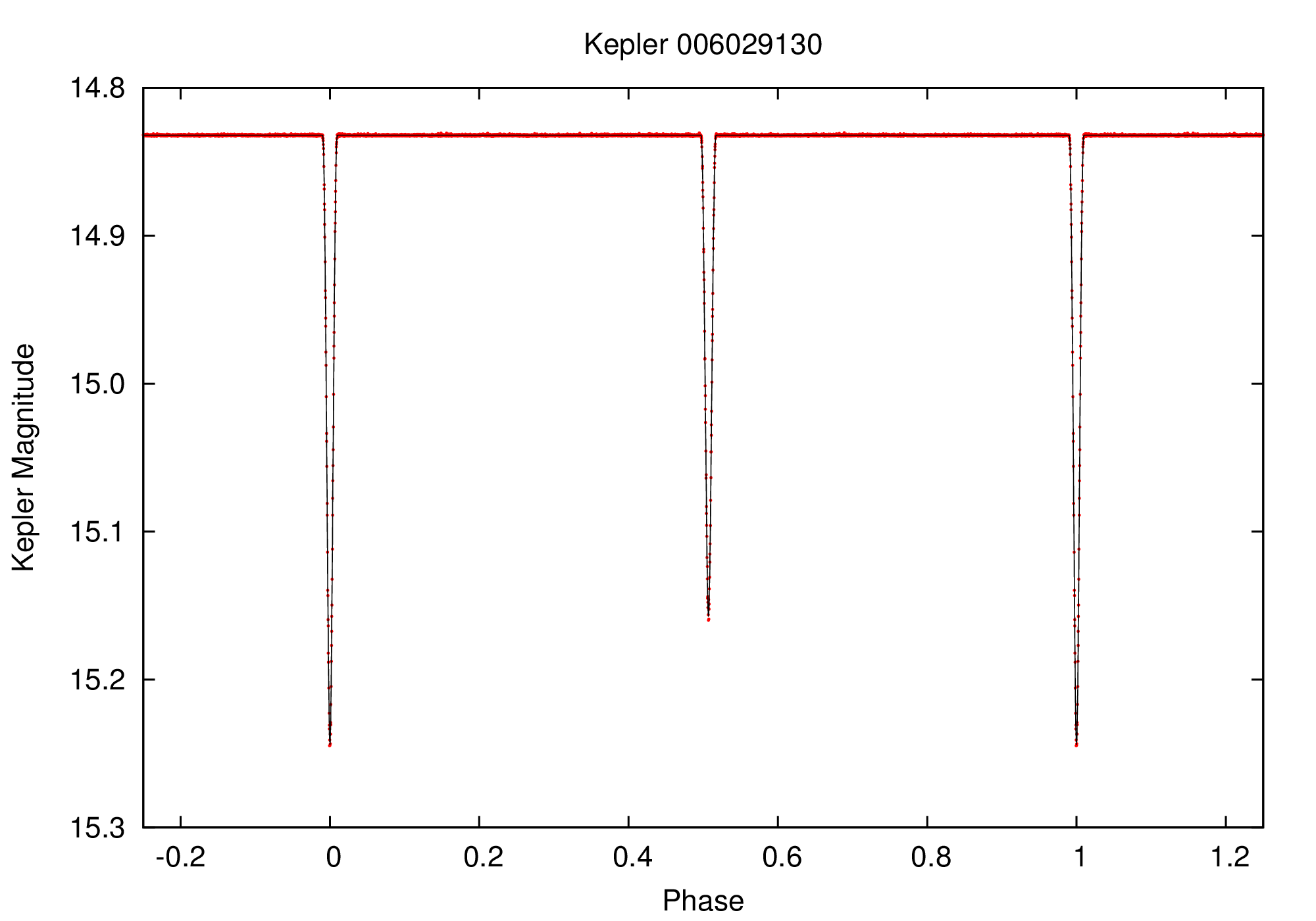}\\
  \epsfig{width=0.9\linewidth,file=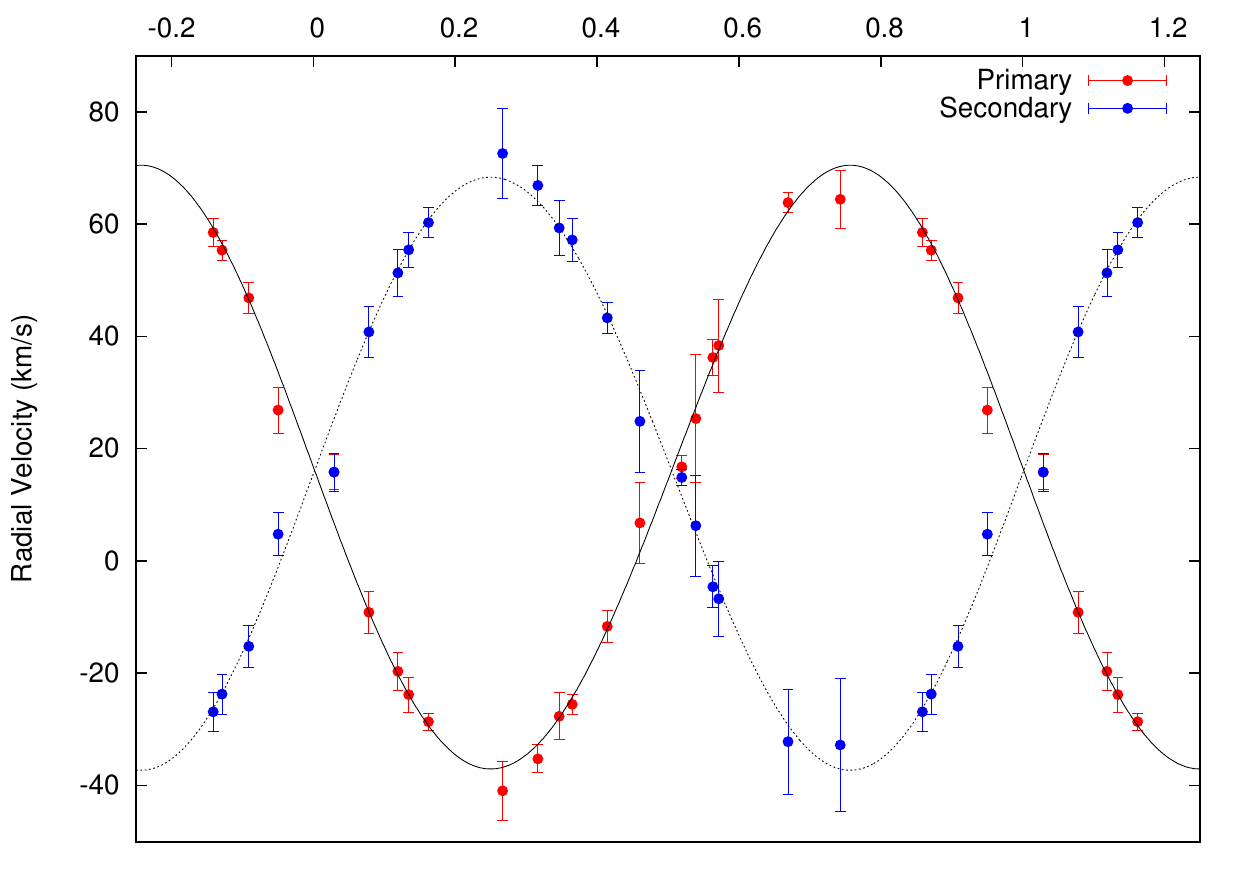}
\end{tabular}
\caption[Light and radial-velocity curves of Kepler 006029130]{$Kepler$ light curve (top panel) and ground-based radial velocity curves (bottom panel) as a function of orbital phase for Kepler 006029130. The best-fit models are shown via black lines.}
\label{Kepler006029130}
\end{figure}

\begin{figure}
\centering
\begin{tabular}{c}
  \epsfig{width=0.9\linewidth,file=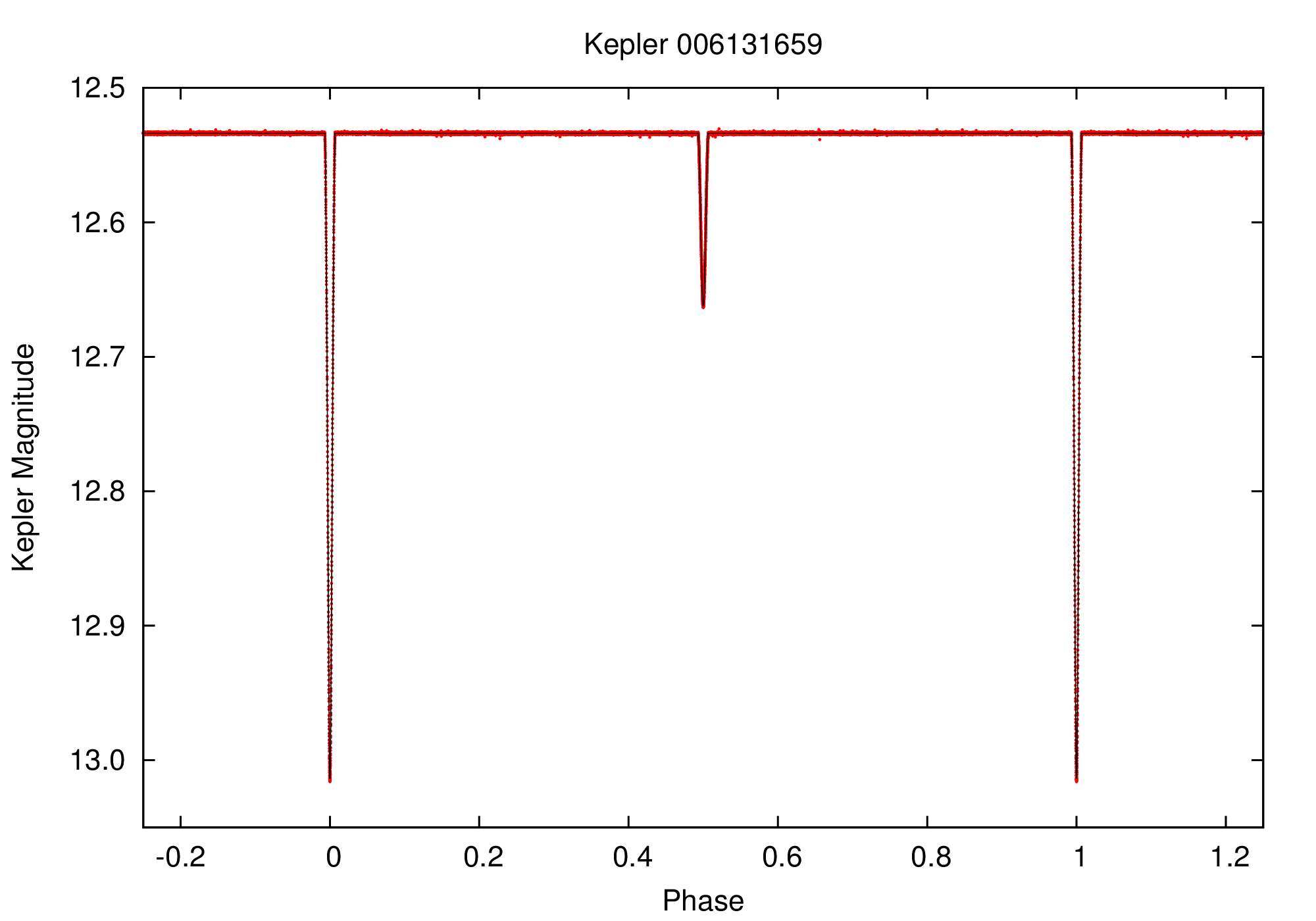}\\
  \epsfig{width=0.9\linewidth,file=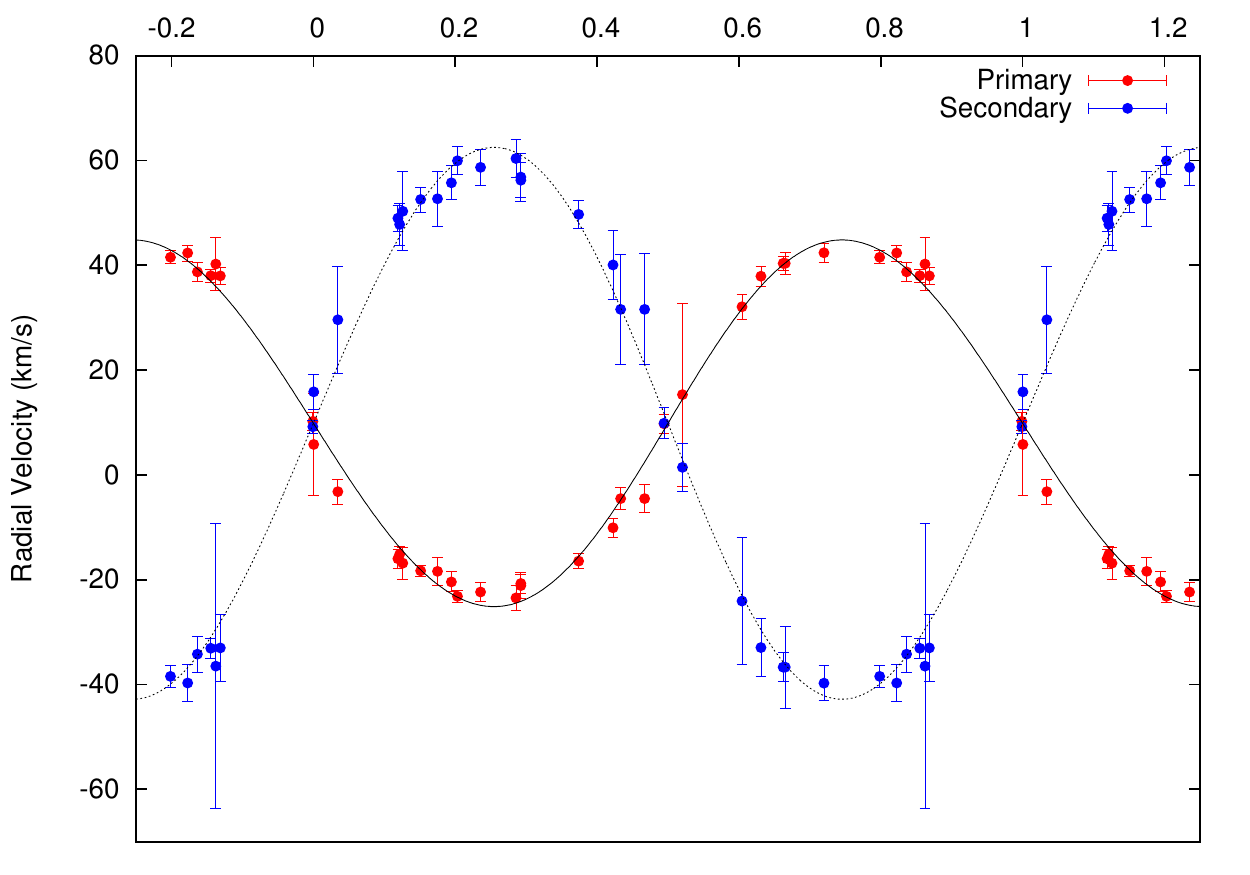}
\end{tabular}
\caption[Light and radial-velocity curves of Kepler 006131659]{$Kepler$ light curve (top panel) and ground-based radial velocity curves (bottom panel) as a function of orbital phase for Kepler 006131659. The best-fit models are shown via black lines.}
\label{Kepler006131659}
\end{figure}

\begin{figure}
\centering
\begin{tabular}{c}
  \epsfig{width=0.9\linewidth,file=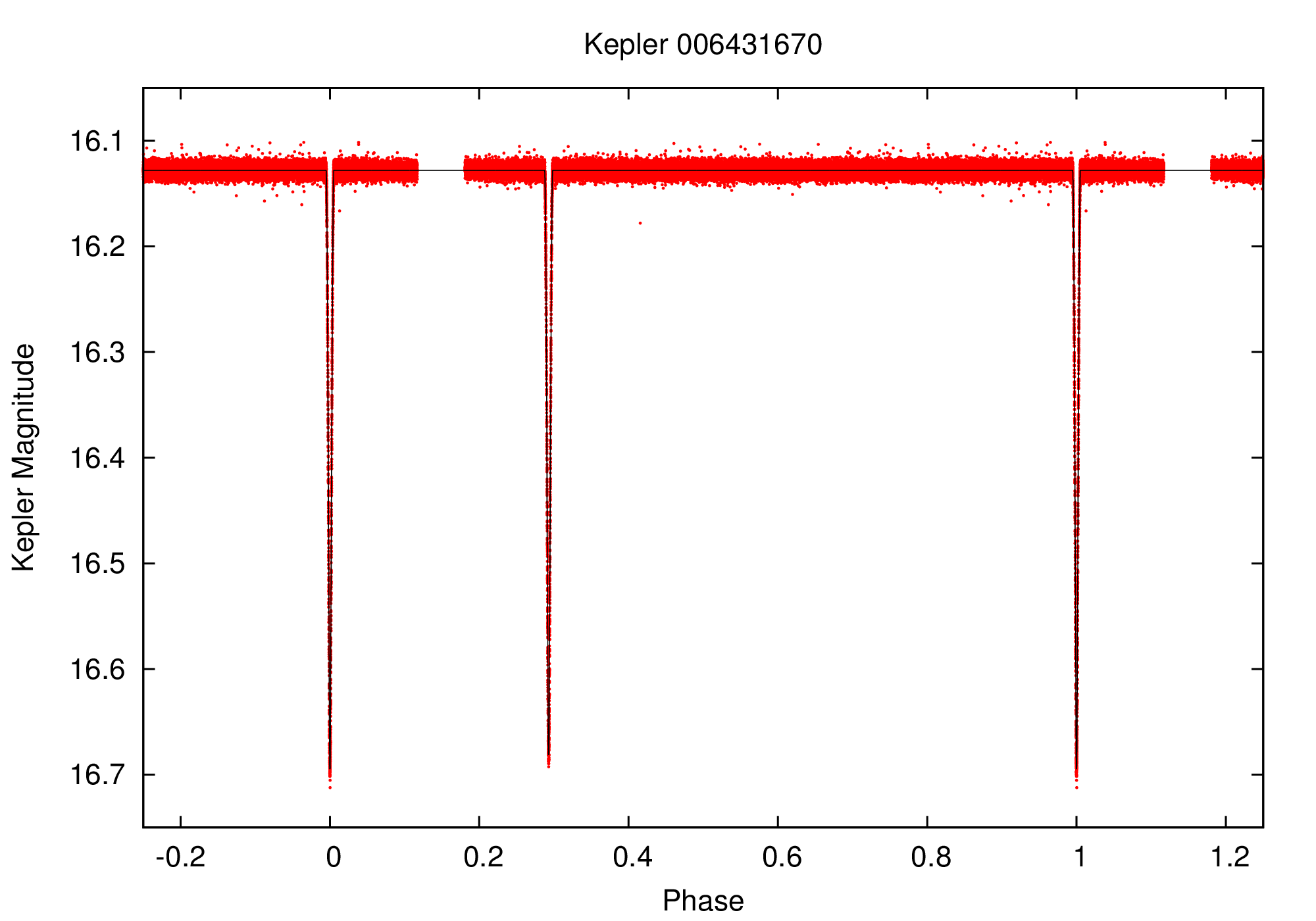}\\
  \epsfig{width=0.9\linewidth,file=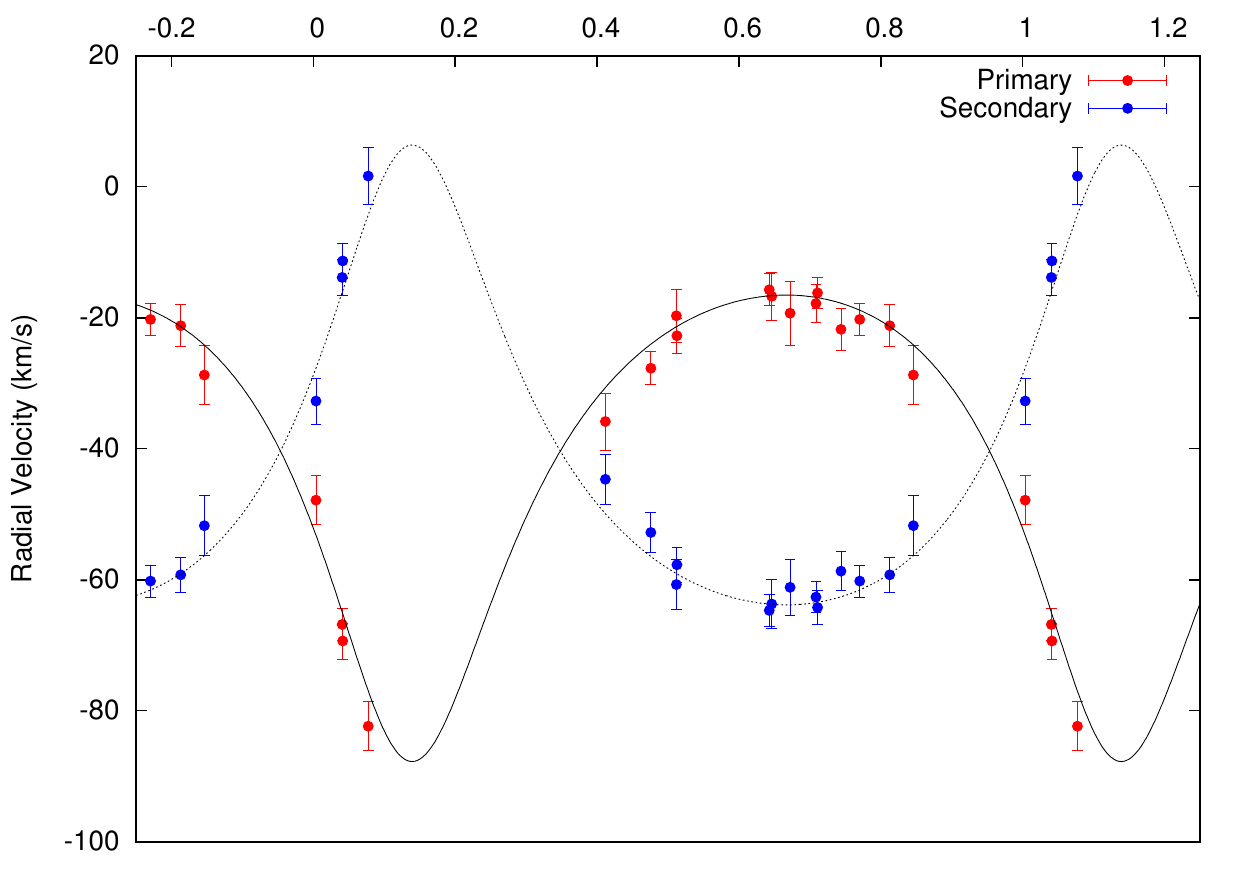}
\end{tabular}
\caption[Light and radial-velocity curves of Kepler 006431670]{$Kepler$ light curve (top panel) and ground-based radial velocity curves (bottom panel) as a function of orbital phase for Kepler 006431670. The best-fit models are shown via black lines.}
\label{Kepler006431670}
\end{figure}

\begin{figure}
\centering
\begin{tabular}{c}
  \epsfig{width=0.9\linewidth,file=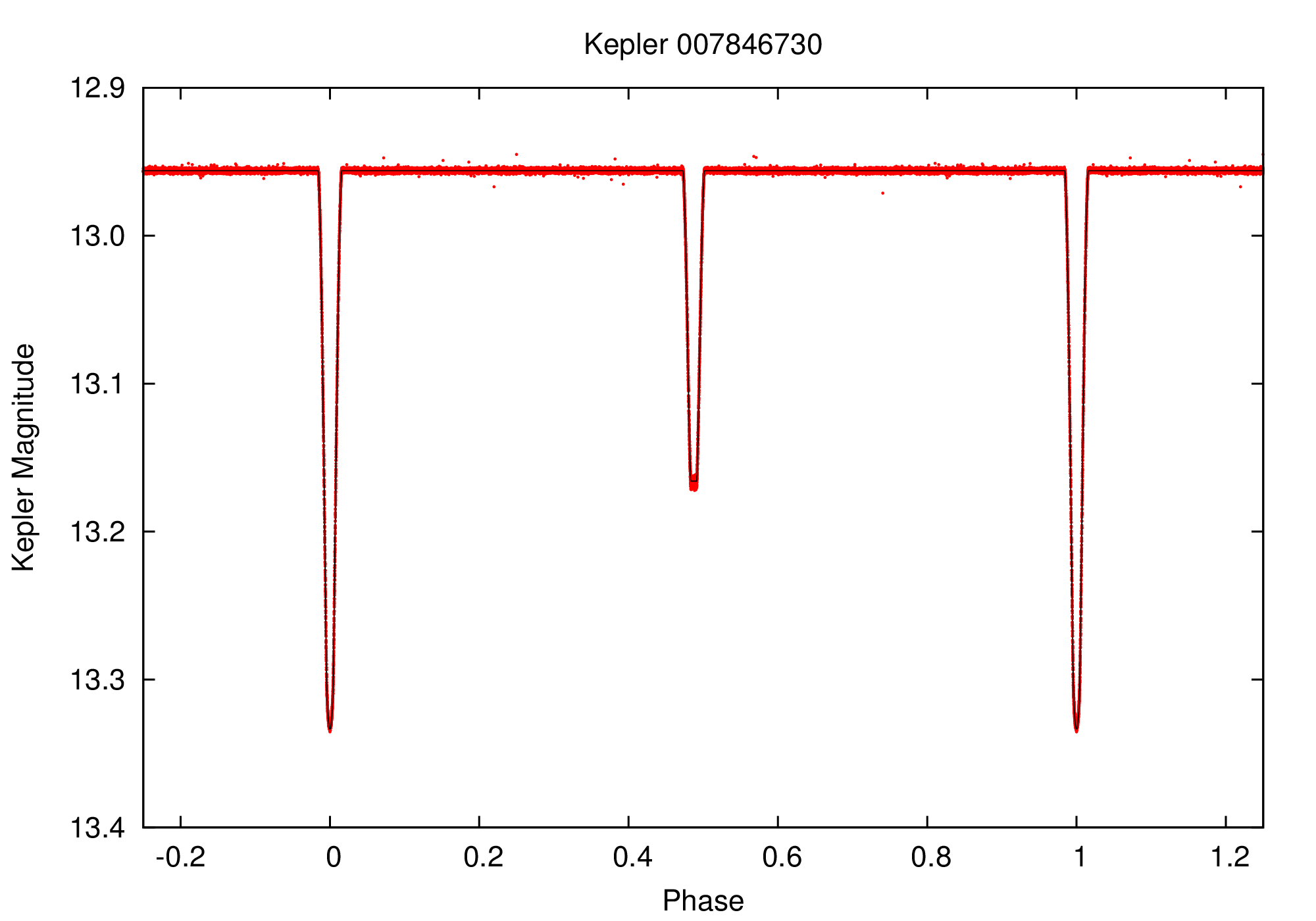}\\
  \epsfig{width=0.9\linewidth,file=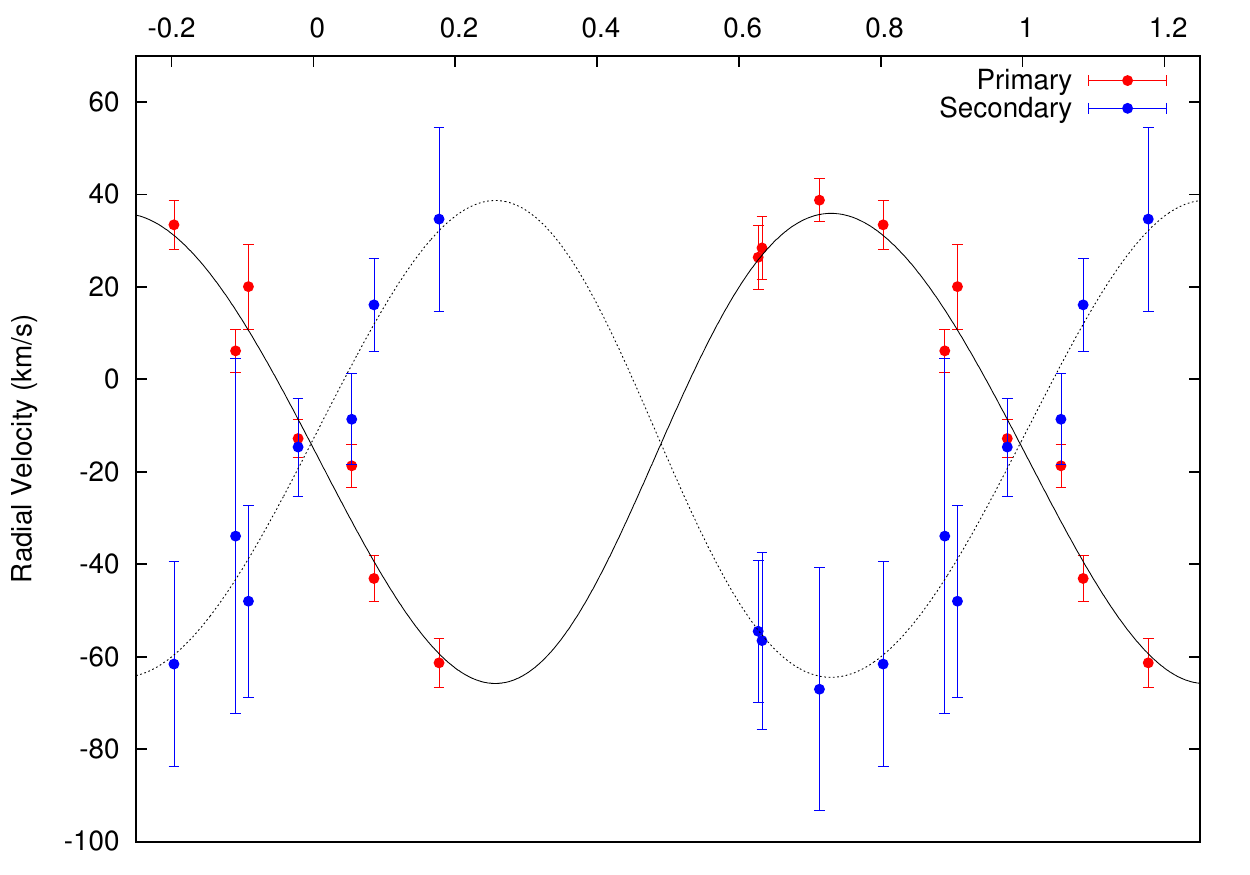}
\end{tabular}
\caption[Light and radial-velocity curves of Kepler 007846730]{$Kepler$ light curve (top panel) and ground-based radial velocity curves (bottom panel) as a function of orbital phase for Kepler 007846730. The best-fit models are shown via black lines.}
\label{Kepler007846730}
\end{figure}

\begin{figure}
\centering
\begin{tabular}{c}
  \epsfig{width=0.9\linewidth,file=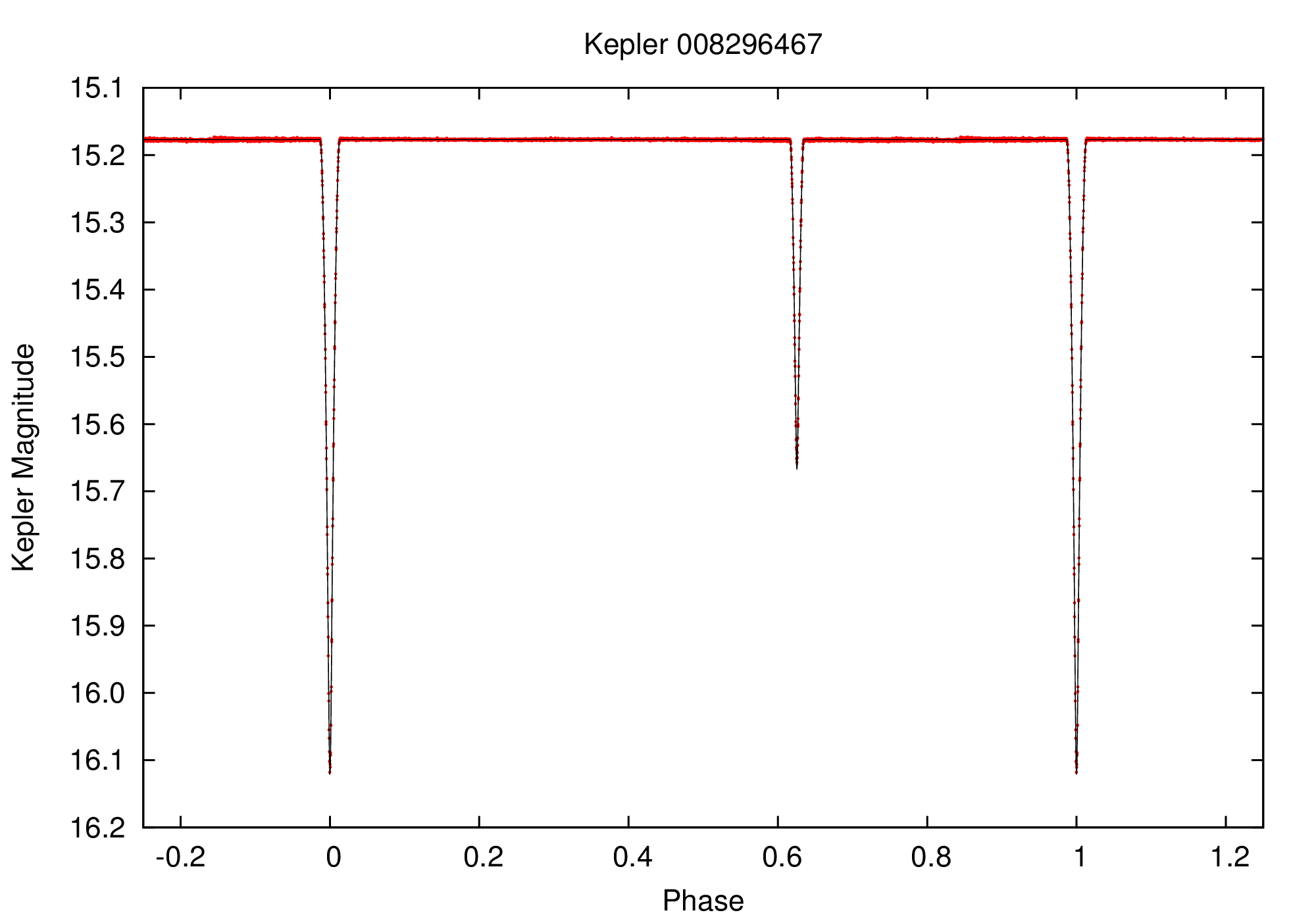}\\
  \epsfig{width=0.9\linewidth,file=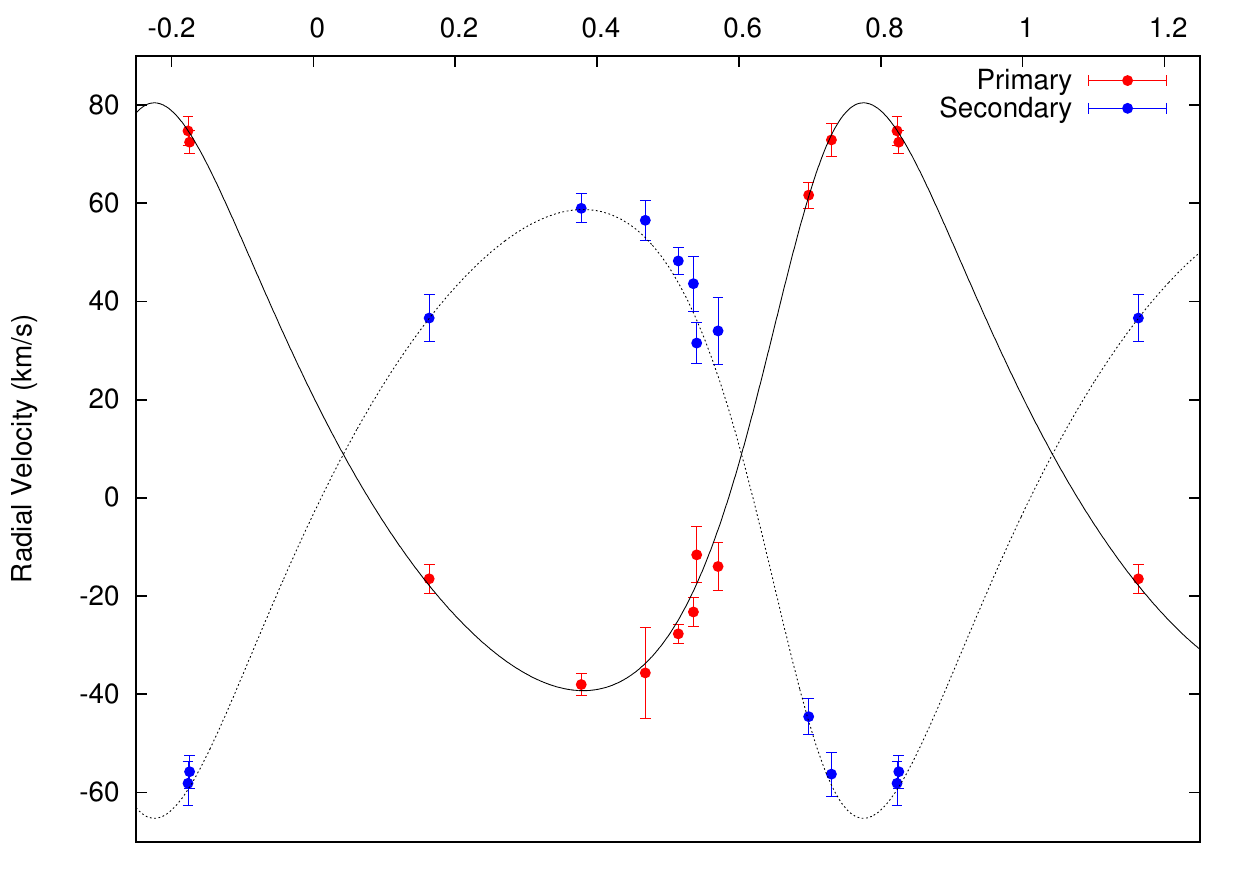}
\end{tabular}
\caption[Light and radial-velocity curves of Kepler 008296467]{$Kepler$ light curve (top panel) and ground-based radial velocity curves (bottom panel) as a function of orbital phase for Kepler 008296467. The best-fit models are shown via black lines.}
\label{Kepler008296467}
\end{figure}

\begin{figure}
\centering
\begin{tabular}{c}
  \epsfig{width=0.9\linewidth,file=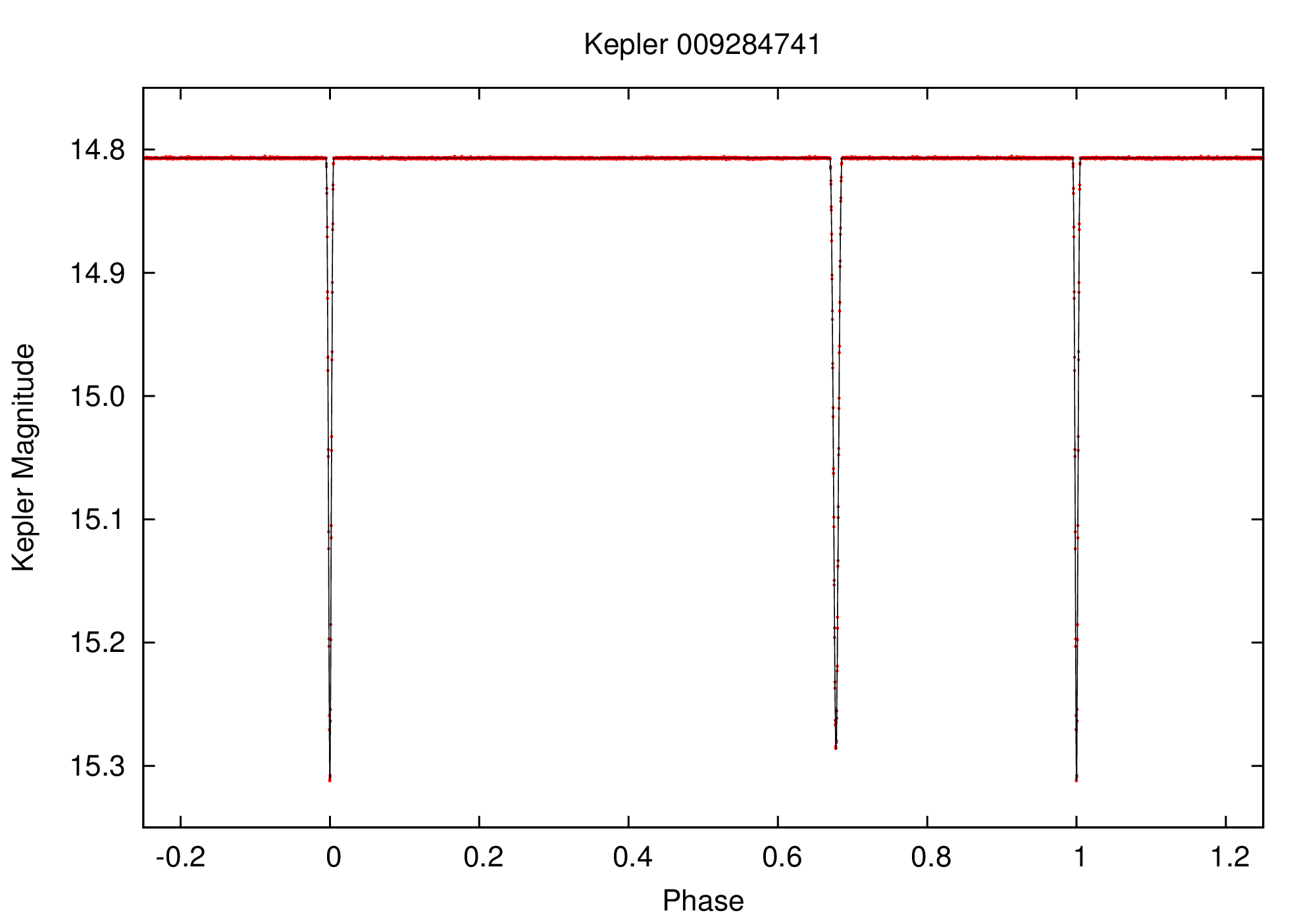}\\
  \epsfig{width=0.9\linewidth,file=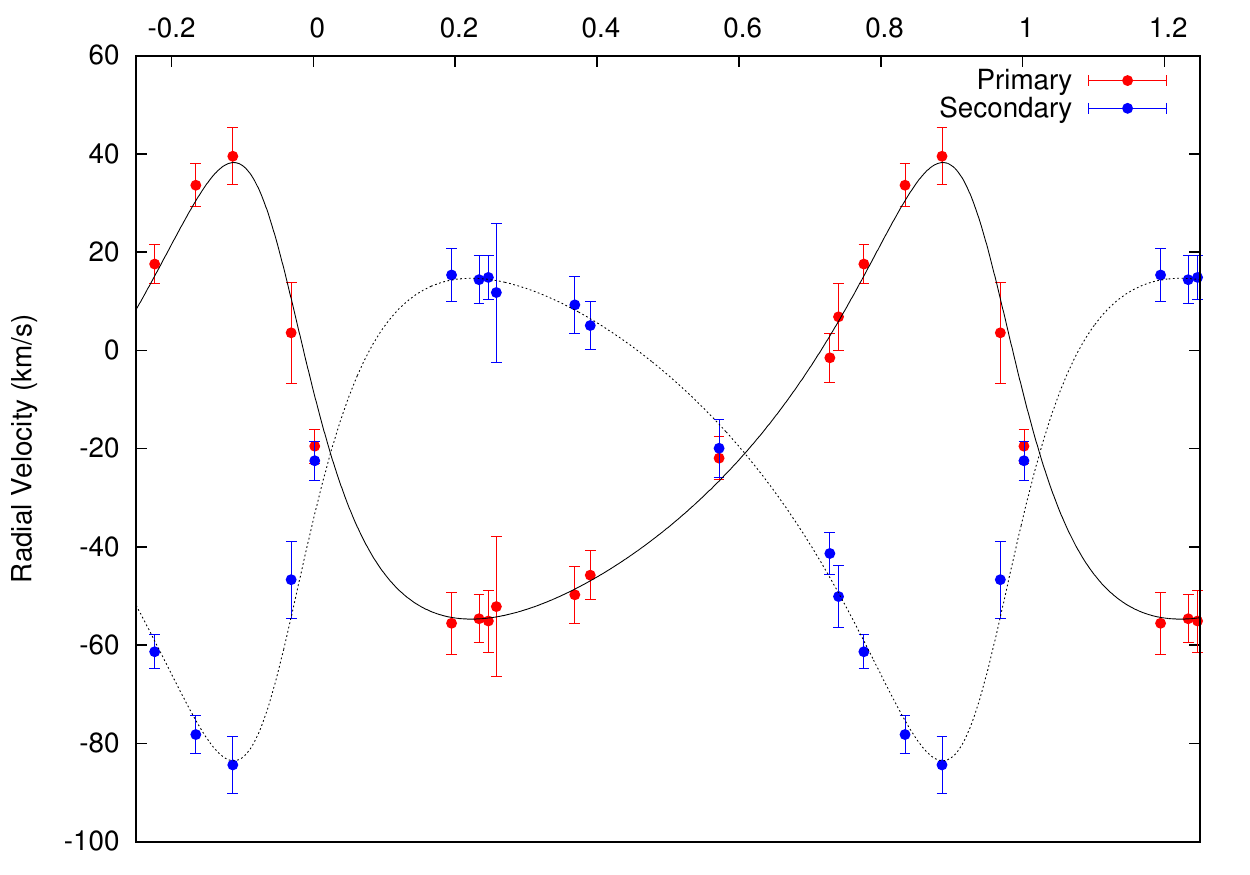}
\end{tabular}
\caption[Light and radial-velocity curves of Kepler 009284741]{$Kepler$ light curve (top panel) and ground-based radial velocity curves (bottom panel) as a function of orbital phase for Kepler 009284741. The best-fit models are shown via black lines.}
\label{Kepler009284741}
\end{figure}

\begin{figure}
\centering
\begin{tabular}{c}
  \epsfig{width=0.9\linewidth,file=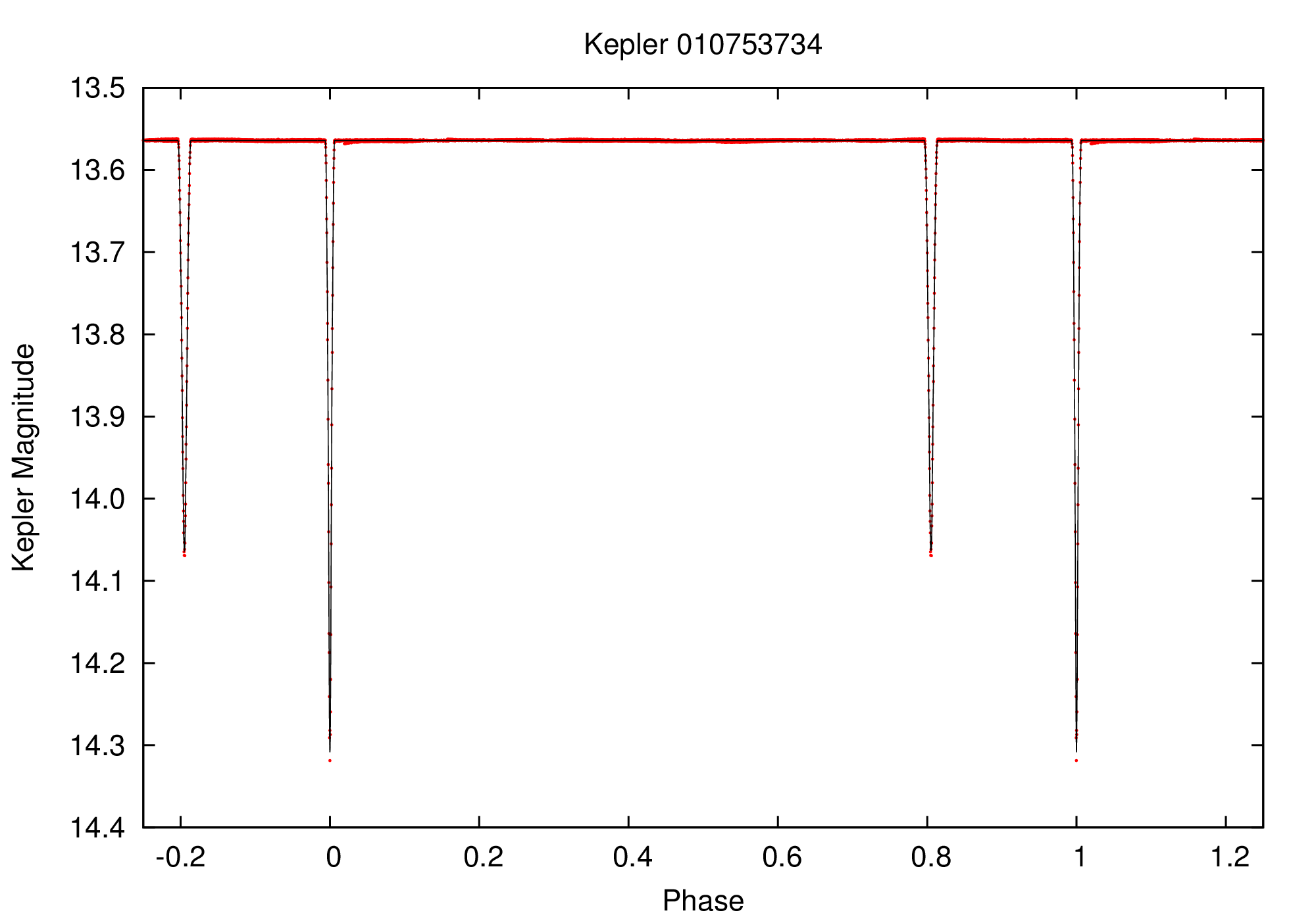}\\
  \epsfig{width=0.9\linewidth,file=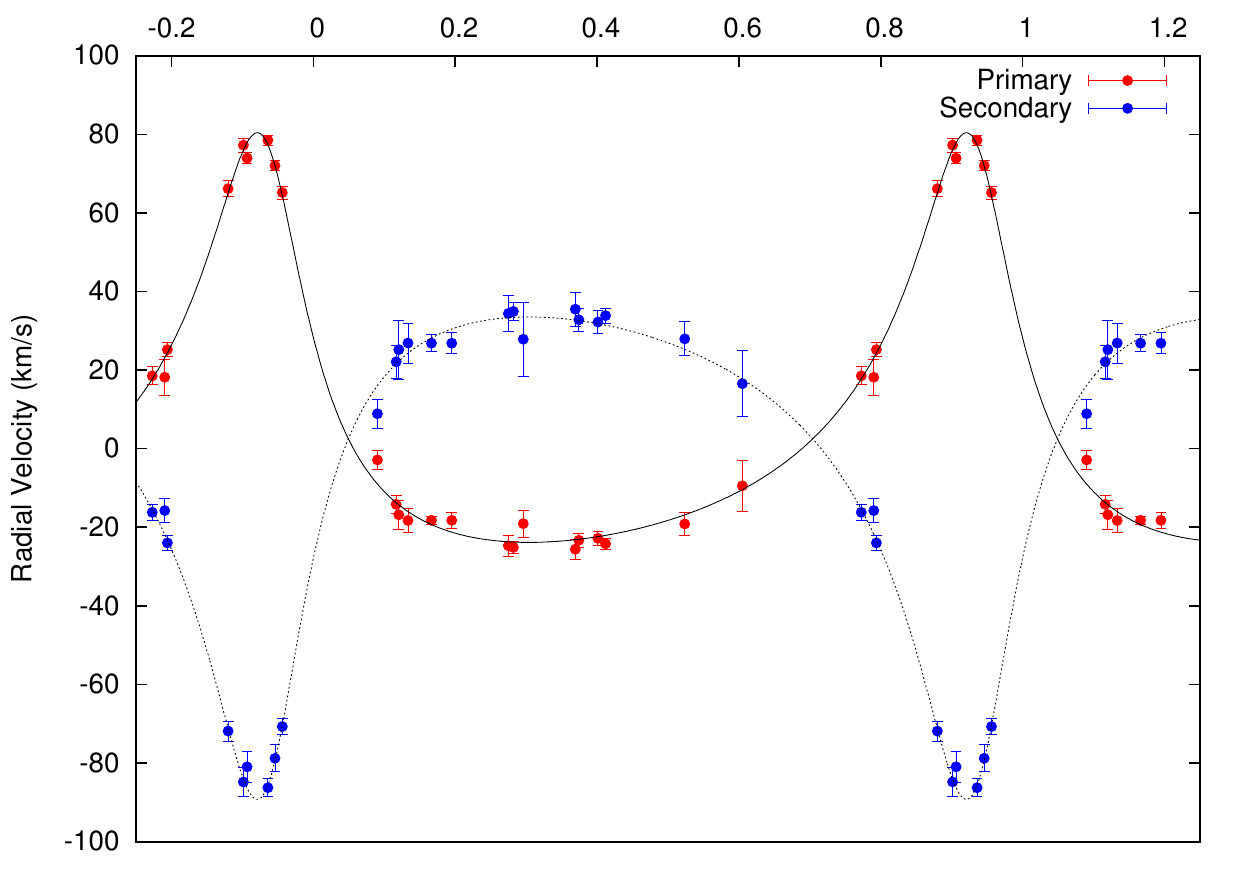}
\end{tabular}
\caption[Light and radial-velocity curves of Kepler 010753734]{$Kepler$ light curve (top panel) and ground-based radial velocity curves (bottom panel) as a function of orbital phase for Kepler 010753734. The best-fit models are shown via black lines.}
\label{Kepler010753734}
\end{figure}

\begin{figure}
\centering
\begin{tabular}{c}
  \epsfig{width=0.9\linewidth,file=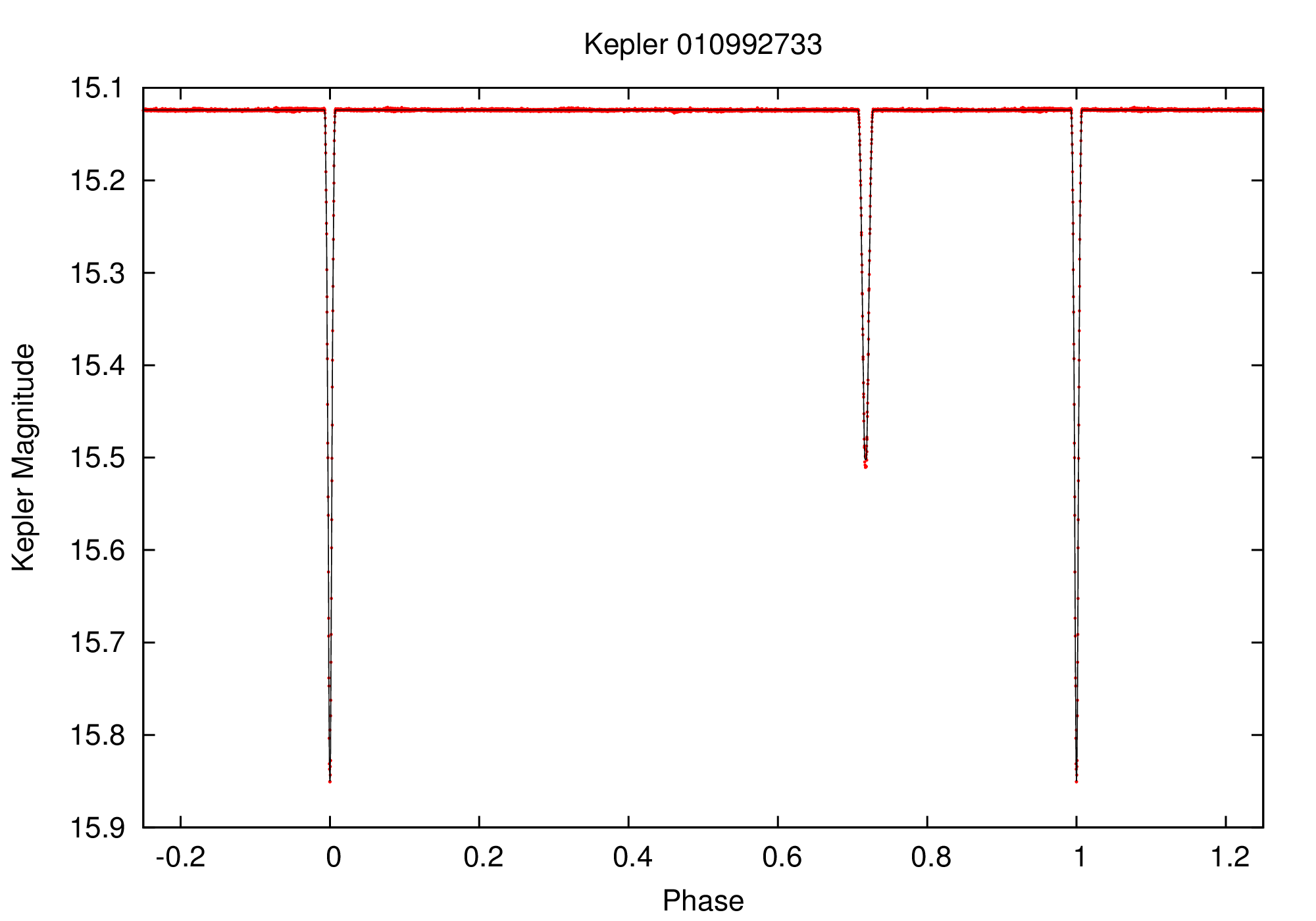}\\
  \epsfig{width=0.9\linewidth,file=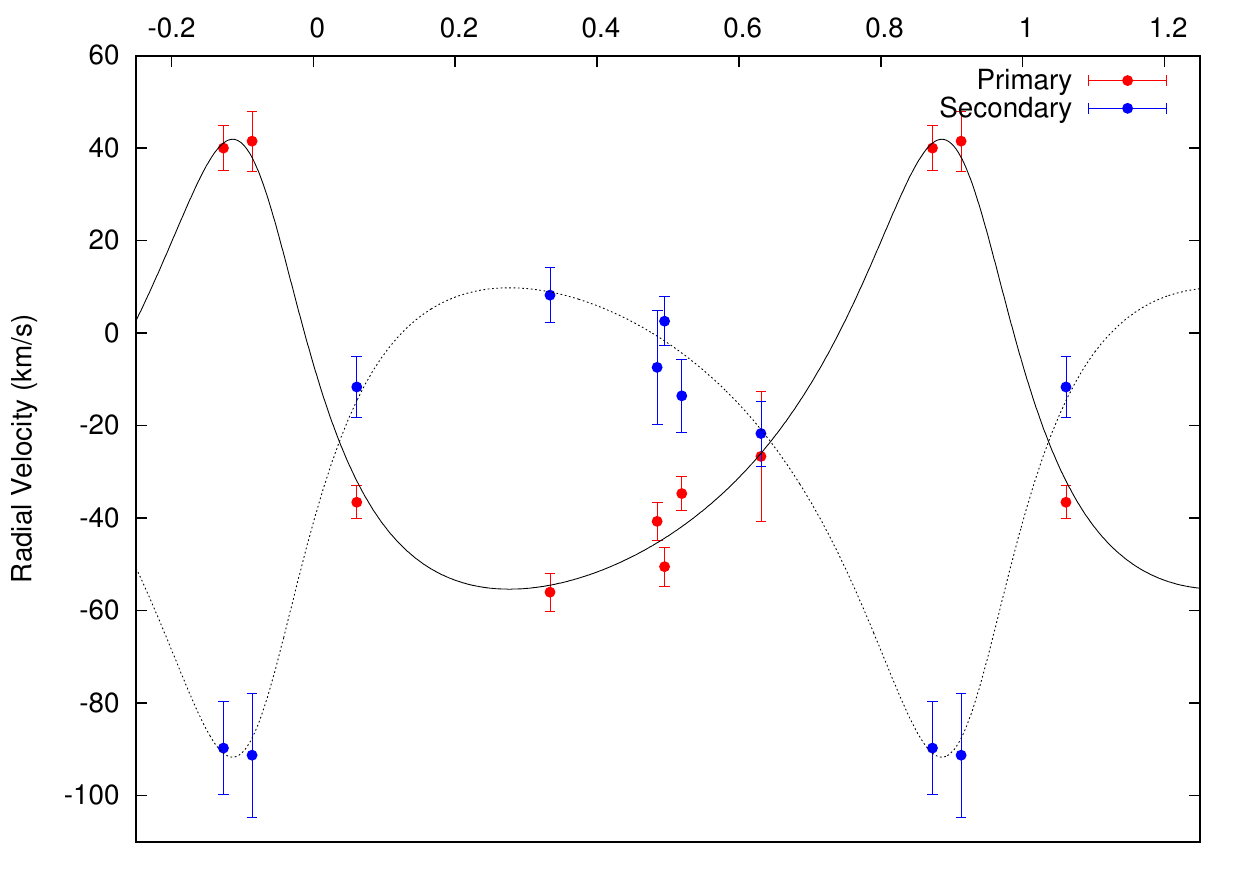}
\end{tabular}
\caption[Light and radial-velocity curves of Kepler 010992733]{$Kepler$ light curve (top panel) and ground-based radial velocity curves (bottom panel) as a function of orbital phase for Kepler 010992733. The best-fit models are shown via black lines.}
\label{Kepler010992733}
\end{figure}

We fit each radial-velocity curve by first fixing the period, time of primary minimum, eccentricity, and longitude of periastron of the system to the values derived from the light curve analysis, as those quantities are extremely well-defined by the light curve alone. We then solve for $K_{1}$, $K_{2}$, and $V_{0}$ via standard $\chi^{2}$ minimization. Plots of the radial velocity curves and best-fit models are shown in Figures~\ref{Kepler003102024}~-~\ref{Kepler010992733}. 

Once the values of $K_{1}$ and $K_{2}$ are obtained from the radial-velocity curves, we can the calculate the mass and radius of each star via the following equations

\begin{equation}
  a = \frac{P \cdot (K_{1} + K_{2}) \cdot \sqrt{1-e^{2}}}{2\pi\sin{i}}
\end{equation}

\begin{equation}
  M_{tot} = \frac{4\pi^{2} a^{3}}{G P^{2}}
\end{equation}

\begin{equation}
  q = \frac{K_{1}}{K_{2}}
\end{equation}

\begin{equation}
  M_{1} = \frac{M_{tot}}{1+q}
\end{equation}

\begin{equation}
  M_{2} = \frac{q \cdot M_{tot}}{1+q}
\end{equation}

\begin{equation}
  R_{1} = r_{1} \cdot a
\end{equation}

\begin{equation}
  R_{2} = r_{2} \cdot a
\end{equation}

\noindent where $P$ is the period of the system, $i$ is the inclination of the binary, $a$ is the semi-major axis of the system, $G$ is the gravitational constant, $q$ is the mass ratio of the system, and $M_{1}$, $M_{2}$, $R_{1}$, $R_{2}$, $r_{1}$, and $r_{2}$ are the physical masses and radii, and fractional radii, of stars 1 and 2 respectively. We again note that $P$, $i$, $r_{1}$, and $r_{2}$ are derived from the light-curve analysis. 

To determine robust errors, we first scaled the error bars of the radial-velocity measurements so that the best fit had a reduced $\chi^{2}$ = 1, as sometimes there were remaining systematics in the radial velocity curves, and we did not want to consequently underestimate errors by ignoring this source of noise. We then performed 10,000 Monte Carlo simulations where new datasets were created by adding noise to the original radial-velocity curves via each point's individual error bars. At each iteration the resulting radial velocity curve was re-fit for the values of $K_{1}$, $K_{2}$, and $V_{0}$, while the values of $P$, $T_{0}$, $i$, $e$, $\omega$, $r_{1}$, and $r_{2}$ were simultaneously perturbed via the errors determined from the light curve modeling, and the resulting quantities of $M_{1}$, $M_{2}$, $R_{1}$, and $R_{2}$ computed. The posterior distributions on all the parameters were then used to compute the median and 1$\sigma$ confidence intervals for all parameters. We list these values for each system in Table~\ref{lmbrvtab}.

\begin{deluxetable}{ccccccc}
\tablewidth{0pt}
\tabletypesize{\scriptsize}
\tablecaption{Physical Masses, Radii, and Temperatures for the Low-Mass Binaries}
\tablecolumns{7}
\tablehead{\emph{Kepler} ID & $M_{1}$ & $M_{2}$ & $R_{1}$ & $R_{2}$ & $T_{1}$ & $T_{2}$\\ & ($M_{\sun}$) & ($M_{\sun}$) & ($R_{\sun}$) & ($R_{\sun}$) & (K) & (K)}
\startdata
\input{lmbrvmodel.tab}
\enddata
\label{lmbrvtab}
\end{deluxetable}

\subsection{Discussion}

\begin{figure}
\centering
\epsfig{width=\linewidth,file=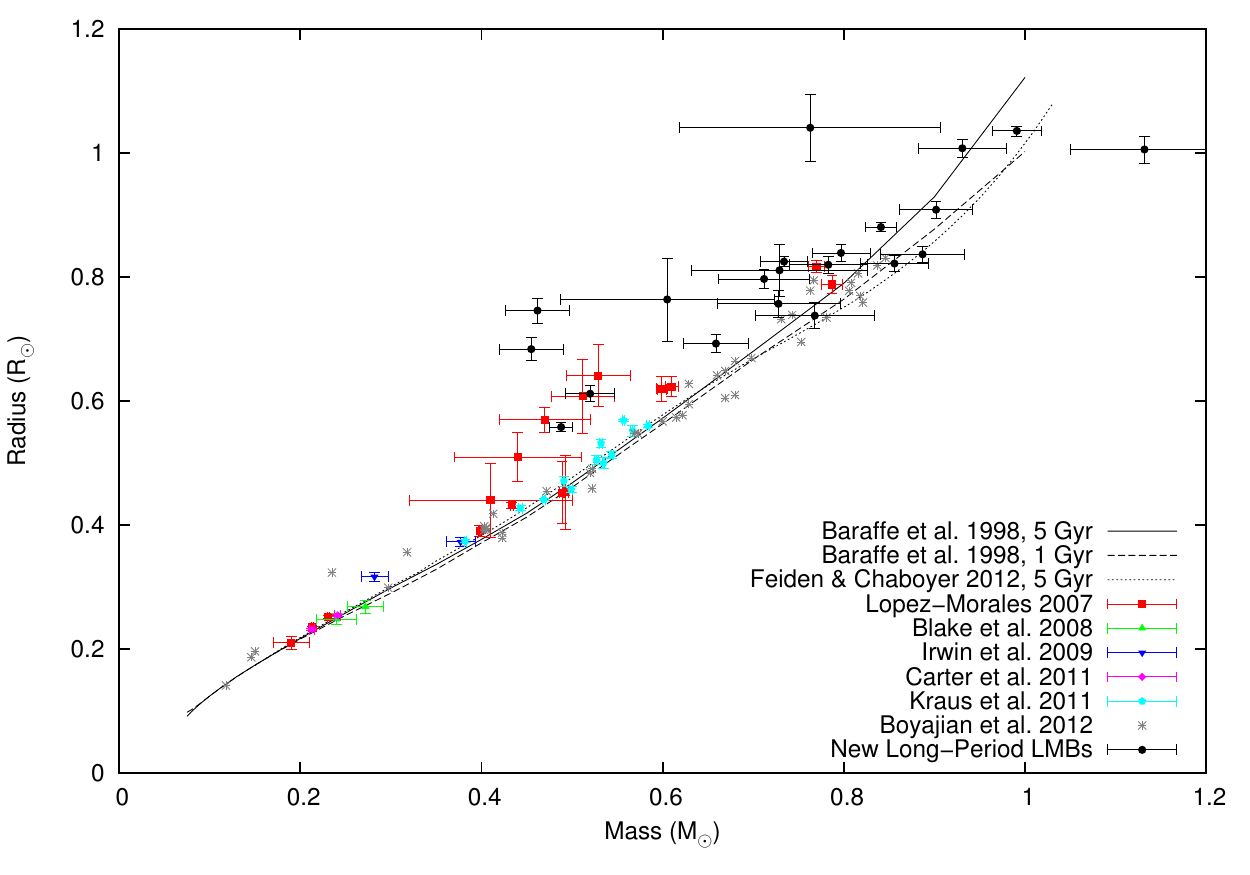}
\caption[Mass-Radius diagram for low-mass stars]{Mass-Radius diagram for low-mass stars. The measurements of the 22 stars in the 11 new long-period low-mass binaries we measured are shown by solid black circles. Measurements from previous studies are shown by other point types and colors. The theoretical mass-radius relations from the stellar models of \citet{Baraffe1998} are shown for [M/H] = 0.0 and ages of both 1 and 5 Gyr via the solid and long-dashed lines, respectively. A more modern mass-radius relationship from the Dartmouth Stellar Evolution Program \citep{Feiden2012} is also shown by a short-dashed line.}
\label{MRdiagram}
\end{figure}

In Figure~\ref{MRdiagram} we plot the mass and radius values of each star of the 11 long-period low-mass eclipsing binaries we modeled, for a total of 22 individual stars. We also plot the canonical theoretical mass-radius relation of \citet{Baraffe1998} given [M/H] = 0.0 and an age of both 1.0 and 5.0 Gyr, and the more recent theoretical relation of \citet{Feiden2012} at 5.0 Gyr, which utilizes the Dartmouth Stellar Evolution Program and tends to predict larger radii at 0.2 $\lesssim$ $M$ $\lesssim$ 0.65 $M_{\sun}$ and smaller radii at 0.65 $\lesssim$ $M$ $\lesssim$ 1.0 $M_{\sun}$ compared to \citet{Baraffe1998}. Furthermore, we plot previously measured low-mass eclipsing binaries from the studies of \citet{LopezMorales2007}, \citet{Blake2008}, \citet{Irwin2009}, \citet{Carter2011}, and \citet{Kraus2011}. Finally, we also plot recent measurements of single low-mass field stars via interferometry by \citet{Boyajian2012}, whose direct measurements are physical size and temperature, but where masses have been inferred from mass-luminosity relations with resulting errors in mass of $\sim$10\% (not plotted for clarity).

Before delving into the analysis, we note that two systems stand out as particularly peculiar: Kepler 006431670 and 007846730. Kepler 006431670 is composed of two stars with values of ($M$, $R$, $T$) = (0.455 $M_{\sun}$, 0.684 $R_{\sun}$, 5012 K) and (0.462 $M_{\sun}$, 0.746 $R_{\sun}$, 5198 K), which are extremely large and hot for their determined masses. Unless there was some systematic error in the radial-velocity measurements of this system, (which is not immediately obvious from Figure~\ref{Kepler006431670}), the only explanation is that this is a pre-main-sequence binary, as the components would not have had enough time to significantly evolve given the age of the universe. The derived masses and radii are fit well by the \citet{Baraffe1998} models at an age of 20 Myr, however there is no sign of $H_{\alpha}$ emission in the spectra, and the temperature should be cooler that observed. A search for excess X-ray, UV, or infrared emission from this system did not turn up any obvious results, however as the system is very faint with $m_{\rm kep}$ = 16.13 and $m_{K}$ = 14.19 such detections may prove difficult with currently available survey data. The components of Kepler 007846730 are ($M$, $R$, $T$) = (0.614 $M_{\sun}$, 1.360 $R_{\sun}$, 5997 K) and (0.605 $M_{\sun}$, 0.764 $R_{\sun}$, 5537 K), which indicates a highly inflated component for the primary, and a moderately inflated component for the secondary. However, the errors on this system are huge, owing in part to the extreme light ratio, (the primary is nearly 5 times more luminous than the secondary), with a $\sim$34\% error on the primary mass and a $\sim$20\% error on the secondary mass. The interesting aspect is that the radial velocity curve shows a mass ratio close to 1, whereas the light curve reveals a ratio of the radii of nearly 2, with the latter measurement extremely precise as the secondary eclipse is flat-bottomed. If there is not a large systematic error resulting from the extreme light ratio, it is likely that, within the large errors, the primary of this system may be a slightly evolved solar-like sub-giant. More precise spectroscopic observations are required though before anything definitive can be claimed.

Examining the rest of the stars, most of the long-period low-mass binaries are inflated compared to the theoretical stellar models, although there is a wide range of individual errors resulting from each system's brightness, luminosity ratio, and temporal coverage, with an apparent intrinsic scatter of $\sim$5\% for the many stars near 0.75 $M_{\sun}$. Of particular interest are the three stars located at ($M$, $R$) = (0.488 $M_{\sun}$, 0.558 $R_{\sun}$), (0.520 $M_{\sun}$, 0.612 $R_{\sun}$), and (0.659 $M_{\sun}$, 0.693 $R_{\sun}$), which are the secondary stars of Kepler 006131659, 003102024, and 004352168 respectively. These stars have low masses, reasonably small errors, are inflated by $\sim$10-15\% compared to the models, and appear to agree well with the previous measurements of \citet{LopezMorales2007} for short-period systems. At face value this argues that low-mass stars in long-period systems, presumed to be at their natural rotation rate, are still in disagreement with theoretical models and thus rotation is not a principal factor in explaining the discrepancy between low-mass stellar models and observations. However, there are a number of factors which could complicate this assessment and deserve attention.

First, it should be noted that with only 3 stars in the critical mass range, we should be wary about any broad conclusions based on small number statistics. Examining some possible factors though, eccentricity may play a role if it is significant enough to cause tidal interactions at periastron. However, the eccentricities for Kepler 006131659, 003102024, and 004352168 are 0.0147, 0.5709, and 0.2146 respectively, and thus do no present any obvious trend or commonality. The periods for the three systems span a wide range of 17.5, 13.8, and 10.64 days respectively as well. It is possible that even though these binaries are in long-period systems, they still have fast rotation rates. Examining the raw PDC light curve for all three systems reveals a possible, (though very weak and possibly residual spacecraft systematics), spot-induced periodic signature of $\sim$15 days for Kepler 006131659 in the out-of-eclipse data, a very clear spot-induced pattern of $\sim$5 days with definite signatures of differential rotation for Kepler 003102024,  and another very clear spot-induced pattern of $\sim$11 days with definite signatures of differential rotation for Kepler 004352168. Thus, from the out-of-eclipse photometric modulations, it is possible that the rotation period of the secondary star in Kepler 003102024 may be at 5 days, and this is fast enough to induce significant inflation, but the fastest rotation periods for Kepler 006131659 and 004352168 are still $\sim$11 and $\sim$15 days respectively, which theoretically means those stars are at their natural rotation rates. A major complication though is the difficulty in determining whether the multiple periods in the out-of-eclipse modulation, which mimics the signature of differential rotation seen in single stars, is actually from differential rotation on one or both of the stars, the result of each stars' individually different rotation rates, or the result of spot patterns changing over time, coupled with the near impossibility of knowing which signals arise from the primary star and which from the secondary in each system. 

As well, a recent study by \citet{Harrison2012} on the rotation rates of single field low-mass stars from $Kepler$ and their spot-induced photometric amplitudes reveal a strong trend of decreasing spot amplitude with rotation period for $P$ $<$ 40 days, with a flat trend beyond 40 days. Relative to the spot amplitudes at 40$^{+}$ days, low-mass stars with periods of 0-5 days, 5-10 days, 10-20 days, 20-30 days, and 30-40 days exhibit approximately 10, 5, 2.5, 2, and 1.5 larger spot amplitudes respectively, though with a decent amount of intrinsic scatter in each period bin. Thus, if one takes spot-induced photometric levels as a proxy for stellar activity, it is possible that one really needs to find stars in eclipsing binaries that have $40^{+}$ day rotation periods in order to truly be measuring ``normal'' non-active stars. However, via the derived levels of photometric variation, one would expect a major trend of decreasing radii versus rotation period by $P$ = 10 days if rotation was a major factor in contributing to activity and radii inflation.

Metallicity has been previously examined as a potential factor in contributing to the discrepancy \citep[e.g.,][]{LopezMorales2007,Boyajian2012}, but it has always been found to be a mostly negligible factor with no observational correlation found, and we are not able to determine metallicity reliably enough from our spectra for our own study. X-ray luminosity has also been examined as a proxy for stellar activity, with \citep{LopezMorales2007} finding a very weak correlation between X-ray flux and the amount of radii inflation for short-period binaries, but \citet{Boyajian2012} finding no correlation for single stars. We were not able to find any X-ray detections for any of ours systems. Finally, it is possible that there is unresolved third light in the PDC aperture that will affect the light curve analysis, but via direct testing we find that adding as much as even 10\% third light into the model only affects the radius of the derived stars to $<$1\%, and is often completely ruled out by the very deep eclipses.

\citet{Boyajian2012} recently found that there was no significant statistical difference between low-mass single stars measured via interferometry and low-mass stars in binary systems, with both groups being inflated relative to the stellar models. The new stars that we have measured appear to be more inflated compared to the single stars in Figure~\ref{MRdiagram}, but it is difficult to perform a rigorous statistical analysis as a result of both our small sample size, and the very large errors on the model-inferred masses of the single stars. If we were able to obtain more accurate temperatures for the eclipsing binaries in the future, then perhaps we could compare more directly to the single stars in the temperature-radius plane. 

Although more observations are always needed, for now, we are thus left with the following conclusions:

\begin{enumerate}

 \item The radii of low-mass main-sequence stars, measured both as components in eclipsing binaries, and interferometrically as single field stars, have radii that are $\sim$5-10\% greater than predicted by both old and new stellar models, with an intrinsic scatter of $\sim$5\%.
 
 \item Low-mass eclipsing binaries with $P$ $>$ 10 days, which should contain components where stars are at their natural rotation rates, analogous to field stars, still present inflated radii relative to stellar models, although more well-studied long-period binaries are still needed to bolster the number of observed stars.
 
 \item No definitive correlation between the amount of radii inflation and a single factor have so far been found.
 
 \item Given the above items, the discrepancy between the observed and predicted radii of low-mass stars appears to be a shortcoming of models of low-mass stars for 0.2 $<$ $M$ $<$ 0.8 $M_{\sun}$. This should not be surprising, given that most stellar models are only 1D or 2D models, and thus fail to capture a lot of complicated physics, including magnetic field creation and its resulting effects.
 
 \item When assuming a radius for a single star from photometric or spectroscopic observations, (e.g., deriving the radius of a transiting exoplanet relative to the radius of its host star), one should use the observational mass-radius-temperature relations, and assume an error of $\sim$5\% in the interpolated radius due to intrinsic star-to-star scatter.
 
\end{enumerate}

%% file: lmblcmodel.tab
003102024 & 13.782617 & 0.05433$\pm$0.00014 & 0.769$\pm$0.007 & 0.414$\pm$0.003 & 88.876$\pm$0.010 & 0.5709$\pm$0.0005 & 298.74$\pm$0.04 & 0.492$\pm$0.014 & 0.523$\pm$0.048\\
004352168 & 10.643735 & 0.06872$\pm$0.00007 & 0.689$\pm$0.002 & 0.328$\pm$0.007 & 89.516$\pm$0.006 & 0.2146$\pm$0.0014 & 224.01$\pm$0.41 & 0.493$\pm$0.031 & 0.741$\pm$0.042\\
004773155 & 25.706040 & 0.04272$\pm$0.00001 & 0.902$\pm$0.003 & 0.833$\pm$0.008 & 89.819$\pm$0.003 & 0.4428$\pm$0.0009 & 306.56$\pm$0.11 & 0.674$\pm$0.016 & 0.750$\pm$0.020\\
006029130 & 12.591616 & 0.06251$\pm$0.00013 & 1.023$\pm$0.018 & 0.798$\pm$0.016 & 88.731$\pm$0.007 & 0.0156$\pm$0.0014 & $ $ 44.92$\pm$5.26 & 0.535$\pm$0.036 & 0.395$\pm$0.042\\
006131659 & 17.527835 & 0.04555$\pm$0.00009 & 0.676$\pm$0.005 & 0.281$\pm$0.002 & 89.161$\pm$0.011 & 0.0147$\pm$0.0005 & 270.28$\pm$0.01 & 0.654$\pm$0.010 & 0.314$\pm$0.024\\
006431670 & 29.911422 & 0.03628$\pm$0.00006 & 1.091$\pm$0.025 & 1.185$\pm$0.052 & 89.519$\pm$0.004 & 0.3336$\pm$0.0008 & 172.91$\pm$1.21 & 0.675$\pm$0.071 & 1.193$\pm$0.057\\
007846730 & 11.028234 & 0.09533$\pm$0.00002 & 0.562$\pm$0.000 & 0.714$\pm$0.002 & 89.482$\pm$0.005 & 0.0462$\pm$0.0002 & 244.38$\pm$0.14 & 0.458$\pm$0.003 & 0.591$\pm$0.008\\
008296467 & 10.303360 & 0.06916$\pm$0.00003 & 0.982$\pm$0.007 & 0.774$\pm$0.013 & 89.772$\pm$0.010 & 0.2540$\pm$0.0017 & 320.43$\pm$0.48 & 0.594$\pm$0.035 & 1.160$\pm$0.025\\
009284741 & 20.729249 & 0.04113$\pm$0.00004 & 1.027$\pm$0.009 & 1.208$\pm$0.018 & 89.416$\pm$0.003 & 0.3722$\pm$0.0011 & $ $ 42.57$\pm$0.21 & 0.498$\pm$0.022 & 0.660$\pm$0.030\\
010753734 & 19.406540 & 0.05156$\pm$0.00004 & 0.850$\pm$0.002 & 0.781$\pm$0.010 & 89.784$\pm$0.016 & 0.5200$\pm$0.0003 & $ $ 18.00$\pm$0.16 & 0.647$\pm$0.029 & 0.557$\pm$0.020\\
010992733 & 18.526059 & 0.05500$\pm$0.00008 & 0.779$\pm$0.005 & 0.696$\pm$0.018 & 89.935$\pm$0.060 & 0.3808$\pm$0.0012 & $ $ 25.30$\pm$0.47 & 0.688$\pm$0.055 & 0.726$\pm$0.040\\

%% file: lmbrvmodel.tab
003102024 & 0.712$\pm$0.050 & 0.520$\pm$0.027 & 0.797$\pm$0.016 & 0.612$\pm$0.013 & 5612$\pm$201 & 4640$\pm$150\\  
004352168 & 1.132$\pm$0.082 & 0.659$\pm$0.036 & 1.006$\pm$0.022 & 0.693$\pm$0.015 & 5547$\pm$188 & 4392$\pm$124\\  
004773155 & 0.931$\pm$0.049 & 0.902$\pm$0.040 & 1.008$\pm$0.015 & 0.909$\pm$0.014 & 5765$\pm$318 & 5523$\pm$294\\  
006029130 & 0.783$\pm$0.043 & 0.797$\pm$0.032 & 0.820$\pm$0.014 & 0.839$\pm$0.014 & 5717$\pm$180 & 5426$\pm$167\\  
006131659 & 0.734$\pm$0.026 & 0.488$\pm$0.013 & 0.825$\pm$0.008 & 0.558$\pm$0.007 & 5493$\pm$129 & 4234$\pm$ 81\\  
006431670 & 0.455$\pm$0.035 & 0.462$\pm$0.035 & 0.684$\pm$0.018 & 0.746$\pm$0.020 & 5012$\pm$207 & 5198$\pm$226\\  
007846730 & 0.614$\pm$0.208 & 0.605$\pm$0.118 & 1.360$\pm$0.119 & 0.764$\pm$0.067 & 5997$\pm$168 & 5537$\pm$146\\  
008296467 & 0.887$\pm$0.046 & 0.856$\pm$0.038 & 0.837$\pm$0.013 & 0.822$\pm$0.013 & 5362$\pm$178 & 5069$\pm$161\\  
009284741 & 0.768$\pm$0.066 & 0.728$\pm$0.068 & 0.738$\pm$0.021 & 0.757$\pm$0.022 & 5082$\pm$252 & 5295$\pm$271\\  
010753734 & 0.991$\pm$0.027 & 0.841$\pm$0.017 & 1.036$\pm$0.008 & 0.881$\pm$0.007 & 5818$\pm$181 & 5492$\pm$164\\  
010992733 & 0.763$\pm$0.144 & 0.729$\pm$0.097 & 1.041$\pm$0.054 & 0.811$\pm$0.042 & 5288$\pm$129 & 4892$\pm$115\\  

%% file: chp7-ld.tex
\begin{singlespace}
\section[\MakeUppercase{Observational Determination of\\Limb-Darkening in the $Kepler$ Bandpass}]{\MakeUppercase{Observational Determination of Limb-Darkening in the $Kepler$ Bandpass}}
\label{chap7}
\end{singlespace}

\subsection{Introduction}
\label{ldintro}

Limb-darkening is the observational phenomenon that stars, when observed as projected disks on the sky, appear to be darker towards their edges, or limbs, when compared to their centers. This effect is typically observed in the near infrared, visible, and ultraviolet wavelengths, with the effect becoming more prominent towards ultraviolet and less prominent towards infrared wavelengths. (Limb-brightening, where the limb appears brighter than the center, can also be observed at both very-short wavelengths, such as extreme ultraviolet and X-ray, and very long wavelengths, such as far-infrared and radio.)

\begin{figure}
\includegraphics[width=\linewidth]{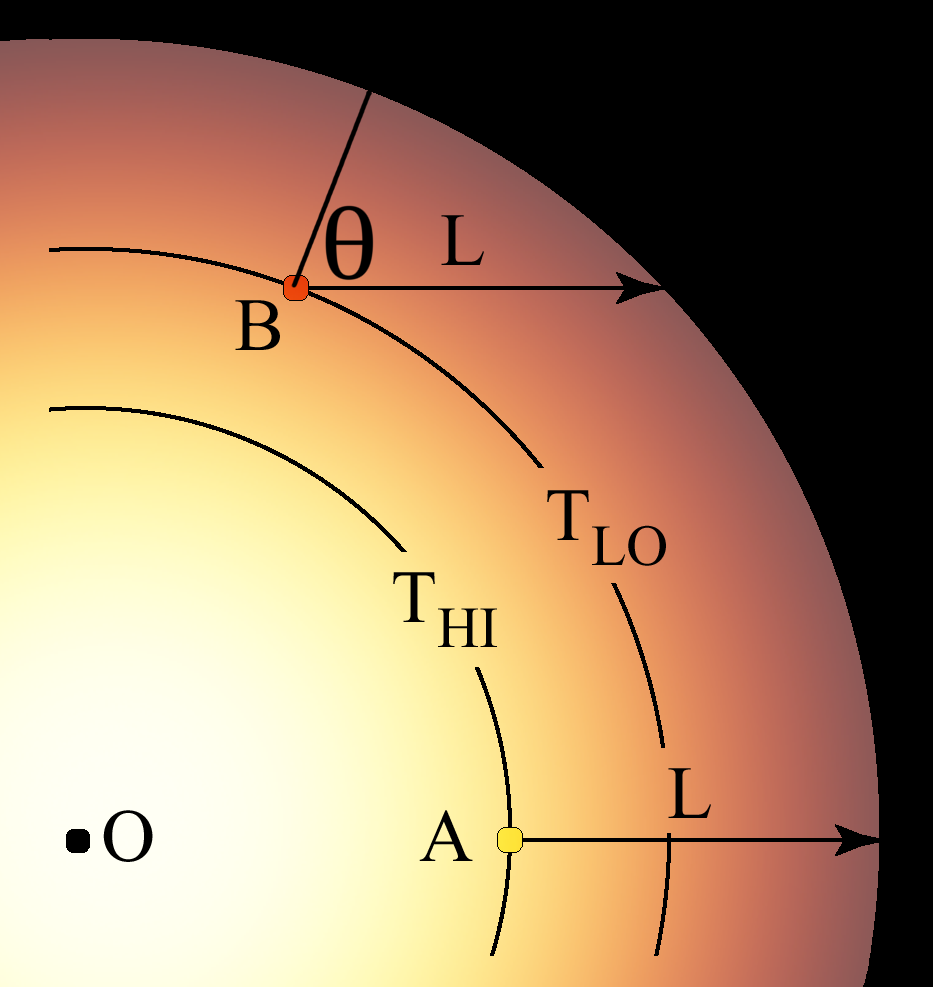}
\caption[An illustration of the effect of limb-darkening]{An illustration of the effect of limb-darkening, not to scale. Light emitted from point $A$, at a temperature $T_{HI}$, travels through a path length $L$ to escape into space to reach the observer situated to the right. Light emitted from the limb of the star, at point $B$, travels a similar path length $L$, but due to spherical geometry point $B$ is located at a higher physical height, and thus radiates at a lower temperature, $T_{LO}$. The resulting difference in temperature causes the limb to appear darker to the observer.}
\label{ldfig}
\end{figure}

The physical cause of limb-darkening results from the fact that a star is a spherical object that contains an extended atmosphere, which has a varying temperature, density, and composition with height, with temperature and density typically decreasing with increasing height. Referring to Figure~\ref{ldfig}, let us assume that one observes a star as shown, and they do so from the right side of the diagram, which is not to scale. If the observer is looking directly towards the center of the star, point $O$, the photosphere, or visible surface of the star, will lie at point $A$, where the optical depth of the atmosphere, $\tau$, reaches approximately $\frac{2}{3}$ at a path length, $L$, from where $\tau$ $\approx$ 0. At this layer the stellar atmosphere will have a temperature, $T_{HI}$, from which most of the observed photons are emitted. However, when one observes towards the limb of the star, away from the center, the layer where $\tau$~=~$\frac{2}{3}$ will still lay at a path length $\sim$$L$ from $\tau$ $\approx$ 0, but due to the spherical geometry, this point, $B$, is physically higher up in the atmosphere. Since point $B$ is higher up, it has a temperature, $T_{LO}$, which is lower than $T_{HI}$. Thus, due to the lower temperature, the amount of flux emitted at point $B$ will be less than point $A$, as seen by the observer, and thus the limb will appear to be darker. Additionally, photons emitted at point $B$ will travel through a different density and composition of material, thus resulting in different spectral absorption features compared to point $A$. Since the relative flux between points $A$ and $B$ mostly depend on the relative difference between their temperatures, limb darkening will strongly vary depending on the wavelength of observation. Limb darkening will be more extreme at shorter wavelengths, where the flux difference between two blackbodies with different temperatures is most extreme, and less extreme at longer wavelengths where blackbodies are in the Rayleigh-Jeans tail. Since stars do not radiate as perfect blackbodies, and since the temperature, density, and composition of the atmosphere is not a linear function of scale height, limb-darkening is an inherently non-linear phenomenon.

\begin{figure}
\includegraphics[width=\linewidth]{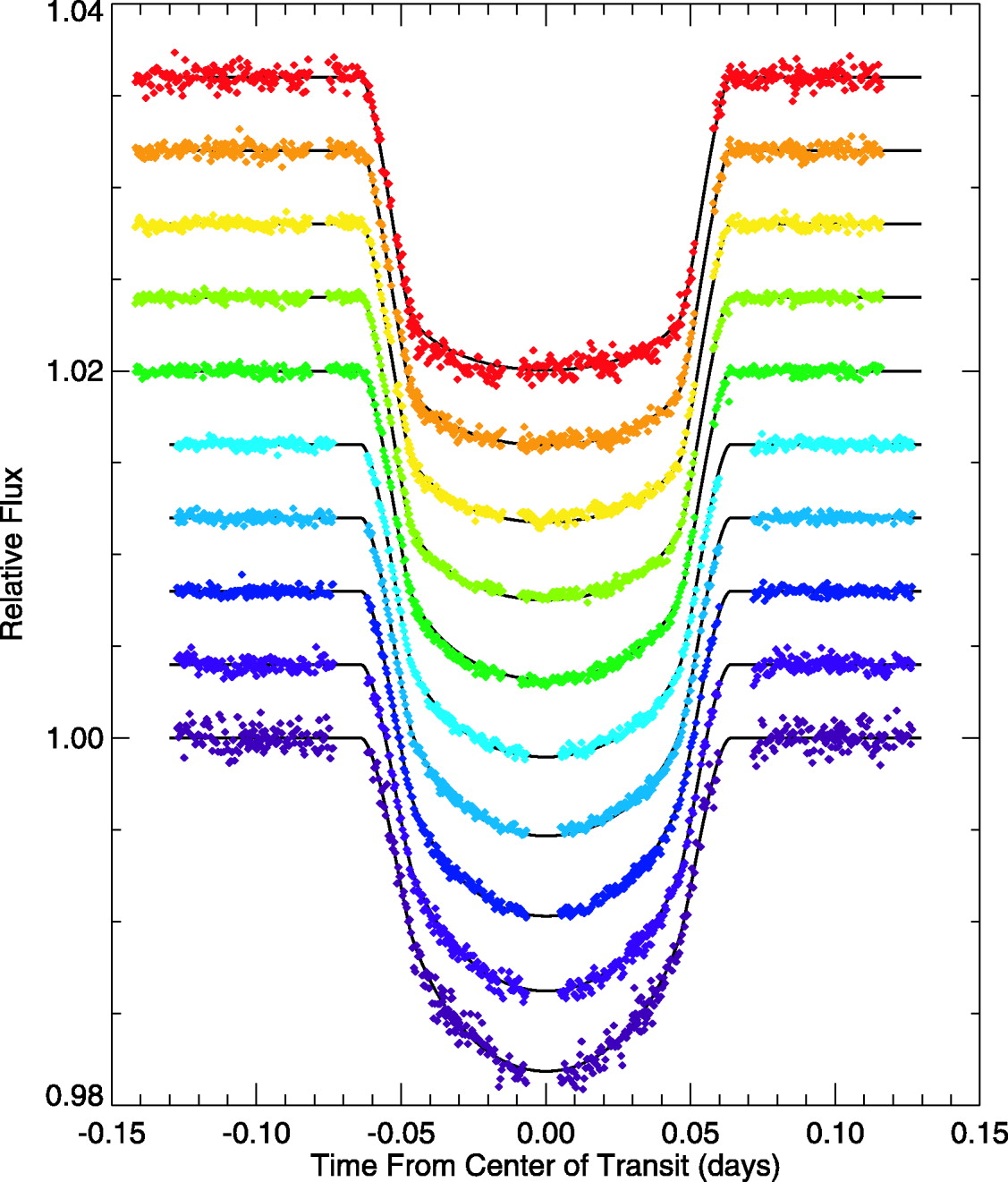}
\caption[Simultaneous multi-wavelength transit observations of HD 209458b, reproduced from \citet{Knutson2007a}]{Simultaneous multi-wavelength transit observations of HD 209458b, reproduced from \citet{Knutson2007a}. The observed data is shown with individual points, and the best-fit limb-darkened model with a solid black line. The top light curve is centered on 970.5 nm, the bottom light curve is centered on 320 nm, with wavelength decreasing from top to bottom. Note how the transit shape transitions from a flat-bottom to a round-bottom shape as wavelength decreases.}
\label{knutson07fig}
\end{figure}

The accurate determination of limb-darkening is important for any science that requires accurately knowing the flux distribution of the projected stellar disk. Of particular relevance, transiting exoplanets can be greatly affected by stellar limb-darkening. In the absence of limb-darkening, between ingress and egress, the observed flux remains constant throughout the transit, and the transit shape is flat-bottomed with a depth exactly equal to the square of the ratio of the radii. However, with limb-darkening, when the planet is towards the limb of the star it will be blocking less total light than when it is at the center of the star. This will result in a transit shape that is deeper towards the center of the transit, and a round-bottom shape. As limb-darkening becomes more severe at shorter wavelengths, transits will be observed to be more round-bottomed at shorter wavelengths, and more flat-bottomed at longer wavelengths. A classic illustration of the effect of limb-darkening in extrasolar planet transits is shown in Figure~\ref{knutson07fig}. This figure is reproduced from \citet{Knutson2007a}, who used the STIS spectrometer on the Hubble Space Telescope to obtain simultaneous transit light curves of HD 209458b in 10 passbands ranging from 290-1030 nm. As can be seen, the transit shape transitions from a flat-bottom to a round-bottom shape as wavelength decreases.

Although limb-darkening is inherently non-linear, many attempts have been made to derive simple analytical formulae to represent it to a reasonable accuracy. Typically these laws express the specific intensity of the stellar disk as a function of $\mu$, where $\mu = \cos{\theta}$, and where $\theta$ is the angle between a line perpendicular to the stellar surface and the line of sight of the observer, (see Fig.~\ref{ldfig}), so that $\mu$ = 1 at the center of the star, and $\mu$ = 0 at the edge. The earliest and simplest expression is known as the linear limb-darkening law, first formulated by \citet{Russell1912} as

\begin{equation}
 \frac{I_{u}}{I_{0}} = 1 - c\cdot(1-\mu)
\end{equation}

\noindent where I$_{\mu}$ is the intensity at a given $\mu$, I$_{0}$ is the intensity at the center of the star, and $c$ is a free coefficient that can be adjusted to best fit the limb-darkening profile for an individual star. In an effort to more accurately represent limb-darkening, the quadratic law was first formulated by \citet{Kopal1950} as

\begin{equation}
 \frac{I_{u}}{I_{0}} = 1 - c_{1}\cdot(1-\mu) - c_{2}\cdot(1-\mu)^{2}
\end{equation}

\noindent where now two free coefficients, $c_{1}$ and $c_{2}$, can be adjusted for each star. Several other 2-parameter have been introduced, notably the logarithmic law \citep{Klinglesmith1970}, the square root law \citep{Diaz-Cordoves1992}, and the exponential law \citep{Claret2003}, expressed as 

\begin{equation}
 \frac{I_{u}}{I_{0}} = 1 - c_{1}\cdot(1-\mu) - c_{2}\cdot\mu\cdot\ln{\mu}
\end{equation}

\begin{equation}
 \frac{I_{u}}{I_{0}} = 1 - c_{1}\cdot(1-\mu) - c_{2}\cdot(1-\sqrt{\mu})
\end{equation}

\begin{equation}
 \frac{I_{u}}{I_{0}} = 1 - c_{1}\cdot(1-\mu) - \frac{c_{2}}{1-e^{\mu}}
\end{equation}

\noindent respectively. As a result of more accurate model stellar atmospheres and the desire to better reproduce their resulting limb-darkening profiles, \citet{Claret2000b} introduced a 4 parameter limb-darkening law as

\begin{equation}
 \frac{I_{u}}{I_{0}} = 1 - c_{1}\cdot(1-\mu^\frac{1}{2}) - c_{2}\cdot(1-\mu) - c_{3}\cdot(1-\mu^\frac{3}{2}) - c_{4}\cdot(1-\mu^{2})
\end{equation}

\noindent which could be extended to more terms as needed based on the general power-law

\begin{equation}
 \frac{I_{u}}{I_{0}} = 1 - \sum^{N}_{n=1}{c_{n}\cdot(1-\mu^{\frac{n}{2}}})
\end{equation}

\noindent where N is the number of terms desired. Finally, \citet{Sing2009b} removed the first term of the \citet{Claret2000b} 4 parameter law to produce a 3 parameter law

\begin{equation}
 \frac{I_{u}}{I_{0}} = 1 - c_{1}\cdot(1-\mu) - c_{2}\cdot(1-\mu^\frac{3}{2}) - c_{3}\cdot(1-\mu^{2})
\end{equation}

\noindent which \citet{Sing2009b} found to produce a smoother and more realistic limb-darkening profile at low values of $\mu$, towards the edge of the star.

\begin{figure}[ht]
\includegraphics[width=\linewidth]{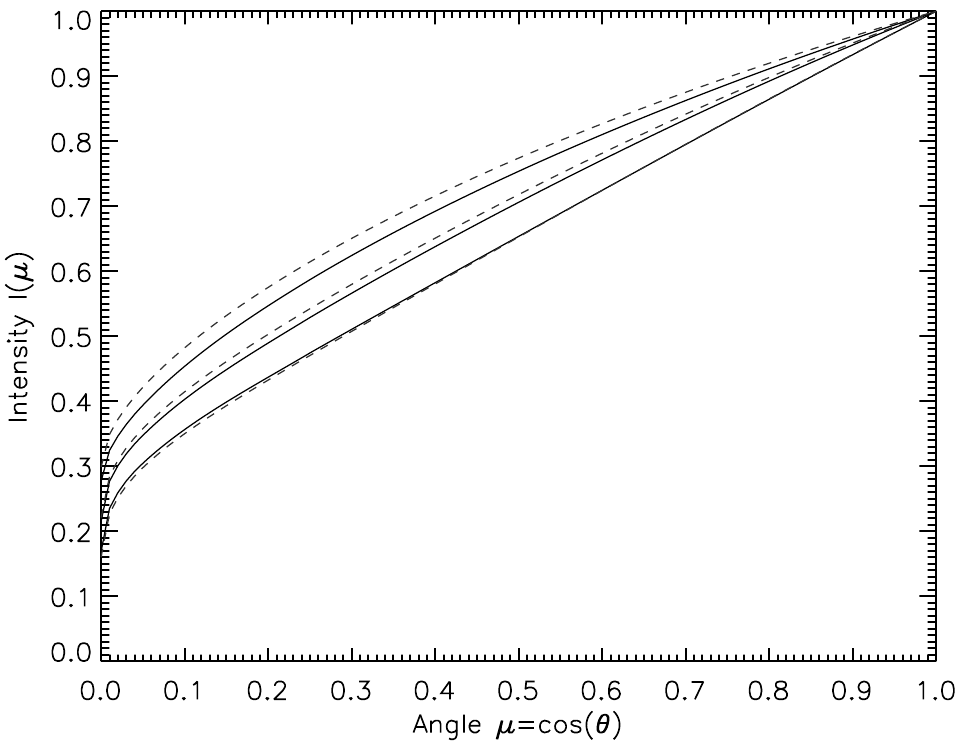}
\caption[Model atmosphere prediction of intensity variation from stellar center to limb, reproduced from \citet{Sing2010}]{The predicted variation of intensity from stellar center ($\mu$ = 1) to limb ($\mu$ = 0) from the ATLAS stellar models, reproduced from \citet{Sing2010}. The solid lines are for the $CoRoT$ bandpass, while the dashed lines are for the $Kepler$ bandpass. From top to bottom, the three sets of lines represent stars with $T_{\rm eff}$ = 6500, 5500, and 4500 K, each with $\log{g}$ = 4.5 and [M/H] = 0.0. Note how the distribution is mostly linear, especially at lower temperatures, with the most non-linearity occurring towards the limb of the star.}
\label{singldfig}
\end{figure}

More precise, and perhaps more importantly, more accurate characterizations of limb-darkening are especially needed today, as the $Kepler$ space telescope is obtaining transit curves of extrasolar planets at unprecedented levels of photometric precision. The physical parameters of extrasolar planets derived from transit light curves are intimately tied to the parameters one assumes for the host star, including its limb-darkening profile. In order to choose appropriate limb-darkening coefficients, typically a limb-darkening law is fit to the stellar disk intensity distribution as predicted via a model stellar atmosphere. Recently, both \citet{Sing2010} and \citet{Claret2011} have derived limb-darkening coefficients for the $Kepler$ bandpass for many of the above laws by fitting them to stellar model predictions. \citet{Sing2010} determined coefficients by performing a least-squares fit (LSF) to the ATLAS models, using only values of $\mu$ $>$ 0.05, as he noted the model atmospheres seem to over-predict limb-darkening at very low values of $\mu$ compared to the Sun. \citet{Claret2011} derived coefficients via both a least-squares fit, as well as a flux conservation method (FSM) that aims to best-fit the model atmospheres while ensuring the overall stellar luminosity from the models is conserved, for both the ATLAS and PHOENIX model atmospheres. Figure~\ref{singldfig} shows the predicted distribution from the ATLAS stellar models for main-sequence stars at 6500, 5500, and 4500 K effective temperatures in both the $CoRoT$ and $Kepler$ bandpasses. As can be seen, the limb-darkening profile is mostly linear, with higher-temperature stars being more non-linear. The largest amount of non-linearity occurs near the limb of the stars.

In addition to computing limb-darkening coefficients from model atmospheres, the observation of eclipsing binaries can directly yield limb-darkening parameters as each star eclipses the other, yielding uniquely shaped eclipse curves. Although limb-darking can be partially degenerate with other parameters, such as the radii of the stars or the inclination of the system, it is not completely degenerate and can be accurately measured with sufficiently precise photometric data. \citet{Claret2008} compared linear limb-darkening coefficients derived from very precise $B$, $V$, and $R$-band eclipsing binary light curves to those computed from model atmospheres, and noted a $\sim$15-20\% average discrepancy, with some systems being up to 40\% discrepant. Such large errors could certainly result in errors of planetary parameters by $\sim$5-10\% or more, which is quite significant compared to the $\lesssim$1\% precision that $Kepler$ can obtain. Thus, it is of great importance to both stellar and planetary astrophysics with $Kepler$ that limb-darkening be observationally determined, and compared to the various model predictions to see which, if any, of the stellar models and limb-darkening fitting techniques provides an accurate representation of reality.

\subsection{Observational Data and Modeling}

We have selected known $Kepler$ eclipsing binaries for modeling that are well-detached, main-sequence systems with deep eclipses, and thus present good cases for determining accurate and precise limb-darkening coefficients. First, we utilize all the eclipsing binaries we have already modeled in Chapter~\ref{chap5} and for which we already have linear limb-darkening coefficients. Second, we selected additional eclipsing binaries with hotter temperatures in order to extend the temperature range of our study, and targeted them for spectroscopic follow-up during our radial-velocity observations as described in \S~\ref{rvobs}. The additional high-temperature systems that we obtained sufficient spectroscopic observations of for temperature determination were Kepler 003327980, Kepler 006610219, Kepler 006766748, and Kepler 009344623. We use the same technique to select and reduce $Kepler$ data for these systems as previously discussed in \S\ref{photobs}. We then model the light curves for these systems using the same technique as described in \S\ref{rvmodeltech} to obtain the linear limb-darkening coefficients, designated $c_{p}$ and $c_{s}$ for the primary and secondary star respectively, and model the spectra using the same technique as described in \S\ref{rvobs}. We list the effective temperatures and linear limb-darkening coefficients for both the low-mass binaries discussed in Chapter~\ref{chap6}, as well as the new higher-temperature binaries, in Table~\ref{ldtab}.

\begin{deluxetable}{ccccc}
\tablewidth{0pt}
\tabletypesize{\scriptsize}
\tablecaption{Effective Temperatures and Linear Limb-Darkening Coefficients of the Stars in the Modeled Eclipsing Binaries}
\tablecolumns{5}
\tablehead{\emph{Kepler} ID & $T_{1}$ & $T_{2}$ & $c_{p}$ & $c_{s}$\\ & (K) & (K) & & }
\startdata
\input{ldmodel.tab}
\enddata
\label{ldtab}
\end{deluxetable}

\subsection{Results and Discussion}

\begin{figure}[ht!]
\centering
\epsfig{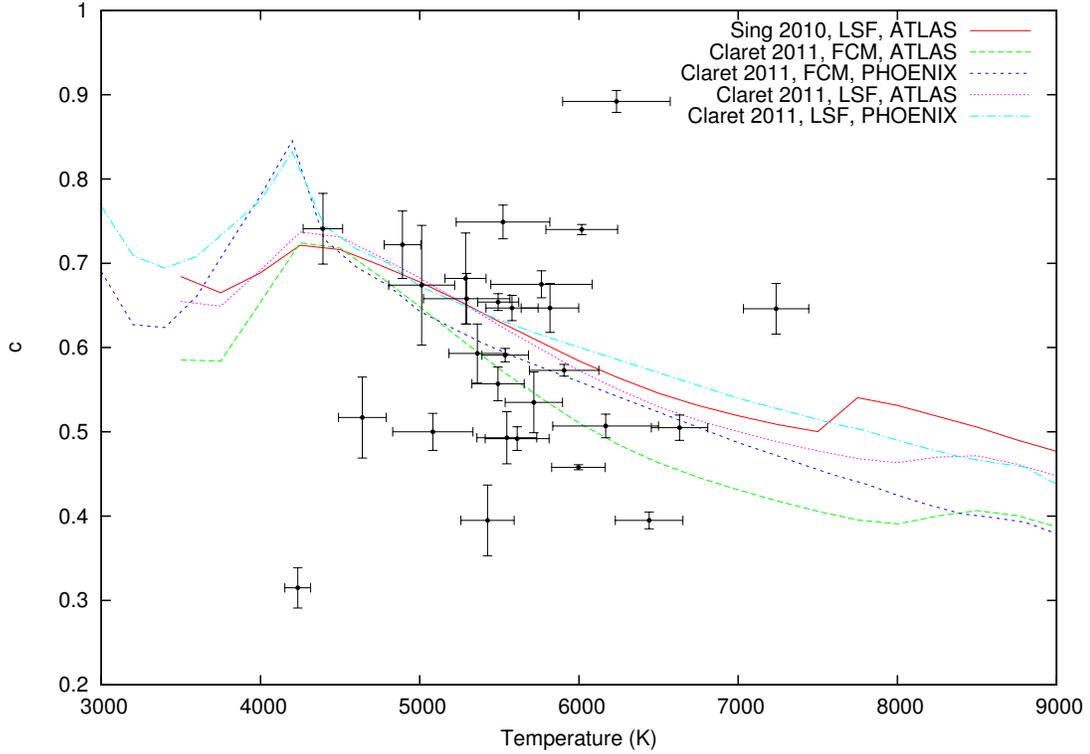}
\caption[Plot of the linear limb-darkening coefficient versus effective temperature]{Plot of the linear limb-darkening coefficient versus effective temperature. Observationally determined coefficients we derived are shown by solid black circles. Theoretical limb-darkening coefficients derived from different model stellar atmospheres and interpolation techniques are shown by lines of varying color and type, all assuming [M/H] = 0.0 and $\log{g}$ = 4.5. ``Sing 2010'' corresponds to the models of \citet{Sing2010}, while ``Claret 2010'' corresponds to the models of \citet{Claret2011}. ``LSF'' corresponds to the least-squares fitting method of interpolation, while ``FCM'' corresponds to the flux conservation method of interpolation. ``ATLAS'' and ``PHOENIX'' correspond to the type of model atmosphere employed. The error-weighted mean and standard deviation for the observed points, as well as the model points, are shown for 3 temperature bins as blue diamonds and red squares respectively.}
\label{LDdiagram}
\end{figure}

In Figure~\ref{LDdiagram} we plot the observationally determined linear limb-darkening coefficients for the modeled eclipsing binaries versus their spectroscopically determined effective temperatures. We also plot the theoretical linear limb-darkening coefficients of \citet{Sing2010} and \citet{Claret2011}, for all combinations of the PHOENIX and ATLAS model atmospheres and the LSM and FCM interpolation techniques they presented, given [M/H] = 0.0 and $\log{g}$ = 4.5.

As can be seen from Figure~\ref{LDdiagram}, the measured linear limb-darkening coefficients have a large amount of scatter compared to the measurement errors. While they tend to cluster around the range of predicted limb-darkening values, no particular theoretical model or interpolation method is clearly favored by the data. As a basic analysis, we divide the data up into three temperature bins that contain enough points for analysis, and as well are common temperatures for exoplanet hosts discovered via $Kepler$: 5250-5500 K, 5500-5750 K, and 5750-6000 K. In each bin we compute the error-weighted mean and error-weighted standard deviation for both the measured points, as well as the mean and standard deviation of all the model predictions. We show these results in Table~\ref{tbindat}. Via these results, we find that the model-predicted linear limb-darkening coefficients have an inherent discrepancy of 4-6\% among models, depending on the temperature bin. The observationally determined linear limb-darkening coefficients have an average discrepancy of 13-25\%, depending on the temperature bin.

\begin{deluxetable}{ccccc}
\tablewidth{0pt}
\tabletypesize{\scriptsize}
\tablecaption{Weighted Mean and Standard Deviation of the Observed and Model Linear Limb-Darkening Coefficients in Three Temperature Bins}
\tablecolumns{5}
\tablehead{$T_{\rm eff}$ Range (K) & Observed Data & Model Data}
\startdata
5250-5500 & 0.625$\pm$0.066 & 0.624$\pm$0.027\\
5500-5750 & 0.581$\pm$0.078 & 0.598$\pm$0.030\\
5750-6000 & 0.519$\pm$0.132 & 0.574$\pm$0.034\\
\enddata
\label{tbindat}
\end{deluxetable}

The first caveat when drawing conclusions from this data is that we still suffer from small-number statistics in each temperature bin, and really require observation of additional systems. Examining the data at-hand however, we must attempt to explain how there could be such large star-to-star scatter in the determined limb-darkening coefficients. One possibility is that the inclinations of these systems are such that only the very edges of the stars are being probed via the eclipses, and thus we are essentially fitting a line to only a small fraction of the stellar disk and thus a small range of $\mu$, so that the determined coefficients are not the average of the entire disk. However, we have computed the fraction of the stellar disk eclipsed for every star in our sample, and find that every star has at least half of its disk eclipsed, and thus every star has its limb-darkening completely probed from $\mu$ = 0 to 1. This is not surprising as we purposely selected systems with very deep eclipses and thus high inclinations. Another possibility is that these stars have a large range of metallicities and surface gravities that result in varying limb-darkening. Examining the models however, for a star with $T_{\rm eff}$ = 5500 K, $\log{g}$ = 4.5, and [M/H] = 0.0, varying the surface gravity by 0.5 dex only results in a change of the linear limb-darkening coefficient by $<$1\%, and varying the metallicity by 0.5 dex only results in a change of $\sim$3\%, which is much less than the observed variation. The last explanation we can think of, and that we deem most likely, is that these stars have a significant amount of star spots and/or plages that result in inhomogeneous stellar disks and affect the derived limb-darkening coefficients. It is well-known that solar-like and low-mass stars can have numerous star spots, sometimes covering a substantial fraction of the star. Although it becomes difficult to model the effect of such spots on the derived limb-darkening coefficients, one telltale sign might be to model the stellar limb-darkening as a function of time. If the limb-darkening coefficients significantly vary for a given system over timescales of months or years, then the most plausible cause would be varying spot patterns, which can change on those time scales. The $Kepler$ mission should definitely be capable of examining this possibility after a few years of data.

Regardless of the cause of the variation, it should be taken as a given that star-to-star linear limb-darkening typically varies by $\sim$15\%. Thus, the recommended prescription when solving extrasolar planet transit curves, or eclipsing binary light curves, is to directly fit for the limb-darkening coefficients simultaneously with other parameters of interest. In the absence of the ability to directly solve for the limb-darkening coefficients, one should then set them to model predictions, but allow them to vary by $\pm$15\% in the error analysis.

%% file: ldmodel.tab
003102024 & 5612$\pm$201 & 4640$\pm$150 & 0.492$\pm$0.014 & 0.517$\pm$0.048\\ 
003327980 & 7239$\pm$205 & 6632$\pm$177 & 0.646$\pm$0.030 & 0.505$\pm$0.015\\ 
004352168 & 5547$\pm$188 & 4392$\pm$124 & 0.493$\pm$0.031 & 0.741$\pm$0.042\\ 
004773155 & 5765$\pm$318 & 5523$\pm$294 & 0.675$\pm$0.016 & 0.749$\pm$0.020\\ 
006029130 & 5717$\pm$180 & 5426$\pm$167 & 0.535$\pm$0.036 & 0.395$\pm$0.042\\ 
006131659 & 5493$\pm$129 & 4234$\pm$ 81 & 0.654$\pm$0.010 & 0.315$\pm$0.024\\ 
006431670 & 5012$\pm$207 & 5198$\pm$226 & 0.674$\pm$0.071 & 1.194$\pm$0.057\\ 
006610219 & 6018$\pm$225 & 5907$\pm$218 & 0.740$\pm$0.006 & 0.573$\pm$0.007\\ 
006766748 & 6441$\pm$212 & 5580$\pm$164 & 0.395$\pm$0.010 & 0.647$\pm$0.015\\ 
007846730 & 5997$\pm$168 & 5537$\pm$146 & 0.458$\pm$0.003 & 0.591$\pm$0.008\\ 
008296467 & 5362$\pm$178 & 5069$\pm$161 & 0.593$\pm$0.035 & 1.160$\pm$0.025\\ 
009284741 & 5082$\pm$252 & 5295$\pm$271 & 0.500$\pm$0.022 & 0.658$\pm$0.030\\ 
009344623 & 6168$\pm$332 & 6236$\pm$338 & 0.507$\pm$0.014 & 0.892$\pm$0.013\\ 
010753734 & 5818$\pm$181 & 5492$\pm$164 & 0.647$\pm$0.029 & 0.557$\pm$0.020\\ 
010992733 & 5288$\pm$129 & 4892$\pm$115 & 0.682$\pm$0.054 & 0.722$\pm$0.040\\ 

%% file: chp8-sim3.tex
\begin{singlespace}
\section[\MakeUppercase{Modeling Multi-Wavelength Stellar\\Astrometry: Determination of the Absolute\\Masses of Exoplanets and Their Host Stars}]{\MakeUppercase{Modeling Multi-Wavelength Stellar Astrometry: Determination of the Absolute Masses of Exoplanets and Their Host Stars}}
\label{chap8}
\end{singlespace}

\subsection{Introduction}
\label{sim3intro}

As part of a Space Interferometry Mission (SIM) Science Study, we previously examined the implications that multi-wavelength microarcsecond astrometry has for the detection and characterization of interacting binary systems, (see  Appendix~\ref{sim1appendix}). We found that the astrometric orbits of binary systems can vary greatly with wavelength, as astrometric observations of a point source only measure the motion of the photocenter, or center of light, of the system. For systems that contain stellar components with different spectral energy distributions, the motion of the photocenter can be dominated by the motion of either component, depending on the wavelength of observation. Thus, with multi-wavelength astrometric observations it is possible to measure the individual orbit of each component, and thus derive absolute masses for both objects in the system. Additionally, we have previously found that multi-wavelength astrometry can be used to directly measure the inclination and gravity darkening coefficient of single stars, as well as the temperature, size, and position of star spots, (see Appendix~\ref{sim2appendix}).

Astrometry has long been used to measure fundamental quantities of binary stars, and more recently has been used to study extrasolar planets. Although no independently confirmed planet has yet been initially discovered via astrometry, many planets discovered via radial-velocity, (which only yields the planetary mass as a function of the system's inclination and host star's mass), have had follow-up astrometric measurements taken in order to determine their inclinations, and thus true planetary mass as a function of only the assumed stellar mass \citep{McArthur2004,Benedict2006,Bean2007,Martioli2010,McArthur2010,Roll2010,Reffert2011}. There are many ground and space-based microarcsecond precision astrometric projects which are either currently operating or on the horizon. The proposed SIM Lite Astrometric Observatory, a redesign of the earlier proposed SIM PlanetQuest Mission, was to be a space-based 6-meter baseline Michelson interferometer capable of 1 $\mu$as precision measurements in $\sim$80 spectral channels spanning 450 to 900 nm \citep{SIM2009}, thus allowing multi-wavelength microarcsecond astrometry. Although the SIM Lite mission has been indefinitely postponed at the time of this writing, it has already achieved all of its technological milestones, and it, or another similar mission, could be launched in the future. The PHASES project obtained as good as 34 $\mu$as astrometric precision of close stellar pairs \citep{Muterspaugh2010}. The CHARA array has multi-wavelength capabilities, and can provide angular resolution to $\sim$200 $\mu$as \citep{Brummelaar2005}. PRIMA/VLTI is working towards achieving $\sim$30-40 $\mu$as precision in the $K$-band \citep{Belle2008}, with GRAVITY/VLTI expected to obtain 10 $\mu$as \citep{Kudryavtseva2010}. The ASTRA/KECK project will be able to simultaneously observe and measure the distance between two objects to better than 100 $\mu$as precision. The GAIA mission will provide astrometry for $\sim$10$^{9}$ objects with 4 - 160 $\mu$as accuracy, for stars with V = 10-20 mag respectively, and does posses some multi-wavelength capabilities \citep{Cacciari2009}. The MICADO instrument on the proposed E-ELT 40-meter class telescope will be able to obtain better than 50 $\mu$as accuracy at 0.8-2.5 $\mu$m \citep{Trippe2010}. Finally, the NEAT mission proposes to obtain as low as 0.05 $\mu$as astrometric measurements at visible wavelengths \citep{Malbet2011}. Thus, astrometric measurements of extrasolar planets are going to become significantly more common in the future.

In this chapter, we examine the multi-wavelength astrometric signature of exoplanets. A star-planet system is a specialized case of a binary system with extreme mass and temperature ratios, and thus the findings in Appendix~\ref{sim1appendix} apply to exoplanets. Specifically, an extrasolar planet has a combination of reflected and thermally emitted light that cause the photocenter to be displaced from the center of mass of the star. Since the planet's temperature is very different from that of the host star, the amount of photocenter displacement due to the planet will greatly vary with wavelength. Although the luminosity ratio between a star and planet is extreme, the planet also lies a much farther distance from the barycenter of the system compared to the star, and thus it has a large ``moment-arm'' with which to influence the photocenter. While conventional single-wavelength astrometric measurements can yield the inclination and spatial orientation of a system's orbital axis, with multi-wavelength astrometry it should be possible to measure the individual orbits of both the star and planet, and thus determine the absolute masses of both.

In Section~\ref{eqsec} we derive analytical formulae for estimating the astrometric motion of a star-planet system at a given wavelength. In Section~\ref{numsec} we perform numerical simulations of the multi-wavelength astrometric orbits of a few systems of interest using the {\sc reflux} code, and examine a few features specific to transiting planets. In both sections we present the most promising systems for future observation and detection of this effect. Finally, in Section~\ref{discusssec} we discuss our results and what future work is needed to achieve these observations.

\subsection{Analytical Formulae for Computing the Reflex Motion}
\label{eqsec}

\begin{figure}[ht]
\centering
\epsfig{width=\linewidth,file=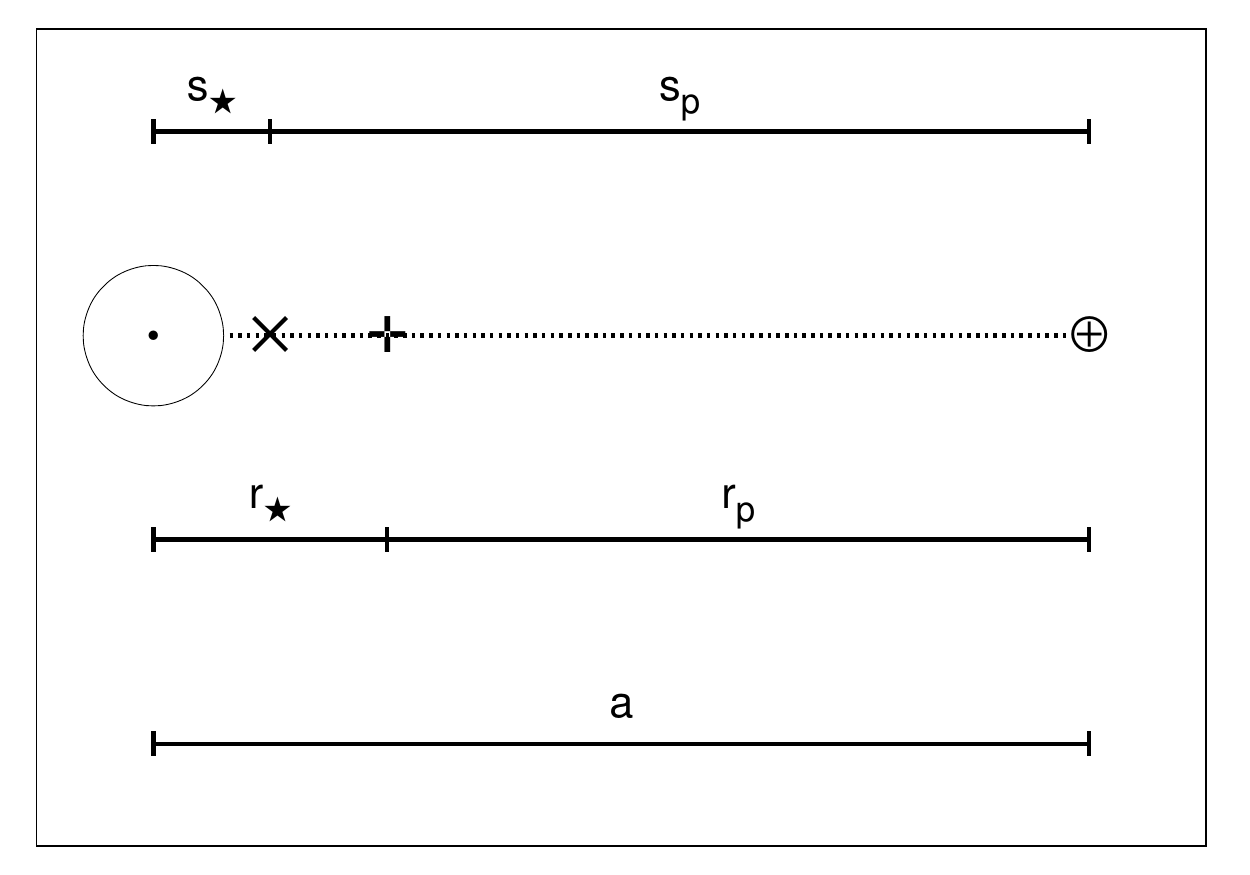}
\caption[Illustration of a star-planet system]{An illustration of a system containing a star, shown on the left, and a planet, shown on the right, separated by a distance $a$, not to scale. The star and planet lie at distances of $r_{\star}$ and $r_{p}$, respectively, from the barycenter of the system, which is marked via a ``+'' symbol. Similarly, the star and planet lie at distances of $s_{\star}$ and $s_{p}$, respectively, from the photocenter of the system, which is marked via a ``$\times$'' symbol. All distances are sky-projected distances along the semi-major axis of the system, and thus are independent of the system's inclination. Note that although in this illustration the photocenter is to the left of the barycenter, it can lie anywhere between the star and planet.}
\label{schematicfig}
\end{figure}

Our objective is to derive an analytical expression for the amplitude of the sky-projected angular astrometric reflex motion of a star-planet system with respect to the wavelength of observation, $\alpha$. In all of the following equations, we are dealing with sky-projected distances measured along the semi-major axis of the system, and thus they are independent of the inclination of the system. We consider the case of a star and single planet in a circular orbit, with masses $M_{\star}$ and $M_{p}$ respectively, separated by an orbital distance, $a$, as illustrated in Figure~\ref{schematicfig}. The system's barycenter, marked via a ``+'' symbol, lies in-between the star and planet, at a distance of $r_{\star}$ from the star, and $r_{p}$ from the planet.

Defining the mass ratio, $q$, as

\begin{equation}
 q = \frac{M_{p}}{M_{\star}} 
\end{equation}

\noindent the values for $r_{\star}$ and $r_{p}$ are then

\begin{equation}
r_{\star} = \frac{a \cdot q}{q+1}
\end{equation}

\begin{equation}
 r_{p} = \frac{a}{q + 1}
\end{equation}

\noindent where by definition $r_{\star}$ + $r_{p}$ = $a$. 

In the case where all the light from the system is assumed to come from the star, i.e., the system's photocenter is the star's center, the wavelength-independent amplitude of the angular astrometric reflex motion of the system, $\alpha_{0}$, is

\begin{equation}
\label{alphaeq1}
\alpha_{0} = \arctan \left(\frac{r_{\star}}{D}\right) = \arctan\left(\frac{a \cdot q}{D\cdot(q+1)}\right)
\end{equation}

\noindent where $D$ is the distance to the system from Earth, and $a$, via Kepler's third law, is

\begin{equation}
\label{aeq}
 a = \left(G(M_{\star} + M_{p})\right)^{\frac{1}{3}}\left(\frac{P}{2\pi}\right)^{\frac{2}{3}}
\end{equation}

\noindent where $G$ is the gravitational constant, and $P$ is the orbital period of the system.

When the planet's luminosity is not negligible, in order to determine the wavelength-dependent value of $\alpha$, the location of the system's photocenter, which varies with wavelength, must be determined. We define $s_{\star}$ and $s_{p}$ to be the distance to the system's photocenter from the star and planet respectively, as shown in Figure~\ref{schematicfig}, where the photocenter is marked with a ``$\times$'' symbol. We define the luminosity ratio at a given wavelength, $L_{r}$, as

\begin{equation}
 L_{r} = \frac{L_{p}}{L_{\star}}
\end{equation}

\noindent where $L_{p}$ is the luminosity of the planet, and $L_{\star}$ is the luminosity of the star. Thus, similar to the previously presented derivations, the values for $s_{\star}$ and $s_{p}$ are

\begin{equation}
 s_{\star} = \frac{a \cdot L_{r}}{L_{r}+1}
\end{equation}

\begin{equation}
 s_{p} = \frac{a}{L_{r} + 1}
\end{equation}

\noindent where by definition $s_{\star}$ + $s_{p}$ = $a$. The observed astrometric motion results from the movement of the system's photocenter around the system's barycenter. Thus, taking into account light from both the star and planet,

\begin{equation}
 \alpha = \arctan\left(\frac{r_{\star} - s_{\star}}{D}\right) = \arctan\left(\frac{s_{p} - r_{p}}{D}\right) 
\end{equation}

\noindent and thus

\begin{equation}
\label{alphaeq2}
 \alpha = \arctan\left(\frac{a \cdot (q - L_{r})}{D \cdot (q + 1) \cdot (L_{r} + 1)}\right)
\end{equation}

\noindent where we have defined $\alpha$ so that $\alpha > 0$ signifies that the star dominates the observed reflex motion, i.e., $L_{r} < q$, and $\alpha < 0$ signifies that the planet dominates the observed reflex motion, i.e., $L_{r} > q$. Note that when the barycenter and photocenter are at the same point, i.e., $L_{r} = q$, and thus $\alpha = 0$, no reflex motion is observable.

We now estimate the value of $L_{r}$ based upon the values of readily measurable system parameters. Light emitted from the planet consists of both thermally emitted light, as well as incident stellar light reflected off the planet. Thus,

\begin{equation}
 L_{r} = \frac{L_{E} + L_{A}}{L_{\star}} = \frac{L_{E}}{L_{\star}} + \frac{L_{A}}{L_{\star}}
\end{equation}

\noindent where $L_{E}$ is the luminosity of the planet from thermal emission, L$_{\star}$ is the luminosity of the star, and $L_{A}$ is the luminosity of light reflected off the planet. To estimate the thermal component, we assume that both the star and planet radiate as blackbodies, and thus

\begin{equation}
 \label{thermaleq}
 \frac{L_{E}}{L_{\star}} = \frac{R_{p}^{2}}{R_{\star}^{2}} \cdot \frac{exp(\frac{hc}{\lambda k T_{\star}}) - 1}{exp(\frac{hc}{\lambda k T_{p}}) - 1}
\end{equation}

\noindent where $R_{p}$ is the radius of the planet, $\lambda$ is a given wavelength, $h$ is Planck's constant, $c$ is the speed of light, $k$ is Boltzmann's constant, $T_{\star}$ is the effective temperature of the star, and $T_{p}$ is the effective temperature of the planet. To derive $T_{p}$ we first assume that the planet is in radiative equilibrium, and has perfect heat re-distribution, i.e. a uniform planetary temperature, and thus

\begin{equation}
\label{tpeq}
 T_{p} = T_{\star} \cdot \left(\frac{(1-A_{B}) \cdot R_{\star}^{2}}{4 a^{2}}\right)^{\frac{1}{4}}
\end{equation}

\noindent where $T_{\star}$ is the temperature of the star, $A_{B}$ is the planetary Bond albedo, and $R_{\star}$ is the radius of the star. 

To estimate the contribution due to reflected light, we first note that the flux received at the planet's surface is $L_{\star}$ divided by the surface area of a sphere at a distance $a$, i.e., $4\pi a^{2}$. The planet intercepts and reflects this light on only one of its hemispheres, which has effective cross sectional area of $\pi R_{p}^{2}$, with an efficiency equal to the albedo. Combining these terms and re-arranging to obtain the luminosity ratio due to reflected light yields

\begin{equation}
\label{refleq}
 \frac{L_{A}}{L_{\star}} = \frac{A_{\lambda} R_{p}^{2}}{4a^{2}}
\end{equation}

\noindent where $A_{\lambda}$ is the planet's albedo at a given wavelength.

Combining the above equations, and assuming values of $A_{B}$ and $A_{\lambda}$, we can estimate $\alpha$ at a given $\lambda$, using only $M_{\star}$, $R_{\star}$, $T_{\star}$, $M_{p}$, $R_{p}$, $P$, and $D$. We note that this assumes that the planet is in radiative equilibrium, but does not account for any additional internal heat sources from the planet, such as gravitational contraction or radioactive decay. While internal heat sources are likely to be negligible for close-in planets, it could significantly contribute to the total luminosity of further out gaseous planets, thus making them even more easily detectable. Our approximation for $\alpha$ also assumes that the planet's luminosity is constant over its orbit as observed from Earth. However, some planets have significant flux differences between their day and night sides due to low day-to-night re-radiation efficiency and/or significant planetary albedos. In these cases, if the inclination of the system is $\neq$ 0$\degr$, then the planet's luminosity will vary with orbital phase as seen by the observer, and the projected astrometric orbit of the photocenter at wavelengths where the planet's luminosity dominates will deviate from an ellipse, with increasing deviation as the inclination approaches 90$\degr$. (This effect is further discussed and illustrated in Section~\ref{numsec}.) As well, we assumed a circular orbit, and thus eccentric planets with varying levels of stellar insolation and temperature would have unique orbital signatures resulting from time-variant planetary flux. Finally, we assumed in this analytical derivation that the star and planet are effectively point sources, but of course in reality they have a physical size. If the star and/or planet have non-symmetric surface features, such as star spots or planetary hot spots, then the star and planet could each influence the location of the photocenter as these features rotated across their surface. The effect on the photocenter would only be a fraction of their physical radii, and would only cause significant deviations to the observed astrometric orbit if the radii of either object was a significant fraction of the object's distance from the system's barycenter. While this would likely be negligible for the planet, it could be significant for the star, e.g., the case of microarcsecond, wavelength-dependent, astrometric perturbations resulting from star spots presented in Appendix~\ref{sim2appendix}.

\begin{figure}[ht!]
\centering
\begin{tabular}{ccc}
\epsfig{width=0.3\linewidth,file=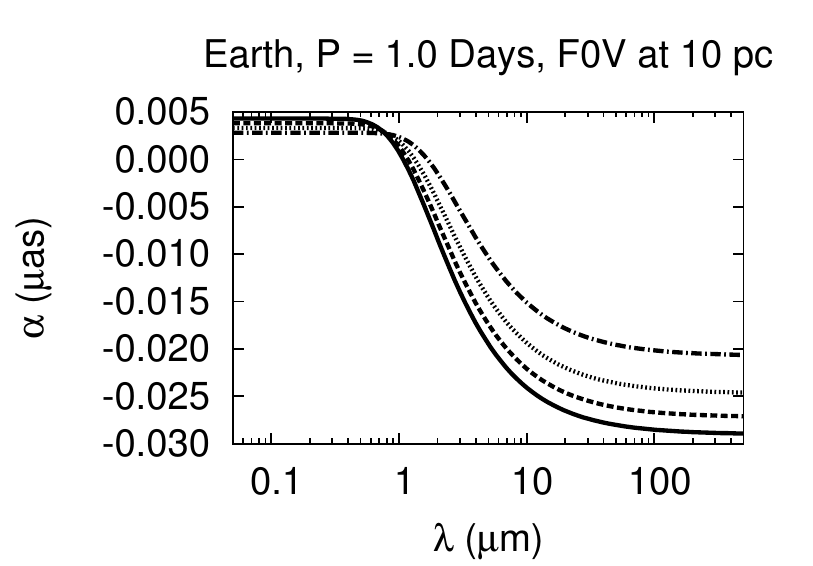} &
\epsfig{width=0.3\linewidth,file=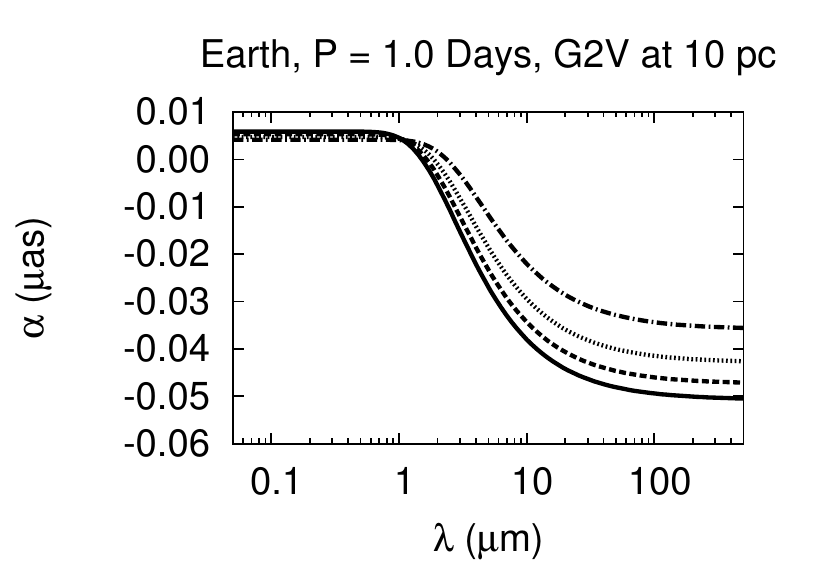} &
\epsfig{width=0.3\linewidth,file=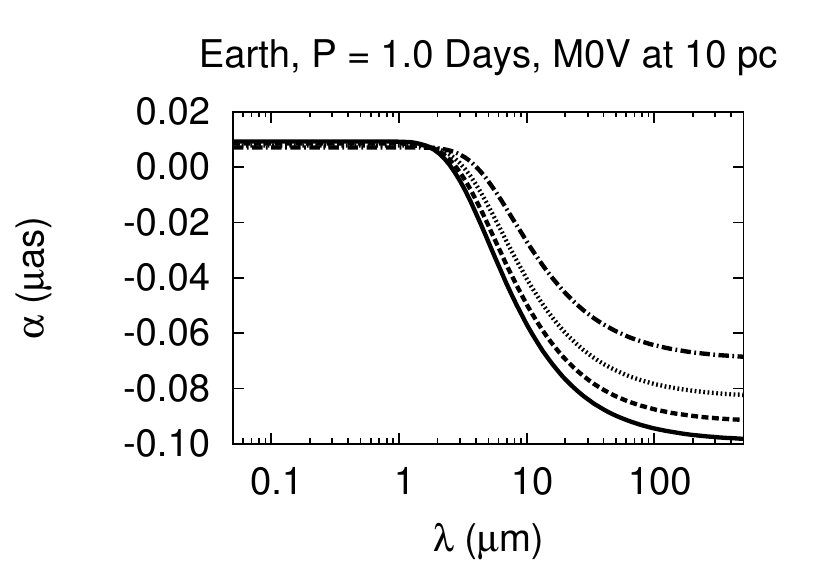} \\
\epsfig{width=0.3\linewidth,file=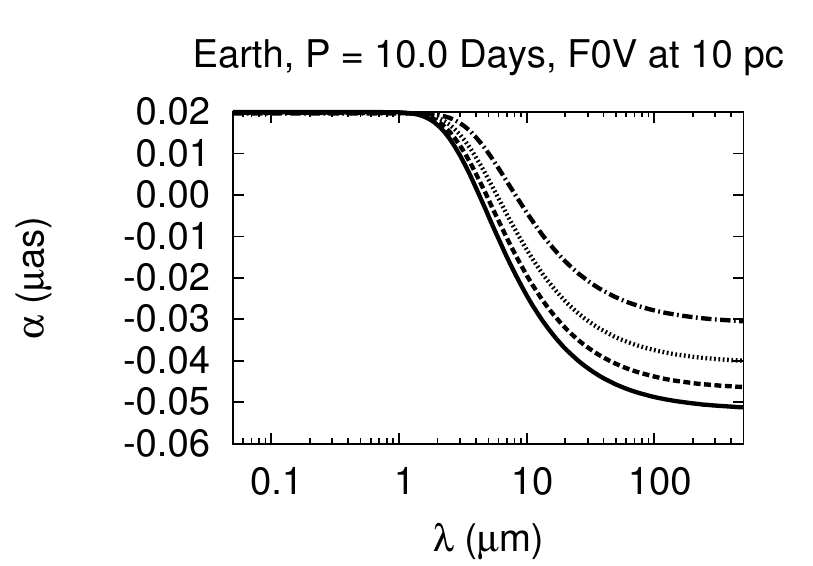} &
\epsfig{width=0.3\linewidth,file=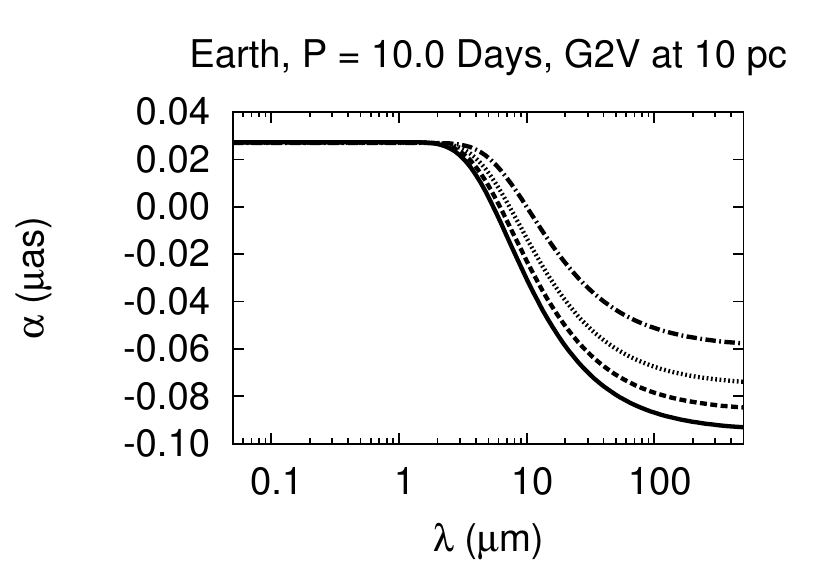} &
\epsfig{width=0.3\linewidth,file=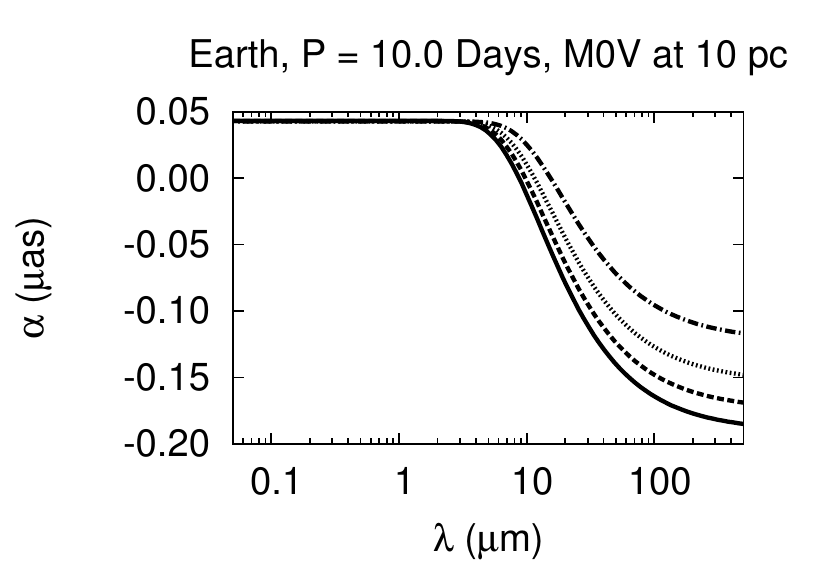} \\
\epsfig{width=0.3\linewidth,file=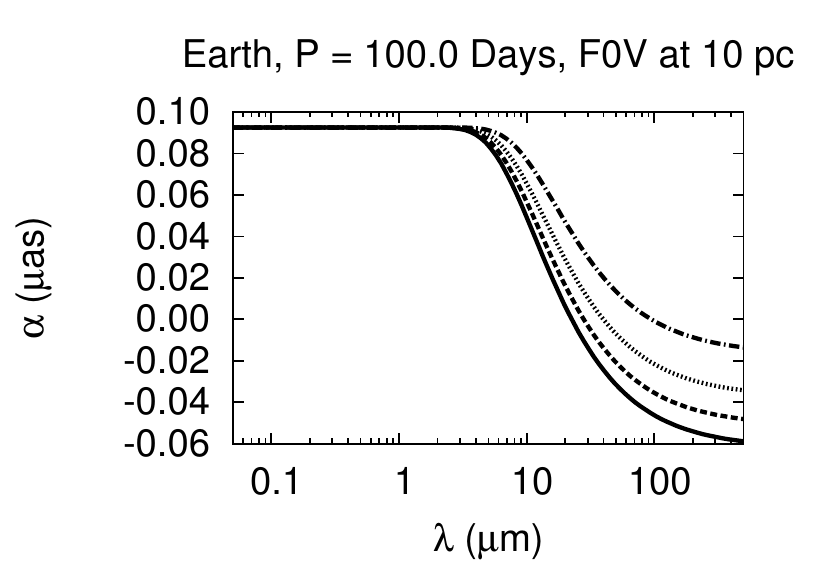} &
\epsfig{width=0.3\linewidth,file=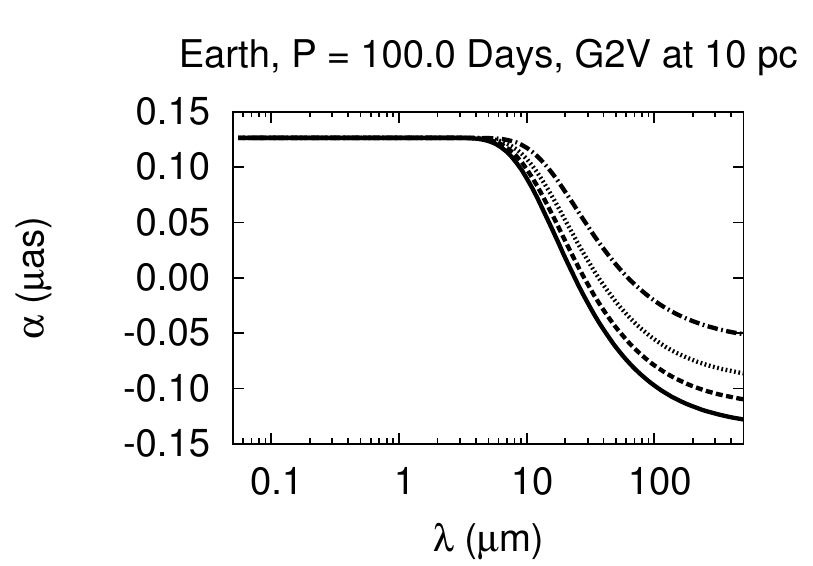} &
\epsfig{width=0.3\linewidth,file=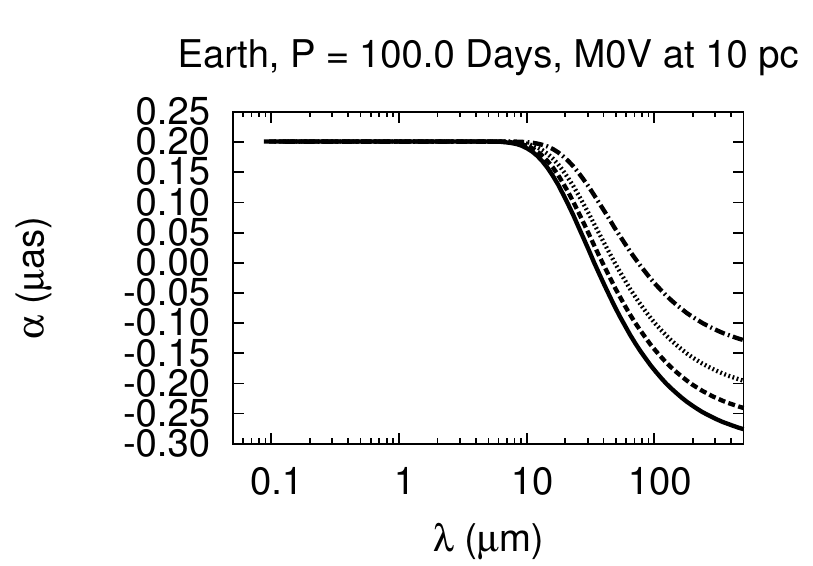} \\
\epsfig{width=0.3\linewidth,file=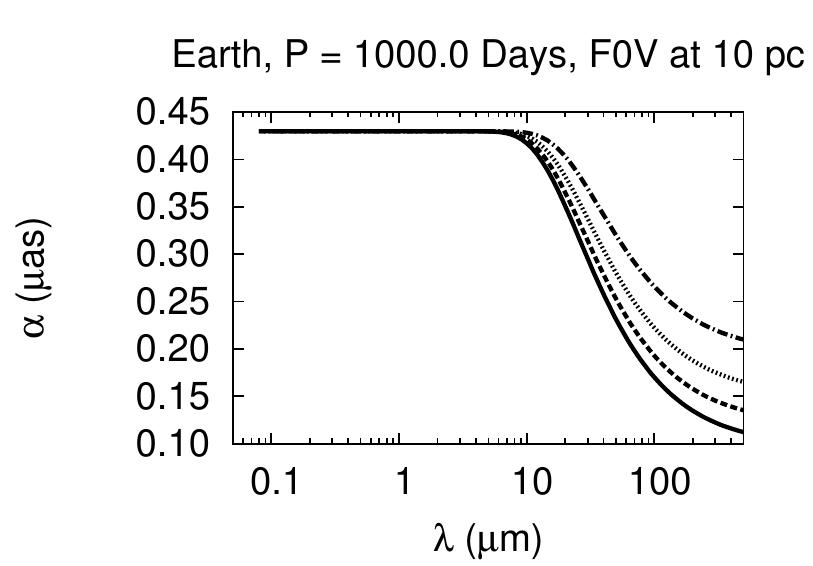} &
\epsfig{width=0.3\linewidth,file=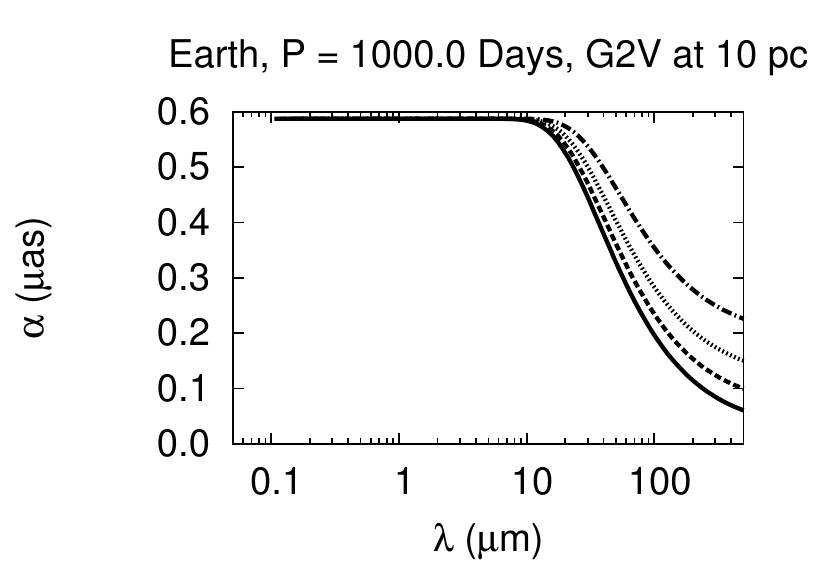} &
\epsfig{width=0.3\linewidth,file=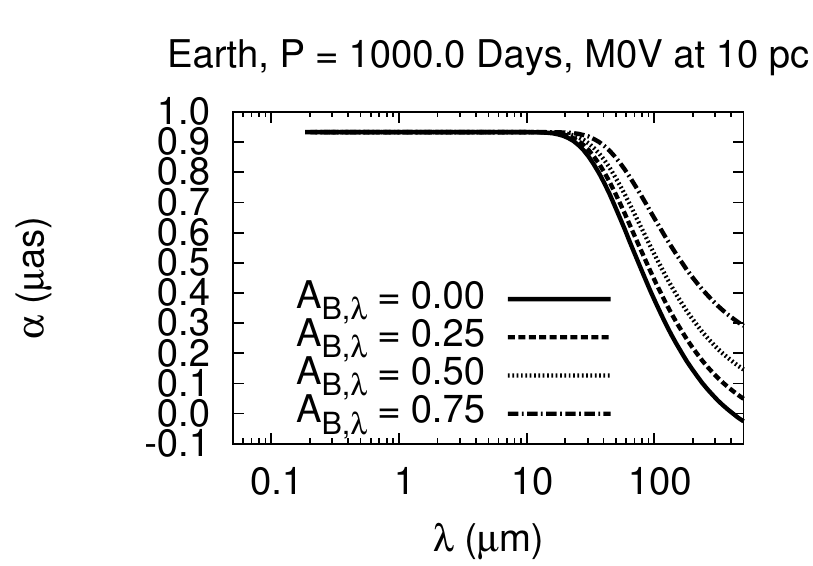} \\
\end{tabular}
\caption[Plots of the reflex motion amplitude versus the wavelength of observations for an Earth-like planet]{Plots of the reflex motion amplitude, $\alpha$, versus the wavelength of observations, $\lambda$, for an Earth-like planet, ($M_{p}$ = 1.0 $M_{\earth}$, $R_{p}$ = 1.0 $R_{\earth}$) around F0V, G2V, and M0V stars at 10 parsecs, (left, middle, and right columns respectively), at periods of 1, 10, 100, and 1000 days, (top to bottom rows, respectively). The solid, dashed, dotted, and dash-dotted lines represent planetary albedos of 0.0, 0.25, 0.5, and 0.75 respectively.}
\label{earthsimfig}
\end{figure}

In Figure~\ref{jupsimfig} we present plots of $\alpha$ versus $\lambda$ for a Jupiter-like planet, ($M_{p}$ = 1.0 $M_{J}$, $R_{p}$ = 1.0 $R_{J}$), around F0V, G2V, and M0V stars at 10 parsecs, with periods of 1, 10, 100, and 1000 days. We also show various planetary albedos, assuming $A_{B}$ = $A_{\lambda}$, of 0.0, 0.25, 0.5, and 0.75. In Figure~\ref{earthsimfig} we do the same for an Earth-like planet, ($M_{p}$ = 1.0 $M_{\earth}$, $R_{p}$ = 1.0 $R_{\earth}$). In general, systems that have large, high-mass stars and large, low-mass planets present the best opportunity to observe negative values of $\alpha$, and thus be able to directly determine their masses. (This is a unique parameter space not covered by other exoplanet characterization techniques such as radial-velocity or the transit method.) Short-period, and thus hot, planets around more massive stars transition to negative values of $\alpha$ at shorter wavelengths, but have lower overall amplitudes compared to long-period, and thus cool, planets around low-mass stars. Reflected light is a fairly minor contribution, only having some significant relevance for planets with very short orbital periods, i.e., $\sim$1 day. For both hot Jupiters and hot Earths, negative values of $\alpha$ can be observed with wavelengths as short as $\sim$2 $\mu$m, i.e., the $K$-band. Considering $\lambda$ $<$ 100 $\mu$m, $\alpha$ $<$ 0 could only be observed for $P$ $\lesssim$ 100 days for a Jupiter-like planet, and for $P$ $\lesssim$ 500 days for an Earth-like planet. Earth itself, (P = 365 days around a G2V star), would have a value of $\alpha$ $\approx$ 0.3 $\mu$as for $\lambda$ $<$ 10 $\mu$m, and $\alpha$ $\approx$ -0.05 $\mu$as at 100 $\mu$m, and thus, theoretically, the absolute mass of an Earth-analogue and its host star could be determined via this technique.

\begin{figure}[ht!]
\centering
\begin{tabular}{ccc}
\epsfig{width=0.3\linewidth,file=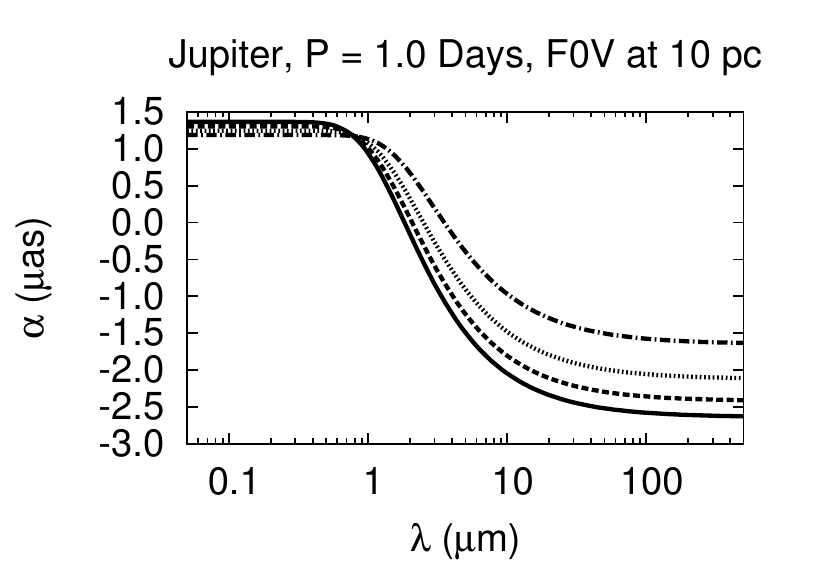} &
\epsfig{width=0.3\linewidth,file=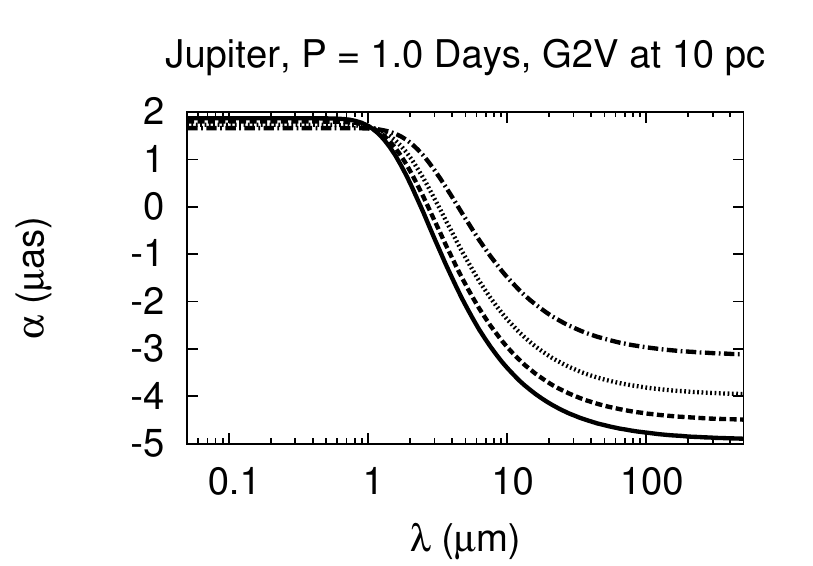} &
\epsfig{width=0.3\linewidth,file=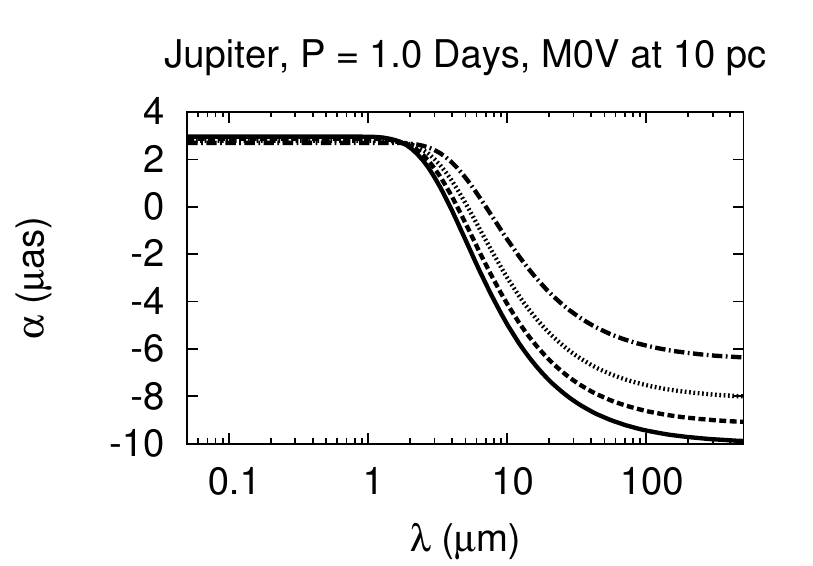} \\
\epsfig{width=0.3\linewidth,file=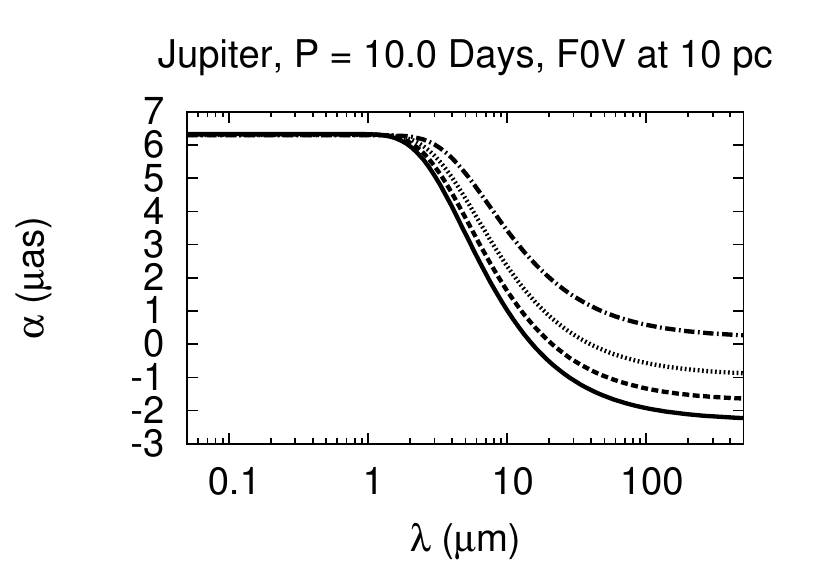} &
\epsfig{width=0.3\linewidth,file=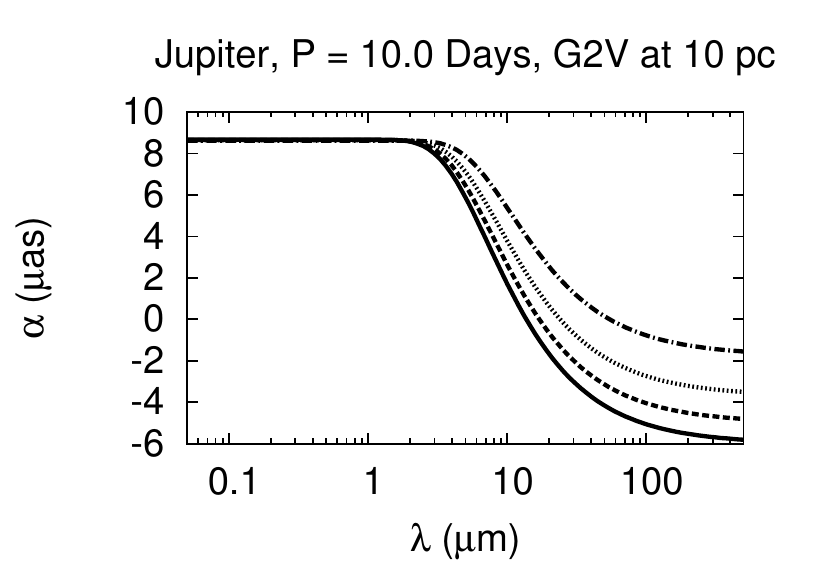} &
\epsfig{width=0.3\linewidth,file=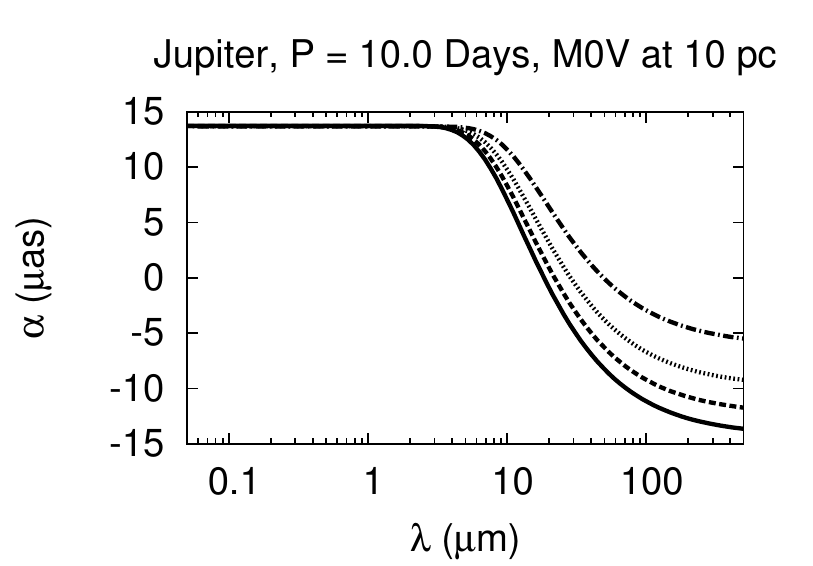} \\
\epsfig{width=0.3\linewidth,file=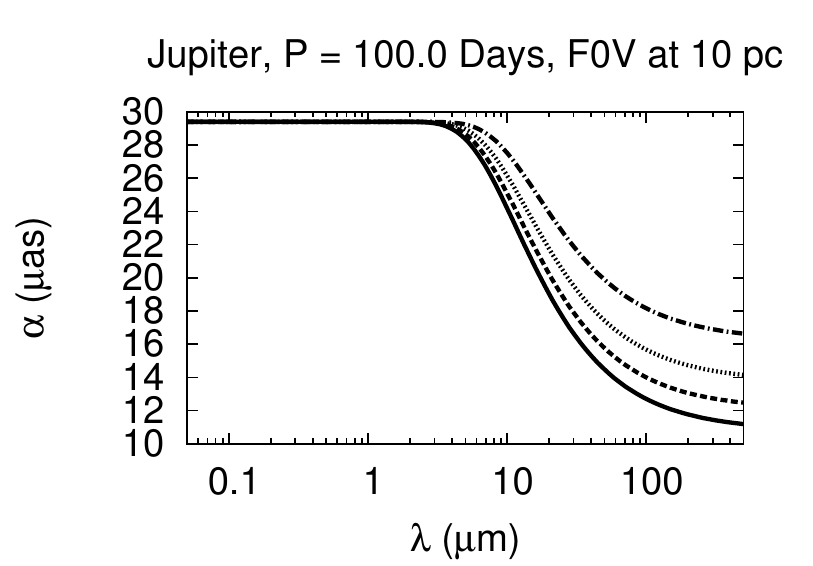} &
\epsfig{width=0.3\linewidth,file=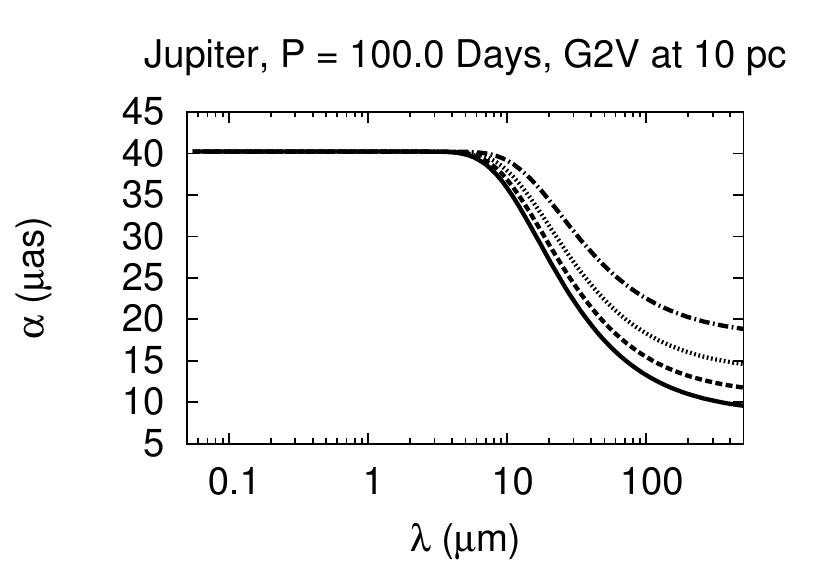} &
\epsfig{width=0.3\linewidth,file=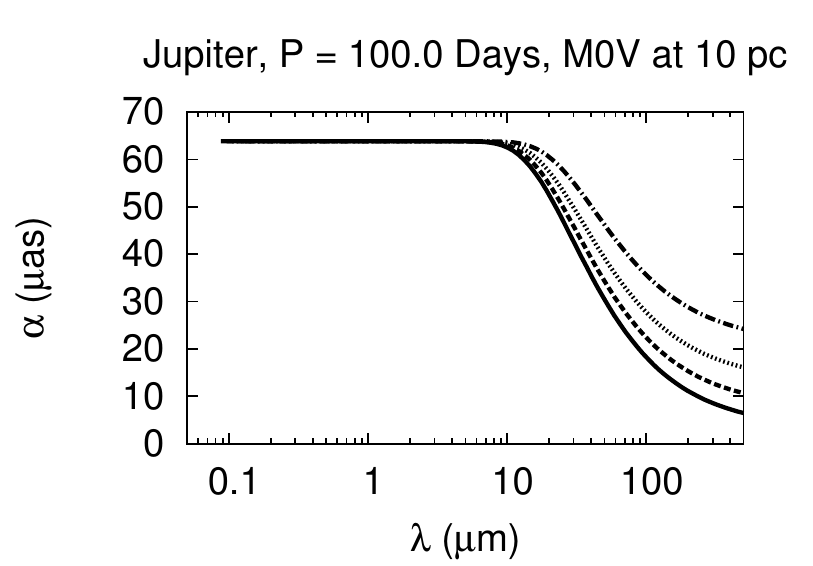} \\
\epsfig{width=0.3\linewidth,file=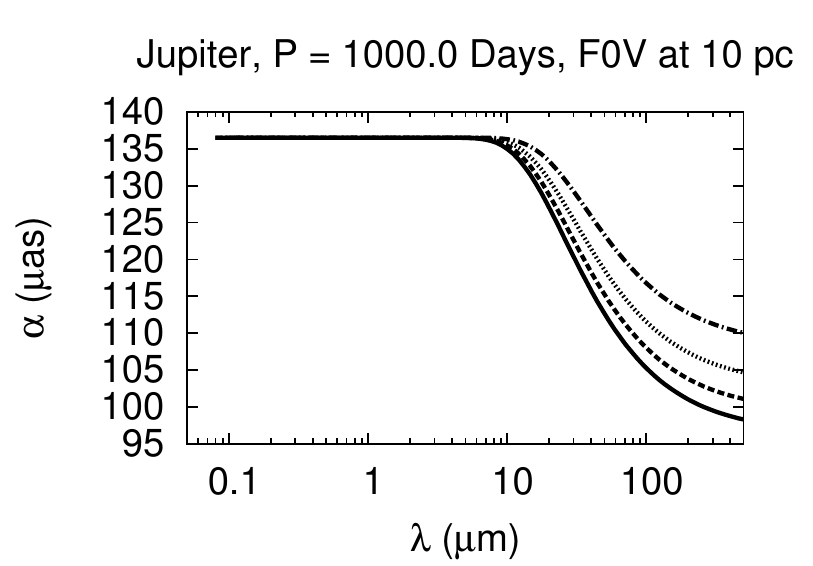} &
\epsfig{width=0.3\linewidth,file=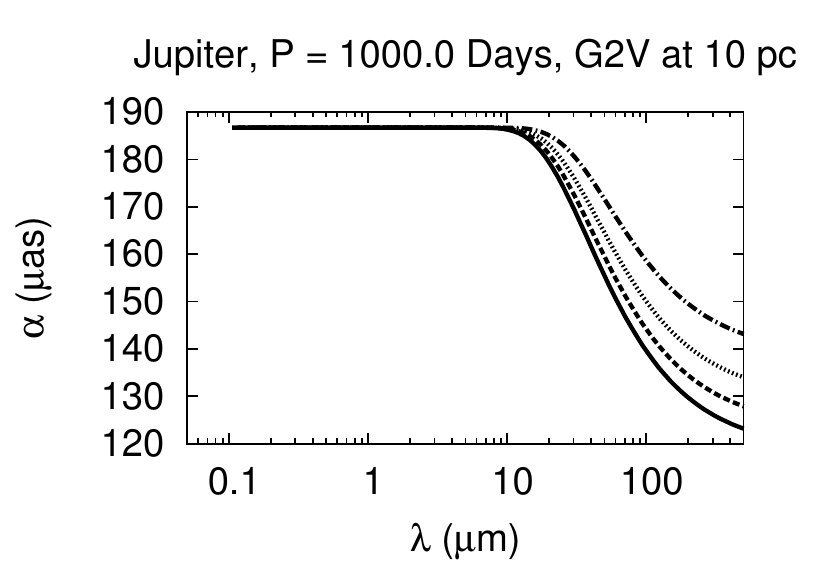} &
\epsfig{width=0.3\linewidth,file=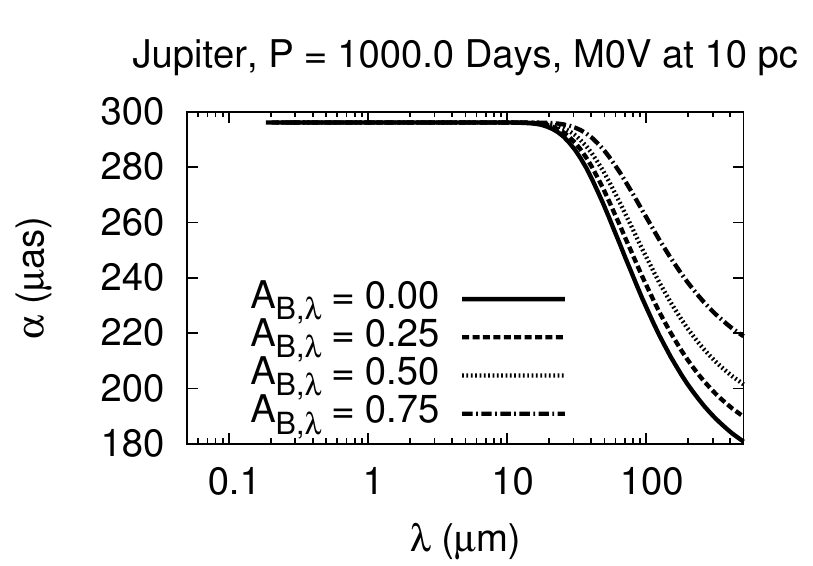} \\
\end{tabular}
\caption[Plots of the reflex motion amplitude versus the wavelength of observations for a Jupiter-like planet]{Plots of the reflex motion amplitude, $\alpha$, versus the wavelength of observations, $\lambda$, for a Jupiter-like planet, ($M_{p}$ = 1.0 $M_{J}$, $R_{p}$ = 1.0 $R_{J}$) around F0V, G2V, and M0V stars at 10 parsecs, (left, middle, and right columns respectively), at periods of 1, 10, 100, and 1000 days, (top to bottom rows, respectively). The solid, dashed, dotted, and dash-dotted lines represent planetary albedos of 0.0, 0.25, 0.5, and 0.75 respectively.}
\label{jupsimfig}
\end{figure}

Utilizing exoplanets.org, we have collected the values for all the previously mentioned system parameters for all currently known exoplanets. Selecting those that have well-determined values of all the needed parameters, in Table~\ref{tab1} we list the top five exoplanets with the largest negative values of $\alpha$ for each of the K (2.19 $\mu$m), L (3.45 $\mu$m), M (4.75 $\mu$m), and N (10.0 $\mu$m) infrared bandpasses, with a total of 11 unique exoplanets. We choose these wavelengths as they are the major ground-based infrared observing windows, and no systems examined had negative $\alpha$ values at wavelengths shorter than $\sim$2 $\mu$m. All of the candidate systems ended up being transiting planets both because they have well-determined values for the planetary radii, and transit surveys are most sensitive to close-in planets. As can be seen, the top candidates for detecting $\alpha$ $<$ 0, and thus measuring the absolute mass of the planet, are WASP-12 b in the $K$-band with $\alpha$ = -0.05 $\mu$as, HD 209458 b in the $L$ and $M$-bands with $\alpha$ = -0.23 and -0.66 $\mu$as respectively, and HD 189733 b in the $N$-band with $\alpha$ = -3.04 $\mu$as. It is interesting that three Neptune and sub-Neptune mass planets, 55 Cnc e, Gliese 436 b, and GJ 1214 b, also make the list, showing that this technique `favors' the characterization of low-mass planets.

\begin{deluxetable}{ccccccccc}
  \tablewidth{0pt}
  \tablecaption{Currently Known Exoplanets with the Most Negative $\alpha$ Values}
  \tabletypesize{\small}
  \tablecolumns{9}
  \tablehead{Name & $D$ & $M_{\star}$ & $R_{\star}$ & $T_{\star}$ & $M_{p}$ & $R_{p}$ & $P$ & $\alpha$ \\ & (pc) & (M$_{\sun}$) & (R$_{\sun}$) & (K) & (M$_{\rm J}$) & (R$_{\rm J}$) & (Days) & ($\mu$as)}
  \startdata
  \cutinhead{$K$-Band (2.19 $\mu$m)}
  \input{sim3-tab1a.tex}
  \cutinhead{$L$-Band (3.45 $\mu$m)}
  \input{sim3-tab1b.tex}
  \cutinhead{$M$-Band (4.75 $\mu$m)}
  \input{sim3-tab1c.tex}
  \cutinhead{$N$-Band (10.0 $\mu$m)}
  \input{sim3-tab1d.tex}
  \enddata
  \label{tab1}
\end{deluxetable}

\subsection{Numerical Modeling via \sc{reflux}}
\label{numsec}

In order to provide a check on our analytical formulae, better illustrate the multi-wavelength astrometric orbits of exoplanet systems, and probe some more subtle effects, we use the \textsc{reflux}\footnote{\textsc{reflux} can be run via a web interface from \url{http://astronomy.nmsu.edu/jlcough/reflux.html}. Additional details as to how to set-up a model are presented there.} code \citep{Coughlin2010a}, which computes the flux-weighted astrometric reflex motion of binary systems at multiple wavelengths, to model a couple known exoplanet systems. We discuss the code in detail in Appendices~\ref{sim1appendix} and \ref{sim2appendix}, but in short, it utilizes the Eclipsing Light Curve (ELC) code, which was written to compute light curves of eclipsing binary systems \citep{Orosz2000}. ELC includes the dominant physical effects that shape a binary system's light curve, such as non-spherical geometry due to rotation and tidal forces, gravity darkening, limb darkening, mutual heating, reflection effects, and the inclusion of hot or cool spots on the stellar surface. The ELC code represents the surfaces of two stars, or a star-planet system, as a grid of individual luminosity points, and calculates the resulting light curve given the provided systemic parameters. \textsc{reflux} takes the grid of luminosity points at each phase and calculates the flux-weighted astrometric photocenter location at each phase, taking into account the system's distance from Earth. Although ELC is capable of using model atmospheres, for this chapter we set the code to calculate luminosities assuming both the star and planet radiate as blackbodies.

We choose to model Wasp-12, HD 209458, and HD 189733, as they are all well-studied systems, and have the most negative $\alpha$ values for the $K$, $L$, $M$, and $N$ bandpasses presented in Table~\ref{tab1}. For each system we set the values for $M_{\star}$, $R_{\star}$, $T_{\star}$, $M_{p}$, $R_{p}$, $P$, $D$, and rotation period of the star to those in the Exoplanets.org database, and set the rotation period of the planet to the orbital period of the system, i.e., assume the planet is tidally locked, and assume a circular orbit. We assume that the spin axes of both the star and planet are perfectly aligned with the orbital axis. We employ the use of spots in the ELC code to simulate a day/night side temperature difference, by assuming a uniform day-side temperature for the planetary hemisphere facing the star, and a uniform night-side temperature for the planetary hemisphere facing away from the star. We employ the values for the day and night side temperatures derived by \citet{Cowan2011}, which were 2939 K for the day-side of Wasp-12 b, 1486 and 1476 K for the day and night sides respectively of HD 209458 b, and 1605 and 1107 K for the day and night sides respectively of HD 189733 b. We adopted a temperature of 1470 K for the night side of Wasp-12 b, i.e., half that of the day side, assuming very little planetary heat redistribution. For all the systems, we set the star's gravity darkening coefficients to those determined by \citet{Claret2000a}, though do not enable gravity darkening for the planet. For both the planet and star, we assume zero albedo, since we are dealing principally with infrared wavelengths where the effect is negligible, and we have already shown that even in the optical reflected light is a minor contribution to the astrometric motions under investigation. Furthermore, the chosen planets are expected to have very low albedos ($A_{B} <$ 0.3) from model atmospheres \citep{Marley1999,Seager2000,Sudarsky2000}, and have even had their albedos constrained to very low values from observations, e.g., \citet{LopezMorales2010} for Wasp-12b, \citet{Rowe2008} for HD 209458b, and \citet{Wiktorowicz2009} for HD 189733b. We also do not assume any limb-darkening since we are dealing principally with infrared wavelengths.

\begin{figure}[h]
\centering
\begin{tabular}{cc}
\epsfig{width=0.475\linewidth,file=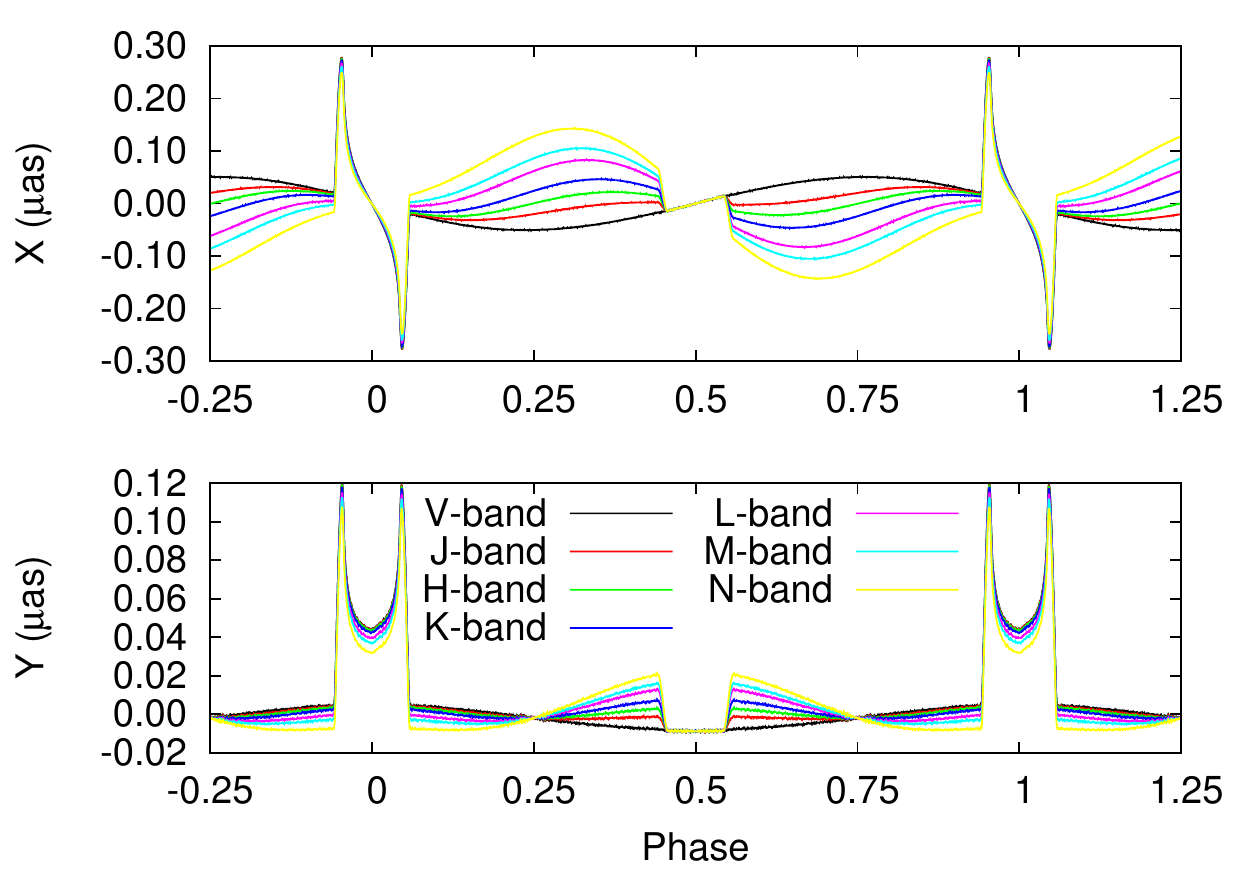} &
\epsfig{width=0.475\linewidth,file=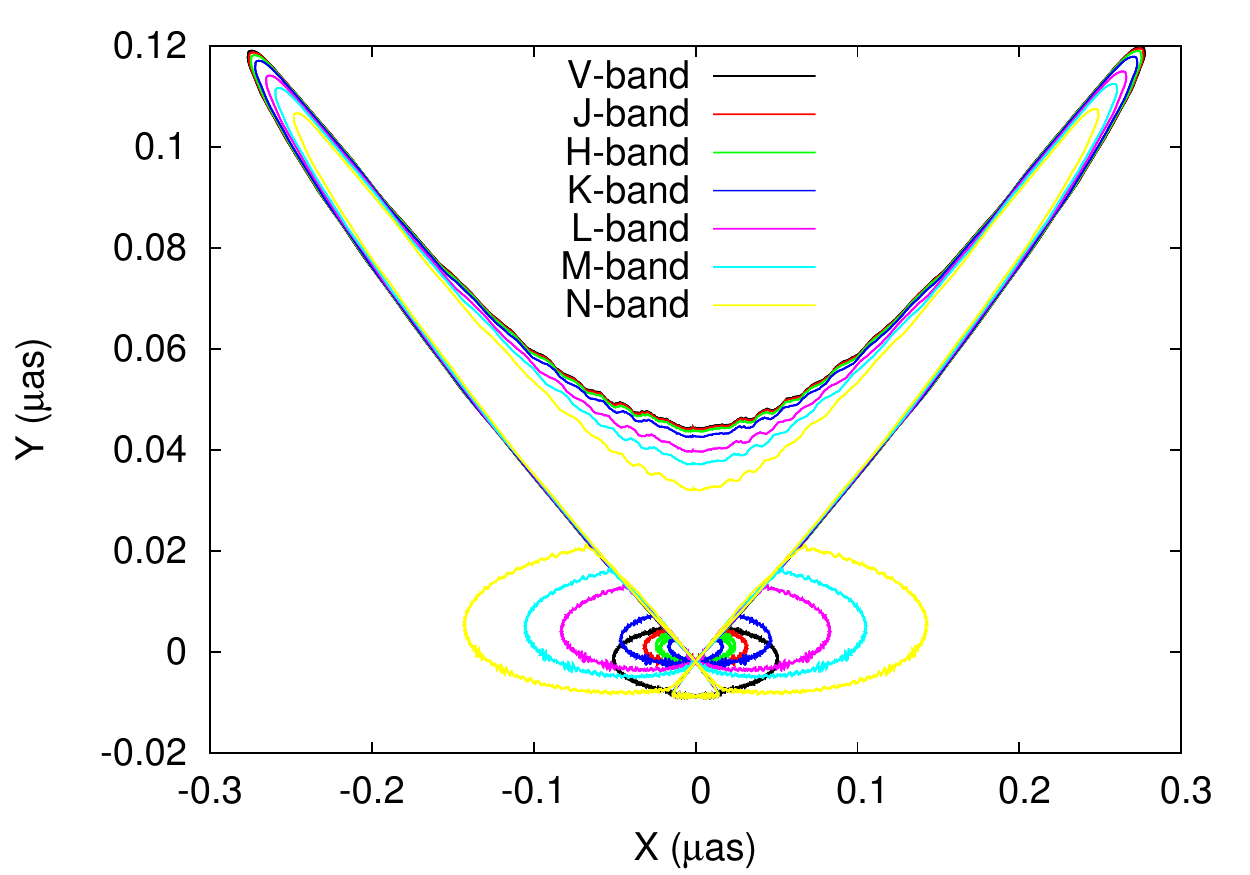} \\
\end{tabular}
\caption[Plots of the multi-wavelength astrometric orbit for the Wasp-12 system]{Plots of the multi-wavelength astrometric orbit for the Wasp-12 system. Left: The X and Y components of motion versus phase. Right: The sky-projected, X-Y, orbit. The point (X,Y) = (0,0) corresponds to the system's barycenter, and the projected orbital rotation axis is parallel to the Y-axis. Phase 0.0 corresponds to the primary transit, when the planet passes in front of the star and is closest to the observer, and phase 0.5 corresponds to the secondary eclipse, when the planet passes behind the star and is farthest away from the observer.}
\label{refluxmodel1}
\end{figure}

\begin{figure}[ht]
\centering
\begin{tabular}{cc}
\epsfig{width=0.475\linewidth,file=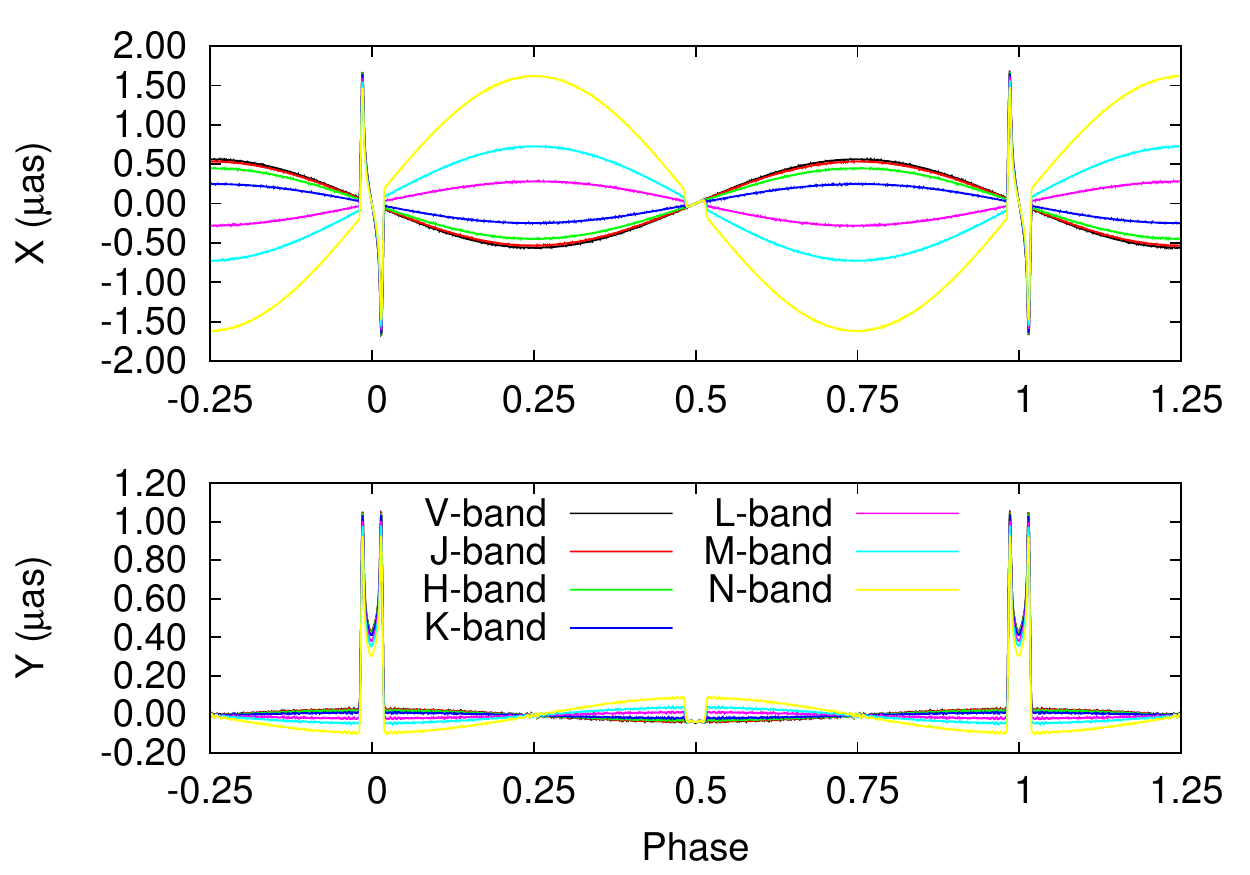} &
\epsfig{width=0.475\linewidth,file=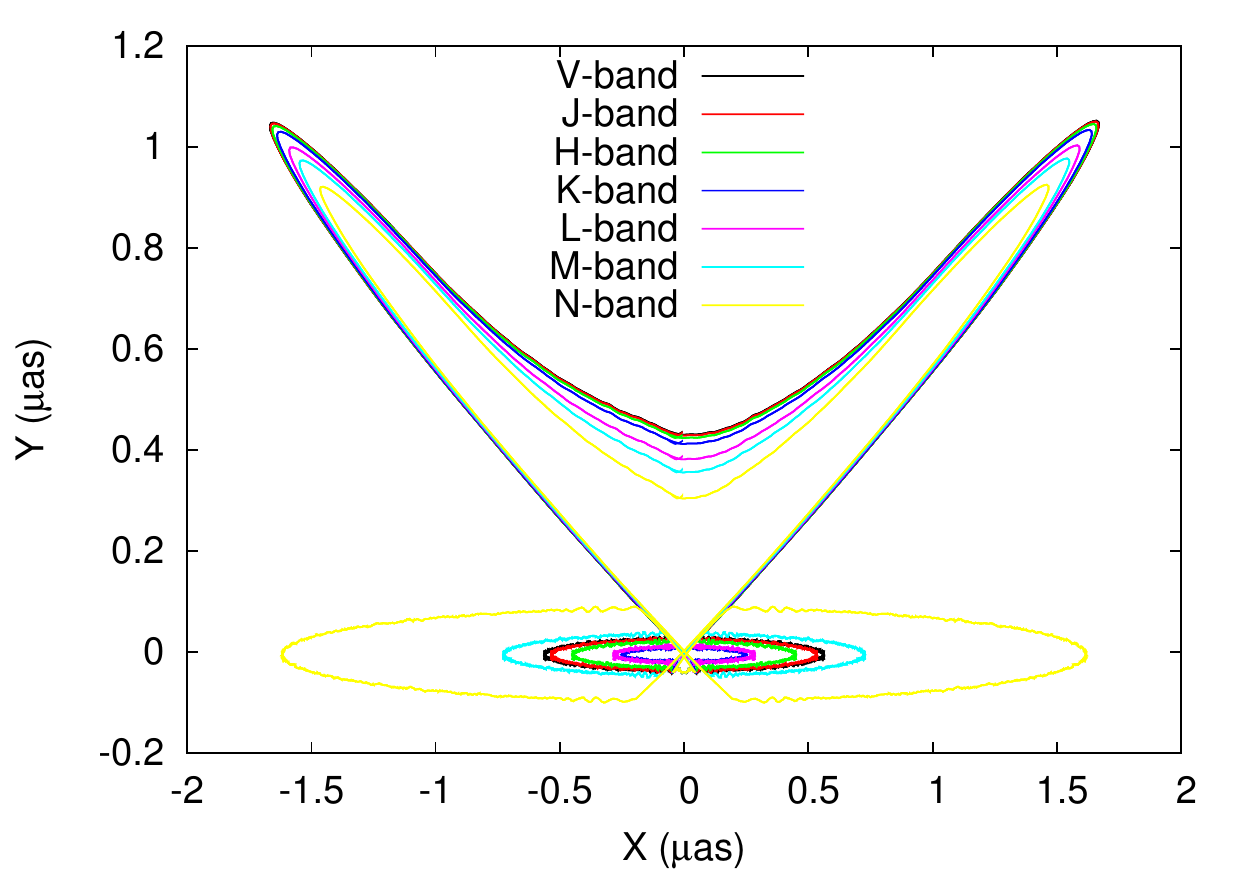} \\
\end{tabular}
\caption[Plots of the multi-wavelength astrometric orbit for the HD 209458 system]{Plots of the multi-wavelength astrometric orbit for the HD 209458 system. Left: The X and Y components of motion versus phase. Right: The sky-projected, X-Y, orbit. The point (X,Y) = (0,0) corresponds to the system's barycenter, and the projected orbital rotation axis is parallel to the Y-axis. Phase 0.0 corresponds to the primary transit, when the planet passes in front of the star and is closest to the observer, and phase 0.5 corresponds to the secondary eclipse, when the planet passes behind the star and is farthest away from the observer.}
\label{refluxmodel2}
\end{figure}

\begin{figure}[ht]
\centering
\begin{tabular}{cc}
\epsfig{width=0.475\linewidth,file=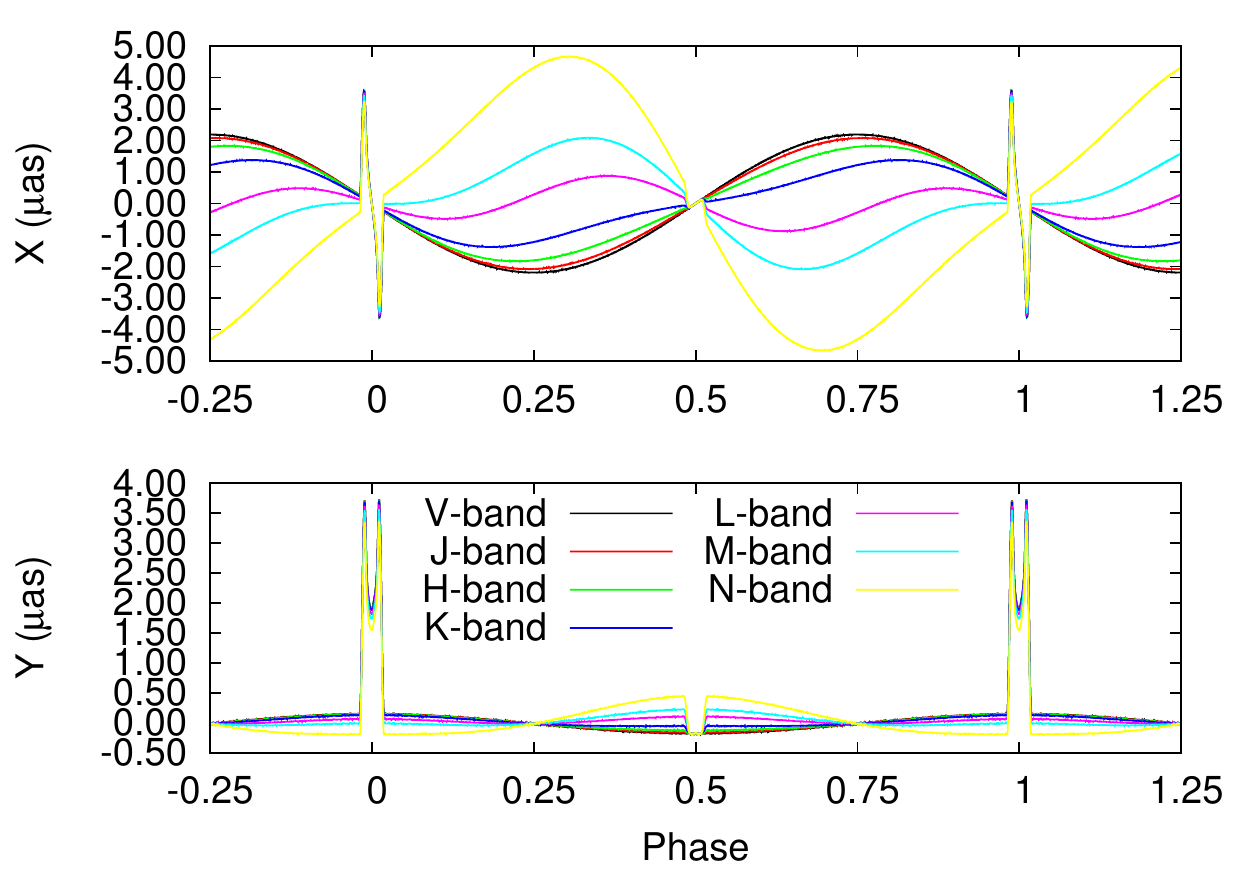} &
\epsfig{width=0.475\linewidth,file=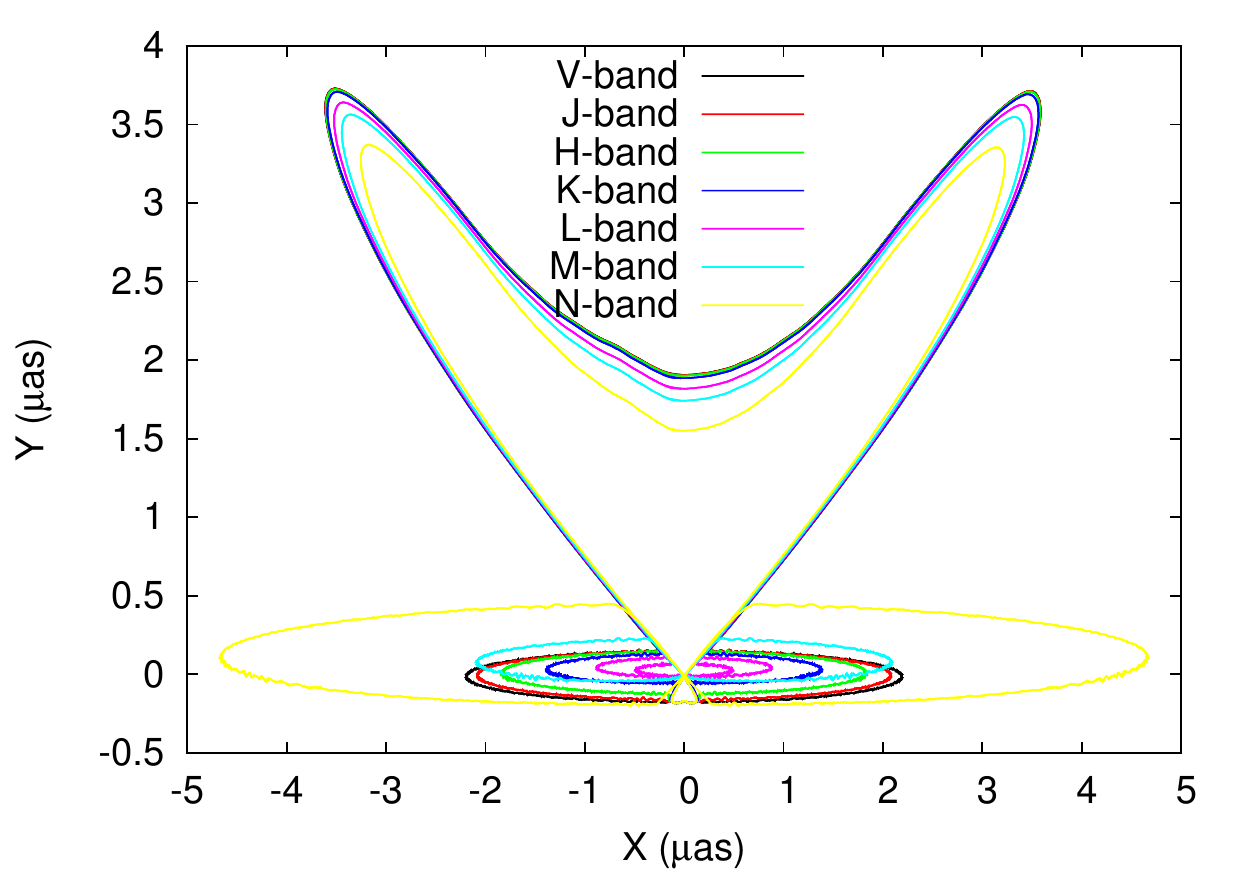} \\
\end{tabular}
\caption[Plots of the multi-wavelength astrometric orbit for the HD 189733 system]{Plots of the multi-wavelength astrometric orbit for the HD 189733 system. Left: The X and Y components of motion versus phase. Right: The sky-projected, X-Y, orbit. The point (X,Y) = (0,0) corresponds to the system's barycenter, and the projected orbital rotation axis is parallel to the Y-axis. Phase 0.0 corresponds to the primary transit, when the planet passes in front of the star and is closest to the observer, and phase 0.5 corresponds to the secondary eclipse, when the planet passes behind the star and is farthest away from the observer.}
\label{refluxmodel3}
\end{figure}

In Figures~\ref{refluxmodel1}, \ref{refluxmodel2}, and \ref{refluxmodel3} we present plots of the X and Y components of the photocenter versus phase, as well as the sky-projected X-Y orbit of the photocenter, in the $V$, $J$, $H$, $K$, $L$, $M$, and $N$ passbands, for Wasp-12, HD 209458, and HD 189733 respectively. The point (X,Y) = (0,0) corresponds to the barycenter of the system, and the projected orbital rotation axis is parallel to the Y-axis. Phase 0.0 corresponds to the primary transit, when the planet passes in front of the star and is closest to the observer, and phase 0.5 corresponds to the secondary eclipse, when the planet passes behind the star and is farthest away from the observer.

Examining the modeling results, the values for $\alpha$ determined via the analytical formulae appear to match the numerical modeling results fairly well. For example, via Table~\ref{tab1}, Wasp-12b was predicted to have $\alpha$ values of -0.05 and -0.10 $\mu$as in the $K$ and $L$-bands respectively, compared to the maximum, out-of-transit, numerical model results of -0.05 and -0.08 $\mu$as. For HD 209458b, expected $\alpha$ values were -0.23, -0.66, and -1.53 $\mu$as for the $L$, $M$, and $N$-bands, compared to -0.30, -0.74, and -1.63 $\mu$as from the numerical models. For HD 189733b, expected $\alpha$ values were -0.47 and -3.04 $\mu$as for the $M$ and $N$-bands, compared to -2.10 and -4.68 $\mu$as from the numerical models. The differences are principally due to the use of observationally determined day and night side temperatures in the numerical models, whereas the analytical formulae assumed perfect radiative equilibrium and a uniform planetary temperature.

Although a transition from positive to negative $\alpha$ appears to occur around the $H$, $K$, and $L$-bands for Wasp-12 b, HD 209458 b, and HD 189733 respectively, a deviation from the visible light signature is clearly visible at shorter wavelengths, and thus it may be possible to disentangle the astrometric motion due to the planet even at shorter wavelengths where it does not dominate the reflex motion of the photocenter. For Wasp-12 b and HD 189733 b the out of transit/eclipse signature deviates from a sinusoid due to the extreme day/night temperature differences on these planets, as discussed in Section~\ref{eqsec}. The different inclinations of the systems are immediately apparent in the X-Y orbit plots, and when actually measured on sky, would directly yield the three-dimensional orbit of the system.

The presence of the primary transit and secondary eclipse is clearly visible in all three cases, with the primary transit dominating the maximum amplitude of the astrometric shift for the visible wavelengths, particularly in the Y-direction. As no limb-darkening was assumed in these models, the variation in the primary and secondary eclipse signatures with wavelength is due to the relative flux of the star and planet in those passbands. As noted by \citet{Gaudi2010}, measuring the astrometric shift of the primary transit directly yields the angular radius of the host star, and if the distance to the system is precisely known, one can directly derive the physical radius of the star. Additionally, if the density of the star is directly determined from the photometric light curve \citep{SeagerOrnelas2003}, then one can also directly derive the mass of the star. We also note, for the first time, that measuring the astrometric signature of the primary transit and, if observing at longer wavelengths, the secondary eclipse, specifically the duration of ingress and egress, similarly directly yields the angular radius of the planet. Since one may directly determine the surface gravity of the planet from the photometric light and radial-velocity curves alone \citep*{Southworth2007}, one may also directly determine the mass of the planet. Thus, for transiting planets, multi-wavelength astrometric measurements yield two independent methods of measuring the physical stellar and planetary masses.

\subsection{Discussion and Summary}
\label{discusssec}

We have shown that the multi-wavelength astrometric measurements of exoplanetary systems can be used to directly determine the masses of extrasolar planets and their host stars, in addition to the inclination and spatial orientation of their orbital axis. If the planet happens to transit the host star, then the angular radius of both the star and planet can be directly determined, and when combined with the trigonometric parallax of the system, the absolute radii of the planet and host star can directly determined via astrometry alone. We found that this technique is best suited, though is certainly not limited to, large, low-mass planets that orbit large, high-mass stars, and thus covers a unique parameter space not usually covered by other exoplanet characterization techniques. 

We have provided analytical formulae and numerical models to estimate the amplitude of the photocenter motion at various wavelengths. We found that, for some systems, the planet can dominate the motion of the system's photocenter at wavelengths as short as $\sim$2 $\mu$m, though the amplitude of the effect is only $\sim$0.05 $\mu$as. If one is able to obtain astrometric measurements at wavelengths up to 10 $\mu$m, then the motion of the photocenter due to the planet could be as high as several microarcseconds, and can often be of a much larger magnitude than seen at optical wavelengths when the photocenter motion is due solely to stellar motion.

We performed numerical modeling of several exoplanet systems via the {\sc reflux} code, and found it to be consistent with the predictions of our analytical model. The numerical modeling revealed that, even at shorter wavelengths where $\alpha$ $>$ 0, the planet has a visible impact on the observed astrometric orbit of the system. As well, deviations from pure sinusoidal motions due to day-night flux differences are clearly visible, and thus multi-wavelength astrometry could probe planetary properties of albedo and heat redistribution efficiency.

One caveat when working to extract the planetary and stellar masses from actual observations is that one will likely need to either precisely know the luminosity ratio of the system, or make assumptions about the luminosity of the planet, e.g., it radiates as a blackbody and is in thermal equilibrium. It may be possible that other observations could yield this information, such as the secondary eclipse depth if the planet happens to transit. The remaining parameters of the system's distance and period should be well determined via other methods such as microarcsecond precision parallax and radial-velocity or photometric light curves.

For the prospects of detection, it is clear that this effect will probably not be detected in the very near-term. Although astrometric measurements are approaching 1 $\mu$as accuracy, they have not yet been performed. Much of the ground-based work is being focused on the optical and $K$-bands, where in the latter the effect is just barely detectable. The development of microarcsecond precision astrometric systems in the mid-infrared, or sub-microarcsecond precision in the near-infrared, are clearly needed, and the methods presented here will serve to preselect the best planetary system candidates to be observed by those systems.

The work presented in this chapter assumed that both the star and planet radiate as blackbodies, however it is known that both can significantly deviate from that assumption, especially in the near infrared \citep[e.g.,][]{Gillon2009,Rogers2009,Gibson2010,Croll2011,deMooij2011,Coughlin2012a}. At the extreme end, \citet{Swain2010} and \citet{Waldmann2012} recently found evidence for a very large non-LTE emission feature around 3.25 $\mu$m in the atmosphere of HD 189733 b\footnote{We note that \citet{Mandell2011} reported a non-detection of a portion of this feature between the publications of \citet{Swain2010} and \citet{Waldmann2012}.}. Although via blackbody approximations we calculate that the planet-to-star flux ratio should be 8.3$\times$10$^{-4}$, \citet{Swain2010} and \citet{Waldmann2012} measure the 3.25 $\mu$m emission feature to be $\sim$8.5$\times$10$^{-3}$ times the stellar flux, or about ten times greater than expected. Assuming blackbody emission, the expected value for $\alpha$ for this system at visible wavelengths is 2.15 $\mu$as, and at 3.25 $\mu$m is 0.83 $\mu$as. If the emission feature is real however, the expected value for $\alpha$ at 3.25 $\mu$m is a very large -11.3 $\mu$as, dominated due to the planetary motion. Thus, the key in performing these types of observations may be to select particular wavelengths where the planets are unusually bright.

Finally, although we did not assume any limb-darkening in our models since we were examining near to mid-infrared wavelengths, limb-darkening will be significant when observed at different optical bandpasses. The astrometric signature of transiting planets will vary greatly due to limb-darkening in the optical regime, and thus multi-wavelength astrometry of transiting planets may be used to explore the limb-darkening profiles of stars, or visa versa, stellar limb-darkening may need to be precisely understood in order to extract planetary and stellar parameters of interest.

%% file: sim3-tab1a.tex
WASP-12 b & 427 & 1.28 & 1.63 & 6300 & 1.35 & 1.79 & 1.091 & -0.05 \\ 
WASP-19 b & 250 & 0.93 & 0.99 & 5500 & 1.11 & 1.39 & 0.789 & -0.05 \\ 
WASP-33 b & 115 & 1.50 & 1.44 & 7430 & 2.05 & 1.50 & 1.220 & -0.04 \\ 
55 Cnc e & 12 & 0.96 & 0.96 & 5234 & 0.03 & 0.19 & 0.737 & -0.01 \\ 
CoRoT-1 b & 480 & 0.95 & 1.11 & 5950 & 1.03 & 1.49 & 1.509 & -0.01 \\

%% file: sim3-tab1b.tex
HD 209458 b & 49 & 1.13 & 1.16 & 6065 & 0.69 & 1.36 & 3.525 & -0.23 \\ 
WASP-33 b & 115 & 1.50 & 1.44 & 7430 & 2.05 & 1.50 & 1.220 & -0.20 \\ 
WASP-19 b & 250 & 0.93 & 0.99 & 5500 & 1.11 & 1.39 & 0.789 & -0.15 \\ 
WASP-17 b & 300 & 1.19 & 1.20 & 6550 & 0.49 & 1.51 & 3.735 & -0.11 \\ 
WASP-12 b & 427 & 1.28 & 1.63 & 6300 & 1.35 & 1.79 & 1.091 & -0.10 \\ 

%% file: sim3-tab1c.tex
HD 209458 b & 49 & 1.13 & 1.16 & 6065 & 0.69 & 1.36 & 3.525 & -0.66 \\ 
HD 189733 b & 19 & 0.81 & 0.76 & 5040 & 1.14 & 1.14 & 2.219 & -0.47 \\ 
WASP-33 b & 115 & 1.50 & 1.44 & 7430 & 2.05 & 1.50 & 1.220 & -0.29 \\ 
WASP-19 b & 250 & 0.93 & 0.99 & 5500 & 1.11 & 1.39 & 0.789 & -0.21 \\ 
WASP-17 b & 300 & 1.19 & 1.20 & 6550 & 0.49 & 1.51 & 3.735 & -0.19 \\ 

%% file: sim3-tab1d.tex
HD 189733 b & 19 & 0.81 & 0.76 & 5040 & 1.14 & 1.14 & 2.219 & -3.04 \\ 
HD 209458 b & 49 & 1.13 & 1.16 & 6065 & 0.69 & 1.36 & 3.525 & -1.53 \\ 
Gliese 436 b & 10 & 0.45 & 0.46 & 3684 & 0.07 & 0.38 & 2.644 & -0.95 \\ 
WASP-34 b & 120 & 1.01 & 0.93 & 5700 & 0.58 & 1.22 & 4.318 & -0.64 \\ 
GJ 1214 b & 12 & 0.16 & 0.21 & 3026 & 0.02 & 0.24 & 1.580 & -0.59 

%% file: appendixA.tex
\begin{singlespace}
\section{\MakeUppercase{Eclipse Phase Dispersion Minimization (EPDM)}}
\label{epdmappendsec}
\end{singlespace}

\begin{figure}[ht!]
\centering
\epsfig{width=\linewidth,file=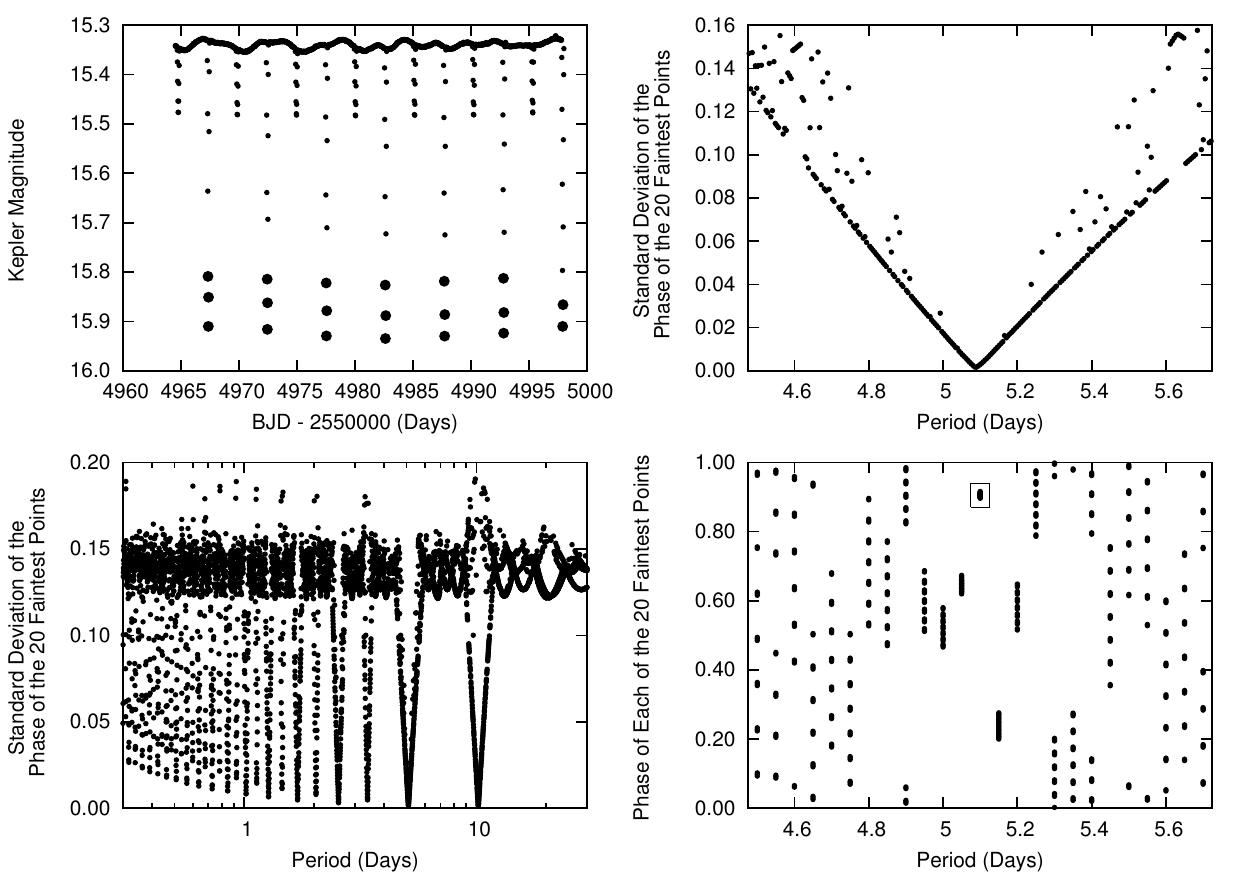}
\caption[Illustration of the EPDM technique]{Illustration of the EPDM technique. Top-left: The unphased light curve of Kepler 006591789, with the 20 faintest points highlighted by using larger point sizes. Bottom-left: Standard deviation of the phase values of the 20 faintest points versus period for this system. As can be seen, the standard deviation approaches 0.0 at $\sim$5.1 days, and integer multiples and fractions thereof. Top-Right: The same plot as in the bottom-left panel, but with the period range restricted to show only the period with the lowest standard deviation, and true period of the system. Bottom-Right: The actual phase values for each of the 20 faintest points at multiple periods, spanning the same period range, (but with a lower period resolution, for clarity), as the plot in the top-right panel. As can be seen, as the examined period approaches the true period of the system, the phase values of the 20 faintest points strongly clump together, producing a very small standard deviation. The best period is highlighted by a box in the lower-right panel.}
\label{epdmfig}
\end{figure}

In this appendix we further explain the EPDM technique introduced in Section~\ref{binaryidentsec}. As mentioned in the text, EPDM finds the period of an eclipsing binary system by seeking the value of the period that best minimizes the dispersion in phase of the faintest N points in a light curve. To illustrate how this method works, we show in Figure~\ref{epdmfig} the period search analysis of the LMMS DDEB candidate Kepler 006591789, which was found to have a period of 5.088435 days via the JKTEBOP model, (see Table~\ref{ddebcandstab}). The unfolded Q1 light curve is shown in the top-left panel of Figure~\ref{epdmfig}. EPDM selects the faintest 20 points of the light curve, which are highlighted by the larger points in that same panel. The number of points should be adjusted based on the quality of the data set. Too few points could result in all the points selected belonging to the same eclipse, if that one eclipse is unusually deep due to systematics or another reason, and thus EPDM will be unable to determine a period. Too many points will cause the results of EPDM to be less precise, as more points are included further away from the center of the eclipses. We have found that 20 points is a good number for \emph{Kepler} data, for which many systems do suffer from moderate systematics, as is evidently the case for Kepler 006591789, as seen by the quasi-sinusoidal variation in the baseline flux.

Having selected the faintest points from the light curve, EPDM then loops over a range of period values. In this case we choose a set of 5,000 period values that range from 0.3 to 30 days, evenly distributed in log space, so that shorter periods are as well-sampled as longer periods. At each period, the phase of each of the 20 faintest points are calculated via the following standard equation,

\begin{equation}
p = \frac{T}{P} - int(\frac{T}{P})
\end{equation}

where $p$ is the phase of a given point, with a time value, $T$, for a given period, $P$, and int() returns the argument rounded down to the nearest integer value. The standard deviation of these 20 phase values is then computed, and we are left with a standard deviation for each trial period. In the bottom-left panel of Figure~\ref{epdmfig}, we plot the standard deviation in phase of the 20 points versus each trial period. The lowest values for the standard deviation indicate the best periods, where the eclipses align in phase-space, while high values indicate bad periods. As can be seen in the bottom-left panel, the standard deviation approaches a value of 0.0 near 10.2 days, 5.1 days, 2.05 days, and decreasing fractions thereof, or period aliases. To determine the three best periods, EPDM first selects the lowest standard deviation, which in this case yields a value of 5.09004 days. It then selects the next lowest value, whose corresponding period value differs from the first by at least 10\%, and yields a value of 10.1747 days. The third period selected via the same method yields a value of 2.54402 days.

To further clarify the technique visually, in the top-right panel of Figure~\ref{epdmfig}, we show the same plot as in the bottom-left panel, but limited in period range to straddle the best period found, 5.09004 days. At the same period range, in the bottom-right panel, we plot the actual values of the phase for each of the 20 points at each period. (For ease of viewing, we use a lower trial period resolution in the bottom-right panel than the top-right panel.) As can be seen in the lower-right panel, when the trial period is far from the true period of the system, the actual phase values have a large dispersion, and range completely from 0.0 to 1.0. As the given period gets closer to the true period, the phase values begin to clump, with their dispersion decreasing as the trial period approaches the true value. Indeed, as highlighted by the box in the bottom-right panel of Figure~\ref{epdmfig}, at the best period, all the phase values are tightly grouped together at P = 5.0904 days, indicating that all the eclipses are extremely well aligned, and the period of the system has been found. 


One complication that can arise is if EPDM encounters an eccentric system with two similarly deep eclipses. In this case, when the algorithm selects the N faintest points, it will be selecting points from both eclipses. Since the system is eccentric, there is a phase offset not equal to 0.5 between primary and secondary eclipse, i.e. the two eclipses occur closer to each other in time compared to the period of the system. In this case, if we were to run EPDM as just described, in a plot like the bottom-right panel of Figure~\ref{epdmfig}, at the true period of the system there would be two groups of points, each by itself having a very small deviation, but separated from each other in phase by a large amount. Thus, the standard deviation calculation will show a much higher value than it should, and the correct period could not be found. Along similar lines, a problem arises when we consider how to calculate the standard deviation of, for example, the distribution of phase points in the bottom-right panel of Figure~\ref{epdmfig} at a period of 4.9 days, which ranges from 0.8 - 1.0, and then jumps to 0.0 - 0.05. It is clear this is a continuous group of points, which simply experiences an abrupt jump from phase 1.0 to 0.0. Although they represent a fairly good period, a calculation of their standard deviation would show a high value, and thus indicate a bad period.

To reconcile both these problems, we insert an additional step into the EPDM technique. At each trial period, EPDM searches for a reflection phase, $p_{r}$, whose value is between 0.0 and 1.0, that will allow the two distinct phase groupings to align. For each value of $p_{r}$, if the phase value of a given point is larger than $p_{r}$, a new value for the phase of the point, $p$, is calculated as

\begin{equation}
  p = p - 2.0\cdot(p-p_{r})
\end{equation}

The value of $p_{r}$ which yields the lowest standard deviation for a given trial period is the correct reflection value, and that corresponding lowest standard deviation should be assigned to that trial period. Thus, in the case of an equal depth, eccentric system, where say the N lowest points group around two phases of 0.2 and 0.4, at a value of $p_{r}$ = 0.3, the two distinct groupings would merge into a single group at phase 0.2, with a very small standard deviation at the correct period of the system. As well, in the case where a group of phase points that range from 0.9 to 1.0 and 0.0 to 0.1, $p_{r}$ allows the points to merge into a single group that only ranges from 0.0 to 0.1. In fact, we have already implemented the use of $p_{r}$ when generating the bottom-left and top-right plots of Figure~\ref{epdmfig}.

In conclusion, because EPDM only utilizes the faintest N points of a light curve, the computations are very quick, especially compared to traditional phase dispersion minimization techniques, which utilize every point in a light curve. This also allows for a more precise determination of the period, as one can apply more computing time towards finer period resolution. As well, for the same reason, EPDM is not affected by systematics or varying star spots, as long as their photometric amplitudes are not on the order of or greater than the amplitude of the eclipses. By selecting the faintest point, or the earliest of the N faintest points, one is also given a good value for the time of primary minimum. We have shown EPDM can be applied to both eccentric and non-eccentric binaries, and since a transiting planet's light curve is similar to an eclipsing binary with only one visible eclipse, the technique works equally well for transiting exoplanets. In theory, EPDM could also be applied to other variables, such as stars with rotating spots, pulsating variables, and contact binaries, although periods for these systems will be less precise than detached eclipsing binaries, due to the broader minima of those systems. In theory though, one may not have to select the faintest points of a light curve, but possibly a very narrow flux range, and achieve the same result.

%% file: appendixB.tex
\begin{singlespace}
\section{\MakeUppercase{Genetic Algorithms for Eclipsing Binaries}}
\label{agaappendix}
\end{singlespace}

As mentioned in the text, in fitting our sample of eclipsing binaries, we have 12 parameters: period, time of primary minimum, inclination, mass ratio, e$\cdot$cos($\omega$), e$\cdot$sin($\omega$), surface brightness ratio, sum of the fractional radii, ratio of the radii, out of eclipse flux level, and the amplitude and phase shift of the sinusoid applied to the luminosity of the primary in order to account for spots. We aim to vary these parameters over their entire range of possible solutions, which if left to a grid search for 10$^{-3}$ precision, would require computing on the order of $\sim$10$^{36}$ light curves; a computationally prohibitive task. Standard steepest descent minimization schemes such as Levenberg-Marquardt have extreme difficulties in large, multi-parameter solution spaces, especially for eclipsing binaries as the solution space is not at all smooth and has many local minima. Thus, we need a minimization technique that is computationally efficient, not adversely affected by a non-smooth solution space, and able to find the global minimum. These criteria are superbly met by the class of optimization schemes known as Genetic Algorithms (GAs).

In a standard GA, \citep[cf.][]{Charbonneau1995}, light curve parameter sets, called individuals, for an initial population of solutions, are randomly generated within a predefined parameter space, and compared to the observational light curve. Their corresponding $\chi^{2}$ value is used as a measure of fitness for natural selection, with parameters from fit individuals bred with each other, (subjected to crossover like chromosomes), to create a second generation of new solutions, and parameters from unfit individuals eliminated. After being subject to random mutations, to maintain parameter diversity and ensure discovery of the global minimum, this second generation is compared to the observational data, and bred into a third generation of solutions. The process continues for a specified number of generations, until a satisfactorily low $\chi^{2}$ is found. \citet{Charbonneau1995} demonstrated the application of GAs to problems in Astronomy and Astrophysics, specifically fitting galactic rotation curves, finding pulsation periods in $\delta$ Scuti stars, and fitting magnetodynamical wind models with multiple critical points, showing how the GA quickly finds the global minimum, regardless of the topography of the solution space. It is this type of GA that has been already been incorporated into the ELC eclipsing binary modeling code, and used with much success \citep{Orosz2000,Orosz2002}.

\citet{Canto2009} recently proposed a new form of GA called an Asexual Genetic Algorithm (AGA). In the AGA, instead of breeding new individuals via crossover, individuals are randomly created within a small predefined parameter space, or breeding box, centered on the fittest members of the previous generation. The size of this breeding box can be shrunk over successive generations to quickly converge to the best-fit solution. As shown by \citet{Canto2009}, the AGA is computationally simpler and more precise since it does not require encoding parameters for crossover, and converges much faster than traditional GAs, without sacrificing any ability to migrate to the global solution, so long as the breeding box size does not decrease too quickly. \citet{Canto2009} first showed that it far outperformed the standard GA in both computational efficiency and final precision by solving one of the exact same problems presented by \citet{Charbonneau1995}. \citet{Canto2009} additionally demonstrated the application of the AGA to fitting the radial-velocities of extrasolar planets and the spectral energy distributions of young stellar objects.

As eclipsing binary solutions have an even larger parameter space with many local minima than most problems, we make a few modifications to the AGA described by \citet{Canto2009} to ensure discovery of the global minimum. First, while we do exactly copy the fittest 10\% of individuals of one generation to the next generation, to ensure forward progress is always made while maintaining parameter diversity, instead of picking the fittest N members of a generation, each of which breeds M offspring, to create a new generation, we randomly select individuals for breeding by weighting them by a factor of (1/$\chi^{2}$)$^{2}$. This ensures that the fittest individuals breed the most offspring, but still allows for a few less fit individuals to breed, maintaining parameter diversity and exploration of the entire parameter space. Second, instead of randomly creating new members within a breeding box of fit individuals, we randomly select a number for each parameter from a Gaussian probability distribution centered on each parameter of a fit individual. Thus, new individuals are not strictly confined to a breeding box, but merely are very likely to be created near a fit individual, and maintain a very small probability that they will be created at many standard deviations away. This mimics mutation in traditional GAs and ensures that the algorithm will not become trapped in a local minimum. Third, as suggested by \citet{Canto2009}, the standard deviation of this normal distribution is chosen for each parameter to be the standard deviation of that parameter in the entire population, times the function $0.1^{(1/\chi_{0}^{2})}$, where $\chi_{0}^{2}$ is the $\chi^{2}$ value of the fittest member of the population. This allows parameters with the greatest impact on the fit, or the smallest range of possible parameters, such as the out of eclipse flux level, to converge rapidly, while allowing parameters that are less certain to converge more slowly and thoroughly explore their parameter space. Furthermore, via this method, the standard deviation is shrunk over successive generations, so that the algorithm converges, but only very slowly initially, rapidly increasing as $\chi_{0}^{2}$ approaches 1.0, i.e. the global minimum has been found. Finally, we take the fittest 10\% of the final generation and perform a standard Levenberg-Marquardt minimization for each member, choosing the member with the resulting lowest $\chi^{2}$ value as our final solution.

\begin{figure}[ht!]
\centering
\begin{tabular}{ccc}
\epsfig{width=0.315\linewidth,file=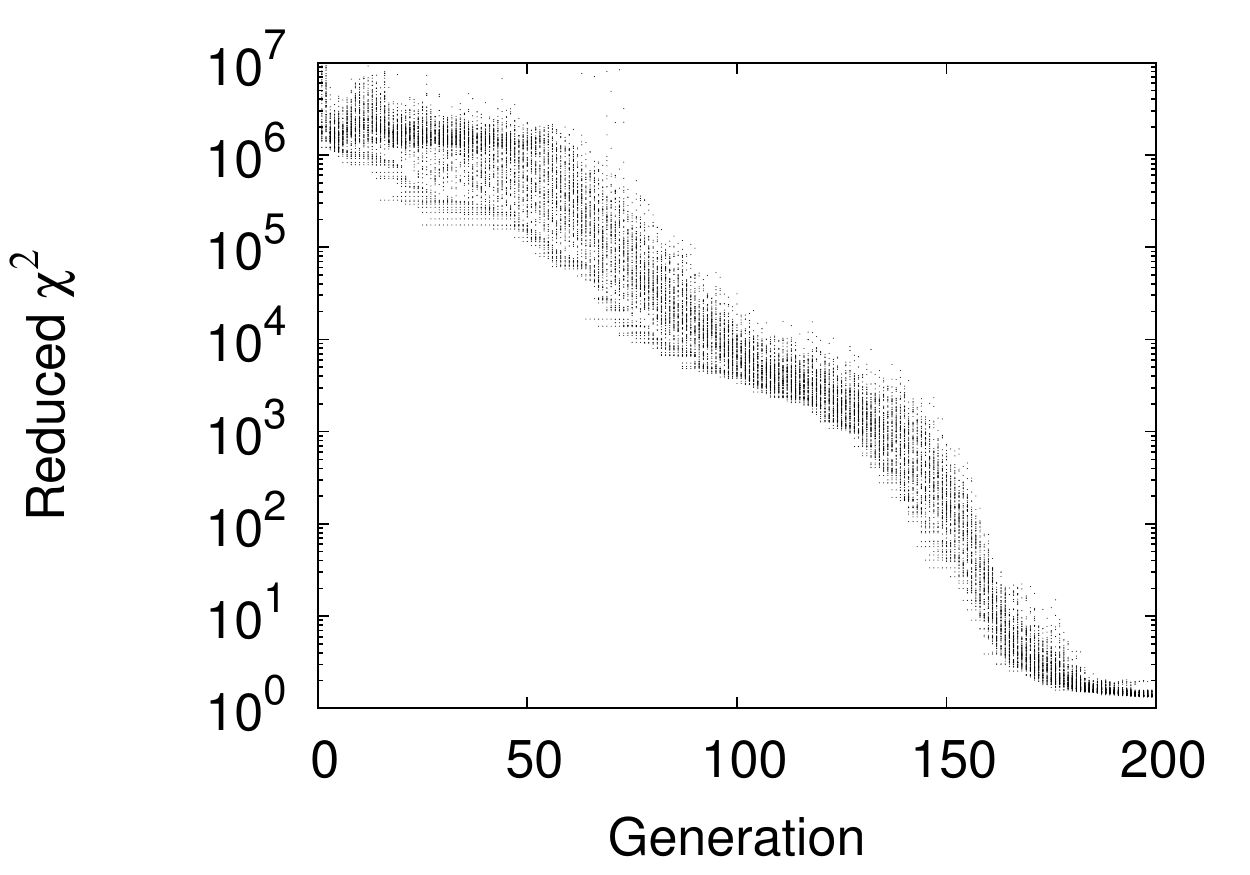} &
\epsfig{width=0.315\linewidth,file=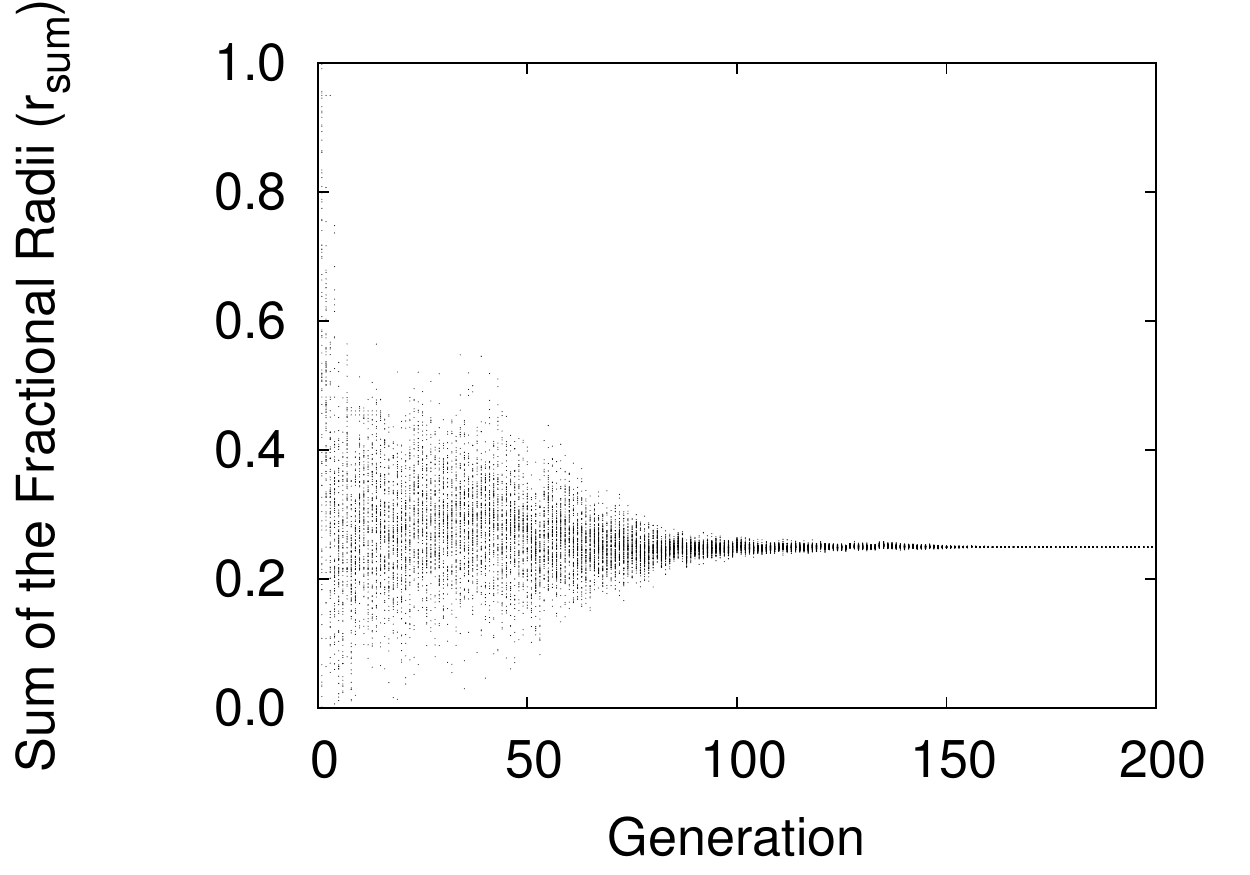} &
\epsfig{width=0.315\linewidth,file=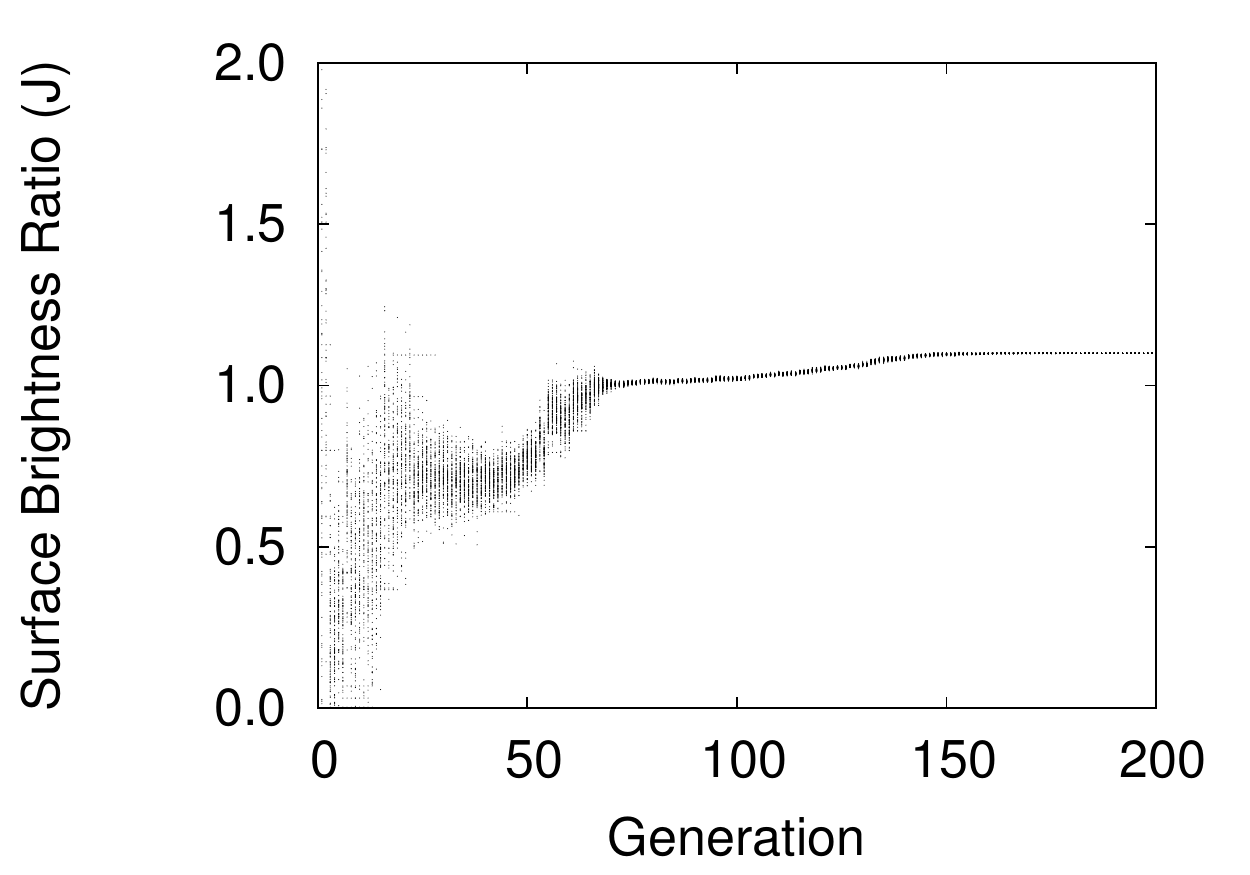} \\
\epsfig{width=0.315\linewidth,file=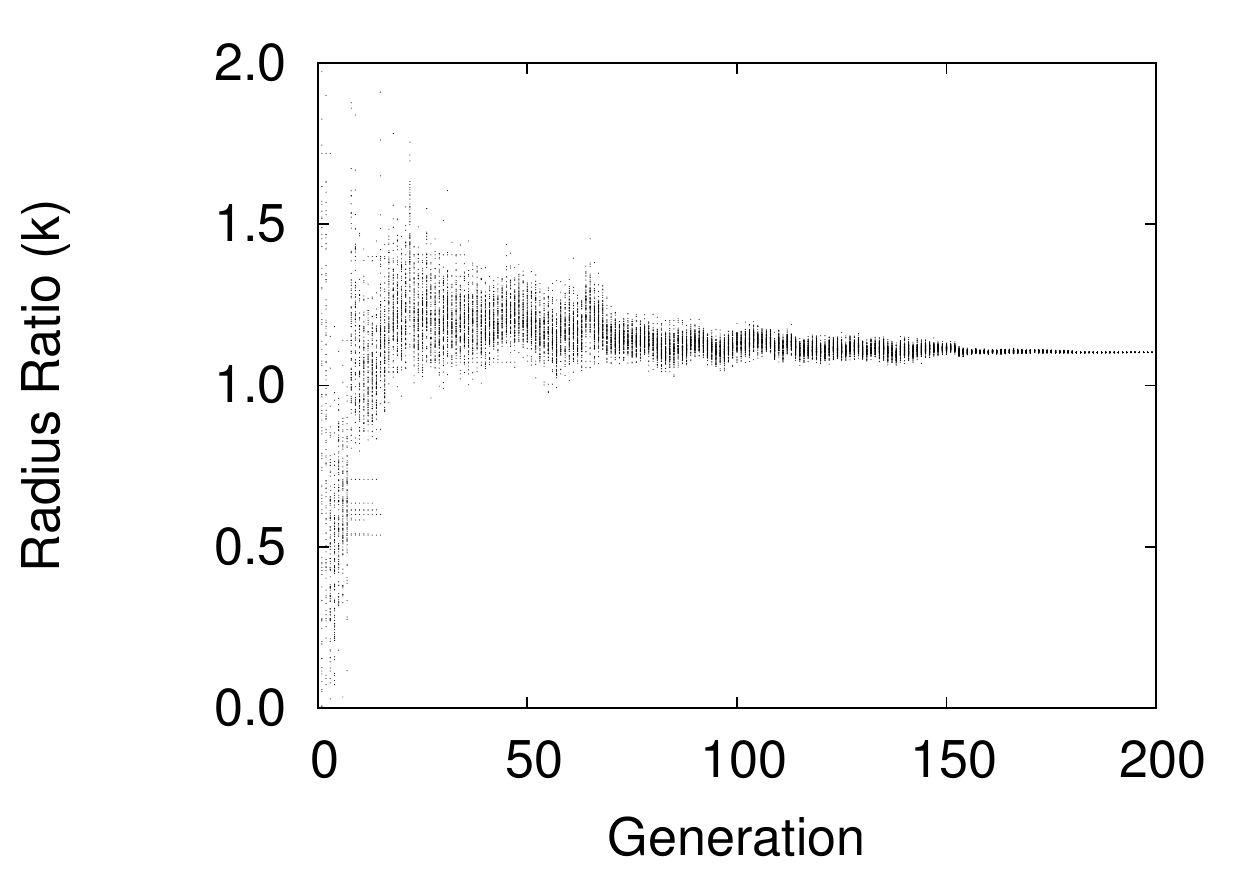} &
\epsfig{width=0.315\linewidth,file=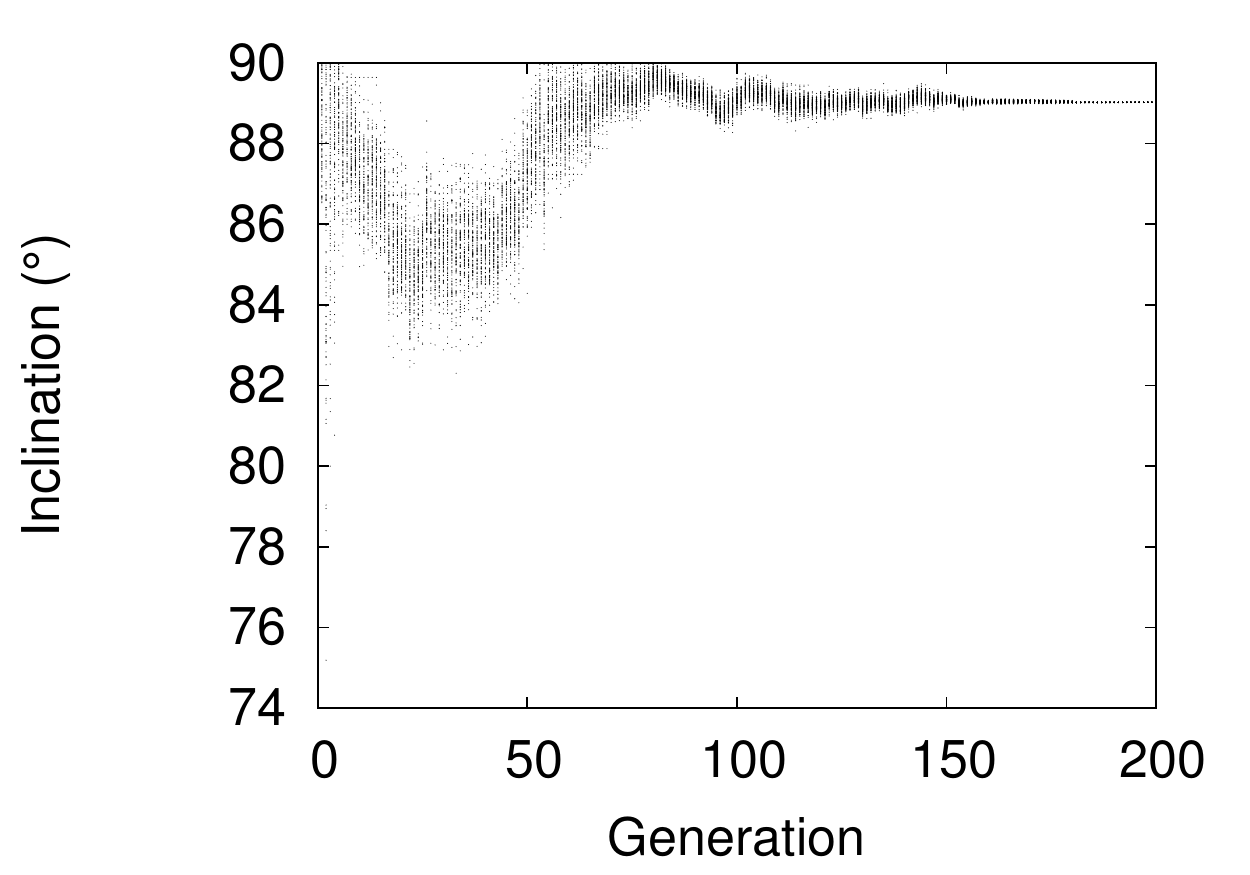} &
\epsfig{width=0.315\linewidth,file=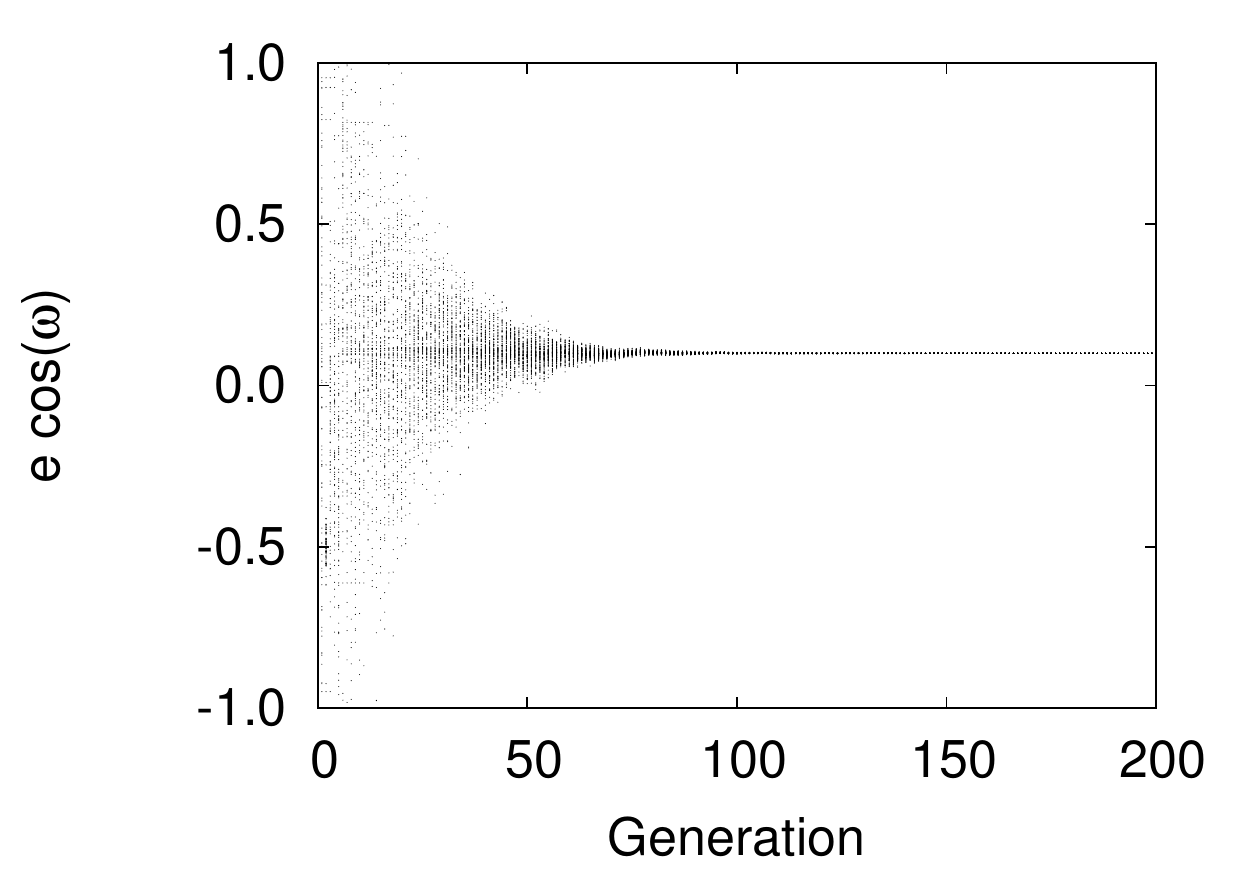} \\
\epsfig{width=0.315\linewidth,file=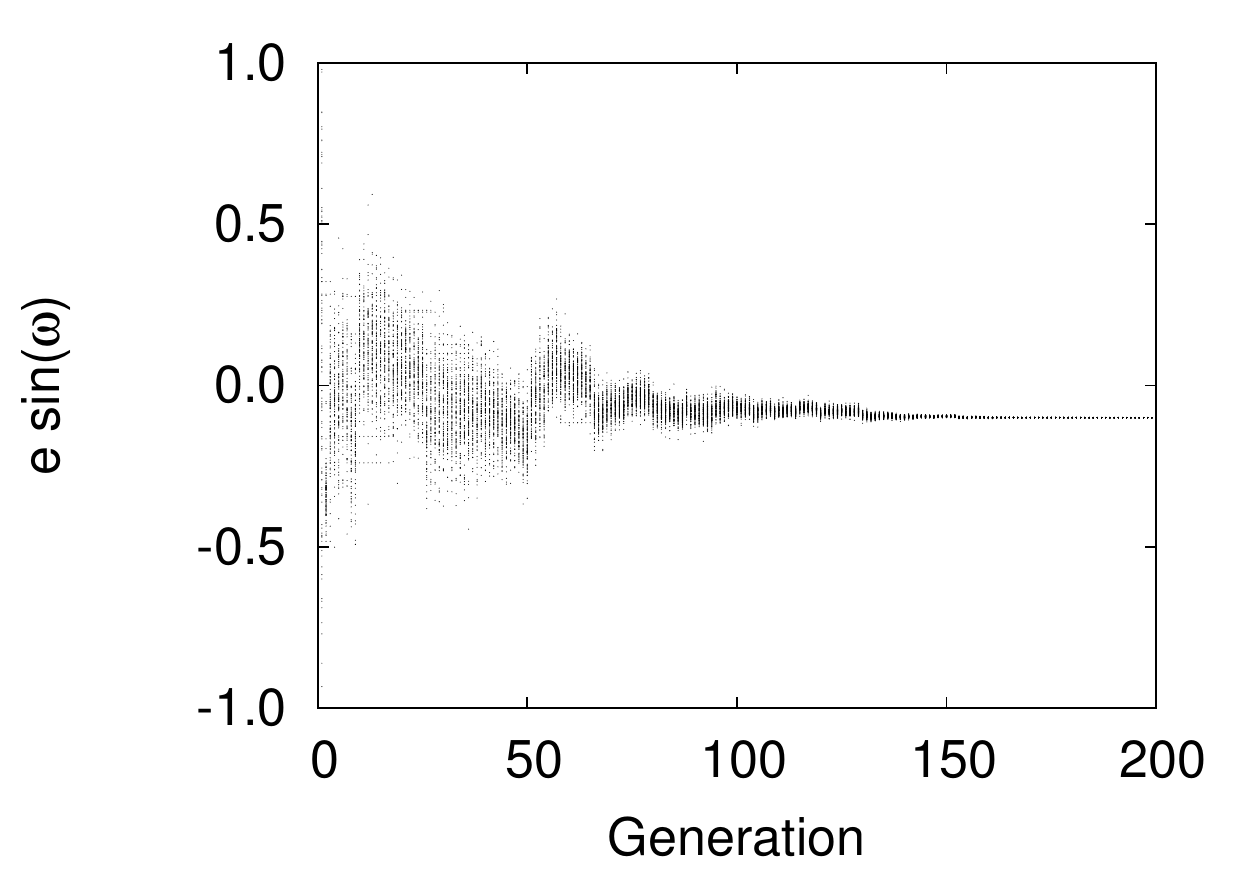} &
\epsfig{width=0.315\linewidth,file=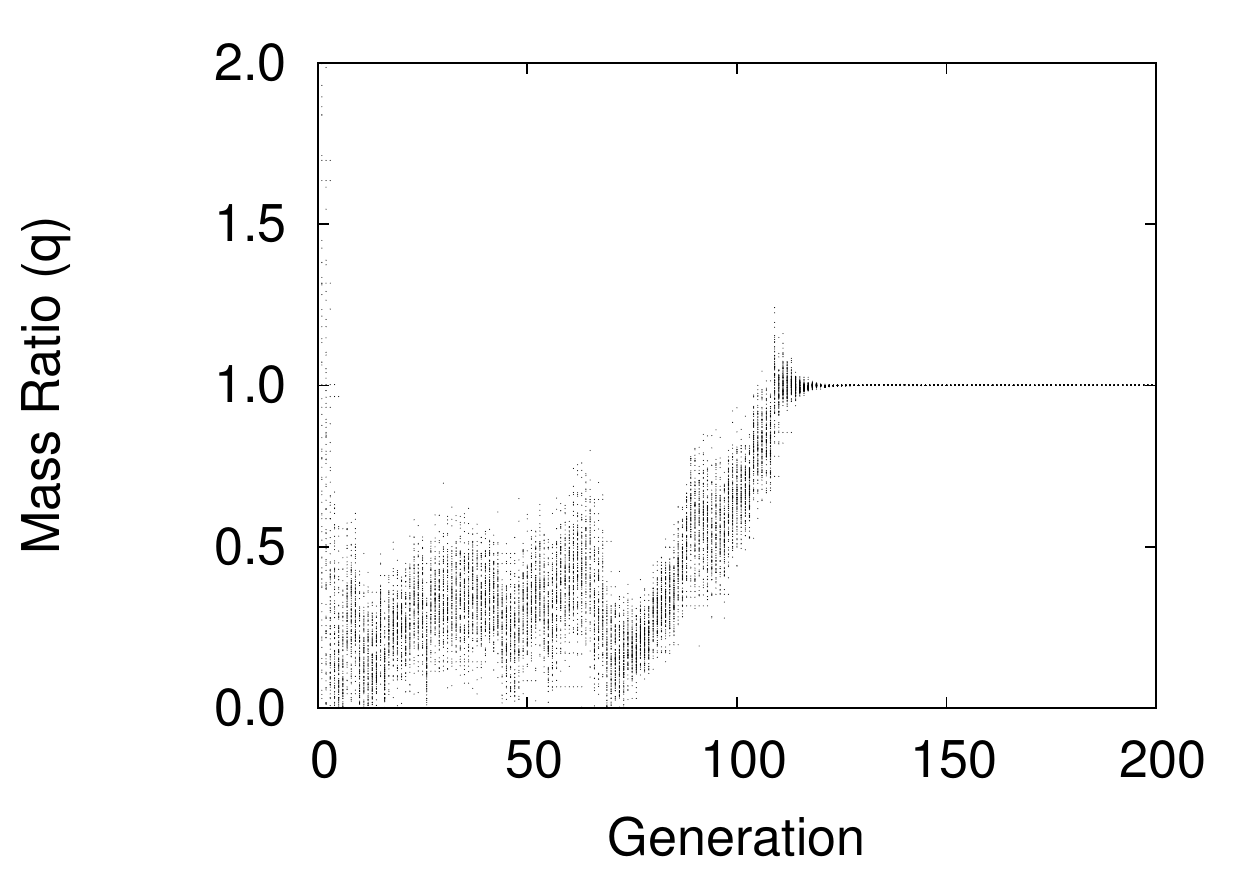} &
\epsfig{width=0.315\linewidth,file=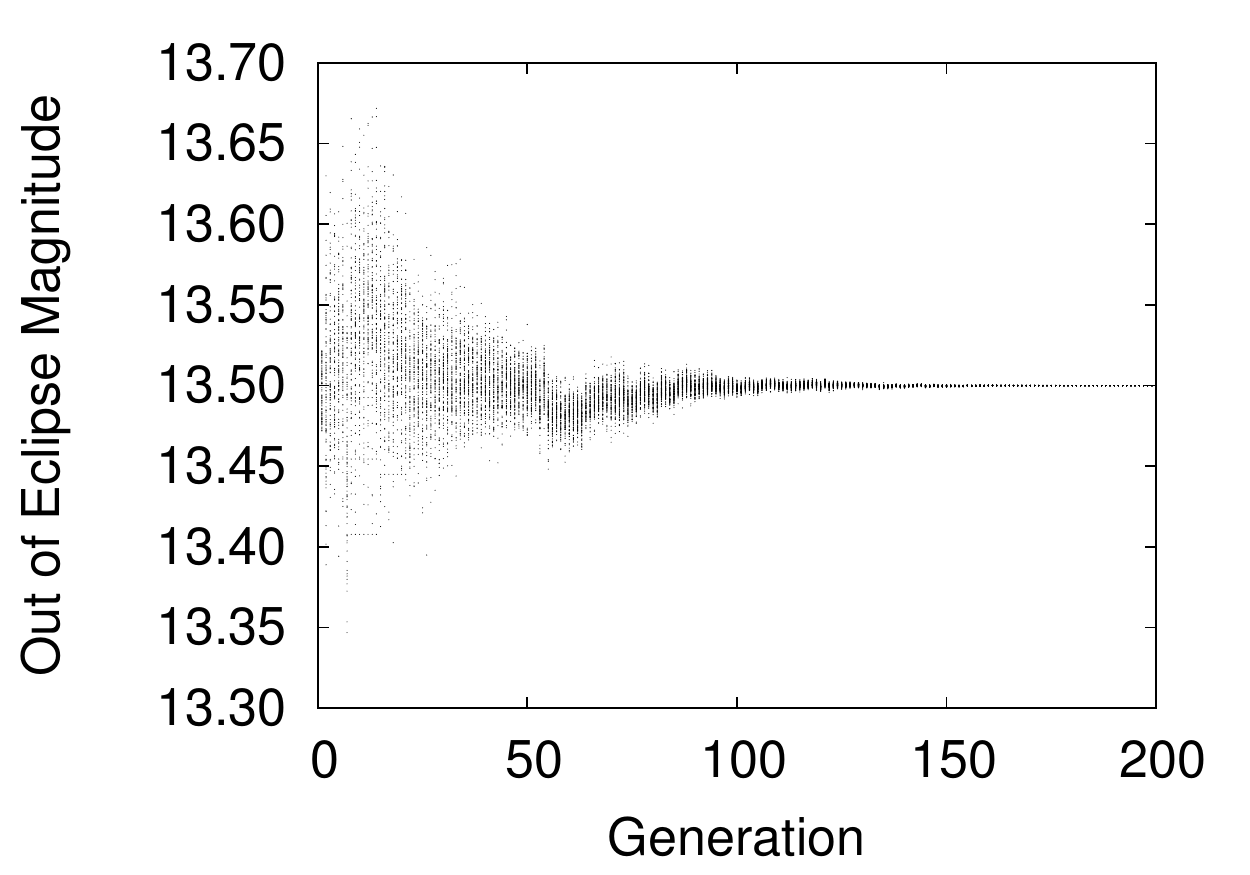} \\
\epsfig{width=0.315\linewidth,file=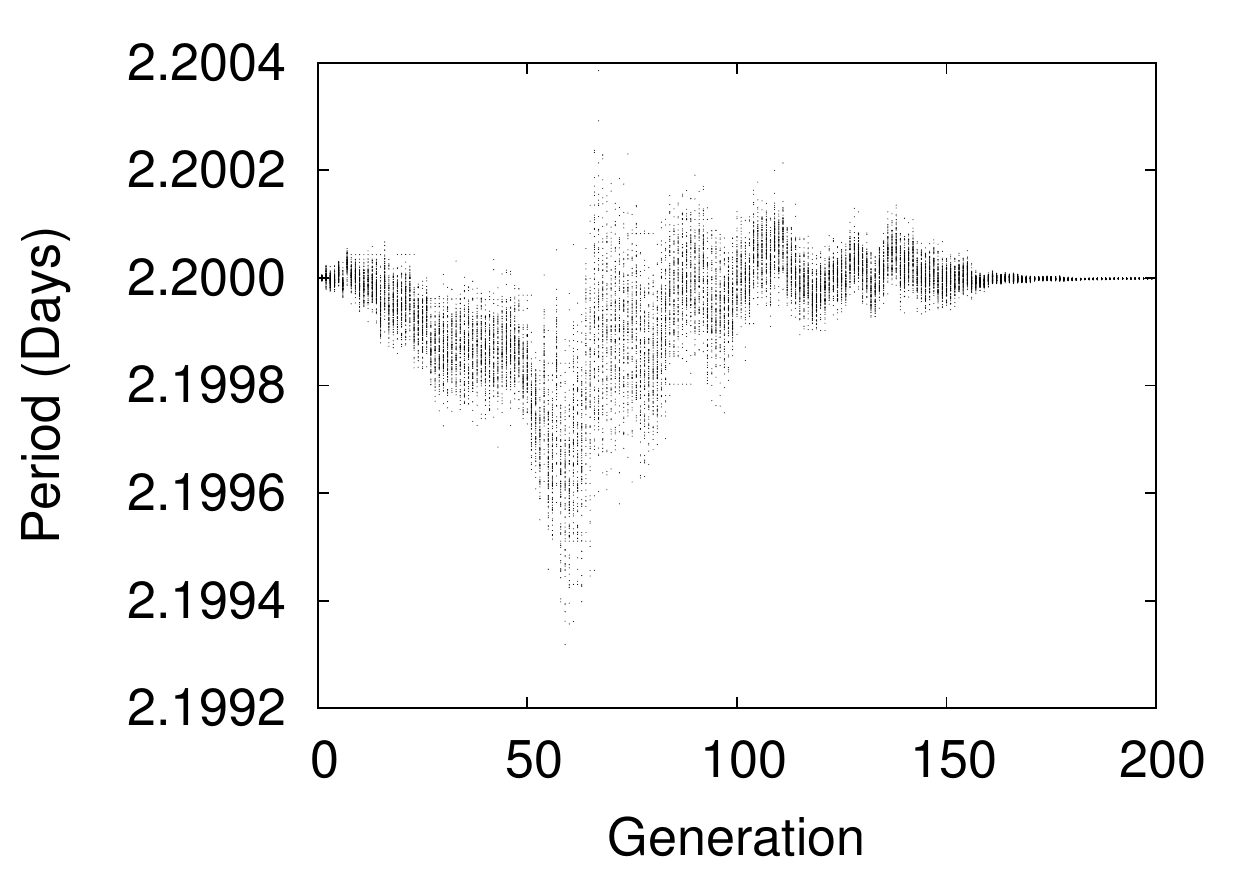} &
\epsfig{width=0.315\linewidth,file=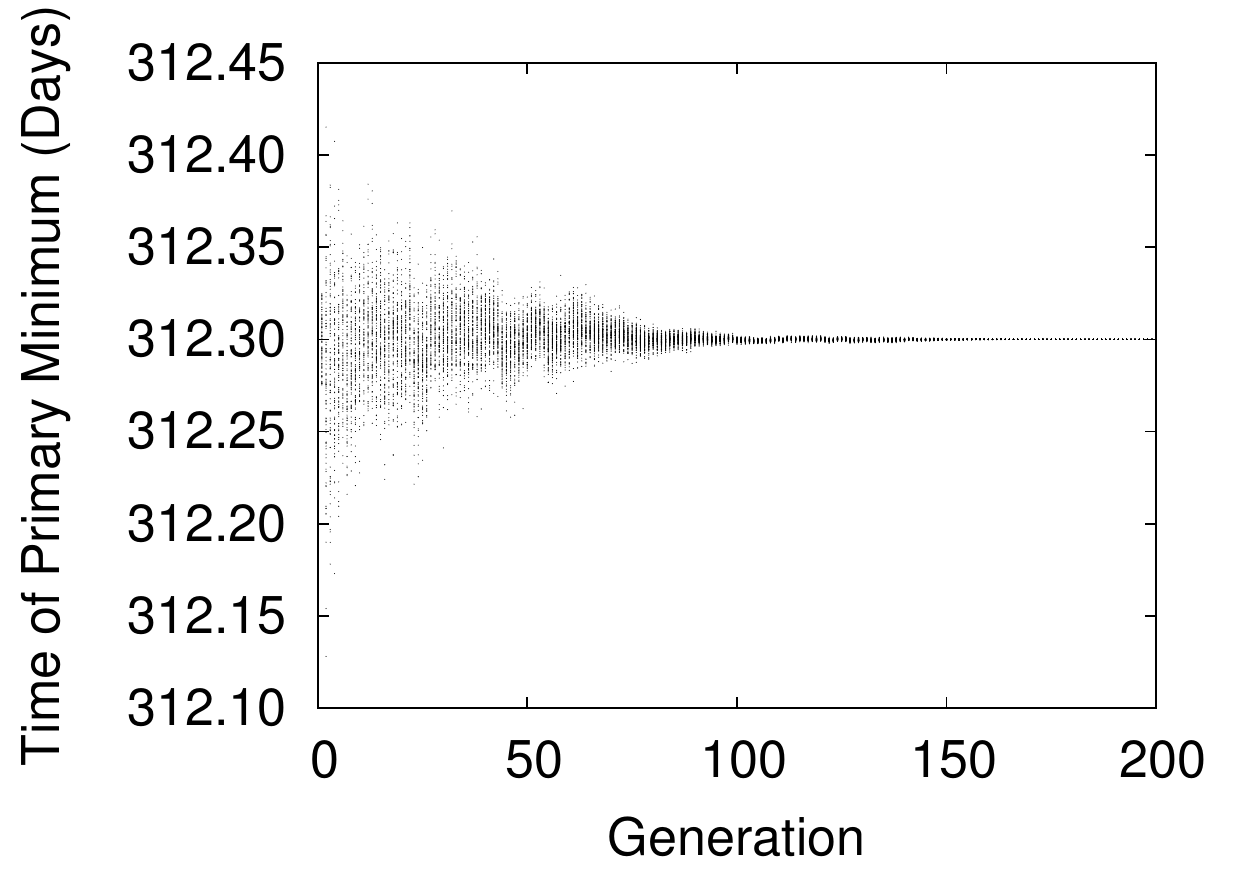} &
\epsfig{width=0.315\linewidth,file=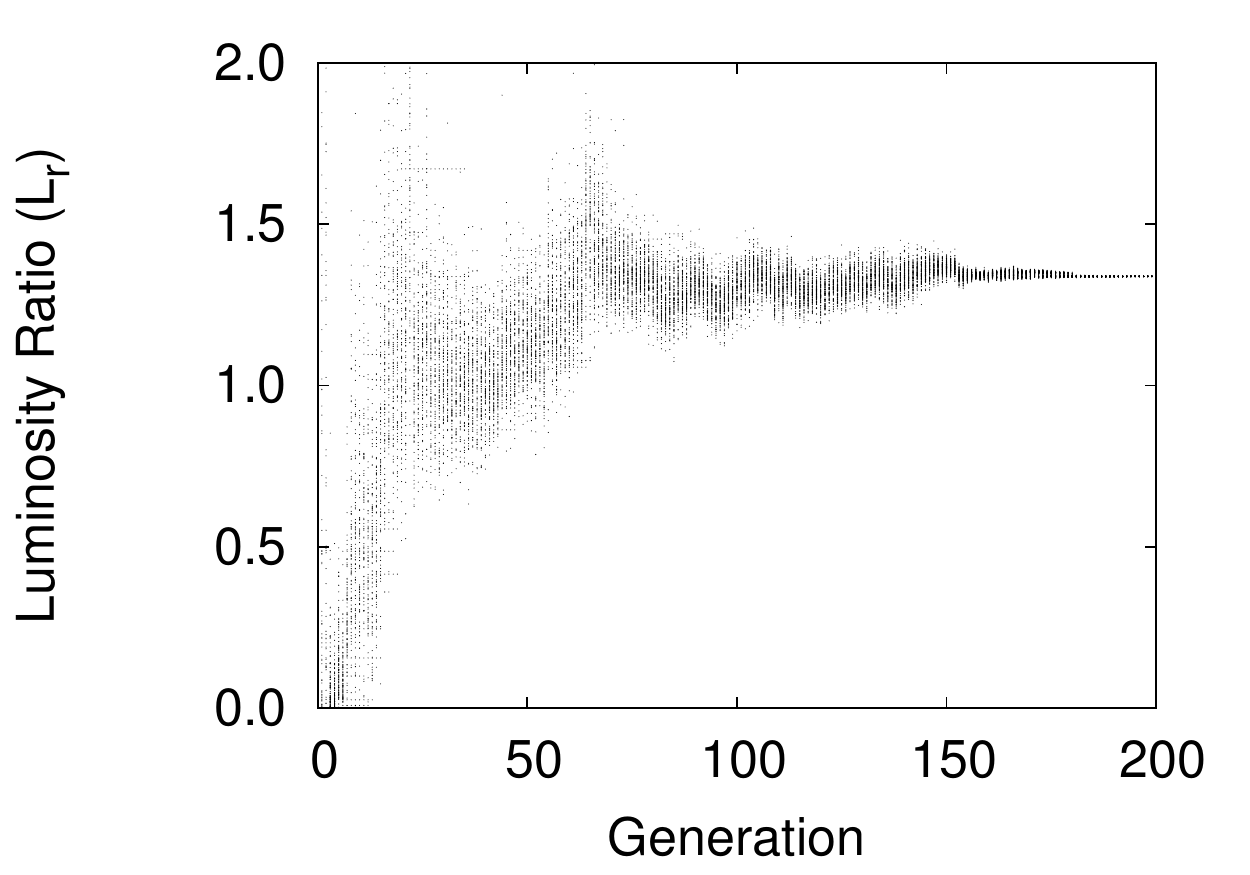} 
\end{tabular}
\newpage
\caption[Illustration of how the AGA converges over subsequent generations]{Illustration of how the AGA converges over subsequent generations (x-axis) by solving an artificially generated light curve, re-binned to the number of data points and error typical for a \emph{Kepler} light curve. Each point represents the parameter of each member in each generation. The parameters of the system are r$_{sum}$ = 0.25, k = 1.1, i = 89.0$\degr$, q = 1.2, e$\cdot$cos($\omega$) = 0.1, e$\cdot$sin($\omega$) = -0.1, J = 1.1, P = 2.20 days, T$_{0}$ = 312.3 days, and out of eclipse magnitude = 13.5. The derived reduced $\chi^{2}$ and luminosity ratio are also plotted. The AGA converges rapidly, decreasing the lowest $\chi^{2}$ value found by an order of magnitude every $\sim$20 generations. It can be seen that the parameters that are most significant to the light curve converge the fastest.}
\label{agafig}
\end{figure}

We nominally found, for the eclipsing binaries in our sample, that a population of 100 individuals, bred for 200 generations, does an excellent job of solving the light curves. This only requires the generation of 20,000 light curves, which with the JKTEBOP code only required $\sim$3 minutes per light curve to solve on a single 2.0 GHz CPU. Of course, some systems may require a smaller or greater number of individuals and/or generations, but it should not be more than a factor of $\sim$2. One may substantially reduce the number of individuals or generations required, and thus the run time, if one can limit the range of parameter space. For example, if one knows, or wants to assume, the orbit is circular or nearly circular, one could constrain $|$e$\cdot$cos($\omega$)$|$ $<$ 0.1 and $|$e$\cdot$sin($\omega$)$|$ $<$ 0.1. Furthermore, the AGA code is extremely parallelizable, and thus with a multi-core computing cluster one could easily use this technique to model thousands of eclipsing binary lightcurves, as is to be expected from Pan-STARRS and other large photometric surveys, in a very reasonable time frame.

To visually demonstrate how the AGA works, we have generated a light curve with the following parameters: r$_{sum}$ = 0.25, k = 1.1, i = 89.0$\degr$, q = 1.2, e$\cdot$cos($\omega$) = 0.1, e$\cdot$sin($\omega$) = -0.1, J = 1.1, P = 2.20 days, T$_{0}$ = 312.3 days, and out of eclipse magnitude = 13.5. We then re-bin this data to match the number of data points in the $Kepler$ Q1 data sets, and add typical Gaussian noise for a bright $Kepler$ star of 0.1 mmag per data point. We then re-solve this light curve with the AGA, varying all the aforementioned parameters, and show in Figure~\ref{agafig} the value of each parameter for every individual in each generation, as well as the values for the derived reduced $\chi^{2}$ and luminosity ratio. One can see how even while searching over the entire global solution space, the AGA rapidly converges to the solution that was used to generate the light curve, with the $\chi^{2}$ decreasing by a factor of $\sim$10 every $\sim$20 generations. Even though the best solution of the 200$^{th}$ generation has $\chi^{2}$ $\sim$ 1.5, if allowed to continue for more generations, this run would eventually converge to $\chi^{2}$ = 1.0, and performing a simple Levenberg-Marquardt minimization from the best solution quickly produces a $\chi^{2}$ = 1.0 fit.

%% file: appendixC.tex
\begin{singlespace}
\section[\MakeUppercase{Modeling Multi-Wavelength Stellar\\Astrometry: SIM Lite Observations of\\Interacting Binaries}]{\MakeUppercase{Modeling Multi-Wavelength Stellar Astrometry: SIM Lite Observations of Interacting Binaries}}
\label{sim1appendix}
\end{singlespace}

\subsection{Introduction}

The field of stellar evolution has matured to the point that we generally understand the entire evolution of an isolated star if we know its initial mass. While many details remain to be solved, it is unlikely that SIM Lite will provide a revolution in our understanding of the evolutionary processes for single stars of low and intermediate mass. The same cannot be said for binary star systems, where there are at present a number of outstanding issues that remain unsolved. Among the most difficult is how mass transfer proceeds during the common envelope stage that occurs in the post-main sequence lives of close binaries. The formation of nearly every mass transferring binary requires a common envelope phase where the secondary star helps to strip the atmosphere of the primary while driving mass from the system, and shrinking the binary's orbit. Current theory \citep[c.f.,][]{Podsiadlowski2003} has a difficult time explaining the creation of systems with massive black hole primaries (M$_{\rm BH}$ $\geq$ 9 M$_{\sun}$) and very low mass secondary stars (M$_{\rm 2}$ $\leq$ 0.7 M$_{\sun}$). Similar problems exist across the mass spectrum of close and interacting binaries (IBs), which for the purposes of the current study include systems with a black hole, neutron star, or white dwarf (WD) primary and a non-degenerate companion. For example, \citet{Howell2001b} explores the well known 2-3 hr period gap observed in cataclysmic variable (CV) systems, which is postulated to be a result of cessation of mass-transfer from the main-sequence, low-mass secondary star to the white dwarf companion for periods between 2-3 hours. \citet{Howell2001b} runs binary population synthesis models that show this theory is only correct if the secondary stars in these systems are up to 50\% oversized, and thus both components are really much less massive than has been typically assumed via standard main-sequence mass-radius-temperature relations. Thus, if accurate masses can be determined for these systems, their formation theories can be directly tested.

Typically in binary star work, radial velocity curves are used to obtain masses for the components. While it can be quite simple to get a radial velocity curve for one component of an IB, (the secondary star if it dominates the luminosity, else typically the disk), it is difficult, or even impossible, to get such data for their black hole, neutron star, or WD primaries. Thus, along with the radial velocity amplitude of the secondary or disk, one must also assume a mass and know the binary orbital inclination in order to solve for the component masses. It is especially difficult to determine the binary inclination. In a large number of IBs the secondary star can be quite prominent at infrared wavelengths, and since the object is distorted, it exhibits ellipsoidal variations as it orbits the primary \citep[c.f.][]{Gelino2001}. The amplitude of these variations are dependent on the orbital inclination, and to a lesser extent, the properties of the secondary star. However, even a small amount of contamination by the accretion disk introduces considerable uncertainty in this method, and nearly all interacting binary systems have some level of contamination from the accretion process. Thus, this technique of using infrared ellipsoidal variations to determine the inclination fails if one cannot ascertain the spectrum and level of the contaminating source.

This paper will show how SIM Lite, with multi-wavelength, microarcsecond astrometry, will be the first mission capable of determining the multi-wavelength astrometric orbits of IBs. The field of optical astrometry has recently been making great progress from both ground-based observations, as well as is posed to make dramatic leaps with the upcoming space-based missions GAIA and SIM Lite. The limits of ground-based astrometry have been recently pushed by both CHARA and PRIMA/VLTI, but neither are capable of the kind of astrometric measurements needed to probe IBs. CHARA has multi-wavelength capabilities, but can only provide angular resolution to $\sim$200 $\mu$as \citep{Brummelaar2005}, which is far larger than the reflex motions to be discussed in this paper, and requires very bright targets. PRIMA/VLTI will achieve $\sim$30-40 $\mu$as precision \citep{Belle2008}, which may be at the limit of usefulness for these systems, but only in $K$-band, which does not help distinguish individual component orbits, as will be shown. The GAIA mission will provide astrometry for $\sim$10$^{9}$ objects with 4 - 160 $\mu$as accuracy, for stars with V=10-20 respectively, and does posses multi-wavelength capabilities \citep{Cacciari2009}. However, GAIA is a scanning satellite and cannot perform pointed or time-critical observations, and in fact only achieves this accuracy by averaging $\sim$80 individual measurements, the individual errors of which range from 36 - 1,431 $\mu$as, again for stars with V=10-20 respectively \citep{Mignard2005}. Thus, GAIA cannot provide the high-precision, time-critical pointed observations needed to study IBs. In contrast, SIM Lite will be able to point at any desired object for any length of time, providing single-measurement accuracy of $\sim$1 $\mu$as \citep{SIM2009}. As well, SIM Lite will have $\sim$80 spectral channels, spanning 450 to 900 nm, thus providing multi-wavelength, microarcsecond astrometry \citep{SIM2009}.

SIM Lite will offer the first chance at measuring astrometric orbits for a large number of IBs, thus directly yielding inclinations and allowing for the precise measurement of the masses of both components in these systems. Determining the inclination of a zero-eccentricity system from astrometry is as straightforward as determining the ratio the semi-major and semi-minor axes of the projected ellipse on the sky. Ideally, one can derive accurate values for the masses of both components from the astrometry of a single component \citep{Benedict2000}, but to do so requires that one knows the orbital period, the semi-major axis of the apparent orbit, the parallax of the system, and the mass ratio of the two components. SIM Lite can provide the first three of these four parameters, but the mass ratios must be estimated for many of the main systems of interest. However, for binaries where there is a significant amount of light from more than one component, \emph{SIM Lite can determine absolute masses for both components}. Since SIM Lite only measures the location of the photocenter of the system, the observed reflex motion of the system will be wavelength-dependent. For example, in a binary where one component is hotter than the other, the measured motion in the blue part of the spectrum will be different from that measured in the red, with shorter wavelengths tracing the motion of the hotter component, and vice-versa. Thus, if one has at least two astrometric orbits at different wavelengths, and one knows the ratio of luminosities in each bandpass, (i.e. the spectra of the components), then one could reconstruct the individual orbits of each component and obtain absolute masses for both. Since SIM Lite is currently designed to have $\sim$80 spectral channels, and since multi-color photometry and spectroscopy already exists for most IBs of interest, this technique should be quite feasible for most IBs.

In the next section we describe the procedure and code used to model the reflex motions of IBs. In section 3 we examine the results for ``proto-typical'' systems, including both X-ray binaries and cataclysmic variables, and estimate the required observing time required by SIM Lite to obtain reasonably accurate parameters for each system. We also briefly investigate how the presence of disk temperature gradients and hotspots in the systems affect the derived orbits. In section 4 we present a full modeling of simulated SIM Lite data for one system, and show the precision and accuracy of recovered astrometric parameters and derived system masses. We summarize our results in section 5.

\subsection{The Modeling Procedure: The {\sc reflux} Code}

\textsc{reflux}\footnote{\textsc{reflux} can be run via a web interface from \url{http://astronomy.nmsu.edu/jlcough/reflux.html}. Additional details as to how to set-up a model are presented there.} is a code that computes the flux-weighted astrometric reflex motions of binary systems. At its core is the Eclipsing Light Curve (ELC) code, which is normally used to compute light curves of eclipsing binary systems (Orosz \& Hauschildt 2000). Besides specifying the primary and secondary stars, ELC allows for the inclusion of an accretion disk.  As with other light curve modeling programs, an array of physical effects are taken into account, such as non-spherical geometry, gravity brightening, limb darkening, mutual heating, and reflection effects. The program can either use a blackbody formula for local intensities of the stellar components, or interpolate from a large grid of NextGen model atmospheres \citep{Hauschildt1999}. ELC also allows for up to two hot or cool spots to be placed on each star, and on the accretion disk. Thus, ELC can reproduce nearly any binary system, including complicated systems such as cataclysmic variables, RS CVn systems, and Low Mass X-ray Binaries (LMXBs).  Additionally, we have modified the ELC code to allow for a mixture of blackbody and model atmosphere intensities for use among different system components, and to allow for the possibility of a free-free, or bremsstrahlung, accretion disk, which follows the form $F_{\lambda} \propto T^{-\frac{1}{2}}e^{-\frac{hc}{\lambda kT}}\lambda^{-2}$, where T is the temperature in Kelvin, and F$_{\lambda}$ is the flux in power per unit area per unit wavelength, $\lambda$.

\textsc{reflux} takes input parameters for a specified system and feeds them to the ELC program, which generates an intensity map of the system at a specified phase and wavelength, composed of N points evenly spaced on a grid around each star and the disk, if present. \textsc{reflux} then computes the system's center of light position using the formula \begin{equation} (X,Y) = \sum_{i=0}^{N}F_{i}\cdot(x,y)_{i}, \end{equation} where ($X,Y$) is the system's center of light (``photocenter''), with the center of mass located at (0,0), and F$_{i}$ is the flux of a grid point $i$, located at ($x,y$). This is done for a complete orbit at 8 different wavelengths, currently chosen to be the standard $UBVRIJHK$ bandpasses, with the astrometric reflex motion output in $\mu$as using an estimate of the system's distance. \textsc{reflux} also calculates the observed spectral energy distribution (SED) of both the sum and the individual components (as seen at quadrature) for comparison to multi-wavelength photometry to help constrain system parameters. \textsc{reflux} will also output a 3D animated gif of the system, the apparent multi-wavelength astrometric orbit, and the $x$ and $y$ components of this motion versus orbital phase. The actual light curves over an orbit are written to a text file for comparison with existing phased-resolved photometric data.

While SIM Lite only operates in the 4500 to 9000 \AA~bandpass, we include $JHK$ photometry into the modeling process to help constrain the SEDs for systems of interest. We do this because it is often possible to detect the secondary star in the near-infrared, allowing one to better quantify its contribution at the wavelengths accessible to SIM Lite. (For consistency and possible comparison to future work in NIR astrometry, we also include the $J$ and $K$-band astrometric motions in the output plots.) Thus, before one can model the expected reflex motions for an IB, it is critical to have both reasonable parameters for the system, as well as multi-color photometry. \textsc{reflux} requires as inputs estimates of the masses of the two stellar components, their temperatures, their radii, the orbital period, and orbital inclination. The eccentricity of the binary is assumed to be zero since in the majority of these systems the companion star fills its Roche lobe, indicating a close orbit with strong tidal forces. In addition, the $V$ magnitude and distance must be input for model normalization. If the system has an accretion disk, the temperature at its inner edge, the power-law index for the radial dependence of its temperature, and the radius of the inner and outer edges must be inputted. Alternatively, one can choose whether the disk follows a blackbody or free-free emission law. In the latter case, one has to specify and adjust a normalization constant for the luminosity of the disk emission to best match the observed SED. To demonstrate the use of \textsc{reflux}, we model several proto-typical IB systems below.

\subsection{Modeling the Reflex Motions of Interacting Binaries}

It is critical that one uses reasonable system parameters to first match the observed SED before relying on the output reflex motions. There are three main scenarios for the visual SEDs of IBs: 1) systems where one stellar component dominates the optical SED, 2) where more than one component contributes to the optical SED, and 3) disk-dominated systems. Besides the stellar components, and symmetric accretion disks, there are a number of other features that are present in IB systems such as accretion disk hot spots, accretion streams, magnetic structures, jets, and other outflows that both affect the SED {\it and} which might have appreciable astrometric signatures. We investigate some of these features below, but to completely cover all of the behavior exhibited by IBs is beyond the scope of the current investigation. In the following we perform case studies for three different IB scenarios, using several well known objects. Additionally, we examine how accretion disk or photospheric ``hotspots'' can distort the reflex motions of an IB. Finally, for each system we also examine the feasibility of observing the system with SIM Lite, providing rough estimates for the amount of time that will be required of SIM Lite to observe the system at a given precision, using the SIM Differential Astrometry Performance Estimator (DAPE) \citep{Plummer2009}.

\subsubsection{IBs with SEDs Dominated by the Primary or Secondary Star}

There are quite a number of IBs where the primary or secondary star completely dominates the SED of the system. For these systems, determining an astrometric orbit is straightforward. We examine two cases: 1) QZ Vul, a LMXB with a black hole primary and a cool, main sequence secondary star, and 2) Cyg X-1, a high mass X-ray binary (HMXB) with an O supergiant primary, and a black hole ``secondary''.

\paragraph{QZ Vul}

\begin{figure}[ht]
\centering
\begin{tabular}{cc}
\epsfig{width=0.475\linewidth,file=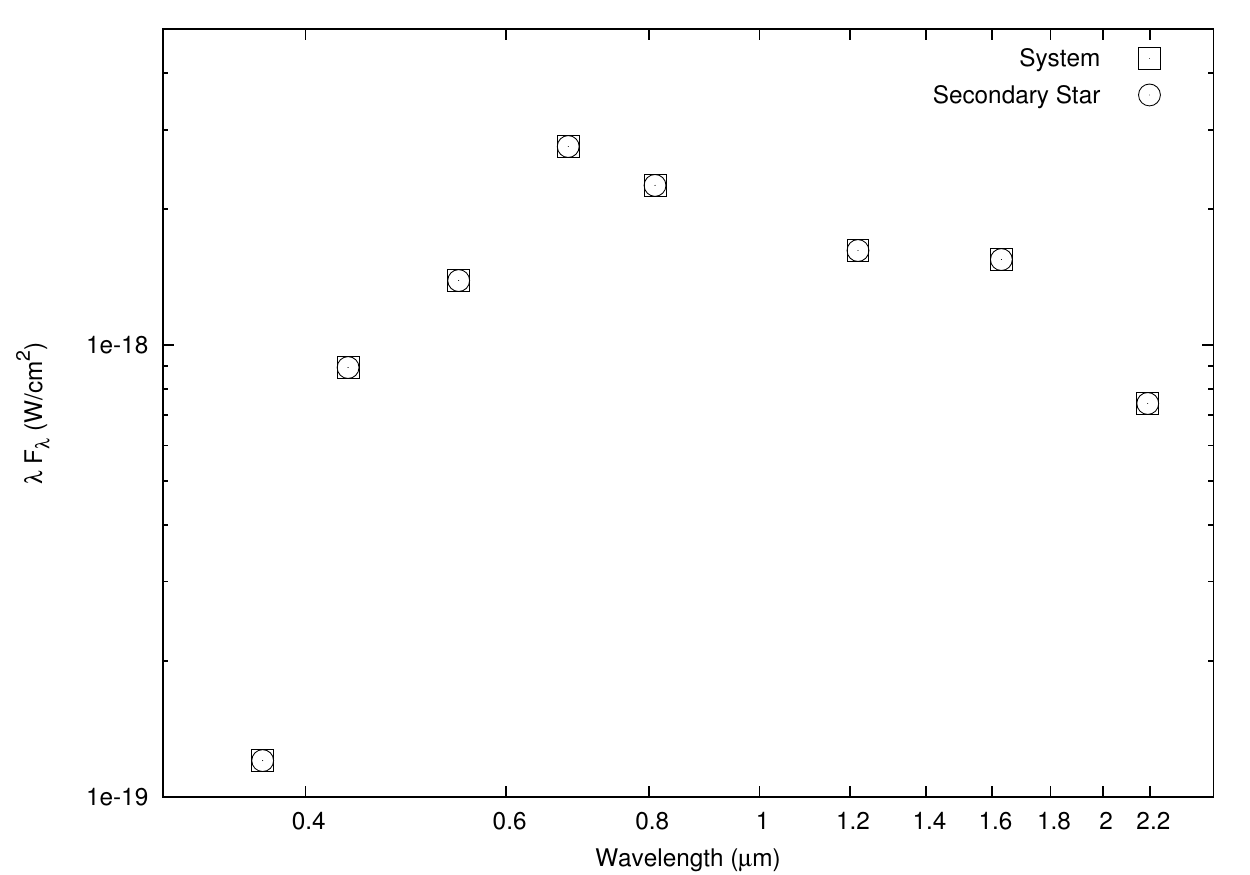} &
\epsfig{width=0.475\linewidth,file=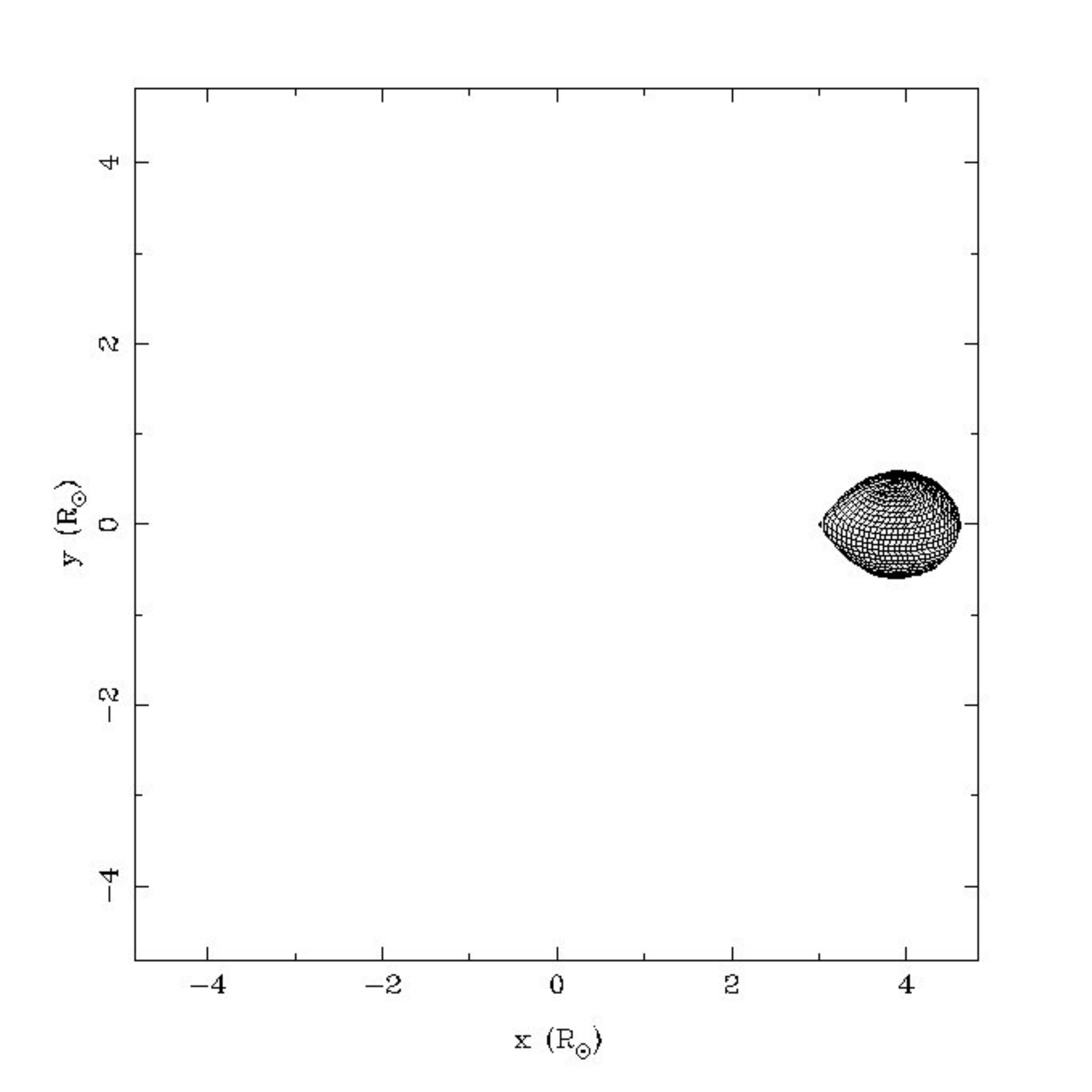}\\
\epsfig{width=0.475\linewidth,file=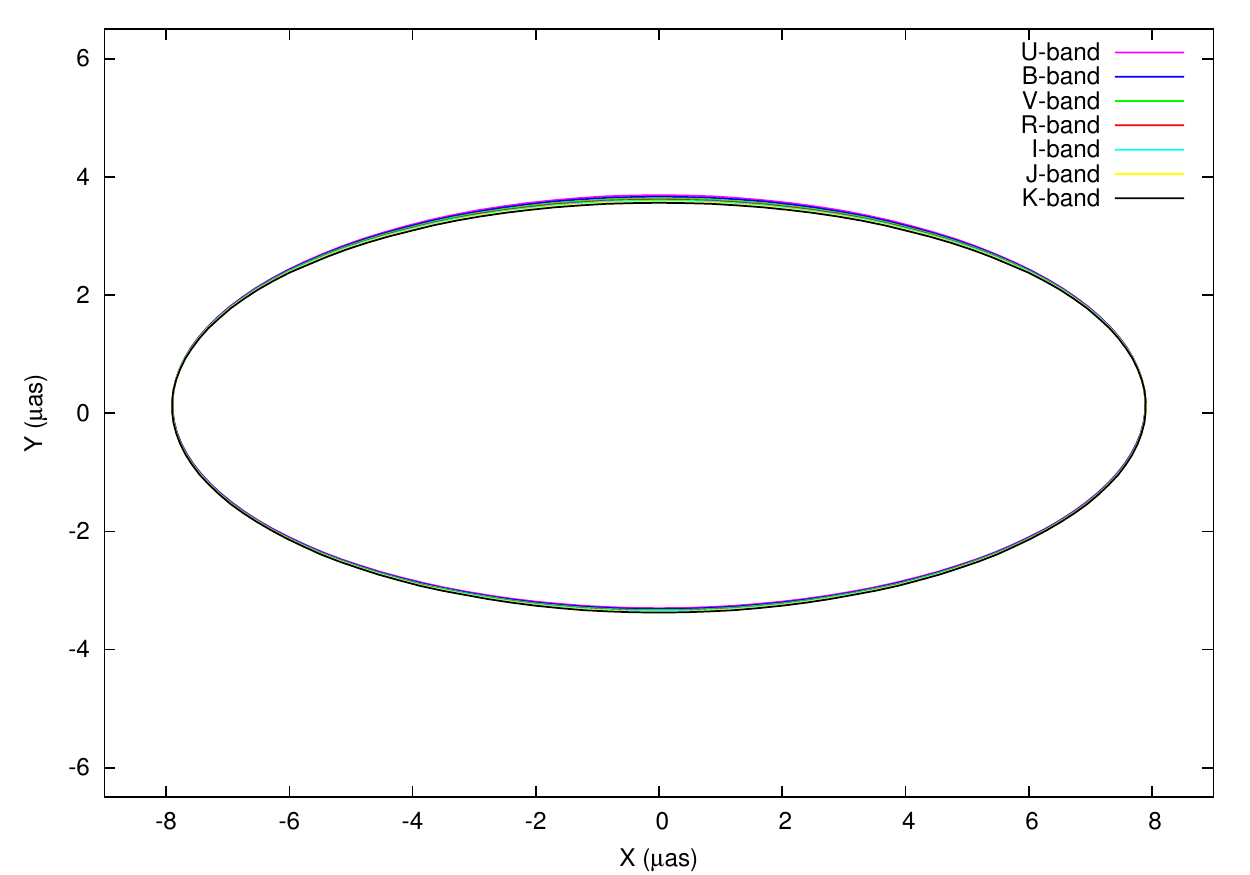} &
\epsfig{width=0.475\linewidth,file=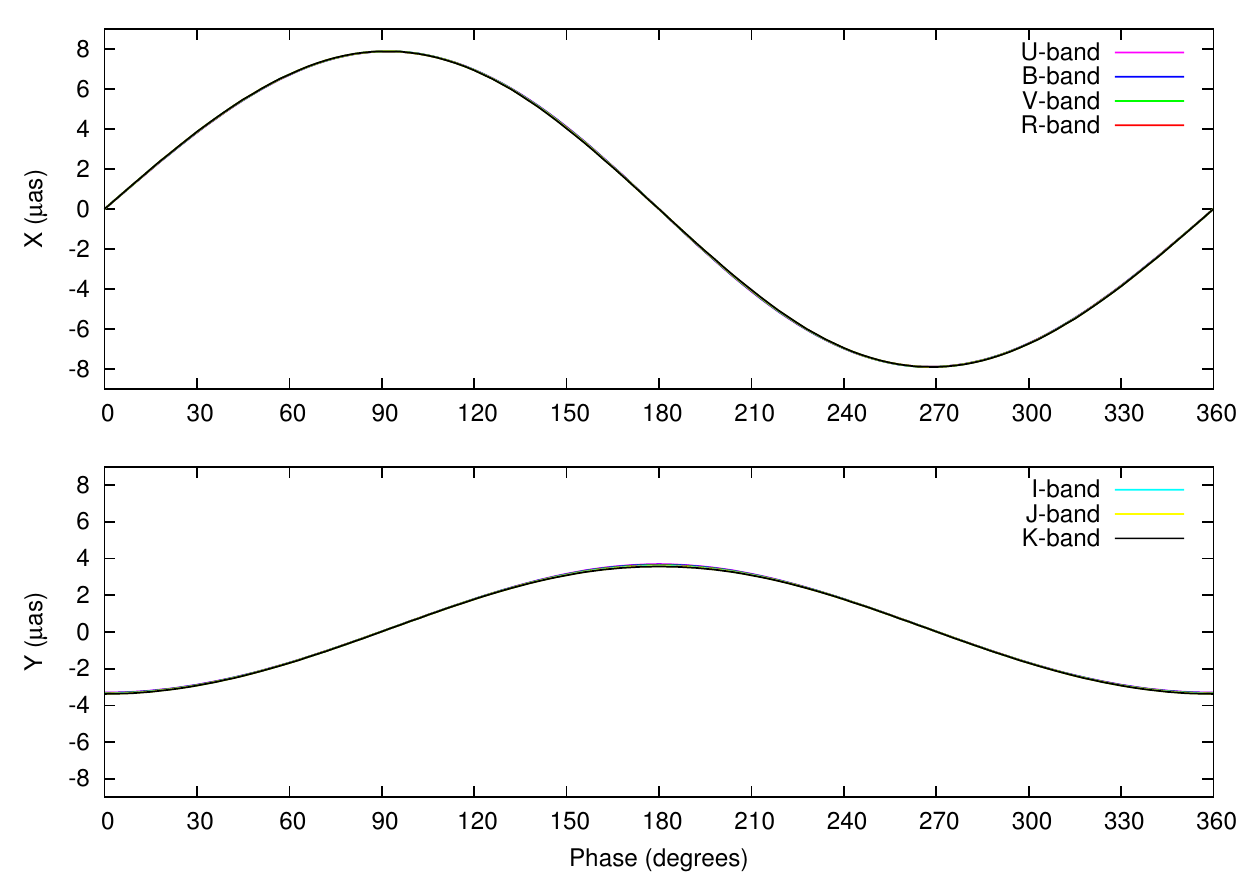} 
\end{tabular}
\caption[Multi-wavelength model of QZ Vul]{\emph{Top Left:} SED plot for QZ Vul. \emph{Top Right:} 3D model of QZ Vul at Phase=90$\degr$. \emph{Bottom Left:} Reflex orbit for QZ Vul. \emph{Bottom Right:} X and Y components of the reflex orbit versus phase for QZ Vul.}
\label{qzvulmainfig}
\end{figure}

As shown in \citet{Gelino2010}, the SED of QZ Vul appears to be that of a reddened K2 dwarf from the optical through the mid-infrared. The system parameters are shown in Table~\ref{qzvultable}. These parameters were input into \textsc{reflux} to model the astrometric motions, with a \textsc{nextgen} model atmosphere used for the secondary star. The model SED, a 3D model of the system, and its wavelength-dependent reflex motions are shown in Fig.~\ref{qzvulmainfig}. Since there is only a single visible component, QZ Vul does not exhibit any discernible wavelength dependency to its reflex motions. However, since the black hole is $\sim$24 times more massive than the secondary star, the K2 dwarf has a large apparent motion, and thus even at a distance of 2.29 kpc, the system's astrometric reflex motion ($\sim$8 $\mu$as), in theory, would be detectable with microarcsecond astrometry. However, in practice, this particular system is so faint (V = 21.2), and has such a short period (0.3342 days), that, according to DAPE, SIM Lite can not reach the needed precision without integrating for longer than the orbital period. Thus, one would need to find a much closer, and thus brighter, LMXB to observe with SIM Lite. The estimated orbital inclination for QZ Vul is 64$^{\circ}$ \citep{GelinoPhD2001}, and the only effect one would see with a QZ Vul type system with a different inclination is a reduction or amplification of the y-component of the astrometric motion, corresponding to an increase or decrease of the inclination respectively.

\begin{deluxetable}{lc}
\tablewidth{0pt}
\tablecaption{Parameters for the QZ Vul System}
\tablecolumns{2}
\tablehead{Parameter & Value\tablenotemark{a}}
\startdata
Magnitude (V) & 21.2\\
Distance (kpc) & 2.29\\
Inclination ($\degr$) & 64.0\\
Period (Days) & 0.3342\\
Eccentricity & 0.0\\
Mass of Star 1 (M$_{\sun}$) & 0.32\\
Mass of Star 2 (M$_{\sun}$) & 7.7\\
Radius of Star 1 (R$_{\sun}$) & 0.397\tablenotemark{b}\\
Radius of Star 2 (R$_{\sun}$) & 0.0\\
T$_{\rm eff}$ of Star 1 (K) & 4500\\
T$_{\rm eff}$ of Star 2 (K) & 0
\enddata
\label{qzvultable}
\tablenotetext{a}{Values from Gelino (2001)}
\tablenotetext{b}{Secondary star fills Roche lobe}
\end{deluxetable}

\paragraph{Cyg X-1}

\begin{figure}[h]
\centering
\begin{tabular}{cc}
\epsfig{width=0.475\linewidth,file=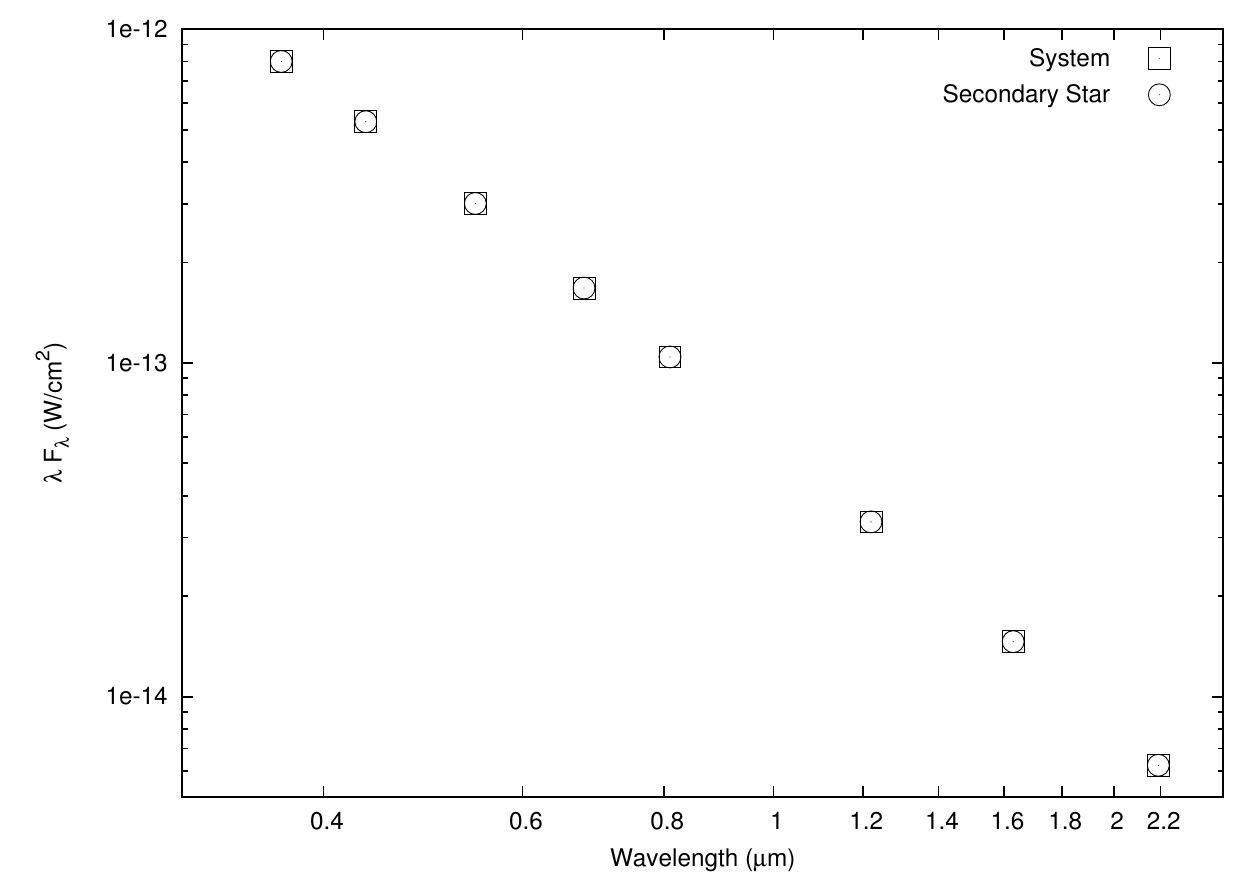} &
\epsfig{width=0.475\linewidth,file=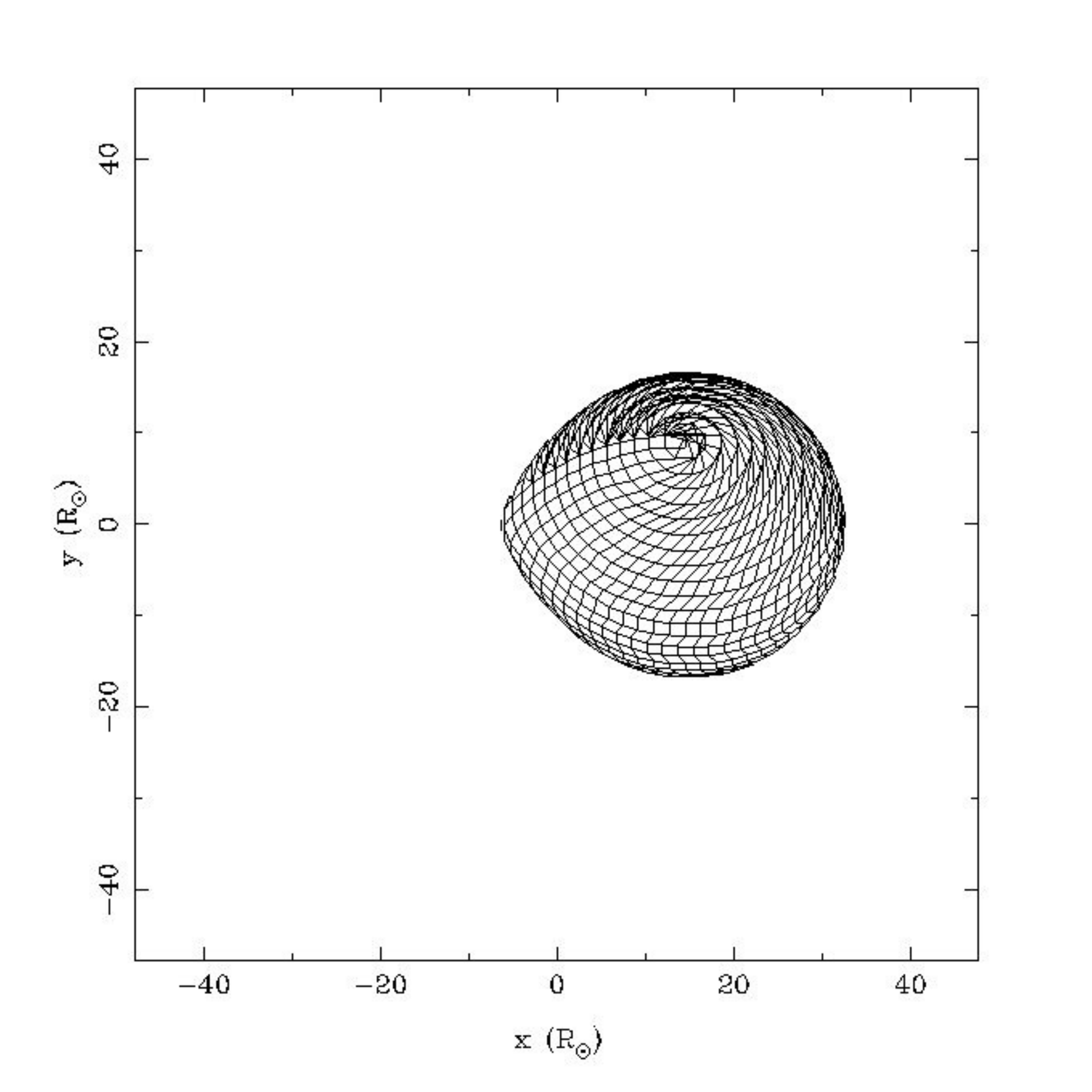} \\
\epsfig{width=0.475\linewidth,file=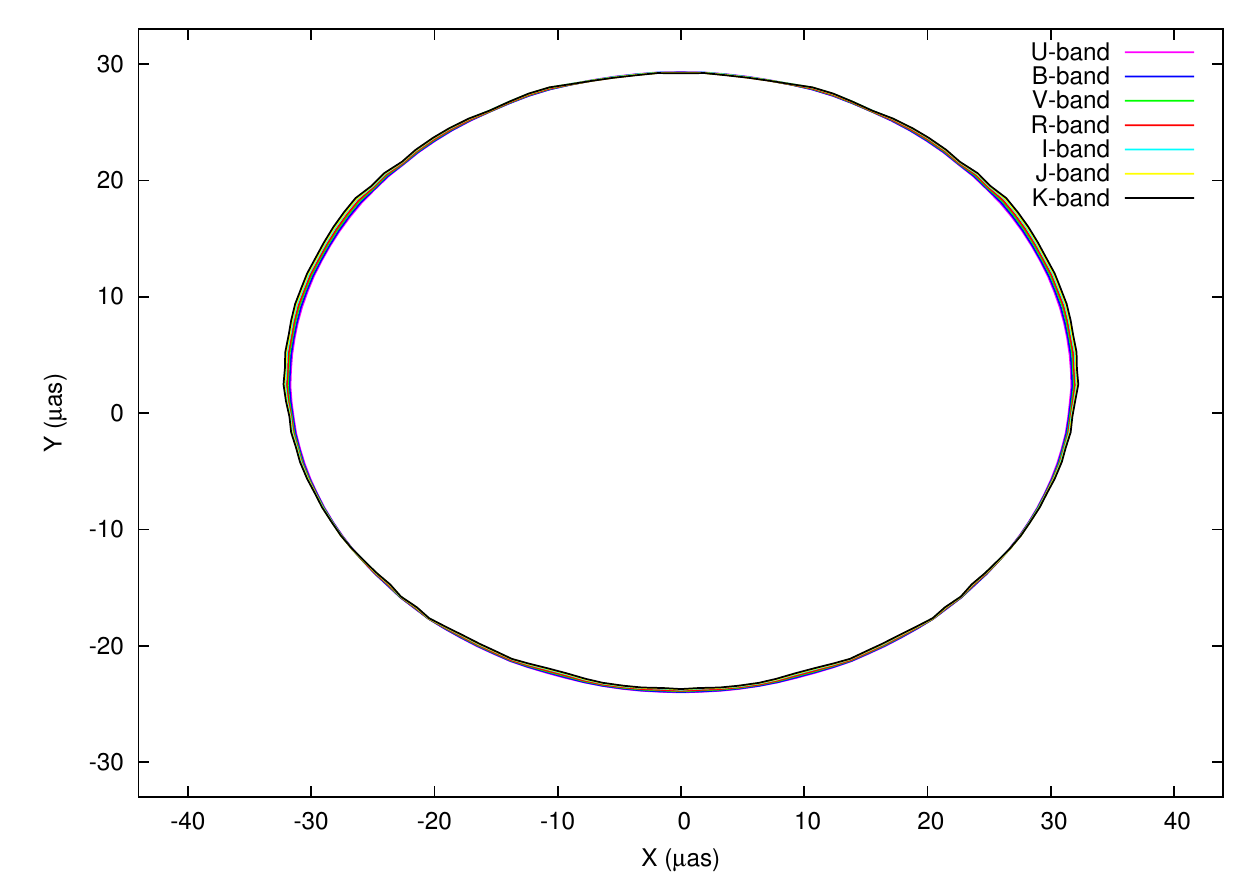} &
\epsfig{width=0.475\linewidth,file=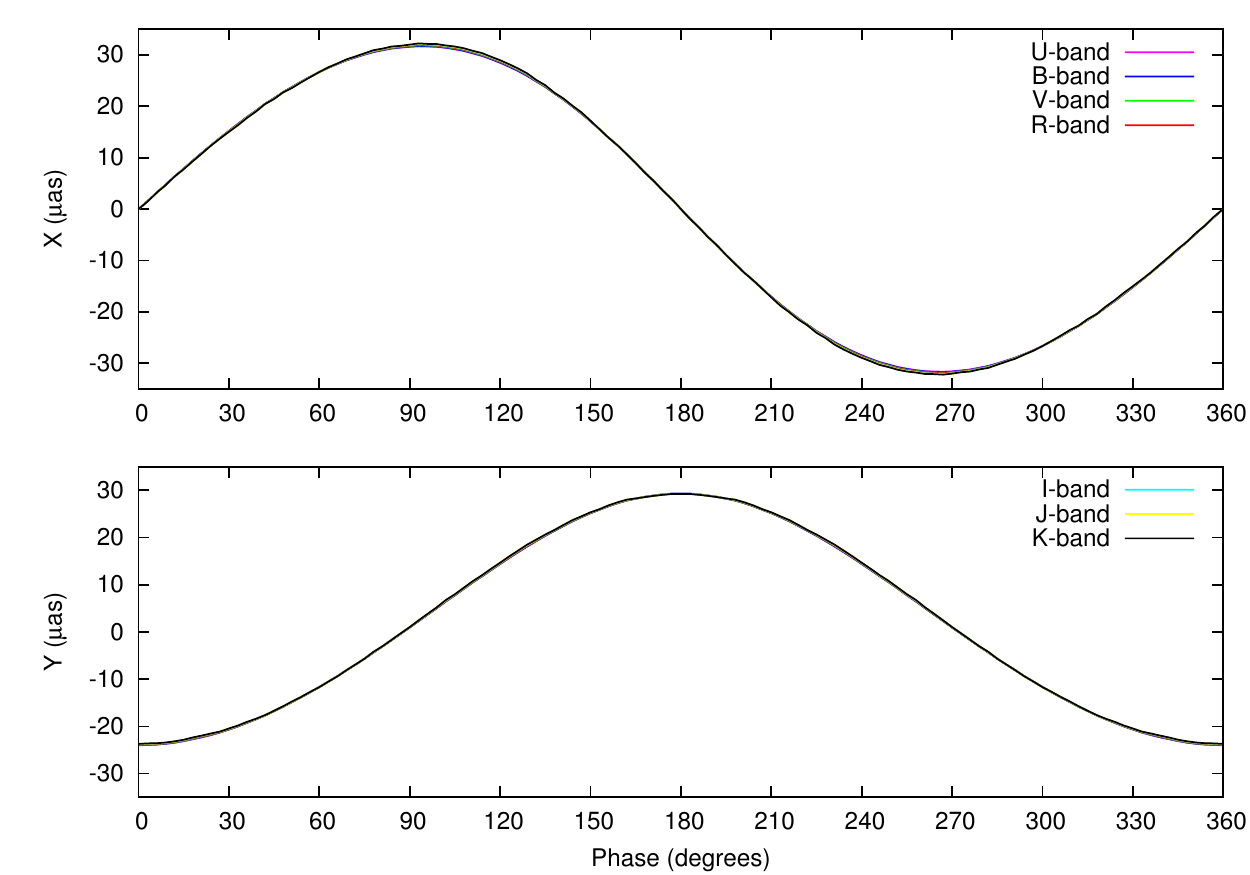}
\end{tabular}
\caption[Multi-wavelength model of Cyg X-1]{\emph{Top Left:} SED plot for Cyg X-1. \emph{Top Right:} 3D model of Cyg X-1 at Phase=90$\degr$. \emph{Bottom Left:} Reflex orbit for Cyg X-1. \emph{Bottom Right:} X and Y components of the reflex orbit versus phase for Cyg X-1.}
\label{cygx1mainfig}
\end{figure}

Cyg X-1 is a well-known HMXB dominated by an O-type supergiant orbited by a 10 M$_{\sun}$ black hole with a period of 5.6 d. Cyg X-1 exhibits a wide variety of behavior, such as highly variable X-ray and radio emission, some of which is presumably due to a relativistic jet. The optical light also varies, but much more weakly ($\Delta$m $\approx$ 0.05 mag). Thus, in contrast to QZ Vul, the primary star will be the visible source in this system. We model this system using the parameters listed in Table~\ref{cygx1table}, using a blackbody for the O star. As can be seen in Fig.~\ref{cygx1mainfig}, there is no wavelength dependence to the astrometric reflex motion, and a single bandpass is sufficient to determine the orbit. Even at a distance of 2.1 kpc, the large separation of the components in Cyg X-1 produces a significant astrometric signature ($\sim$25 $\mu$as), even though it is the more massive component that is responsible for the visible astrometric motion in this system. Given the brightness of the system (V = 8.95), and the long orbital period, this system is easy to observe with SIM Lite. According to DAPE, 10 individual measurements, each with 2.5 $\mu$as precision, could be obtained in only a total of 1 hour and 40 minutes of mission time, (given 5 minutes of target integration time per visit, broken into 5, 1-minute chops between the target and reference star.) This would provide more than sufficient high-precision measurements to obtain a good solution for this system, directly yielding the inclination of the system and the semi-major axis of the primary star's orbit, and thus indirectly the masses of the components. For systems in which one component completely dominates the visual SED, having accurate photometry for the system is unnecessary in interpreting the astrometric data.

\begin{deluxetable}{lc}
\tablewidth{0pt}
\tablecaption{Parameters for the Cyg X-1 System}
\tablecolumns{2}
\tablehead{Parameter & Value\tablenotemark{a}}
\startdata
Magnitude (V) & 8.95\\
Distance (kpc) & 2.10\\
Inclination ($\degr$) & 35\\
Period (Days) & 5.566\\
Eccentricity & 0.0\\
Mass of Star 1 (M$_{\sun}$) & 17.8\\
Mass of Star 2 (M$_{\sun}$) & 10.1\\
Radius of Star 1 (R$_{\sun}$) & 16\\
Radius of Star 2 (R$_{\sun}$) & 0.0\\
T$_{\rm eff}$ of Star 1 (K) & 32000\\
T$_{\rm eff}$ of Star 2 (K) & 0
\enddata
\label{cygx1table}
\tablenotetext{a}{Values from \citet{Herrero1995}}
\end{deluxetable}

\subsubsection{IBs Where the SED is Not Fully Dominated by Either the Primary or Secondary Star}

\paragraph{SS Cygni}

SS Cyg is a bright, well-known, cataclysmic variable. In quiescence, SS Cyg has an apparent magnitude of $V$ = 12.2. During outbursts, it brightens to $V$ $\sim$ 8.8. It has an orbital period of 6.603 hrs, and the secondary star has a spectral type of K5 \citep{Harrison2004}. \citet{Dubus2004} obtained near-simultaneous multi-wavelength photometry of SS Cyg. As shown in Figure 5 of \citet{Harrison2007} the white dwarf and accretion disk dominate the blue end of the SED, while the secondary star becomes prominent in the red and near-infrared. \citet{Harrison2007} modeled the SED of SS Cyg as a combination of the two stellar components plus a free-free accretion disk around the white dwarf. Due to the availability of simultaneous $UBVRIJHK$ photometry, and well-constrained stellar parameters, SS Cyg is an ideal system to investigate what happens when more than one component in the binary system is visible. As we demonstrate, SS Cyg has strongly wavelength-dependent astrometric reflex motions. We only consider the reflex motion in quiescence, since during outburst, the luminosity of the system is completely dominated by the hot (10,000~K), optically thick accretion disk.

\begin{figure}[ht]
\centering
\begin{tabular}{cc}
\epsfig{width=0.475\linewidth,file=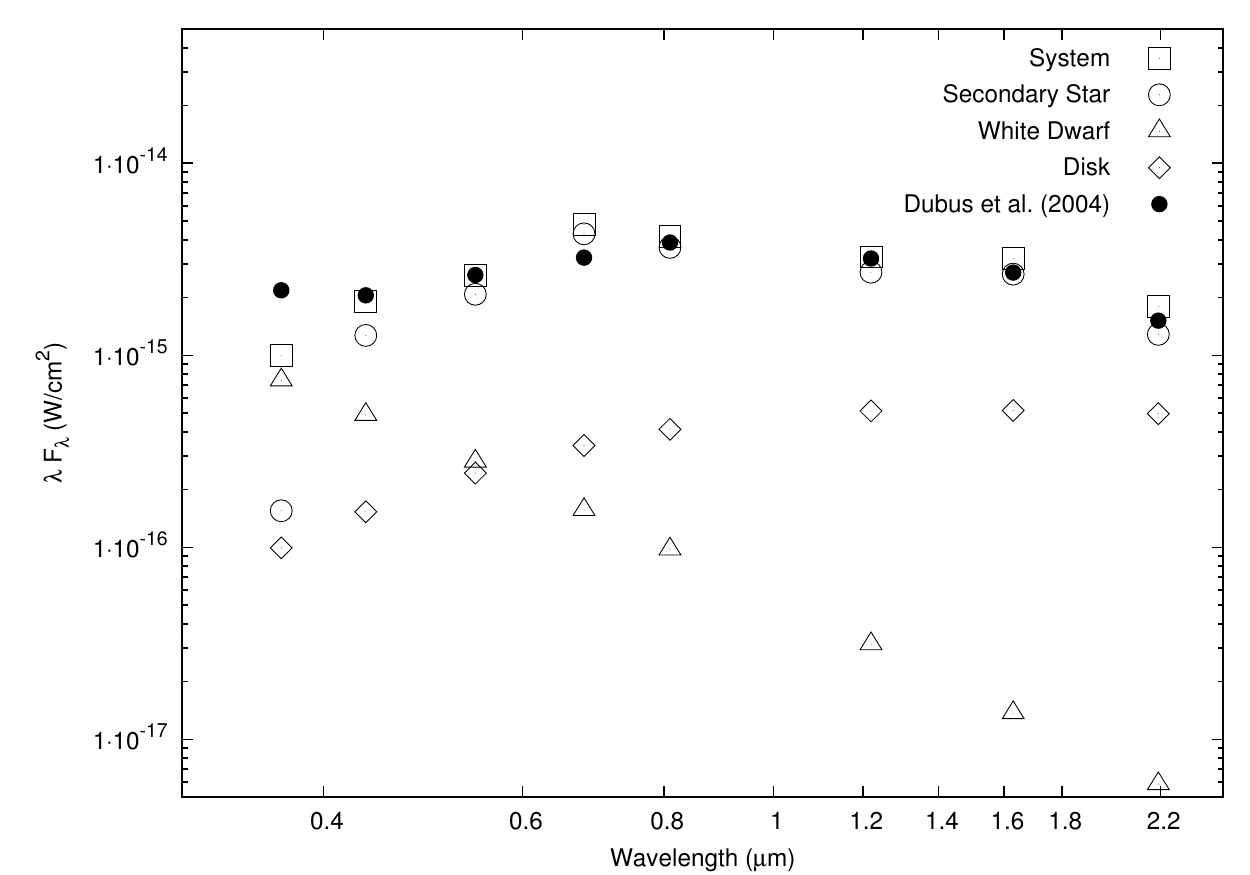} &
\epsfig{width=0.475\linewidth,file=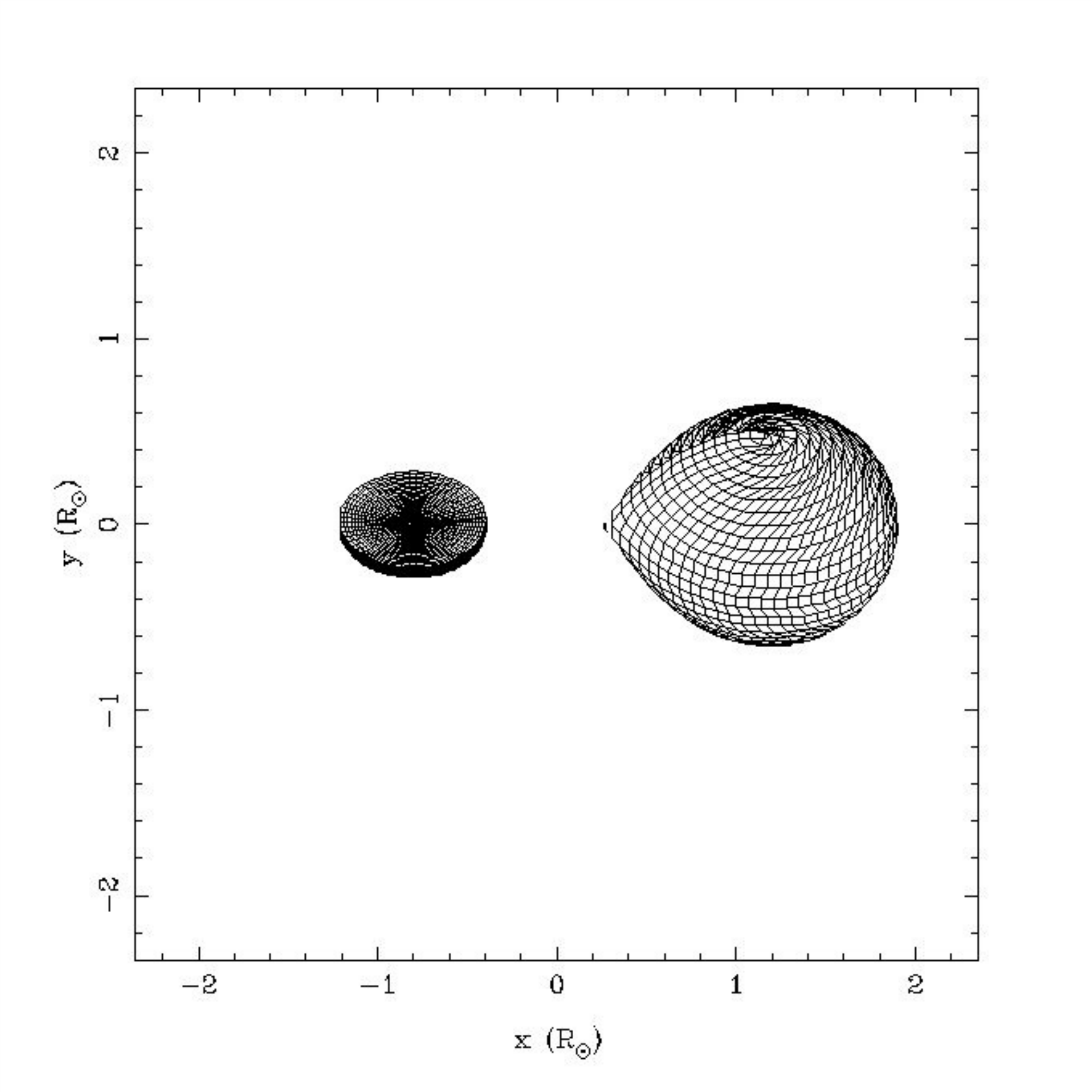} \\
\epsfig{width=0.475\linewidth,file=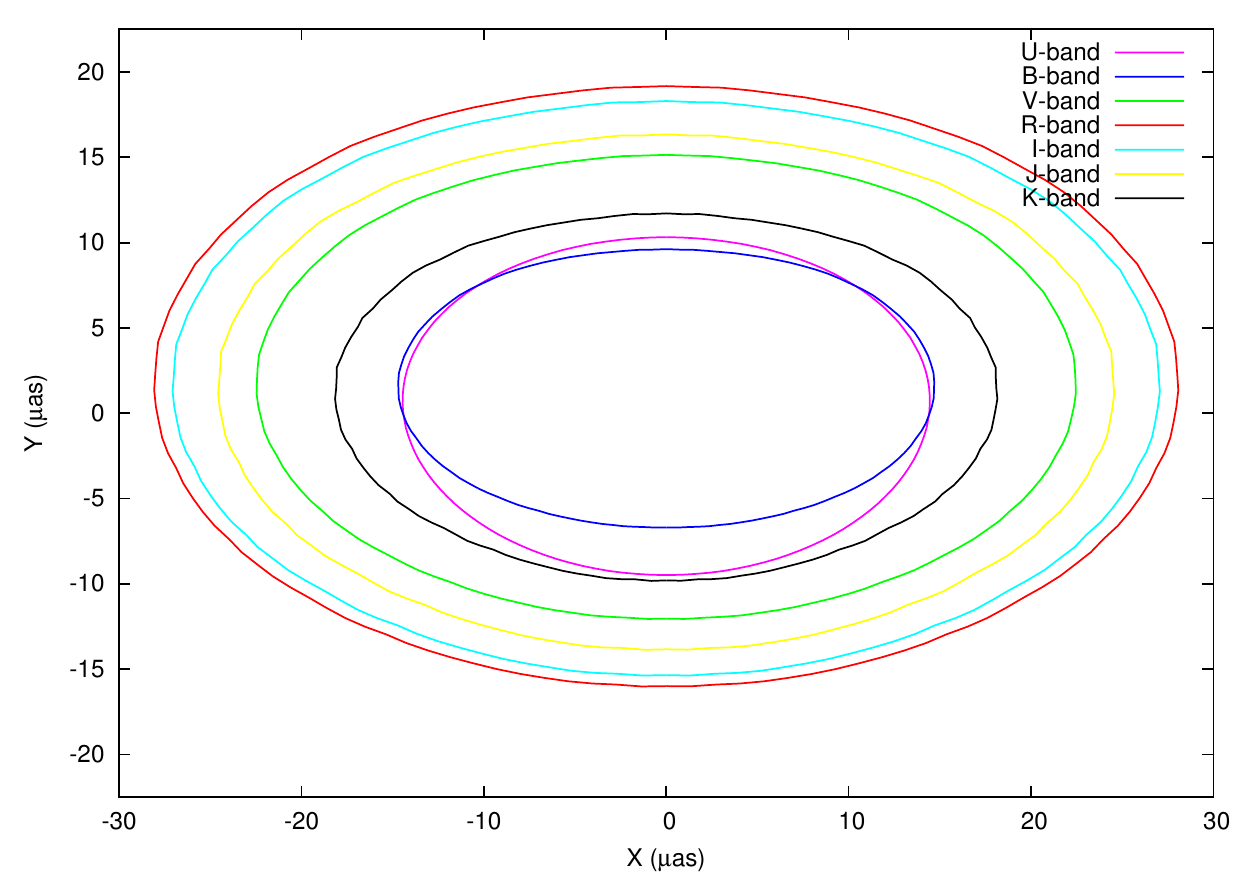} &
\epsfig{width=0.475\linewidth,file=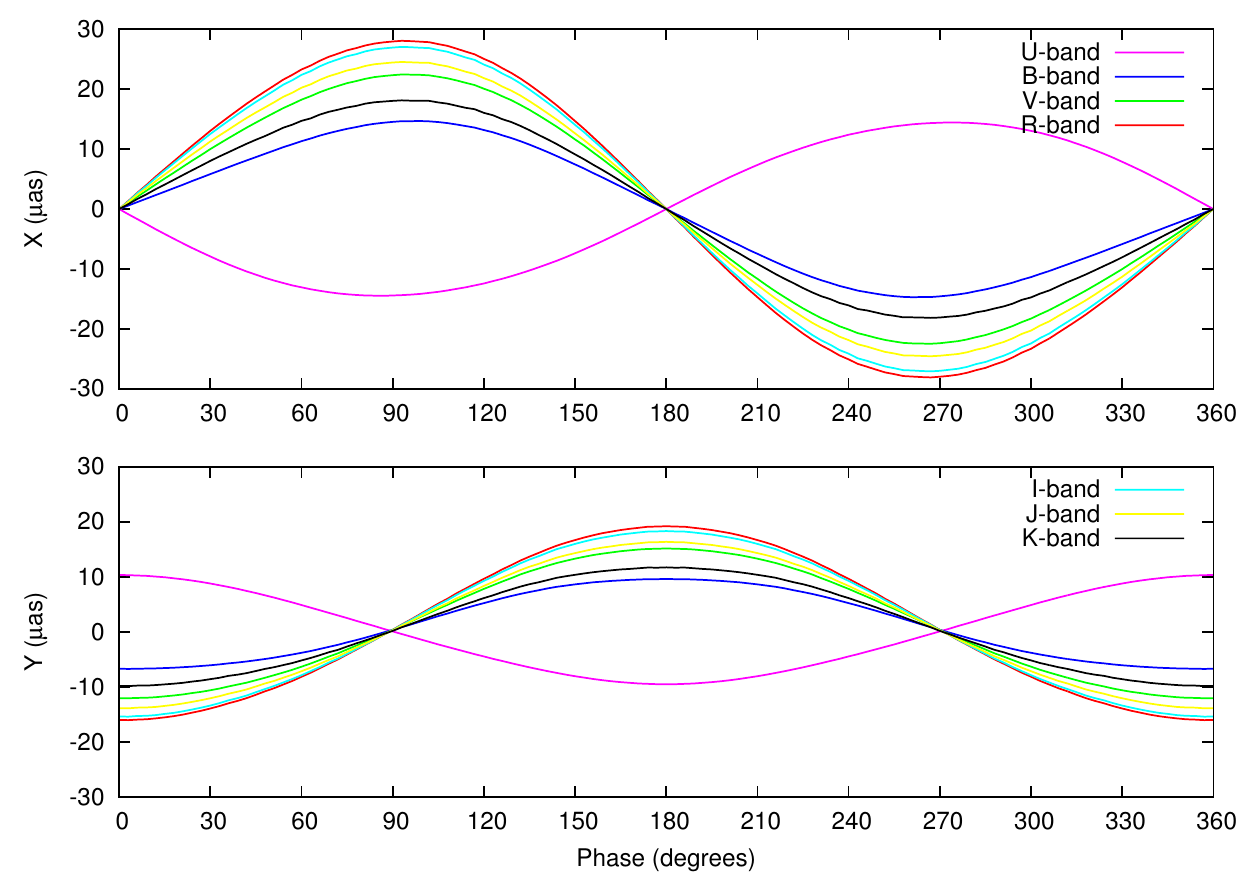}
\end{tabular}
\caption[Multi-wavelength model of SS Cyg]{\emph{Top Left:} SED plot for SS Cyg compared to photometry from \citet{Dubus2004}. \emph{Top Right:} 3D model of SS Cyg at Phase=90$\degr$. \emph{Bottom Left:} Reflex orbit for SS Cyg. \emph{Bottom Right:} X and Y components of the reflex orbit versus phase for SS Cyg.}
\label{sscygmainfig}
\end{figure}

We have generated a model system with the parameters listed in Table~\ref{sscygtable}, where the temperatures and radii of the stellar components have been derived from the literature. In the case of SS Cyg, we used a \textsc{nextgen} model atmosphere for the secondary star, and assumed a blackbody for the white dwarf primary. We attempted both blackbody and free-free accretion disk models, and found that the free-free model provides the best match to the photometry. The results are shown in Fig.~\ref{sscygmainfig}, where the model SED is compared to the photometry, with the contributions from each of the three components shown with separate symbols. As can be seen from these plots, the white dwarf dominates the flux contribution in the $U$-band, and fades thereafter, with the secondary dominating from $B$ through $K$ (though the accretion disk contribution becomes quite important in the infrared). A wire grid representation of the SS Cyg system at phase 0.25 is also shown in Fig.~\ref{sscygmainfig}. As shown in the bottom-left panel, the amplitudes of the reflex motions smoothly increase from $B$ (semi-major axis $a$ = 12 $\mu$as) to $R$ ($a$ = 28 $\mu$as), and then decline at longer wavelengths due to the contribution of the accretion disk. The reflex motion in the $U$-band is larger than in $B$, as it is almost completely dominated by the WD primary. In the $B$-band, the two stars have similar luminosities, and thus the motions are a mixture of the individual astrometric orbits. If there was no accretion disk, and the white dwarf was invisible, the total amplitude of the reflex motions for the secondary star in this system would be $a$ = 34.1 $\mu$as. Thus, the true reflex motions are diluted, and one cannot determine the actual astrometric orbit without modeling the system. However, with proper modeling, one can deconvolve the motion of the white dwarf + accretion disk and the secondary star, thus directly yielding the semi-major axis of each component's orbit, which when combined with the well-known period and distance to the system, directly yields absolute masses for each component, (see \S4).

\begin{deluxetable}{lc}
\tablewidth{0pt}
\tablecaption{Parameters for the SS Cyg System}
\tablecolumns{2}
\tablehead{Parameter & Value\tablenotemark{a}}
\startdata
Magnitude (V) & 8.8 (max) 12.20 (min)\\
Distance (pc) & 159.5\\
Inclination ($\degr$) & 50.5\\
Period (Days) & 0.275130\\
Eccentricity & 0.0\\
Mass of Star 1 (M$_{\sun}$) & 0.555\\
Mass of Star 2 (M$_{\sun}$) & 0.812\\
Radius of Star 1 (R$_{\sun}$) & 0.684\tablenotemark{b}\\
Radius of Star 2 (R$_{\sun}$) & 0.015\\
T$_{\rm eff}$ of Star 1 (K) & 4400\\
T$_{\rm eff}$ of Star 2 (K) & 35000\\
Disk Inner Radius (R$_{\sun}$) & 0.022\\
Disk Outer Radius (R$_{\sun}$) & 0.407\\
Disk Inner Temperature (K) & 10000\tablenotemark{c}\\
Disk Temp Power-Law Exponent & 0.0
\enddata
\tablenotetext{a}{Values from \citet{Bitner2007}}
\tablenotetext{b}{Star fills its Roche Lobe}
\tablenotetext{c}{Disk is free-free}
\label{sscygtable}
\end{deluxetable}

Careful examination of the wavelength-dependent astrometric orbits reveals that the actual center about which the motion occurs is wavelength dependent. Note that the $U$-band reflex motion is centered on (0,0), but at the other wavelengths, the motions are offset from this position. This subtle effect has two causes: 1) the secondary star in SS Cyg fills its Roche lobe, and therefore has a teardrop shape, and 2) gravity brightening for non-degenerate stars is significant. These two effects combine to offset
the center of light of the secondary star from its center of mass. This creates an offset in the astrometric motions in the bandpasses where the secondary star dominates the systemic luminosity. Since the WD is spherical, and does not have gravity brightening effects, the $U$-band motion remains centered at (0,0). For more detailed modeling on the wavelength dependence of the astrometric center due to gravity brightening, see \citet{Coughlin2010b}.

One of the important consequences of these effects is that the actual orbital inclination is harder to derive. Naively, one would assume that the orbital inclination can be calculated from the ratio of the minor and major axes of the orbital ellipse. If we do this in the $R$-band [$i$ = cos$^{\rm -1}$(17.5/28.0)] we derive an orbital inclination of 51.4$^{\circ}$, instead of the input value of 50.5$^{\circ}$. Depending on the bandpass, the orbital inclinations derived for our simulation of SS Cyg range from a minimum of 47$^{\circ}$ in the $U$-band, to a maximum of 51.4$^{\circ}$ in the $R$-band! While it is likely that these differences will be lost in the astrometric noise for most IB systems, there will be a small number of IBs (and many non-interacting binary systems) where this difference should be detectable, and proper modeling is required to accurately extract the system's inclination.

With respect to observing this system with SIM Lite, according to DAPE, even in its low state with V = 12.1, a total mission time of 6 hours and 40 minutes would provide 10 measurements with respect to an astrometric standard star, each with 1.7 $\mu$as precision and an individual integration time of 20 minutes per measurement, each composed of 20 chops between the target and reference stars, with 1 minute exposures on the target and 30 second exposures on the reference. As a minimum of 7 points are needed to solve for a full astrometric orbit, (given that 7 parameters describe an orbit), and as the orbital period of SS Cyg is $\sim$6.6 hours, this would successfully measure the full astrometric orbit with limited orbital smearing, (see \S\ref{solvesec}).

SS Cyg can be used as a template for cases of CVs and LMXBs where the secondary star is clearly visible in the optical. It simply requires changing the input values of the stellar components, the relative prominence (and nature) of the accretion disk, and altering the orbital inclination. Other changes, such as orbital period, mass ratio, and distance, act to scale the amplitude of the reflex motions. However, it remains critically important to have simultaneous multi-wavelength photometry, and a reasonable handle on the system properties to derive accurate astrometric orbits for these types of objects. As one would expect, the wavelength dependent behavior of the astrometric motions for these types of systems is more complex than for systems with only a single, visible component, but the potential scientific payoff is much higher. When varying the inclination angle for a SS Cyg type system, the y component of the astrometric motion decreases with increasing inclination angle for all wavelengths as expected, but at large inclination angles eclipse effects produce significant deviations from simple sinusoidal reflex motions.

\subsubsection{IBs with Accretion Disk Dominated SEDs}

The majority of LMXBs, and many CVs, have their optical light dominated by the accretion disk. The accretion disk surrounds the compact object, and thus the observed astrometric motion is that of the compact object, although there are some complications arising from the extended structure of the disk. Here we examine two cases, the short period CV system V592 Cas, and the well-known LMXB Sco X-1.

\paragraph{V592 Cas}

\begin{figure}[ht]
\centering
\begin{tabular}{cc}
\epsfig{width=0.475\linewidth,file=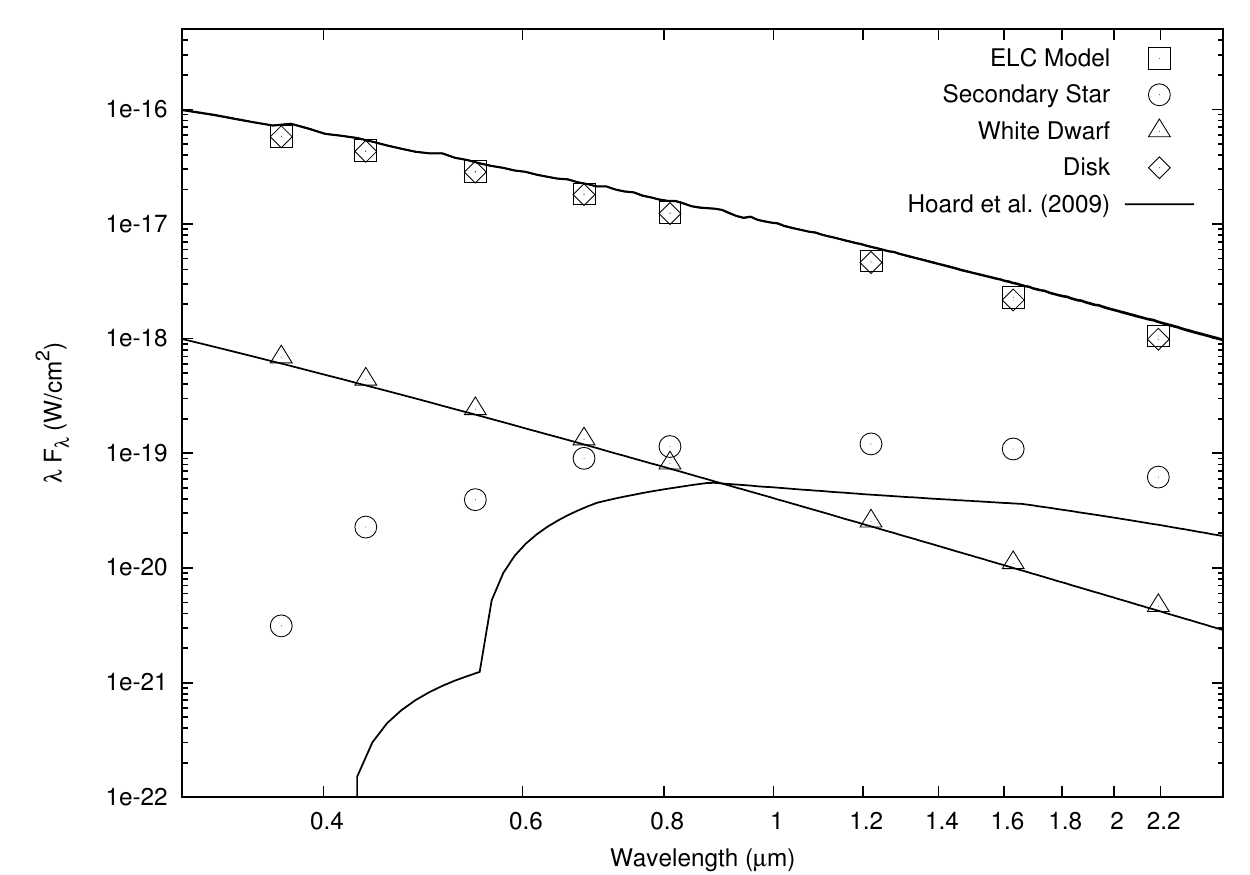} &
\epsfig{width=0.475\linewidth,file=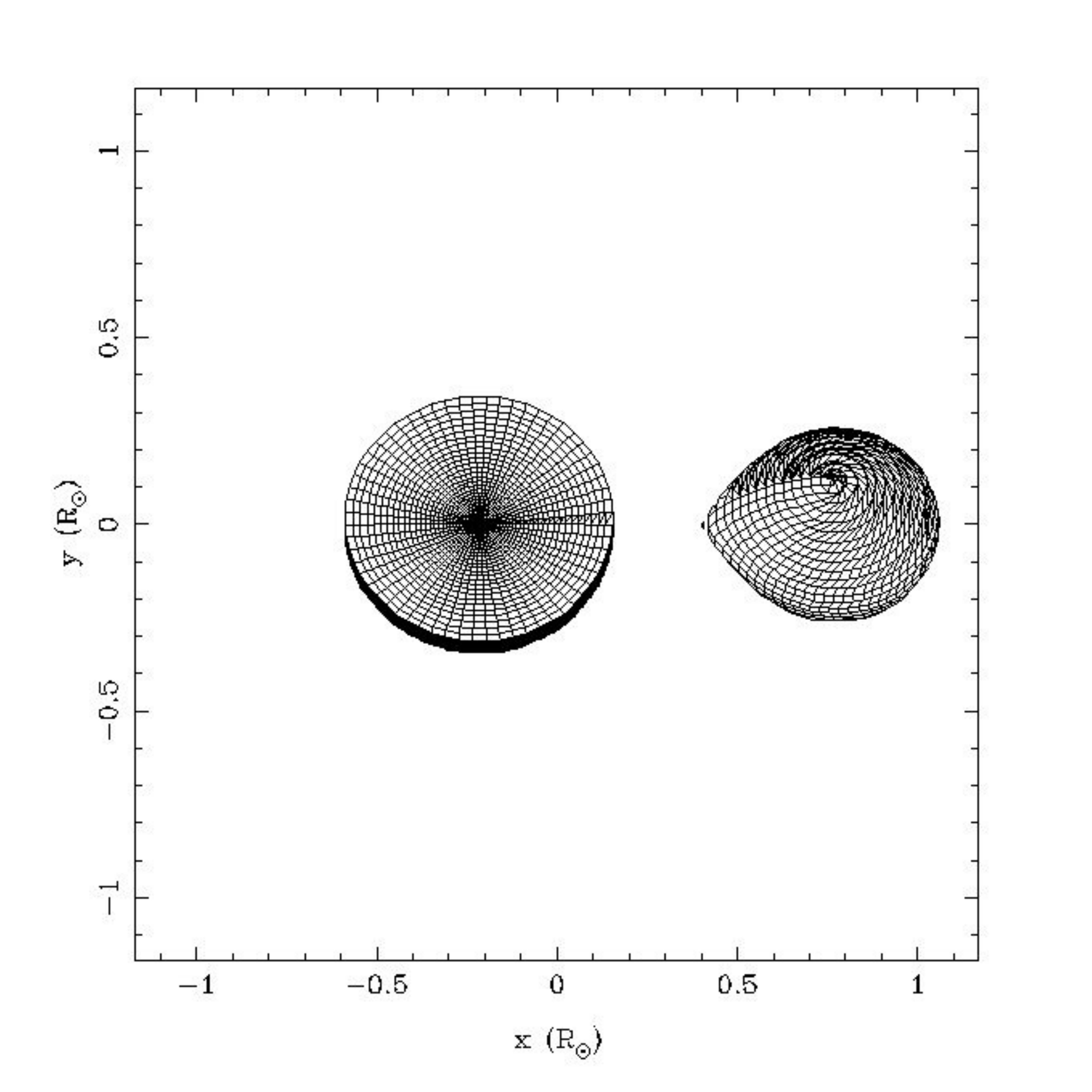} \\
\epsfig{width=0.475\linewidth,file=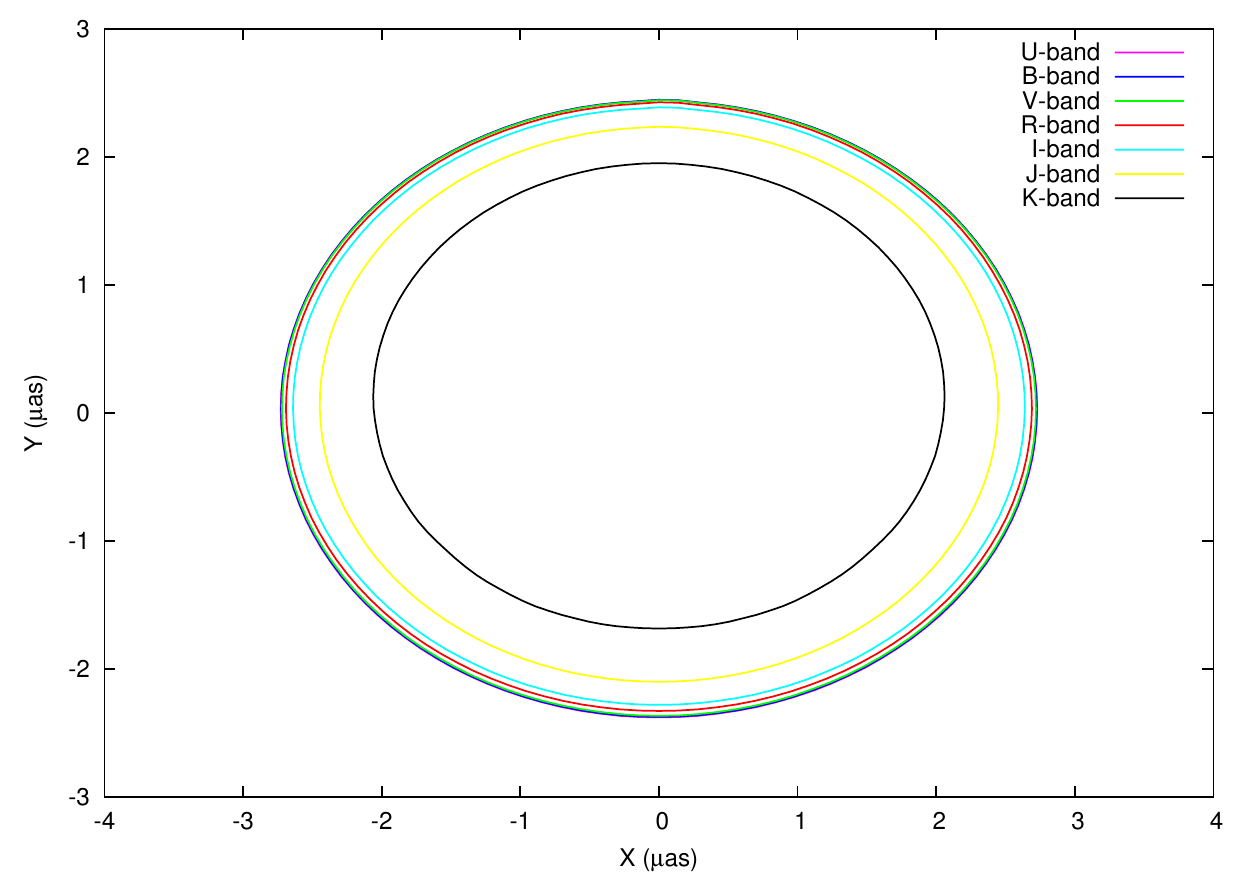} &
\epsfig{width=0.475\linewidth,file=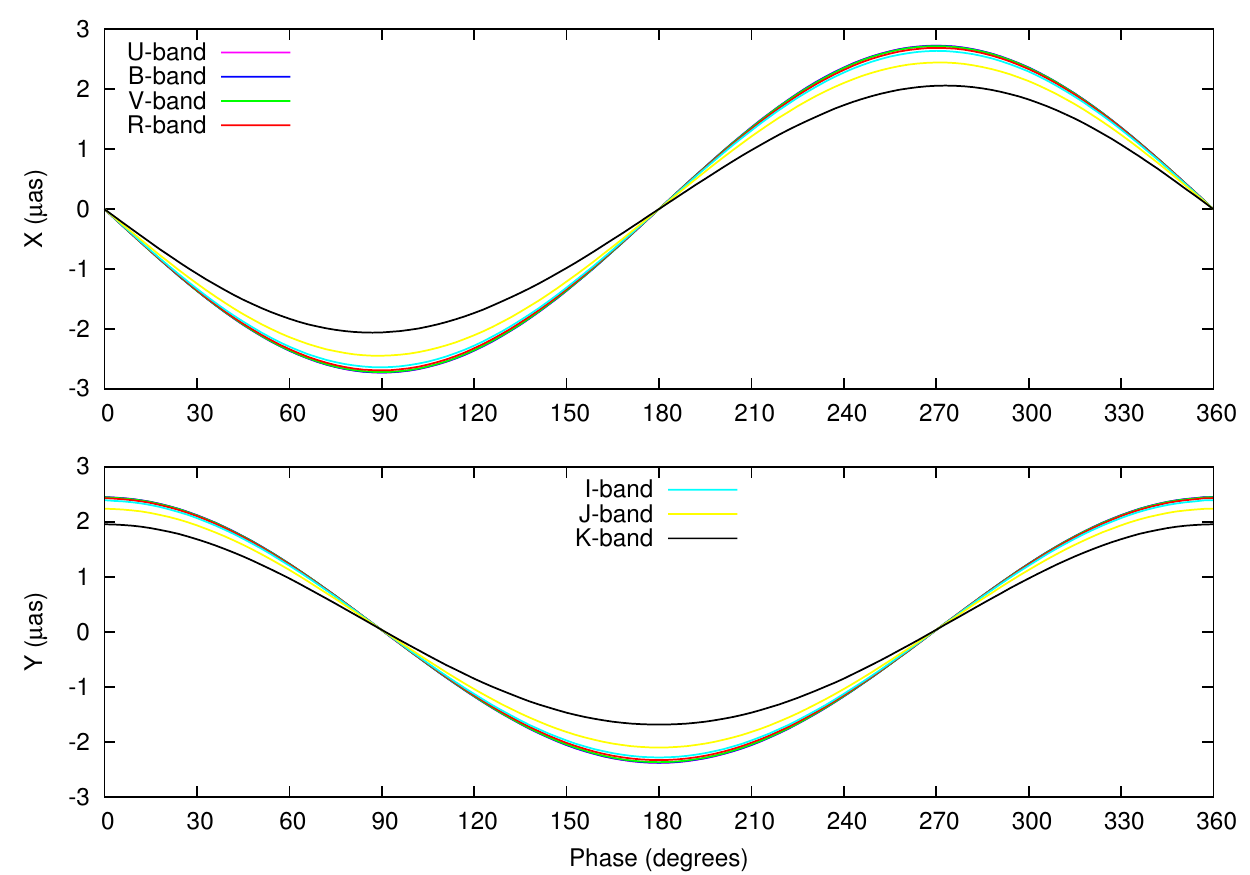}
\end{tabular}
\caption[Multi-wavelength model of V592 Cas]{\emph{Top Left:} SED plot for V592 Cas compared to models from \citet{Hoard2009}. \emph{Top Right:} 3D model of V592 Cas at phase 0.5. \emph{Bottom Left:} Reflex orbit for V592 Cas. \emph{Bottom Right:} X and Y components of the reflex orbit versus phase for V592 Cas.}
\label{v592casmainfig}
\end{figure}

V592 Cas is a short-period (P$_{\rm orb}$ = 2.76 hr) CV  system consisting of a Roche lobe-filling M dwarf, a very hot white dwarf, and a disk that dominates the luminosity at optical and near-infrared wavelengths. The majority of short period CVs closely resemble V592 Cas, and thus it can act as a prototype for disk-dominated CVs. The SED of the system has been investigated by \citet{Hoard2009}, who reproduced the observed photometry using a model that included a red dwarf, white dwarf, and a two component blackbody disk around the white dwarf consisting of an inner flat component and an outer flared component. To model the mid-infrared excess, a circumbinary dust disk was also included. This dust disk is only important at mid-infrared wavelengths, and thus can be ignored in this present study. We model the reflex motions of V592 Cas using the parameters from \citet{Hoard2009} shown in Table~\ref{v592castable}, while using a single component blackbody disk model with a flare inclination that is the average of the angles in their two component disk model. The SED produced by \textsc{reflux} is compared to that of \citet{Hoard2009} in Fig. \ref{v592casmainfig}, where the contributions from each component are also shown. We find a greater contribution of flux from the M dwarf compared to \citet{Hoard2009}, by a factor of $\sim$1.75. This is most likely due to the fact that \citet{Hoard2009} used the spectrum of a field M5.0V dwarf, while our model, which incorporates full Roche geometry, produces a star with the same effective temperature, but a larger surface area by the same factor of $\sim$1.75, adopting the radius for an isolated M5.0V as 0.21 R$_{sun}$, as shown by recent measurements of low-mass stars \citep[c.f.,][]{LopezMorales2007}. Either way, the M dwarf is a few percent of the total system flux, and thus does not greatly affect the derived astrometric measurements.

\begin{deluxetable}{lc}
\tablewidth{0pt}
\tablecaption{Parameters for the V592 Cas System}
\tablecolumns{2}
\tablehead{Parameter & Value\tablenotemark{a}}
\startdata
Magnitude (V) & 12.8\\
Distance (pc) & 364.0\\
Inclination ($\degr$) & 28.0\\
Period (Days) & 0.115063\\
Eccentricity & 0.0\\
Mass of Star 1 (M$_{\sun}$) & 0.210\\
Mass of Star 2 (M$_{\sun}$) & 0.751\\
Radius of Star 1 (R$_{\sun}$) & 0.270\tablenotemark{b}\\
Radius of Star 2 (R$_{\sun}$) & 0.014\\
T$_{\rm eff}$ of Star 1 (K) & 3030\\
T$_{\rm eff}$ of Star 2 (K) & 45000\\
Disk Inner Radius (R$_{\sun}$) & 0.0106\\
Disk Outer Radius (R$_{\sun}$) & 0.371\\
Disk Inner Temperature (K) & 109,700\tablenotemark{c}\\
Disk Temp Power-Law Exponent & $-$0.75
\enddata
\tablenotetext{a}{Values from \citet{Hoard2009}}
\tablenotetext{b}{Star fills its Roche Lobe}
\tablenotetext{c}{Disk is blackbody}
\label{v592castable}
\end{deluxetable}

As seen in Figure~\ref{v592casmainfig}, due to the domination by the disk, the system shows less wavelength dependent motion than SS Cyg. However, there is still a discernible effect towards longer wavelengths as the secondary becomes a more significant source of flux. Thus, with enough measurements, one could in theory derive the individual component masses. The influence of the orbital inclination is readily apparent, and is uncontaminated by the secondary star gravity effects that were visible in our models for SS Cyg. With respect to SIM Lite observing time, this system would require a moderate amount of SIM Lite mission time, given its small amplitude and short period. According to DAPE, if we constrain individual integration times to 15 minutes, (0.1 of the orbital period), to prevent orbital smearing, one could obtain 2.3 $\mu$as precision per visit (composed of 20 target-reference chops of 45 second target integration each), which is on par with the amplitude of the observed reflex motions. Thus, in comparison to SS Cyg and Cyg X-1, discussed above, one might need more than 10 measurements to build up signal-to-noise and really nail down the system parameters. 30 measurements would take a total mission time of 17.5 hours, and thus we can say with reasonable confidence that this system would require less than a day of total mission time to observe so that the astrometric parameters could be obtained with reasonable precision.

We decided to use this system to investigate the possibility of using the multi-wavelength astrometric capability of SIM Lite to discern two more subtle properties of disk-dominated systems: the disk temperature gradient and disk hotspots. The disk model employed in ELC allows for an inner disk temperature to be specified, along with a power law exponent, so that the disk temperature will vary radially according to the formula \begin{equation} T(r) = T_{inner}\cdot(\frac{r}{R_{inner}})^{\chi}, \end{equation} where T is the temperature at a given distance from the compact object, $r$, given a temperature T$_{inner}$ at the innermost radius, R$_{inner}$, and a power law exponent $\chi$ \citep{Orosz2000}. Usually a value of $\chi$ = $-$0.75 is assumed for a steady-state disk \citep{Pringle1981}, but we wanted to test if changing $\chi$ to $-$0.5 or $-$1.0, (which physically reflect a centrally irradiated disk \citep{Friedjung1985,Vrtilek1990,Bell1999} and its corresponding extrema), would have a discernible effect on the wavelength dependent astrometric reflex motion. Thus, we produced two models, one with $\chi$ = $-$0.5 and the other with $\chi$ = $-$1.0, while also adjusting the inner disk temperature so that the resulting SED best matched that with $\chi$ = $-$0.75. Neither the SEDs nor the reflex motions differed significantly from the models that used $\chi$ = $-$0.75. Thus, it will not be possible to constrain the overall temperature gradient in accretion disks using astrometry.

\paragraph{Sco X-1}

In many CV systems, the optical light is orbitally modulated due to the presence of a hotspot on the outer edge of the accretion disk. This hotspot is where the accretion stream from the secondary impacts the disk. These spots typically have temperatures of 20,000 K, and can produce modulations on order of $\pm$50\% in visible bandpasses \citep{Mason2002}. Because this feature is on the outer edge of the disk, it has the potential to distort the reflex motions in systems where it is prominent. In addition, due to its higher temperature than the surrounding disk, the hotspot could be the source of strong line emission, especially in HI or HeI lines \citep{Skidmore2002}. Thus, its detection might be isolated, and the reflex motions amplified, using very narrow bandpasses centered on the strongest emission lines. To investigate the effects of a hotspot, we used the same V592 Cas model as above, but added a hotspot (20,000 K) on the outer edge of the accretion disk that leads the secondary by 30$\degr$ in phase, and that is 30\% of the $V$-band flux.  The differences between a model with a spot, and one without are extremely slight, with a maximum difference in models with and without a spot of 0.15 $\mu$as. Increasing the inclination of the system does not have a significant effect. To be even remotely detectable, V592 Cas would have to be $\sim$3 times closer, making the spot signature $\sim$0.5 $\mu$as, although it would still only be $\sim$5\% of the reflex amplitude signature. Unless an IB has a hotspot that is much more dominant, such features will remain undetectable.

Sco X-1 is the proto-type LMXB consisting of a neutron star accreting from a low-mass companion. The X-ray luminosity of Sco X-1 is very close to the Eddington limit for a 1.4 M$_{\sun}$ neutron star. The secondary star in Sco X-1 has never been directly detected, though an orbital period of 0.787 days was detected through the analysis of 85 years of visual photometry \citep{Gottlieb1975}, and from radial velocity measurements \citep{LaSala1985}. \citet{Steeghs2002} detected narrow HI emission lines that they showed are from the irradiated secondary star. From these data they estimate masses for the components in this system of M$_{\rm 1}$ = 1.4 M$_{\sun}$ and M$_{\rm 2}$ = 0.42 M$_{\sun}$, thus making the secondary star a significantly evolved subgiant. Since the accretion disk dominates the spectrum, we have simply assumed a Roche lobe-filling M type secondary star corresponding to the observed mass, and an invisible neutron star. The stellar and accretion disk parameters are listed in Table~\ref{scox1table}, and the accretion disk model we employ is similar to that for V592 Cas.

\begin{deluxetable}{lc}
\tablewidth{0pt}
\tablecaption{Parameters for the Sco X-1 System}
\tablecolumns{2}
\tablehead{Parameter & Value\tablenotemark{a}}
\startdata
Magnitude (V) & 11.1 (max) 14.1 (min) \\
Distance (kpc) & 2.80\\
Inclination ($\degr$) & 50.0\\
Period (Days) & 0.7875\\
Eccentricity & 0.0\\
Mass of Star 1 (M$_{\sun}$) & 0.42\\
Mass of Star 2 (M$_{\sun}$) & 1.4\\
Radius of Star 1 (R$_{\sun}$) & 1.473\tablenotemark{b}\\
Radius of Star 2 (R$_{\sun}$) & 0.000\\
T$_{\rm eff}$ of Star 1 (K) & 4500\\
T$_{\rm eff}$ of Star 2 (K) & 0\\
Disk Inner Radius (R$_{\sun}$) & 0.026\\
Disk Outer Radius (R$_{\sun}$) & 1.011\\
Disk Inner Temperature (K) & 100 000\tablenotemark{c}\\
Disk Temp Power-Law Exponent & $-$0.75
\enddata
\tablenotetext{a}{Values from \citet{Steeghs2002}, \citet{Gottlieb1975}, and \citet{LaSala1985}}
\tablenotetext{b}{Star fills its Roche Lobe}
\tablenotetext{c}{Disk is blackbody}
\label{scox1table}
\end{deluxetable}

\begin{figure}[ht]
\centering
\begin{tabular}{cc}
\epsfig{width=0.475\linewidth,file=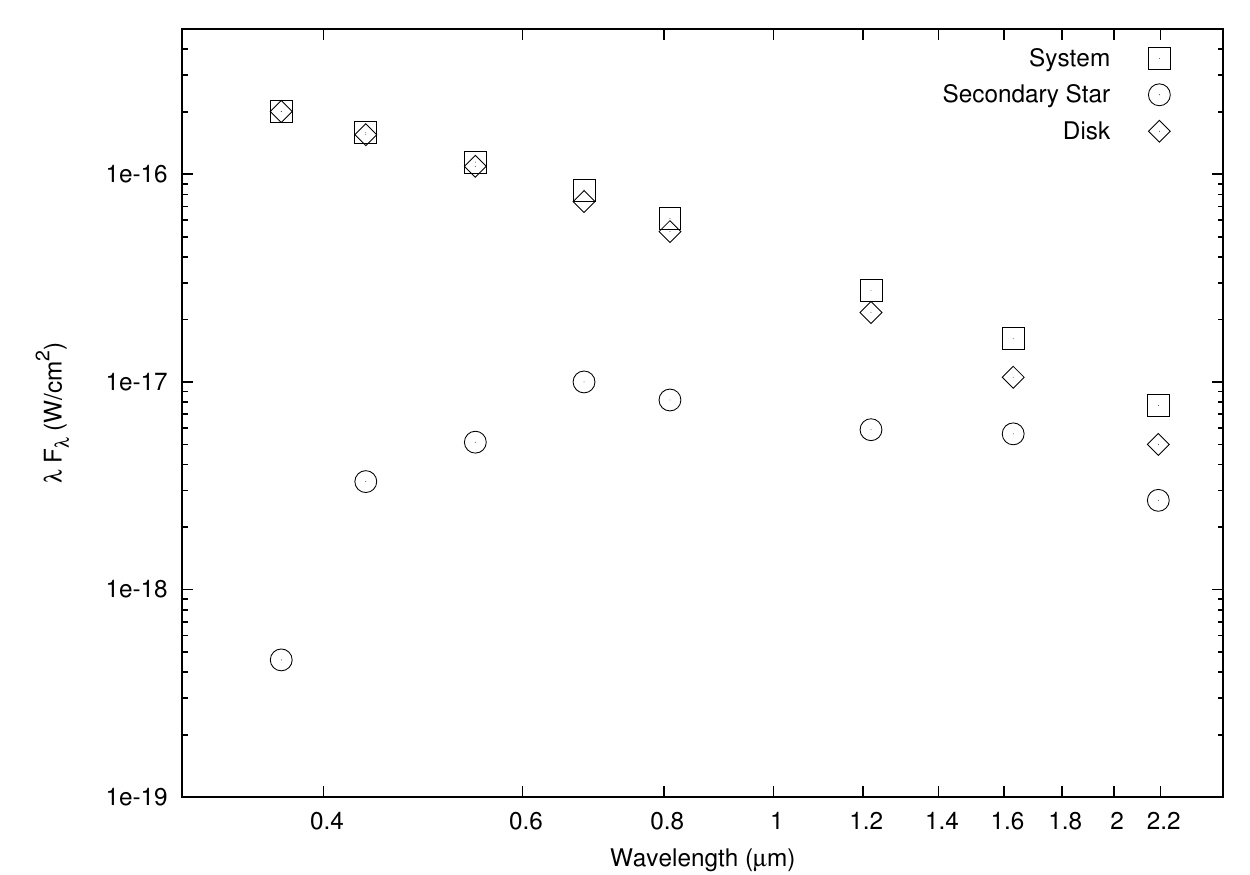} &
\epsfig{width=0.475\linewidth,file=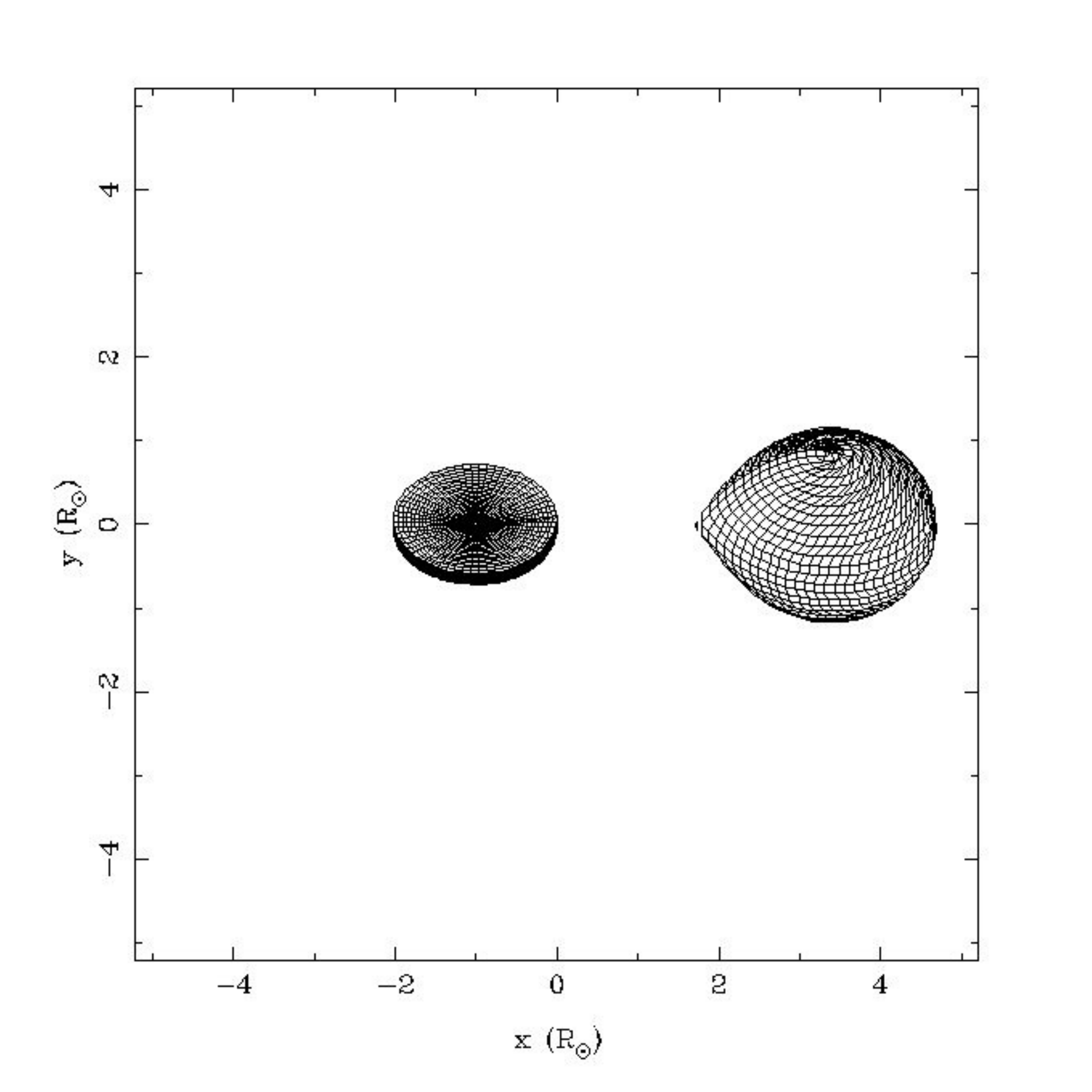}\\
\epsfig{width=0.475\linewidth,file=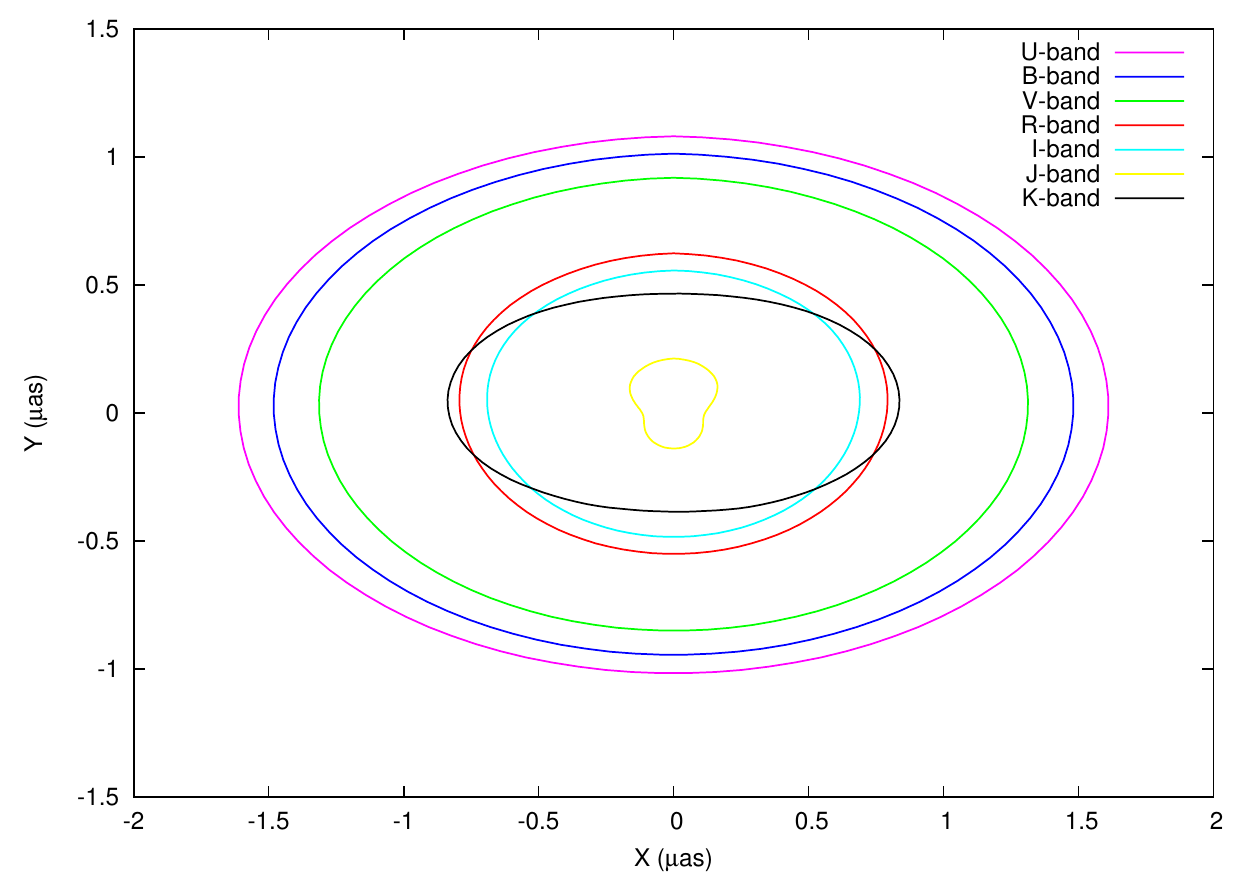} &
\epsfig{width=0.475\linewidth,file=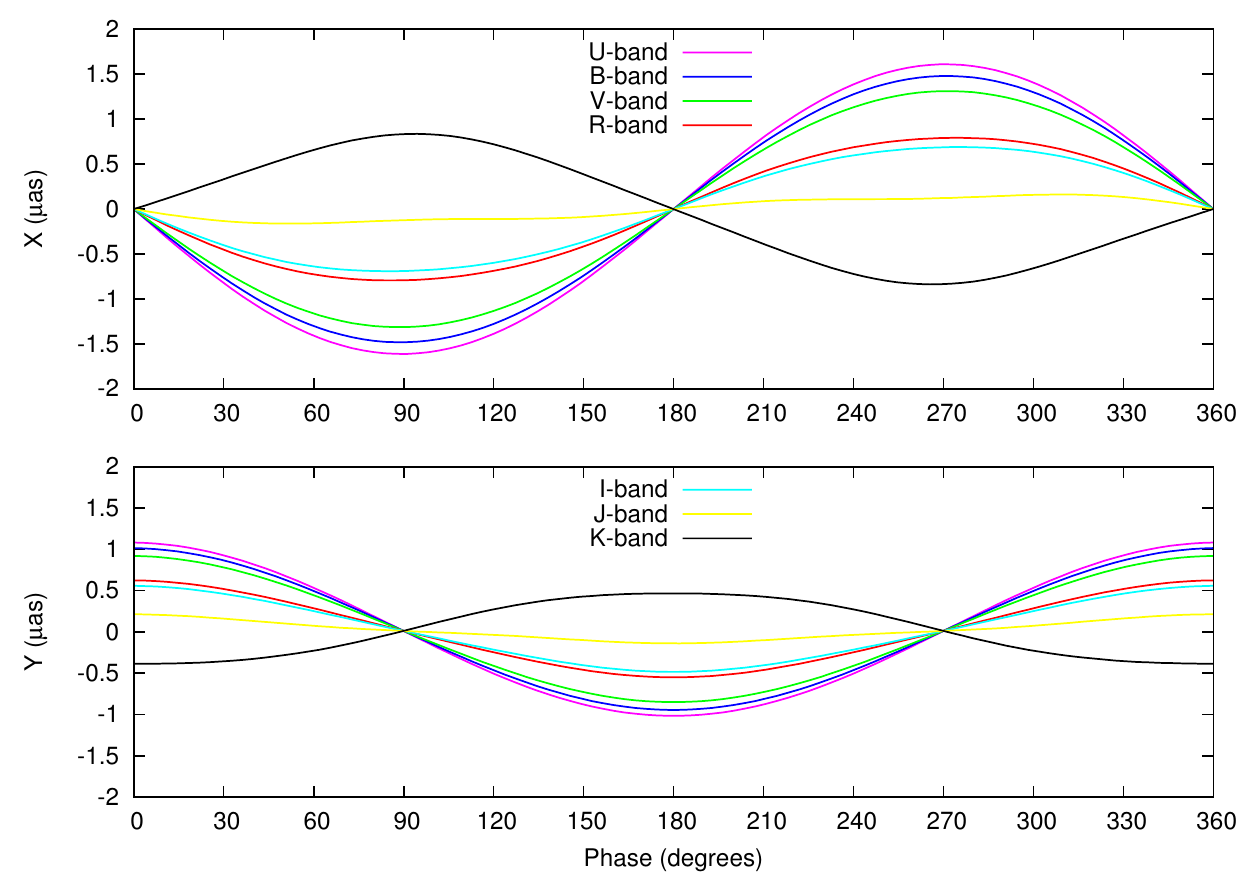}
\end{tabular}
\caption[Multi-wavelength model of Sco X-1]{\emph{Top Left:} SED plot for Sco X-1. \emph{Top Right:} 3D model of Sco X-1 at Phase=90$\degr$. \emph{Bottom Left:} Reflex orbit for Sco X-1. \emph{Bottom Right:} X and Y components of the reflex orbit versus phase for Sco X-1.}
\label{scox1mainfig}
\end{figure}

As seen in Figure~\ref{scox1mainfig}, the accretion disk dominates until the near-infrared when the secondary star begins to contribute. Thus, as also seen in Figure~\ref{scox1mainfig}, the reflex motion has a weak wavelength dependence, but SIM Lite cannot observe in the near infrared where the reflex motion reverses, and thus it will be difficult to disentangle the components with such a weak wavelength dependence in the optical. However, it might be possible to recover the contribution to the astrometric reflex motion for systems with slightly more prominent secondary stars (e.g., Cyg X-2). As seen in Figure~\ref{scox1mainfig}, Cyg X-1 has a very small reflex motion, only $\sim$1.25 $\mu$as, and can get relatively faint, (V = 14.1 at its faintest), and thus this system is just at the limit of SIM Lite's capabilities. Adopting the faintest magnitude of 14.1 of the system to be conservative, and limiting ourselves to observations of 110 minutes, (0.1 of the orbital period), to prevent orbital smearing, we find via DAPE that we can achieve individual measurements with precisions of 2.2 $\mu$as, almost twice the semi-major axis of the system. Thus, assuming we would need at least 50 measurements to begin to derive a reasonable orbit for this system, this system would require at least 4 days of SIM Lite mission time, which may prove more expensive than the scientific payoff justifies. If we are able to observe in its high state when V = 11.1, we could achieve 30 measurements with 1.5 $\mu$as precision in only 20 hours of mission time, which is far more reasonable.  When varying the inclination of a Sco X-1 type system, the contribution from the secondary star becomes more significant as more of the disk becomes self-eclipsed and its total luminosity decreases, and thus alters the individual wavelength reflex motions accordingly.

\subsubsection{Polars and the Astrometric Signature of Magnetic Threading Regions}

\paragraph{AR UMa}

AR UMa is a ``polar'', a CV with a highly magnetic WD primary and a Roche lobe filling M5/6 dwarf \citep{Harrison2005}. As shown in \citet{Szkody1999}, the $V$-band light curve is complex, being dominated by phase-dependent cyclotron emission. The near-IR light curve shows normal ellipsoidal variations, allowing \citet{Howell2001a} to estimate an orbital inclination of 70$\degr$. The magnetic field strength in AR UMa has been estimated to be as high as B = 240 MG, though \citet{Howell2001a} argue for a lower value of 190 MG. In these systems, at some point after material from the secondary star has passed through the inner Lagrangian point, it is captured by the magnetic field of the white dwarf, producing a ``magnetic threading region'' that can be a significant source of luminosity in H$_{\alpha}$. We investigate the possibility of astrometrically detecting this region via the following modeling process. We model the system as a white dwarf + red dwarf system without an accretion disk, using the parameters shown in Table~\ref{arumatable}, but for \emph{only} the  H$_{\alpha}$ bandpass add a narrow ``accretion'' disk halfway between the inner Lagrangian point and the WD. This disk then has a hot spot leading the inner Lagrangian point by 45$\degr$ in phase. The actual contribution from the disk has been made to be completely negligible, but the spot itself makes the system 25\% brighter in H$_{\alpha}$. In effect, it is an isolated emission spot located between the WD and secondary star, mimicking a threading region.

\begin{deluxetable}{lc}
\tablewidth{0pt}
\tablecaption{Parameters for the AR UMa System}
\tablecolumns{2}
\tablehead{Parameter & Value\tablenotemark{a}}
\startdata
Magnitude (V) & 14.50 (max) 18.00 (min)\\
Distance (pc) & 85.0\\
Inclination ($\degr$) & 65.0\\
Period (Days) & 0.0805\\
Eccentricity & 0.0\\
Mass of Star 1 (M$_{\sun}$) & 0.70\\
Mass of Star 2 (M$_{\sun}$) & 1.40\\
Radius of Star 1 (R$_{\sun}$)\tablenotemark{b} & 0.322\\
Radius of Star 2 (R$_{\sun}$) & 0.013\\
T$_{\rm eff}$ of Star 1 (K) & 3200\\
T$_{\rm eff}$ of Star 2 (K) & 35000
\enddata
\tablenotetext{a}{Values from \citet{Howell2001a} and \citet{Harrison2005}}
\tablenotetext{b}{Star fills its Roche Lobe}
\tablenotetext{c}{Disk is free-free}
\label{arumatable}
\end{deluxetable} 
\clearpage

\begin{figure}[ht]
\centering
\begin{tabular}{cc}
\epsfig{width=0.475\linewidth,file=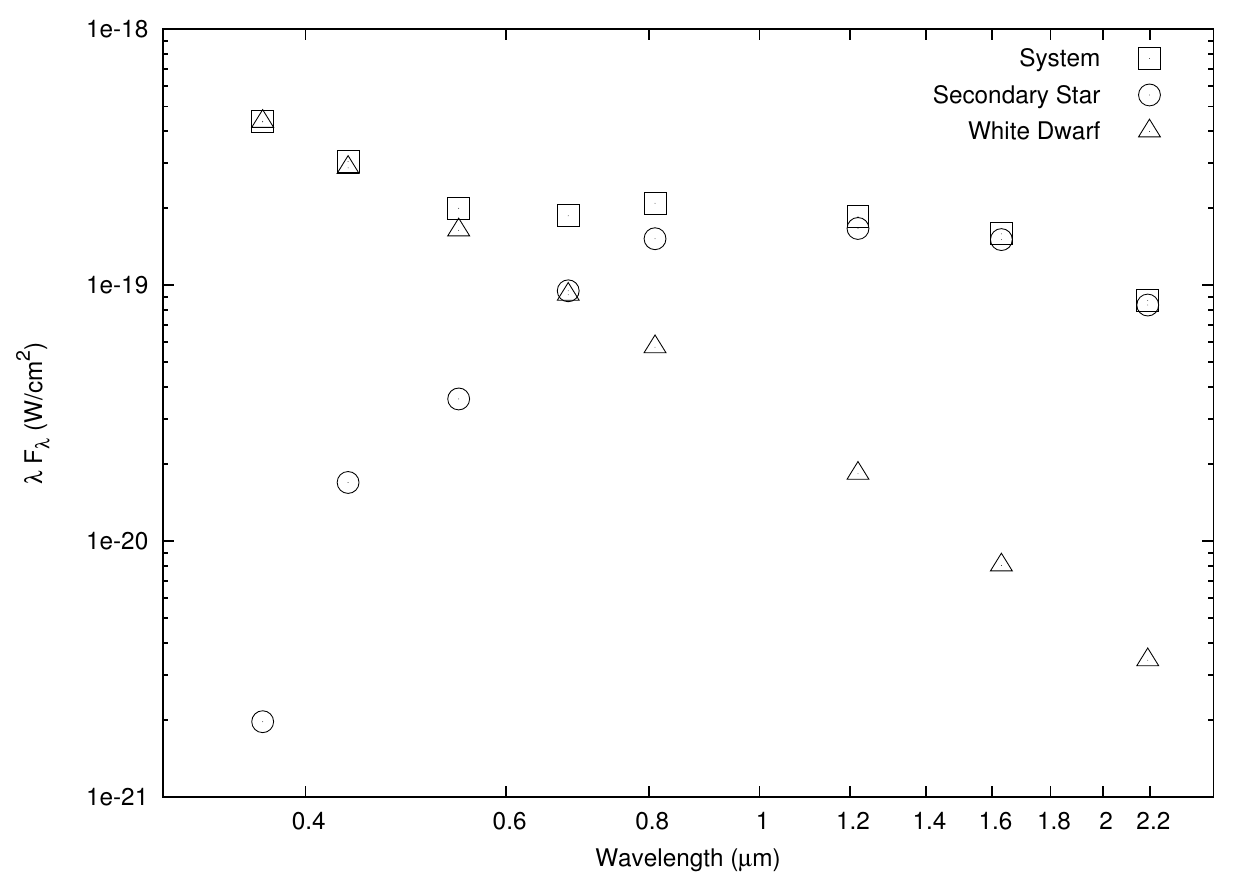} &
\epsfig{width=0.475\linewidth,file=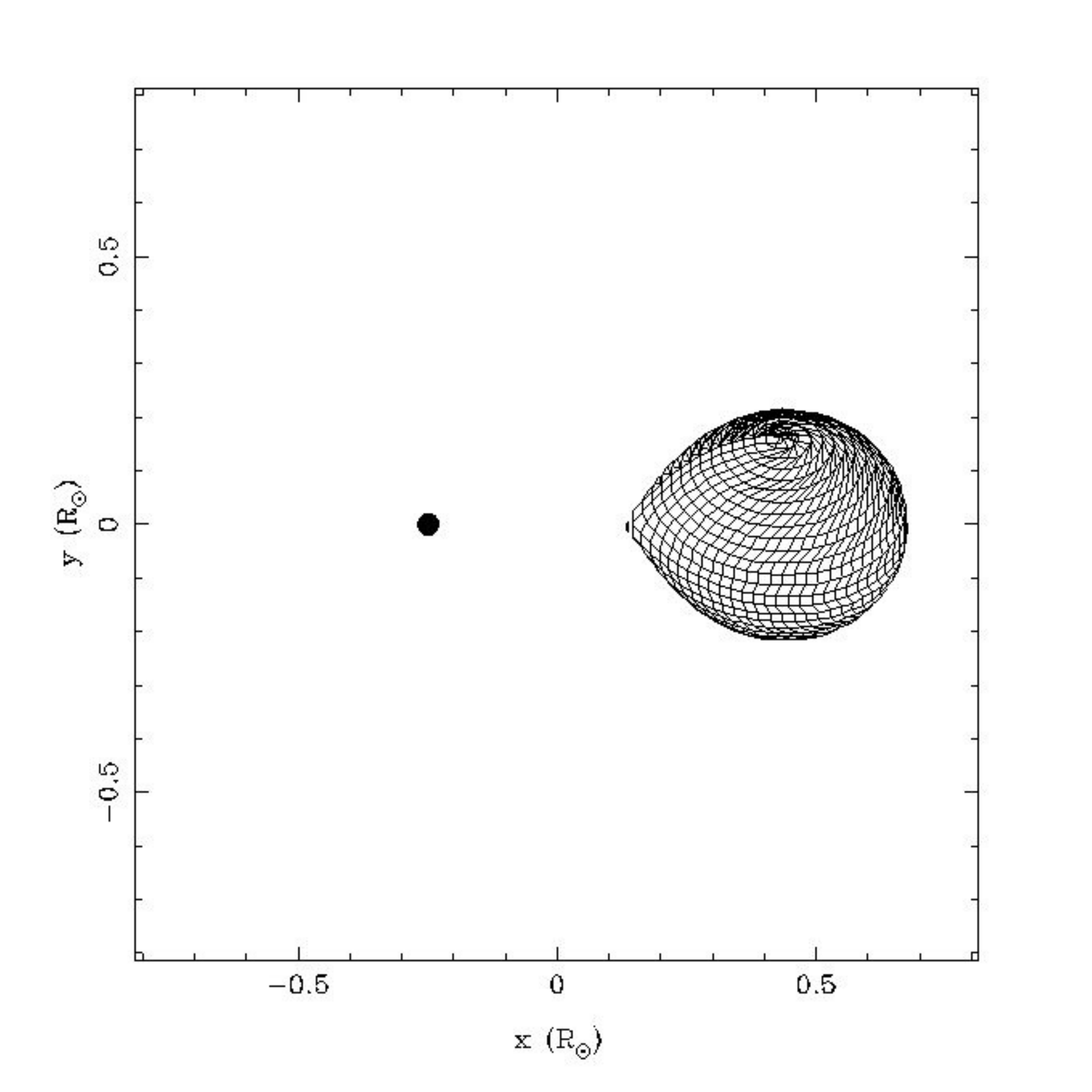} \\
\epsfig{width=0.475\linewidth,file=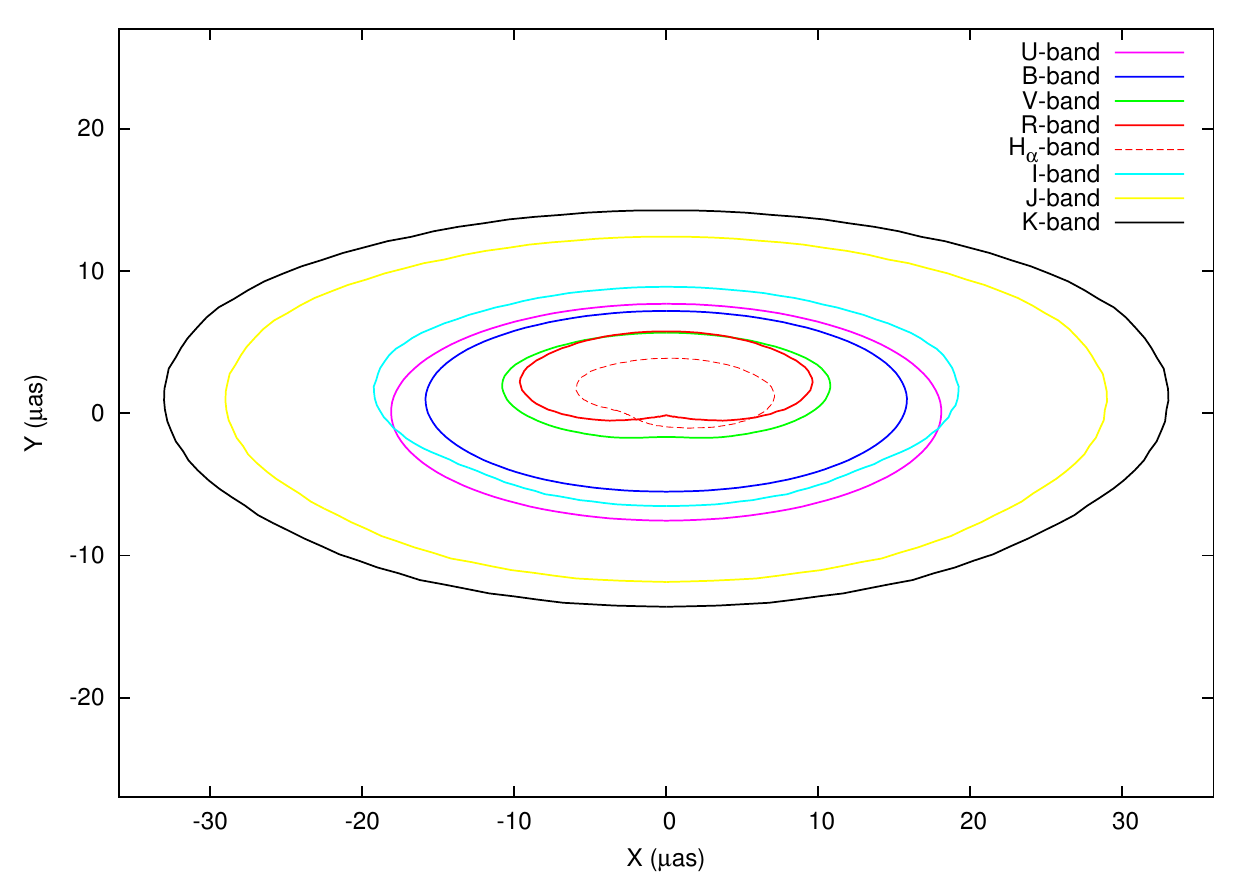} &
\epsfig{width=0.475\linewidth,file=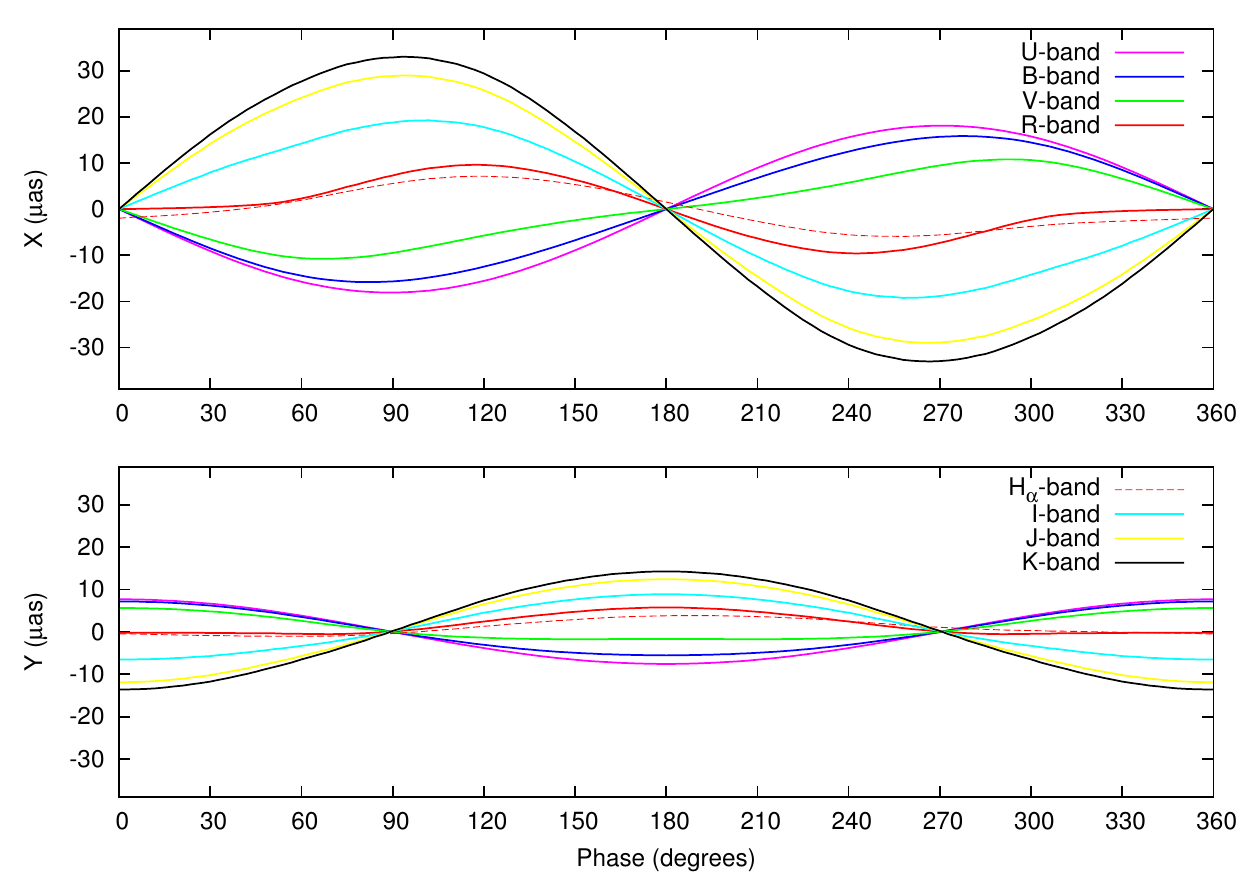}
\end{tabular}
\caption[Multi-wavelength model of AR UMa]{\emph{Top Left:} SED plot for AR UMa. \emph{Top Right:} 3D model of AR UMa at Phase=90$\degr$. \emph{Bottom Left:} Reflex orbit for AR UMa. \emph{Bottom Right:} X and Y components of the reflex orbit versus phase for AR UMa.}
\label{arumamainfig}
\end{figure}

As can be seen in Figure~\ref{arumamainfig}, there is a large amount of  wavelength dependent astrometric reflex motion as expected, ranging from $\sim$10-20 $\mu$as in the optical, with the WD dominating at short wavelengths and the secondary star dominating at long wavelengths. This should easily allow the determination of both component masses with orbital coverage at a few passbands. According to DAPE, limiting individual observations to 12 minutes, (0.1 of the orbital period), to prevent orbital smearing, one could obtain individual measurements of $\sim$6 $\mu$as precision at the system's brightest, (V = 14.5), or $\sim$48 $\mu$as at its faintest, (V = 18.0). Thus, in its high state one could obtain 30, $\sim$6 $\mu$as precision, measurements in as little as 9 hours of total mission time, but in its low state, assuming 100, $\sim$48 $\mu$as precision, measurements would yield a meaningful solution, a total mission time of 30 hours would be required, which is still a reasonable amount of time for the scientific payoff in our opinion.

Of particular interest is that in H$_{\alpha}$ the threading region produces a detectable astrometric signature that has a phase offset of $\sim$15$\degr$ from the broadband wavelengths, due to the original 45$\degr$ offset of the hot spot diluted by the light from the rest of the system. Thus, through use of narrow-band astrometry, one could recover the location of the threading accretion regions in these types of systems. When varying the inclination angle for a AR UMa type system, the y component of the astrometric motion decreases with increasing inclination angle for all wavelengths as expected, but at large inclination angles eclipse effects produce significant deviations from simple sinusoidal reflex motions.

\subsection{Simulated SIM Observations}
\label{solvesec}

Although it is beyond the scope of this paper to perform a full simulation of exactly what parameters and to what precision one could extract with a given amount of SIM Lite time at specific wavelengths for each system, we can present a basic analysis for the SS Cyg system. To test the accuracy and precision to which we can recover system parameters, we generated simulated astrometric data for SS Cyg in the $U$ and $V$-bands, using the parameters listed in Table~\ref{sscygtable}, but with an inclination of 54.0$\degr$, comprised of 10 measurements at random phases with 1.7 $\mu$as Gaussian noise added to each measurement. We then simultaneously solved for the astrometric parameters of period, semi-major axis, inclination, the position angle of the line of nodes, and the angle in the plane of the true orbit between the line of nodes and the major axis, while fixing the eccentricity and the epoch of passage through periastron to be zero, using the \textsc{GAUSSFIT} program \citep{Jefferys1988}, with a procedure described in \citet{Benedict2001}. We recover inclinations of 55.6 $\pm$ 5.1 and 54.0 $\pm$ 3.8 degrees, and semi-major axes of 13.58 $\pm$ 0.94 and 23.58 $\pm$ 1.28 $\mu$as, each for the $U$ and $V$-bands respectively. Given the distance to the system of 159.5 parsecs \citep{Dubus2004}, and assuming that 58\% of the $U$-band astrometric motion results from the white dwarf, and 69\% of the $V$-band flux results from the secondary star, as per the modeling, (see \S3.2.1), this yields masses of 0.559 $\pm$ 0.070 and 0.815 $\pm$ 0.096 M$_{\sun}$ for the secondary star and white dwarf respectively, compared to the input masses of 0.555 and 0.812 M$_{\sun}$. Thus, one will be able to disentangle individual component masses with reasonable uncertainties for this system with only $\sim$8 hours of SIM mission time.

\subsection{Conclusion}

We have presented a modeling code, \textsc{reflux}, that is capable of modeling the multi-wavelength astrometric signature of interacting binaries. We have presented models for multiple IB systems, and find that for primary or secondary star dominated systems, the contamination of the photocenter is minimal and SIM Lite will provide good astrometric orbits if the system is bright enough, yielding absolute inclinations and indirect masses. For mixed component systems, SIM Lite will be able to directly determine absolute masses for both components when multi-wavelength astrometric curves are obtained and combined with existing spectroscopy and multi-color photometry. For disk-dominated systems, SIM Lite will only be able to obtain astrometric orbits for a few systems, due to the small relative motion of the compact object that the disk surrounds, but these should yield accurate inclinations. We find that while multi-wavelength SIM Lite data will be unable to distinguish between various disk temperature gradients and disk hotspots, it should be able to determine the location of magnetic threading regions in polars, and thus similar effects in other systems, via narrow wavelength astrometry. In total, SIM Lite should contribute greatly to our understanding of interacting binary systems, especially if its multi-wavelength capability is maintained in the final design.

%% file: appendixD.tex
\begin{singlespace}
\section[\MakeUppercase{Modeling Multi-Wavelength Stellar\\Astrometry: Determining Absolute\\Inclinations, Gravity Darkening\\Coefficients, and Spot Parameters of Single\\Stars with SIM Lite}]{\MakeUppercase{Modeling Multi-Wavelength Stellar Astrometry: Determining Absolute Inclinations, Gravity Darkening Coefficients, and Spot Parameters of Single Stars with SIM Lite}}
\label{sim2appendix}
\end{singlespace}

\subsection{Introduction}
\label{introsection}

SIM Lite is currently expected to have $\sim$80 spectral channels \citep{SIM2009}, spanning 450 to 900 nm, thus allowing multi-wavelength microarcsecond astrometry, which no current or planned ground or space-based astrometric project, (GAIA, CHARA, VLT/PRIMA, etc.) is able to match. We showed in Appendix~\ref{sim1appendix} the implications multi-wavelength microarcsecond astrometry has for interacting binary systems. In this paper, we discuss an interesting effect we encountered while modeling binary systems, namely that gravity darkening in stars produces a wavelength dependent astrometric offset from the center of mass that increases with decreasing wavelength. It is possible to use this effect to derive both the inclination and gravity darkening exponent of a star in certain cases.

Determining the absolute inclination of a given star has many practical applications. There is much interest in the formation of binary stars, where whether or not the spin axis of each star is aligned with the orbital axis provides insight into the formation history of the system \citep{Turner1995}. The mutual inclination between the stellar spin axes and orbital axis can greatly affect the rate of precession, which is used to probe stellar structure and test general relativity \citep{Sterne1939a,Sterne1939b,Sterne1939c,Kopal1959,Jeffery1984}. \citet{Albrecht2009} recently reconciled a 30-year-old discrepancy between the observed and predicted precession rate of DI Herculis through observations which showed the stellar spin axes were nearly perpendicular to the orbital axis. Along similar lines, extrasolar planets discovered via the radial velocity technique only yield the planetary mass as a function of the inclination of the orbit \citep{Mayor1995, Noyes1997, Marcy2000}, and thus, if one assumes the planetary orbit and stellar rotation axes are nearly parallel, determining the absolute inclination of the host star yields the absolute mass of the planet. If the stellar spin axis is found not to be parallel to the planetary orbital axis, this provides valuable insights into the planet's formation, migration, and tidal evolution histories \citep{Winn2006,Fabrycky2009}. A final example is the study of whether or not the spin axes of stars in clusters are aligned, which both reveals insight into their formation processes, as well as significantly affects the determination of the distances to those clusters \citep{Jackson2010}.

Our proposed technique can also be used in conjunction with other methods of determining stellar inclination to yield more precise inclination values and other stellar parameters of interest. \citet{Gizon2003} and \citet{Ballot2006} have shown that one can derive the inclination of the rotation axis for a given star using the techniques of astroseismology given high-precision photometry with continuous coverage over a long baseline, such as that provided by the $CoRoT$ and $Kepler$ missions. This technique is sensitive to rotation rates as slow as the Sun's, but becomes easier with faster rotation rates. \citet{Domiciano2004} discuss how spectro-interferometry can yield both the inclination angle and amount of differential rotation for a star, parameterized by $\alpha$. For both eclipsing binaries and transiting planets, the observation of the Rossiter-McLaughlin (RM) effect can yield the relative co-inclination between the two components \citep{Winn2006,Albrecht2009,Fabrycky2009}. The technique we propose in this paper would be complementary to these techniques in several ways. First, it would provide an independent check on the derived inclination axis from each method, confirming or refuting the astroseismic models and spectro-interferometric and RM techniques. Second, in principle the astroseismic technique is not dependent on the gravity darkening coefficient $\beta_{1}$, and the spectro-interferometric technique is correlated with the value for $\alpha$; combining techniques would yield direct and robust observationally determined values for $i$, $\alpha$, and $\beta_{1}$. Finally, the accurate, observational determination of $\alpha$ and $\beta_{1}$, (along with stellar limb-darkening), is critical to accurately deriving the co-inclination from the RM effect, as well as other quantities in stellar and exoplanet astrophysics.

In this paper, we also present models for and discuss the determination of spot location, temperature, and size on single stars, which produce a wavelength-dependent astrometric signature as they rotate in and out of view. Star spots are regions on the stellar surface where magnetic flux emerges from bipolar magnetic regions, which blocks convection and thus heat transport, effectively cooling the enclosed gas, and thus are fundamental indicators of stellar magnetic activity and the internal dynamos that drive it. \citet{Isik2007} discuss how the observation of spot location, duration, stability, and temperature can probe the stellar interior and constrain models of magnetic flux transport. Through the observation of the rotation rates of starspots at varying latitudes, one is able to derive the differential rotation rate of the star \citep{Collier2002}, which may be directly related to the frequency of starspot cycles. Mapping spots in binary star systems provides insight into the interaction between the magnetic fields of the two components, which can cause orbital period changes \citep{Applegate1992}, radii inflation \citep{LopezMorales2007,Morales2008}, and may possibly explain the $\sim$2-3 hour period gap in cataclysmic variable systems \citep{Watson2007}. Detecting and characterizing star spots via multi-wavelength astrometry would be complementary to other existing techniques, namely optical interferometry \citep{Wittkowski2002}, tomographic imaging \citep{Donati2006,Auriere2008}, photometric monitoring \citep{Alekseev2004,Mosser2009}, and in the future, microlensing \citep{Hwang2010}.

We present the details of our modeling code, \textsc{reflux}, in Section~\ref{refluxsection}, discuss the inclination effect and present models for multiple stars in Section~\ref{incsection}, discuss the spot effects and present models in Section~\ref{spotsection}, and present our conclusions in Section~\ref{conclusionsection}.

\subsection{The {\sc reflux} Code}
\label{refluxsection}

\textsc{reflux}\footnote{\textsc{reflux} can be run via a web interface from \url{http://astronomy.nmsu.edu/jlcough/reflux.html}. Additional details as to how to set-up a model are presented there.} is a code that computes the flux-weighted astrometric reflex motion of binary systems. We discussed the code in detail in Appendix~\ref{sim1appendix}, but in short, it utilizes the Eclipsing Light Curve (ELC) code, which was written to compute light curves of eclipsing binary systems (Orosz \& Hauschildt 2000). The ELC code represents the surfaces of two stars as a grid of individual luminosity points, and calculates the resulting light curve given the provided systemic parameters. ELC includes the dominant physical effects that shape a binary's light curve, such as non-spherical geometry due to rotation, gravity darkening, limb darkening, mutual heating, reflection effects, and the inclusion of hot or cool spots on the stellar surface. For the work in this paper we have simply turned off one of the stars, thus allowing us to probe the astrometric effects of a single star. To compute intensity, ELC can either use a blackbody formula or interpolate from a large grid of NextGen model atmospheres (Hauschildt et al. 1999). For all the simulations in this paper, we have used the model atmosphere option, and will note now, and discuss more in detail later, that the calculation of limb-darkening is automatically included in NextGen model atmospheres. These artificially derived limb-darkening coefficients have recently been shown to be in error by as much as $\sim$10-20\% in comparison to observationally derived values \citep{Claret2008}, and thus their uncertainties must be included, although for this work, due to symmetry, we find the introduced error is negligible. For all our simulations, we model the $U$, $B$, $V$, $R$, $I$, $J$, $H$, and $K$-bands for completeness and comparison to future studies, though we note that SIM Lite will not be able to observe in the $U$, $J$, $H$, or $K$ bandpasses.

\subsection{Inclination and Rotation}
\label{incsection}

\begin{figure}[h]
\centering
\begin{tabular}{cc}
\epsfig{file=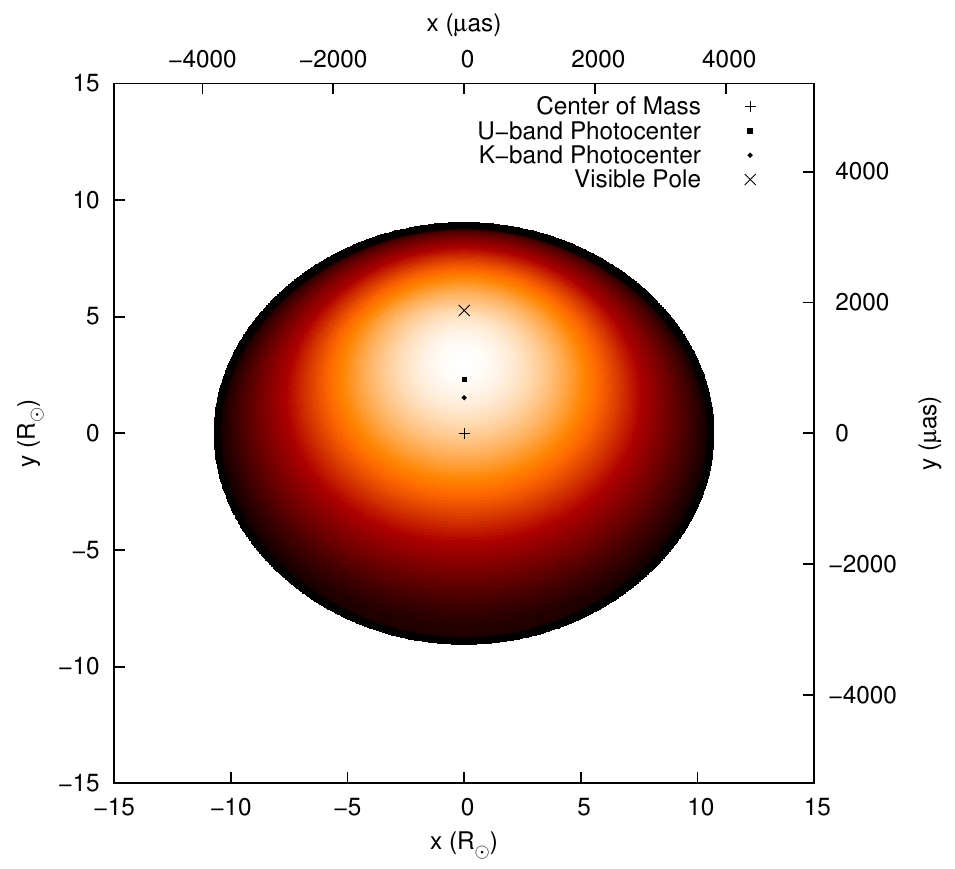, width=0.475\linewidth} &
\epsfig{file=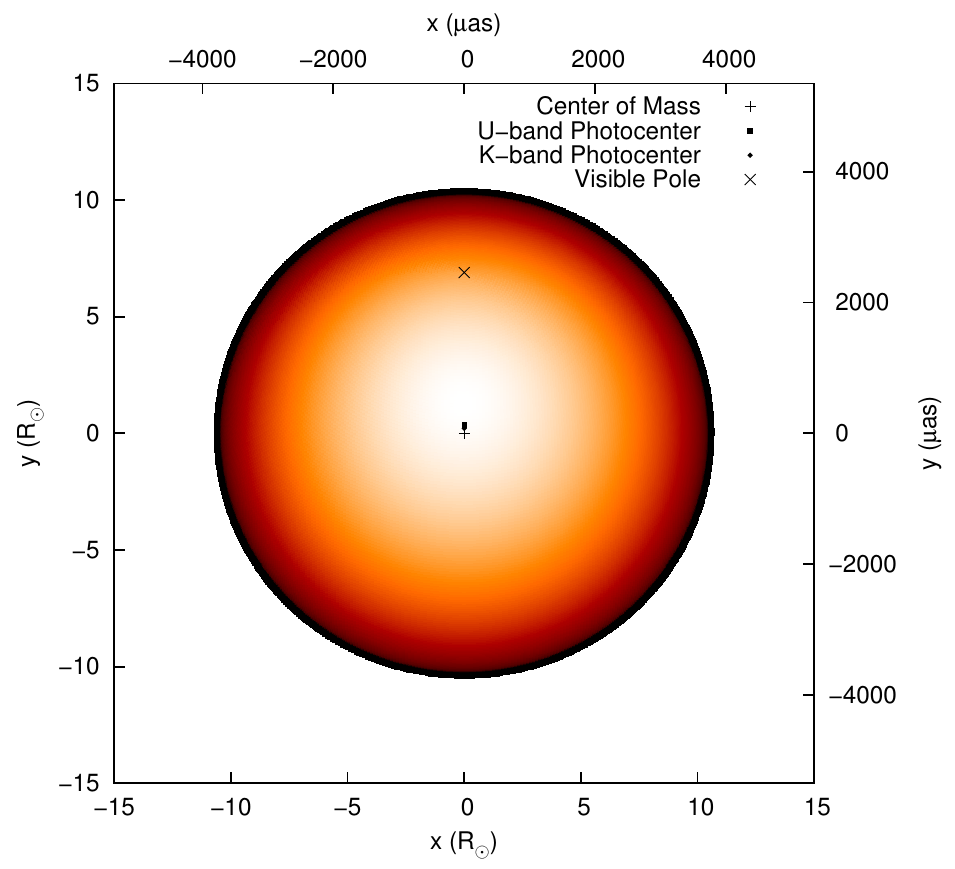, width=0.475\linewidth} \\
\end{tabular}
\caption[Illustration of the astrometric-inclination effect]{Illustration of the inclination effect using brightness maps of Capella Ab, which has an inclination of 42.788$\degr$ \citep{Torres2009}. Left: Capella Ab artificially spun-up to near its break-up speed, to accentuate the gravity darkening effect for ease of viewing. As can be seen, the photocenter of the system is dramatically shifted away from the center of mass, towards the visible pole, which is brighter than the rest of the star due to gravity darkening. Furthermore, since the pole is physically hotter, the $U$-band photocenter is shifted more than the $K$-band photocenter, and in the direction of the projected rotation axis, or y-axis. Right: Capella Ab at its actual rotation period. As can be seen, the actual effect is small compared to the angular size of the star on the sky, but still large compared to the 1 $\mu$as benchmark of SIM Lite. The presence of gravity darkening is clearly visible, which causes a decrement in flux towards the limb of the star. Note that the broad wavelength coverage of SIM Lite will only cover the $B$, $V$, $R$, and $I$ bandpasses.}
\label{incfig}
\end{figure}

The astrophysical phenomenon of gravity darkening, also sometimes referred to as gravity brightening, is the driving force behind the ability to determine the inclination of a single star using multi-wavelength astrometry. A rotating star is geometrically distorted into an oblate spheroid, such that its equatorial radius is greater than its polar radius, and thus the poles have a higher surface gravity, and the equator a lower surface gravity, than a non-rotating star with the same mass and average radius. This increased surface gravity, $g$, at the poles results in a higher effective temperature, T$_{\rm eff}$, and thus luminosity; decreased $g$ at the equator results in a lower T$_{\rm eff}$ and luminosity. This temperature and luminosity differential causes the star's center of light, or photocenter, to be shifted towards the visible pole, away from the star's gravitational center of mass. Since the inclination determines how much of the pole is visible, the amount of displacement between the photocenter and the center of mass is directly related to the inclination. Furthermore, since the luminosity difference effectively results from a ratio of blackbody luminosities of differing temperatures, the effect is wavelength dependent, with shorter wavelengths shifted more than longer wavelengths. Thus, the amount of displacement between the measured photocenter in two or more wavelengths is directly related to the inclination. See Figure~\ref{incfig} for an illustration of the effect.

An additional complicating factor is the exact dependence of temperature on local gravity. \citet{Zeipel1924} was the first to derive the quantitative relationship between them, showing that $T_{\rm eff}^{4} \propto g^{\beta_{1}}$, where $\beta_{1}$ is referred to as the gravity darkening exponent. The value of $\beta_{1}$ has been a subject of much study and debate; for a complete review, see \citet{Claret2000a}, who presents both an excellent discussion of past studies, as well as new, detailed computations of $\beta_{1}$ using modern models of stellar atmospheres and internal structure that encompass stars from 0.08 to 40 M$_{\sun}$. Since the value of $\beta_{1}$ affects the temperature differential between equator and pole, the multi-wavelength displacement will also be dependent on the value of $\beta_{1}$. The total amplitude of the effect will be scaled by the angular size of the star, which depends on both its effective radius and distance. Thus, in total, the components of this inclination effect are the effective stellar radius, distance, effective temperature, rotation rate, $\beta_{1}$, and inclination of the star. In principle, one is able to determine the effective stellar radius, effective temperature, rotation rate, and distance of a target star using ground-based spectroscopy and space-based parallax measurements, including from SIM Lite. Thus, when modeling the multi-wavelength displacement of the stellar photocenter, the only two components that need to be solved for are the inclination and $\beta_{1}$, with $\beta_{1}$ already having some constraints from theory.

A good trio of stars for modeling and testing this inclination effect are the components of the binary system Capella, (Aa and Ab), and the single star Vega. \citet{Torres2009} has very recently published an extremely detailed analysis of both the binary orbit of Capella and the physical and evolutionary states of the individual components, providing both new observations, as well as drawing from the previous observations and analyses of \citet{Hummel1994} and \citet{Strassmeier2001}. Vega, in addition to being one of the most well-studied stars in the sky, has recently been discovered to be a very rapid rotator seen nearly pole-on \citep{Aufdenberg2006,Peterson2006,Hill2010}. In total, these three stars represent both slow and rapid rotators for giant and main-sequence stars at a range of temperatures, as Capella Aa is a slow-rotating K-type giant, Capella Ab is a fast-rotating G-type giant, and Vega is a very fast-rotating A-type main-sequence star. With many ground-based interferometric observations to compare with, and being bright and nearby, these stars also present excellent targets for SIM Lite.

\begin{figure}[ht]
\centering
\epsfig{file=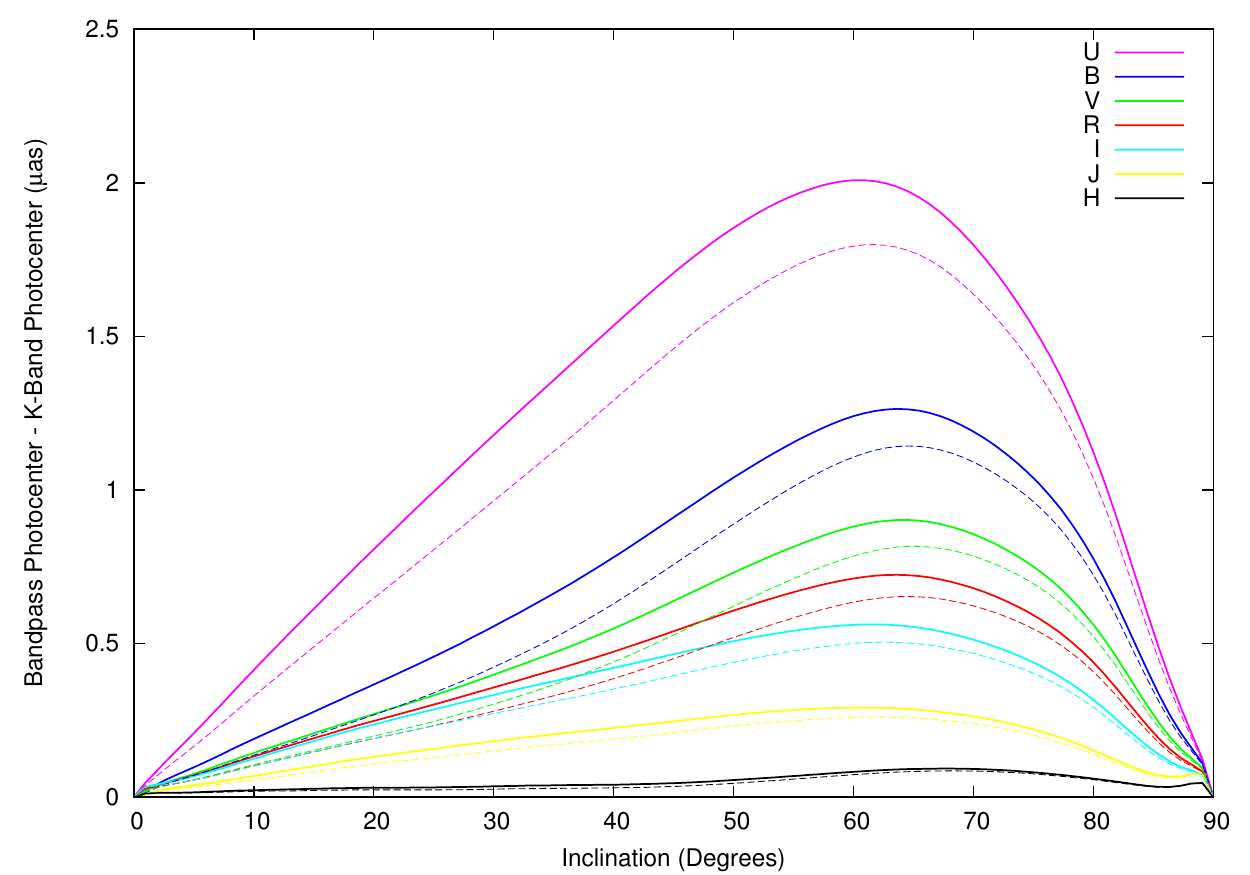, width=\linewidth}
\caption[The astrometric displacement of each bandpass with respect to $K$-band versus inclination for Capella Aa]{The astrometric displacement of each bandpass with respect to $K$-band versus inclination for Capella Aa. Dashed lines are a model with $\beta _{1}$ decreased by 10\%. Due to the slow rotation rate of Capella Aa, the effect is limited to a maximum of $\sim$2.0 microarcseconds between $U$ and $K$-band, and only a maximum of $\sim$0.7 $\mu$as between $B$ and $I$-band, where SIM Lite will operate. This puts the detection of this effect for Capella Aa at the very edge of SIM Lite's capability.}
\label{CapellaAaFig}
\end{figure}

\begin{figure}[ht]
\centering
\epsfig{file=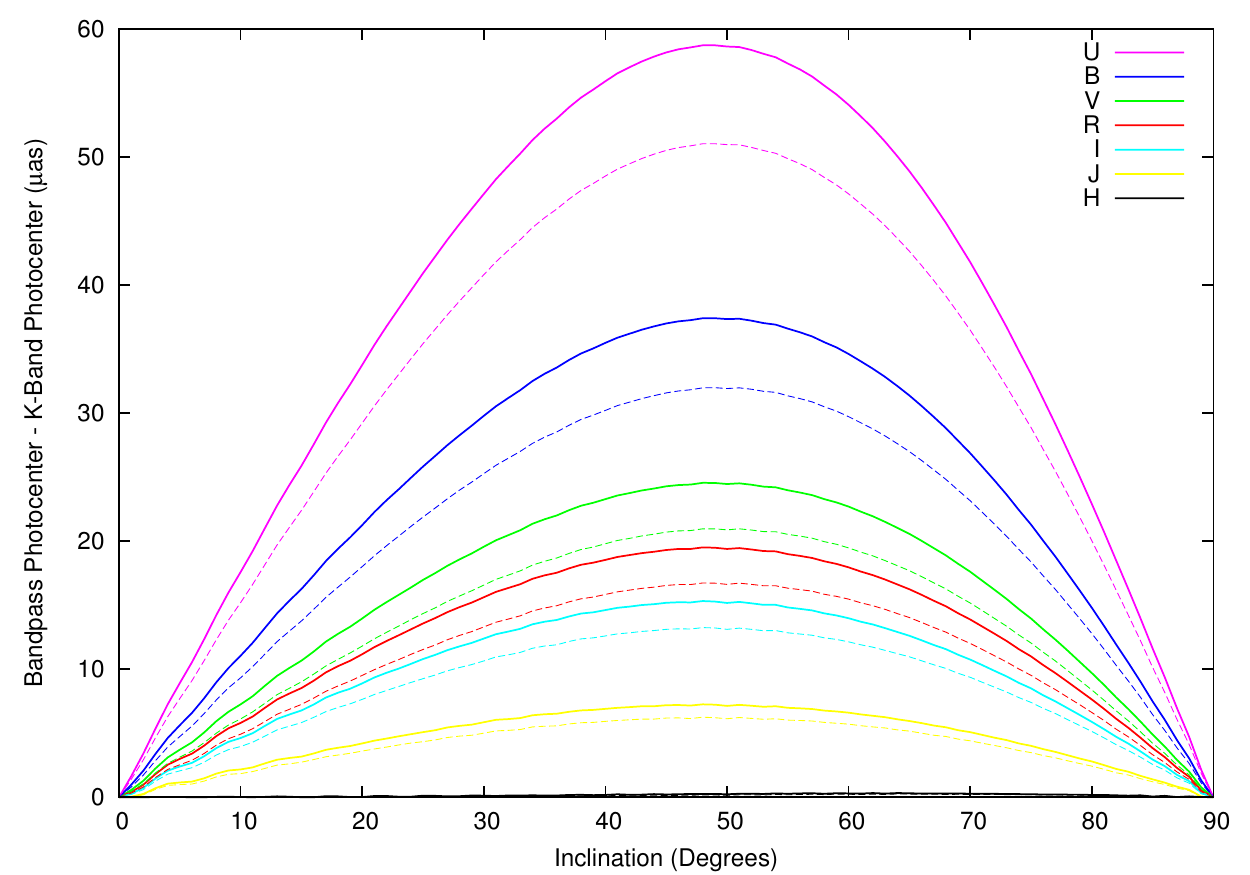, width=\linewidth}
\caption[The astrometric displacement of each bandpass with respect to $K$-band versus inclination for Capella Ab]{The astrometric displacement of each bandpass with respect to $K$-band versus inclination for Capella Ab. Dashed lines are a model with $\beta_{1}$ decreased by 10\%. Due to the fast rotation rate of a Capella Ab like star, the effect is moderate with tens of microarcseconds of displacement, and thus these types of stars are excellent targets for SIM Lite. The actual inclination of Capella Ab is 42.788$\degr$ \citep{Torres2009}, and thus Capella Ab itself should show a large shift of the photocenter with wavelength. Note that the broad wavelength coverage of SIM Lite will only cover the $B$, $V$, $R$, and $I$ bandpasses.}
\label{CapellaAbFig}
\end{figure}

\begin{figure}[ht]
\centering
\epsfig{file=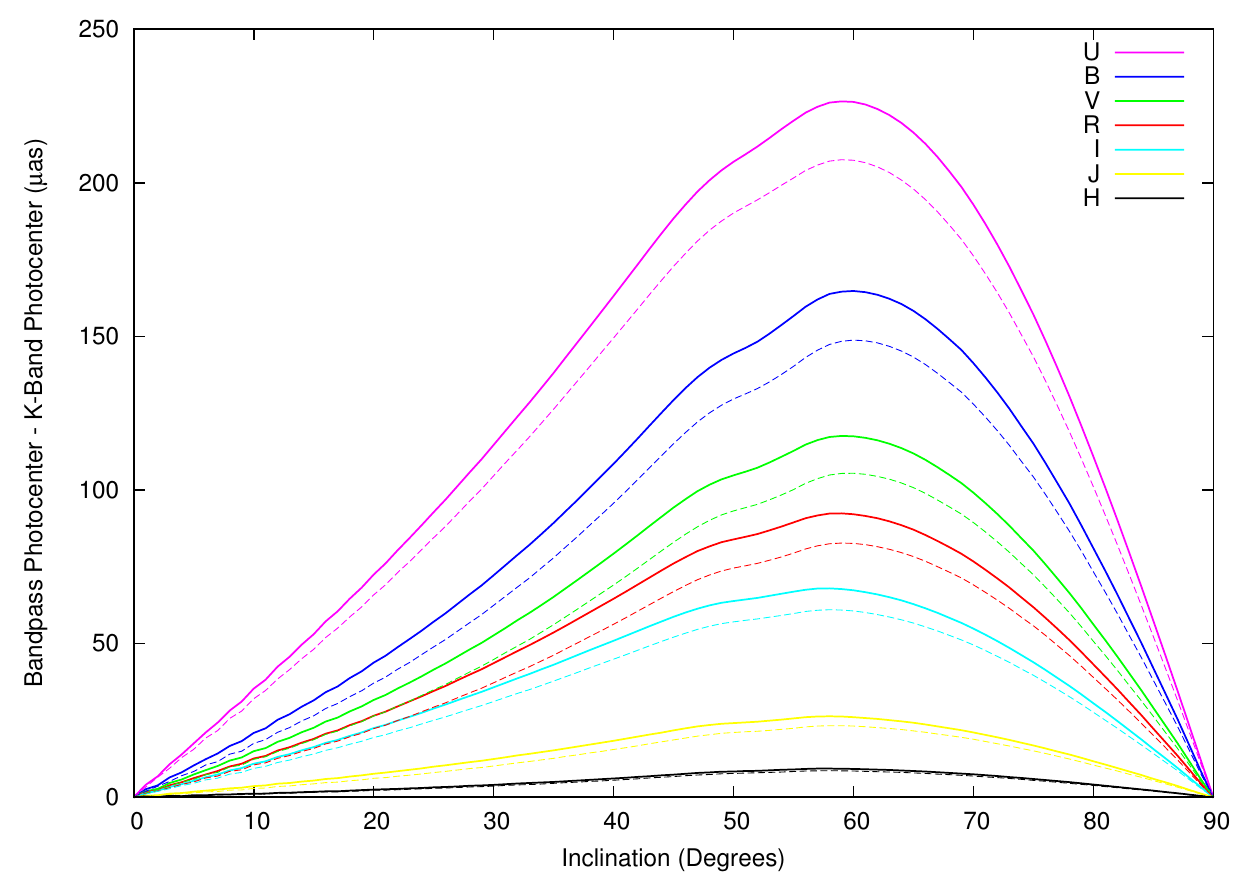, width=\linewidth}
\caption[The astrometric displacement of each bandpass with respect to $K$-band versus inclination for Vega]{The astrometric displacement of each bandpass with respect to $K$-band versus inclination for Vega. Dashed lines are a model with $\beta_{1}$ decreased by 10\%. Note that due to the very fast rotation of Vega, along with a high value of $\beta_{1}$, the effect can be quite large, at a couple hundred micro-arcseconds. For a Vega-like star, SIM Lite observations would yield very accurate values for $\beta_{1}$ and the inclination. For Vega itself, which is known to be nearly pole on, with an inclination of 5.7$\degr$ \citep{Hill2010}, there should be a $B$-band minus $I$-band displacement of 6.0 $\mu$as, still detectable by SIM Lite. Note that the broad wavelength coverage of SIM Lite will only cover the $B$, $V$, $R$, and $I$ bandpasses.}
\label{VegaFig}
\end{figure}

We use the \textsc{reflux} code to generate models of the astrometric displacement from $U$-band to $H$-band, with respect to the $K$-band photocenter, for inclinations from 0 to 90$\degr$, for each star, as shown in Figures~\ref{CapellaAaFig},~\ref{CapellaAbFig}, and~\ref{VegaFig}. We use systemic parameters given by \citet{Torres2009} for Capella Aa and Ab, and by \citet{Aufdenberg2006} and \citet{Peterson2006} for Vega, listed in Tables~\ref{CapellaAaTable},~\ref{CapellaAbTable}, and~\ref{VegaTable} respectively. We employ the model atmospheres incorporated into the ELC code, as well as automatically chosen values for $\beta_{1}$ based on Figure 1 of \citet{Claret2000a}. Additionally, in each figure we show a dashed line to indicate the effect of decreasing the gravity darkening coefficient by 10\% to simulate the uncertainty of the models \citep{Claret2000a} and explore the correlation with other parameters.

\begin{deluxetable}{lc}
\tablewidth{0pt}
\tablecaption{Parameters for Capella Aa}
\tablecolumns{2}
\tablehead{Parameter & Value\tablenotemark{a}}
\startdata
Distance (pc) & 12.9\\
Rotation Period (Days) & 106.0\\
Mass (M$_{\sun}$) & 2.70\\
Radius (R$_{\sun}$) & 12.2\\
Effective Temperature (K) & 4940\\
$\beta_{1}$ & 0.43
\enddata
\label{CapellaAaTable}
\tablenotetext{a}{Values from \citet{Torres2009}}
\end{deluxetable}

\begin{deluxetable}{lc}
\tablewidth{0pt}
\tablecaption{Parameters for Capella Ab}
\tablecolumns{2}
\tablehead{Parameter & Value\tablenotemark{a}}
\startdata
Distance (pc) & 12.9\\
Rotation Period (Days) & 8.64\\
Mass (M$_{\sun}$) & 2.56\\
Radius (R$_{\sun}$) & 9.2\\
Effective Temperature (K) & 5700\\
$\beta_{1}$ & 0.39
\enddata
\label{CapellaAbTable}
\tablenotetext{a}{Values from \citet{Torres2009}}
\end{deluxetable}

\begin{deluxetable}{lc}
\tablewidth{0pt}
\tablecaption{Parameters for Vega}
\tablecolumns{2}
\tablehead{Parameter & Value\tablenotemark{a}}
\startdata
Distance (pc) & 7.76\\
Rotation Period (Days) & 0.521\\
Mass (M$_{\sun}$) & 2.11\\
Radius (R$_{\sun}$) & 2.5\\
Effective Temperature (K) & 9602\\
$\beta_{1}$ & 1.02
\enddata
\label{VegaTable}
\tablenotetext{a}{Values from \citet{Aufdenberg2006} and \citet{Peterson2006}}
\end{deluxetable}

As can be seen from these models, we find that the effect is quite large for a Capella Ab-like or Vega-like fast rotator, but only marginally detectable for a slower-rotating system like Capella Aa. This also implies that this effect would not be detectable for a slow-rotating, main-sequence star like our Sun. Our modeling confirms this, showing a total U-K amplitude of $\ll$ 0.1 $\mu$as for a 1.0 M$_{\sun}$, 1.0 R$_{\sun}$ star with a rotation period of 30.0 days at 10.0 parsecs. These conclusions on detectability are made with the assumption that, for bright stars like these, SIM Lite can achieve its microarcsecond benchmark. We show this is possible in narrow angle (NA) mode by employing the SIM Differential Astrometry Performance Estimator (DAPE) \citep{Plummer2009}. For a target star with magnitude V$=$5, and a single comparison star with V$=$10 located within a degree of it on the sky, by integrating 15 seconds on the target, and 30 seconds on the reference, for 10 visits at 5 chop cycles each, a final precision of $\pm$1.01 $\mu$as is achieved in only 1.04 hours of total mission time. For a fainter target with $V$=10, this precision is only reduced to $\pm$1.32 $\mu$as in the same amount of mission time. In utilizing NA mode, one must be careful in choosing the reference star(s), to ensure that they are not stars with a substantial wavelength dependent centroid. Given the only constraints on reference stars are that they need to have V $\gtrsim$ 10 and are within one degree on the sky, one could easily choose a slow-rotating, main-sequence star, determined as such via ground-based observations, as a wavelength-independent astrometric reference star. We also note that wide angle SIM Lite measurements, with a precision of $\sim$5 $\mu$as, may not detect the wavelength dependent photocenter of a system like Capella, but will have no difficulty detecting it in stars like Capella Ab or Vega.

The effect of decreasing the gravity darkening exponent is to decrease the total amplitude of the effect in each wavelength, with shorter wavelengths affected more than longer wavelengths. Thus, the choice of gravity darkening exponent is intimately tied to the derived inclination. If one were to model observed data with a gravity darkening exponent that was $\sim$10\% different than the true value, they would derive an inclination that would also be $\sim$10\% different from the true inclination. However, the two combinations of inclination and gravity darkening exponent do not produce identical results, and can be distinguished with a sufficient precision at a number of wavelengths. For example, if one were to adopt the nominal value for $\beta_{1}$ and derive an inclination of 40 degrees for a Vega-like star, then adopt a $\beta_{1}$ value that was 10\% lower, one would derive an inclination of 43 degrees, a 7.5\% change. In this case though, with the lower $\beta_{1}$ value, the measured photocenter in the $U$, $B$, $V$, $R$, $I$, $J$, and $H$ bandpasses, with respect to the $K$-band photocenter, would differ from the nominal $\beta_{1}$ model by $\sim$0.5, -1.0, -2.0, -2.0, -1.6, -1.0, and 0.2 $\mu$as respectively. Note that for B, V, R, and I, where SIM Lite can observe, these discrepancies, on the order of $\sim$1.0 $\mu$as, should be large enough to be distinguished in NA mode. Thus, a unique solution exists for the values of \emph{i} and $\beta_{1}$ if the photocenter is measured in three or more wavelengths. (The photocenter of one wavelength is used as a base measurement that the photocenters of other wavelengths are measured with respect to, as we have chosen $K$-band as the base measurement in our models. With the photocenter measured in three or more wavelengths total, there are two or more photocenter difference measurements, with two unknown variables for which to solve.) Another complication is the possibility of having equally good fitting high and low solutions for $\emph{i}$. For example, if one observed and determined a best-fit inclination of 70 degrees for a Vega-like star, one could obtain a reasonably good fit as well at 46 degrees, (see Fig~\ref{VegaFig}). However, just as in the case of the uncertainty in the value of $\beta_{1}$, discernible discrepancies would exist. In this case, the discrepancies in the measured photocenter in the $U$, $B$, $V$, $R$, $I$, $J$, and $H$ bandpasses, with respect to the $K$-band photocenter, would be $\sim$0.1, -9.0, -2.0, 1.5, 6.0, 1.0, and 0.2 $\mu$as respectively. Just as in the case of the uncertainty in the value of $\beta_{1}$, this discrepancy between equally good fitting high and low inclination solutions can be resolved if one has three or more wavelengths obtained in NA mode.

As mentioned in Section~\ref{introsection}, we note that the limb-darkening function, which was automatically chosen by the ELC code as incorporated into the model atmospheres, can differ from actual observed values by $\sim$10\% \citep{Claret2008}. We have tested how changing the limb-darkening coefficients by 10\% affects the resulting astrometric displacements, and find that the result is less than 0.5\% for all wavelengths, and thus is negligible in the modeling. The reason is that limb-darkening is symmetric, and thus while increased limb-darkening damps the visible pole, it also damps the rest of the star, and thus the relative brightness between regions is maintained. 

Additionally, this inclination technique yields the orientation of the projected stellar rotation axis on the sky, which is parallel to the wavelength dispersion direction. When coupled with the derived inclination, this technique thus yields the full 3-dimensional orientation of the rotation axis. This could be a powerful tool in determining the overall alignment of stellar axes in the local neighborhood and in nearby clusters.

\subsection{Star Spots}
\label{spotsection}

Another area of astrophysical interest to which multi-wavelength astrometric measurements from SIM Lite can contribute is the study of star spots. As the cause of star spots are intense magnetic fields at the photosphere, they are typically found in stars with convective envelopes, especially rapidly rotating stars. Thus, both low-mass, main-sequence K and M dwarfs, as well as rapidly rotating giant and sub-giant stars, are known to host large spots on their surface. The study of the distribution, relative temperature, and size of these spots would greatly contribute to the study of magnetic field generation in stellar envelopes. A starspot that rotates in and out of view will cause a shift of the photocenter for a single star, which has been a subject of much recent discussion in the literature \citep[e.g.][]{Hatzes2002,Unwin2005,Eriksson2007,Catanzarite2008,Makarov2009,Lanza2008}, especially in light of its potential to mimic, or introduce noise when characterizing, an extrasolar planet. However, there has been no mention in the literature of the multi-wavelength astrometric signature of stellar spots, where, just as in the case of the gravity darkening inclination effect, we are looking at essentially two blackbodies with varying temperatures, and thus shorter wavelengths will be more affected by a spot than longer wavelengths.

\begin{figure}
\centering
\begin{tabular}{cc}
\epsfig{file=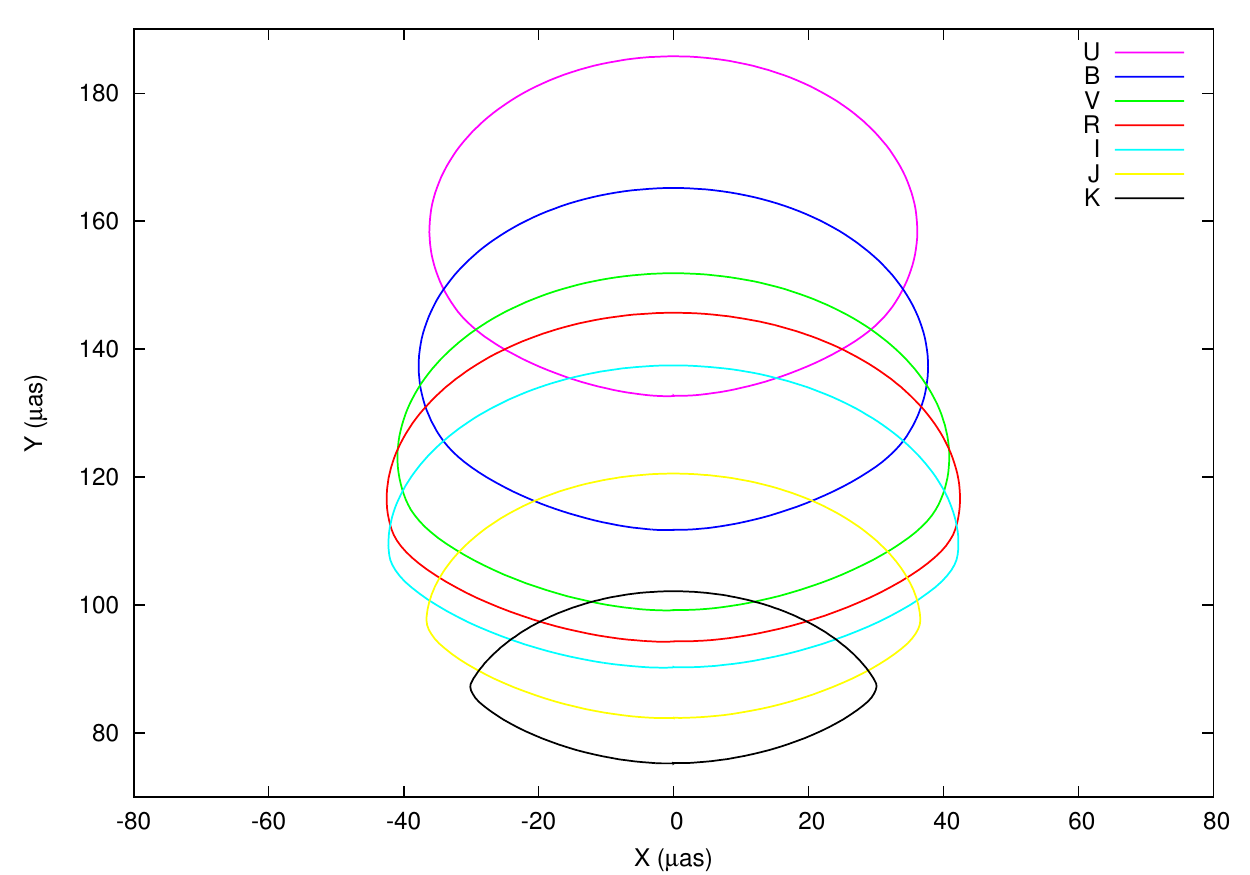, width=0.45\linewidth} &
\epsfig{file=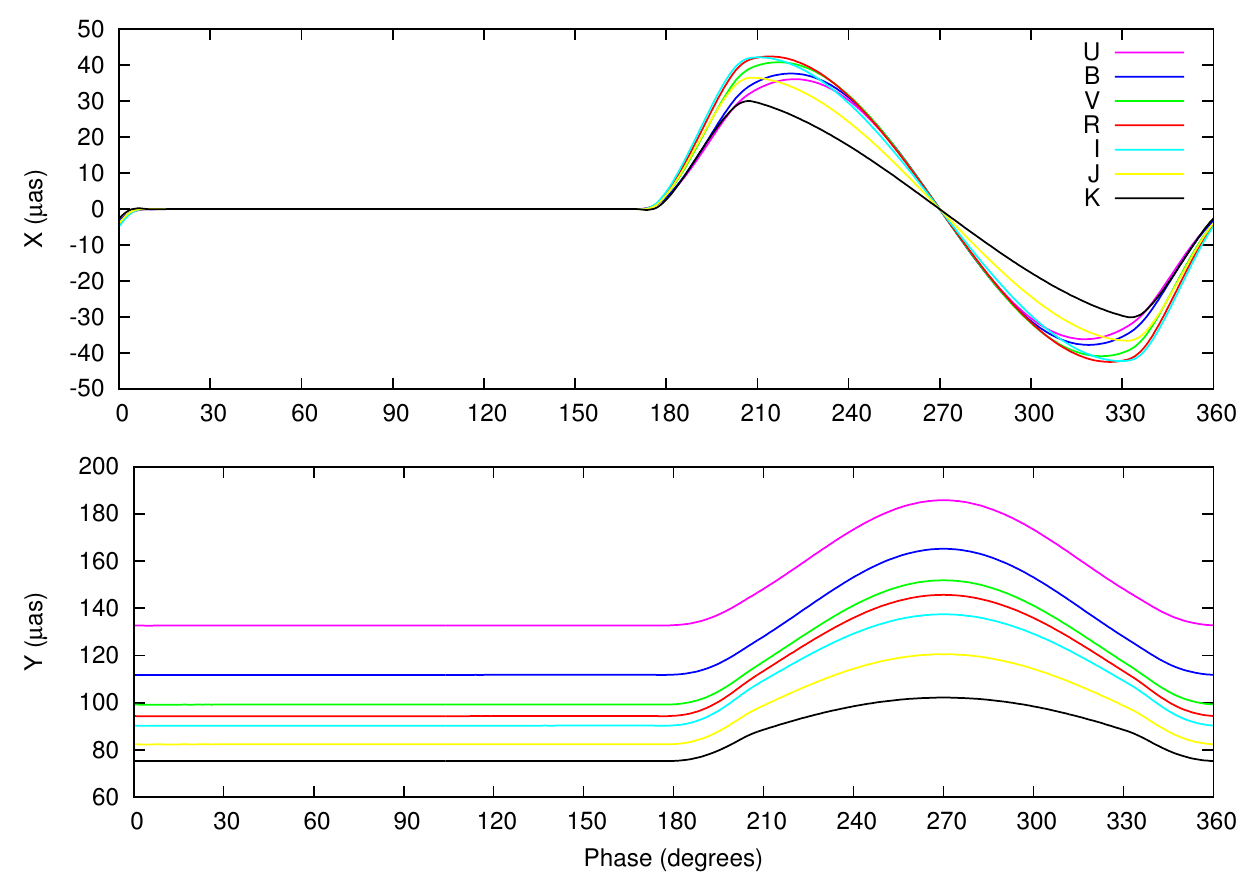, width=0.45\linewidth} \\
\end{tabular}
\caption[A simulated cool Spot on CapellaAb]{A simulated cool Spot on CapellaAb. The spot is located on the equator, with a longitude such that it is seen directly at phase 270$\degr$. The strong presence of the gravity darkening effect, discussed in Section~\ref{incsection}, dominates the wavelength spread in the y direction. Note that the broad wavelength coverage of SIM Lite will only cover the $B$, $V$, $R$, and $I$ bandpasses.}
\label{CapellaAbSpotFig}
\end{figure}

\begin{figure}[ht]
\centering
\begin{tabular}{cc}
\epsfig{file=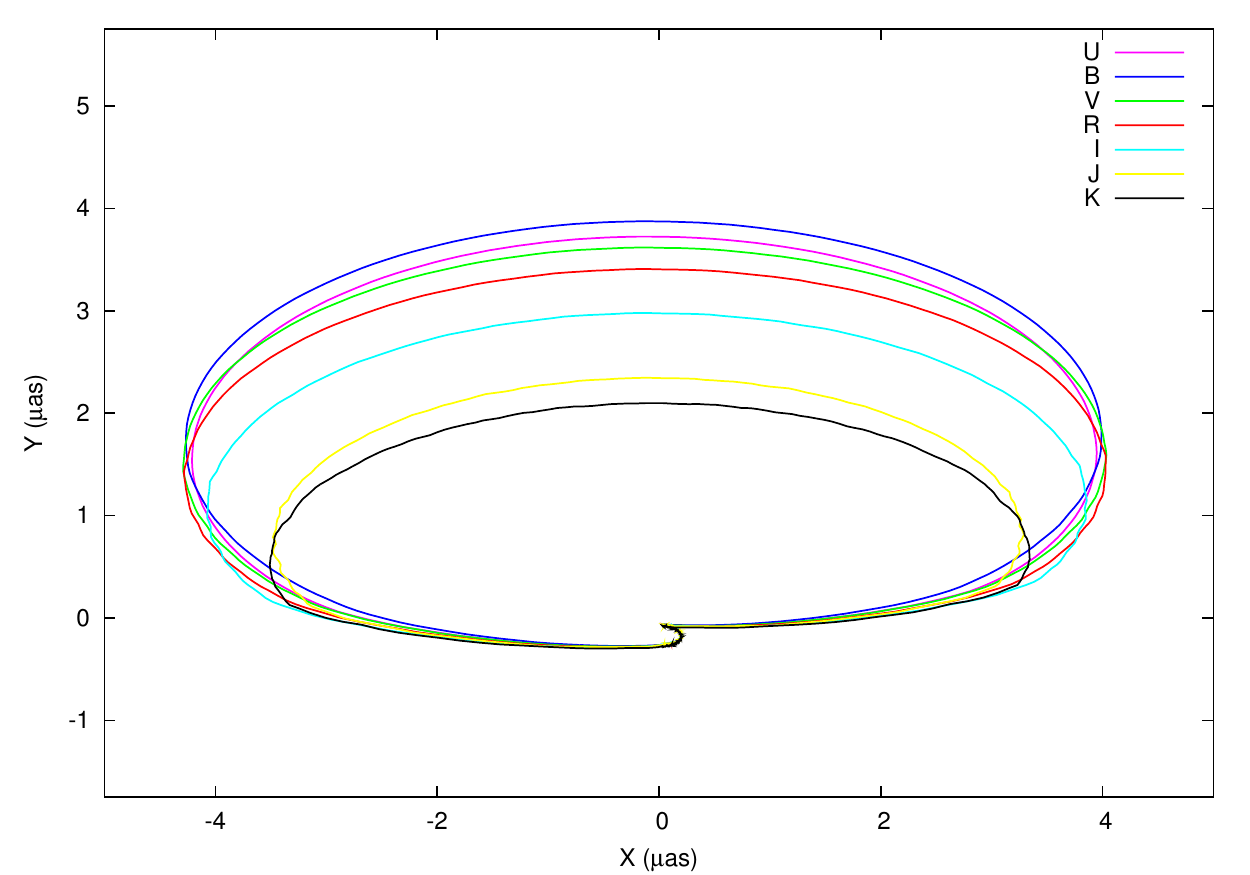, width=0.45\linewidth} &
\epsfig{file=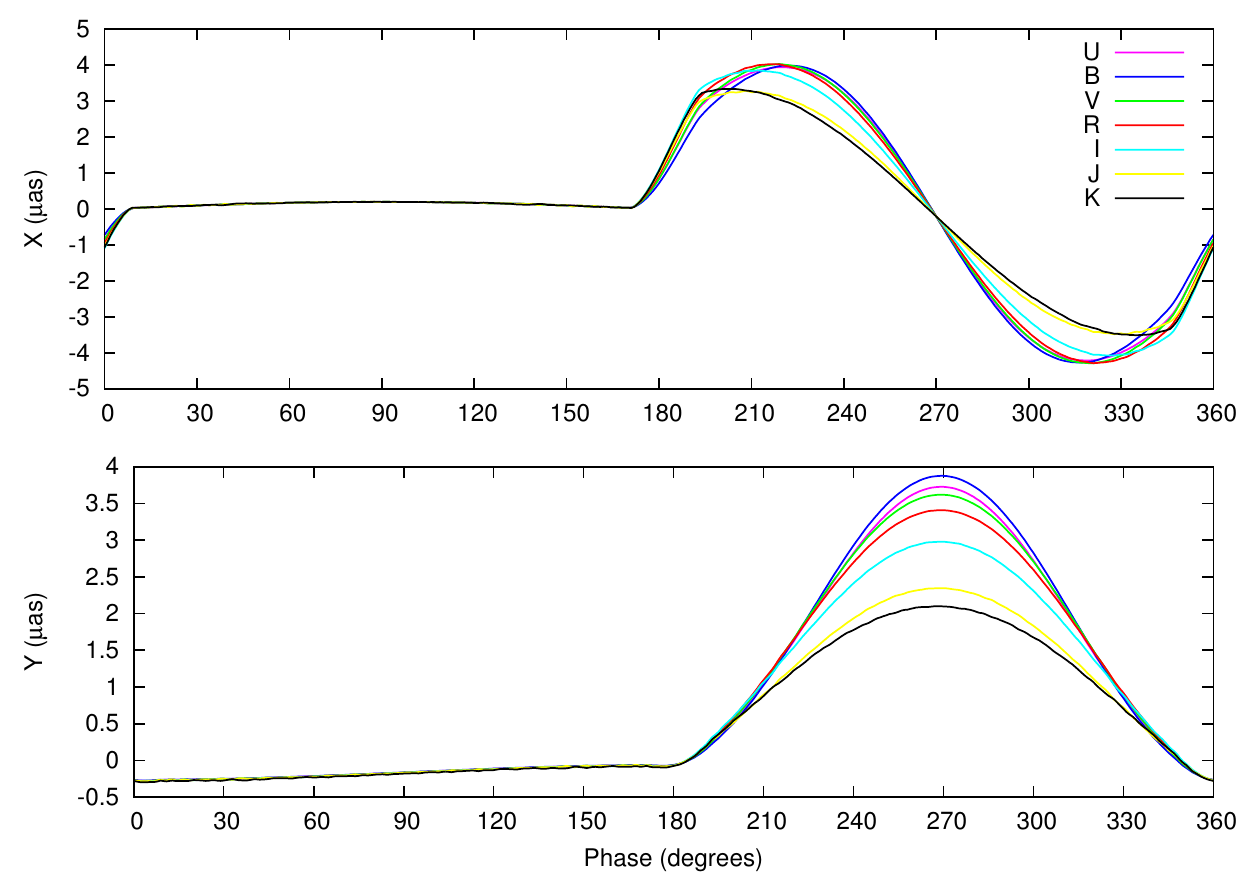, width=0.45\linewidth} \\
\end{tabular}
\caption[A simulated cool spot on a nearby K dwarf star]{A simulated cool spot on a nearby K dwarf star with an inclination of 60$\degr$, whose parameters are given in Table~\ref{kdwarftable}. The spot is located on the equator, with a longitude such that it is seen directly at phase 270$\degr$. Note that the broad wavelength coverage of SIM Lite will only cover the $B$, $V$, $R$, and $I$ bandpasses.}
\label{SingleStarCoolSpot}
\end{figure}

\begin{figure}[ht]
\centering
\begin{tabular}{cc}
\epsfig{file=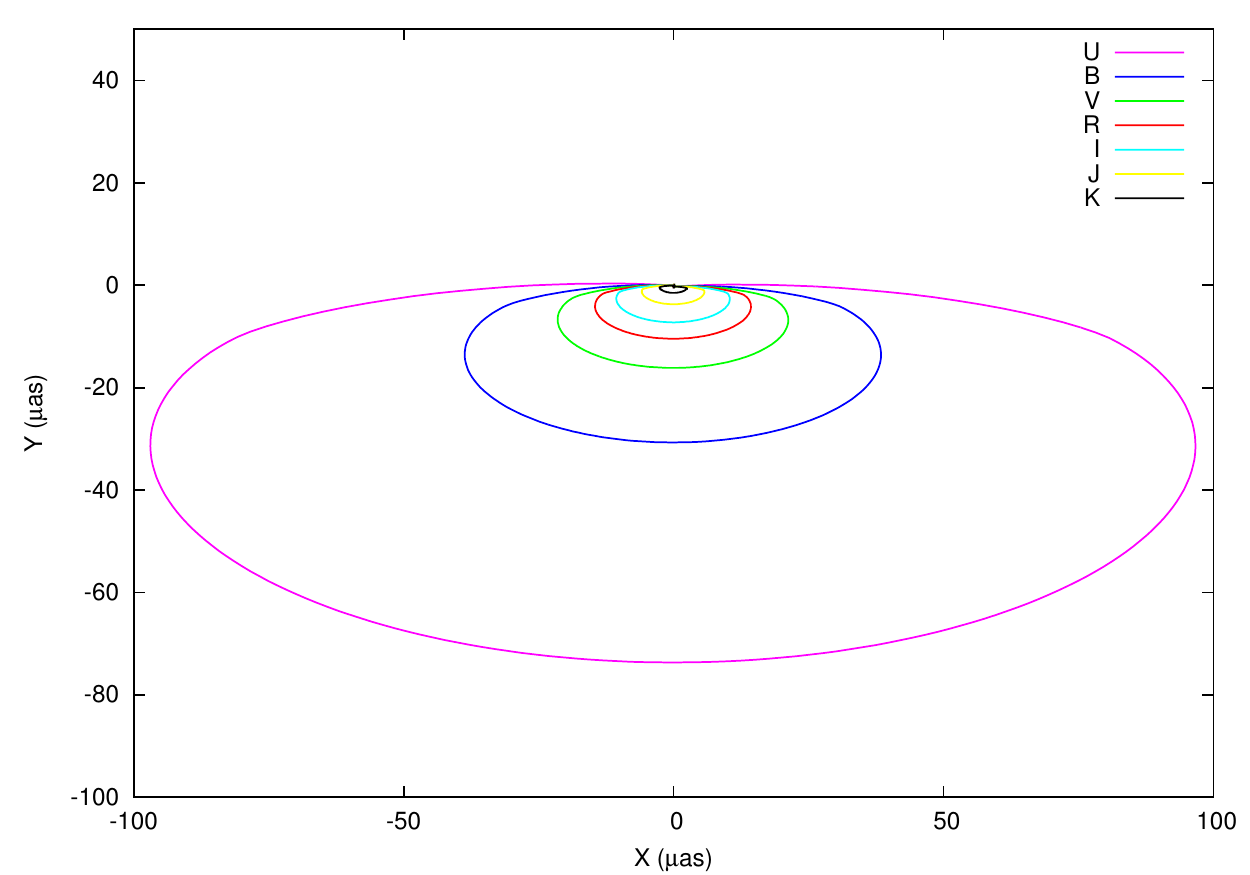, width=0.45\linewidth} &
\epsfig{file=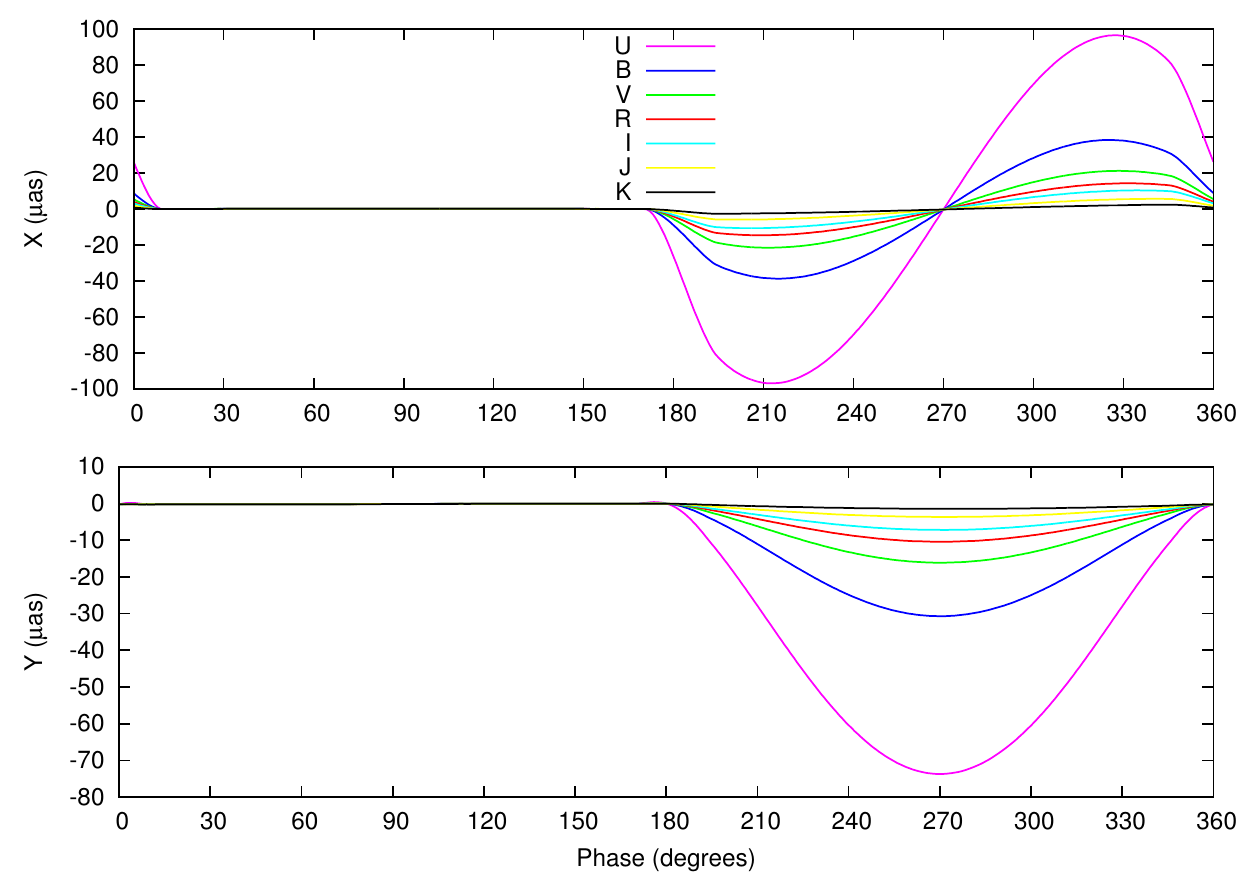, width=0.45\linewidth} \\
\end{tabular}
\caption[A simulated hot spot or flare on a nearby K dwarf star]{A simulated hot spot or flare on a nearby K dwarf star with an inclination of 60$\degr$, whose parameters are given in Table~\ref{kdwarftable}. The spot is located on the equator, with a longitude such that it is seen directly at phase 270$\degr$. Note that the broad wavelength coverage of SIM Lite will only cover the $B$, $V$, $R$, and $I$ bandpasses.}
\label{SingleStarHotSpot}
\end{figure}

To characterize the multi-wavelength astrometric signature of stellar spots, we model two spotty systems, again using the {\sc reflux} code. We model Capella Ab, which shows evidence of large spots and is suspected of being a RS CVn variable \citep{Hummel1994}, and a typical main-sequence K dwarf. For Capella Ab, we use the parameters listed in Table~\ref{CapellaAbTable}, along with the star's determined inclination of 42.788$\degr$ \citep{Torres2009}, and add a cool spot that has a temperature that is 60\% of the average surface temperature, located at the equator, at a longitude such that it is seen directly at phase 270$\degr$, and having an angular size of 10$\degr$, (where 90$\degr$ would cover exactly one half of the star). For the K dwarf system, we use the physical parameters listed in Table~\ref{kdwarftable}, simulating a typical K Dwarf at 10 parsecs, and add a cool spot with the same parameters as we do for Capella Ab. Additionally, to investigate the effects of cool versus hot spots or flares, we also run a model with a hot spot by changing the spot temperature to be 40\% greater than the average surface temperature. We present our models in Figures~\ref{CapellaAbSpotFig},~\ref{SingleStarCoolSpot}, and~\ref{SingleStarHotSpot}.

\begin{deluxetable}{lc}
\tablewidth{0pt}
\tablecaption{Parameters for the K Dwarf System}
\tablecolumns{2}
\tablehead{Parameter & Value}
\startdata
Distance (pc) & 10.0\\
Inclination ($\degr$) & 60.0\\
Period (Days) & 20\\
Mass (M$_{\sun}$) & 0.6\\
Radius (R$_{\sun}$) & 0.6\\
Effective Temperature (K) & 4500\\
Latitude of Spot ($\degr$) & 90\\
Longitude of Spot ($\degr$) & 270\\
Angular Size of Spot ($\degr$) & 10\\
Cool Spot Temperature Factor & 0.6\\
Hot Spot Temperature Factor & 1.4
\enddata
\label{kdwarftable}
\end{deluxetable}

As can be seen for CapellaAb, the gravity darkening inclination effect presented in Section~\ref{incsection} dominates the spread of colors in the y-direction, the direction parallel to the stars' projected rotation axis. However, the amplitude of the spot motion is quite large, with a total amplitude of $\sim$40 $\mu$as in all bandpasses, which would be easily detectable by SIM Lite. For the K dwarf with a cool spot, we see a much smaller, but still detectable shift of amplitude $\sim$5-8 $\mu$as, depending on the wavelength. In the case of a hot spot or flare, we see a much larger displacement, on the order of $\sim$10-200 $\mu$as, depending on the wavelength, which would be easily detectable by SIM and provide extremely precise values in deriving the spot parameters. 

In general, the temperature of the spot, in relation to the mean stellar surface temperature, is related to the spread in observed wavelengths, with a larger spread indicating a larger temperature difference. The duration of the astrometric displacement in phase, coupled with the overall amplitude of the astrometric displacement, yields the size of the spot, as larger spots will cause larger displacements and be visible for a larger amount of rotational phase. The latitude of the spot can also affect the total duration. Finally, the amplitude of the astrometric displacement in the x versus the y direction is dependent on both the latitude of spot as well as the inclination of the star. Thus, when modeled together, one is able to recover these parameters. This work can also be combined with our work in Appendix~\ref{sim1appendix} to derive the location of spots in binary systems, as the astrometric signature of the spot is simply added to the astrometric signature of the binary system.

The astrometric motion induced upon a parent star by a host planet does not have a wavelength dependence. Spots however, as we have shown via our modeling, have a clear wavelength dependence. Thus, if one has a candidate planetary signal from astrometry, but it shows a wavelength-dependent motion, it must then be a false positive introduced from star spots at the rotation period of the star, (assuming that the planet's emitted flux is negligible compared to the star.) Furthermore, when SIM is launched, there will likely be many cases where a marginally detectable signal due to a planetary companion is found at a very different period than the rotation period of the star. However, starspots will still introduce extra astrometric jitter which will degrade the signal from the planetary companion. Multi-wavelength astrometric data can be used to model and remove the spots, which will have a wavelength dependence, and thus strengthen the planetary signal, which will not have a wavelength dependence.

\subsection{Discussion and Conclusion}
\label{conclusionsection}

We have presented detailed models of the multi-wavelength astrometric displacement that SIM Lite will observe due to gravity darkening and stellar spots using the \textsc{reflux} code. We find that SIM Lite observations, especially when combined with other techniques, will be able to determine the absolute inclination, gravity darkening exponent, and 3-dimensional orientation of the rotational axis for fast and slow rotating giant stars, and fast-rotating main-sequence stars. This technique will be especially useful in probing binary star and exoplanet formation and evolution, as well as the physics of star forming regions. Direct observational determination of the gravity darkening exponent has direct applications in both stellar and exoplanet astrophysics. This technique is also relatively inexpensive in terms of SIM Lite observing time, as one need only to observe a given star once, as opposed to binary stars and planets, which require constant monitoring over an entire orbit. It should be noted that this effect should be taken into account when constructing the SIM Lite astrometric reference frame, such that fast-rotating giants should be excluded so as not to produce a wavelength-dependent astrometric reference fame.

We also have presented models of star spots on single stars, and find that SIM Lite should be able to discern their location, temperature, and size. Combined with other techniques, this will provide great insight into stellar differential rotation, magnetic cycles and underlying dynamos, and magnetic interaction in close binaries. From this modeling, it should especially be noted that multi-wavelength astrometry is a key tool in the hunt for extrasolar planets, either by ruling out false signals created by spots, or simply removing extra astrometric jitter introduced by spots. Thus, it remains critical that SIM Lite maintains a multi-wavelength astrometric capability in its final design.